\documentclass[11pt]{article}
\usepackage{graphicx}
\usepackage{color}
\usepackage{titlesec}
\usepackage{amssymb}
\usepackage{hyperref}
\usepackage{subcaption}
\usepackage{sidecap}
\usepackage{eurosym}

\usepackage{longtable,tabulary}

\def\ltabulary{%
\def\endfirsthead{\\}%
\def\endhead{\\}%
\def\endfoot{\\}%
\def\endlastfoot{\\}%
\def\tabulary{%
  \def\TY@final{%
\def\endfirsthead{\LT@end@hd@ft\LT@firsthead}%
\def\endhead{\LT@end@hd@ft\LT@head}%
\def\endfoot{\LT@end@hd@ft\LT@foot}%
\def\endlastfoot{\LT@end@hd@ft\LT@lastfoot}%
\longtable}%
  \let\endTY@final\endlongtable
  \TY@tabular}%
\dimen@\columnwidth
\advance\dimen@-\LTleft
\advance\dimen@-\LTright
\tabulary\dimen@}

\makeatletter
\def\ltabulary{%
\def\endfirsthead{\\}%
\def\endhead{\\}%
\def\endfoot{\\}%
\def\endlastfoot{\\}%
\def\tabulary{%
  \def\TY@final{%
\def\endfirsthead{\LT@end@hd@ft\LT@firsthead}%
\def\endhead{\LT@end@hd@ft\LT@head}%
\def\endfoot{\LT@end@hd@ft\LT@foot}%
\def\endlastfoot{\LT@end@hd@ft\LT@lastfoot}%
\longtable}%
  \let\endTY@final\endlongtable
  \TY@tabular}%
\dimen@\columnwidth
\advance\dimen@-\LTleft
\advance\dimen@-\LTright
\tabulary\dimen@}

\makeatother

\definecolor{hbl}{rgb}{0.153,0.498,0.835} 
\definecolor{dbl}{rgb}{0,.15,0.46}              

\titleformat{\chapter}[display]
{\color{hbl}\bfseries\boldmath\huge}{\chaptertitlename\
\thechapter}{20pt}{\Huge}
\titleformat{\section}
{\color{hbl}\bfseries\boldmath\Large}{\thesection}{1em}{}

\titleformat{\paragraph}
{\normalfont\normalsize\bfseries}{\theparagraph}{1em}{}
\titlespacing*{\paragraph}
{0pt}{3.25ex plus 1ex minus .2ex}{1.5ex plus .2ex}

\newcommand{\muJy}{$\mu$Jy}

\newcommand{\hi}{H\mbox{\,\sc i}}
\newcommand{\hii}{H\mbox{\,\sc ii}}

\newcommand{\msol}{\mbox{${\rm M}_\odot$}}

\newcommand{\kms}{\mbox{$\rm km\, s^{-1}$}}

\def\sun{\hbox{$\odot$}}
\def\degree{\nobreak\ifmmode{^\circ}\else{$^\circ$}\fi}
\def\arcsec{\hbox{$^{\prime\prime}\,$}}

\def\grtsim{\mathrel{\hbox{\rlap{\hbox{\lower2pt\hbox{$\sim$}}}\raise2pt\hbox{$>$}}}}
\def\lesssim{\mathrel{\hbox{\rlap{\hbox{\lower2pt\hbox{$\sim$}}}\raise2pt\hbox{$<$}}}}

\newcommand{\etal}{{et~al.}}

\newcommand{\noi}{\noindent}

\newcommand{\mnras}{MNRAS}
\newcommand{\apj}{ApJ}

\newcommand{\apjl}{ApJL}

\newcommand{\aap}{A\&A}

\newcommand{\boldvec}[1]{\vec{\mbox{\boldmath{$#1$}}}}
\def\,{\thinspace}
\def\lsim{\mathrel{\raise .4ex\hbox{\rlap{$<$}\lower 1.2ex\hbox{$\sim$}}}}
\def\gsim{\mathrel{\raise .4ex\hbox{\rlap{$>$}\lower 1.2ex\hbox{$\sim$}}}}

\newcommand{\ergs}{\mathrm{ erg\,s^{-1}}}
\def\nh{N_{\rm H}}
\def\cmmoinsdeux{\mbox{ cm}^{-2}}

\newcommand{\nancay}{Nan\c{c}ay}
\newcommand{\enancay}{EMBRACE@\nancay}

\usepackage{xcolor,soul,lipsum}
\newcommand{\myul}[2][black]{\setulcolor{#1}\ul{#2}\setulcolor{black}}

\newcommand{\lagrange}{Universit\'e C\^ote d'Azur, Observatoire de la C\^ote d'Azur, CNRS, Laboratoire Lagrange, Nice, France}
\newcommand{\cnrs}{CNRS-INSU, Institut National des Sciences de l'Univers}
\newcommand{\lpsc}{Laboratoire de Physique Subatomique et de Cosmologie, Universit\'e Grenoble Alpes, CNRS, Grenoble, France}
\newcommand{\irap}{IRAP, Universit\'e de Toulouse, CNRS, UPS, CNES, Toulouse, France}
\newcommand{\cral}{Universit\'e Lyon, Universit\'e Lyon1, Ens de Lyon, CNRS, Centre de Recherche Astrophysique de Lyon UMR5574, France}
\newcommand{\lam}{Aix Marseille Univ, CNRS, LAM, Laboratoire d'Astrophysique de Marseille, Marseille, France}
\newcommand{\gepi}{GEPI, Observatoire de Paris, PSL Research University, CNRS, Meudon, France}
\newcommand{\apc}{Astroparticule et Cosmologie APC, Universit\'e Paris Diderot, CNRS/IN2P3, CEA/IRFU, Observatoire de Paris, Sorbonne Paris Cit\'e, Paris, France}
\newcommand{\artemis}{Universit\'e C\^ote d'Azur, Observatoire de la C\^ote d'Azur, CNRS, Laboratoire Artemis, Nice, France}
\newcommand{\ias}{Institut d'Astrophysique Spatiale, CNRS, UMR 8617, Universit\'e Paris-Sud, Universit\'e Paris-Saclay, Orsay, France}
\newcommand{\irfu}{IRFU, CEA, Universit\'e Paris-Saclay, Gif-sur-Yvette, France}
\newcommand{\irfuSAp}{CEA, IRFU/SAp, Universit\'e Paris-Saclay, Gif-sur-Yvette, France}
\newcommand{\lermasorb}{LERMA, Observatoire de Paris, \'Ecole normale sup\'erieure, PSL Research University, CNRS, Sorbonne Universit\'es, UPMC Univ. Paris 06, Paris, France}
\newcommand{\iapsorb}{Sorbonne Universit\'es, UPMC Univ Paris 6 et CNRS, UMR 7095, Institut dÕAstrophysique de Paris, Paris, France}
\newcommand{\stras}{Universit\'e de Strasbourg, CNRS, Observatoire astronomique de Strasbourg, UMR 7550, Strasbourg, France}
\newcommand{\lal}{Laboratoire de l'Acc\'el\'erateur Lin\'eaire, IN2P3-CNRS, Universit\'e de Paris-Sud, Orsay Cedex, France}
\newcommand{\usn}{Station de Radioastronomie de Nan\c{c}ay, Observatoire de Paris, PSL Research University, CNRS, Univ. Orl\'eans, Nan\c{c}ay, France}
\newcommand{\ucb}{Universit\'e Claude Bernard Lyon I, Institut de Physique Nucl\'eaire, Lyon, France}
\newcommand{\colfr}{LERMA, Observatoire de Paris, Coll\`ege de France, \'Ecole normale sup\'erieure, PSL Research University, CNRS, Sorbonne Universit\'es, UPMC Univ. Paris 06, Paris, France}
\newcommand{\lupm}{Laboratoire Univers et Particules de Montpellier, Universit\'e de Montpellier, CNRS/IN2P3, Montpellier, France}
\newcommand{\aim}{Universit\'e Paris-Diderot, AIM, Sorbonne Paris Cit\'e, CEA, CNRS, Gif-sur-Yvette, France}
\newcommand{\iram}{Institut de Radioastronomie Millim\'etrique (IRAM), Saint Martin dÕH\`eres, France}
\newcommand{\univgren}{Universit\'e Grenoble Alpes, CNRS, IPAG, Grenoble, France}
\newcommand{\lab}{Laboratoire dÕastrophysique de Bordeaux, Universit\'e de Bordeaux, CNRS, Pessac, France}
\newcommand{\unib}{Universit\'e Bordeaux, CNRS/IN2P3, Centre dÕ\'Etudes Nucl\'eaires de
Bordeaux Gradignan, Gradignan, France}
\newcommand{\lpcee}{LPC2E, CNRS, Universit\'e d'Orl\'eans, Orleans, France}
\newcommand{\luth}{LUTh, Observatoire de Paris, PSL Research University, CNRS UMR 8109,
Universit\'e Pierre et Marie Curie, UniversitŽ\'e Paris Diderot, Meudon, France}
\newcommand{\subatech}{SUBATECH, \'ecole des Mines de Nantes, CNRS-IN2P3, Universit\'e de Nantes, France}
\newcommand{\iaa}{Institut d'Astronomie et d'Astrophysique, CP-226, Universit\'e Libre de Bruxelles (ULB), Brussels, Belgium}
\newcommand{\ganil}{Grand Acc\'el\'erateur National d'Ions Lourds (GANIL), CEA/DRF Ð CNRS/IN2P3, Caen, France}
\newcommand{\lpc}{Universit\'e de Caen, ENSICAEN and CNRS, UMR6534, LPC, Caen, France}
\newcommand{\usa}{Institute for Nuclear Theory, University of Washington, Seattle, Washington, USA}
\newcommand{\ipn}{Institut de Physique Nucl\'eaire de Lyon, CNRS/IN2P3, Universit\'e de Lyon, Universit\'e Claude Bernard Lyon 1, Villeurbanne, France}
\newcommand{\insarennes}{IETR, INSA Rennes, CNRS UMR 6164, Rennes, France}
\newcommand{\leme}{LEME, EA 4416, Universit\'e Paris-Ouest, Ville d'Avray, France}
\newcommand{\satie}{SATIE, UMR 8029, \'Ecole Normale Sup\'erieure de Cachan, Universit\'e Paris-Saclay, Cachan, France}
\newcommand{\lss}{L2S, UMR 8506, CentraleSup\'elec/CNRS/Universit\'e Paris-Sud, Universit\'e Paris-Saclay, Gif-sur-Yvette, France}
\newcommand{\lesia}{LESIA, Observatoire de Paris, PSL Research University, CNRS, Sorbonne Universit\'es, UPMC Universit\'e Paris 06, Universit\'e Paris Diderot, Meudon, France}
\newcommand{\syrte}{SYRTE, Observatoire de Paris, PSL Research University, CNRS, Sorbonne Universit\'es, UPMC Universit\'e Paris 06, LNE, Paris, France}
\newcommand{\ceadam}{CEA Commissariat \`a l'Energie Atomique DAM, Arpajon Cedex, France}
\newcommand{\stcruz}{Santa Cruz Institute for Particle Physics and Department of Physics, University of California at Santa Cruz, Santa Cruz, CA, USA}
\newcommand{\iuf}{Institut Universitaire de France, Paris, France}
\newcommand{\ariane}{ArianeGroup, Paris, France}
\newcommand{\callisto}{Callisto, Villefranche de Lauragais, France}
\newcommand{\tas}{Thales Alenia Space, Domaine Observation \& Sciences, Toulouse, France}
\newcommand{\atos}{Bull, an ATOS company, Bezons, France}
\newcommand{\airliquide}{Air Liquide, Grenoble, France}
\newcommand{\ddn}{DDN Storage, Paris, France}
\newcommand{\ipgp}{Institut de Physique du Globe de Paris, Paris, France}
\newcommand{\lpnhe}{Sorbonne Universit\'es, UPMC Universit\'e Paris 06, Universit\'e Paris Diderot, Sorbonne Paris Cit\'e, CNRS, Laboratoire de Physique Nucl\'eaire et de Hautes Energies (LPNHE), Paris Cedex 5, France}
\begin{document}
%
%
%
\sffamily
\vspace{5cm}
\pagecolor{white}
\color{black}
\pagestyle{empty}

\parbox{\textwidth}{
{\bfseries\boldmath \Huge \color{hbl} \vspace{1cm}  French SKA White Book \vspace{0.5cm}}\\
\noi {\LARGE \color{hbl} The French community towards the Square Kilometre Array}\\
}
\parbox{\textwidth}{\centering\includegraphics[scale=0.94,angle=0]{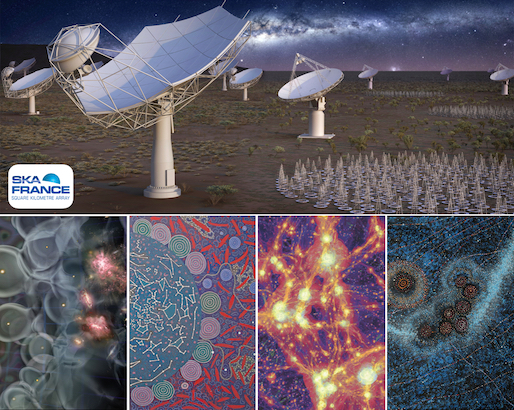}}

\vspace{1cm}
\parbox{\textwidth}{
{\large
\noi \underline{\bf Editor in Chief:}\\
\noi  C. Ferrari 

\vspace{0.3cm}
\noi \underline{\bf Editors:}\\
\noi G. Lagache, J.-M. Martin, B. Semelin --- {\color{hbl}Cosmology and Extra-galactic astronomy}\\
\noi M. Alves, K. Ferri\`ere, M.-A. Miville-Deschenes, L. Montier --- {\color{hbl} Galactic Astronomy} \\
\noi E. Josselin, N. Vilmer, P. Zarka --- {\color{hbl}Planets, Sun, Stars and Civilizations} \\
\noi S. Corbel, S. Vergani --- {\color{hbl} Transient Universe} \\
\noi S. Lambert, G. Theureau --- {\color{hbl}Fundamental Physics} \\
\noi S. Bosse, A. Ferrari, S. Gauffre --- {\color{hbl}Technological Developments} \\
\noi G. Marquette --- {\color{hbl}Industrial Perspectives and Solutions} \\
}}

\newpage
\pagecolor{white}
\parbox{0.8\textwidth}{\vspace{17.5cm}

\noi Published by the SKA France Coordination in collaboration with AS SKA-LOFAR 
\bigskip

\noi We acknowledge financial support of Universit\'e Paris-Saclay and AS SKA-LOFAR for the first ``French SKA White Book'' organisation meeting. We are grateful to the MPIfR colleagues, editors of the German ``White Paper'', who provided the adopted latex macro 

\bigskip

\noi Cover images show: (a) an artist's impression of the Square Kilometre
Array (SKA) and of its different antenna types, (b) an artist's impression of the Epoch of Re-ionisation, (c) numerical simulations of the magnetised cosmic web, (d) two paintings of the ``Shared Sky'' SKA Indigenous Art/Astronomy Exhibition

\smallskip

\noi Image courtesy: (a, d) SKAO; (b) A. Loeb (CfA) (c) F. Vazza (INAF)

\bigskip

\noi This White Book is available on-line at the SKA France web page:\\
\noi \href{https://ska-france.oca.eu/images/SKA-France-Media/FWB\_051017.pdf}{\color{blue} \myul[blue] {https://ska-france.oca.eu/images/SKA-France-Media/FWB\_051017.pdf}}}











%
%
%

\tableofcontents
 
\newpage
\pagestyle{plain}
\renewcommand{\thepage}{\Roman{page}}


\newpage
\pagestyle{empty}
\noi{\bf }

\newpage
\pagestyle{plain}





\newpage
\setcounter{page}{1}
\pagestyle{plain}
\cleardoublepage\phantomsection\addcontentsline{toc}{section}{R\'esum\'e  ex\'ecutif}{}{}

\noi {\bfseries\boldmath\LARGE \color{hbl} R\'esum\'e  ex\'ecutif}\\

\noi {\bfseries\boldmath \color{hbl} Introduction}

\smallskip
\noi Le ``Square Kilometre Array'' (SKA) est un projet de radiot\'elescope g\'eant, de surface collectrice \'equivalente de un kilom\`etre carr\'e comme son nom l'indique, constitu\'e de plusieurs r\'eseaux interf\'erom\'etriques dans les longueurs d'onde m\'etriques et centim\'etriques. Il est pr\'evu de d\'eployer SKA sur deux sites, en Afrique du Sud et en Australie. Le d\'eploiement se d\'eroulera en deux phases s\'epar\'ees dans le temps:

\begin{itemize}

\item La Phase 1, dont le co\^ut estim\'e est de 674 M\euro, le d\'ebut de construction pr\'evu pour 2020 pour une mise en service \`a l'horizon 2024+, consiste \`a installer environ 10\% du r\'eseau final, sous forme d'environ 200 antennes paraboliques en Afrique du Sud et 130 000 antennes phas\'ees fixes travaillant aux basses fr\'equences dans l'ouest australien. Dans cette configuration, SKA1 repr\'esentera un saut qualitatif immense par rapport aux instruments existants, et permettra des avanc\'ees d\'ecisives dans toutes les th\'ematiques de l'astrophysique et de la physique modernes, comme la cosmologie,  l'origine des champs magn\'etiques cosmiques, le milieu interstellaire, la formation des \'etoiles aux diff\'erentes \'epoques de l'univers, les ondes gravitationnelles, \dots

\item La Phase 2 est envisag\'ee pour les ann\'ees 2030+. Dans cette configuration finale, SKA2 sera l'instrument ultime de la radioastronomie basse-fr\'equence du 21\`eme si\`ecle. 

\end{itemize}

\noi D\`es la phase 1 SKA sera l'une des plus formidables machines jamais d\'eploy\'ees par l'homme, et de loin la plus impressionnante en termes de d\'ebit de donn\'ees et de puissance de calcul engag\'ee.

\smallskip
\noi Le projet est pour l'instant pilot\'e par un ``project office'' (SKA Organisation, SKAO), auquel doit succ\'eder une organisation intergouvernementale (IGO), l'Observatoire SKA, qui devrait se mettre en place progressivement \`a partir de fin 2017. Dans la phase initiale de mise en place de l'IGO les \'el\'ements de gouvernance et les r\`egles de fonctionnement seront n\'egoci\'es par les membres.\\

\noi {\bfseries\boldmath \color{hbl} Communaut\'e concern\'ee en France}

\smallskip
\noi La France, membre fondateur de SKA, a quitt\'e fin 2011 l'organisation charg\'ee de pr\'eparer sa construction, pour des raisons budg\'etaires et programmatiques. Les activit\'es scientifiques autour du projet ne se sont pas arr\^et\'ees cependant. 
La communaut\'e astronomique fran\c{c}aise a r\'eaffirm\'e son int\'er\^et majeur pour le projet SKA lors de son exercice quinquennal de prospective en 2014, et en a organis\'e depuis 2016 la pr\'eparation scientifique et technique autour de la structure de coordination nationale SKA~France. Cette structure est pilot\'ee par cinq \'etablissements (CNRS-INSU, Observatoires de Paris et de la C\^ote d'Azur, Universit\'es de Bordeaux et d'Orl\'eans).

\smallskip
\noi La publication de ce livre blanc, avec la participation de presque 200 auteurs fran\c{c}ais et de plus de 40 laboratoires de recherche, d\'emontre bien le fort investissement de notre communaut\'e astronomique, et celui, rapidement croissant, d'acteurs scientifiques et technologiques majeurs des domaines des Big Data et du calcul tr\`es haute performance.

\smallskip
\noi Les actions initi\'ees par la coordination SKA~France, relevant tout autant de la pr\'eparation scientifique de SKA que des activit\'es de R\&D n\'ecessaires \`a son d\'eveloppement, sont donc men\'ees aussi bien dans les \'etablissements qui pilotent SKA~France qu'en dehors de ceux-ci. L'exploitation d'instruments \'eclaireurs comme LOFAR et NenuFAR en France et en Europe, ou pr\'ecurseurs comme MeerKAT, ASKAP, MWA en Afrique du Sud et Australie, d\'emontre tr\`es clairement que la radioastronomie du 21\`eme si\`ecle n'est plus l'affaire exclusive des radioastronomes "classiques", mais implique toute la communaut\'e astronomique, exploitant des donn\'ees r\'eduites et calibr\'ees mises \`a disposition. Ainsi une estimation prudente du nombre de chercheurs concern\'es par l'exploitation de SKA1 s'\'el\`eve \`a 400 personnes en France, et plus de 4000 au niveau mondial. Cette estimation recouvre les chercheurs de la communaut\'e astronomique fran\c{c}aise, de la communaut\'e HPC et d'autres domaines applicatifs, ainsi que de l'industrie qui est pr\^ete \`a investir des efforts \`a long terme dans SKA. \\

\newpage
\noi {\bfseries\boldmath \color{hbl} Retour industriel}
\smallskip

\noi Des r\`egles de retour industriel seront d\'efinies au moment de la mise en place de l'IGO. Dans ce contexte, la France a une place essentielle \`a revendiquer, notamment dans des domaines o\`u SKA pr\'esente des d\'efis technologiques majeurs : production et stockage d'\'energie renouvelable ; infrastructures de calcul ; r\'ecepteurs dans les diff\'erentes bandes de fr\'equences ; traitement du signal et des donn\'ees ; ing\'enierie syst\`eme. Notre ambition est que le succ\`es de SKA soit enrichi par l'expertise fran\c{c}aise.

\smallskip
\noi De grands groupes industriels \`a forte composante fran\c{c}aise ont d\'ej\`a manifest\'e leur int\'er\^et pour ces d\'eveloppements; parmi ces groupes, la coordination SKA~France a d\`ej\`a re\c{c}u des lettres de soutien d'Air Liquide, Ariane Group, ATOS-Bull, Callisto, DDN Storage, ENGIE, FEDD, NVIDIA, Thales Alenia Space. Les laboratoires fran\c{c}ais sont \'egalement tr\`es bien plac\'es dans les domaines des r\'ecepteurs et du traitement du signal et des donn\'ees.

\smallskip
\noi Ainsi, ce positionnement permettra de mettre en avant la haute technologie fran\c{c}aise, ce qui facilitera potentiellement l'acc\`es \`a d'autres march\'es pour les industriels fran\c{c}ais, en particulier dans les domaines tr\`es comp\'etitifs de l'\'energie intelligente, de l'\'electronique et du calcul tr\`es haute performance, ainsi que celui de l'exploitation des masses de donn\'ees \`a l'\'echelle dite {\em Exa}.\\

\noi {\bfseries\boldmath \color{hbl} Conclusion}
\smallskip

\noi Suite \`a la recommandation r\'ecente du Haut Conseil des Tr\`es Grandes Infrastructures de Recherche (HC TGIR) de poursuivre la pr\'eparation du projet SKA en France, la communaut\'e astronomique nationale a redoubl\'e d'efforts sous la coordination de SKA~France. La r\'einscription de SKA sur la feuille de route fran\c{c}aise des TGIR d\`es 2018, et le retour de la France dans SKAO, pr\'eparant son entr\'ee dans l'IGO, nous placeraient dans une position favorable pour recueillir les fruits du fort investissement intellectuel consenti ces derni\`eres ann\'ees, en nous permettant de participer \`a des avanc\'ees scientifiques majeures pendant les 50 prochaines ann\'ees.

\smallskip
\noi Dans ce cadre, la coordination SKA France, avec ses partenaires priv\'es, a d\'ecid\'e d'\'evoluer vers la ``Maison SKA France'', destin\'ee \`a \^etre non seulement un forum pour ses membres afin d'organiser leur participation aux travaux pr\'eparatoires de SKA et de son IGO, mais aussi un pr\'ecurseur d'un nouveau paradigme pour les relations entre les mondes de l'industrie et de la recherche, ayant le m\^eme calendrier et les m\^eme objectifs, avec cependant diff\'erentes perspectives d'utilisation finale. La dimension transversale et multi-usage de SKA en font un cas d'\'etude parfait pour une approche nouvelle et innovante des modalit\'es de financement des TGIR.

\pagestyle{plain}

\cleardoublepage\phantomsection\addcontentsline{toc}{section}{Executive summary}{}{}

\noi {\bfseries\boldmath\LARGE \color{hbl} Executive summary}\\

\noi {\bfseries\boldmath \color{hbl} Introduction}

\smallskip
\noi The ``Square Kilometre Array'' (SKA) is a large radio telescope project characterised, as suggested by its name, by a total collecting area of approximately one square kilometre, and consisting of several interferometric arrays to observe metric and centimetric wavelengths. The deployment of the SKA will take place in two sites, in South Africa and Australia, and in two phases separated in time:

\begin{itemize}

\item Phase 1, with an estimated cost of \euro 674 million, the start of construction planned for 2020 and of commissioning by 2024+, consists of installing approximately 10\% of the final arrays. It will include about 200 dishes in South Africa and slightly more than 130,000 low-frequency simple antennas organised in phased arrays in Western Australia. In this configuration, SKA1 will represent a huge qualitative leap with respect to existing instruments, and will allow decisive advances in all the domains of modern astrophysics and physics, such as cosmology, the origin of cosmic magnetic fields, the physics of the interstellar medium, the formation of stars at different epochs of the universe, the detection of gravitational waves, \dots

\item Phase 2 is envisaged for 2030+. In this final configuration, SKA2 will be the ultimate instrument of low-frequency radio astronomy of the 21st century.

\end{itemize}

\noi From its Phase 1, the SKA will be one of the most formidable machines ever deployed by mankind, and by far the most impressive in terms of data throughput and required computing power.

\smallskip
\noi The project is currently being led by a project office (SKA Organisation, SKAO), to be succeeded by an intergovernmental organisation (IGO), the SKA Observatory, which is expected to be gradually implemented from the end of 2017. In the initial phase of the implementation of the IGO, the members will negotiate governance elements and operating rules.\\

\noi {\bfseries\boldmath \color{hbl} Community in France}

\smallskip
\noi France, a founding member of the SKA, left the organisation in charge of preparing its construction at the end of 2011, for budgetary and programmatic reasons. Scientific activities around the project did not stop, though. 
The French astronomical community restated its major interest in the SKA project in 2014, during its five-year perspective exercise, and, since 2016, has organised the French scientific and technical preparation to the SKA through the national coordination SKA France. This structure is managed by five institutions (CNRS National Institute for Earth Sciences and Astronomy (CNRS-INSU), Observatories of Paris and of C\^ote d'Azur, Universities of Bordeaux and Orl\'eans).

\smallskip
\noi The publication of this white paper, with the participation of almost 200 French authors and more than 40 research institutes, demonstrates the strong investment of our astronomical community and of a rapidly growing number of major scientific and technological players in the fields of Big Data and high performance computing.

\smallskip
\noi The actions initiated by the SKA France coordination, concerning both the scientific preparation of the SKA and the R\&D activities necessary for its development, are therefore carried out both within and outside the research institutes founding of SKA France. The exploitation of pathfinder instruments such as LOFAR and NenuFAR in France and Europe, or precursors as MeerKAT, ASKAP, MWA in South Africa and Australia, shows very clearly that radio astronomy of the 21st century is no longer the exclusive business of ``classical'' radio astronomers, but involves the whole astronomical community that is interested to exploit available reduced and calibrated data. Based on this, a conservative estimate of the number of researchers interested to SKA1 is of about 400 in France and more than 4000 in the world. These numbers include researchers from the French astronomical community, from the HPC community, as well as from private companies that are ready to lasting efforts in the SKA project. \\

\noi {\bfseries\boldmath \color{hbl} Industrial return}
\smallskip

\smallskip
\noi Rules of industrial return will be defined when the IGO will become operational. In this context, France has an essential place to claim, particularly in those areas where the SKA presents major technological challenges: production and storage of renewable energy; computing infrastructures; receivers in various frequency bands; signal and data processing; system engineering. Our ambition is that the success of the SKA will benefit from the French expertise.

\smallskip
\noi Large industrial groups with a strong French component have already expressed interest in these developments; SKA France has received letters of support from Air Liquide, Ariane Group, ATOS-Bull, Callisto, DDN Storage, ENGIE, FEDD, NVIDIA, Thales Alenia Space. French laboratories have also a wide expertise in the domains of receivers and signal and data processing.

\smallskip
\noi French high technology will also strongly benefit from the participation to the SKA, which will facilitate potential access to other markets for French manufacturers, particularly in the highly competitive fields of intelligent energy, electronics and very high performance computing, as well as that of the exploitation of Exa-scale Big Data.\\

\noi {\bfseries\boldmath \color{hbl} Conclusion}
\smallskip

\noi Following the recent recommendation by the Haut Conseil des Tr\`es Grandes Infrastructures de Recherche (HC-TGIR, High level committee for large research infrastructures) to continue the French participation to the preparation of the SKA instrument, the French astronomical community has redoubled efforts under the coordination of SKA France. Having the SKA on the French roadmap of large research infrastructures in 2018 and France joining the SKA Organisation, preparing for its entry into the IGO, would put us in a favourable position for exploiting the strong intellectual investment made in recent years, and enable France to participate in major scientific breakthroughs over the next 50 years.

\smallskip

\noi In this framework, the SKA France Coordination, together with the companies engaged, has decided to evolve towards the ``Maison SKA France'', intended to be not only a forum for its members to organise their participation to the preparatory work for the SKA and its IGO, but also a precursor of a new paradigm for industry and research relations, sharing the same calendar and performance agenda, but with different final use perspectives. The multi-usage, transversal domain dimension of the SKA makes it a perfect case study for such new and innovative financial approach of large research infrastructures.

\pagestyle{empty}
\renewcommand{\thepage}{\arabic{page}}

\newpage
\pagestyle{plain}


\section{Introduction}
\setcounter{page}{1}
\smallskip

\noi {\sffamily \scriptsize
{\sffamily\bf C. Ferrari} [\lagrange],
{\bf G.~Marquette} [\cnrs],
{\bf M.~P\'erault} [\lermasorb]}

\subsection{The Square Kilometre Array}
\vspace{0cm}

\subsubsection{The telescopes}
\vspace{0cm}

\begin{figure*}[t]
\centering
\includegraphics[width=0.3\linewidth]{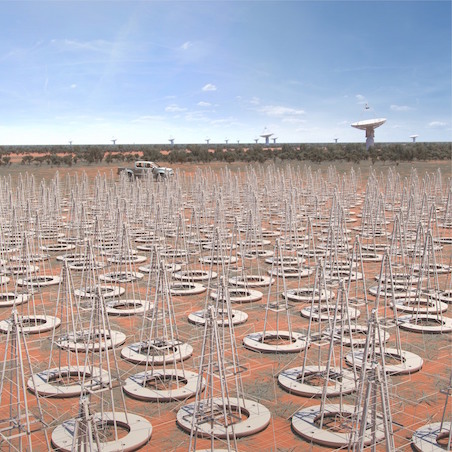}
\includegraphics[width=0.3\linewidth]{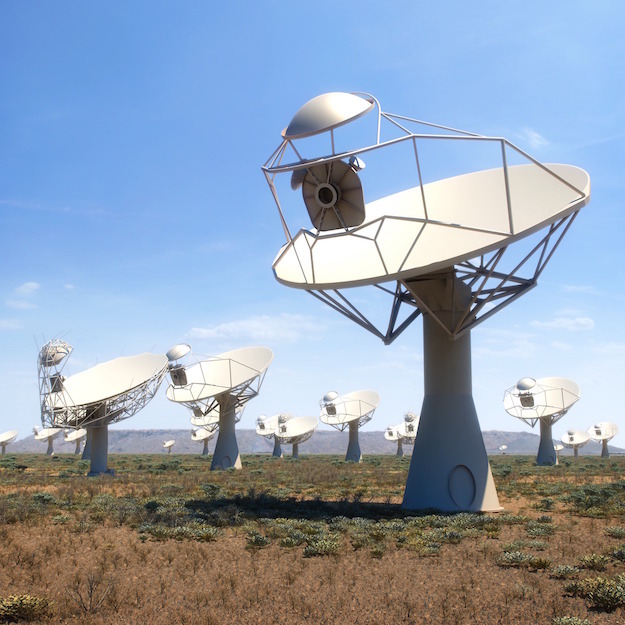}
\includegraphics[width=0.6\linewidth]{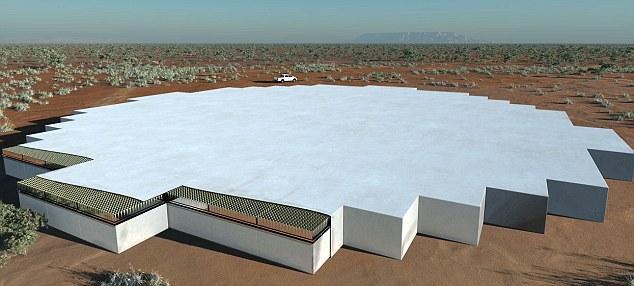}
\caption{\label{fig:ant} Artist's rendition of the SKA low- and mid-frequency aperture arrays ({\em top left} and {\em bottom}) and SKA dishes ({\em top right}). Image courtesy: SKA Organisation.}
\end{figure*}

\begin{figure}[t]
\centering
\includegraphics[width=0.8\linewidth]{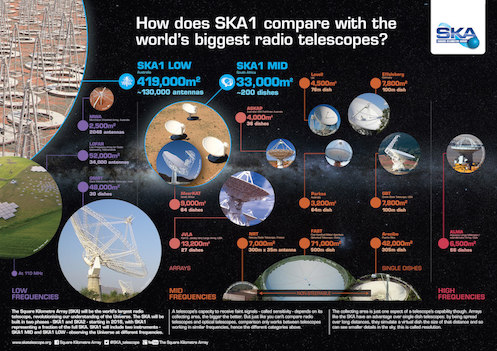}
\caption{\label{fig:skatec1} Image summarising how the SKA compares to currently operating radio telescopes. Image courtesy: SKA Organisation.}
\end{figure}

\begin{figure}[t]
\centering
\includegraphics[width=0.8\linewidth]{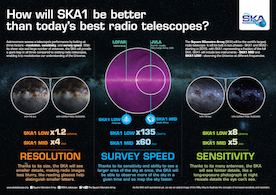}
\caption{\label{fig:skatec2} Expected gain for the SKA in angular resolution, survey speed (i.e. the velocity at which the telescope is able to map the sky, given resolution and sensitivity limits) and sensitivity. Image courtesy: SKA Organisation.}
\end{figure}

\noi The Square Kilometre Array (SKA) is an ambitious international project to develop the world's largest radio telescope that, when complete, is planned to reach 1 million square metres of collecting area (hence its name). The full array will be built over two sites in Australian and African deserts, as far as possible from the origin of disturbing radio-frequency interferences (RFI), mostly of human origins. The frequency range that the SKA is expected to cover is unprecedented, going from approximately 50 MHz to 15 GHz (and, possibly, more in its final phase - up to about 30 GHz), thus requiring a variety of antenna designs (see Fig.\,\ref{fig:ant}). This, together with the expected gain in survey speed and image quality compared to current radio telescopes (Figs.\,\ref{fig:skatec1} and \ref{fig:skatec2}), will enable transformational science at centimetre and metre wavelengths. Its status as an ESFRI Landmark Project recognises the SKA as a major research infrastructure for Europe. 

\smallskip
\noi In its lowest frequency part ($\sim$50 to 350 MHz), the telescope will be made up by hundreds of thousands of simple antenna elements (e.g. dipoles or log--periodic elements), which will be arranged in hundreds of stations of a few metres in diameters. The signal of all elements within one station will be combined numerically and all the stations will work together forming a so called ``aperture array''. The station separation will go from a few tens of meters in a central core area, to several tens (up to hundreds) of kilometres in the outer distribution, which will include a few spiral arms. This low-frequency part of the telescope is going to be built in the Murchison desert of Western Australia. At the higher frequencies covered by the SKA (above 350 MHz), the array will consist of hundreds of 15 m diameter dishes, to be initially distributed within the Karoo desert (about 500 km North of Cape Town) and subsequently extending to different states in central up to northern Africa, going from maximum baselines of hundreds to thousands of kilometres. Further technical developments are planned, aiming to cover the intermediate frequency part of the SKA also with dense aperture arrays.

\smallskip
\noi These numbers alone prove to the reader with an idea of  why the SKA is considered as a fully ``Big Data'' project. These numerous elements generate a huge data rate of several Tb/s already in the first phase of the project (called ``SKA1'', see Sect.\,\ref{sect:project}), when the total collecting area will be approximately one tenth of the final expected array. The SKA1 data rate is  expected to exceed the total global internet traffic at present day rate. Even after data reduction, the archived data rate for astronomical exploitations will be of the order of 50 to 300 Pbytes per year. Data processing and storage will require by early 2020's super-computers about 10 times more powerful than the fastest machines available today. 

\subsubsection{The project organisation}\label{sect:project}

\smallskip
\noi A so big and ambitious international project is the result of a quite long development history. The SKA project, originally conceived in the late 1980's / early 1990's (see Ekers 2012 for a complete overview), formally began with signing a first Memorandum of Agreement in 2000, followed by the establishment of the SKA International Project Office with R. Schilizzi as the first Project Director (2003). An early design study (called ``SKADS'') started in 2005, and a preparatory work (``PrepSKA'') in 2008, paving the way to the birth of the SKAO legal entity in 2011. Since 2011, a UK Company Limited by Guarantee, called SKA Organisation (SKAO), has been adopted as a temporary solution to enable the SKA project to proceed. In order to provide a long-term government commitment and funding stability, the SKAO is currently evolving towards an intergovernmental organisation (IGO), similar to other big research infrastructures (such as ESO, ESA and CERN). 

\smallskip
\noi SKAO includes today ten formal members (Australia, Canada, China, India, Italy, New Zealand, South Africa, Sweden, the Netherlands and the United Kingdom), with in addition several countries that have expressed their potential interest in joining the Organisation. Without entering in a detailed description of the SKA Governance, a General Director is appointed by a Board of Directors, which includes voting representatives of the 10 member countries (as well as one person per observer countries) and which is called to take all relevant decisions for the development of the project. 

\smallskip
\noi In 2012 the SKA board proceeded to the selection of the SKA construction sites. Both Australia and South African deserts, sites of the three technological and scientific precursors telescopes (MWA, ASKAP and MeerKAT, see Sect.\,\ref{intro:consortia}), were considered as excellent locations for building the arrays covering the low- and mid-frequency part of the e.m. spectrum. At that phase, in particular, both the precursors going to GHz frequencies (ASKAP and MeerKAT) were planned to be integrated to the future SKA antennas. In 2015, however, a redefinition of the design needed to be developed (known as ``re-baselining'') due to cost issues. This lead to the definition of the design Baseline for the first phase of SKA (generally referred to as ``SKA1'') that,  within the set cost cap of Û674M (2016 euros), will consist of two arrays:

\begin{itemize}

\item {\bf SKA1-LOW} in Australia, including $\sim$131,000 simple antennas covering the frequency range from $\sim$50 MHz to $\sim$350 MHz. The array will be in the same region as ASKAP and will have a maximum baseline of approximately 65 km;

\item {\bf SKA1-MID} in South Africa, including 133 15\,m diameter dishes and observing frequencies from 350 MHz to 15.5 GHz, divided in five frequency bands (from Band 1 to Band 5). The instrument will integrate the MeerKAT antennas, for a total number of 197 dishes, separated by a maximum distance of 150 km. 

\end{itemize}

\noi After the Cost Control Project (CCP) initiated at the November 2016 Board of Directors meeting, in July 2017 the Board decided that the Design Baseline remains the long-term ambition of the SKA1 project and the focus should be on the Critical Design Review of the project, expected in 2018. At that occasion, the Board approved though the definition of a ``Deployment Baseline'', which corresponds to the telescopes currently deliverable at that funding level and takes advantage of the scalable nature of interferometers\footnote{All information in the \href{http://newsletter.skatelescope.org/wp-content/uploads/2017/04/SKAO-eNewsletter-Edition-35-August-2017.pdf}{\color{blue} \myul[blue] {SKA Organisation eNewsletter edition 35 Ð August 2017}}.}.

\smallskip
\noi Based on these decisions, the construction of SKA1, which corresponds to $\lesssim$10\% of the full final instrument, is planned to start in 2019, with the first science operations beginning in early 2020's. SKA1 will be a single observatory built over three sites, among which one will host the headquarters (at Jodrell Bank, in UK, site decision taken in 2015) and the others two telescopes, SKA1-LOW in Australia and SKA1-MID in South Africa. After 2025, the instrument is planned to be further developed towards the full square kilometre total collecting area. This will of course require the necessary preparation from the technological and scientific point of view, for which SKA1 will play a crucial role, but also a likewise important budget and managing strategy.

\subsection{General overview of the SKA science and technology} \label{intro:techno}
\vspace{0cm}

\subsubsection{An instrument designed for a wide range of astronomical and general physics studies}
\vspace{0cm}

\begin{figure}
\centering
\includegraphics[width=0.5\linewidth]{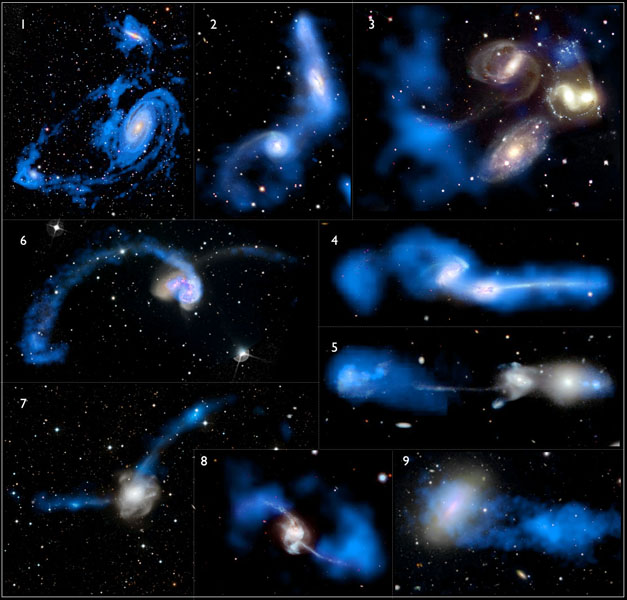}
\caption{\label{fig:HI} Sample of merging galaxies in different phases of the collision between two or more systems. The yellow/white and blue colours indicate, respectively, the distribution of stars observed at visible wavelengths and of the hydrogen atomic gas detected in the radio (at 21\,cm). From Duc \& Renaud 2013.}
\end{figure}

\noi The SKA is a long-standing project, whose original conception dates back to the late 1980's -- early 1990's. The telescope was originally conceived as an optimised instrument to measure the 21\,cm rest-frame emission of neutral hydrogen (\hi) gas at cosmological distances (see Ekers 2012 and references therein; Fig.\,\ref{fig:HI}). Indeed, the SKA is still a wonderful ``\hi~machine'', which will provide us with the possibility to study how galaxies acquire and loose their hydrogen gas, how it is transformed into stars, how it is related to the possible presence of an active nucleus and to the density of the environment. Going from current facilities to SKA1, we will be able to perform this kind of studies along the history of the Universe: we will measure the \hi~content of hundreds of thousands galaxies, up to look-back times of approximately 5 to 6 billion years, while with present telescopes we are limited to a very few gas-rich galaxies observed, at the maximum, about 2-2.5 billion years ago (Staveley-Smith \& Oosterloo 2015). 

\begin{figure}
\centering
\includegraphics[width=0.8\linewidth]{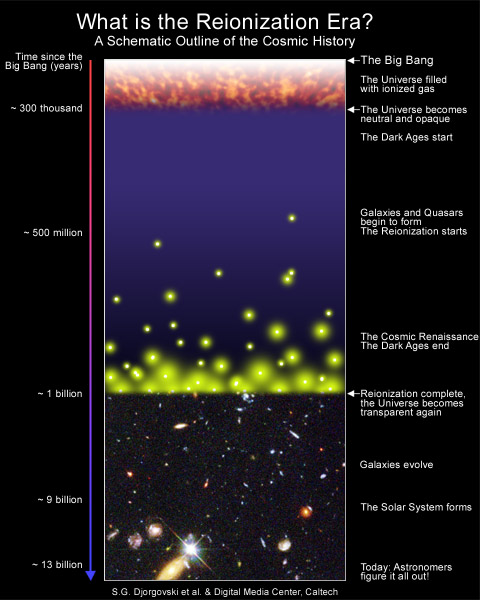}
\caption{\label{fig:history}Artist's conception of the evolution of structures within the history of the universe, starting from neutral hydrogen, to the appearance of the first objects (stars and galaxies) that re-ionise the gas, until the formation of all the astrophysical sources that we observe today. The SKA will be an exquisite, and often unique, tool to study most of the phases shown here. Image courtesy: Djorgovski et al. (Caltech).}
\end{figure}

\smallskip
\noi In the early 2000's, another exceptional scientific application of SKA \hi~measurements started to emerge: the study of the Cosmic Dawn (CD) and the Epoch of Re-Ionisation (EoR). These phases of the Universe started around 100 and 280 million years after the Big Bang, respectively, when, after the Dark Ages in which the matter of the Universe was completely dominated by neutral hydrogen, the first sources (stars, galaxies, \dots) began to form (see Fig.\,\ref{fig:history}). These luminous objects, intrinsically faint and suffering absorption by their medium, were able to ionise their surrounding gas, but at the same time they are extremely difficult to detect. The SKA will instead be able to map the structure of the \hi~gas from which these sources formed, whose distribution will therefore be characterised by voids (i.e. the bubbles of ionised medium surrounding luminous objects). This kind of measurements, which require exquisite sensitivity and the capacity to get rid of all kind of foreground sources, are extremely challenging. However, it is definitely worth investing telescope time and resources, since we will have a unique access to the phases during which the very initial seeds of structures observed by other telescopes (Planck, being the latest large mission) in the Cosmic Microwave Background (CMB) are transformed into the panoply of sources that we observe in the more local Universe (see e.g. Koopmans et al. 2015). 

\begin{figure*}[h]
\centering
\includegraphics[angle=-90,width=0.6\linewidth]{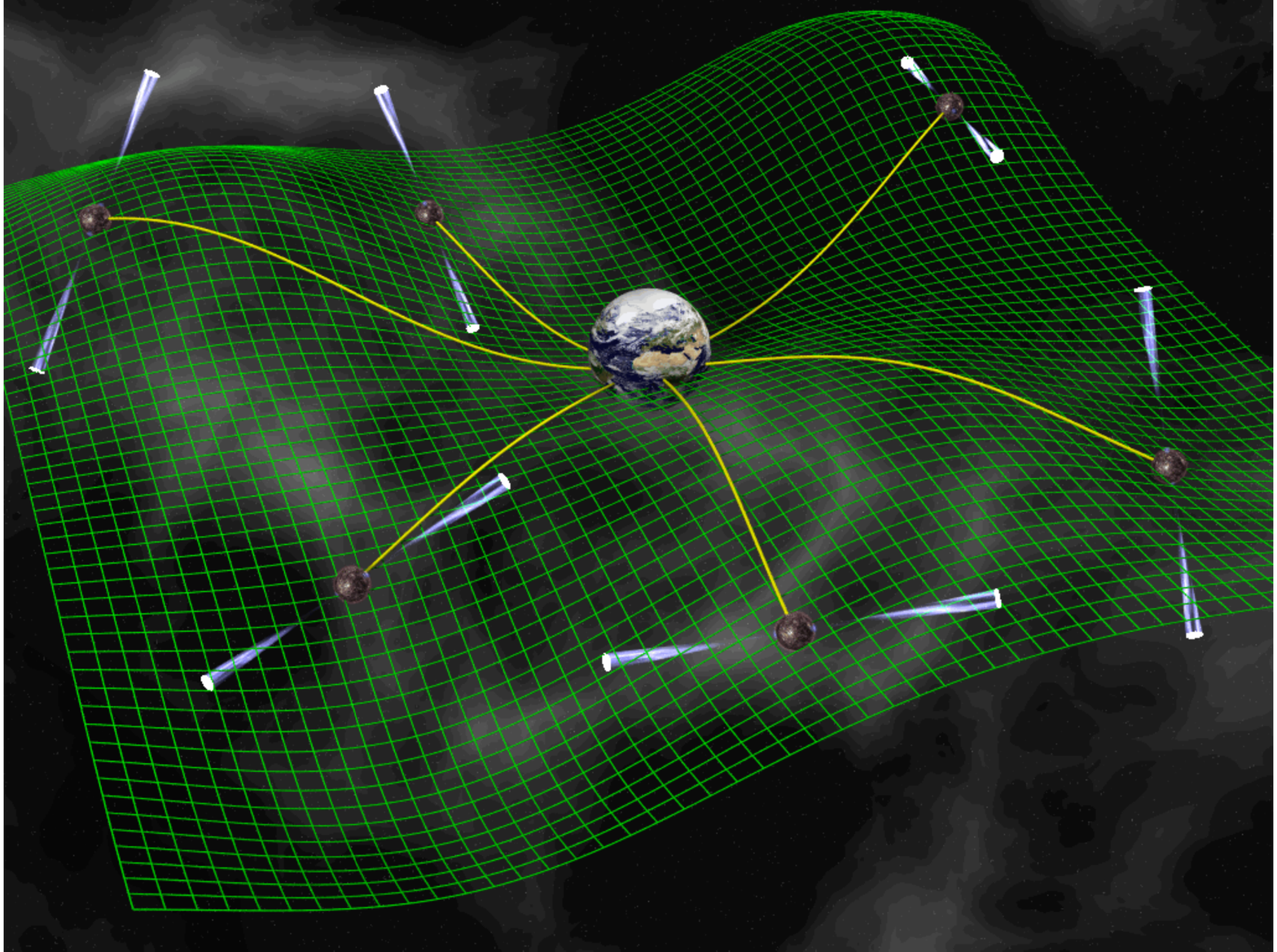}
\caption{ArtistÕs conception of a pulsar timing array (PTA): pulsars form the arms of a cosmic gravitational wave detector. Image courtesy: David J. Champion.}
\label{fig:pta}
\end{figure*}

\smallskip
\noi A third major science case for the present SKA project is the study of pulsars, highly magnetised, rotating neutron stars, whose beamed radio emission is detected as a pulse of radiation each time the beam sweeps across our line-of-sight; a pulsar is therefore a ``cosmic light-house'', whose light blinks on and off to our view with a constant and extremely short pulse period (observed values for different pulsars go from $\approx$ 1.4 ms to $\lesssim$ 10 seconds). Pulsar research has acquired more and more weight over time. The first reason is that they are extreme, and thus exquisite, physics laboratories, with properties very far from what can be available on Earth (deep gravitational potentials, densities exceeding nuclear densities, magnetic field strengths as high as $B \sim 10^{14} - 10^{15}$ Gauss, see e.g. the excellent NRAO ``Essential Radio Astronomy'' series of lectures by Jim Condon and collaborators, Condon \& Ransom 2016). The second reason, whose weight can be fully understood today, after the first direct detection of Gravitational Waves (GW; LIGO Scientific Collaboration \& Virgo Collaboration 2016), is that a network of pulsars can be used as a giant GW detector in space. Pulses from these neutron stars are extremely regular (accuracies approach 1 part in $10^{16}$). This means that even small perturbations in the fabric of space-time, as those created by the propagation of a GW, can be detected through pulsar distance measurements: when they propagate through our network of observed pulsars, perturbations can indeed increase our distance from some of these neutron stars and decrease the distance from others, thus slightly increasing or decreasing the pulse arrival time. Extremely precise measurements of the detection time of the different signals can therefore be used to detect GW propagation in space. This is the work performed by researchers involved in Pulsar Timing Array (PTA) studies (e.g. Janssen et al. 2015; see Fig.\,\ref{fig:pta}).

\begin{figure*}[ht!]
\begin{center}
\includegraphics[width=0.6\textwidth]{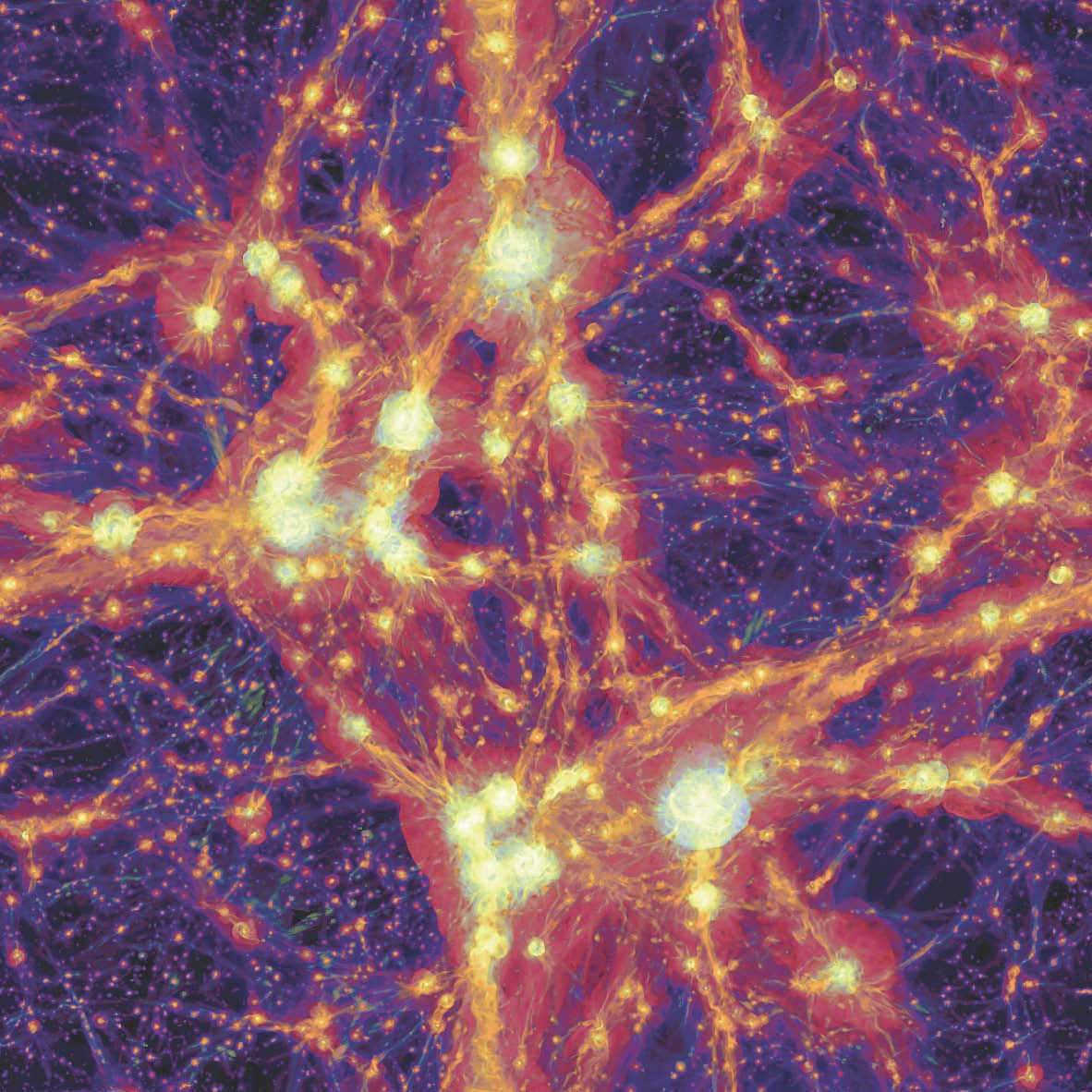}
\end{center}
\caption{\label{fig:web} Numerical simulations of the ``cosmic web'': purple and red colours show the gas distribution at temperatures respectively lower and higher than 10$^5$ degrees. Magnetic fields are represented in yellow (for intensities higher than 10\,nGauss) and orange (for intensities lower than 10\,nGauss). Filaments of cold and low-magnetised gas, which connect hotter and higher magnetised structures (galaxy clusters), are currently undetected. The SKA could allow to reveal them through their expected synchrotron radiation. From Vazza et al. 2015.}
\end{figure*}

\smallskip
\noi The SKA will also give us the unique opportunity to study magnetic fields in all kinds of sources, on spatial scales ranging from the few millions of kilometres of heliospheric coronal mass ejections to the dozens of Mpc ($\approx 10^{21}$ km, 1 parsec (pc) corresponding to $\sim 3 {\times} 10^{13}$ km) of the cosmic filaments that connect galaxies and galaxy clusters within the Universe (Fig.\,\ref{fig:web}). This will be possible mostly through Faraday Rotation Measure (RM) studies. In a very simplified way, the synchrotron signal from radio sources is linearly polarised and its polarisation direction is rotated when it passes through a foreground magnetised plasma before reaching our telescopes on Earth. Very usefully, this rotation scales with the square of the observed wavelength and with a quantity (the RM) that depends on the intensity of the traversed magnetic field. Basically, if we detect a huge amount of background radio sources, multi-wavelength observations allows us to trace the 3D structure of magnetic fields in their foreground. The big advantage of the SKA compared to current state-of-the-art radio telescopes is that we will increase the number of detected background radio sources with measured RM from about 40\,000 to several millions (Johnston-Hollitt et al. 2015).

\smallskip
\noi If these are the major science drivers for the SKA in terms of transformational science, which can be achieved {\em only} with this instrument, other astrophysical sectors will enormously benefit from it. The evolution of galaxies will not only be addressed through \hi-line studies, but also thanks to the analysis of diffuse radio emission. This is a powerful tracer of active galaxies, in terms of both on-going star-formation and presence of a massive central black-hole accreting mass. In both cases, we detect continuum synchrotron radiation, related to the presence of relativistic electrons and magnetic fields. In star-forming galaxies, supernovae explosions are responsible for cosmic ray acceleration, while in Active Galactic Nuclei (AGN) the relativistic plasma is ejected by the central compact nucleus hosting a black-hole. Through radio observations, we can therefore unveil magnetic fields within galaxies, study star formation from within our own Milky Way to high-z ($\lesssim$ 4) galaxies, as well as analyse the role of BH in galaxy evolution up to cosmological distances (Prandoni \& Seymour 2015; Johnston-Hollitt et al. 2015; Umana et al. 2015). 

\smallskip
\noi In the very local Universe, one of the most exciting science cases of the SKA, complementary to other instruments, is the so called ``Cradle of Life''. The SKA is expected to be able to examine planet-forming disks around young stars, the organic chemistry of Earth-like terrestrial planets, and even detect signs of possible extra-terrestrial life from planets in nearby solar systems (Search for Extraterrestrial Intelligence (SETI) experiment).

\smallskip
\noi New scientific discoveries, as well as technological developments, have therefore enormously widened the scientific ambitions of the SKA, which will address open questions covering a variety of astronomy and physics fields, across a huge range of scales and over a big fraction of the history of the Universe, schematically represented in Fig.\,\ref{fig:history}. 

\smallskip
\noi We can interestingly follow the evolution of SKA related science cases through major comprehensive publications, starting with the first science case document (Taylor \& Braun 1999), followed by the ``Science with the Square Kilometre Array'' book which appeared in 2004 (Carilli \& Rawlings 2004). Ten years after, in summer 2014, the SKA Organisation, in collaboration with institutes and researchers participating to the SKA effort, has organised the meeting \href{http://pos.sissa.it/cgi-bin/reader/conf.cgi?confid=215}{\color{blue} \myul[blue] {``Advancing Astrophysics with the Square Kilometre Array''}} (AASKA14, in the rest of this Volume), held in Giardina Naxos, Sicily. This resulted in an impressive two volume book, containing 135 chapters from 1,231 contributors spread over 31 countries (Braun et al. 2015).

\smallskip
\noi The SKA Office is organised to take full profit of this enthusiastic international participation to the preparation of the SKA. In order to conceive the best instrument to address the open questions mentioned above, a number of Science Working Groups (SWG) have been settled. They are listed below, together with the reference papers written, generally, by their chairs\footnote{In order to bring new ideas and have a very open approach, the chairs of the SWG change approximately every four years.} and published in the AASKA14 book:

\begin{itemize}

\item \href{http://skatelescope.org/the-science/cosmicdawn/}{\color{blue} \myul[blue] {Epoch of Reionisation}} (Koopmans et al. 2015);

\item \href{http://skatelescope.org/the-science/galaxyevolution/dark-energy/}{\color{blue} \myul[blue] {Cosmology}} (Maartens et al. 2015);

\item \href{http://skatelescope.org/the-science/challenging-einstein/}{\color{blue} \myul[blue] {Fundamental Physics with Pulsars}} (Kramer \& Stappers 2015);

\item \href{http://skatelescope.org/the-science/radio-transients/}{\color{blue} \myul[blue] {Radio Transients}} (Fender et al. 2015);

\item \href{http://skatelescope.org/the-science/continuum-surveys/}{\color{blue} \myul[blue] {Extragalactic Continuum}} (Prandoni \& Seymour 2015);

\item \href{http://skatelescope.org/the-science/magnetism/}{\color{blue} \myul[blue] {Cosmic Magnetism}} (Johnston-Hollitt et al. 2015);

\item \href{http://skatelescope.org/the-science/cradle-life/}{\color{blue} \myul[blue] {Cradle of Life/Astrobiology}} (Hoare et al. 2015);

\item \href{http://skatelescope.org/the-science/galaxyevolution/dark-energy/galaxy-evolution-more/}{\color{blue} \myul[blue] {H I Galaxy Science}} (Staveley-Smith \& Oosterloo 2015);

\item \href{http://skatelescope.org/dark-energy/galaxy-evolution-more/}{\color{blue} \myul[blue] {Extragalactic Spectral Line}} (Beswick et al. 2015);

\item \href{http://skatelescope.org/galaxyevolution/}{\color{blue} \myul[blue] {Our Galaxy}} (Umana et al. 2015);

\item \href{http://skatelescope.org/shi/}{\color{blue} \myul[blue] {Solar and Heliospheric Physics}} (Nakariakov et al. 2015).

\end{itemize}

\noi To be noted that the last three SWG were part of other groups before and have been added after the Giardini Naxos meeting in 2014, based on the need of specific science cases that emerged during that extremely interesting and useful conference. It is also worth stressing that the SKA will be developed in parallel to other major astronomical facilities. Interesting sections both of the 2014 SKA book (see Sect. 9 of Braun et al. 2015) and of this Volume (Sect.\,\ref{science:synergies}) are therefore devoted to the developing synergies and contact points with other projects.

\smallskip
\noi The main activities of the different SWG include providing advice on the science requirements for the SKA, collaborating with the technical consortia described in Sect.\,\ref{intro:consortia} to achieve an instrument optimised for the expected goals, as well as communicating with the rest of the astronomical community about SKA developments. Scientists interested to have more information about the different SWG and the possibility to join them can visit the \href{https://www.skatelescope.org/swg-terms-of-reference/}{\color{blue} \myul[blue] {SKA Organisation web page}}. In addition to the SWG, a Science and Engineering Advisory Committee (SEAC) was established by the SKA Board to advise the Director-General and the Board on matters related to scientific and technical issues.

\subsubsection{Design of SKA components}\label{intro:consortia}
\vspace{0cm}

\noi Similarly to other Earth and space science large projects, the SKA design has been broken down in various work packages (WPs) managed by international consortia responsible for specific elements of the final observatory. 

\smallskip
\noi These consortia were formed after 2013, when a \href{http://skatelescope.org/request-for-proposals/}{\color{blue} \myul[blue] {request for proposal}} (RfP) was sent out by SKAO to all research institutes and potential industrial partners. The leading organisation or entity in all proposals answering to this call had to be in a member country of the SKA Organisation; no restrictions were placed on the other partners of a proposing consortium nor in any of the subcontracting. While the SKA Board and SKAO played a key role in selecting and, still now, coordinating the consortia, the latter operate independently and are responsible to fullly funding their work. Since 2014, a yearly SKA Engineering Meeting with all consortia participating is organised, with the aim of providing a global overview of the status, progress and way forward for the project in terms of science, governance, engineering and management. 

\smallskip
\noi At the moment of the RfP, eleven WPs were defined by SKAO, which are listed below, grouped into two main categories. To be noted that, with a few exceptions, the name of each consortium is the same as the one of the corresponding WP.

\begin{itemize}

\item {\bf Elements of the SKA1 programme}, including the following WPs:

\begin{itemize}

\item Dish (\href{http://skatelescope.org/dish/}{\color{blue} \myul[blue] {DSH}})\footnote{At the time of the RfP, this WP included the design of phased array feeds, which is now part of a separate WP of the Advanced Instrumentation programme, as described below.};

\item Low frequency aperture array (\href{http://skatelescope.org/lfaa/}{\color{blue} \myul[blue] {LFAA}})\footnote{In this case, the name of the consortium in charge of the WP is ``Aperture Array Design and Construction'' (AADC).};

\item Telescope manager (\href{http://skatelescope.org/tm/}{\color{blue} \myul[blue] {TM}});

\item Science data processor (\href{http://skatelescope.org/sdp/}{\color{blue} \myul[blue] {SDP}});

\item Central signal processor (\href{http://skatelescope.org/csp/}{\color{blue} \myul[blue] {CSP}});

\item Signal and data transport (\href{http://skatelescope.org/sadt/}{\color{blue} \myul[blue] {SaDT}});

\item Assembly Integration Verification (\href{http://skatelescope.org/aiv/}{\color{blue} \myul[blue] {AIV}});

\item Infrastructure, both in Australia and in South Africa (\href{http://skatelescope.org/infra/}{\color{blue} \myul[blue] {INFRA~AUS \& INFRA~SA}}).

\end{itemize}

\item {\bf Advanced Instrumentation programme}, including:

\begin{itemize}

\item Mid-frequency aperture array (\href{http://skatelescope.org/mfaa/}{\color{blue} \myul[blue] {MFAA}})\footnote{In this case, the name of the consortium in charge of the WP is ``Aperture Array MID frequency'' (AAMID).};

\item Wide-band single pixel feed (\href{http://skatelescope.org/wbspf/}{\color{blue} \myul[blue] {WBSPF}}).

\end{itemize}

\end{itemize}

\begin{figure*}[ht!]
\begin{center}
\includegraphics[width=0.9\textwidth]{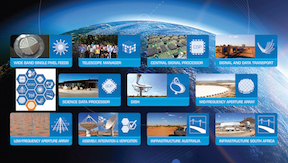}
\end{center}
\caption{\label{fig:logosWP} Logos of the different SKA WPs and consortia. The new PAF consortium is indicated in orange in the inset. Images courtesy: SKA Organisation.}
\end{figure*}

\smallskip
\noi Two stages were identified for each WP: Stage 1, a requirements analysis and preliminary design stage, and Stage 2, a detailed design stage. While Stage 1 for the Elements of the SKA1 programme included both to derive the functional and performance requirements, and to elaborate the Element preliminary designs, in order to prepare for the detailed designs and procurement specifications necessary for the construction of the SKA telescope in Stage 2, the final aim for the second WP category (Advanced Instrumentation programme) was to elaborate preliminary design for future developments, beyond SKA1. 

\smallskip
\noi After having passed the PDR (Preliminary Design Review) between the end of 2015 and the first half of 2016, the consortia are now in the detailed definition phase, preparing the Critical Design Reviews (CDR), which are expected in 2018. We will summarise in the following their main tasks in the framework of the Design Baseline set in March 2015.
 
\smallskip
\noi Several consortia are responsible for the construction of two telescopes, SKA1-LOW and SKA1-MID. The design of the different types of antennas, their structure, receivers and verification is the responsibility of the LFAA and DSH consortia respectively. In the case of SKA1-LOW, the 131\,000 antennas that will constitute the telescope will be grouped into 512 stations (of which about half in a central core of 1 km and the rest on spiral arms up to a maximum distance of 65 km between stations). The LFAA consortium is also responsible for the signal processing chain and hardware to combine the antenna signal within each station to form the equivalent of a single larger receiving system (phased arrays). The use of phased arrays for higher frequency observations (above 350 MHz) is currently limited mostly by energy consumption issues, but its implementation, planned in a second phase of the SKA project, is the subject of the developments within the MFAA consortium. The other consortium that is conducting studies for the SKA Advanced Instrumentation Program is the WBSPF, involved in the design of very broadband receivers, covering ranges of four to eight in frequency instead of the factors of two of conventional (``octave'') radio astronomy receivers.

\smallskip
\noi All infrastructures (roads, buildings, vehicles, \dots) will be delivered by the two INFRA consortia in the host countries of the SKA (Australia and South Africa), while the AIV consortium is responsible for planning all activities necessary to incorporate SKA elements into existing infrastructure and to verify their compliance with the SKA specifications.

\smallskip
\noi The data rates produced by SKA1 will be enormous (several Tbit/s). The SaDT consortium is in charge of the hardware and software necessary for the data transmission between the elements of the telescopes. It is also responsible for providing the time reference, which is essential for an interferometer.

\smallskip
\noi The biggest challenge in terms of computing performance and storage capacity is related to the steps between the acquisition of raw data at the telescopes, the production of multi-dimensional images of the sky and their distribution to the astronomical community. The two main consortia in charge of data processing are the CSP and the SDP. Very briefly, the first is the SKA ``central brain''. It is responsible for a) removing terrestrial radio-frequency interferences (RFIs) from data, b) detecting and providing accurate timing of pulsars and  c) converting the electrical signals received by SKA antennas into interferometric visibilities. At this point, visibilities will be transported from the telescope sites to big data centres in Perth and Cape Town, where it is the responsibility of the SDP consortium to build up the hardware platforms, software and algorithms needed to produce and distribute to the community scientific products ready for the analysis of the scientists (image cubes, catalogs of sources, time-domain spectral data, \dots). Typical processing powers of 50 and 100 Pflop will be requested for CSP and SDP activities, with an increasing rate of archival capacity of a few hundreds of TBytes/yr. 

\smallskip
\noi Importantly, consortia have to face the ``Big Data'' challenge of the SKA not only in terms of computing power and data rate, but also concerning energy consumption, with a proposed ceiling of about 5MW per telescope. Concerning the organisation of clever solutions for the crucial energy issue of the SKA project, a ``Power Supply Options Workgroup'' (PSOW) has been set up in 2016. This SKAO-managed advisory group is in charge of investigating energy supply (preferentially green) solutions, both in Australia and in South Africa. The members of this international team are directly interacting with industry in Member and Observer countries.

\smallskip
\noi To complete our panorama of the SKA WPs and consortia defined at the time of the RfP, the Telescope Manager, in charge of all hardware and software required to control the telescope and associated infrastructure, is particularly important for planning all SKA operations (from observations to data storage) and to evaluate their online performance.

\smallskip
\noi At the time of the RfP, the development of Phased Array Feeds (PAF) was part of the DSH WP. After the Design Baseline decision in 2015, which does not include the use of the PAF technology in Phase 1 of the SKA, a separate PAF consortium has been created for developing current designs of these receiver arrays placed at the focus of each parabolic dish, replacing single receiver elements employed by most current radio telescopes. This technology, which can enhance the telescope field of view by a factor of a few dozens, is currently used in the case of two pathfinder/precursor instruments: ASKAP (AUS) and APERTIF (NL). 

\smallskip
\noi Before concluding this quick overview of the huge technical work behind the construction of such an amazing ``machine'', we stress that, as in the case of PAF, the technical consortia work is largely based on the construction and operation of precursors and pathfinders instruments. The SKA is going to become a reality since, in the last years, through the huge progress in antenna design and wide bandwidth feeds, as well as in the ability to transport and process massive amounts of data, a new generation of radio telescopes is being built. 

\smallskip
\noi At low frequencies, instruments such as the Low-Frequency Array (LOFAR, in Europe, with its major extension NenuFAR, in France) and the Murchison Widefield Array (MWA, in Australia) employ several thousands of inexpensive dipole antennas arranged within stations without moving parts; the signals from these stations are digitised, transported to a central processor and combined to emulate a conventional dish antenna. These telescopes process the signal on sufficiently short time scales to correct for the ionospheric changes, severely affecting radio observations below a few hundreds MHz. At higher frequencies (from hundreds of MHz to a few GHz), two major arrays are being built, which are formed by tens of big ($\sim$12-15\,m diameter) dishes: ASKAP in Australia and MeerKAT in South Africa. In addition, several major radio facilities have been recently upgraded to improve their performance, including, for instance the Jansky Very Large Array (JVLA, in New Mexico), the Giant Meterwave Radio Telescope (GMRT, in India) and the Westerbork Synthesis Radio Telescope (WSRT, in the Netherlands). 

\smallskip
\noi The reader should keep in mind that SKA precursors are the three instruments that are located on the future SKA sites (MWA, ASKAP and MeerKAT), while the pathfinder status is given to radio telescopes engaged in SKA related technology and science studies (a census of these instruments is available at the \href{https://www.skatelescope.org/precursors-pathfinders-design-studies/}{\color{blue} \myul[blue] {SKA web page}}). 

\subsubsection{Beyond the SKA Observatory: the regional data centres}
\vspace{0cm}

\noi The data archive operated directly by the SKA Observatory will only have the capacity to manage a portion of the enormous scientific output from the SKA telescopes. In addition, within the current construction costs, no provision is made for the distribution of data to users, nor for computational facilities to enable data analyses beyond the SDP standard data products. 

\smallskip
\noi A Data Flow Advisory Panel (DFAP) was thus initiated by the SKA Board in July 2015 to advise SKAO on how to optimise the telescopes data flow. Based on DFAP recommendations, other research infrastructures are being organised all over the world, called ``Regional Data Centres'', which will include public and private organisations and will operate in agreement with the SKA Observatory but not under its control. In this case, ``regions'' are widely identified as large areas in the world.  At the moment, six Regional Centres are being planned, in Europe (see Sect.\,\ref{tech:interoperability} for more information), Canada, Africa, India, China and Australia. Each regional data centre will probably be a network of several nodes in different countries. 

\smallskip
\noi The centres will provide the necessary additional massive archiving capability, as well as the computational power to help researchers with the scientific analysis. 
In particular, three components have been identified by the DFAP that, being not provisioned within the current scope of the SKA Observatory, are the main responsibilities of regional data centres:

\begin{itemize}

\item computational capacity for re-processing and science analysis;
\item long-term storage capacity for archiving of standard SKA and derived data products;
\item local user support for post-processing and science analysis.
\end{itemize}

\subsection{Overview of the SKA situation in France}
\vspace{0cm}

\subsubsection{The organisation of the French community towards the SKA}
\vspace{0cm}

\noi Despite being amongst the founding partners in the first stages of SKA development (see Sect.\,\ref{PrepSKA}), for programmatic and financial reasons France did not join the group of countries which formed the SKA Organisation at the end of 2011. The SKA project is thus completing the pre-construction phase without France as a formal partner, which is nevertheless invited as an observer to the SKA Board meetings. 

\smallskip
\noi French scientific and technological implications in the project did not stop after 2011, assisted amongst others by the initiatives of the \href{https://as-ska-lofar.oca.eu}{\color{blue} \myul[blue] {Action Sp\'ecifique SKA-LOFAR}} (AS SKA-LOFAR), a structure set up in 2009 by CNRS-INSU to promote activities related to low-frequency radio astronomy in France. During 2014, the French community worked on the \href{http://www.insu.cnrs.fr/files/documentcomplet.pdf}{\color{blue} \myul[blue] {2015-2020 CNRS-INSU astronomical roadmap}} and an expression of the undeniable scientific and technological interest for the SKA emerged. As a result, in 2016 the CNRS-INSU, together with national institutes already involved in the SKA project (Paris and C\^ote d'Azur Observatories, and Bordeaux and Orl\'eans Universities), created the \href{http://ska-france.oca.eu/}{\color{blue} \myul[blue] {SKA France Coordination}}, in charge of organising the technological and scientific activities preparing a future French participation to the SKA project, within both research institutes and potentially interested industrial partners.

\smallskip
\noi The current organisation of SKA France includes:

\begin{itemize} 

\item a {\bf steering committee}, responsible for the strategic coordination of activities related to SKA between the various institutes involved in France;

\item a {\bf coordinator}, who chairs the SKA~France Advisory Committee (see below), coordinates the technical and industrial initiatives towards the SKA, facilitates the connections between the various French scientific groups working on the SKA project, reports to the steering committee on the status of the SKA and on the progress of activities of SKA France;

\item an {\bf advisory committee}, in charge of facilitating the activities and communication between the various scientific groups working in France on the SKA. 

\end{itemize}

\smallskip
\noi The establishment of SKA France as a national project is intended to complement the work done by the AS SKA-LOFAR, which has a mandate to encourage scientific involvement in all current radio astronomical facilities. SKA France has the specific responsibility to prepare and coordinate the French contributions to the international SKA project in the period before the SKA construction phase. This includes the communication of technical, strategic and political information amongst all the SKA participants in France. 

\subsubsection{French participation to the SKA preparation}
\vspace{0cm}

\noi Several institutes in France are involved at various levels in the design and pre-construction phase of the SKA, with engineers and researchers participating in five of the eleven technical design consortia and in all the science working groups (SWG) established by the SKA Organisation. 

\smallskip
\noi The ramping scientific interest in the SKA exploitation is demonstrated by the fact that, while at the end of 2016 approximately 30 French researchers were officially participating to the SKA SWG, today about 170 astrophysicists, from more than 40 different research institutes, are co-authors of this Volume. Their fields of interest cover the whole topics of the international SWG, going from the far Universe through cosmology and extra-galactic astronomy, to the study of our own Galaxy, of its stars and planets, without forgetting the investigation of the transient universe and the fascinating and extreme objects that allow fundamental physics studies (Sect.\,\ref{science}). 

\smallskip
\noi Some of the contributions of this White Book reflects the yearly implication of French researchers in the international definition of the key scientific performances of the SKA: sixteen French researchers were authors or co-authors of more than 30 out of the 135 chapters of the AASKA14 SKA Science Book published in 2015, with five of them being in the core-teams of SWG. Fran\c{c}oise Combes (Observatoire de Paris and Coll\`ege de France) has recently been nominated co-chair of the Extragalactic spectral line SKA Science Working Group (SWG), together with Robert Beswick (Manchester University). This group was created after the 2014/2015 SKA re-baselining related activities in order to address the SKA capability for galaxy evolution studies through (non-\hi) extragalactic spectral lines, a topic of great interest for the French community, in particular due to the synergy with higher-frequency radio telescopes. A French researcher has also served as chair for one of the \href{https://astronomers.skatelescope.org/science-assessment-teams/}{\color{blue} \myul[blue] {Science Assessment Teams}} organised during the recent Cost Control Project in spring 2017.

\smallskip
\noi ``Synergy'' has been the keyword that has allowed to make the interest of French astronomers exploding in the last months. Through seminars, workshops and discussions organised in France, researchers non-expert of radioastronomical techniques have realised that this will not be a handicap at all in the SKA-era, since images, spectra and catalogs will be provided to the whole community that wants to derive physical informations about the sky from radio photons. It is this aspect that, together with the fact that the SKA will nicely complement most of the main present and future astronomical projects, has significantly increased the number of French astronomers interested to fully exploit its observations. To be noted that, in addition to the co-authors of this White Book, an expression of interest for the SKA has been signed by about one hundred researchers, who, even if not yet ready to contribute to this Volume, are conscious of the expected impact of this project for their future work. Their names are listed at the end of the White Book. 

\smallskip
\noi Next sections will allow the reader to discover interesting and sometimes innovative ideas related to the four fields on which the SKA will be a {\em unique} instrument (i.e. Epoch of Reionisation, \hi~studies, cosmic magnetic fields and fundamental physics with pulsars), but also a very rich overview of what the SKA will be able to do together with other projects in which France is deeply involved, both space missions and telescopes on Earth, covering the whole electromagnetic spectrum. This is clearly the main object of the very rich ``Scientific synergies with other instruments'' chapter (Sect.\,\ref{science:synergies}), but it emerges too from some topics on which the French community has particularly innovative ideas compared to international SKA partners, such as on the study of the Interstellar Medium (Sect.\,\ref{science:galaxy}).

\smallskip
\noi The technical participation of France to the SKA preparation is also of great value and described in this Volume (Sect.\,\ref{technology}). The French community is actively involved on pathfinders and prototypes of the world's largest project in radio astronomy for the 21st century, namely NenuFAR and EMBRACE (Sect.\,\ref{tech:pp}).

\smallskip
\noi As detailed in Sect.\,\ref{tech:preconstr}, several French institutes are officially involved in five of the eleven SKA design consortia (Sect.\,\ref{intro:consortia}). The Nan\c{c}ay Radio Observatory ({\it Unit\'e Scientifique de Nan\c{c}ay}, USN, a department of Paris Observatory) has a leading role in the design of Application Specific Integrated Circuits (ASIC) for the conception of dense aperture arrays within the Advanced Instrumentation WP MFAA and is thus part of the AAMID consortium. Within the Advanced Instrumentation programs, it is important to mention the participation of the {\it Laboratoire d'Astrophysique de Bordeaux} (LAB) on the design of wideband receivers (4.6-24\,GHz). Based on its expertise developed on the ALMA project, LAB has also been solicited to work on Analog-to-Digital conversion technology for the DSH consortium, in particular for the highest-frequency band of SKA1 (Band 5). Two French institutes, Paris and C\^ote d'Azur Observatories, are involved in the AADC consortium. Paris Observatory, through the USN expertise, has made several development proposals for  realisation of Low Noise Delay to decrease the power consumption, with the same performances. {\it Observatoire de la C\^ote d'Azur} works on the development of algorithms for image reconstruction. The signal processing expertise of the French community is of great interest for the SKA data challenge. At the moment, Paris Observatory is involved in the SDP consortium for the development of algorithms for the mitigation of Radio Frequency Interferences, imaging and calibration. 

\smallskip
\noi It is important to stress that the French activity related to the optimisation of data processing is ramping up, both at academic and industry level in France, through a joint approach coordinated by SKA France (see Sect.\,\ref{texh:TdS}). A major achievement of the SKA France coordination has indeed been the interest raised in big companies and small/medium size enterprises. As developed in Sect.\,\ref{industry}, the SKA~France coordination has performed a joint comprehensive analysis of both the expertise available in France and the needs of the SKA project not yet covered by the partnering countries. Based upon this analysis, an ensemble of workshops grouping industry and academic experts has led to the identification of four key axes of interest for French companies, including high-frequency digitalisation technologies and HPC, signal processing, renewable energy production, distribution and storage, and system engineering. This joint public-private research approach has recently lead to discuss the evolution of SKA France towards a contractual structure, called ``Maison SKA France'', which is described in Sect.\,\ref{conclusions}.

\smallskip
\noi France participates to several other key fields of the SKA project, through the implications of internationally recognised experts. The French SKA Industry Liaison Officer (ILO), Gabriel Marquette (General Manager of \href{http://www.eurogia.com}{\color{blue} \myul[blue] {``EUROGIA''}}, a Cluster for Low carbon energy technologies developed within the \href{http://www.eurekanetwork.org}{\color{blue} \myul[blue] {``EUREKA''}} IGO that groups 44 countries), is one of the eight members of the SKA Power Solutions Options Workgroup. France is also participating through CNRS-INSU to the organisation of the European SKA data centre within the H2020 AENEAS project (Sect.\,\ref{tech:interoperability}), and, despite the fact of being a non-member partner\footnote{The institutes or organisations participating to AENEAS are considered {\em full member} or {\em non-member partners} if they are from countries that are current SKA members or non-members of the SKA Organisation, respectively. In the first case they lead the AENEAS work-packages and are responsible for their delivery. In the second case they are keen to develop their technological experience and may host a node of the European data centre. These partners are funded at a lower level than the full members.}, some key roles in the project are played by French researchers: C. Ferrari, SKA France coordinator, is the elected chair of the AENEAS General Assembly, and J.-P. Vilotte, coordinator of HPC activities within CNRS-INSU, is one of the members of the External Advisory committee of AENEAS, together with Ian Bird (LHC Computing Grid Project Leader at CERN) and Martin Zwaan (Head of the ALMA Regional Centre at ESO).

\smallskip
\noi We now leave the reader to discover more details about the wide scientific and technological interests of the French community in the SKA project that we have shortly summarised above. \\

\parbox{0.9\textwidth}{
\noi{References:}\\
\noi{\scriptsize  
Beswick, R.~J., \etal, 2015, AASKA14, 70;
Braun, R. \etal\ 2015, ``Proceedings of Advancing Astrophysics with the Square Kilometre Array'' (AASKA14), 9-13 June, 2014, Giardini Naxos, Italy, PoS, id.174
Carilli, C. \& Rawlings, S., 2004, Science with the Square Kilometer Array, New Astronomy Reviews, 48 (Nos 11-12), 979 (Elsevier, Amsterdam), eds. C. Carilli \& S. Rawlings;
Condon, J.~J. \& Ransom, S.~M., 2016, Essential Radio Astronomy, Princeton University Press;
Duc, P.-A. \& Renaud, F., 2013, LNP, 861, 327;
Ekers, R., 2012, in Proceedings of the meeting "Resolving The Sky - Radio Interferometry: Past, Present and Future". April 17-20,2012 Manchester, UK. Published online at \url{http://pos.sissa.it/cgi-bin/reader/conf.cgi?confid=163}, id.7;
Fender, R., \etal, 2015, AASKA14, 51;
Hoare, M., \etal, 2015, AASKA14, 115;
Janssen, G., \etal,  2015, AASKA14, 37;
Johnston-Hollitt, M., \etal, 2015, AASKA14, 92;
Koopmans, L. \etal, 2015, AASKA14, 1;
Kramer, M. \& Stappers, B., 2015, AASKA14, 36;
LIGO Scientific Collaboration \& Virgo Collaboration, 2016, PRL, 116, 061102;
Prandoni, I. \& Seymour, N., 2015, AASKA14, 67;
Maartens, R., \etal, 2015, AASKA14, 16;
Nakariakov, V., \etal, 2015, AASKA14, 169;
Staveley-Smith, L. \& Oosterloo, T., 2015, AASKA14, 167;
Taylor, A.~R. \& Braun, R., 1999, Science with the Square Kilometer Array, eds. A. R. Taylor \& R. Braun;
Umana, G., \etal, 2015, AASKA14, 118;
Vazza, F., \etal, 2015, A\&A, 580, 119
}}

\newpage

\section{Science}\label{science}

\subsection{Early Universe, cosmology and large scale structures}

\smallskip

\noindent {\normalsize Contributors of this section in alphabetic order: }

\smallskip

\noi {\sffamily \scriptsize
{\sffamily\bf R.~Ansari} [\lal],
{\bf D.~Aubert} [\stras],
{\bf M.~Bucher} [\apc],
{\bf J.~Cohen-Tanugi} [\lupm],
{\bf F.~Combes} [\colfr],
{\bf H.~Courtois} [\ucb],
{\bf C.~Ferrari} [\lagrange],
{\bf K.~Ferri\`ere} [\irap],
{\bf G.~Lagache} [\lam],
{\bf M.~Langer} [\ias],
{\bf G.~Lavaux} [\iapsorb],
{\bf M.~Limousin} [\lam],
{\bf C.~Magneville} [\irfu],
{\bf P.~Noterdaeme} [\iapsorb],
{\bf P.~Ocvirk} [\stras],
{\bf M.~Pandey-Pommier} [\cral],
{\bf C.~Peroux} [\lam],
{\bf J.~Richard} [\cral],
{\bf F.~Rincon} [\irap],
{\bf C.~Schimd} [\lam],
{\bf B.~Semelin} [\lermasorb],
{\bf S. Torchinsky} [\apc]
}

\subsubsection{Constraining astrophysical models of the Cosmic Dawn and the Epoch of Reionisation with the 21\,cm signal}
\vspace{0cm}



\noi 
The reionisation of the universe starts with the formation of the first stars and ends around $z=6$ (as revealed by the observation of the Gun-Peterson trough in Quasar spectra, e.g. Fan et al. 2006). During this period, bubbles of ionised intergalactic medium (IGM) grow around primordial galaxies until they overlap. All the while, regions of the IGM that have not been reionised yet emit in the 21\,cm line, tracing a distinctive pattern on the sky. The intensity of the signal against the background of the CMB is:

$$
\delta T_b \,=\, 27. (1-x_{\mathrm{H_{II}}}) \,(1+\delta)  \left( T_S - T_{\mathrm{CMB}} \over T_S \right) \left( 1 + {1 \over H(z)} {d v_{||} \over dr_{||}} \right)^{-1}   \,\left( {1+z \over 10} \right)^{1 \over 2}  \mathrm{mK}
$$
\noi 
where $x_{\mathrm{H_{II}}}$ is the local ionised fraction of hydrogen (fluctuations on scales smaller than $1$ kpc are smoothed by a convolution with the line profile along the line of sight), $\delta$ is the overdensity of the gas, $T_S$ is the local spin temperature of hydrogen, $T_{\mathrm{CMB}}$ the CMB temperature at that redshift, $H(z)$ the Hubble parameter, $d v_{||} \over dr_{||}$ the velocity gradient along the line of sight and $z$ the redshift. The $27$ mK typical amplitude depends on the cosmology, see e.g. Mellema (2013) for more details. Astrophysical processes mainly have an impact on $x_{\mathrm{H_{II}}}$ and $T_S$. A loose definition of the Epoch of Reionisation (EoR) vs Cosmic Dawn (CD) is whether the fluctuations of the signal are dominated by the fluctuations of $x_{\mathrm{H_{II}}}$ (late) or the fluctuations of $T_S$ (early on). The redshift of the transition between these two periods is very uncertain based on current observations, but likely lies between $8$ and $12$.
The SKA is expected to measure the signal between $50$ Hz ($z \sim 27$) and $200$ Hz ($z \sim 6$) statistically through the power spectrum of the fluctuations but also by building a full tomography with a typical sensitivity of 1 mK at a resolution of $5^\prime$. Koopmans et al. (2015) discuss in detail the impact of the SKA design on the expected capability to detect the signal. In this contribution we focus on some of the physical processes that determine the strength of the signal and how to constrain these processes using the upcoming observations. 

\smallskip

\noi 
{\bf Three parameters regulating the signal during the Cosmic Dawn}

\smallskip

\noi 
During the CD, the strength of the signal depends mainly on the spin temperature, that is itself determined mainly by the local temperature of the gas and the flux in the Lyman-alpha line. Through the Wouthuysen-Field mechanism, the Lyman-$\alpha$ photons couple $T_S$ to the kinetic temperature of the gas, pulling it away from $ T_{\mathrm{CMB}}$ (see, e.g., Furlanetto et al. 2006).
However, since cosmological distances are involved between the sources of the Lyman-$\alpha$ photons and the location where the 21\,cm signal is emitted in the IGM, these Lyman-$\alpha$ photons were actually emitted in the whole band between Lyman-$\alpha$ and the Lyman limit and underwent cosmological redshifting. Thus the local strength of the coupling is decided by the  source formation efficiency (fraction of  halo mass converted into sources within a Hubble time) and the sources emissivity in the full Lyman band.
The source formation efficiency is the main parameter that connects the growth of structures dictated by cosmology and the actual production of photons. This fundamental parameter is crucial both during the CD and the EoR. On the other hand, the emissivity in the Lyman band is relevant only during the CD. Later on, the Lyman-$\alpha$ flux becomes large enough so that $T_S=T_K$ and $\delta T_b$ becomes insensitive to the exact value of the local Lyman-$\alpha$ flux.
We explored the influence of actually including the effect of the Lyman-$\alpha$ coupling with detailed Lyman lines radiative transfer simulations in several works (Baek et al. 2009; Baek et al. 2010; Vonlanthen et al. 2011).
Once the transfer is performed correctly in a model where the source formation history is fixed, only the ratio $f_\alpha$ (see Semelin et al. 2017)  between the emissivity in the Lyman band and the emissivity above the ionisation threshold is still a free parameter. $f_\alpha$ is an effective parameter that depends in the initial mass function of the stellar populations. It can hopefully be constrained by the observations (see next paragraph).

\smallskip
\noi
We mentioned that $T_S$ is driven to $T_K$ by the Wouthuysen-Field coupling. But what determines the local value of $T_K$? If the local evolution of $T_K$ was the result of a simple adiabatic evolution, it would be fully determined by the cosmology and initial conditions, without the need of additional astrophysical parameters.
However, as soon as the first sources are formed, there is the possibility that X-ray are emitted (by X-ray binaries, SNs, AGNs) and travel far into the neutral IGM where they heat up the gas. Thus a parameterisation of the production of X-rays is required. A simple parameter $f_X$ has been widely used. It quantifies how much energy is emitted in the form of X-ray per unit of stellar mass formed (see Furlanetto et al. 2006).
However, X-rays can be emitted in a large range of energies and their mean-free path in the neutral IGM is proportional to $E^3$. X-rays with energies of several KeV actually have little chance to interact with the IGM before the end of reionisation. Thus the spectral properties of the X-ray emission is important in determining its heating efficiency (Fialkov et al. 2014). In Semelin et al. (2017) we define $r_{H/S}$, the ratio of the energy emitted as Hard X-rays by X-ray binaries (using the template spectra of Fragos et al. 2013) to the energy emitted as Soft X-rays by QSOs.
We show that varying this ratio between $0$ and $1$ is similar, in terms of effective heating, to decreasing $f_X$ by a factor of 3. This however applies to the spatially averaged heating. If we increase $r_{H/S}$ and $f_X$ to keep the average heating similar, the spatial fluctuations of the heating decrease in amplitude. Thus $f_X$ and $r_{H/S}$ cannot be collapsed in a single parameter when we study the 21\,cm signal fluctuations.
Thus $f_\alpha$, $f_X$ and $r_{H/S}$ form a basic 3-parameter set that regulates the 21\,cm signal during the CD, {\sl for a fixed source formation history}. The ability of SKA1 to constrain these parameters characterising CD is still under investigation: preliminary results can be found in Semelin et al. (2017). Due to its large low frequency coverage, SKA1 is however the only instrument likely to yield constraints. More detailed prediction are available for other parameters characterising the EoR (e.g. Grieg et al. 2015; Grieg et al. 2017; Shimabukuro and Semelin 2017).
The modelling could be improved further by including more detailed physics: for example adding a parameter to describe the efficiency of H$_2$ formation in primordial gas.

\smallskip
\noi 
{\bf Machine Learning as a method to derive constraints on the model parameters}

\begin{figure}[!t]
\begin{subfigure}{.66\textwidth}
  \centering
  \includegraphics[width=0.95\linewidth]{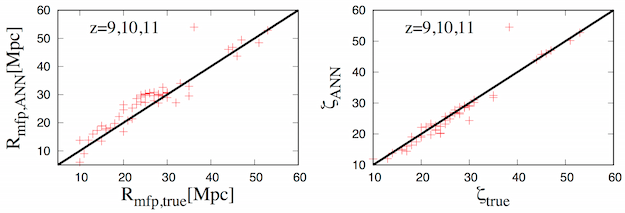}
\end{subfigure}
\begin{subfigure}{.32\textwidth}
  \centering
  \includegraphics[width=0.95\linewidth]{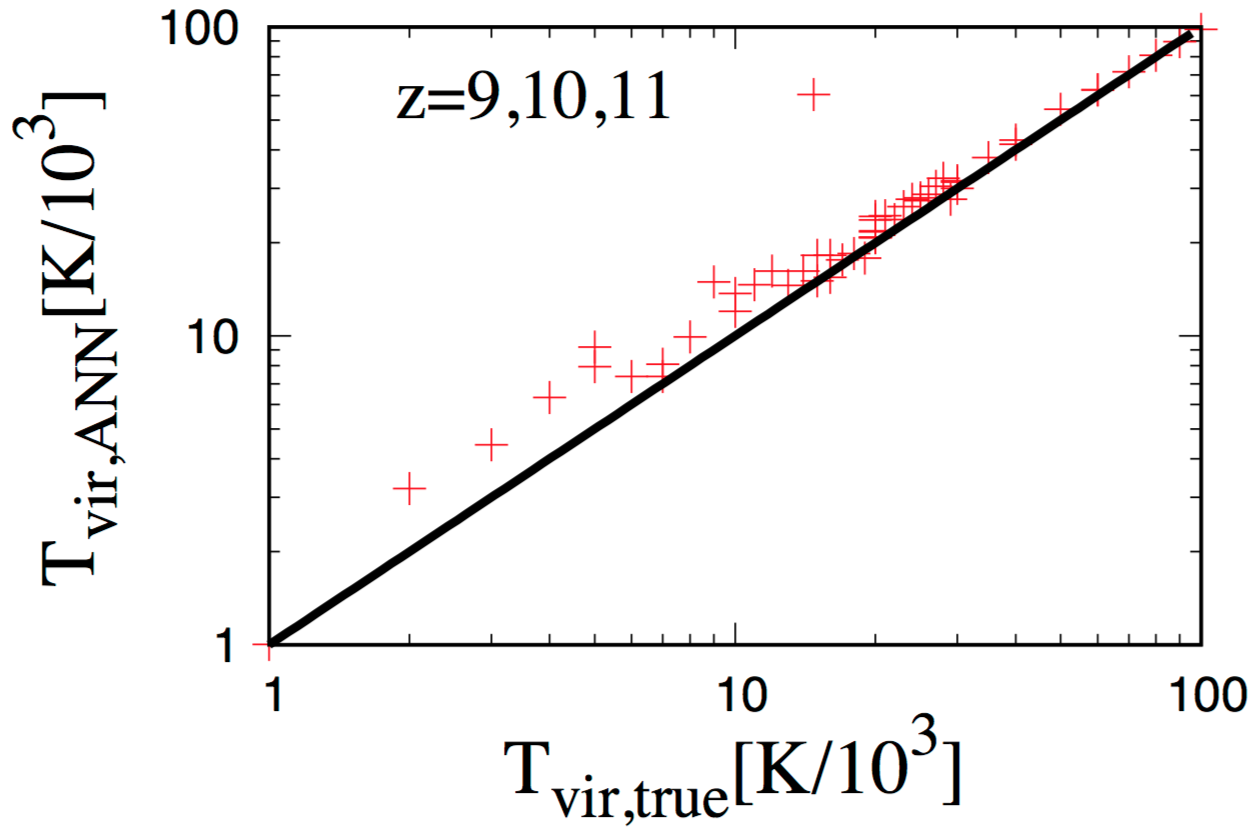}
\end{subfigure}
  \caption{\label{fig:EoR_CD} Values for 3 parameters reconstructed by a trained Neural Network. R$_{\mathrm{mfp}}$ is the mean free path of ionising photons in ionised regions due to unresolved Lyman limit systems, $\zeta$ is the ionising photon production efficiency and T$_{\mathrm{vir}}$ is the minimum virial temperature of halos producing photons. The NN uses the power spectrum at redshifts 9, 10 and 11 as an input and the effect of SKA-like thermal noise and of sample variance are included. From Shimabukuro and Semelin (2017).}
\end{figure}

\smallskip
\noi 
Semi-numerical and full numerical simulations predicting the 21\,cm signal have been used mainly to explore the range of possible signals. This has been an important step in formulating scientific requirements for the SKA EoR project and adjust the instrument design accordingly. In the future however, the main task of both kind of simulations will be to extract astrophysical information from the observed signal.
This will concretely mean putting constraints on astrophysical parameters such as defined in the previous section, in the same way the CMB was used to put constraints on the cosmological parameters. Preliminary work has been done in this direction. Greig and Mesinger (2015, 2017) have developed a MCMC method using the 21\,cm power spectrum as a metric and a fast semi-numerical code to sample the parameter space.
The advantage of the method is that it relies on a well established Bayesian formalism and provides local confidence contours in the parameter space. The limitation is that it has to use fast semi-numerical simulations because it requires generating the signal for a large number ($\sim 10^5$) of points in the parameter space. An alternative method is to use Neural Networks (NN) to either, in a forward modelling, predict the power spectrum for a given set of parameter (work in progress, J. Pritchard, private communication) or, in backward modelling, predict the values of the parameters for an observed power spectrum (Shimabukuro and Semelin, 2017).
The parameters used in Shimabukuro and Semelin (2017) characterise the EoR rather than the CD: they are the same as in Greig and Mesinger works to allow comparison of the efficiency of the two methods. We have shown that using a learning sample limited to a few tens of points in the chosen 3-parameter space we can train a simple NN to reconstruct the parameters values from the power spectrum (see Fig.\,\ref{fig:EoR_CD}) with an accuracy only slightly worse than with the MCMC method (but with a factor of a thousand less computing time).  This makes it possible to use full simulations to build the training sample and thus, hopefully, benefit from a more robust modelling. Both approaches and probably new ones will most likely be the subject of intense investigation in the years to come.\\

\parbox{0.9\textwidth}{
\noi{References:}\\
\noi{\scriptsize 
Baek S. et al., 2009, A\&A, 495, 389;
Baek S. et al., 2010, A\&A, 523, 4;
Fan H. X. et al. 2006, ARAA, 44, 415;
Fialkov A. et al., 2014, Nature, 506, 197;
Fragos T. et al., 2013, ApJ, 776, 31;
Furlanetto S. et al., 2006, Phys Rep, 433, 181
Greig B. and Mesinger A., 2015, MNRAS, 449, 4246;
Greig B. and Mesinger A., 2017, arXiv170503471;
Koopmans L. et al., 2015, AASKA14, 1;
Mellema G. et al., 2013, Experimental Astronomy, 36, 235;
Semelin B. et al., 2017, in prep.;
Shimabukuro H. and Semelin B., 2017, MNRAS, 468, 3869;
Vonlanthen P. et al., 2011, A\&A, 532, 97
}}\\

\subsubsection{21\,cm signal from the EoR}
\vspace{0cm}

\begin{figure}[!ht]
  \centering
  \includegraphics[width=0.75\linewidth]{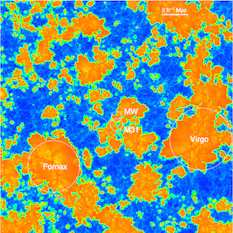}
  \caption{Temperature distribution in a 45 $h{_{70}}^{-1}$\,Mpc x 45 $h{_{70}}^{-1}$\,Mpc x 0.03 $h{_{70}}^{-1}$\,Mpc slice of the CODA simulation at half-reionisation (Ocvirk et al. 2016). Orange regions show photo-heated, ionised material, while the cold, still neutral medium appears in blue.}
  \label{fig:CODA}
\end{figure}

\smallskip

\noi Once the first sources of UV radiation appear in the Universe, the cosmic hydrogen experiences a global transition known as the Epoch of Reionisation (EoR). During this transformation a network of hot and ionised 'bubbles' is established (see Fig.\,\ref{fig:CODA}), with a topology dictated by the distribution of large scale structures and a redshift evolution set by the photons production history of the first galaxies and AGNs. Growing \hii~regions eventually merge, resulting in a fully ionised inter-galactic medium. This scenario is supported by the evolution of the IGM opacity as measured in quasar spectra (see e.g. Fan et al 2006; Becker et al. 2015) with a full completion of hydrogen reionisation by $z\sim 6$. Planck latest estimates of Thompson optical depth also favour a late Reionisation with typical half-reionisation redshifts $z\sim 8$ (Planck Collaboration 2016). This timing is also supported by less conclusive probes such as the reconstruction of star formation histories in dwarf galaxies (Brown et al. 2014) or the rapid evolution of Ly$\alpha$ emitters (Konno et al. 2014) at $z=7$. SKA will offer first insights on the transition through direct 3D (2D+time) tomography of the 21\,cm signal (Madau et al. 1997). 

\smallskip

\noi These data will be used to investigate the connection between the sources (first galaxies and AGNs) and the IGM. For instance the evolving geometries of a Reionisation powered by a few bright galaxies or abundant fainter ones are expected to be different (see e.g. McQuinn et al. 2007; Chardin et al. 2012).  Imaging will also provide a better view of the actual Reionisation history experienced by biased regions (and therefore galaxies): structures tend to be reionised earlier than the Universe as a whole and often on large durations (up to $\sim 200$ Myrs, see e.g. Li et al 2014).  The details of these timings could be important to investigate the suppression of star formation in subhaloes by radiation (Ocvirk \& Aubert 2011; Ocvirk et al. 2014).

\smallskip

\noi Likewise powerful AGNs are expected to create large \hii~regions  with radii of 10s of Mpc (see e.g. Datta et al. 2012, 2016): detections with SKA1-LOW should be possible with 1000 hrs of observation and will provide constraints on AGNs or IGM high-z properties. Such prospects open the possibility of linking the population of \hii~regions with e.g. the luminosity functions (LF) of galaxies and AGNs : first constraints on LFs are already available (e.g. Bouwens et al. 2015; Giallongo et al. 2015) and will be further refined with future experiments such as JWST or ATHENA. SKA will provide an alternate view of the first sources by putting constraints on their influence on their environment : as such, SKA will be essential component to deepen our understanding of structure formation in its earliest stages. It would also provide key insights on the relative contribution of star forming galaxies and quasars to the Reionisation of the Universe.

\smallskip

\noi Remains the question of the best way to quantify the statistical properties of the Reionisation process from 21\,cm maps. For instance, a large number of studies focus on the well-established power-spectrum estimator (e.g. Santos et al. 2008; Baek et al. 2010) : the frequency structure of the 21\,cm signal is shown to be quite sensitive to the nature of the dominant source of UV photons and well within the capability of SKA1.  Furthermore, power spectra measurements are within the capabilities of some SKA precursors. However, the power spectrum is well suited to gaussian signals whereas the structure of the ionised gas is expected to be highly non-gaussian at all scales (see e.g. Mellema et al. 2006) : 21\,cm maps with the same 2-pt statistics can present quite different features (see e.g. Mellema et al. 2015).  As a consequence, power spectra are unlikely to be sufficient to break degeneracies between different reionisation scenarios : imaging and 3D tomography, only achievable with SKA, appear as necessary. Maps could be analysed using techniques such as the distribution of \hii~regions sizes (e.g. Zahn et al. 2007; Iliev et al. 2006; Chardin et al. 2012) or the calculation of the Euler-characteristics (Friedrich et al. 2011) : such estimators were shown to relate to the properties of the underlying sources.

\smallskip

\noi Finally, 21\,cm imaging could also be used to target specific objects, such as luminous quasars. Properties such as the mass or the age of these objects or even the anisotropy of their emission could be obtained (e.g. Wyithe 2008; Datta et al. 2012 or Mellema et al. 2006), providing valuable informations on the formation and the growth of supermassive blackholes at these early times. 
In principle, the same kind of investigations could be conducted with galactic sources. Taking Koopmans et al. (2015) as a reference, detection of ionised bubbles with radii in the 2-10 comoving Mpc range could be envisioned at z$\sim9 $ with SKA1-LOW. According to studies based on semi-analytic and radiative hydrodynamics simulations, such sizes are expected for masses typical of progenitors of groups of galaxies and clusters. The \textit{average} corresponding masses are similar to $10^{12+}$ M$_\odot$ during the Reionisation, leading to $10^{13-14}$\,M$_ \odot$  objects at $z=0$ (e.g. Alvarez et al. 2009; Chardin et al. 2014).\\

\parbox{0.9\textwidth}{
\noi{References:}\\
\noi{\scriptsize
Alvarez, M., et al., 2009, ApJL, 703, 167;
Baek, S., et al., 2010, A\& A, 523, A4;
Becker, G. D., et al., 2015, MNRAS, 447, 3402;
Bouwens, R. J., et al., 2015, ApJ, 803;
Brown, T. M., et al., 2014, ApJ, 796, 91;
Chardin, J., et al., A\& A, 2012, 548, 9;
Chardin, J. et al., A\& A, 2014, 568, 52;
Datta, K. K., et al., 2012, MNRAS, 424, 762;
Fan X., et al., 2006, AJ, 132, 117; 
Friedrich, M. M., et al., 2011, MNRAS, 413, 1353;
Giallongo, A., et al., 2015, A\&A, 578, 83;
Iliev, I. T., et al. MNRAS, 371, 2006;
Konno, A., et al. 2014, ApJ, 797, 16;
Li, T., et al., 2014, 785, 134;
Lidz A., et al., 2008, ApJ, 680, 962;
Madau P., et al., 1997, ApJ, 475, 429;
Mellema G., et al., 2006, MNRAS, 372, 679;
Mellema, G., et al., 2015, AASKA14, 10; 
Ocvirk, P. \& Aubert, D., MNRAS, 2011, 417, 93;
Ocvirk, P., et al., ApJ, 2014, 794;
Planck Collaboration, 2016,	A\& A, 596, 108;
Santos, M. G., et al., 2008, ApJ, 689, 1;
Wyithe, J. S. B., 2008, MNRAS, 387, 469;
Zahn, O., et al., 2007, ApJ, 654, 12
}}\\

\subsubsection{Cross-correlating cosmic fields in the EoR}
\vspace{0cm}

\noi SKA1 will deliver \hi~intensity maps over a broad range of frequencies (and thus redshifts) and a substantial fraction of the sky.
Cross-correlating \hi~intensity maps with galaxy surveys, Cosmic Backgrounds (CMB, near- and far-IR backgrounds), or line surveys (Ly$_{\alpha}$, CO or CII) can provide, e.g., high precision clustering measurements, cosmological constraints (e.g.\ on the nature of primordial fluctuations), the statistical measure of average reionisation bubble size and the HI content of early galaxies.
Cross-correlation studies are advantageous since the measurable statistics do not suffer in the same way from foregrounds and systematic effects as auto-correlation function measurements. What can be learnt from cross-correlations in the EoR is potentially huge, and we evoke hereafter only a few examples in which the French community intends to play a leading role.

\smallskip

\noi {\bf The cross-correlation with CMB temperature anisotropies} 

\smallskip

\noi Through Thomson scattering, the CMB is sensitive to the amount of free electrons in the Universe since Recombination. On the contrary, 21\,cm line fluctuations trace the evolution of the amount of neutral hydrogen. Both observables are thus natural probes of reionisation. Complementary information can be in principle extracted from their cross-correlation. Several authors have studied the cross-correlation of CMB temperature anisotropies and cosmological 21\,cm line fluctuations both on large (e.g.\ Alvarez et al. 2006) and small (e.g.\ Cooray, 2004) angular scales (see Tashiro et al. 2010, for additional references). On large scales, the 21\,cm line fluctuations correlate with the Doppler CMB anisotropies. For a given redshift of reionisation $z_\mathrm{re}$ (corresponding to an ionised fraction of $50 \%$), the resulting cross-power spectrum is sensitive to the duration of reionisation $\Delta z$. Adapting the results of Tashiro et al. (2010) to the current specification from the SKA1 System Baseline Description documents, this signal could be detectable with a signal-to-noise ratio of $\sim 2$ (for a total observing area of $1000\,\mathrm{deg}^2$), or even up to $\sim 6$ (for $10000\,\mathrm{deg}^2$), given the limits on the duration of reionisation provided by the Planck Collaboration (2016b; note that the relation between the $\Delta z$ in Tashiro et al. 2010, and the $\Delta z$ defined in the Planck Collaboration paper is not trivial).

\smallskip

\begin{figure}[!ht]
 \centering
 \includegraphics[width=0.5\linewidth]{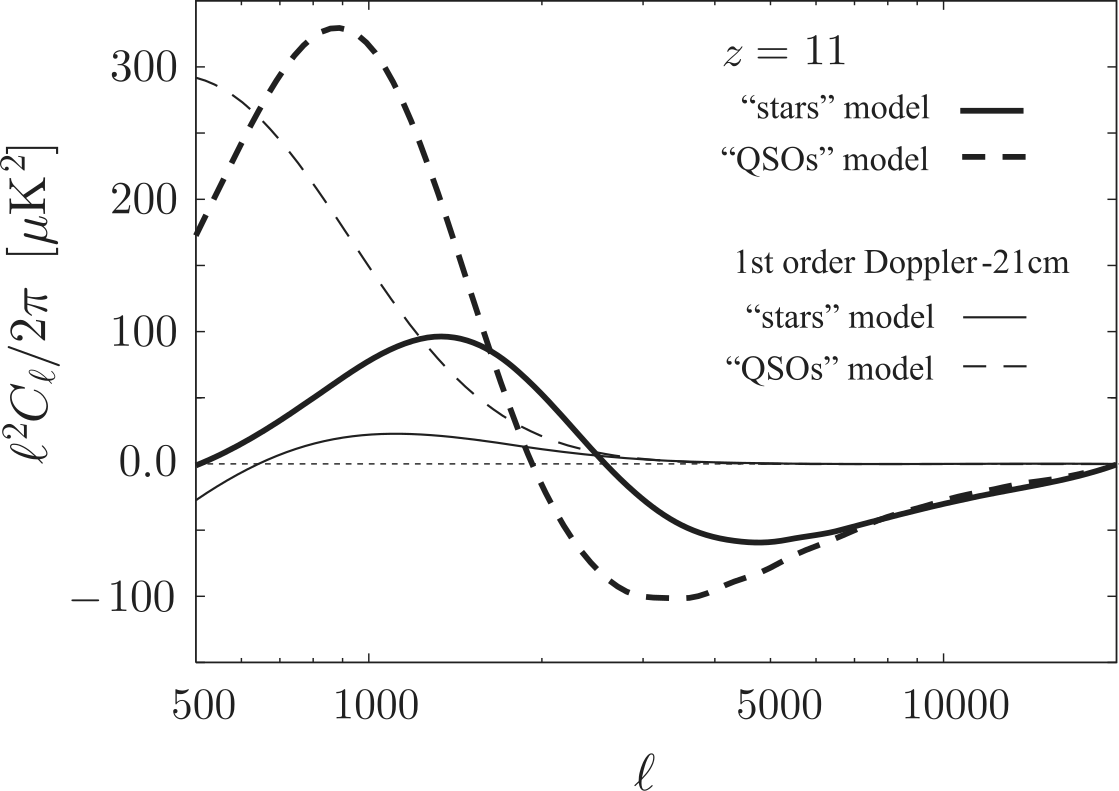}
 \caption{\label{fig:totocross} Dependence of the cross-correlation of the CMB kSZ signal with the 21\,cm fluctuations on the sources of reionisation. Thick solid and dashed lines correspond to star- or quasar-driven reionisation respectively. Thin lines corresponding to the CMB Doppler-21\,cm cross-correlation are given for reference. Extracted from Tashiro et al. (2011) who assumed $z_\mathrm{re} = 11$.}
\end{figure}

\noi On the other hand, before reionisation is complete, small scale secondary CMB temperature anisotropies are produced by the Sunyaev-Zeldovich (SZ) effect due to the inverse Compton scattering off the free electrons contained in the cosmological \hii~regions surrounding sources of ionising photons (Aghanim et al. 1996). The kinetic SZ (kSZ) effect, due to bulk motions of those electrons, is dominant over the thermal effect during reionisation. The cross-correlation of the kSZ with the 21\,cm line fluctuations produces characteristic shapes in the cross-power spectrum that may in principle be used to extract information on the size of the \hii~regions, and therefore on the nature of the ionising sources (see Fig.\,\ref{fig:totocross} for illustration). However, depending on the dominant ionising sources and on $z_\mathrm{re}$, the detection of this signal will be difficult, even with the SKA (Tashiro et al. 2011).

\smallskip

\noi {\bf The cross-correlation with CMB polarisation}

\smallskip

\noi The precise measurement of the CMB E-mode polarisation by the Planck-HFI has allowed the Planck Collaboration (2016a,b) to reduce significantly the uncertainty on the redshift of reionisation and to show that $z_\mathrm{re}$ is substantially lower than previously thought. The CMB E-mode polarisation, as a powerful probe of reionisation, can also be cross-correlated with the 21\,cm line fluctuations. As Tashiro et al. (2008) have shown, both the amplitude and the shape of the cross-power spectrum are sensitive to $z_\mathrm{re}$ and $\Delta z$, and might also be used to test for more exotic reionisation scenarios. It is interesting to note that the expected signal-to-noise ratio of this signal is essentially independent of the duration $\Delta z$ (Tashiro et al. 2010).

\smallskip

\noi {\bf The cross-correlation with galaxy surveys}

\smallskip

\noi Lidz et al. (2009) and Wiersma et al. (2013) have undertaken theoretical studies of the cross-correlation between 21\,cm measurements and galaxies detected with high-redshift surveys, in particular Ly$-\alpha$ emitters. This cross-correlation promises to be an excellent tool for inferring the topology of reionisation. On large scales, the 21\,cm and galaxy fields are anti-correlated during most of the EoR. However, the two fields become roughly uncorrelated on scales smaller than the size of the \hii~regions around detectable galaxies. Hence, the 21\,cm galaxy cross-power spectrum provides a tracer of bubble growth during reionisation, with the signal turning over on progressively larger scales as reionisation proceeds. Measuring the turnover scale as a function of galaxy luminosity constrains the characteristic bubble size around galaxies of different luminosities (Lidz et al. 2009). At lower redshifts, the cross-correlation of 21\,cm intensity maps with galaxy surveys is a measure of the \hi~content of galaxies and comoving \hi~density, $\Omega_{HI}$ - if the \hi~bias, b$_{HI}$, can be estimated externally (Wolz et al. 2016; Masui et al. 2013).

\smallskip

\noi {\bf The cross-correlation with cosmic near- and far-IR backgrounds}

\smallskip

\noi The lower value of the optical depth $\tau$ of Thomson scattering reported by the Planck collaboration (2016a,b) strengthens the likelihood that early star-forming galaxies dominated the reionisation process (e.g.\ Bouwens et al. 2015; Robertson et al. 2015). Consequently, any attempt at understanding reionisation is paired with an understanding of primeval star formation. In the current view, a very steep faint-end slope of the UV luminosity function is required, implying the existence of plenty of faint galaxies. These galaxies may be too faint to be detected individually but their signal is embedded in the near- and far-IR backgrounds that measure the cumulative light from all galaxies in the near-IR -- which probes the redshifted starlight (e.g.\ Cooray et al. 2012; Kashlinsky et al. 2012; Fernadez \& Zaroubi 2013) -- and in the far-IR/sub-millimeter -- which probes the light reprocessed by dust (e.g.\ Lagache et al. 2005; Planck Collaboration 2014). The \hi~and IR backgrounds result from different, mutually exclusive regions, resulting in a strong anti-correlation (Fernandez et al. 2014). Indeed, the cross-correlation coefficients are the most negative during the mid-stages of reionisation.
There are many free parameters that can change the intensity of the high-redshift component of the IR backgrounds. However, Fernandez et al. (2014) have shown that the cross-correlation is not sensitive to the detailed physics of high-redshift galaxies, but to the reionisation history, such as $z_\mathrm{re}$ and $\Delta_z$. This cross-correlation will also help answering the question of which types of galaxies are mostly responsible for reionisation (Jeli\'c et al. 2015).

\smallskip

\noi {\bf The cross-correlation with CII intensity mapping experiments} 

\smallskip

\noi Instrumental concepts for reionisation CII experiments are emerging, with the Tomographic  Ionized-Carbon  Mapping  Experiment  (TIME-Pilot,  Crites  et  al., 2014),  and  CONCERTO  project (CarbON  CII line in post-reionisation and ReionisaTiOn epoch,  Lagache et al., in prep). These intensity mapping experiments will measure the 3D fluctuations of the [CII] line emission at redshifts $4.5<z<8$. The atomic [CII] line is one of the most valuable tracers of star formation at high redshift. Like the aforementioned probes (IR background, galaxy surveys), CII and \hi~fluctuations are strongly anti-correlated at large angular scales. As shown by Gong et al. (2012),  the CII-21\,cm cross-power spectrum will capture the topology of the end of EoR, including the ionised bubble sizes and the mean ionisation fraction. It will help us understand how reionisation proceeded.\\

\parbox{0.9\textwidth}{
 \noi{References:}\\
 \noi{\scriptsize
  Aghanim, N., et al., 1996, A\&A, 311, 1;
  Alvarez, M. A., et al., 2006, ApJ, 647, 840;
  Bouwens, R.J., et al., 2015, ApJ, 811, 140;
  Cooray, A. 2004, Phys. Rev. D, 70, 63509;
  Cooray, A., et al., 2012, ApJ, 756, 92;
  Crites, A.T., et al., 2014, SPIE, 9153, 1;
  Fernandez, E.R. \& Zaroubi S., 2013, MNRAS, 433, 2047;
  Fernandez, E.R., et al., 2014, MNRAS, 440, 298;
  Gong, Y., et al., 2012, ApJ, 745, 49;
  Jeli\'c, V., et al., 2015, AASKA14, 8;
  Kashlinsky, A., et al., 2012, ApJ, 753, 63;
  Lagache, G., et al., 2005, ARAA, 43, 727;
  Lidz, A., et al., 2009, ApJ, 690, 252;
  Masui, K.W., et al., 2013, ApJ, 763, 20;
  Planck collaboration, 2014, A\&A, 571, 30;
  Planck collaboration, 2016a, A\&A, 596, 107;
  Planck collaboration, 2016b, A\&A, 596, 108;
  Robertson, B.E., et al., 2015, ApJ, 802, 19;
  SKA1 System Baselinev2 Description (SKA-TEL-SKO-0000308);
  SKA1 Level 0 Science Requirements (SKA-TEL-SKO-0000007);
  Tashiro, H., et al., 2008, MNRAS, 389, 469;
  Tashiro, H., et al., 2010, MNRAS, 402, 2617;
  Tashiro, H., et al., 2011, MNRAS, 414 3424;
  Wiersma, R.P.C., et al., 2013, MNRAS, 432, 2615;
  Wolz, L., et al., 2016, MNRAS, 458, 3399
  }}\\

\subsubsection{Cosmic magnetism} \label{science:cosmicB}
\vspace{0cm}

Magnetic fields seem ubiquitous in the Universe. Their presence has
been detected not only at present times, but also in the distant past,
up to a redshift $\sim$ 4. The origin of these fields remains
unconstrained, and their role in the formation and evolution of cosmic
structures, from sub-parsec to cosmological scales, is still
a matter of intense investigation. Future generation radio-telescopes, and especially the SKA,
will offer wonderful capabilities to answer many of the outstanding
questions pertaining to cosmic magnetism.

\smallskip

\noi {\bf Introduction} 

\smallskip
\noi Magnetic fields are a familiar component
of our everyday life. Quantum and statistical physics explain
magnetism in materials on small scales. In
contrast, the problem of the origin of magnetic fields in cosmology
and astrophysics on large scales is still not settled. What are the
mechanisms that generated magnetic fields in the Universe? Were the
first magnetic fields created at high energies in the primordial era,
or were they generated later on by astrophysical processes? Are
present-day magnetic fields, measured in galaxies and galaxy clusters,
the result of turbulent dynamos, or are they persistent relics of the
Big Bang? What role did they play in the formation of cosmic
structures? SKA1 and (some of) its precursors will provide crucial
elements for answering those and related questions. Indeed, the origin
and evolution of cosmic magnetism is one of the main science drivers
of the SKA, and Cosmic Magnetism is already Key Science for LOFAR and
the ASKAP. A very short review of the topic is presented here, and the
reader is referred for instance to Beck et al. (2013),
Johnston et al. (2008), Beck (2015) and Johnston-Hollitt et al. (2015) for more information.

\smallskip

\smallskip

\noi {\bf Origin of large-scale magnetic fields} 

\smallskip
\noi While the presence of large-scale magnetic fields in cosmic structures has been
established for a long time, their origin is still unclear. Many
possibilities have been put forward and examined in detail in the
literature. The possible magnetogenesis mechanisms can be
roughly divided into two kinds. Those of the first kind rely on the
high-energy physics of the primordial, pre-recombination universe,
namely inflation or phase transitions (e.g. Long et al. 2014; Bamba 2015). Those of the second kind rely on astrophysical batteries that occurred in the classical,
post-recombination universe within formed structures or during their
formation (e.g. Gnedin et al. 2000; Durrive \& Langer 2015). For a review, we refer the reader to Widrow et al. (2012)
and Ryu et al.= (2012). The only consensus reached so far is that cosmic magnetic
fields must have been generated with relatively weak strengths
($10^{-21}-10^{-16}~\mathrm{G}$, and up to the nanoGauss, according to current estimates; e.g.\ Durrer \& Neronov, 2013), 
otherwise they would have
noticeably affected the structure formation process itself. These
initial weak fields have then evolved to reach the observed
magnitudes, scales, and shapes, during and after
the formation of structures, by mechanisms that we evoke below.

\smallskip

\noi {\bf Evolution of cosmic magnetic fields} 

\smallskip
\noi  Due to the high
conductivity of the cosmic plasma and to the large scales involved,
magnetic fields are frozen in matter. This ensures that, on the
largest scales, magnetic field strengths decay as the inverse of the
scale factor squared. On smaller, gravitationally unstable scales, the
initial weak magnetic fields are amplified by several concurring
effects. First, there is an adiabatic compression that operates during
the gravitational collapse of structures. Second, further
amplification can be provided by the stretching of field lines by shearing, swirling motions and compression in localised shocks.  In particular, small-scale turbulence can amplify magnetic fields on small scales through a dynamo. A turbulent dynamo mechanism amplifying and sustaining magnetic fields up to $\mu\mathrm{G}$ levels on cosmologically fast timescales may notably be at work in the intra-cluster medium (ICM; e.g.\ Rincon et al. 2016). The cosmic history of this process, as well as the dynamics of field-amplifying magnetised convective processes relevant to the ICM, such as the magnetothermal and heat-flux driven buoyancy instabilities, are far from fully understood. The SKA provides the opportunity to learn a lot about magnetised dynamics in the ICM.
Within rotating galaxies, large-scale differential rotation
and small-scale turbulence combine in order to amplify, order and
sustain large-scale magnetic fields through a large-scale dynamo. A
review of the current status of the dynamo theory is available in
Brandenburg et al. (2012). The flip-side is that present observations
do not allow us to distinguish between magnetogenesis mechanisms of
the primordial universe and astrophysical batteries as the
characteristics of the initial fields are essentially erased by the
powerful turbulent dynamo mechanisms acting upon them. There is
one place, though, where the latter is possibly absent, or
weak at most, and where we should look for clues
about the origin of magnetic fields: the intergalactic space, and
notably cosmic voids. The detection of intergalactic magnetic fields
(IGMFs) is therefore one of the major objectives of the SKA.

\smallskip

\noi {\bf Magnetic fields of the Milky Way} 

\smallskip
\noi Interstellar
magnetic fields in the Milky Way exist over wide ranges of sizes and
strengths. They play a crucial role in the Galactic environment, where
they govern the structure and dynamics of the interstellar gas,
regulate the process of star formation, affect the evolution of
supernova remnants and superbubbles, accelerate cosmic rays and
channel their trajectories\dots Observation methods include dust
polarisation, Zeeman splitting, Faraday rotation, and synchrotron
emission, all of which are complex and indirect. Our current
observational picture is that of a global spiral magnetic field with a
significant turbulent component (e.g.\ Ferri\`ere 2015). However, many open questions remain regarding both the global field (azimuthal structure, vertical parity, number and locations of radial reversals\dots) and the turbulent field
(energy spectrum, structure functions, outer scale\dots). The reader
may consult Haverkorn (2015) for reference. Using the SKA1 to improve
our knowledge of magnetic fields in the Milky Way (e.g.\ Haverkorn et al. 2015) will not only lead to a better understanding of our own Galaxy(see, e.g., Sects.\,\ref{sci:FarTom} and \ref{sci:zeem}), but it will also help us define and remove Galactic foregrounds for a multitude of
extragalactic and cosmological studies (e.g.\ Planck Collaboration, 2016).

\smallskip
\noi {\bf Future observations with the SKA}

\smallskip
\noi Currently, we know that magnetic fields are present in galaxies of all
types (Beck \& Wielebinski, 2013) as well as in galaxy clusters
(Feretti et al. 2012). There are hints indicating that intergalactic
filaments, too, may be magnetised, and recent analyses of high-energy
gamma-ray experiments suggest that so could the entire intergalactic
space (e.g.\ Haverkorn, 2015). Future magnetic field radio measurements
with the SKA1 and its precursors rely on Faraday rotation, synchrotron
radiation and full-sky polarisation measurements. The construction of
rotation measure (RM) grids of both Galactic pulsars and unresolved
polarised extragalactic sources will make it possible to unravel the
3D structure of magnetic fields in the Milky Way, on sub-parsec to
Galactic scales. The extragalactic-source RM
grid will also be essential for the detection and characterisation of
IGMFs. Both RM and extended synchrotron polarimetry measurements with the SKA1 will provide new information on the spatial distribution, structure and evolution of magnetic fields of the ICM (Bonafede et al. 2015), and possibly in the cosmic web (Giovannini et al. 2015). Such measurements, possibly used in synergy with velocity-field measurements with future X-ray observatories such as Athena, will be extremely valuable to diagnose dynamical field amplification processes in clusters. In addition, Faraday tomography (or RM synthesis) will allow us to reconstruct the 3D structure of magnetic fields in nearby galaxies. Finally, deep polarisation surveys will make it possible to
detect the magnetic fields present in the filamentary structure of the
Warm Hot Intergalactic Medium and measure their properties (Taylor et al. 2015, see also Sect.\,\ref{science:clustersNT}).\\

\parbox{0.9\textwidth}{
\noi{References:}\\
\noi{\scriptsize Bamba, K., 2015, Phys. Rev. D, 91, 043509;
Beck, R., 2015, in \textit{Magnetic Fields in the Universe: From Laboratory and Stars to Primordial Structures}, Eds.; A. Lazarian, E. M. de Gouveia dal Pino, C. Melioli, 3--17 (Springer);
Beck, R., et al., 2013, Astron. Nach., 557, 548;
Beck, R. \& Wielebinski, R., 2013, in \textit{Planets, Stars Stellar Systems -- Vol. 5: Galactic Structure and Stellar Populations}, Eds. T. D. Oswalt \& G. Gilmore, 641--723 (Springer);
Brandenburg, A., et al., 2012, SSRv, 169, 123;
Durrer, R. \& Neronov, A., 2013, A\&ARv, 21, 62;
Durrive, J.-B. \& Langer, M., 2015, MNRAS, 453, 345;
Feretti, L., et al., 2012, A\&ARv, 20, 54;
Gnedin, N., et al., 2000, ApJ, 539, 505;
Haverkorn, M., 2015, in \textit{Magnetic Fields in Diffuse Media}, Astrophysics and Space Science Library, Eds. A Lazarian, E. M. de Gouveia Dal Pino \& C. Melioli, 483--506 (Springer);
Haverkorn, M., et al., 2015, AASKA14, 96;
Johnston, S., et al., 2008, Exp. Astron., 22, 151;
Johnston-Hollitt M., et al., 2015, AASKA14, 92;
Ferri\`ere, K., 2015, J. Phys.: Conf. Ser., 577, 012008;
Giovannini, G., et al., 2015, AASKA14, 104;
Long, A. J., et al., 2014, JCAP, 2014, 036;
Planck Collaboration 2016, \aap, 596, A105;
Rincon, F., et al., 2016, PNAS, 113, 3950;
Ryu, D., et al., 2012, SSRv, 166, 1;
Taylor, A. M., et al., 2011, \aap, 529, A144;
Taylor. A. R., et al., 2015, AASKA14, 113;
Widrow, L. M., et al., 2011, SSRv, 166, 37
}}\\

\subsubsection{Cosmological Evolution of the Neutral and Cold Gas Mass Density}
\vspace{0cm}

\begin{figure}[!ht]
  \centering
  \includegraphics[width=\hsize]{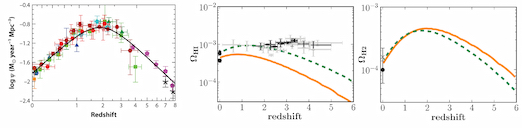}
\vspace{-1.0cm}\caption{\label{fig:MassDensities}History of cosmic star-formation ({\em left}, adapted from Madau \& Dickinson 2014) compared to that of the overall neutral gas (centre, models
  from Popping et al. 2014 and data from 21\,cm measurements at $z=0$ and high-$z$ DLA surveys - Noterdaeme et al. 2012; Zafar et al. 2013) and with that of
  molecular hydrogen (model from Popping et al. 2014). We clearly lack measurements of the total (warm+cold) neutral gas cosmic density at z$<$2 and for the cold gas at all redshifts z$>$0.}
  \end{figure}

\noi The physical processes driving the dramatic change in star formation rate between $z=1$ and today remain to be identified (Fig~\ref{fig:MassDensities}). In order to address this issue, we need to target the fuel for star formation, i.e., the neutral and cold phase of the interstellar medium in high-redshift galaxies which account for the bulk of all star formation. Such studies are essential to understand the formation of galaxies and the growth of structures.

\smallskip

\noi Instead of starlight, SKA will use the 21\,cm line emission of neutral hydrogen to trace the matter.
The expected number of accurate \hi~21\,cm redshift measurements in the SKA survey exceeds that from planned optical and infrared 
surveys by at least an order of magnitude. Yet these approaches are complementary
to study the relationship between the luminous stellar populations and the gas from which they form. 

\smallskip

\noi Observations of the 21\,cm line of HI, either in absorption against background radio sources or in emission, probe neutral gas, and trace the gravitationally dominated galaxy dynamics rather well.
Neutral gas constitutes most of the total interstellar gas mass in galaxies and tends to segregate in a warm ($T \sim$ a few $10^3$~K) and a cold ($T \sim$ 100~K) component.

\smallskip

\noi {\bf \hi~Emission Line Surveys}

\smallskip

\noi In the optically thin regime, the 21\,cm emission traces the neutral gas
independently of its temperature. 
The SKA will be able to map the \hi~distribution and kinematics at ISM cloud-size resolution in many galaxies, 
and will be be complementarity with ALMA and the E-ELT to improve our understanding of star formation, 
gas flows into and out of galaxies. 
The roadmap towards achieving the observation of \hi~at cosmological redshifts and the imaging of more nearby objects at increasing spatial resolution and sensitivity involves the SKA Precursor instruments, ASKAP and MeerKAT, plus APERTIF in the northern hemisphere, SKA1 and ultimately SKA2. During the SKA Precursor phase, ASKAP and APERTIF will perform two concerted all-sky \hi~surveys (WALLABY and WNSHS), thanks to their large FoV (8-30 deg$^2$), with an angular resolution of 8-30 arcsec and a velocity resolution of a few km/s. It is expected that a million galaxies will be detected, with an average redshift $z \sim 0.1$, and 3D \hi~maps produced for thousands of objects. At the same time, MeerKAT, which has a smaller FoV than ASKAP (1 deg$^2$) but higher sensitivity, will perform a high-sensitivity survey (MHONGOOSE) of 60 selected nearby galaxies, aimed at probing low-density halo gas and a very high-sensitivity survey of part of the Fornax cluster, aimed at detection of cosmic web filaments. It will also perform an ultra-deep \hi~survey (LADUMA) aimed at detecting $\sim$ 6000 galaxies at the highest achievable redshifts, on average $z \sim 0.4$ (but going out to $z \sim1$ for the most massive objects).
  For the SKA1 phase, observations will include the imaging of several millions of galaxies at a few arcsec resolution, at an average redshift of 0.4. 
   The ultimate goal of SKA2 is the observation of a billion galaxies with an average redshift of $z \sim 1$ at a resolution of 0.1 arcsec or (much) less.

\smallskip

\noi {\bf 21\,cm Intensity mapping}

\smallskip

\noi The SKA \hi~emission line surveys discussed above has the advantage of determining the source redshift $z$ by a direct comparison of the observed radio frequency with the intrinsic emission 
frequency ($\sim 1420 \, \mathrm{MHz}$). However, the relatively low radio brightness of these objects has limited their detection 
to the cosmological neighbourhood of our galaxy with the current instrument. Even with SKA, the detection 
of \hi~galaxies at higher redshifts, $z \sim1$ and beyond would remain a challenge, 
and be only for the brightest \hi~galaxies would be detected by SKA at $z \sim 1$.   
Intensity mapping provides a way to economically map huge cosmic volumes in three dimensions.
Indeed, the total \hi~emission from 3D cosmic volume cells of $\sim \mathrm {5-10~Mpc}^3$ , from the combined  
emission of  hundreds of galaxies and \hi~clouds would be large enough to be detected by an instrument 
with a collecting area of few 10 000 $\mathrm{m^2}$, and with few hours integration time.
Intensity mapping up to redshifts $z \sim 2-3$ will thus require an instrument with a wide 
field of view $\gtrsim 10 \, \mathrm{deg^2}$, large collecting area $ \sim 10 000 \, \mathrm{m^2}$ 
and high sensitivity, but with a modest angular resolution ($\sim 5-10$ arcmin). SKA aperture arrays will be well suited to carry intensity mapping surveys, covering a wide redshift range, 
up to $z \sim 3$. Intensity mapping technique and expected results are discussed in Sect.\,\ref{science:IM} of this book.

\smallskip

\noi {\bf \hi~Absorption Line Surveys} 

\smallskip

\noi Neutral gas is efficiently traced by Damped Lyman-$\alpha$ Absorption systems (DLAs)
against bright background sources such as quasars. Since the
Lyman-$\alpha$ line is insensitive to the temperature of the gas, DLAs can arise from
both the warm and cold neutral interstellar medium (WNM and CNM, respectively). While the WNM dominates
the absorption cross-section,
it is possible to estimate the CNM fraction by deriving the physical conditions through e.g.
the excitation of different species (e.g. Neeleman et al. 2015) or use tracers of cold gas,
such as H$_2$ (e.g. Noterdaeme et al. 2008).
Despite the fact that about thirty thousand DLAs have been detected so far (e.g. Noterdaeme et al. 2012),
these are mostly limited to $z>2$, where the Lyman-$\alpha$ lines becomes easily observable from the ground.
In addition, current DLA samples are likely biased against the high metallicity/dusty gas which is more likely to be cold. 

\smallskip

\noi In the radio domain, not only flux-limited radio samples are unaffected by the presence of dust but there is a strong
weighting of the CNM to the detectability of the 21\,cm line. Indeed, 
the observed 21\,cm optical depth ($\tau$) of a homogeneous cloud in thermal equilibrium is related to both
its column density, $N$(H\,{\sc i}) and its spin temperature ($T_s$) as
\begin{equation}
N({\rm HI}) = 1.823\times10^{18}~T_s \int \tau dv
\end{equation}

\smallskip

\noi However, the search of 21\,cm absorption has currently been limited to
small sub-samples of optically-selected absorbers, thereby adding
selection effects to the dust bias inherited from the optical selection.
Indeed, only a small fraction of the quasars are bright enough in the
radio for current facilities that also have limited frequency coverage and
poor RFI environment. Only a small number of 21\,cm absorption systems have been detected to date (e.g. Gupta et al. 2012) and the evolution of the cold gas in galaxies is poorly constrained. The extreme sensitivity of SKA will change the situation dramatically, allowing to carry out blind (i.e. without optical selection) sensitive 21\,cm absorption measurements against thousands of background radio sources and over the the full range of redshifts from $z=0$ out to $z=4$. 

\begin{itemize}

\item[$\bullet$] Incidence of cold gas: The undergoing MeerKAT Absorption Line Survey (PIs Gupta/Srianand) should already detect several hundred intervening
21\,cm absorbers at $z<1.5$ towards ~1000 compact background sources, constraining the evolution of the amount of cold gas in the Universe at 10\% accuracy in 5 redshift bins. 

\item[$\bullet$] Gas kinematics and physical conditions: The 21\,cm line is intrinsically very narrow (a few km/s) so that different velocity components in a given intervening galaxy are easily resolved, allowing for detailed studies of the gas kinematics (in the optical, only the total \hi~column density is available due to the saturation of the Lyman-$\alpha$ line). The width of the each line component also provides a direct measure of the gas temperature.

\item[$\bullet$] Spatial mapping: An exciting prospect will be to use extended background radio sources (i.e. galaxies) to spatially map the intervening cold gas (Biggs et al. 2016).

\end{itemize}

\smallskip

\noi We finally note that associated 21\,cm absorbers, i.e. at the redshift of the background source will allow us to study the presence and properties of cold gas close to the central engine of quasars and in their host galaxy.\\

\parbox{0.9\textwidth}{
\noi{References:}\\
\noi{\scriptsize 
Biggs, A., \etal , 2016, MNRAS, 462;
Gupta N., \etal , 2012, A\&A, 544, 21;
Neeleman M., \etal , 2015, ApJ, 800, 7;
Noterdaeme \etal , 2008, A\&A, 481, 327;
Noterdaeme \etal , 2012, A\&A, 547, L1;
Popping, G., \etal , 2014, MNRAS, 442;
Zafar, T., \etal , 2013, A\&A, 556
}}\\

\subsubsection{Large scale structure through \hi~velocity fields} 
\vspace{0cm}

\noi For the past 30 years considerable effort has been put into improving the quality of numerical simulations and galaxy redshift surveys to unveil the nature of the Dark Matter and its role in the clustering of cosmic structures. However, the dynamics of this process is not totally understood yet; we are still unable to successfully reproduce even the well-known structures and empty regions in the nearby Universe, let alone determine their masses. 

\smallskip

\begin{figure}[t]
  \centering
  \begin{subfigure}[b]{0.24\textwidth}
        \includegraphics[width=\textwidth]{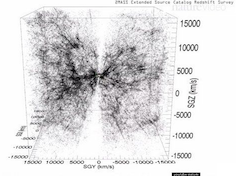}
        \caption{}
  \end{subfigure}
  \begin{subfigure}[b]{0.24\textwidth}
        \includegraphics[width=\textwidth]{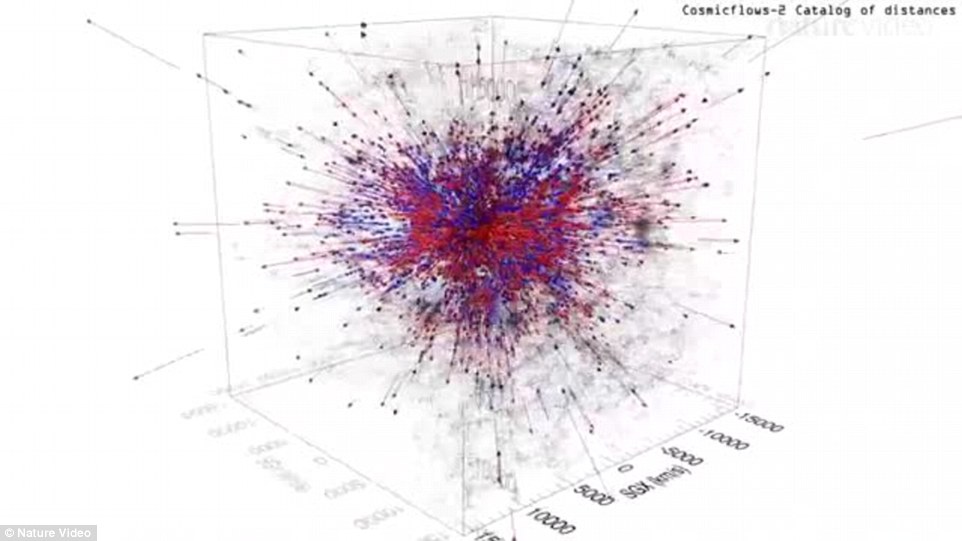}
        \caption{}
  \end{subfigure}  
  \begin{subfigure}[b]{0.24\textwidth}
        \includegraphics[width=\textwidth]{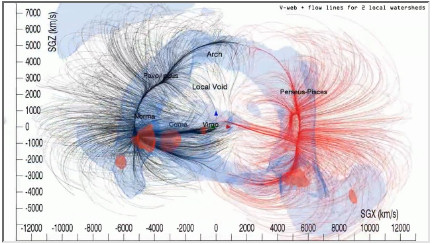}
        \caption{}
  \end{subfigure}  
  \begin{subfigure}[b]{0.20\textwidth}
        \includegraphics[width=\textwidth]{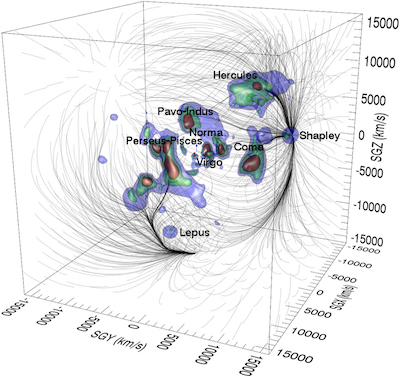}
        \caption{}
  \end{subfigure}  
  \caption{\label{fig:wf_cflow}
The full process from observational data to models: ({\em left to right}) from the static picture of the large-scale structure charted by galaxies (a)  and their peculiar velocities (b) one can reconstruct the 3D model of galaxy flows and dark matter distribution (c) and finally identify dynamical basins e.g. by watershed computations (d) (Courtois et al. 2013).}
\end{figure}

\noi A global effort is recently pursued to improve the detailed modelling of these processes using new analysis techniques (Libeskind et al. 2015; Carrick et al. 2015; Lavaux \& Jasche 2016) and observational programs in optical and near-infrared bands, e.g. Cosmicflows (Courtois \& Tully 2014; Tully et al. 2016) and TAIPAN (Kuehn et al. 2014). In Fig.\,\ref{fig:wf_cflow}, we illustrate the inference procedure from observations to gravitational and structuration models of our Local Universe. Using SKA capabilities we would enter into a new era of extragalactic velocity surveys, permitting to reconstruct the largest cosmographic maps of dark matter at an unprecedented level of accuracy. These will allow us the study of the accurate measurement of the expansion rate of the universe and the growth-rate of structures the standard methods based on the baryonic acoustic oscillations (BAO) and redshift space distortions (RSD) statistics, two of the best probes to test the cosmological model and General Relativity on large scales.

\smallskip

\noi {\bf Observations}

\smallskip

\noi We will measure BAO and RSD by a \hi~redshift survey, the distinctive profile of the \hi~line allowing very accurate redshift measurement ($\delta z < 10^{-4}$). With the radio-telescope today having the largest collecting area, Arecibo, \hi~redshifts can only be measured in the local universe, $z\lesssim 0.14$ (Freudling et al. 2011). Cosmological applications will require detecting enough galaxies to beat shot-noise and over a large enough area to reduce cosmic variance. As shown in Raccanelli et al. (2015), with 10,000~hours of observation time and a large optimal survey area of at least 5,000 deg$^2$, SKA1 will produce $\sim 10^7$ redshifts using band-2 from SKA1-MID, while only 10$^4$ redshifts at higher redshifts ($0.4<z<3$) can be observed with SKA1-MID Band 1. SKA2 would increase quite drastically those numbers with about $10^9$ galaxies detected in a survey of 30,000 deg$^2$ up to $z\sim 2$, which would be the largest unprecedented redshift survey.

\smallskip

\noi The upcoming pathfinders \hi~(imaging) surveys with ASKAP, WSRT/APERTIF and MeerKAT will improve the estimate of galaxy distances in the local universe by reducing the scatter in the Tully-Fisher relation (TFR), establishing the statistical properties of the TFR at modest redshifts ($z<0.2$) using spatially resolved HI, and up to $z<0.5$ using global \hi~line widths.  With SKA one can analyse the slope, scatter, and zero-point of the TFR as a function of redshift, cosmic environment, morphological type, and (specific) star formation rate of galaxies. This requires adequate angular resolution and column density sensitivity to image the \hi~kinematics in the outer gaseous disks of galaxies at higher redshifts, and to survey volumes that encompass the cosmic environments with sufficient statistical significance to mitigate cosmic variance. A spatial resolution of 10\,kpc, corresponding to an angular resolution of 1.6~arcsec at a frequency of 950\,MHz, would be sufficient to assess the kinematic state of the largest \hi~disks in $M_*$-galaxies at $z=0.5$. Based on studies of nearby galaxies, the required column density sensitivity at this angular resolution is within reach of an ultra-deep survey using SKA1-MID. The volume surveyed by a single SKA1-MID pointing at $z=0.5$ at 950~MHz is large enough to escape the cosmic-variance limit. Pushing TFR studies based on resolved \hi~kinematics to even higher redshifts would require SKA2 capabilities.

\smallskip

\noi Because TFR is a relation between the apparent brightness (from optical photometry observations) and the intrinsic galaxy luminosity (derived from \hi~line-width radio observations, accurate distances derived from the TFR require accurate photometric calibration (Courtois et al. 2011). In practice, to obtain usable peculiar velocities, galaxy distances must be more accurate than 20\%, translating in 0.2~mag accuracy in photometry. Two ground-based optical surveys, PanSTARRS in Hawaii and SkyMapper in Australia, will cover the entire sky in $g$, $r$, $i$, $z$ bands. During the SKA timeline, these surveys will generate increasingly deep images of the entire sky with rigorous photometric quality control. By that time we will also have near-infrared imaging from Spitzer and/or Wide-Field Infrared Survey Explorer (WISE) satellites (WISE has now completed the imaging of the entire sky). We have prepared successful proof-of-concepts for the calibration and use of the TFR at Spitzer1 and WISE2 bands for a few hundreds of galaxies (Sorce et al. 2014; Neill et al. 2014). 

\smallskip

\noi {\bf Dark Matter, Dark Energy, and test of the gravitational theory by the large-scale dynamics}

\smallskip

\noi From the independent measurement of redshift ($z$) and distance ($d$) of sources, the latter obtained e.g. by the Tully-Fisher relation (TFR), their radial peculiar velocity $V_\mathrm{pec}$ can be inferred in the local universe according to the first order relation $cz = H_0 d + V_\mathrm{pec}$, where $H_0$ is the Hubble constant. Strong constraints on cosmological parameters can be obtained by using SKA peculiar velocities to recover the primeval BAO signal before it was blurred by non-linear clustering (e.g. Kazin et al. 2014; Anderson et al. 2015), and to model the RSD (Koda et al. 2014). Combined with the underlying redshift distribution, the reconstructed velocity field can also be used for density-density or velocity-velocity comparisons (e.g. Erdogdu et al. 2006; Lavaux et al. 2010; Carrick et al. 2015), or for cross-correlation with the dipole moment of the galaxy distribution as done by Nusser (2017). On sufficiently large scales, where the growth rate can be approximated by $f\approx\Omega_\mathrm{m}^\gamma$ with $\gamma \simeq 0.55$ in General Relativity (GR), the low-redshift observations like the one in SKA-MAPS (Santos et al 2014) will deliver the stringent test of GR by measuring the parameter $\gamma$ at the percent level. This measurement will complement the constraints achieved by high-redshift galaxy surveys like WiggleZ, VVDS, VIPERS, and in the future eBOSS, PFS-SuMIRe, DESI, Euclid, or WFIRST, which do not have the information on velocities.

\smallskip

\noi Besides, it will be interesting to combine SKA velocities with cosmic shear measurements obtained e.g. by Euclid or internally by SKA (see the SKA German proposal by P. Schneider) to test the gravitational instability paradigm on very large scales measuring the so-called gravitational slip (Dor\'e et al. 2003; Reyes et al. 2010). Indeed, weak-lensing by large scale structure provides complementary information since it depends on both the spatial and temporal scalar perturbations of the metric while velocities depend on the temporal part only.

\smallskip

\noi Finally, a remarkable application of SKA velocities will be to probe the gravitational theory
in unexplored gravitational field regimes such as at the galactic scale (Desmond et al. 2017).

\smallskip

\noi It is worth to mention that the SKA pathfinders, with accurate peculiar radial velocities for more than 50,000 galaxies within a sphere of radius 600\,Mpc, will prepare the field for the best direct measurement of the present-day expansion rate of the Universe, the largest maps of the density and velocity fields of local structures, and unique new tests of large-scale gravitational physics using galaxy motions, including tests of modified theories of gravity.
For this purpose, it will be crucial to realise realistic cosmological simulations that can be compared directly to the observed velocity and density fields in the Universe, eventually considering not only the standard $\Lambda$CDM model but also different kinds of Dark Matter and different theories of gravity with the appropriate gravity solvers. Indeed, the motion of galaxies is most directly in response to the clumped distribution of Dark Matter, which is directly related to the divergence of the cosmic velocity field. More generally cosmic flows depends on the physics of the Dark Sector, which is one of the main SKA science drivers.\\

\parbox{0.9\textwidth}{
\noi{References:}\\
\noi{\scriptsize 
Becker, R.H., \etal, 2001, AJ, 122, 2850;
Carrick, J., \etal, 2015, MNRAS, 450, 317;
Courtois, H. \etal, 2011, MNRAS, 415, 1935;
Courtois, H. \etal, 2013, AJ, 146, 69;
Courtois, H. \& Tully, R. B., 2014, MNRAS, 447, 1531;
Desmond, H., \etal, 2017, arXiV:02420;
Dor\'e, O. \etal, 2003, ApJL, 2, 81;
Erdogdu, P., \etal, 2006, MNRAS, 373, 45;
Freudling, W., \etal, 2011, ApJ, 727, 40;
Kazin, E. A. , \etal 2014, MNRAS, 441, 3524;
Koda, J., \etal, 2014, MNRAS, 445, 4267;
Kuehn, K., \etal, 2014, SPIE, 9147, 10;
Lavaux, G., \etal, 2010, ApJ, 709, 483;
Lavaux, G. \& Jasche J., 2016, MNRAS, 455, 3169;
Libeskind, N., \etal, 2015, MNRAS, 453, 108;
Neill, J.D., \etal, 2014, AJ, 792, 129;
Nusser, A., 2017, arXiv:1703.05324;
Racanelli, A., \etal, 2015, arXiV:1501.03821;
Reyes, R., \etal, 2010, Nature, 464, 256;
Santos, M., \etal, 2014, AASKA14; arXiV:1501.03989
Sorce, J., \etal, 2014, MNRAS, 444, 527;
Tully, R.B., \etal, 2016, AJ, 152, 50}
}\\

\subsubsection{Large Scale Structure and BAO with Intensity Mapping} \label{science:IM}

\vspace{0cm}

\noi The statistical properties of the matter distribution in the
Universe encodes invaluable information about the cosmological
parameters, such as the mean cosmic matter and energy densities, the
physics underpinning the Standard Cosmological Model, especially
General Relativity, Modified Gravity, and Dark Energy.  Most of this
information can be extracted from the large-scale statistical
properties of the matter distribution on comoving length scales of
10\,Mpc and above, corresponding to apparent angular
scales larger than a few arcminutes.  The cosmological matter
distribution, dominated by dark matter, can be probed through tracers
such as galaxies, which are often detected and observed through their
optical emission, their position being determined by imaging and their
redshifts by spectroscopy.

\smallskip
\noi Since the first astronomical observation of 21\,cm emission by
Ewen and Purcell (1951) following the suggestion by van~de~Hulst
(1945), galaxies are known to harbour neutral hydrogen (\hi) that is
destined eventually to form stars. Galaxies can thus be detected and
observed in radio through the 21\,cm line emission of this neutral
atomic hydrogen. Moreover, the redshifted 21\,cm feature pinpoints
directly the galaxy positions in redshift space.  In contrast, optical
surveys cannot obtain accurate redshifts easily.  Optical surveys rely
on photometric redshifts which are not very accurate and are prone to
error due to misidentification of the object type for distant objects.
Alternatively, more accurate redshifts can be obtained through costly
spectroscopic follow-up on a subset of the catalogue.

\smallskip
\noi The SKA \hi\ line survey provides a method similar 
to optical surveys to map matter distribution through compact
\hi\ cloud detection in radio, through the 21\,cm hydrogen hyperfine
transition.  This method has the advantage of determining the source
redshift $z$ by a direct comparison of the observed radio frequency
with the intrinsic emission frequency of 1420\,MHz (Abdalla \& Rawlings et al. 2005).  The expected number of such accurate redshift
measurements in the SKA survey exceeds that from planned optical and
infrared surveys by at least an order of
magnitude (Rawlings et al. 2004). Yet these two approaches are complementary.  Having
both will help to clarify a number of astrophysical
questions, such as the relationship between the luminous stellar
populations and the gas from which they form, and the star formation
history. 

\smallskip
\noi A large galaxy survey such as expected from the SKA \hi\ line
survey will provide precision measurements of the statistical properties of Large Scale Structure (LSS) and the corresponding power spectrum $P(k)$.  In particular, it is possible
to extract the precise measurement of the angular diameter
subtended by the $\sim$ 110\,Mpc/h baryon acoustic oscillation (BAO)
feature as a function of redshift. This feature, a remnant of the
tight coupling between photons and baryons in the early Universe, is
imprinted on the galaxy distribution during the epoch of decoupling of
photons and baryons at $z\sim1100.$ It is one of the most robust
cosmological probes used to constrain Dark Energy properties. On
intermediate angular scales $(\lesssim \mathrm{deg})$
the survey will resolve the cosmic web (filaments, walls, and the
voids they enclose), which is emerging as a rich target for defining
cosmological observables.

\smallskip
\noi By using the BAO signature alone to map the expansion history of
the Universe during the acceleration epoch, one can establish the
onset and shape of dark energy domination (Blake \& Glazebrook 2003;
Seo \& Eisenstein 2003). Using the \hi \, galaxy survey and BAO,
SKA1 will be able to measure the expansion rate $H(z)$ and the
angular diameter distance $D_A(z)$ with a relative precision of $1-2
\%$ for redshifts up to $z \lesssim 0.5$.  The relative precision on
$H(z), D_A(z)$ measurements will reach $0.2-0.5 \%$ over the redshift
range $ 0.5 \lesssim z \lesssim 1.5$ with the second phase of SKA
(Yahya et al. 2015).  This is a conservative estimate since it uses
only the information from this one scale. Analysing the entire
correlation function and modelling the way galaxies trace the
underlying matter distribution would further reduce the expected error
bar on the Dark Energy equation of state parameters.

\smallskip
\noi However, the relatively low radio brightness of these
\hi \, clumps has limited their detection to the
cosmological neighbourhood of our galaxy with current instruments.
SKA will significantly extend the detection redshift range thanks to
its enormous collecting area combined with its large field of
view. Despite the huge increase in sensitivity brought by SKA, the
detection of \hi\ galaxies at higher redshifts (i.e., $z\sim1$ and
beyond) will remain a challenge, and only the brightest \hi\ galaxies
would be detected by SKA at $z\sim1$.

\smallskip
\noi A technique to further enhance the cosmological signal is to
image directly the 3D structure of the cosmic \hi\ distribution by
mapping the integrated 21\,cm brightness.  An early suggestion of this
method (Hogan \& Rees 1979) and the idea to use 21\,cm line
observations for the determination of cosmological parameters (Scott
\& Rees 1990) was later proposed as a radio astronomy technique
called Intensity Mapping (Battye et al. 2004; Peterson et al. 2006)
and further developed for radio interferometers (Ansari et al. 2008;
Ansari et al. 2012).
The Intensity Mapping technique eliminates the intermediate step of
constructing a galaxy catalog since all photons are used without
selecting only those that come from objects detected with high
significance. By eliminating the necessity to reach detection
thresholds for individual galaxies, the Intensity Mapping approach
will allow using early stages of SKA to measure cosmological
parameters before completion of the ``full instrument'' (SKA Phase 2).  A number of
instruments are currently in development (see Bull et al. 2015 for a
comparison), and in particular the TIANLAI project (Chen 2012), 
as well as HIRAX (Newburgh et al. 2016), has significant input
from France.  The proposed project called MANTIS (Cappellen et al.
2016) would be an early version of the SKA Mid Frequency Aperture
Array, and ideally suited for Intensity Mapping. Using the Intensity Mapping 
technique, SKA1 should be able to measure the Dark Energy equation of state parameters 
$w_0$ and $w_a$ with $5-10 \%$ precision, close to the expected precision 
from Euclid  (Santos et al. 2015).

\smallskip
\noi Intensity mapping provides the most economical way to map huge cosmic
volumes in three dimensions. The total \hi\ emission from 3D
cosmic volume cells of $\sim5-10\,{\rm Mpc}^3$, arising from the combined
emission of hundreds of galaxies and \hi\ clouds, would be large enough
to be detected by an instrument with a collecting area of a few
10\,000\,m$^2$ within a few hours integration time.  Intensity
Mapping up to redshifts $z\sim2-3$ will thus require an instrument
with a wide field of view on the order of tens of square degrees, a
collecting area of $\sim10000 {\rm m}^2$ and high sensitivity,
but with a modest angular resolution ($\sim5-10$\,arcmin).

\smallskip
\noi SKA aperture arrays will be well suited to perform Intensity Mapping
surveys, covering a wide redshift range, up to $z\sim3$. The SKA Intensity
Mapping surveys will be used for competitive cosmological
studies, including, but not limited to, precise determination of the
BAO scale.\\

\parbox{0.9\textwidth}{
\noi{References:}\\
\noi{\scriptsize
Abdalla, F. \& Rawlings, S., 2005, MNRAS, 360, 27; 
Ansari, R., et al., 2008, arXiv:0807.3614;   
Ansari, R., et al., 2012, \aap , 540 , 129; 
Battye, R., et al., 2004, \mnras, 355, 1339;  
Blake, C. \& Glazebrook, K., 2003, \apj , 594, 665; 
Bull, P., et al., 2015, \apj , 803, 21; 
van Cappellen, W., et al., 2016, arXiv:1612.07917; 
Chen, X., 2012, Int. J. Modern Physics Conference Series, 12, 256; 
Ewen, H. I. \& Purcell, E.M., 1951, Nature, 168, 356; 
Hogan, C. J. \& Rees, M. J., 1979,  \mnras , 188, 791;  
Newburgh, L. B., et al., 2016, Proc.SPIE Int.Soc.Opt.Eng. 9906, 99065X; 
Peterson, J. B., et al., 2006, arXiv:0606104; 
Santos, M., et al., 2015, AASKA14, 19; 
Scott, D. \& Rees, M. J., 1990,  \mnras , 247, 510;  
Seo, H.-J. \& Eisenstein, D. J., 2003, \apj , 598, 720;  
van~de~Hulst, H. C., 1945, Nederl. Tij.  Naturkuunde, 11, 201;  
Yahya, S. et al., 2015, \mnras , 450, 2251}}\\

\subsubsection{Distant Universe with Gravitational Lensing in the SKA Era}
\vspace{0cm}

\begin{figure}[!ht]
  \centering
  \includegraphics[width=0.9\linewidth, height=160px]{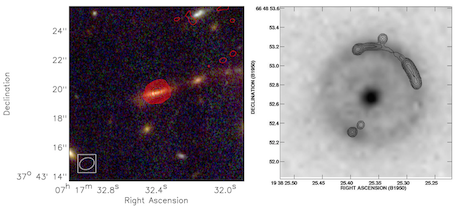}
  \caption{\label{fig:lensing1} Lensing effect - {\em Left}: cluster-galaxy lensing: lensed galaxy at z=2.32 seen in merging cluster 
MACSJ$0717.5+3745$ with radio contours (red) overlaid on HST optical background image (adapted from van Weeren et al. 
2016). {\em Right}: galaxy-galaxy lensing: Einstein's ring seen in JVAS B1938+666 with radio contours (black) overlaid on 
HST image (adapted from King et al. 1997).  
}
\end{figure}

\noi Gravitational lensing is an unique and efficient technique to
observe faint distant galaxies and measure the mass distribution
(including dark matter) locally as well as up to cluster-scale and
beyond, thus providing a major insight into galaxy formation and
evolution over cosmic timescale. Strong-gravitational lens systems, in
which the gravitational field of a foreground galaxy and/or cluster of
galaxies multiply images of a background distant galaxy, gives rise to
a rare Einstein's ring or giant arc-like structures respectively, as
seen e.g. in the HST, VLA and e-Merlin radio observations (Biggs et
al. 2004; Estrada et al. 2007; Gladders et al. 2002; Hewitt, J. et
al. 1988; Hennawi et al. 2008; Scarpine et al. 2006).  However, strong
lensing events are rare and found around clusters, where each
square-degree of sky contains only about one cluster of sufficient
mass to produce a giant arc. Nevertheless, lensing clusters are
preferred to study high redshift galaxies, as their weak amplification
factor helps to reconstruct the remote galaxy profile with
ease. The first big surveys for the identification of lens systems and
lensed galaxies were carried out on galaxy-galaxy strong lensing
systems (e.g. quasars) by the ESO Hamburg survey (Wisotzki et
al. 1996) and CfA-Arizona-ST-LEns-Survey (CASTLES by Munoz et
al. 1998; Richard et al. 2016) at optical wavelengths. In the radio
domain a similar search was carried out by the Jodrell Bank-VLA
Astrometric survey (Browne et al. 1998; Patnaik et al. 1992) and
Cosmic Lens All Sky Survey (CLASS; Jackson et al.1995;
Myers et al. 1995).  Additionally, with the advent of deep and wide
imaging at radio wavelengths (with the GMRT and JVLA) combined with
large spectroscopic HST and MUSE survey, many more new lensed galaxies
around lensing clusters have been recently discovered (see e.g. Fig.\,\ref{fig:lensing1},
Pandey-Pommier et al. 2017; van Weeren et al. 2016; Lagattuta et
al. 2016). In spite of these successful attempts, the number of
gravitational lenses (galaxies and clusters) known is still very small,
with only a few hundreds discovered as of now (McKean et al. 2015).  

\smallskip

\noi Thanks to its high sensitivity, SKA will transform our understanding of the Universe
with gravitational lensing, particularly at radio wavelengths where
the number of gravitational lenses will increase to $\sim$10$^5$
helping to probe the distant lensed galaxy population in the Universe
(McKean et al. 2015). A similar increase is expected from the future
Optical/IR surveys, notably EUCLID and LSST. In this chapter we
highlight the technical capability and main scientific results that
will be achieved via gravitational lensing technique in the SKA era.

\begin{figure}[!ht]
  \centering
  \includegraphics[width=1.02\linewidth, height=150px]{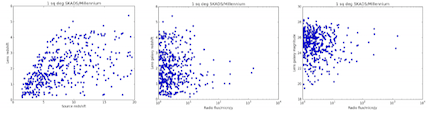}
  \caption{
    \label{fig:lensingstat} SKA gravitational lensing
    simulations. {\em Left}: Lens redshift versus lensed galaxy
    redshift. {\em Middle}: Lens redshift versus lensed galaxy flux density.
    {\em Right}: Lens magnitude versus lensed galaxy flux density (adapted
    from McKean et al. 2015).  }
\end{figure}

\smallskip
\noi Compared to current instruments (such as the JVLA or eMERLIN), SKA1-MID will be the first radio telescopes with the required survey speed {\it and} sub-arcsec angular resolution for large-scale lensing searches. McKean et al. (2015) derived radio gravitational lensing estimates based on semi-analytical models for the number density of lenses, CLASS survey lensing statistics (Myers et al. 2003) and the SKADS (SKA Design Study) database for the population of background radio sources (Wilman et al. 2008). They found that a shallow wide field SKA-MID final survey with an rms noise level of $\mu$Jy/beam and arcsec-scale resolution for gravitational lenses will be most efficient to increase the number of lens systems as well as lensed galaxies (AGN and starbursts) up to redshift z=1-5 (see e.g. Fig.\,\ref{fig:lensingstat}). In addition, thanks to the magnification provided by lensing (factors 5-100), a deeper and smaller area survey with SKA1-MID would further detect a fainter population of sub-$\mu$Jy level exotic high redshift objects, thus providing an opportunity to study their detailed radio properties - an unexplored region so far (Prandoni and Seymour 2015). These objects are expected to be detected routinely with SKA2, which will go up to a factor 5-10 deeper than SKA1-MID (McKean et al. 2015).

\smallskip
\noi The SKA, in combination with the optical/infrared missions such as Euclid, 4MOST and LSST, will discover a whole new class of rare gravitational lenses allowing us to explore the lensed distant galaxies in the Universe. The main cosmological results expected for galaxy formation and evolution as well as lenses themselves with SKA lensing studies are:

\begin{itemize}

\item The statistical survey data will place a firm constraint on mass
distribution models of lenses in the Universe and confirm if the
low-mass, early-type galaxies are dark-matter dominated. It will be
also possible to confirm if all lensing galaxies have quasar-type
properties

\item An increase in the number count of lensed galaxies by factors of
$>$10$^3$ will provide a new population of rare and cosmologically
important objects at the low mass end (Impellizzeri et al. 2008).
Lensed galaxies with star-formation rates (SFR) of $>$ 1 M$_{\odot}$
yr$^{-1}$ at z = 3-4 and $>$5 M$_{\odot}$ yr$^{-1}$ at z = 6 (near
epoch of reionisation) will be detected and used to probe the
evolution of SFR w.r.t. redshift. Further, spectral properties of high
redshift AGNs (discovered due to lensing) derived via broad band
continuum observations will be used to study the SFR and radio-FIR
luminosity correlation.

\item Deep \hi~imaging with the SKA will map the neutral hydrogen gas
content in the lensed galaxies that are usually early/or late-type
gas-rich galaxies.  Moreover, the wide-field SKA \hi~spectral line
survey on galaxies with \hi~masses of $>$ 12 M$_{\odot}$, will be able
to find new gravitational lens systems (approx. 0.5-5 gravitational
lenses per square-degree) whose redshift will be confirmed via
upcoming optical/IR survey with EUCLID, LSST (Serjeant et al. 2014;
Negrello et al. 2010; Abdalla et al. 2010).

\end{itemize}

\parbox{0.9\textwidth}{
\noi{References:}\\
\noi{\scriptsize Abdalla, F. B., et al. 2010 MNRAS, 401, 743;
Browne, I.,W.,A., 1998, ASSL, 226, 323B;
Biggs, A.,D., et al. 2004, MNRAS, 350, 949B;
Estrada, J., et al. 2007, ApJ, 660, 1176E;
Gladders, M., et al. 2002, AAS, 201, 5906G;
Hennawi, J. F., et al. 2008, AJ, 135, 664H;
Impellizzeri, C., et al. 2008, Nature, 456, 927;
Jackson, N., et al. 1995, MNRAS, 274L, 25J;
King, L., et al. 1997, MNRAS, 298, 450;
Lagattuta, D., et al. 2016, arXiv161101513L;
Mckean, J., et al. 2015, AASKA14, 84;
Munoz, J., et al., 1998, ApJ, 492L, 9M;
Myers, S., et al. 1995, ApJ, 447L, 5M;
Myers et al., 2003, MNRAS, 341, 1M;
Negrello et al., 2010, Sci, 330, 800N;
Pandey-Pommier, M., et al., 2017, A$\&$A, submitted;
Patnaik, A., et al., 1992, MNRAS, 259P, 1P;
Richard, J., et al., 2016, IAUFM, 29B.764R;
Scarpine, V., et al., 2006, AAS, 20921507S;
Serjeant, S., et al., 2014, ApJL, 793,10S;
van Weeren, R., et al., 2016, ApJ, 817,98V;
Wisotzki, L., et al., 1996, A$\&$A,315L,405W;
Wilman, R. et al., 2008, MNRAS, 388, 1335W
}}

\newpage
\subsection{Extra-galactic astronomy}

\noindent {\normalsize Contributors of this section in alphabetic order: }

\smallskip

\noi {\sffamily \scriptsize
{\sffamily \bf {\bf R.~Adam}} [\lagrange],
{\bf N.~Aghanim} [\ias],
{\bf B.~Ascaso} [\apc],
{\bf E.~Athanassoula} [\lam],
{\bf M.~B\'ethermin} [\lam],
{\bf S.~Boissier} [\lam],
{\bf A.~Boselli} [\lam],
{\bf A.~Bosma} [\lam],
{\bf V.~Buat} [\lam],
{\bf L.~Ciesla} [\irfu;\aim]
{\bf F.~Combes} [\colfr],
{\bf B.~Comis} [\lpsc],
{\bf E.~Daddi} [\irfu;\aim],
{\bf M.~Douspis} [\ias],
{\bf W.~van Driel} [\gepi],
{\bf C.~Ferrari} [\lagrange],
{\bf A.~Hughes} [\irap],
{\bf O.~Ilbert} [\lam],
{\bf G.~Lagache} [\lam],
{\bf M.~Langer} [\ias],
{\bf M.~Lehnert}  [\iapsorb],
{\bf J.~Mac\'{i}as-P\'{e}rez} [\lpsc],
{\bf N.~Nesvadba} [\ias],
{\bf M.~Pandey-Pommier} [\cral],
{\bf G.~W.~Pratt} [\irfu;\aim],
{\bf F.~Rin\c{c}on} [\irap],
{\bf C.~Tasse} [\gepi]
}

\subsubsection{Clusters}

\paragraph{Non-thermal emission from galaxy clusters}\label{science:clustersNT}
\vspace{0cm}

\begin{figure}[!ht]
  \centering
  \includegraphics[width=0.95\linewidth]{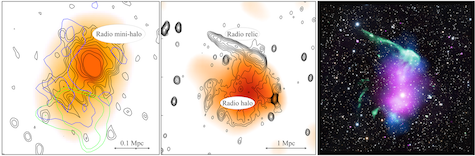}
    \caption{\label{fig:clusters2} {\em Left and middle panels:} contours (black) of synchrotron radiation observed in the radio band are overlaid on the images of the thermal bremsstrahlung X-ray emission of the galaxy clusters RX\,J1347-1145 (adapted from Ferrari et al. 2011; Chandra and GMRT pointed observations) and RX\,J0603.3+4214 (adapted from van Weeren et al. 2012; WSRT and ROSAT surveys respectively). In the left panel, blue and green contours indicate a high pressure region in the ICM detected through Sunyaev-Z'eldovich (SZ) observations with the millimeter-wave instruments MUSTANG (blue; Mason et al. 2010) and NIKA (green; Adam et al. 2014). Shock related to cluster mergers seem to be the favourite hypothesis to explain the NT emission of radio relics, while relativistic electrons observed in radio halos and mini-halos are generally attributed to turbulence in the ICM. The elongated radio and X-ray emissions towards the high-pression region of  RX\,J1347-1145, though, suggests shock-induced cosmic-rays in addition to stochastic turbulence acceleration. {\em Right panel:} A spectacular multi-wavelength view of RX\,J0603.3+4214 (Image credit: X-ray: NASA/CXC/SAO/R. van Weeren et al; Radio: LOFAR/ASTRON; Optical: NAOJ/Subaru. Adapted from van Weeren et al. 2016). The cluster radio and X-ray emissions (green and purple, respectively) are overlaid on the visible image (white) that also allowed to map the cluster mass distribution from gravitational lensing (blue).}
\end{figure}

\noi Galaxy clusters are the most massive objects in the Universe that have had the time to collapse under the influence of their own gravity; their number density as a function of mass and redshift is used to constrain cosmological parameters (Voit 2005). Located at the nodes of the cosmic web, galaxy clusters are not isolated systems: mass is accreted from the surrounding filamentary structure either at a smooth rate or through mergers with galaxy groups and (sub-)clusters. After a few seminal works, mostly in the 90s (Feretti 1999), in the last decade we have discovered an increasing number (from less than twenty to several dozens) of massive merging clusters that ``light up'' at radio wavelengths due to the presence of a non-thermal (NT) intracluster component, i.e. weak magnetic fields ($\sim$ $\mu$G) and cosmic ray electrons (CRes) in the ICM, giving rise to Mpc-scale synchrotron radiation detected at $\mu$Jy/arcsec$^2$ level around 1.4 GHz. Even if all clusters present some level of radio emission associated to active galaxies, in the following we will refer to ``radio-loud'' clusters as those systems hosting this kind of diffuse radio sources, which are usually classified as ``halos'', ``relics'' or ``mini-halos'' (see Fig.\,\ref{fig:clusters2}), depending on their position, size, morphology and polarisation properties (e.g. Ferrari et al. 2008).

\smallskip

\noi In recent years, increasing attention has been paid to study the link between the NT component of the ICM and the dynamical state and evolution of galaxy clusters. Up to now, most of the detected radio-loud clusters show signatures of recent mergers, with a radio power (P$_{radio}$) that correlates to the mass (M$_{Cl}$) of the cluster (e.g. Martinez-Aviles et al. 2016). These results agree with theoretical models predicting that ICM shocks and turbulence that develop during cluster mergers are responsible for electron acceleration and magnetic field compression (e.g. Dolag et al. 2008): the higher is the mass of colliding clusters, the more gravitational energy is released within the thermal and NT components of the ICM. Other theoretical models for the origin of the NT intracluster component have been proposed (e.g Pfrommer et al. 2008), which are in less good agreement with the observed properties of clusters, in particular due to the lack of gamma-ray emission from the ICM that is predicted by those models. However, with a few dozens of radio-loud clusters at low redshift discovered so far (Feretti et al. 2012), we are most likely observing just the ``tip of the iceberg'' of diffuse cluster radio emission. Deeper observations of richer cluster samples, in particular with SKA pathfinders (such as LOFAR, see Fig.\,\ref{fig:clusters2}), are starting to show the NT emission in exquisite details, often opening new questions on points that seemed well established (e.g. Bonafede et al. 2015a; Sommer et al. 2017). 

\smallskip 

\noi SKA1 will allow us to study diffuse radio emission in statistical cluster samples and in a totally new range of masses and redshifts: the discovery of thousands of radio-loud clusters (up to z$\approx$1 and down to M$_{Cl}\approx10^{14}$ \msol, including $\sim$1000 ultra-steep diffuse radio sources, Ogrean et al. 2015) is predicted by current estimates (Cassano et al. 2015; Ferrari et al. 2015). Thanks to its sensitivity, polarisation purity and resolution, SKA will also permit a step forward in magnetic field measurements. Hundreds of sources in the background of massive clusters are expected to be detected by SKA1, hence entering a new era for the study of the magnetic field through Faraday Rotation Measure studies (Bonafede et al. 2015b). Predictions indicate that SKA1 will be able to recover tiny differences in the magnetic field properties of the ICM, which are far beyond the capabilities of the present instruments. 
\smallskip 

\noi SKA1 surveys will thus allow us to detect clusters through their synchrotron radiation, which will be complementary to more classical cluster detection methods (SZ signal, X-ray emission, galaxy over-densities). Interestingly, simulations predict that accretion and merging shocks taking place around galaxy clusters and within cosmic filaments could accelerate particles and compress magnetic fields. As already mentioned in Sect.\,\ref{science:cosmicB}, low-frequency observations with SKA1-LOW could thus be key also for the detection of intracluster filaments through their synchrotron radiation. This is one of the hottest astrophysical topics: most of the baryons in the Universe are expected to reside in the filaments that connect galaxy clusters, but the warm-hot intergalactic medium (WHIM) is currently largely unconstrained from the observational point of view. In the case of efficient amplification of magnetic fields within filaments, SKA1 should be able to detect a significant portion of emission from the cosmic web at low redshifts (z$\lesssim$0.1). For a better imaging of their full extent or for their detection up to z=0.5, SKA1-LOW sensitivity should be improved by a factor $\gtrsim$3. This can be achieved during the Phase 2. Remarkably enough, even the non detection of radio emission from filaments will offer a way of limiting the magnetisation properties of the cosmic web, thereby providing important clues on the efficiency of the turbulent amplification of early weak cosmological fields during structure formation (Vazza et al. 2015).

\smallskip 

\noi The Square Kilometre Array is expected to revolutionise our knowledge of the NT component of the ICM, similarly to what happened in the early 00's with the study of its thermal component  thanks to the NASA and ESA satellites Chandra and XMM.\\

\parbox{0.9\textwidth}{
\noi{References:}\\
\noi{\scriptsize Adam, R., et al., 2014, A\&A, 569, 66;
Bonafede, A., et al., 2015a, MNRAS, 454, 3391;
Bonafede, A., et al., 2015b, AASKA14, 95;
Cassano, R., et al., 2015, AASKA14, 73;
Dolag, K., et al., 2008, SSRv, 134, 311;
Feretti, L., 1999, dtrp.conf, 3;
Feretti, L., et al., 2012, A\&ARv, 20, 54;
Ferrari, C., et al., 2008, SSRv, 134, 93;
Ferrari, C., et al., 2011, A\&A, 534, 12;
Ferrari, C., et al., 2015, AASKA14, 75;
Mason, B. S., et al., 2010, ApJ, 716, 739;
Martinez-Aviles, G., et al., 2016, A\&A, 595, 116;
Ogrean, G. A., et al., 2015, ApJ, 812, 153;
Pfrommer, C., 2008, MNRAS, 385, 1211; 
Sommer, M. W., et al., 2017, MNRAS, 466, 996;
Vazza, F., et al., 2015, AASKA14, 97;
Voit, M., 2005, RvMP, 77, 20;
van Weeren, R.~J., et al., 2016, ApJ, 818, 204
}}\\

\paragraph{SKA SZ studies of galaxy clusters}
\vspace{0cm}

\noi With the last generation of CMB experiments, many studies have
been done on Galaxy Clusters using the inverse Compton effect of CMB
photons on the hot gas (Sunyaev Zeld'ovich effect, SZ). From blind
detections on the whole sky to particular investigations of shocks in
the intra-cluster medium (ICM), SZ offers many possibilities. This effect
being furthermore insensitive to redshift, it theoretically allows us
to probe high redshift warm-hot gas in structures. The sensitivity and resolution of SKA1-MID Band~5 brings a new window to probe the SZ effect, complementary to many
other observations. Recent observations of SZ clusters have taken
advantage of multi-wavelength observations and the peculiar shape of
the spectral function of SZ at high frequencies: indeed, the spectral
function goes from negative for frequency lower than 217 GHz, null at
217 GHz, to positive above 217 GHz, offering a unique signature. Note
however that the first detections and images of SZ were made at 15 GHz
(Birkinshaw 1984, Jones 1993).

\smallskip

\noi Recent studies based on high resolution X-ray and SZ observations have shown the
complexity of gas density and temperature in the ICM. High resolution
studies of the ICM state by SKA1-MID will allow us to access the
pressure distribution and thus the dynamical state of
clusters. Pressure discontinuities will reveal shocks and/or cold
fronts, signatures of cluster mergers, on a large range of redshifts
allowing us to study the ICM and cluster evolution. These observations
will bring to the next step the current work initiated with Planck
(Planck collaboration 2013b) and done with NIKA and NIKA2 (Adam et al. 2014). In the
2030s, Athena will offer an X-ray complementarity to SKA observations,
bringing the possibility of combined analyses of high resolution, high
sensitivity observations of warm-hot gas not only in clusters but also
in filaments and possibly more diffuse environments. While SKA and SZ
will bring the thermal and pressure properties of clusters, X-ray
spectroscopy from Athena will offer the dynamical state of the
substructures, giving a full and powerful view of the different
physical processes at play in clusters.

\begin{figure}[!ht]
  \centering
  \includegraphics[width=0.55\linewidth]{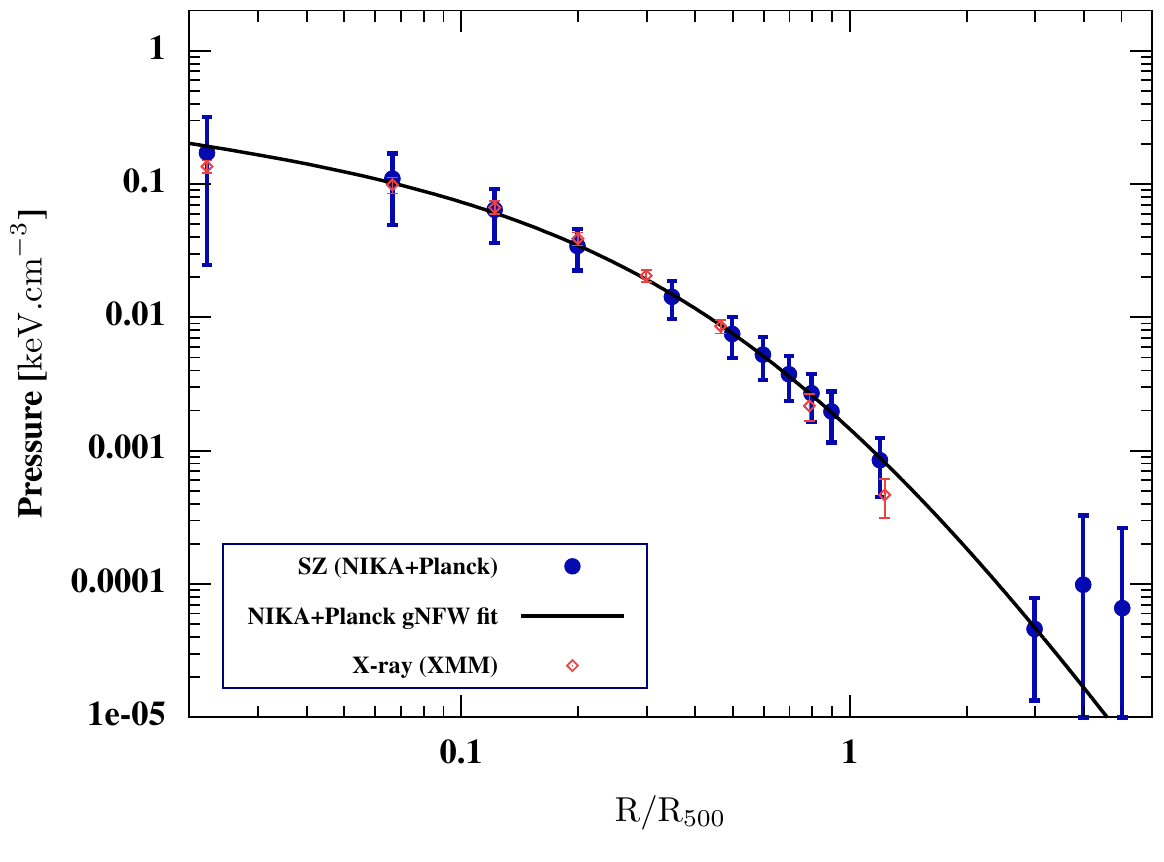}
    \caption{\label{fig:clustersSZ1} Non-parametric pressure profile (blue) deprojected from the NIKA tSZ surface brightness map and the Planck Compton parameter map. A generalised NFW pressure profile model has been fitted on the NIKA+Planck deprojected pressure points (black solid line). XMM-Newton estimated pressure profile (red) based on the deprojected density profile and the temperature estimation from spectroscopic observations. The NIKA/Planck and XMM-Newton estimates are compatible within error bars. Taken from Rupin et al. 2017.}
\end{figure} 

\smallskip 

\noi In addition to thermal and dynamical probes, non-thermal
signatures have shown to be crucial to understand physical properties
of the ICM. Combination of GMRT and MUSTANG (Ferrari et al. 2011) showed a
strong correlation between the high pressure seen in SZ and the excess
of radio surface brightness of the central radio source in
RXJ1347-1145 (see also Fig.\,\ref{fig:clusters2}). This result brought a hint of the presence of cosmic rays
contributing to the excess radio emission coming from shock fronts in
the ICM. SKA1 will thus offer a complete view of the interaction
between thermal and non thermal physical processes through the high
sensitivity of detection of both the synchrotron and SZ emissions.

\smallskip 

\noi With its high sensitivity and high resolution, SKA1-MID will map
the SZ effect from the centre to the more diffuse external regions,
even for relatively high redshift clusters, allowing to reconstruct
the pressure profile on a large range of scales. While combination of
SZ and X-ray was needed until recently (Planck collaboration 2013b), current SZ
observations are improving (see Fig.\,\ref{fig:clustersSZ1}), but SKA1-MID will be a
precious tool to explore the evolution of global properties of
clusters over time. In the coming years, optical (Euclid - Sect.\,\ref{science:euclid} - LSST) and
X-ray (eRosita - Sect.\,\ref{science:athena-erosita}) will detect high redshift clusters that SKA1-MID will
be able to follow up, providing information on the SZ profile and
flux. Combining observations from low and high redshifts of the
complementary experiments will thus allow for cluster property
evolution studies. In addition, as the SZ measurement is a good mass
proxy, SKA1-MID will provide cluster mass estimates in the high
redshift domain, where lensing reconstruction from Euclid and X-ray
estimates from eRosita will be limited. These SKA1-MID estimated
masses will complement the low redshift ones, allowing us to build a
unique cosmological sample over a large range of redshifts and thus to
provide stronger cosmological constraints.

\smallskip

\noi SZ measurements with SKA1-MID will not be free from contaminations
but the versatility of SKA will allow for cleaning. At small scales,
the SKA1-MID (Band~5) SZ signal is contaminated by the radio emission
of active galaxies in the field of view. By using the SKA1-MID long
baseline, the position and flux of the point sources can be measured
with high precision and the radio sources contamination removed. On
larger scales, synchrotron emission of our own Galaxy can also be
removed by observations at different lower frequencies.

\smallskip 

\noi The Square Kilometre Array is thus an efficient probe of the hot-warm gas in structures, provides unique information on high redshift clusters and  complements many other wavelengths surveys in which the French community is involved.\\

\parbox{0.9\textwidth}{
\noi{References:}\\
\noi{\scriptsize
Adam, R., et al., 2014, A\&A, 569, 66;
Birkinshaw, M., et al., 1984, Nature, 309, 34;
Ferrari, C., et al., 2011, A\&A, 534, 12;
Jones, M. E. et al., 1993, Nature, 365, 320;
Planck collaboration, 2013, A\&A, 550, id.A131, 24;
Planck collaboration, 2013, A\&A, 554, id.A140, 19;
Rupin et al., 2017, A\&A 597, A110 
}}\\

\paragraph{Environmental effects on galaxy evolution}
\vspace{0cm}

\noi The environment is a key parameter in galaxy evolution. Since the seminal work of Dressler (1980),
it is now well established that rich environments such as clusters and compact groups are dominated  
by early-type galaxies, while late-type objects are the main galaxy population within the field.
Multifrequency observations have also indicated that the gas content and the activity of star 
formation of late-type systems in rich environments are significantly reduced with respect to their 
field counterparts (e.g. Boselli \& Gavazzi 2006).

\smallskip

\noi Two main families of physical processes have been proposed to explain these differences: gravitational
interactions between galaxies and the potential well of the cluster (Merritt 1983; Byrd \& Valtonen 1990;
Moore et al. 1998), and the dynamical interaction of galaxies moving at high velocity within a hot ($T_{ICM}$ $\sim$ 10$^7$-10$^8$ K)
and dense ($\rho_{ICM}$ $\sim$ 10$^{-3}$ atoms cm$^{-3}$) intra-cluster medium (ICM) (Gunn \& Gott 1972;
Cowie \& Songaila 1977; Larson et al. 1980). All these mechanisms are able to stop the infall of fresh gas, to 
partly or totally remove the cold gas component from the disc, and thus to reduce the activity of star formation of the
perturbed galaxies.

\smallskip

\noi If the importance of the environment in shaping galaxy evolution is now well established, it
is still unclear which, among the proposed mechanisms is the dominant in different density regions, from loose groups to rich clusters,
and at different epochs. Indeed, observations and simulations indicate that 
the densest regions observed in the local universe have been formed through the accretion of smaller structures, where
the different mechanisms might have been already at place (pre-processing, Dressler 2004).

\smallskip

\noi 
The radio spectral domain offers a unique opportunity in the study of the role of the environment on galaxy evolution.
The presence of bent head tail radio galaxies in rich clusters (Rudnick \& Owen 1976) have been the first strong observational evidence of ram pressure.
Radio continuum observations have also revealed the presence of $\sim$ 50 kpc long tails associated to three 
late-type galaxies in the periphery of the cluster A1367 (Gavazzi et al. 1995), witnessing a ram pressure stripping process.
It has been also shown that cluster late-type galaxies do not follow the far-IR radio correlation since have, 
on average, a stronger radio continuum emission than their field counterparts (Gavazzi et al. 1991; Gavazzi \& Boselli 1999; Andersen \& Owen 1995). 
This result has been interpreted as due to the compression of the magnetic field in cluster galaxies 
interacting with the surrounding environment.

\smallskip

\noi 

\begin{figure}[!ht]
  \centering
  \includegraphics[width=0.95\linewidth]{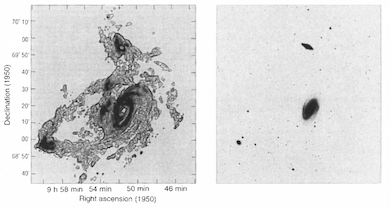}
  \caption{\label{fig:M81} The \hi~gas distribution within the M81 galaxy group derived using the 21 cm emission line ({\em left}) 
  is compared to the stellar distribution obtained in the visible ({\em right}) (from Yun et al. 1994). 
  The \hi~gas component is heavily perturbed by the gravitational interactions between the different group members, 
  whose stellar morphology does not show any evident sign of an ongoing perturbation.}
\end{figure}

\smallskip

\begin{figure}[!ht]
  \centering
  \includegraphics[width=0.8\linewidth]{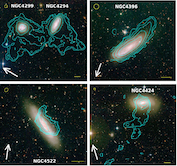}
  \caption{\label{fig:viva} VLA \hi\ contours superposed on NGVS optical images (Boissier et al. 2012),
from the VIVA sample of Virgo galaxies (Chung et al. 2009). The white bar is 1 arcmin = 6 kpc. The arrows indicate the direction of the cluster centre. The \hi\ beam is shown as an ellipse in the {\em top-left} corner of each panel. Observations with SKA1 will be able to see such deformations due to the
environment up to z=1.}
\end{figure}

\noi 
The most spectacular results on the perturbations induced by the environment on galaxy evolution, however, come from
21 cm \hi~observations. The atomic gas, which is generally distributed on discs more extended by a factor of $\sim$ 2 than the stellar disc,
and is thus loosely bounded to the gravitational potential well of galaxies, is easily perturbed during any kind of interaction (Fig.\,\ref{fig:M81}).
There are indeed several examples of galaxies in nearby clusters and groups with \hi~tidal or cometary tails, indicating 
an ongoing stripping process (Yun et al. 1994; Chung et al. 2007; Scott et al. 2012) (cf. Fig.\,\ref{fig:viva}).
It has been also shown that galaxies in groups (Catinella et al. 2013) and clusters (Cayatte et al. 1990; 
Solanes et al. 2001; Chung et al. 2009; Gavazzi et al. 2013) have, on average, less \hi~gas than similar objects in the field.
There is also evidence of molecular gas depletion (Casoli et al. 1998;
Vollmer et al. 2005, 2008; Fumagalli et al. 2009; Scott et al. 2013; Boselli et al. 2014; Jachym et al. 2014).
The lack of gas, principal feeder of star formation, has strong implication in the quenching process and 
thus on the evolution of the perturbed galaxies (Boselli et al. 2006, 2016a).

At intermediate redshifts (z$\le$0.5)
CO detections suggest that diffuse molecular gas may be stripped before the SFR is affected
(Jablonka et al. 2013).
The star formation-density relation is
perhaps reversed at some redshift close to 1 (Elbaz et al. 2007) and
such a reversal should be observed in the gas content of individual and
in the ensemble of galaxies in proto-clusters in the early universe.
Mapping of the \hi\ line in clusters and
proto-clusters will be possible with SKA1 up to z$\sim$1 at low resolution,
and up to z=2 at higher resolution with SKA2.

\medskip

\smallskip

\noi 
The SKA will be a unique opportunity for understanding the role of different environments on galaxy evolution. 
Thanks to its sensitivity, angular resolution, and area coverage (4 $\mu$Jy/beam at 0.5-2 arcsec resolution over 30.000 deg.$^2$
for the proposed all sky continuum survey - Prandoni \& Seymour 2015, $\sim$ 10$^{20}$ cm$^{-2}$ at 15 arcsec resolution over 20.000 deg$^2$
for the \hi~all sky survey - Staveley-Smith \& Oosterloo 2015)
we will be able to map the the radio continuum and \hi~distribution within galaxies
for large samples of local and high-$z$ objects, selected according to similar criteria, and belonging to different environments. 
The different programmed extragalactic wide field surveys will provide large samples of galaxies spanning a wide range in stellar mass 
and morphological type (from giant to dwarfs) and galaxy density (from voids, to the general field, loose and compact groups, and rich clusters), 
and will thus be ideally suited for this purpose. The exquisite sensitivity of SKA, able to rich \hi~column densities of $\sim$ 10$^{-18}$ cm$^{-2}$
for pointed observations, will allow the detection of the stripped gas in perturbed galaxies down to still unreached levels. 
Combined with observations of other gas phases (hot gas from X-rays - Sun et al. 2007, ionised gas from
H$\alpha$ - Boselli et al. 2016b, molecular gas from CO - Jachym et al. 2017) and with tuned simulations (e.g. Roediger \& Hensler 2005, Tonnesen \& Bryan 2012) 
we will be able to study for the first time in a coherent and consistent way the fate of the stripped material in an hostile environment.

\smallskip

\noi 
The study of the role of environment on galaxy evolution is already a key aspect in several precursor projects under way with APERTIF, MeerKAT, and ASKAP, in which the French community is deeply involved.
The Fornax cluster survey (PI P. Serra, MeerKAT) is indeed dedicated to the study of a nearby cluster of galaxies at unprecedented sensitivity, while EMU in radio continuum (R. Norris) and Wallaby in \hi~(PI B. Koribalski and L. Staveley-Smith; both with ASKAP) will provide a sample 
of $\sim$ 500.000 nearby galaxies, including those in rich environments such as the nearby Virgo cluster. Wallaby and EMU will also provide ideal reference samples of isolated galaxies, necessary for every comparison.
These samples are crucial for determining the typical scaling relations of unperturbed objects, to which those derived for galaxies
in different density regions will be confronted.\\

\parbox{0.9\textwidth}{
\noi{References:}\\
\noi{\scriptsize Andersen, V., \& Owen, F.~N., 1995, AJ, 109, 1582;
Boissier, S., \etal, 2012, A\&A 545, A142;
Boselli, A., \& Gavazzi, G., 2006, PASP, 118, 517;
Boselli, A., et al., 2006, \apj, 651, 811;
Boselli et al. 2014, A\&A, 564, 67;
Boselli, A., et al., 2016a, A\&A, 596, A11;
Boselli, A., et al., 2016b, A\&A, 587, A68;
Byrd, G., \& Valtonen, M., 1990, ApJ, 350, 89;
Casoli F., \etal, 1998, A\&A 331, 451;
Catinella, B., et al., 2013, MNRAS, 436, 34; 
Cayatte, V., et al., 1990, AJ, 100, 604;
Chung, A., et al., 2007, ApJL, 659, L115;
Chung, A., \etal, 2009, AJ 138, 1741;
Cowie, L., \& Songaila, A., 1977, Nature, 266, 501;
Dressler A., 1980, ApJ, 236, 351;
Dressler, A., 2004, in Clusters of Galaxies: Probes of Cosmological Structure and Galaxy Evolution, 206;
Elbaz, D., \etal, 2007, A\&A 468, 33;
Fumagalli et al. 2009, ApJ, 697, 1811;
Gavazzi, G., \& Boselli, A., 1999, A\&A, 343, 93;
Gavazzi, G., et al., 1991, AJ, 101, 1207;
Gavazzi, G., et al., 1995, A\&A, 304, 325;
Gavazzi, G., et al., 2013, A\&A, 553, A89;
Gunn, J.~E., \& Gott, J.~R., III 1972, ApJ, 176, 1;
Jablonka, P., \etal, 2013, A\&A 557, A103;
J{\'a}chym, P., \etal, 2014, ApJ 792, 11;
J{\'a}chym, P., et al., 2017, arXiv:1704.00824;
Larson, R., et al., 1980, ApJ, 237, 692;
Merritt, D., 1983, ApJ, 264, 24;
Moore, B., et al., 1998, ApJ, 495, 139;
Prandoni, I., \& Seymour, N., 2015, AASKA14, 67;
Roediger, E. \& Hensler, G: 2005, A\&A, 433, 875;
Rudnick, L., \& Owen, F.~N., 1976, ApJL, 203, L107;
Scott, T.~C., et al., 2012, MNRAS, 419, L19;
Scott, T. C., \etal, 2013, MNRAS 429, 221;
Solanes, J.~M., et al., 2001, ApJ, 548, 97;
Staveley-Smith, L., \& Oosterloo, T., 2015, AASKA14, 167; 
Sun, M., et al., 2007, ApJ, 671, 190;
Tonnesen, S., \& Cen, R., 2012, MNRAS, 425, 2313;
Vollmer, B., \etal, 2005, A\&A 441, 473;
Vollmer, B., \etal: 2008, A\&A 491, 455;
Yun, M.~S., et al., 1994, Nature, 372, 530
}}\\

\subsubsection{Formation and evolution of galaxies}

\paragraph{Unobscured star formation and cosmic star formation history}\label{science:unobscured}
\vspace{0cm}

\noi Several surveys of SKA1 are dedicated to the measure of the star formation rate (SFR) from radio continuum in order to trace the star formation history of the Universe (SKA1-MID). Radio continuum observations give access to the only SFR measurements unaffected by dust. The ultraviolet (UV) and optical recombination lines are directly linked to the emission from young stars but suffer from dust extinction. On the opposite side the infrared (IR) emission from dust misses the unobscured starlight, leading to an underestimate of the SFR which can be severe in the earliest phasesTable of galaxy formation, or in metal and dust poor environments. These issues have been identified for a long time and hybrid star formation tracers combining UV-optical and IR luminosities are used at the expense of complex selection functions (e.g. Kennicutt and Evans, 2012).
\smallskip

\noi Deep radio continuum surveys with SKA1 will provide a very powerful alternative with the detection of synchrotron and free-free continuum emissions up to very high redshifts (Mancuso et al. 2015). The continuum surveys in Band 2 will be sensitive to  the synchrotron emission whereas the deep surveys in Band 5 will sample a frequency range where the free-free emission from the gas ionised by young stars is expected to dominate the synchrotron component in strong starbursting objects and provide a very direct measure of the current SFR (Murphy et al. 2015, Fig.\,\ref{fits} below). 

\smallskip

\begin{figure}[!ht]
  \centering
  \includegraphics[width=0.60\linewidth]{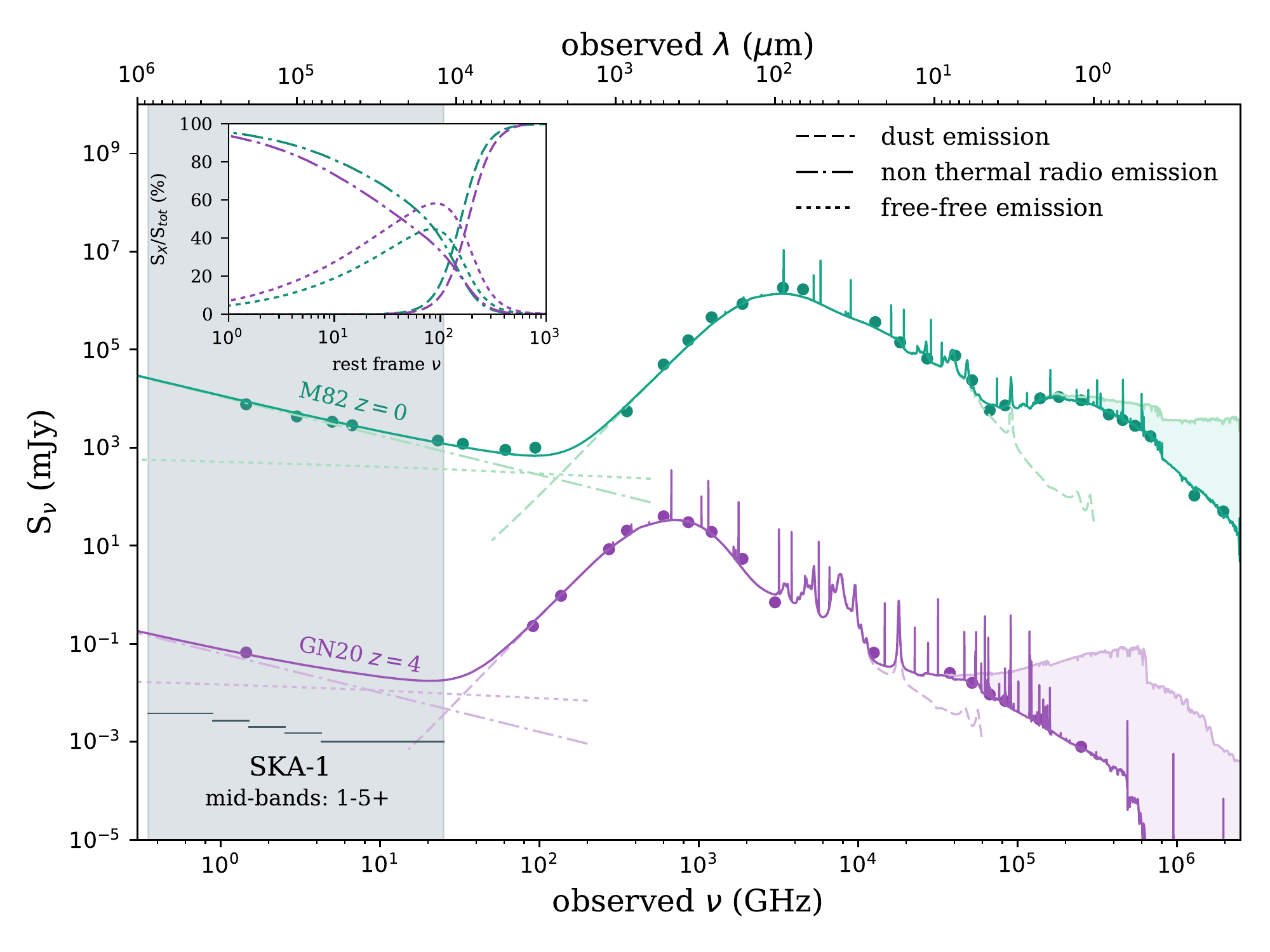}
  
  \caption{\label{fits} SED of the  starburst galaxies Messier 82 ($z$$=$0) and GN20 ($z$$=$4) fitted with the code \href{http://cigale.lam.fr/}{\color{blue} \myul[blue] {CIGALE}}. The best model (solid line), the unattenuated SED (shaded area on the right), and the observed fluxes (dots) are displayed in green and purple for M82 and GN20, respectively. The different  components are detailed in the legend inside, the contribution of each of them as a function of rest frame frequency is plotted in the insert panel. The stellar mass, SFR, star formation history, dust mass or energy budget are output parameters of the fitting analysis. The observed data for Messier 82 are taken from the NASA/IPAC extragalactic database, the data for GN20 are courtesy of E.~Daddi. The grey shaded area indicates the frequency range probed by the SKA1 mid-Bands 1 to 5+.}
\end{figure}

\noi The targeted fields of these surveys will be selected in synergy with other imaging and spectroscopic surveys from optical to submm (Euclid, LSST, DES, Prandoni and Seymour 2015; see also Sect.\,\ref{science:synergies}):  redshift determinations are required for any study of galaxy evolution. The analysis of the entire electromagnetic spectrum from UV to radio will yield to the measure of physical parameters like the stellar mass, the recent star formation history,  the dust mass or the amount of obscuration. Continuum and line emissions from the ionised gas trace the very recent star formation, in the last few Myrs, when UV, synchrotron emission, and dust emissions are sampling larger timescales  ($\sim$100\,Myrs, Kennicutt and Evans, 2012, their Table 2): the comparison of the different emissions will provide key information on the age of the starburst event. The shape of the radio continuum itself results of the combination of different emissions: the synchrotron, the free-free, and the cold dust emissions, each of them related to star formation activity. The link of each of these emissions to the SFR depends on the star formation history and ISM properties of the galaxies (Oti-Floranes et al. 2010; Boquien et al. 2014). By combining shallow and deep SKA1 surveys, the main sequence of star-forming galaxies (SFR-stellar mass correlation) and its dispersion can be  studied up to high redshifts and/or low mass for individual targets with a very robust SFR estimate. The best way to derive such physical characteristics, and their evolution with redshift, is  to fit  the full spectral energy distributions (SEDs) with models including all the physical processes at work. In Fig.\,\ref{fits} the SED of two dust-obscured starburst galaxies, M82 at $z$$=$0 and GN20 at $z$$=$4, are plotted with the best fits obtained by combining stellar, dust, nebular, and synchrotron emission. AGN emission is also a possible contributor to the radio continuum and a careful modelling of this emission over the whole SED will open the field of the co-evolution between star formation and black hole growth. The addition of  X-ray measurements  from the future \textit{Athena} observatory  will dramatically improve further this research field (Sect.\,\ref{science:athena-erosita}).

\smallskip

\noi The  MeerKAT deep survey MIGHTEE Tier 2 will cover 35 deg$^2$ at 1.4\,GHz down to a 5$\sigma$ limit of 5\,$\mu$Jy for unresolved sources  similar to  the wide survey of 1000 deg$^2$ planned for SKA1, but with an 8 times higher spatial resolution: the SKA1 survey will dramatically improve the statistics and the quality of  the cross-identification with sources at other wavelengths. The deepest SKA1 survey will detect SFR of the order of 100 $\rm M_{\sun} yr^{-1}$ at $z$$\sim$7. 
A direct outcome from the unique characteristics of the SKA1 survey will be the ability to constrain the average SFR along cosmic time in a statistically robust way, tremendously improving the situation thanks to its efficiency in surveying the millimeter sky, compared to
ALMA, for instance (e.g., Dunlop et al. 2017).

Since the seminal works by Lilly et al. (1996) and Madau (1996),
intensive efforts have been carried on in establishing the SFR Density
(SFRD) evolution with all possible SFR tracers (Madau \& Dickinson
2014). Compilation of numerous works shows that the evolution of the
SFRD with redshift exhibits a bell shape with a peak possibly located
at 1$<$$z$$<$3. While the increase of the SFRD at 0$<$$z$$<$1 is well
established, the position of the SFRD peak varies from $z$$\sim$1 in
FIR (Gruppioni et al. 2013) to $z$$\sim$3 in radio (Novak et
al. 2017).  Despite an apparent agreement in the SFRD measurements at
3$<$$z$$<$8-9, the situation is even more uncertain given the difficulty to
correct the UV light for dust attenuation (Burgarella et al. 2013),
with UV being currently the only SFR tracers for the bulk of the
galaxy population at $z$$>$3.


The radio emission has been already used to derive the SFRD out to $z$$\sim$5
with 6000 individual detected galaxies using VLA over 2 deg$^2$ (Novak
et al. 2017). Given the 35 deg$^2$ which will be covered with the deep
survey of MeerKAT with a rms twice better, we will be able to reduce
significantly any uncertainties linked to cosmic variance and
improve the constraint on the SFRD peak. Figs.\,2 and 3 of Jarvis et
al. (2014) demonstrate the capability of SKA1 to constrain the radio luminosity function
well below the characteristic peak (reaching SFR$\sim$10\,M$_{\sun}$
yr$^{-1}$ for $z$$\sim$3 with SKA1 deep) with an incredible sample of
detected galaxies (up to $>$50 millions with SKA1 Wide) improving the
situation by two orders of magnitude, both in term of depth and sample
size. Therefore, SKA1 deep will produce an unique picture of the SFRD
evolution, sampling efficiently the location of the SFRD peak, and
establishing the mass assembly rate in the first half of the Universe
life using dust-free tracers, which has never been done.

Given the high resolution of radio data, allowing for an easy comparison 
with ancillary data, stacking analysis is a powerful tool in radio to push the 
limits of studies to lower SFR and higher redshifts. For instance, 
Karim et al. (2011) stacked VLA 1.4\,GHz radio data proving it to 
be extremely efficient to constrain the evolution of the SFRD at $z$$<$4 
with only 2deg$^2$ observed with a sensitivity of 10\,$\mu$Jy.
SKA1 deep will allow a gain better than
two order of magnitude in luminosity over a similar area.  Given the
richness of the multi-colour data which will be available (e.g. LSST,
Euclid), stacking will become a standard and essential tool. Combining stacking and
direct detections, we will get one of the most accurate view of the cosmic
star formation history.\\
 
\parbox{0.9\textwidth}{
\noi{References:}\\
\noi{\scriptsize
Boquien, M., \etal., 2014, A\&A, 571, A72;
Burgarella, D., \etal, 2013, A\&A, 554, 70;
Dunlop, J., \etal, 2017, MNRAS, 466, 861;
Gruppioni, C., \etal, 2013, MNRAS, 432, 23;
Karim, A., \etal, 2011, ApJ, 730, 61; 
Kennicutt, R. C. \& Evans, N.J., 2012, ARA\&A, 50, 531;
Lilly, S.J., \etal, 1996, ApJ, 460, 1;
Madau, P., \etal, 1996, MNRAS, 283, 1388;
Madau, P. \& Dickinson, M.E., 2014, ARA\&A, 52, 415;
Mancuso , C., \etal., 2015, ApJ 810, 72;
Murphy, E.J.,  etal, 2015, AASKA14, 85;
Novak, M., \etal, 2017, arXiv170309724;
Oti-Floranes, H. \& Mas-Hesse, J.M., 2010, A\&A, 511, A61;
Prandoni, I. \& Seymour, N., 2015, AASKA14, 67
}}\\

\paragraph{Large samples of \hi\ galaxies}
\vspace{0cm}

\noi The total amount and internal distribution and dynamics of neutral gas in galaxies are essential features to understand galaxy evolution. The \hi\ 21\,cm line has been instrumental in discovering the physics of nearby galaxies, in particular the influence of environment (Chamaraux et al. 1980; Chung et al. 2009) and the importance of dark matter (Bosma 1981; Noordermeer et al. 2007; McGaugh 2012). However, the weakness of the \hi\ line makes it very difficult to detect its emission at redshifts above z$\sim$0.2, and our knowledge of neutral gas at high z is only based on its molecular phase, traced by CO emission, thanks to the rotational ladder of the molecule (e.g. Tacconi et al. 2013; Carilli \& Walter 2013). 
Due to ill-determined conversion factors, the exact relative fraction of the two gas phases,
\hi\ and H$_2$, is not known at z$>$0.3 and indications from damped Ly$\alpha$ absorption lines (Noterdaeme et al. 2012) are in contradiction with semi-analytical models (Lagos et al. 2012). 

\begin{figure}[!ht]
  \centering
  \includegraphics[width=0.90\linewidth]{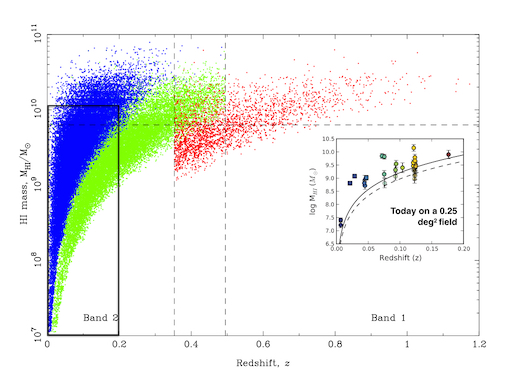}
  \caption{\label{fig:HISurveys}  A simulation showing the distribution with redshift of \hi~masses of galaxies likely to be detected with SKA1 in the representative tiered set of 1,000 hr surveys discussed by Staveley-Smith \& Oosterloo (2015). Blue corresponds to the ``medium wide'' survey, green to the ``medium deep'' and red to the ``deep''. The turn-over of the \hi~mass Schechter distribution function M(HI)* at z=0 is displayed as the dashed horizontal line. The diagram inserted at {\em right}, corresponding to the black rectangle inserted at {\em left}, shows what is possible to do now, in 50 hours of observations with the JVLA, from the pilot survey of Fern\`andez et al. (2013).}
\end{figure}

\smallskip
\noi 
In recent decades, large 21\,cm surveys have been performed to derive a statistical knowledge of \hi\ in nearby galaxies (z $\leq$0.05) such as HIPASS (\hi\ Parkes All Sky; cf. Barnes et al. 2001) and ALFALFA (Arecibo Legacy Fast ALFA; cf. Giovanelli et al. 2005). To extend our knowledge to higher redshifts, stacking techniques are used to estimate mean galaxy \hi\ contents (e.g. Lah \etal\, 2007; Delhaize et al. 2013).
In the near future, \hi\ in tens of thousands of galaxies will be detected at higher redshifts, up to z=0.2-0.5, with SKA precursors such as the Australian Square Kilometre Array Pathfinder (ASKAP, Johnston
et al. 2008) and MeerKAT (de Blok et al. 2009). With the first phase of the Square Kilometre Array, SKA1, it will be possible to detect galaxies up to z=1.7 (see Fig.\,\,\ref{fig:HISurveys} from the simulations by Staveley-Smith \& Oosterloo, 2015), and to obtain reasonable maps for a large fraction of them, to investigate outflows, environmental deformation and dark matter distributions. Eventually, with the higher sensitivity of SKA2, it will be possible to determine the \hi\ content of a myriad of galaxies up to z$\sim$2, and to map Milky-Way type galaxies up to z$\sim$1 (Blyth et al. 2015). 
This corresponds to a look-back time covering a large fraction of the age of the Universe, up to 3 Gyr after the Big Bang, when galaxies assemble their mass, and therefore will provide significant, new insights in galaxy formation and evolution.

\smallskip
\noi 
{\bf \hi\ mass functions: accretion and winds} 

\smallskip

\noi If early galaxies are formed with a higher surface density, due to the high
density of the Universe at the beginning of its expansion, one can expect a more
efficient transformation of atomic into molecular gas in galaxy centres. However,
outside the optical disk where stars form efficiently, there must exist huge
reservoirs of atomic gas, linked to cosmic filaments, fuelling the mass assembly
of galaxies. Cosmological simulations by tying the growth of dark matter structure to simple gas physics, show that much of the gas in galaxies is accreted
cold in filamentary structures (Keres et al. 2005; Dekel et al. 2009).
To understand and quantify these processes, it is important to establish
the \hi\ mass function of galaxies as function of redshift, and follow the fate of 
the gas, by comparison with the molecular evolution. The detailed balance, including emission and
absorption, will show the importance of gas accretion and outflowing winds
(e.g. Bouch\'e et al. 2012).

\medskip

\smallskip
\noi 
{\bf Rotation curves and dark matter content } 

\smallskip

\noi In the nearby universe, it is well known that the dark matter (DM) fraction in galaxies
decreases with stellar mass (e.g. Salucci \& Burkert 2000). Galaxies like the Milky
Way or more massive are not dominated by DM in their optical disk, contrary to dwarfs.
Only \hi\ gas is able to reveal their rotation curve outside the optical disk,
for an accurate determination of their dark matter content. At high redshift, limited
spatial resolution and sensitivity prevent the study of \hi\ rotation curves for dwarfs, and 
H$\alpha$ or CO velocity fields, confined to the optical radius, are unable to reveal the dark matter content of massive 
galaxies. Detailed studies of \hi\ kinematics with the SKA will make a breakthrough here.
What is the fraction of baryons in galaxies as a function of redshift? Is this
fraction closer to the universal fraction of 17\%, while supernovae and AGN feedback
have been implicated in reducing it to less than 4\% today?
Obtaining rotation curves extending to the outskirts is time consuming at
high z, but much will be learned from a well-selected sample of $\sim$ 100
galaxies. A deep integration of 100h with SKA1-MID, with 15kpc spatial resolution 
(2.4 arcsec at z=0.5 and 1.8 arcsec at z=1 and beyond), will be
required to reach \hi\ column densities of 4.1 and 7.5 10$^{20}$ cm$^{-2}$ respectively,
at 5$\sigma$, in 20km/s channels.  The study
of a more significant sample could be done only at lower resolution and
higher sensitivity with SKA1 and only then by exploiting the relation
between luminosity and rotation speed (i.e., Tully-Fisher relation).
Studying galaxies at even higher redshifts, up to z=2, will have to wait
for the sensitivity provided by SKA2.

\smallskip
\noi
{\bf The growth of angular momentum in galaxies }
\smallskip

\noi The angular momentum of nearby disk galaxies is about
that predicted for their dark matter halos (Steinmetz \& Navarro 1999).
How can a dissipative and non-dissipative matter have the same specific
angular momentum, when one is found on much smaller scales than the
other?  Models of the cosmological accretion of gas driven by
the growth of large scale structure have made significant progress in
solving this issue (Danovich et al. 2015) but even then, these models do
not follow the gas down to galaxy (kpc) scales. 
The angular momentum as a function of redshift, should vary as
$j(z) \propto (1+z)^{3/2}$
(Mo et al. 1998). This is something we can test rather uniquely with SKA
observations of many galaxies in \hi. 

\smallskip
\noi \hi\ is observed in galaxies with a small range of column
densities, from about 0.1 to 10 M$_\odot$ pc$^{-2}$, it extends
well beyond the optical disk, and obeys a
size-\hi\ luminosity relation 
(e.g., Broeils \& Rhee 1997). The cosmic \hi\ density
may not evolve significantly with redshift (Zafar et al. 2013) suggesting
that at a minimum, the \hi\ content of galaxies does (Lehnert et al. 2015).
SKA, in a deep integration of a well-studied field with galaxy
morphologies, stellar masses, and estimated scale lengths, will not only
provide \hi\ luminosities, but also rotation speeds and a constraint on
the total angular momentum of a significant number of galaxies across
a wide range of redshifts (up to z$\sim$1 for SKA1 and z$\sim$2 for SKA2). 

\smallskip
\noi
{\bf Low Surface Brightness and dwarf galaxies }

\smallskip

\noi Low Surface Brightness galaxies (LSBs), with disk central surface
brightness typically below 23 mag arcsec$^{-2}$ in the optical, could
represent a large fraction of galaxies. Their evolution and role in the
big picture of the formation and evolution of galaxies is still under debate however. 
The class of ``giant'' LSBs is characterised by huge
\hi\ masses (above 10$^{10}$ M$_\odot$, e.g. Matthews et al. 2001) and large sizes
(over hundreds of kpc).

\smallskip 

\noi Recent progress in optical detectors and telescopes have allowed us to
discover extended stellar disks around ''normal'' galaxies pertaining to
this class (e.g. Hagen et al. 2016), and to re-observe the prototypical
giant LSB Malin 1 in the nearby universe (Boissier 
et al. 2016). Disseau \etal\, (2017) found giant LSBs at redshifts of about 0.5.
Giant LSBs are usually gas-rich; \hi\ measurements at different redshifts
will be necessary to understand their evolution. Malin 1 has an \hi\
column density varying from 5 to 2 10$^{10}$ atoms/cm$^2$ from the centre to a radius
of 100 kpc (Lelli et al. 2010) and a \hi\ mass of 10$^{10.66}$ M$_\odot$.
According to the simulations of Staveley-Smith \& Oosterloo (2015), this
mass can be detected up to redshift 0.5 in wide/medium deep surveys, and
up to 1.2 in a deep survey with SKA1 (adopting a column density
threshold of 2 to 0.2 x 10$^{20}$ atoms/cm$^2$, a resolution between 2 and 10
arcsec).
Malin 1 being extremely large, its radius at this column density threshold
corresponds to the resolution of the aforementioned wide/medium surveys proposed
at redshift about 0.5/1.35. SKA1 will allow us to quantify the presence of massive 
\hi\ disks and their evolution up to these intermediate redshifts. In the meantime, the
WALLABY survey on ASKAP will detect Malin 1-like galaxies up to 1 Gpc.\\

\parbox{0.9\textwidth}{
\noi{References:}\\
\noi{\scriptsize
Barnes, D. G., \etal, 2001, MNRAS, 322, 486;
Blyth, S., \etal, 2015, AASKA14, 128;
Boissier, S., \etal, 2016, A\&A 593, A126;
Bosma, A., 1981, AJ 86, 1825;
Bouch\'e, N., \etal, 2012, MNRAS 419, 2;
Broeils, A. H. \& Rhee, M.-H, 1997, A\&A 324, 877;
Carilli, C. L. \& Walter, F., 2013, ARAA 51, 105;
Chamaraux, P., \etal, 1980, A\&A 83, 38;
Danovich, M., \etal, 2015, MNRAS 449, 2087;
de Blok, E., \etal, 2009, in Panoramic Radio Astronomy: 
Wide-field 1-2 GHz Research on Galaxy Evolution;
Dekel, A., \etal, 2009, Nature 457, 451
Delhaize, J., \etal, 2013, MNRAS, 433, 1398;
Disseau, K., \etal, 2017, MNRAS 466, 2337;
Fern\`andez, X., \etal, 2013, ApJ 770, L29;
Giovanelli, R., \etal, 2005, AJ, 130, 2598;
Hagen, L. M. Z., \etal, 2016, ApJ 826, 210;
Johnston, S., \etal, 2008, Experimental Astronomy, 22, 151;
Keres, D., \etal, 2005, MNRAS 363, 2;
Lagos, C. d. P., \etal, 2012, MNRAS, 426, 2142;
Lah, P., \etal, 2007, MNRAS, 376, 1357;
Lehnert M. D., \etal, 2015 A\&A 577, 112;
Lelli, F., \etal, 2010, A\&A 516, A11
Matthews, L. D., \etal, 2001, A\&A 365, 1;
McGaugh, S. S., 2012,  AJ 143, 40;
Mo, H. J., \etal, 1998, MNRAS 295, 319;
Noordermeer, E., \etal, 2007, MNRAS 376, 1513;
Noterdaeme, P., \etal, 2012, A\&A, 547, L1;
Salucci, P. \& Burkert, A.: 2000, ApJ 537, L9;
Staveley-Smith, L. \& Oosterloo, T: 2015, AASKA14, 167;
Steinmetz, M., \& Navarro, J., 1999, ApJ 513, 555;
Tacconi, L. J., \etal, 2013, ApJ, 768, 74;
Zafar, T., \etal : 2013 A\&A 556, A141
}}\\

\paragraph{Molecular gas content at high redshift from the CO lines}\label{science:CO}
\vspace{0cm}

\noi Taking the census of gas reservoirs in galaxies across cosmic times is essential to understand the star formation in the Universe. In addition to the atomic gas (\hi) observed via the 21\,cm line, SKA will probe the molecular gas (H$_2$) through the redshifted $^{12}$CO J=1-0 transition ($\nu_{\rm rest}$ = 115\,GHz), which is a standard tracer of molecular gas reservoirs. Characterising these cold gas reservoirs is particularly important to understand the evolution of galaxies, as they are the fuel for star formation.

\smallskip

\noi It is now well established that high-redshift galaxies have much higher star formation rate at fixed stellar mass than low-redshift galaxies (e.g. Daddi et al. 2007; Elbaz et al. 2011; Schreiber et al. 2015), and that ultra luminous infrared galaxies (with SFR$>$100\,M$_\odot$) have a strong contribution to the star formation density at z$>$2 (e.g. Caputi et. al. 2007; Magnelli et al. 2013). The origin of their intense star formation could be due either to large gas reservoirs or to a temporary increase of the star formation efficiency induced by mergers (e.g. Daddi et al. 2010; Genzel et al. 2010). 

\smallskip

\noi At z$<$3, massive follow-up observations of optically or near-IR selected objects in CO were performed with millimetre interferometers (e.g. Tacconi et al. 2013). These observations revealed a quick rise of the gas fraction with increasing redshift.  At higher redshift, no large systematic CO searches were performed. However, observations of extreme starbursts (e.g. Riechers et al. 2013) or lensed galaxies (e.g. Aravena et al. 2016) showed that intensely star-forming galaxies at high redshift are a mix of extreme starbursts and gas-rich objects. Some pioneering deep line search observations were performed with PdBI and ALMA
(e.g. Decarli et al. 2014; Walter et al. 2016) and detected $\sim$10 sources in CO mostly below z=3. 

\smallskip

\noi Most of these studies were not based on the fundamental (1-0) transition, which is the most reliable transition to derive a gas mass, but on higher J (2-1, 3-2...) transitions. The ratio between the fundamental and the higher-J transitions can vary significantly depending on the galaxy type (Carilli et al. 2013). Consequently, gas mass estimates based on these higher-J transitions are not fully reliable. SKA1 will be able to target directly the CO 1-0 transition above z=7.3. SKA2 with its coverage of higher frequencies will be able to probe this line at z$>$3.8. 

\smallskip

\begin{figure}[!ht]
  \centering
	\begin{tabular}{cc}
	 \includegraphics{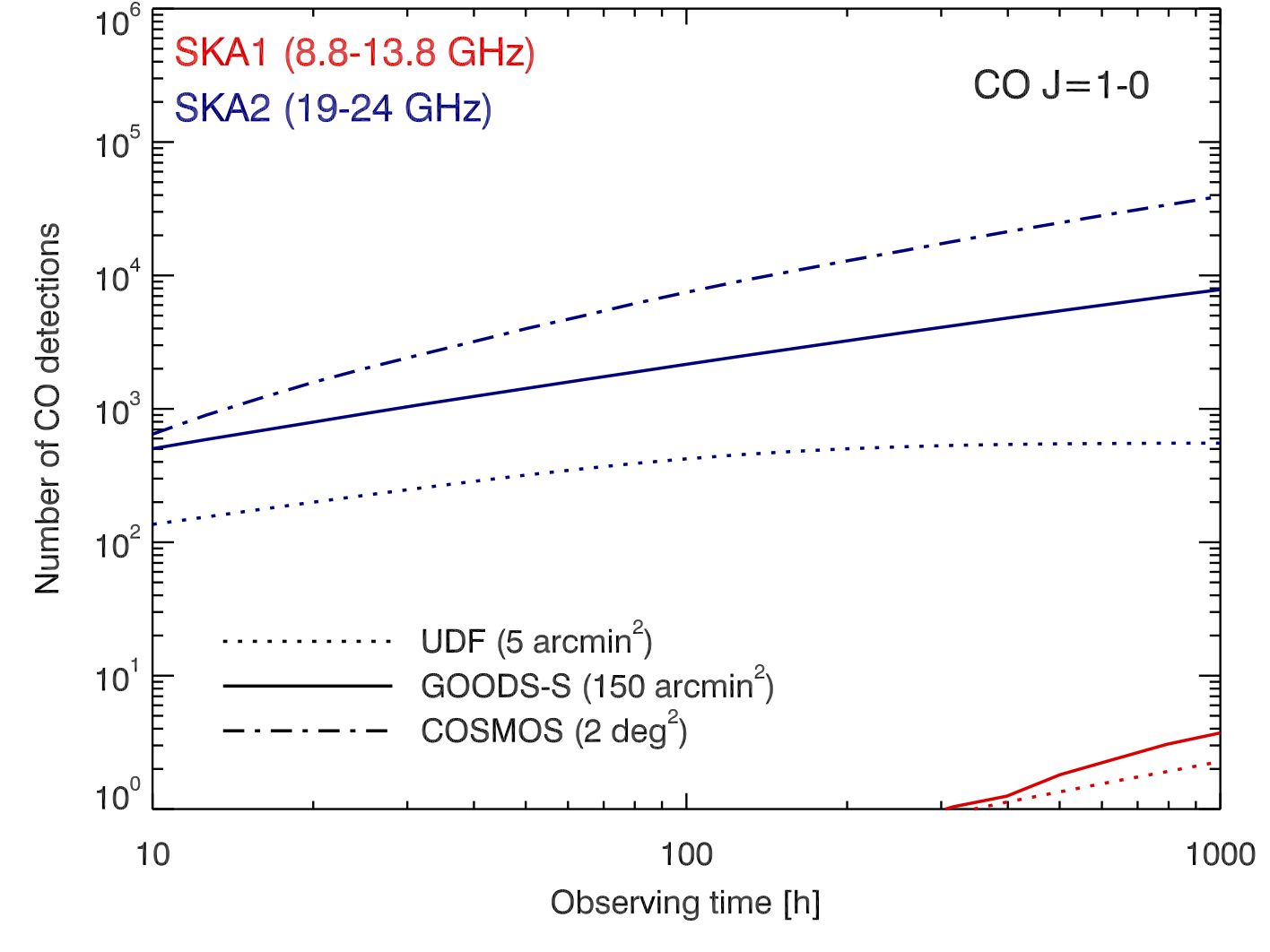} & \includegraphics{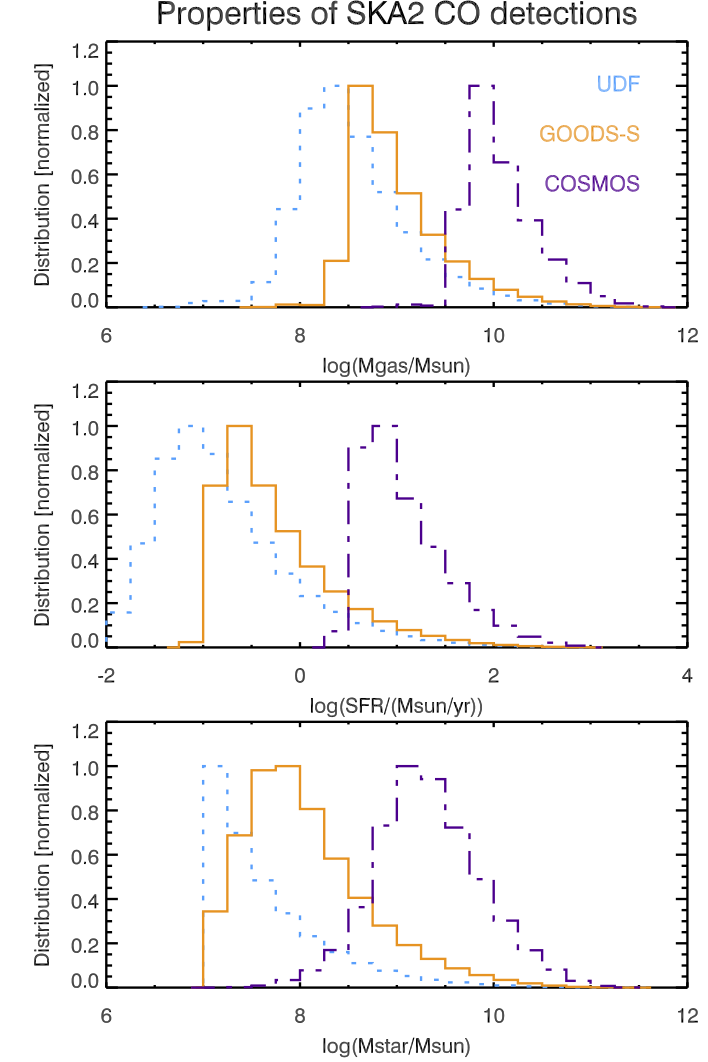} \\
	\end{tabular}	
  \caption{\label{fig:predictionsCO} {\em Left panel}: Number of expected CO 1-0 detections with SKA1 (8.3-13.3\,GHz, 7.3$<$z$<$12.1, red) and SKA2 (19-24\,GHz, 3.8$<$z$<$5.1, blue) as a function of observing time. We performed predictions for 3 different survey sizes: UDF (5\,acrmin$^2$, dotted line), GOODS-S (150\,acrmin$^2$, solid line), and COSMOS (2\,deg$^2$, dot-dash line). There is less than 1 detection expected in average in 1000\,h in COSMOS with SKA1 and it is thus not visible in this plot. {\em Right panels}: properties ({\em top}: gas mass, {\em middle}: SFR, {\em bottom}: stellar mass) of CO-detected galaxies for 3 different 1000-h survey sizes. }
\end{figure}

\smallskip

\noi We present here some predictions of the number of CO detections expected with SKA for various survey strategies. They are based on the Simulated Infrared Dusty Simulated Sky (SIDES, B\'ethermin et al. 2017). The gas content and the CO line flux of the simulated galaxies are derived following Sargent \etal (2014). Line widths are derived using Harris et al. (2012) relation between L'$_{\rm CO}$ and the FWHM. From these predictions and the SKA1 science requirements, we estimate the number of CO 1-0 detections at 5$\sigma$ for different observational strategies. These predictions do not take into account the effect of the CMB on the observed line flux, which can reduce the observed line flux by a factor of 2 (Da Cunha et al. 2013; Zhang et al. 2016).

\smallskip

\noi SKA1 will be a powerful instrument for observing CO 1-0 in high-z galaxies identified by ALMA (probably with [OIII] and [CII], see e.g. Laporte et al. (2017) galaxy at z=8.3) or by JWST. For instance, the z=8.3 source (SFR = 20\,M$_\odot$/yr, FWHM = 50\,km/s, magnified by a factor of 1.8) will be detectable in only $\sim$1\,h. However, as shown in Fig.\,\ref{fig:predictionsCO} ({\em left}, red curves), SKA1 will not be able to perform competitive deep spectral scans, since it can detect less than one source per 100\,h. This is caused by the absence of high frequency coverage, which forces to target to be at z$>$7.3, where the probed volume and the number density of gas-rich galaxies are both very small.

\smallskip 

\noi In contrast, with its coverage of higher frequencies and its sensitivity improved by more than one order of magnitude (from 190 to 2500 antennae), SKA2 will be a powerful CO 1-0 survey machine able to detect tens of thousands of galaxies in 1000\,h (Fig.\,\ref{fig:predictionsCO}). The exact number of detections varies slightly with the strategy and a field size similar to the COSMOS survey (2\,deg$^2$) is predicted to be the most efficient. In a 1000\,h survey, we could thus detect $\sim$40\,000 galaxies over 2\,deg$^2$ between z=3.8 and z=5.1 with a typical stellar mass of 10$^{9.5}$\,M$_\odot$ and a star formation rate of 10\,M$_\odot$/yr (see Fig.\,\ref{fig:predictionsCO}, {\em right} panels). These unprecedented performances will open a new era for statistical CO studies above z=4. In addition, deeper fields covering $\sim$100\,arcmin$^2$ could detect progenitors of Milky-Way-like galaxies at z$\sim$5. Finally, the size of these spectroscopic samples will be sufficient to perform 3-dimension clustering measurements and to constrain their typical mass of the host dark-matter halos, which is a critical information for models of galaxy evolution.

\smallskip

\noi SKA will have strong synergies with ALMA (see also Sect.\,\ref{science:alma}), which will be able to probe complementary tracers of interstellar medium (ISM) at higher frequencies. It is particularly important to calibrate the $\alpha_{\rm CO}$ conversion factor between the CO 1-0 line luminosity and the gas mass. Combining the CO 1-0 from SKA with higher-J transitions, we will construct the CO SLED (Spectral Line Energy Distribution, e.g. Weiss et al. 2007; Daddi et al. 2015) and put better constrain on the properties of molecular gas in z$>$3 galaxies. Robust constraints on the various phases of the ISM will be obtained by combining CO with other ISM tracers: [NII] and [OIII] from ALMA for the ionised gas, [CII] from ALMA for the cold neutral medium and photo-dominated regions, \hi\ from SKA and [CI] from ALMA for the atomic gas, and HCN and CS from SKA (and ALMA for high-J transitions) for the dense molecular gas (see Sect.\,\ref{science:otherlines}).\\

\parbox{0.9\textwidth}{
\noi{References:}\\
\noi{\scriptsize Aravena, M., et al.\ 2016, \mnras, 457, 4406;
Bethermin, M., et al.\ 2017, arXiv:1703.08795, sub. to \aap;
Caputi, K.~I., et al.\ 2007, \apj, 660, 97;
Carilli, C.~L., \& Walter, F.\ 2013, ARA\&A, 51, 105;
da Cunha, E., et al.\ 2013, \apj, 766, 13;
Daddi, E., et al.\ 2007, \apj, 670, 156;
Daddi, E., et al.\ 2010, \apjl, 714, L118;
Daddi, E., et al.\ 2015, \aap, 577, A46;
Decarli, R., et al.\ 2014, \apj, 782, 78;
Elbaz, D., et al.\ 2011, \aap, 533, A119;
Genzel, R., et al.\ 2010, \mnras, 407, 2091;
Harris, A.~I., et al.\ 2012, \apj, 752, 152;
Laporte, N., et al.\ 2017, \apjl, 837, L21;
Magnelli, B., et al.\ 2013, \aap, 553, A132;
Riechers, D.~A., et al.\ 2013, Nature, 496, 329;
Sargent, M.~T., et al.\ 2014, \apj, 793, 19 ;
Schreiber, C., et al.\ 2015, \aap, 575, A74;
Tacconi, L.~J., et al.\ 2013, \apj, 768, 74;
Walter, F., et al.\ 2016, \apj, 833, 67;
Wei{\ss}, A., et al.\ 2007, \aap, 467, 955;
Zhang, Z.-Y., et al.\ 2016, RSOS, 3, 160025}}\\


\paragraph{Extragalactic spectral lines but \hi\, and CO}\label{science:otherlines}
\vspace{0cm}

\noi The SKA will be able to cover high frequencies in the cm range, up to 16 GHz in its first phase SKA1, and up to 24 GHz in the second phase SKA2. In this range, one of the main lines to observe is the CO(1-0), redshifted from z=3.8 and higher, since the first line of the ladder is required to better estimate the CO-to-H$_2$ conversion ratio in galaxies. CO lines are discussed in Sect.\,\ref{science:CO}, and here we concentrate on other molecular lines, such as HCN, HCO$^+$, HNC, CS, useful as dense gas tracers in redshifted galaxies (e.g. Wagg et al. 2015), H$_2$O, OH, which maser lines help to trace nuclear regions at very high spatial resolution, either AGN or intense starbursts (e.g. Braatz et al. 2010; Darling \& Giovanelli 2002), and recombination lines in nearby galaxies, tracing the physics of diffuse cold gas (e.g. Oonk et al. 2015). Absorption lines in front of background radio sources will yield precious clues on the dynamics and chemistry of the intervening galaxies, and help to constrain the variations of fundamental constants (e.g. Wiklind \& Combes 1996; Combes \& Wiklind 1999; Muller et al. 2014).

\smallskip
\noi 
{\bf Dense gas tracers in redshifted galaxies}

\smallskip
\noi 
In the last decade, lots of efforts have been done to investigate the gas content of main sequence galaxies, forming stars at a sustained rate, with a depletion time of the order of 1-2 Gyr (e.g. Daddi et al. 2010; Tacconi et al. 2013). The observation of several CO lines has been used to derive the gas excitation, and estimate the CO-to-H$_2$ conversion ratio. However, these estimations are still quite uncertain, and one way to progress is to observe dense gas tracers, such as HCN, HCO$^+$ or CS, to probe the physical conditions (density, temperature) of the gas. This has been possible in rare cases of huge starbursts, or lensed objects with high amplification factors (e.g. Carilli \& Walter 2013 for a review). Due to their high dipole moment, the critical density to excite these molecules is high (10$^5$ cm$^{-3}$ or higher for the more excited levels), therefore they are difficult to detect at high J with ALMA. The first lines at J=1-0 are thus more promising. Their emission is a better tracer of star formation, since giving directly the amount of dense clumps in the molecular medium (Gao \& Solomon 2004). With SKA1, HCN(1-0) will be observable for z$>$4.5, and CS(1-0) for z$>$2. For SKA2, these will be z$>$2.7 and z$>$1, respectively. The expected lines are weak, though, about an order of magnitude less than the CO lines for starbursts, and even less for normal galaxies. Wagg \etal\, (2015) estimate that the SKA1 deep survey (5000h) could detect about 1 source per arcmin in CS and HCN, and SKA2 (1000h) about 10 times more.

\smallskip
\noi 
{\bf Water masers and AGN}

\smallskip
\noi 
In the sub-pc regions around AGN, the gas is so dense and excited that H$_2$O masers can be excited (e.g. the prototype NGC 4258, Greenhill et al. 1996). The emission is so strong that long baselines (VLBI) can be used, bringing a lot of information on the gas dynamics and mass of the central black hole. The rest-frame frequency being 22 GHz, nearby galaxies could be observed with SKA2, and z$>$0.5 masers could be found with SKA1. Up to now, more than 100 type-2 AGN have been detected (e.g. Braatz et al 2010). Beyond the study of the accreting disk itself, measuring its thickness, and its warping in some cases, and the accurate determination of black hole masses, the high resolution imaging and proper motions can determine accurate distances, and the value of the local Hubble constant. SKA will bring a strong, independent and complementary method to determine H$_0$ at different redshifts with a precision better than 5\% (e.g. Braatz et al. 2015), a valuable advantage when the other methods are in tension (e.g. Planck collaboration 2016).

\smallskip
\noi 
{\bf OH megamasers and starbursts}

\smallskip
\noi 
Starbursting galaxies are strong infrared emitters (LIRG and ULIRGs), and provide the conditions to excite OH mega-masers, 2 to 4 orders of magnitude more luminous than OH galactic masers, at the 1.6 GHz frequency (Baan \& Haschick 1983). The powerful lines are broad ($\sim$1000km/s) and can be multiple. More than 100 OH megamasers have been detected now, in particular with the Arecibo survey (e.g. Darling \& Giovanelli 2002), and their redshifts are between 0.1 and 0.23. Since the star formation rate increases with redshift by an order of magnitude until z$\sim$2, the number of actively star-forming galaxies are expected to increase by several orders of magnitude, given the much larger volume explored at high z. The detection of OH megamasers at any redshift will be a side product of the deep SKA surveys of redshifted \hi\, line. This will give insight in the merger evolution with redshift. Based on the correlation between OH megamasers and IR luminosity, the OH masers could be detected up to z=3-5 (cf Roy \& Kl\"ockner 2012). In addition to yield the rate of mergers as a function of redshift and the OH mega-maser luminosity function, this research will allow to statistically investigate the nuclear regions of starbursts, and propose targets to follow up at various wavelengths, in particular with ALMA.

\smallskip
\noi 
{\bf Radio Recombination lines}

\smallskip
\noi 
The \hi\, 21\,cm line does not inform about the physical conditions of the atomic gas, and in particular its density and temperature. A complementary tracer are the radio recombination lines of hydrogen and carbon, which correspond to dense and clumpy media at high frequency (corresponding to SKA-MID) and to a diffuse medium at low frequency, $<$350 MHz, corresponding to SKA-LOW (e.g. Thompson et al. 2015). The diffuse carbon radio recombination lines (CRRL) are sensitive to the cold medium, or CNM, and have narrow line-widths, while that of hydrogen (HRRL) are found in the warm medium (WIM), with broader widths. In the Milky-Way, these lines are observed both in emission and absorption in front of strong continuum sources, with typical optical depth of 0.5 to 2 10$^{-3}$. Diffuse CRRL have been observed towards external galaxies by LOFAR in absorption in M82 (Morabito et al. 2014). The lines are issued from atomic gas or PDR, where the carbon is ionized. The non-LTE models necessary to interpret the data yield the electronic temperature and density and the abundance of carbon. By comparison of \hi-21\,cm lines, it is possible to derive the fraction of CNM, WNM and WIM along the line of sight. With SKA1, about 100 nearby galaxies could be detected in the CRRL, while 400 will be detected and mapped with SKA2 (Oonk et al. 2015).

\begin{figure}[!ht]
  \centering
  \includegraphics[width=0.99\linewidth]{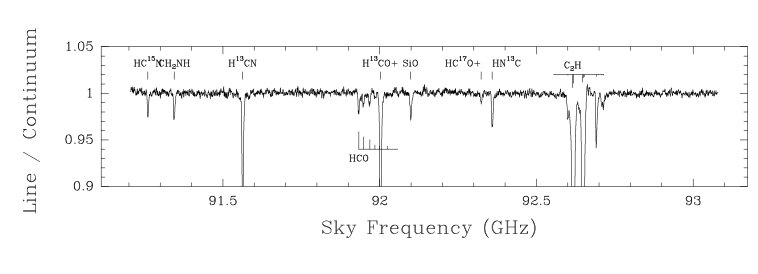}
  \caption{\label{fig:pks} Part of the molecular line survey, done in absorption
in front of the quasar PKS1830-21 with ALMA (Muller et al. 2014). }
\end{figure}

\smallskip
\noi 
{\bf Absorption lines in intervening galaxies}
\smallskip
\noi 
Molecular absorption lines from cold medium in intervening galaxies is a powerful tool to study high-redshift galaxies with high resolution and high sensitivity, since the optical depth can rise up to saturation, and only the pencil beam of the strong background radio core is involved. The detection power is then only due to the continuum strength, and only small masses of molecular medium are necessary for detection (Wiklind \& Combes, 1996). Five absorbing systems have been detected with mm telescopes, and  three of them correspond to lensed quasars, where the absorption comes from the lensing galaxy. In PKS1830-21, a molecular line survey was even possible with ALMA, as shown in Fig.\,\ref{fig:pks} (Muller et al. 2014). Since several lines of sight cross the galaxy, its dynamic can be probed. The absorption lines from the cold medium are so narrow, that a high precision on their frequency/velocity is obtained, thus these systems are ideal targets to constrain the variation of fundamental constants, by comparing various elements (\hi, CI, HCN, CS, CH$_3$OH, e.g. Murphy et al. 2001). With SKA, many more absorbing systems are expected to be detected, since the strength of continuum sources increases at low frequency (synchrotron emission). These detections will be a side product of wide redshifted \hi\, surveys.\\

\parbox{0.9\textwidth}{
\noi{References:}\\
\noi{\scriptsize
Baan, W.A.,  \& Haschick, A.D., 1983, AJ 88, 1088;
Braatz J.A., \etal, 2010, ApJ, 718, 657;
Braatz J.A., \etal, 2015, IAUGA 2255730;
Carilli, C., \& Walter, F., 2013, ARA\&A 51, 105;
Combes, F., \& Wiklind, T., 1999, in Highly Redshifted Radio Lines, ASP Conf. Series Vol. 156, p. 210
Daddi, E., \etal, 2010, ApJ, 713, 686;
Darling J. \& Giovanelli R., 2002, ApJ, 572, 810;
Gao, Y. \& Solomon, P. M., 2004, ApJS, 152, 63;
Greenhill, L. J., \etal, R., 1996, ApJ 472, L21;
Morabito, L. K., \etal, 2014, ApJ  795, L33;
Muller, S., \etal, 2014 A\&A, 566, 112;
Murphy, M. T., \etal, 2001, MNRAS 327, 1244;
Oonk, J.B.R., \etal, 2015, AASKA14, 139;
Planck collaboration, 2016, A\&A 594, A13
Roy, A., Kl\"ockner, H.R., 2012, in "SKA German White Paper";
Tacconi, L. J., \etal, 2013, ApJ, 768, 74;
Thompson, M., \etal, 2015, AASKA14, 126;
Wagg, J., \etal, 2015, AASKA14, 161;
Wiklind, T, Combes, F, 1996 Nature 379, 139
}}\\

\paragraph{The role of AGN}\label{sci:AGN}
\vspace{0cm}

\noi Active galactic nuclei (AGN), powered by accretion of baryonic
matter onto the supermassive black holes that reside in the centres of
most galaxies, are amongst the most luminous objects in the
Universe. The most powerful AGN release the equivalent of the binding
energy of a massive galaxy within their short activity periods of few
ten to hundred Million years (e.g., Silk \& Rees, 1998), separated by
much longer epochs of relative quiescence. The injection of a small
amount of this energy is sufficient to regulate baryon cooling and
cosmic structure formation from the sizes of individual galaxies to
galaxy groups and clusters (Dubois et al. 2016), the largest
gravitationally bound structures in the Universe today. Most galaxies
host dormant AGN, whose masses are tightly correlated with the mass and other
structural properties of their host galaxy, pointing out a close
evolutionary link. AGN emit varying fractions of their kinetic energy in
form of relativistic particles, producing radio jets and lobes, whose
sizes can range from few parcsec to the Mpc sizes of entire galaxy
clusters. In the most powerful high-redshift AGN, these jets take the
leading role in stirring up the gas of the host galaxy and expelling
large fractions of it, thereby quenching further star formation, and
playing a decisive role for the evolution of massive galaxies (e.g.,
Nesvadba et al. 2006, 2017). It is probably also the radio source
which keeps massive galaxies gas-poor over the 10 Billion years
following their initial growth phase, and until the present day (e.g.,
Best et al. 2006; Nyland et al. 2016).

\smallskip

\begin{figure}[!ht]
  \centering
  \includegraphics[width=0.63\linewidth]{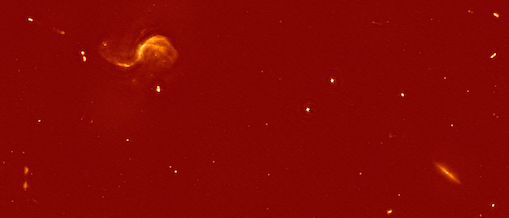}
  \includegraphics[width=0.35\linewidth]{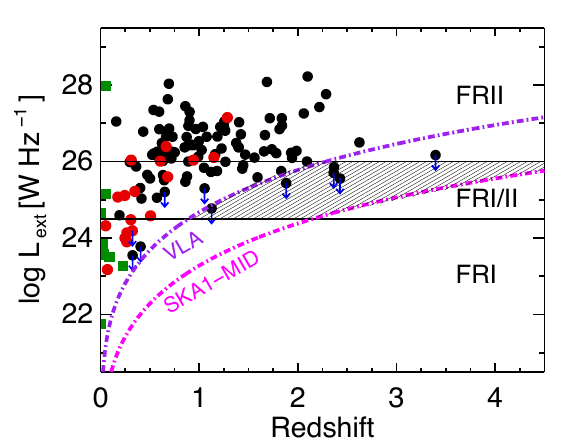}
  \caption{\label{fig:agn_figure} {\em Left:} zoom into a very small
    region of one of the extragalactic fields of the SKA precursor
    LOFAR at 150~MHz, showing a multitude of AGN and star
    forming galaxies, including the nearby Seyfert galaxy M~106 ({\em top
    left}), and many relativistic radio jets produced by supermassive
    black holes at redshifts up to z$\gtrsim$1. This images was
    produced with algorithms for post-processing adaptive optics
    developped at Observatoire de Paris (credits: LOFAR Surveys Key
    Science Project). {\em Right:} radio power vs. redshift
    distribution for radio-loud AGN expected via SKA1-MID survey. The
    solid black line denotes the FRI/FRII divide (adapted from Kharb et al. 2016).}
\end{figure}

\noi SKA will revolutionise our understanding of AGN in many ways. For
the first time, we will have comprehensive surveys of the AGN
population, including radio-quiet sources, out to redshifts z$\sim6$,
well beyond the peak of cosmic star formation at z$\sim2$. This will
provide us with the definite answer on the global role of AGN
feedback, its occurrence during different key phases of the cosmic
star formation history, galaxy evolution (e.g., starburst, galaxy
interactions and mergers, passive evolution), and as a function of
galaxy mass and environment. In addition SKA, by exceeding the spatial
resolution and sensitivity of current data sets by more than an order
of magnitude, will allow us to follow the expansion of individual
radio sources through the interstellar gas of their host galaxies and
surrounding intergalactic or intracluster environments, and provide us
with the physical parameters necessary to unify the rich morphologies
and spectral shapes of these sources within a single coherent physical
and evolutionary framework (Pandey-Pommier et al. 2014, 2016a), and to
quantify the energy and momentum transfer from the AGN to the gas.

\smallskip

\noi As a result of the survey work and detailed studies of individual
galaxies or small samples, SKA will trace the co-existence of radio
jets and star formation within the nuclei of the most intensely
star-forming galaxies and galaxy mergers across all of cosmic
star-formation history, including systems, which are optically thick
even in the far-infrared and millimetre regimes. We will probe
directly, whether the pressure enhancement produced by the expanding
jets in the surrounding gas inhibits, or perhaps even boosts star
formation (e.g., Salom\'e et al. 2017; Elbaz et al. 2009; Bieri et
al. 2016), and constrain the number of black-hole
binaries in galaxy mergers, and the physics and timescales of their
merging (Capelo et al. 2017), a major hypothesised source of
gravitational waves and perhaps the origin of black-hole spin and
radio-loudness. Maser emission from black-hole accretion disks will
provide us with accurate black-hole mass estimates out to high
redshifts, tracing the co-evolution of black holes and their hosts
across cosmic time (Nesvadba et al. 2011), and also constraining our
understanding of the first black holes in galaxies (Volonteri et
al. 2009).

\smallskip

\noi SKA will give us direct access to the low-frequency radio
spectrum of AGN out to the highest redshifts, which constrains the
kinetic power and momentum carried by the radio lobes, and is the
central parameter that determines the power of AGN feedback. For the
first time, SKA will let us establish a kinetic luminosity function
for AGN across cosmic history, constraining their global potential in
shaping galaxies and small-scale cosmic structure. What fractions of
the AGN energy output are injected into the gas, as a function of
black hole and host properties, and of environment? What fraction of
the rest-mass energy equivalent of the supermassive black holes does
this correspond to? The energy transfer efficiency from jet to gas,
and the rest-mass energy equivalent are the scaling parameters that
are predicted by hydrodynamic and cosmological models of AGN feedback,
and are observationally constrained only to about one order of
magnitude today -- a major limitation in our understanding of AGN
feedback (Nesvadba et al. 2006, 2017).

\smallskip

\noi 
SKA will also show whether AGN really fall into two classes of
``radio-loud'' and ``radio-quiet'' sources, with a clear dichotomy in
black-hole properties or accretion rate (Cadolle Bel et al. 2007), or
whether these are extreme ends of a continuous sequence of black-hole
and host galaxy properties. SKA will be the first facility to trace
AGN at the divide between Fanaroff \& Riley (1974) luminosity class I
and II out to the peak of cosmic star-formation at $z\sim2$, and to
probe even radio-quiet, powerful quasars out to $z\sim6$, across
nearly all of cosmic star-formation history. This will show whether
AGN activity was common enough to shape galaxies at all cosmic times,
a basic, and unproven ingredient of our most advanced scenarios of
galaxy evolution today.

\smallskip

\noi 
Another major limitation in our understanding of the AGN-host galaxy
connection is the episodic nature of AGN activity. With the exception
of rare peculiar cases, the period of quiescence between major
activity cycles, and the relative timing of AGN activity and star
formation are today essentially unconstrained. With SKA, we will close
the feedback cycle, by probing the low-frequency radio spectrum out to
high redshifts, including the past radio activity of galaxies, over
timescales when the radio synchrotron spectrum of these sources has
already faded at higher frequencies (Pandey-Pommier et al. 2016b). This
will provide direct measures of the timing and duration of AGN
activity and feedback cycles, including for low-power radio
sources. The fate of the gas at the end of AGN activity could
determine whether sporadic AGN activity alone can maintain low gas
fractions and star formation rates in previously quenched galaxies
over cosmic time (Nesvadba et al. 2011), or whether additional
mechanisms are necessary (Martig et al. 2009).

\smallskip
\noi SKA will also turn radio sources across cosmic history into
beacons to probe the 21\,cm line of atomic hydrogen in their massive
host galaxies and surrounding gaseous halos {\it in absorption}. This
includes in particular the diffuse, low-density gas outside of massive
star-forming regions, that is otherwise unobservable (Nesvadba et
al. 2016). \hi~absorption line studies will provide sensitive limits of
turbulence and fragmentation of the ambient gas in high-z AGN host
galaxies, and provide us with accurate limits of gas accretion, as
well as the terminal velocities and gas escape fractions reached by
AGN and starburst-driven outflows.\\

\parbox{0.9\textwidth}{
\noi{References:}\\
\noi{\scriptsize 
Best, P., \etal, 2006, MNRAS, 368, L67;
Bieri, R., \etal, 2016, MNRAS, 455, 4166;
Cadolle Bel, M., \etal, 2007, ApJ 659, 549;
Capelo, P. R., \etal, 2017, MNRAS, 469, 4437;
Dubois, Y., \etal, 2016, MNRAS, 463, 3948;
Elbaz, D., \etal, 2009, A\&A, 507, 1359;
Fanaroff, B. L. \& Riley, J. M., 1974, MNRAS, 167, 31; 
Kharb, P., \etal, 2016, JApA, 37, 34; 
Martig, M., \etal, 2009, ApJ, 707, 250;
Nesvadba, N., \etal, 2006, ApJ 650, 693;
Nesvadba, N., \etal, 2016, A\&A 593, L2;
Nesvadba, N., \etal, 2017, A\&A 600, 121;
Nyland, K., \etal, 2017, MNRAS, 464, 1029; 
Pandey-Pommier, M., \etal, 2014, in \href{http://nenufar.obs-nancay.fr/IMG/pdf/nenufar-science-case-v5_2014_10_10_pz.pdf}{\color{blue} \myul[blue]{NenuFAR : instrument description and science case}};
Pandey-Pommier, M., \etal, 2016b, Proceedings of the SF2A, 373;
Pandey-Pommier, M., \etal, 2016a, Proceedings of the SF2A, 379;
Salom\'e, P., \etal, 2016, A\&A, 595, 65;
Silk, J. \& Rees, M. J., 1998, A\&A 331, L1;
Volonteri, M. \& Natarajan, P., 2009, MNRAS, 400, 1911
}}%


\subsubsection{Nearby resolved galaxies}

\paragraph{Galaxy dynamics and \hi~distribution}
\vspace{0cm}

\noi 
\hi\ studies played a major role in the 1970s and 1980s in establishing the presence of dark matter
in gas rich galaxies, necessary to keep up the extended flat or rising rotation curves at large radii,
beyond the optical image. It became readily clear, however, that additional dynamical methods
are needed to establish accurately the amount of dark matter, due to the well known degeneracy
between models with different stellar mass-to-light ratios of the visible parts. Even though some
work argues for nearly maximum disks in Milky Way type galaxies, based on considerations of
spiral structure or the strength of shocks in a few barred spirals, other work, primarily using stellar
velocity dispersions, indicates non-maximum disks. This problem will continue to be addressed in
the near future, with extensive surveys using integral field unit spectrographs (IFUs) on dedicated
optical telescopes. 

\smallskip

\noi 
More recent work on late type low surface brightness galaxies showed convincingly that the
current $\Lambda$CDM models are in disagreement with observations, the so-called core-cusp controversy
(de Blok 2010). Recent attempts to solve this problem concentrate on introducing more violent feedback in
the star formation recipes used in the numerical simulations of galaxy formation (Oh et al. 2015), so that an
initially cuspy dark matter distribution can be modified sufficiently. Research on this problem can
be expected to continue to progress, until a better understanding of these feedback mechanisms has
been reached. At present, the predictions from these simulations seem to begin to work for small
galaxies, but for larger ones the answer is still open. This is an active area of research, and further
improvements of the galaxy formation models can be expected (e.g. Tollet et al. 2016). \smallskip

\noi 
For bright edge-on galaxies, the star formation intensity is such that a thick \hi\ disk is set up,
whose kinematics shows rotation speeds which are smaller than those in the plane of the disk (e.g.
Oosterloo et al. 2007 for NGC 891). This greatly complicates the analysis of flaring \hi\ disks, 
which are expected to
exist in the outer parts of galaxies if the dark halo is near spherical, and which can help constrain
the shape of the dark halo. Perhaps this latter problem can still be studied for a number of small galaxies,
which have a more quiescent star formation activity.
\smallskip

\noi 
Further \hi\ observations can be expected in the near future, since the all-sky surveys planned
with the SKA precursors enable the selection of an adequate sample of relatively unperturbed
galaxies, for which far deeper \hi\ observations can be fruitfully done with the SKA, and combined
with more extensive diagnostics of the kinematics of the visible parts using IFUs. Such data will
contain crucial information for the kinematics of the gas, and hence can be used to study the dark
matter problem. Improvement in constraining the stellar mass-to-light ratios using multi wavelength 
data can also be expected. Moreover, these observations can also be used for the purposes of 
studying the star formation and gas accretion in disk galaxies, as well as the connection to the 
cosmic web. Modern star formation laws relate primarily the molecular gas to star formation, but
the role of the \hi\ and total gas reservoirs, and its dynamics, play a role in shaping the ISM in which
these molecular clouds appear, a role which needs further investigation.
\smallskip

\begin{figure}[t]
  \centering
  \includegraphics[width=0.95\linewidth]{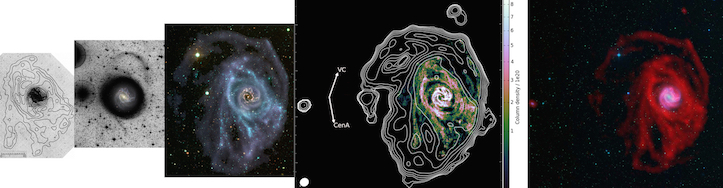}
  \caption{\label{fig:m83} Changing perspectives in the \hi\ imaging of M83. The {\em left} panel shows the
 \hi\ image from Rogstad et al. (1974) - angular resolution 2', outer contour 1.37 10$^{20}$ cm$^{-2}$ -, 
 and the panel next to it a deep optical image on the same scale by 
 Malin \& Hadley (1997). The panel next to it is a composite image by Koribalski (2015), obtained with the
 ATCA (blue is \hi\, superimposed on a colour image produced from a GALEX UV image in light blue, a DSS R-band image in green and a 2MASS J band in red red, credit A.R. L\'opez-S\'anchez). The next panel shows deep KAT-7 \hi\ observations by Heald et al. (2016), with the THINGS \hi\ image (Walter et al. 2008) as inset. The outer contour of the KAT-7 image is 5.6 10$^{18}$ cm$^{-2}$, and the angular resolution 3.75 x 3.42 arc min.  
 The {\em rightmost} panel is the recently obtained image with a MeerKAT-16 element configuration, using seven
 50 minutes exposures (see \href{http://www.ska.ac.za}{\color{blue} \myul[blue] {media release 16/05/2017}}). The \hi\ images have
 progressive better resolution and/or sensitivity, while a wider field of view permits the detection of close
 dwarf companions. 
 The link with a wider perspective w.r.t. the cosmic web is indicated in the KAT-7 picture (VC means Virgo cluster, while CenA indicates the direction of Centaurus A).}
\end{figure}

\smallskip
\noi 
Another field whose exploration will flourish in the near future is the study of faint stellar
extensions around the primary optical image of galaxies, such as extended UV disks 
(e.g. Espada et al. 2011) or faint optical emission (e.g. Martinez-Delgado et al. 2010). 
The spatial coincidence of these features with the \hi~is not always
unique (as in e.g. M33, Lewis et al. 2013), and deep high resolution imaging with the SKA will 
contribute to a better understanding of this phenomenon, which may be related to interactions with 
and/or minor mergers of dwarf satellites. In Fig.\,\ref{fig:m83} we show the evolution of the \hi\ imaging of M83 
over the last $\sim$40 years, with from {\em left} to {\em right} a low resolution image from 1974, a deep optical image 
from 1997, multi wavelength imaging next, then an extended image at 
low resolution but high sensitivity coupled with a wider field of view from 2016, and a taste of things to come 
with a 2017 MeerKAT image.
The \hi\ studies
will be accompanied by studies of the metallicity, stellar populations and star formation of galaxy
outskirts in general, as in e.g. Boissier et al. (2016). 

\smallskip
\noi 
Present plans for work in this area with the SKA pathfinders/precursors, in which members of the French astronomical community are involved, are 
as follows: i) with the WSRT and the Apertif instrument a large shallow survey and a medium deep
survey will commence early next year. Follow-up optical
data will be collected for selected galaxies with the WEAVE IFU-spectrograph on the 4.2m WHT at La Palma,
planned to be ready in 2018.
ii) with ASKAP, a large \hi\ survey of the southern sky and part of the northern sky (up to declinations of 
30$^\circ$), named WALLABY, is planned to commence later this year, and iii) with MeerKAT, deep observations
of 30 nearby galaxies in the project MHONGOOSE are planned to commence in 2018, when the telescope is ready.

\smallskip
\noi 
With the SKA1 and SKA2 
telescopes, the sensitivity of the observations will be further improved, and better image quality 
at higher spatial resolution will be obtainable. Currently under discussion are: a medium wide
survey (400 deg$^2$ at resolution 10 arcsec) out to z $\sim$ 0.3, a medium deep survey (20 deg$^2$ at resolution 5 arcsec)
out to z $\sim$ 0.5, a deep survey (single pointing at resolution 
2 arcsec) out to z $\sim$ 1, deep imaging (100h/galaxy) of 
30 nearby galaxies for the study of the interstellar medium, and another 30 galaxies for gas accretion. 
Hence, in addition to deep observations of local galaxies, a 
 major push will be made to get 
 resolved \hi\ data of spiral galaxies as function of redshift. The depth of the observations will allow the study of the 
connection of the galaxies to the Cosmic Web. The SKA data will help refine 
the models of galaxy formation and evolution significantly.\\

\parbox{0.9\textwidth}{
\noi{References:}\\
\noi{\scriptsize 
Boissier, S., \etal,, 2016, A\&A, 593, A126;
de Blok, W.J.G. 2010, Advances in Astronomy, 2010, article id. 789293;
Espada, D., \etal, 2011, ApJ, 736, 20;
Heald, G., \etal, 2017, MNRAS, 462, 1238;
Koribalski, B.S., 2015, IAUS, 309, 39;
Lewis, G.F., \etal,  2013, ApJ, 763, 4;
Malin, D., Hadley, B.,1997, PASA, 14, 52;
Martinez-Delgado, D., \etal, 2010, AJ, 140, 962;
Oh, S.-H., \etal,  2011, AJ, 142, 24;
Oosterloo, T., \etal, 2007, AJ, 134, 1019;
Rogstad, D.H., \etal, 1974, ApJ, 193, 309;
Tollet, E., \etal,  2016, MNRAS, 456, 3542;
Walter, F., \etal,  2008, AJ, 136, 2563
}}

\paragraph{The Interstellar Medium in Nearby Galaxies}

\vspace{0cm}

\noi {\bf Introduction}

\smallskip
\noi Galaxy evolution is driven by the accretion of gas onto galaxies,
the cycling of matter within galaxies between the interstellar medium
(ISM) and stars, and the ejection of gas from the galactic
disk. Direct detailed observations of these physical processes are
essential for understanding the astrophysics of galaxy evolution, to
interpret unresolved observations of galaxy populations, and to
implement ``sub-grid'' physics into numerical simulations. Nearby
galaxies provide the best opportunity to study these processes in
their galactic context across a representative range of galaxy
properties and local ISM conditions. Here we briefly outline three
science themes that will be advanced by SKA1-MID surveys of galaxies
in the local Universe, focussing on topics relating to the neutral ISM
and galactic-scale star formation. SKA1 studies of accretion onto
supermassive black holes, galaxy dynamics, and AGN activity in nearby
galaxies are described elsewhere in this volume. Time and
sensitivity estimates assume the characteristics of SKA1-MID presented
in Dewdney \etal\ (2013).

\smallskip
\noi {\bf Calibrating the Star Formation Rate in Galaxies}

\smallskip
\noi Radio continuum (RC) observations are a powerful diagnostic of
the star formation rate (SFR), allowing a direct measurement of the
local star formation activity even in highly obscured environments
(see e.g. Condon 1992). At GHz frequencies, RC emission consists of
two components: thermal free-free emission due to ionizing photons
from high-mass ($M \geq 8\,{\msol}$) stars, and non-thermal
synchrotron emission from relativistic electrons gyrating in a
magnetic field. The thermal RC is directly proportional to the
instantaneous SFR, while the non-thermal RC depends on both the
magnetic field strength and the cosmic ray electron density. The
physics that produces a link between a galaxy's global radio
synchrotron luminosity and its SFR are not yet fully understood (see
e.g. Bell 2003; Lacki, Thompson \& Quataert 2010), and radio-based SFR
estimates are usually calibrated via the RC-to-infrared (IR)
correlation (e.g. Yun, Reddy \& Condon, 2001). Since the non-thermal
RC in normal galaxies is typically an order of magnitude brighter than
the thermal RC at GHz frequencies, there is great interest in
calibrating the non-thermal RC as a SFR diagnostic that can be used
from the local Universe out to intermediate redshifts (e.g. Murphy
2009; see also Sect.\,\ref{science:unobscured}).

\smallskip
\noi SKA1-MID studies of the RC emission in nearby galaxies will make
important progress on this topic. First, SKA1-MID will permit a
spatially resolved imaging RC survey of galaxies across the entire
frequency range between 1.6 and 10\,GHz. As an example, observations
with 3\arcsec\ resolution and 1\,hr per band (4\,hrs per target) would
achieve $\sim1$\,\muJy\ per beam sensitivity and linear resolutions of
300\,pc for galaxies within 20\,Mpc. Whereas existing facilities are
restricted to studying individual targets with this combination of
resolution, frequency coverage and sensitivity, SKA1-MID thus opens
the possibility of a deep RC survey of $\sim100$ nearby galaxies in
less than 500\,hrs of observing. These data would enable radio
spectral index mapping and a robust pixel-by-pixel separation of the
thermal and non-thermal RC across a wide range of galaxy types. The
former will reveal how cosmic ray electrons propagate and cool within
galactic disks, while the latter is needed to investigate how the
non-thermal RC fraction depends on local ISM conditions (e.g. gas
density, interstellar radiation field) and global galaxy processes
(e.g. cosmic ray escape). Existing studies of individual targets with
the VLA and Westerbork have reported variations in the relationship
between the non-thermal RC emission and the SFR as a function of
galactic environment (e.g. Dumas \etal\ 2011; Tabatabaei \etal\
2013), but the larger samples and better sensitivity to diffuse
emission of SKA1-MID are essential to translate these first results
into a physical explanation for how the non-thermal RC traces the SFR
across the galaxy population.

\smallskip
\noi Second, with sensitive, sub-arcsecond imaging of individual
nearby galaxies, it will become possible to obtain a complete census
of the compact radio sources that trace both star formation and the
final stages of stellar evolution, e.g. compact \hii~regions, super
star clusters, X-ray binaries, planetary nebulae, supernovae (SNe) and
supernova remnants (SNR). An inventory of these populations can be
used to infer the level of current and recent SF within the host
galaxy, independent of the mechanisms that tie a galaxy's non-thermal
radio luminosity to its SFR. For example, SKA1-MID observations with
0.5 arcsec resolution and $\sim$\muJy\ sensitivity at 1\,GHz would
detect all long-lived radio SNR within $\sim20$\,Mpc. Combined with
information regarding the SNR size and age, these SNR number counts
track the number of recent core collapse SNe due to high-mass stars,
providing a direct estimate of a galaxy's recent SF activity and the
shape of the high-mass end of the stellar initial mass function
(e.g. Fenech \etal\ 2008). Comparing SFR estimates based on the
population statistics of SF phenomena with common multi-wavelength SFR
proxies (e.g. IR, UV, H$\alpha$) will place significant new
constraints on the calibration and interpretation of extragalactic SFR
diagnostics.

\smallskip 

\noi {\bf Tracing the Gas Cycle in Galaxies}

\smallskip
\noi A key result from recent multiwavelength studies of nearby
galaxies is that there is a tight, approximately linear correlation
between the molecular gas and SFR surface densities on $\sim$kpc
scales within star-forming disks (e.g. Bigiel \etal\ 2008; Leroy
\etal\ 2008). This correlation can be interpreted as constant star
formation efficiency in the molecular gas reservoir. However, the
existing data do not access the spatial scales at which cool atomic
gas collapses to form giant molecular clouds (GMCs), or the subsequent
formation of individual stars. The physical processes that regulate
the rate and efficiency of star formation therefore remain poorly
understood. Equally puzzling is that the observed $\sim$\,kpc-scale
relation implies that nearby galaxies should deplete their molecular
gas reservoirs within a few Gyr. Numerical simulations show that gas
in galactic disks can be replenished by accreting gas from the
intergalactic medium (IGM, e.g. Kere\u{s} \etal\ 2005), but there is
little direct observational evidence for this process. Therefore, two
major outstanding questions for galactic-scale theories of star
formation are, How do extragalactic star formation scaling relations
from the small-scale physics of cloud collapse and star formation? and
How do galaxies sustain their star formation activity throughout
cosmic time?

\smallskip
\noi Targeted SKA1-MID observations of the \hi\ emission in nearby
galaxies will make significant progress towards answering both of
these questions. First, the SKA will have sufficient sensitivity to
image the atomic gas reservoir in nearby galaxies on the scale of
individual GMCs and star-forming complexes. Assuming a channel width
of 5\,\kms, SKA1-MID observations with 3\arcsec\ resolution and 10\,hr
per galaxy would achieve a column density sensitivity limit of
$4\times10^{20}$\,cm$^{-2}$ and linear resolutions of 150\,pc for
galaxies within 10\,Mpc. A survey of 50 to 100 galaxies would open the
possibility of a systematic study of the phase balance between the
warm and cold atomic gas components, turbulence in the neutral ISM,
and the transition from atomic to molecular gas on GMC scales across a
wide range of ISM conditions and galactic environments. Combined with
high resolution ALMA observations of the molecular gas in galaxies,
such a survey would establish the timescales and requisite physical
conditions for GMC formation, the first step -- and perhaps a critical
bottleneck -- for star formation.

\smallskip
\noi Empirical estimates for molecular gas depletion times indicate
that star-forming galaxies must acquire fresh gas from the IGM to
maintain their star formation activity.
Simulations favour a cold gas accretion process in which the accreting
gas is not shock heated as it enters a galaxy's halo. However
\hi\ studies of the Milky Way and local galaxies consistently obtain
cold gas accretion rates that are an order of magnitude below the
SFR. This measurement is not straightforward, since accreting
primordial gas must be distinguished from `galactic fountain' gas that
has been expelled from the disk by supernovae and stellar winds
(e.g. Fraternali \& Binney 2008).  As shown by surveys such as
HALOGAS (Heald \etal\ 2011), a thick halo of diffuse neutral gas is
present around many nearby galaxies, but a significant population of
\hi\ clouds and streams transporting cold gas mass directly onto
galactic disks appears to be absent. One possibility is that the
accretion occurs at column densities lower than current observational
limit of $\sim10^{19}$\,cm$^{-2}$. With its excellent sensitivity, the
SKA will be a powerful machine to characterize extraplanar
\hi\ features and to put strong observational limits on cold gas
accretion rates. For example, with an angular resolution of
30\arcsec\ and 5\,\kms\ channel width, SKA1-MID will detect a column
density of $3\times10^{18}$\,cm$^{-2}$ with $5\sigma$ significance in
10\,hours. This opens the prospect to obtain a robust statistical
description of cold gas accretion activity in the local Universe by
surveying $\sim100$ gas-rich galaxies within the Local Volume
($<10$\,Mpc). Looking further ahead, detailed studies of individual
targets with SKA2 would reach column densities of less than
$10^{17}$\,cm$^{-2}$ in 100\,hours, accessing the regime of Lyman
Limit Systems. Both modes of observations will yield strong tests of
predictions from cosmological models of cold gas accretion and
galactic fountain models.

\smallskip
\noi {\bf Revealing the Relativistic Interstellar Medium}

\smallskip
\noi In addition to gas and dust, the ISM consists of magnetic fields
and relativistic cosmic rays (CRs). These non-thermal components of
the ISM are dynamically important, likely playing a significant role
in the regulation of star formation, the launching of galactic
outflows and winds, and in the hydrostatic balance and stability of
galaxy gas disks (e.g. Cox 2005). Most of our knowledge of the
magnetic field properties in nearby galaxies comes from radio
observations (e.g. Beck 2016): the intensity of the synchrotron
component measures the product of the magnetic field strength in the
sky plane and the number density of CRs, while the field direction
projected in the sky plane can be determined from the polarisation of
the synchrotron emission. To date, radio polarisation maps of nearby
galaxies have mostly been obtained with a linear resolution of
$\geq0.5$\,kpc or greater, due to surface brightness sensitivity
limitations of existing telescopes. Polarisation maps with 1-100\,pc
resolution are needed to investigate important physical properties of
interstellar magnetic fields, e.g. the correlation scale and degree of
anisotropy of turbulent magnetic fields, the universality of
large-scale field reversals and magnetic helicity, and the interaction
between magnetic fields and the density and kinematics of the neutral
ISM.

\smallskip
\noi SKA1-MID will be a powerful machine for mapping the polarised
synchrotron emission in nearby galaxies. For polarisation
observations, SKA1-MID Band 4 (2.8 to 5.2\,GHz) or the low frequency
end of Band 5 (4.6 to 13.8\,GHz) offer the best compromise between
synchrotron intensity (typically declining as $\nu^{-0.8}$ in spiral
galaxies) and Faraday depolarisation, which increases at lower frequencies. 
As an example, observations using 2.5\,GHz of SKA1-MID
Band 5 with 5\arcsec\ resolution would achieve a sensitivity of
$\sim0.2$\,\muJy\ per beam in 10\,hours of integration time, $\sim10$\ times
more sensitive than can be achieved with the EVLA using a similar
observing configuration. This resolution is sufficient to access
spatial scales of $\sim1$\,pc in the Magellanic Clouds, and 0.5\,kpc
or better for all galaxies within 20\,Mpc. With only 10\,hours
required to detect even faint polarised emission in galaxy disks,
SKA1-MID therefore presents an opportunity to obtain sensitive
polarization maps for a significant number of galaxies, enabling the
first systematic characterization of the strength and structure of the
magnetic field in galaxies in the local Universe.

\smallskip 

\noi In addition to the references quoted above, the interested reader can refer to: Beswick \etal\ (2015), Murphy \etal\ (2015) and P\'{e}rez-Torres \etal\ (2015) to know more about the calibration of galaxy SFR; de Blok \etal\ (2015), McClure-Griffiths \etal\ (2015) and Popping \etal\ (2015) about tracing the gas cycle in galaxies; Beck \etal\ (2015) concerning how to reveal the relativistic interstellar medium.\\

\parbox{0.9\textwidth}{
\noi{References:}\\
\noi{\scriptsize
Beck, R., \etal, 2015, AASKA14, 94;
Beck, R., 2016, A\&AR, 24, 4;
Bell, E., 2003, ApJ, 586, 794;
Beswick, R., \etal, 2015, AASKA14, 70;
Bigiel, F., \etal, 2008, AJ, 136, 2846;
de Blok, E., \etal, 2015, AASKA14, 129;
Condon, J., 1992, ARA\&A, 30, 575;
Cox, D., 2005, ARA\&A, 43, 337;
Dumas, G., \etal, 2011, AJ, 141, 41;
Fenech, D., \etal, 2008, MNRAS, 391, 1384;
Fraternali, F. \& Binney, J., 2008, MNRAS, 386, 935;
Heald, G., \etal, 2011, A\&A, 526, 118;
Kere\u{s}, \etal, 2005, MNRAS, 363, 2;
Lacki, B., \etal, 2010, ApJ, 717, 1;
Leroy, A., \etal, 2008, AJ, 136, 2782;
McClure-Griffiths, N., \etal, 2015, AASKA14, 130;
Murphy, E., 2009, ApJ, 706, 482;
Murphy, E., \etal, 2015, AASKA14, 85;
P\'{e}rez-Torres, \etal, 2015, AASKA14, 60;
Popping, A., \etal, 2015, AASKA14, 132;
Tabatabaei, F., \etal, 2013, A\&A, 552, 19;
Yun, M., \etal, 2001, ApJ, 554, 803
}}

\newpage
\subsection{Galactic astronomy}\label{science:galaxy}

\noindent {\normalsize Contributors of this section in alphabetic order: }

\smallskip

\noi {\sffamily \scriptsize
{\sffamily \bf {\bf F.~Acero}} [\irfu;\aim],
{\bf M.~Alves} [\irap],
{\bf A.~Araudo} [\lupm],
{\bf A.~Bacmann} [\univgren],
{\bf J.-P.~Bernard} [\irap],
{\bf S.~Bottinelli} [\irap],
{\bf F.~Boulanger} [\lermasorb;\ias],
{\bf A.~Bracco} [\ias;\irfu;\aim],
{\bf J.-M.~Casandjian} [\irfu;\aim]
{\bf L.~Cambresy} [\stras]
{\bf E.~Caux} [\irap],
{\bf I.~Cognard} [\lpcee],
{\bf E.~Dartois} [\ias],
{\bf K.~Demyk} [\irap],
{\bf E.~Falgarone} [\lermasorb],
{\bf K.~Ferri\`ere} [\irap],
{\bf I.~A.~Grenier} [\irfu;\aim],
{\bf J.~M.~Griessmeier} [\lpcee;\usn],
{\bf L.~Guillemot} [\lpcee;\usn],
{\bf A.~Gusdorf} [\lermasorb],
{\bf E.~Habart} [\ias],
{\bf P.~Hennebelle} [\irfu;\aim],
{\bf F.~Herpin} [\lab],
{\bf R.~Lallement} [\gepi],
{\bf B.~Lefloch} [\univgren],
{\bf M.~Lemoine-Goumard} [\unib],
{\bf F.~Levrier} [\lermasorb],
{\bf D.~Lis} [\lermasorb],
{\bf A.~L\'opez-Sepulcre} [\iram],
{\bf A.~Marcowith} [\lupm],
{\bf D.~J.~Marshall} [\irfu;\aim],
{\bf M.-A.~Miville-Deschenes} [\ias],
{\bf L.~Montier} [\irap],
{\bf J.~Pety} [\iram;\lermasorb],
{\bf M.~Renaud} [\lupm],
{\bf I.~Ristorcelli} [\irap],
{\bf M.~Schultheis} [\lagrange],
{\bf G.~Theureau} [\lpcee;\usn;\luth],
{\bf C. Vastel} [\irap],
{\bf L. Verstraete} [\ias],
{\bf N.~Ysard} [\ias]
}

\subsubsection{The nearby interstellar medium}
\vspace{0cm}

\noi {\bf Spectral line mapping} 

\smallskip
\noi Different evolutionary processes involving the interstellar medium (ISM)  dictate how our Galaxy evolves. From molecular formation through to star formation and the enrichment provided by evolved stars, detailed knowledge of these processes is only available in our Galactic backyard, yet they allow us to better understand how external galaxies evolve as well.
The nearby ISM  provides a particular appeal with respect to SKA observations as it will be possible to study the different phases of the ISM in exquisite detail.

\smallskip
\noi 
Data from the SKA will provide us with homogeneous coverage in angular resolution of the three main phases of the ISM (H, H$_2$, H$^+$) for the first time. Spectral line mapping at the resolution of 0.22'' corresponds to 30/90 AU at the $\sim$ 140/400 pc  of the Taurus/Orion complex, respectively. SKA will provide information on the neutral ISM through the {\hi} hyperfine line at 1.4 GHz, on the diffuse molecular phase through four OH lines at 1.612, 1.665, 1.667 and 1.720 GHz, as well as on the ionised phase through several radio recombination lines of both hydrogen and helium. Due to the limited wavelength range where these lines fall they can be observed simultaneously by the SKA, providing increased mapping speed over current observatories. 
Using line information in emission and in absorption from the SKA, the nearby ISM can be studied from the appearance of molecules through star forming clouds and the feedback of energetic processes as traced by the ionised phase.

\smallskip

\begin{figure}[!ht]
  \centering
  \includegraphics[width=0.65\linewidth, viewport=20 0 1000 500 ]{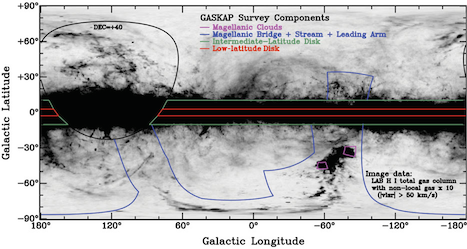}
  \caption{\label{fig:gaskap} Survey of the Milky Way using ASKAP. Different components are part of the survey such as the Galactic plane, high latitude Galactic clouds, the Magellanic clouds and the Magellanic stream.}
\end{figure}

\smallskip

\noi 
As a taster of things to come, observations are currently underway using the Australian SKA Pathfinder (ASKAP) to map the Galactic plane (GASKAP, Dickey et al. 2013). Fig.\,\ref{fig:gaskap} shows the different areas of the survey which, beyond the Galactic disk and Magellanic clouds, overlap nearby molecular clouds with different levels of stellar activity (e.g., Chamaeleon and Pegasus) and translucent clouds in different dynamical states (e.g., cirri and intermediate velocity clouds). The survey totals over 13000 square degrees in roughly 8000 hours of observing time. The angular resolution will reach down to 20'' with $\sim$ 1mJy sensitivity. The sampling of the atomic and molecular phases through {\hi} absorption as well as {\hi} and OH  emission will highlight the interplay between these two phases, specifically the production and shielding of molecular gas, and the influence of the Galactic environment on this interface. Information on this transition, the conditions under which it happens, and the mass locked up in this optically-thick {\hi} and CO-dark H$_2$ transition will provide valuable insight into the differences in evolution of molecular clouds. In the SKA1-MID phase this study can be extended to other local complexes at medium and high Galactic latitudes. It will finally be possible to combine {\hi} and OH observations (from the SKA) with CO data in absorption and emission (from ALMA) and with total gas tracers (such as dust and $\gamma$ rays). This comparison is essential if we wish to improve the interpretation of {\hi} and CO line emission along with dust emission which are the only available diagnostics in external galaxies.

\smallskip
\noi
{\bf 3D Exploration of the ISM} 

\smallskip
\noi The large amount of quality stellar data in the last two decades (2MASS, SDSS, PanStarrs ... ) has allowed the construction of three dimensional maps of dust density (Lallement et al. 2014; Marshall et al. 2006; Green et al. 2014).
In many environments, dust and {\hi} trace the same underlying material (Planck \& Fermi collaborations, 2015), therefore, combined with a SKA diffuse {\hi} survey, the aforementioned 3D dust maps provide a tool to explore the distribution of gas in 4D phase space. This will help us trace the flow of gas from the halo through the ISM where it collapses and feeds star formation. With this description of the distance and kinematics of the gas, we can tackle the essential question of ``where does the gas come from?'' (McClure-Griffiths et al. 2015).

\smallskip
\noi 
The Gaia mission (Gaia collaboration, 2016) will help a lot with this effort by providing parallax and extinction measurements to a billion stars. However, Gaia will not provide a full description of nearby molecular clouds as it will not ``see'' stars embedded within the clouds due to the high extinction suffered there. In contrast, the SKA is able to pick up these embedded stars, such as young stellar objects, and through accurate astrometry will also provide robust distances and proper motions to these sources. The radial velocity can be obtained through line emission, thus providing an exceptional complement to Gaia, and enabling the full characterisation of all local star-forming complexes, even in the SKA1-MID phase. These studies could be extended to local spiral arms using the full SKA2 array. 

\smallskip
\noi
Additionally, information on the sites and processes of formation/destruction of the complex organic molecules responsible for the hundreds of optical and IR Diffuse Interstellar absorption Bands (DIBs) will benefit from the combination of new multiplex stellar surveys and high-resolution SKA maps of the three ISM phases. Ground based surveys and Gaia provide massive amounts of DIB strengths and radial velocities towards targets seen through all kinds of interstellar phases and cloud opacities. Cross matching this data with SKA lines will allow deeper studies of the link between DIBs and the physical state and the history of interstellar clouds. Subsequently, in combination with dust emission, it will be possible to explore the link with the dust grain distributions in the clouds. 

\smallskip
\noi
{\bf  Probing the ionised ISM} 

\smallskip
\noi Radio recombination lines (RRLs) of many atomic species will be detected by the SKA. Each species can produce many different RRLs and their combination can be used to deduce many properties of the ionised phase of the ISM (Thompson et al. 2005, and references therein). %
In contrast to present day surveys, the SKA will detect RRLs both at both high resolution and high sensitiviy.
These lines are excellent probes of the WIM and are essential to complete our picture of the nearby ISM with its shells, bubbles, and cavities. In addition, MeerGAL is a proposed Galactic plane survey using the 64-dish MeerKAT telescope, a SKA precursor and the most sensitive cm-wave telescope in the southern hemisphere. Observations are to be collected from 10-14 GHz over a total of 3300 hours. These observations will map out young \hii~regions as well as RRLs. 

\smallskip
\noi 
The ionised ISM can also be probed using dispersion measures of the radio pulses of Galactic pulsars in order to reconstruct  a 3D model of the free electron density in the Galaxy (NE2001, Cordes \& Lazio, 2003). This 
model is very poorly defined in the nearby ISM due to low numbers of pulsars off the Galactic plane and to the complexity of the 3D distribution of nearby ionised bubbles.
SKA1 will  increase the number of observed pulsars by more than an order of magnitude (Keane et al. 2014), including many at high latitude. The combination of these dispersion measures, of the RRLs that the SKA will detect, as well as observations of free-free emission (e.g., Planck collaboration, 2016), will enable a major improvement of the NE2001 model in the nearby ISM. \\

\parbox{0.9\textwidth}{
\noi{References:}\\
\noi{\scriptsize 
Cordes \& Lazio, 2003, arXiv:astro-ph/0207156v3; 
Gaia collaboration, 2016, A\&A, 595, A1; 
Dickey, J.M., et al., 2013, PASP, 30, 3; 
Green et al., 2014, ApJ, 783, 114; 
Lallement, R. et al., 2014, A\&A, 561, A91;
Keane, E.F. et al., 2015, AASKA14, 40;
Marshall, D.J. et al., 2006, A\&A, 453, 635;
McClure-Griffiths, N. M. et al., 2015, AASKA14, 130;
Planck collaboration, 2016, A\&A, 594, A25;
Planck \& Fermi collaborations, 2015, A\&A, 582, A31;
Thompson, M. et al., 2015, AASKA14, 126
}}

\subsubsection{Turbulent cascade}
\vspace{0cm}

\noi {\bf Basics of turbulence}

\smallskip
\noi Turbulence is the occurrence of chaotic motions in a fluid, preventing any deterministic description of its velocity field $\vec{v}$ on long timescales, so that only statistical analyses are appropriate. This impredictability stems from the non-linearity and non-locality of the equation describing the time evolution of $\vec{v}$, through which instabilities lead to the fragmentation of large eddies into ever smaller ones, until the viscous damping time is equal to the eddy turnover time. That picture led to the K41 description of a "turbulent cascade" (Kolmogorov 1941), which is based on the assumption that the energy transfer rate $\epsilon$ from scale to scale is constant, such that the solution is statistically time-invariant. Key results from this K41 theory are that the velocity fluctuations $\delta v_l$ at a scale $l$ are approximately given by $\delta v_l\approx(\epsilon l)^{1/3}$, that the velocity power spectrum integrated over directions scales with wavenumber $k$ as $P_v(k)\propto \epsilon^{2/3}k^{-5/3}$, and that the structure functions of the velocity field\footnote{These are defined through $S_p(l)=\langle\left[\vec{v}(\vec{r}+\vec{l}).\vec{u}-\vec{v}(\vec{r}).\vec{u}\right]^p\rangle$, where  $\vec{u}=\vec{l}/l$ and the average is over lag vectors $\vec{l}$ at given $l$.} scale as $S_p(l)\propto (\epsilon l)^{p/3}$ for any positive integer $p$. The main flaw of the K41 theory, despite its successes, is that it cannot explain phenomena such as {\it intermittency}, which is the fact that energy dissipation occurs in intense bursts, highly localised in both time and space. In experimental studies, intermittency manifests itself through the presence of non-Gaussian wings in the probability distribution functions (PDFs) of $\delta v_l$ at small scales. Several improvements of this turbulent cascade picture have been proposed to take into account effects of 1/ compressibility (see, e.g., Falceta-Gon\c{c}alves \etal\ 2014), 2/ magnetisation in both isotropic (Iroshnikov 1964; Kraichnan 1965a,b) and anisotropic (Sridhar \& Goldreich 1994; Goldreich \& Sridhar 1995) frameworks, and 3/ both compressibility and magnetisation (Kritsuk \etal\ 2007).
 
\smallskip
\noi {\bf Interstellar turbulence}

\smallskip
\noi The idea of interstellar turbulence was first suggested in 1951 by Von Weisz\"acker and Von H\"orner. It is supported by observations of power-law power spectra over a large range of scales for a number of tracers, such as the local free electron density (Armstrong \etal\ 1995) or atomic hydrogen ({\hi}) density and velocity fluctuations (e.g., Miville-Desch\^enes \etal\ 2003a). This self-similarity is expected to span up to eight orders of magnitude in size: most of the energy injection, from massive stellar feedback, supernov{\ae}, and Galactic shear, occurs on 100\,pc to 1\,kpc scales, and the turbulent cascade trickles energy down to dissipation scales of a few AU ($10^{-5}\,\mathrm{pc}$). Although the spectral indices found in different studies reveal a large scatter (see the review by Hennebelle \& Falgarone 2012), this may be due to projection effects (Lazarian \& Pogosyan 2000; Miville-Desch\^enes \etal\ 2003b; Levrier 2004) and to the large panel of tracers being used. Other observational diagnostics are suggestive of turbulence, such as the Larson (1981) scaling between the sizes of molecular clouds and their suprathermal linewidths, $\sigma_v\propto l^\alpha$ with $\alpha\sim 0.4$, or the constancy of $\rho\sigma_v^3/l$ observed over a wide range of scales (Hennebelle \& Falgarone 2012), in agreement with theoretical findings of Kritsuk \etal\ (2007) for compressible MHD. Hints of the intermittent dissipation of turbulence at small scales are also apparent : the non-Gaussian wings of centroid velocity increments in CO observations by Hily-Blant \etal\ (2008) are linked to the plane-of-the-sky projection of the vorticity (Lis \etal\ 1996), and they are shown to arise from coherent, milliparsec-scale elongated structures exhibiting large velocity shears (Falgarone \etal\ 2009).

\smallskip
\noi {\bf The promise of SKA for interstellar turbulence}

\smallskip
\noi The SKA is purposefully built as a "gas survey machine", with a particular focus on the 21\,cm line from atomic neutral hydrogen (\hi), a ubiquitous component of the interstellar medium (ISM) which exhibits structure on a wide range of scales, from $\sim 1\,\mathrm{kpc}$ to a few tens of AU. By measuring this spectral line with high brightness temperature sensitivity, high angular resolution, wide field-of-view, and high spectral resolution, SKA will provide tremendous advances in our understanding of the distribution and kinematics of the atomic gas, and therefore of the turbulent cascade over the entire range of relevant scales (Staveley-Smith \& Osterloo 2015). 

\smallskip
\noi
The {\hi} emission is mostly optically thin, so its brightness temperature is proportional to the column density of atomic hydrogen, $T_b \propto N_\mathrm{H}$, and includes contributions from the warm neutral medium (WNM) and the cold neutral medium (CNM), in proportions that vary depending on the line-of-sight probed. The spectral line cubes of {\hi} emission provided by state-of-the art surveys such as GASS (McClure-Griffiths \etal\ 2009; Kalberla \& Haud 2015), EBHIS (Winkel \etal\ 2016) and GALFA-\hi~(Peek \etal\ 2011) allow one to characterise the power spectrum of turbulent motions in the WNM, provided one can remove the CNM part. These surveys reach angular resolutions of $4'$ to $16'$, spectral resolutions of $0.2$ to $1.5\,\mathrm{km\,s^{-1}}$, and typical column density sensitivities of a few $10^{18}\,\mathrm{cm^{-2}}$. The high angular and spectral resolutions, and the high sensitivity provided by SKA will extend these studies to much smaller scales. The CNM itself is best traced by {\hi} absorption measurements, either as self-absorption features (HISA) yielding maps of its spatial distribution, or using strong continuum background sources, which allows one to derive its spin temperature. These absorption measurements will probe the smallest scales and therefore constrain turbulence dissipation mechanisms (Dickey \etal\ 2004). For instance, by targeting extended continuum background emission such as supernova remnants, variations of a few percent in $\tau$ can be mapped over scales of a few $0.1\arcsec$ (corresponding to 10\,AU at the 100\,pc distance to nearby clouds).

\smallskip
\noi
While it is difficult to determine the density and temperature structure of the atomic gas solely from 21\,cm {\hi} data, low-frequency narrow carbon radio recombination lines (RRL, probed by the $50-350\,\mathrm{MHz}$ SKA1-LOW), which are good tracers of the cold atomic gas, will help disentangle contributions from CNM and WNM (Oonk \etal\ 2015). Broader RRLs from hydrogen will likely arise from the ionised gas (WIM) and will trace its own kinematics (Thompson \etal\ 2015). The distinction between cold and warm components of {\hi} will also help constrain models of turbulent mixing by measuring the amount of thermally unstable gas, thought to be transitioning between the CNM and WNM stable phases. The SKA will thus be able to compare turbulent motions in various phases of the ISM and address important topics such as, e.g., 1/ determining whether turbulence in the CNM is "frozen-in" as cold atomic gas transitions to the molecular phase; 2/ assessing the dominant mode of turbulent energy injection (solenoidal vs compressive), which can be estimated from spectral line cubes using the method of Brunt \& Federrath (2014), as demonstrated with the molecular line survey of the Orion B molecular cloud (Orkisz \etal\ 2017); and 3/ providing constraints for global turbulent ISM models, using dust reddening measurements from PAN-STARRS and LSST and distance measurements from GAIA, which together with SKA data will provide a true tomography of the atomic ISM, with 3 spatial dimensions and one velocity component (McClure-Griffiths et al. 2015).

\smallskip
\noi
Moreover, SKA Galactic surveys aimed at the {\hi} line will simultaneously provide maps of the synchrotron continuum at 1.4\,GHz, both in total intensity and linear polarisation. As demonstrated by Gaensler \etal\ (2011) using ATCA, this data will help probe turbulence in the magneto-ionic medium through polarisation gradients. Regarding the interplay between matter structures and magnetic fields in the ISM, this will complement similar Faraday tomography studies of this medium with LOFAR (Zaroubi \etal\ 2015) and the analysis of thermal dust polarisation by Planck (see, e.g., Planck Int. XX 2015; Planck Int. XXXV 2016; Soler \etal\ 2016).
\\

\parbox{0.9\textwidth}{
\noi{References:}\\
\noi{\scriptsize 
Armstrong,  J. W., \etal\, 1995, ApJ, 443, 209;
Brunt, C. M. \& Federrath, C., 2014, MNRAS, 442, 1451;
Dickey, J. M., \etal\, 2004, New Astr. Rev., 48, 1311;
Falceta-Gon\c{c}alves, D., \etal\, 2014, Nonlinear Processes in Geophysics, 21, 587;
Falgarone, E., \etal\, 2009, A\&A, 507, 355;
Gaensler, B. M., \etal\, 2011, Nature, 478, 214;
Goldreich, P. \& Sridhar, S., 1995, ApJ, 438, 763;
Hennebelle, P. \& Falgarone, E., 2012, ARAA, 20, 55;
Hily-Blant, P., \etal\, 2008, A\&A, 481, 367;
Hily-Blant, P. \& Falgarone, E., 2009, A\&A, 500, L29;
Iroshnikov, P. S., 1964, Sov. Ast., 7, 566;
Kalberla, P. M., W. \& Haud, U., 2015, A\&A, 578, A78;
Kolmogorov, A. N., Akademiia Nauk, 1941, SSSR Doklady, 30, 301;
Kowal, G. \& Lazarian, A., 2007, ApJL, 666, L69;
Kraichnan, R. H., 1965, Phys. of Fluids, 8, 575;
Kraichnan, R. H., 1965, Phys. of Fluids, 8, 1385;
Kritsuk, A. G., \etal\, 2007, ApJ, 665, 416;
Larson, R. B., 1981, MNRAS, 194, 809;
Lazarian, A. \& Pogosyan, D., 2000, ApJ, 537, 720;
Levrier, F., 2004, A\&A, 421, 387;
Lis, D. C., \etal\, 1996, ApJ, 463, 623, 1996;
McClure-Griffiths, N. M., \etal\, 2009, ApJS, 181, 398;
McClure-Griffiths, N. M., \etal\, 2015, PoS, AASKA14, 130;
Miville-Desch\^enes, M.-A., \etal\, 2003, A\&A, 411, 109;
Miville-Desch\^enes, M.-A., \etal\, 2003, ApJ, 593, 831;
Oonk, R., \etal\, 2015, AASKA14, 139;
Orkisz, J. H., \etal\, 2017, A\&A, 599, 99;
Peek, J. E. G.,  \etal\, 2011, ApJS, 194, 20;
Planck collaboration Int. XX, 2015, A\&A, 576, 105; 
Planck collaborarion Int. XXXV, 2016, A\&A, 586, 138; 
Soler, J. D., \etal\, 2016, A\&A, 596, 93;
Sridhar, S. \& Goldreich, P., 1994, ApJ, 432, 612;
Staveley-Smith, L. \& Osterloo, T., 2015, AASKA14, 167;
Thompson, M., \etal\, 2015, AASKA14, 126;
von H\"orner, S., 1951,  Zeit. Ast., 30, 17;
von Weizs\"acker, C. F., 1951, ApJ, 114, 165;
Winkel, B., \etal\, 2016, A\&A 585, A41;
Zaroubi, S., \etal\, 2015, MNRAS, 454, 46
}}\\

\subsubsection{The formation of cold atomic structures}
\vspace{0cm}

\noi{\bf Introduction}

\smallskip
\noi Most structures in the Universe are formed through the action of self-gravity. However, it appears that some structures, like atomic (\hi) interstellar clouds, have an internal energy far greater than their gravitational energy. These cold clouds (CNM - $T\sim 20-200$\,K) are formed from the diffuse and warm \hi\ (WNM - $T\sim 10^4$\,K) through a condensation process that does not involve gravity but the thermal instability (Field 1965). 
The efficiency of the transition of the gas from the WNM to the CNM phases is an important aspect of the regulation of the star formation rate in galaxies (Ostriker et al. 2010). How exactly atomic gas transits from diffuse warm gas into denser, colder clouds in which the hydrogen is mostly in molecular form and in which stars form, is still not fully understood.
What seems clear is that the time it takes for \hi\ to condensate is similar to the dynamical time of the turbulent motions. This is attested by the fact that a significant fraction of the \hi\ mass is observed at temperatures between 500 and 5000\,K corresponding to the thermally unstable regime (Heiles \& Troland 2003). 

\smallskip
\noi
The current scenario by which the CNM forms involves an increase of the pressure of the WNM by a large scale process, like a converging flow, sending it out of thermal equilibrium. Sonic turbulence in the pressured WNM then provides the required velocity and density fluctuations to trigger the transition to the thermally stable cold phase (Hennebelle \& Audit, 2007; Saury et al. 2014). One current hypothesis is that the formation of molecular gas and stars may be the result of gravitational contraction starting in the atomic phase (e.g., Heitsch et al. 2008). Large molecular clouds may then develop from an agglomeration of small cold clouds, collected together by the sweeping action of spiral arms (Dobbs et al. 2012) or through the compressive action of H\mbox{\,\sc ii} regions or superbubbles that can trigger the phase transition and collect \hi\ clouds (Ntormousi et al. 2011; Clark et al. 2012).
The combined action of photons and mechanical energy from young stars ultimately destroys the molecular clouds, and a large fraction of the gas is returned to either ionized or atomic form. Part of understanding how a galaxy evolves is understanding how much hydrogen exists in its various states and the time it takes for gas to flow between states and scales. 

\smallskip
\noi{\bf The HI-H2 transition} 

\smallskip
\noi Linking the phase transition of the \hi\ to the star formation cycle requires estimating the amount and properties of gas in atomic and in molecular form. %
The nature of the \hi\ associated with a molecular cloud (e.g., its temperature distribution, the velocity field of the \hi) is needed to inform on how molecular clouds form and on what timescale. 

\smallskip
\noi
Since the first observations of the 21\,cm line in the 1950s, the combined analysis of absorption and emission observed in neighbouring lines of sight has been used to separate the contributions of the two phases (CNM and WNM) to the emission (e.g., Heiles \& Troland 2003). This leads to the estimate of the temperature, the column density and the velocity dispersion of the cold phase. This type of analysis is demanding and has been done only on a few hundred of lines of sight so far, often away from molecular clouds (Heiles \& Troland 2003; Strasser \& Taylor 2004; Dickey et al. 2009; Murray et al. 2017). 
In the last 10 years we have entered a period where the level of details reached by numerical simulation exceeds by far what has been obtained with observational data. There is now a lack of information from the data to constrain the numerical simulation results (on the CNM fraction for instance) and determine what processes dominate in different environments.
With the important gain in sensitivity, SKA key programs will provide $\sim 2 \times 10^5$ absorption measurements against radio sources (McClure-Griffiths et al. 2015). Having several of these sources per square degree, combined with the fully sampled imaging of the 21\,cm emission line, will allow us for the first time to map the structure of the CNM without ambiguity. Dedicated absorption observations will also allow us to map the \hi\ change of state through the molecular interface. 

\smallskip
\noi
In denser environments, self-absorption signature has been detected in the 21\,cm line emission where cold \hi\ gas is located in front of a warmer \hi\ background. These cold structures are often not seen in molecular tracers (Gibson et al. 2005). Therefore the self-absorption structures can be used to map the structure of the cold gas in and around molecular clouds, but because the temperature of the background emission is unknown, the temperature of the cold gas can not be estimated directly. Only the combination of \hi\ emission and absorption towards continuum sources gives access to the spin temperature of the gas, $T_s$. The combination of the two types of measurements provided by the SKA will allow us to use the full richness of the 21\,cm emission (including the self-absorption features) to map the spatial and thermal structure of cold \hi. 
Extracting the contribution of the cold gas from the emission data will then permit to estimate the properties of the warm and thermally unstable phases and finally understand the exact scenario that triggers the phase transition leading to the formation of cold and dense structures in the ISM. In particular, this SKA absorption grid will allow us to estimate the amount of thermally unstable gas over a large range of physical conditions and therefore evaluate the timescales of the formation process. One important goal of future studies will then be to characterise observationally the association between atomic and molecular gas.

\smallskip
\noi
Many difficulties hinder the estimate of the association of CNM and molecular structures. The combination of turbulent motions and thermal broadening from CNM and WNM structures makes the 21\,cm line complex and significantly wider ($\sim 7-10$\,km\,s$^{-1}$) than molecular lines ($\sim 1$\,km\,s$^{-1}$). This difference in spectral characteristics is the main reason why the atomic to molecular transition in the ISM has not been mapped systematically, in spite of the existence of both large scale \hi\ maps and large scale CO maps for more than three decades. 
{SKA \hi\ absorption measurements, combined with high surface brightness sensitivity data at high Galactic latitude and sub-arcminute resolution will generate a substantial catalogue of isolated \hi\ clouds. These clouds can be observed with matching angular ($\sim 20''$) and velocity ($0.3$\,km\,s$^{-1}$) resolutions in molecular tracers, such as OH, CO and HCO$^+$. The matching of molecular and atomic tracers will enable quantitative analysis of important atomic to molecular transitions in many well-defined systems. }

\smallskip
\noi
The SKA will enable essential complementary surveys of the hydroxyl (OH) molecule, whose emission is very weak, despite its probable ubiquity and large observed column densities. Recent evidence that a large fraction of molecular gas exists in the so-called "CO-dark molecular gas" form (Grenier et al. 2005; Planck Collaboration 2011) is an incentive to look for  other tracers of cold H$_2$ than CO. 
Due to its low excitation temperature, large scale maps of OH emission have not been available. This will become possible with the SKA. Combined with dust tracers and gamma ray measurements, the combination of \hi\ and molecular tracers should lead to better constraints on the \hi-H$_2$ transition, on the molecular content of the ISM, and on the evolution of the fraction of CO-dark H$_2$ with Galacto-centric radius and location with respect to the spiral arms.

\smallskip
\noi{\bf Link with star formation} 

\smallskip
\noi Current theoretical expectations are that the fraction of the \hi\ in the CNM phase ($f_{\rm CNM}$), as well as the star formation rate, should scale with the pressure of the WNM (Ostriker et al. 2010; Saury et al. 2014). The exact value of $f_{\rm CNM}$ is still a matter of debate, even more its variation across the Galactic plane.
Absorption measurements in the Galactic plane seem to indicate a constant value in the outer Galaxy ($R_{\rm gal} > 8.5$\,kpc) of $f_{\rm CNM} \sim 10-25$\,\% (Dickey et al. 2009). There the scale height of the CNM and WNM are similar. In the inner galaxy, where star formation rate is higher, the CNM is much more concentrated towards the plane with a scale height of $\sim 100$\,pc, a third of the scale height of the WNM. The difference in scale heights is attributed to the different pressure regime which puts the \hi\ in physical conditions that favour the phase transition (Wolfire et al. 2003). To make progress on this topic, more absorption measurements are needed, in the Milky Way and in nearby galaxies. 
For now there is only a few tens of absorption measurements in the SMC and LMC. The sensitivity of the SKA will allow us to extend this type of measurements dramatically which will open the possibility to study the efficiency of the phase transition as a function of the local physical conditions (metallicity, pressure, star formation).\\


\parbox{0.9\textwidth}{
\noi{References:}\\
\noi{\scriptsize Clark, P.~C., et al., 2012, MNRAS, 424, 2599;
Dickey, J.~M., et al., 2009, ApJ, 693, 1250; 
Dobbs, C.~L., et al., 2012, MNRAS, 425, 2157;
Field, G.~B., 1965, ApJ, 142, 531;
Gibson, S.~J., et al., 2005, ApJ, 626, 195;
Grenier, I.~A., et al., 2005, Science, 307, 1292;
Heiles, C. \& Troland, T.~H., 2003, ApJ, 586, 1067;
Heitsch, F., et al., 2008, ApJ, 683, 786;
Hennebelle, P. \& Inutsuka, S.-i., 2006, ApJ, 647, 404;
Hennebelle, P. \& Audit, E., 2007, A\&A, 465, 431;
McClure-Griffiths, N.~M., et al. 2015, AASKA14, 130;
Murray, C.~E., et al., 2017, ApJ, 837, 55;
Ntormousi, E., et al., 2011, ApJ, 731, 13;
Ostriker, E.~C., et al., 2010, ApJ, 721, 975;
Planck Collaboration, et al., 2011, A\&A, 536, A19;
Saury, E., et al., 2014, A\&A, 567, A16;
Strasser, S. \& Taylor, A.~R., 2004, ApJ, 603, 560;
Wolfire, M.~G., et al., 2003, ApJ, 587, 278
}}\\

\subsubsection{Molecular complexity in cold cores and hot corinos}
\vspace{0cm}

\noi A tremendous advance in instrumentation for spectroscopy of the interstellar medium took place during past decades. Major discoveries have occurred because of the greater sensitivities of existing telescopes such as IRAM (30m and NOEMA), ALMA and SOFIA. The high sensitivity as well as the spectral resolution of the instruments led to the discovery of many species, and the spatial resolution was the key point to uncover the spatial distribution of these species. Complex organic molecules (COMs), such as as methyl formate (HCOOCH$_3$), dimethyl ether (CH$_3$OCH$_3$) and formamide (NH$_2$CHO), should form in every step through ice mantle desorption as well as gas-phase reactions. Formamide has recently been suggested to be a central species in the synthesis of metabolic and genetic molecules (Saladino et al. 2012), the chemical basis of life, such as amino acids, nucleic acids and bases, acyclonucleosides, sugars, amino sugars, and carboxylic acids. Glycine (NH$_2$CH$_2$COOH) is one of the simplest amino acids and a key constituent of living organisms. So far it has been detected in a comet (Altwegg et al. 2016) and remains a holy grail for the detection of COMs in space. 

\begin{figure}[!ht]
  \centering
  \includegraphics[angle=270,width=0.4\linewidth]{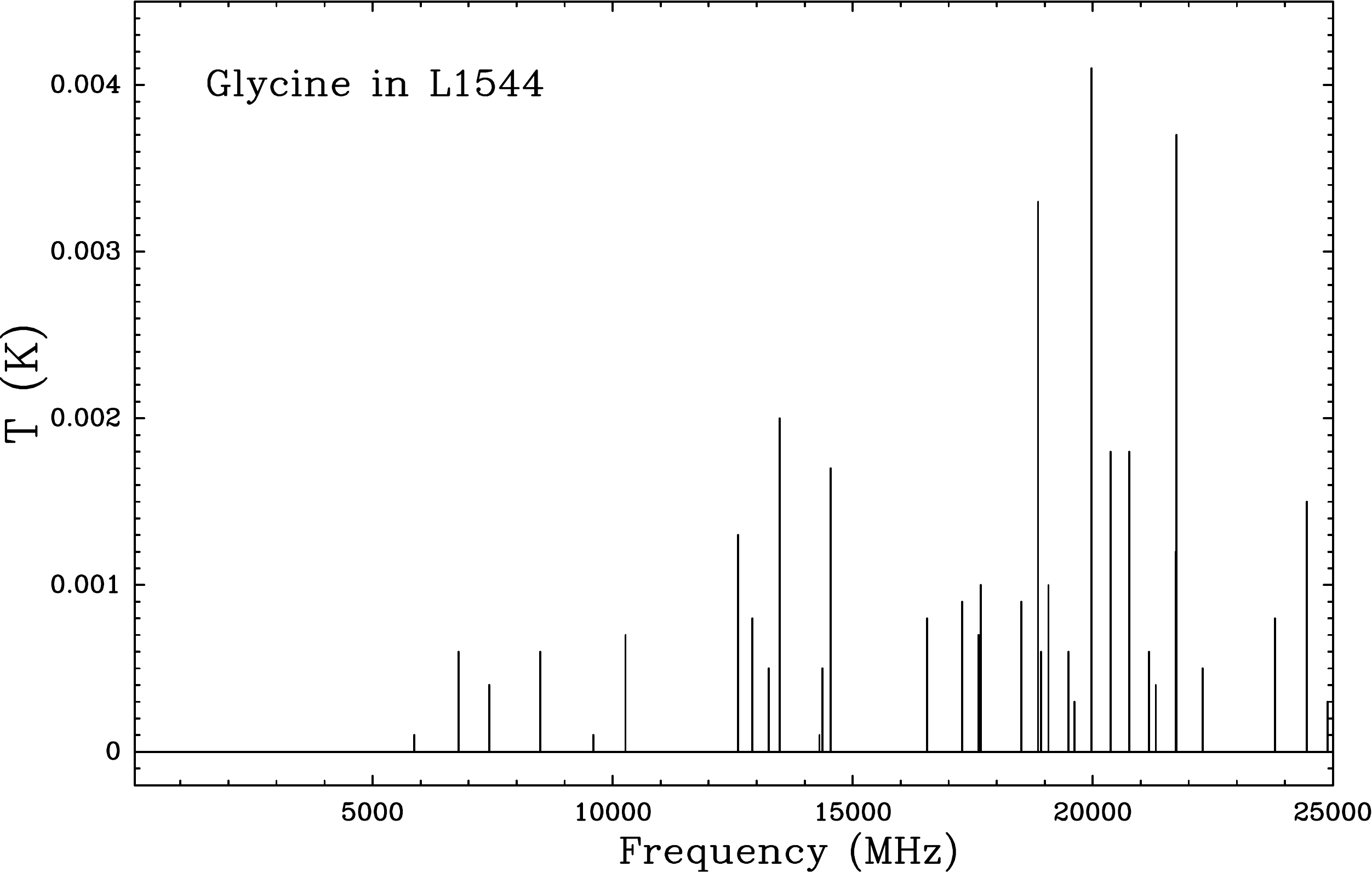}%
  \includegraphics[angle=270,width=0.4\linewidth]{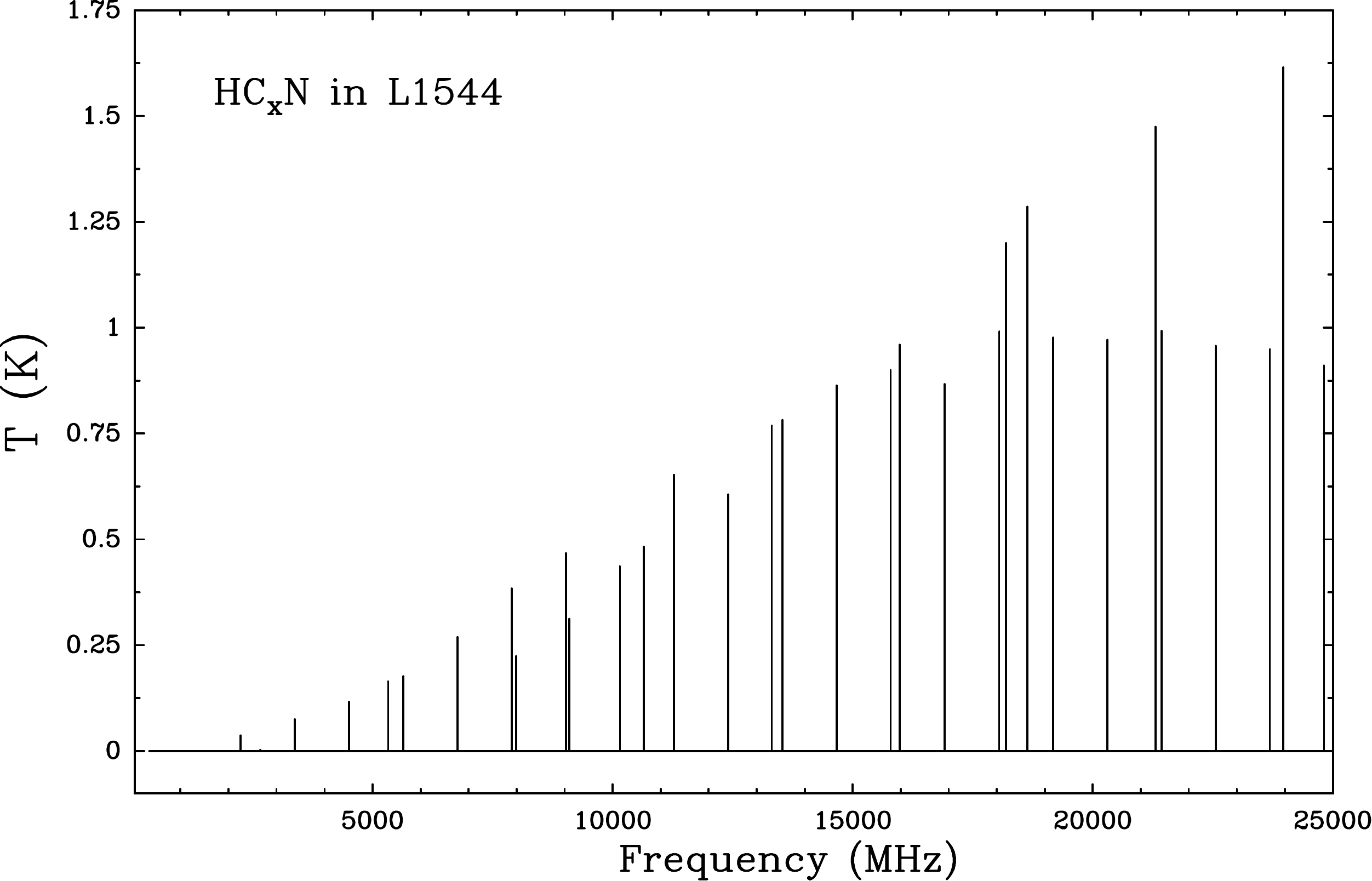}
  \caption{\label{fig:l1544} CASSIS simulation for glycine (conformer I) and HC$_n$N (where $n$=3,5,7, ...) obtained between 350~MHz and  25~GHz.}
\end{figure}

\begin{figure}[!ht]
  \centering
  \includegraphics[width=0.4\linewidth]{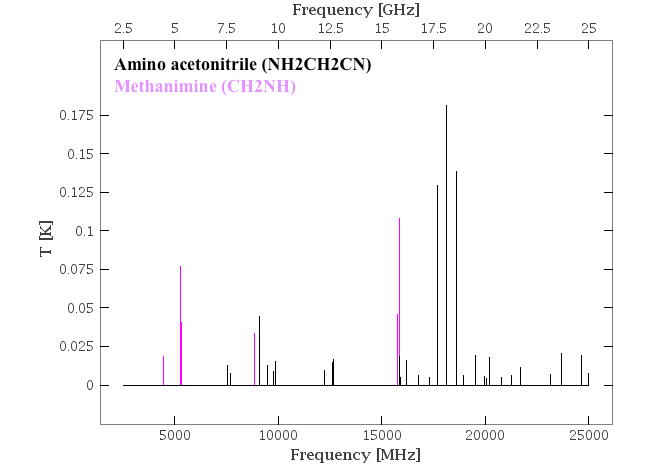}
  \caption{\label{fig:I16293} CASSIS simulation for two precursors of glycine in IRAS 16293-2422. Note that for methanimine, each apparent line actually corresponds to at least three transitions.}
\end{figure}

\smallskip

\noi COMs have been detected in dark clouds (Marcelino et al. 2007), prestellar cores (Cernicharo et al. 2012; Bacmann et al. 2012; Vastel et al. 2014) and hot corinos (Cazaux et al. 2003; Bottinelli et al. 2004; Jorgensen et al. 2012; Kahane et al. 2013). Low mass star forming regions are best suited for the detection of COMs, especially in the earliest phases represented by prestellar cores. The level of confusion is indeed lower in those cores, considering the low gas temperatures ($\sim$ 10 K), leading to a smaller excitation of the lines compared to hot cores and hot corinos. The spectra are also expected to suffer less from line blending, with lower linewidths ($\le$ 0.5 km/s) leading to a more precise identification. Recent spectral surveys at IRAM-30m such as TIMASS and ASAI revealed a high level of molecular complexity (Caux et al. 2011; Vastel et al. 2016 and references within). Upper limits on the glycine abundance have been estimated in a prestellar core (Jimenez-Serra et al. 2016) and a low-mass protostar (Ceccarelli et al. 2000) with observed transitions higher than 100~GHz. \\
We present in Fig.\,\ref{fig:I16293} a LTE CASSIS (http://cassis.irap.omp.eu/) simulation (between 350 MHz and 25 GHz) for glycine in the prototypical prestellar core L1544, using N=6$\times10^{12}$ cm$^{-2}$, T$_{ex}$=5~K and a 0.3~\kms\ linewidth, based on the upper limits estimated by Jimenez-Serra et al. (2016). We show that the brightest lines lie around 20 GHz and are around 4~mK, corresponding to a 0.2~mJy intensity considering a 12$^{\prime\prime}$ beam. For baselines corresponding to this beam size, we expect to find $\sim$100 SKA1 dishes. Using a sensitivity per SKA1 dish at 20~GHz of 2~m$^2$\,K$^{-1}$ (Fig.\,5 of the SKA1 system baseline V2 description SKA-TEL-SKO-0000308), an integration time of 1000 hours and a spectral resolution of 0.2~\kms\ (sufficient to detect the line), we expect a rms noise level of 63 $\mu$Jy (see equations 9.36 from Tools of Radioastronomy, 5th edition):\\

\begin{equation}
\Delta S_{\nu} (Jy) = 1.02\frac{1}{(A_e/T_{sys}) \sqrt{Nt\Delta\nu}} \hspace{2cm}
\Delta T_{b} (K) = 13.58\frac{\lambda^2}{(A_e/T_{sys})\theta^2 \sqrt{Nt\Delta\nu}}
\end{equation}

\smallskip

\noi where $A_e/T_{sys}$ is the dish sensitivity (m$^2$/K), N=n(n-1)/2 the simultaneous pair-wise correlations (n being the number of antennas), $t$ the integration time in hours, $\Delta\nu$ the resolution in kHz, $\lambda$ in mm and $\theta$ in arcsec.
Hence, the three brightest transitions at $\sim$~20~GHz should be marginally detected, with a S/N of 3.2. The sensitivity at 20 GHz is a crucial parameter for the detection of glycine and other COMs such as formamide in prestellar cores and increasing its value by a factor 2 will increase the S/N by the same factor.
\smallskip

\noi 
Nonetheless, other species of interest can be observed towards these sources with the frequency coverage currently scheduled for deployment. In particular, cyanopolyynes (HC$_n$N where $n$=3,5,7, ...) are a key family of molecules that can be used as probes of physical and chemical conditions. We show that cyanopolyynes have much brighter lines than glycine, leading to quicker detections (less than a minute).
Another species of interest is the NH$_3$ low energy transition (1$_{1,1} - 1_{1,0}$) at 23.7 GHz (1.3 K for a column density of at least 3.5$\times10^{14}$ cm$^{-2}$: Crapsi et al. 2007). For all these observations, line blending is not a problem as reported in spectral surveys of prestellar cores in the 70--110 GHz range and we checked with our CASSIS simulation, taking into account many species (already detected in L1544), that it will be the same in the 350 MHz--25 GHz range. 
\smallskip

\noi 
Despite the unprecedented planned performances of the SKA, the identification of glycine in a prestellar core will remain very challenging. Unfortunately, the search for this amino acid is not expected to be easier in young protostellar sources such as IRAS16293-2422. Indeed, although the column density may be a bit higher than in prestellar cores ($\lesssim 10^{13}$~cm$^{-2}$ in a $11^{\prime\prime}$ beam -- Ceccarelli et al. 2000), the higher $T_{\rm ex}$ implies that the glycine spectrum is shifted to higher frequencies, and transitions at $\sim$~20~GHz are modelled to be weaker than 0.2~mK. Better targets for hot corinos would be glycine precursors such as amino acetonitrile (NH$_2$CH$_2$CN) and methanimine (CH$_2$NH) as they are expected to be $\sim$~100 times more abundant (Belloche et al. 2008). The CASSIS simulation for these species is shown in Fig.\,\ref{fig:I16293} ($N=10^{15}$~cm$^{-2}$, $T_{\rm ex}$~=~100~K, in a $11^{\prime\prime}$ beam). We checked with CASSIS that the transitions in Fig.\,\ref{fig:I16293} would not suffer blending with other species. These glycine precursors could be identified in 1000 hours with SKA1 (sensitivity per dish of 4~m$^2$~K$^{-1}$ at 10~GHz, spectral resolution of 0.2~\kms, adequate to resolve the $\gtrsim$~1~\kms\ lines), whereas only 10 to 50 hours would be enough with SKA2 at 20 and 15~GHz. Also, as for L1544, cyanopolyynes would be detected in a very short integration time.

\smallskip

\noi The SKA  will give access to a new spectral range, with an unprecedented spatial resolution and sensitivity. SKA1 Low and SKA1 Mid will cover a frequency range from 50 MHz to 14 GHz, giving access to low energy levels of numerous COMs. The study of COMs in Sun-like star-forming regions would highly benefit from SKA2, since it would give access to the frequency band at $\sim$20 GHz, where the emission of glycine is expected to be the brightest. SKA appears as a good complement to ALMA for the study of molecular complexity in hot corinos and starless cores.\\

\parbox{0.9\textwidth}{
\noi{References:}\\
\noi{\scriptsize Altwegg, K., et al., 2016, Science Advances, 2, 5; Bacmann, A., et al., 2012, A\&A, 541, 12; Belloche, A., et al., 2008, A\&A, 482, 179; Bottinelli, S., et al., 2004, ApJ, 615, 354; Caux, E., et al., 2011, A\&A, 532, 23; Cazaux, S., et al., 2003, ApJ, 593, L51; Ceccarelli, C., et al., 2000, A\&A, 362, 1122; Cernicharo, J., et al., 2012, ApJ, 759, 43; Crapsi, A., et al., 2007, A\&A, 470, 221; Jimenez-Serra, I., et al., 2016, ApJ, 830, L6; Jorgensen, J. K.,  et al., 2012, ApJ, 757, 4; Kahane, C., et al., 2013, ApJ, 763, 38; Marcelino, N., et al., 2007, ApJ, 665, 127; Saladino, R., et al., 2012, cospar, 39, 1657; Vastel, C., et al., 2014, ApJ, 795, L2; Vastel, C., et al., 2016, A\&A, 591, L2}}\\

\subsubsection{Interstellar dust}
\vspace{0cm}

\noi Dust grains have a decisive role for many physical and chemical processes acting in the various environments of the interstellar medium. A good knowledge of their size, structure, and composition is thus required first to characterise their properties in all interstellar phases and how they do evolve from one phase to the other, and second to provide critical inputs for other areas of Galactic astronomy (see the other sections of this chapter). We will outline two key science topics for which SKA1 will advance our understanding of dust: the Anomalous Microwave Emission and the grain growth in protoplanetary disks towards the formation of the rocky cores of planets.

\begin{figure}[!ht]
  \centering
  \includegraphics[width=0.95\linewidth]{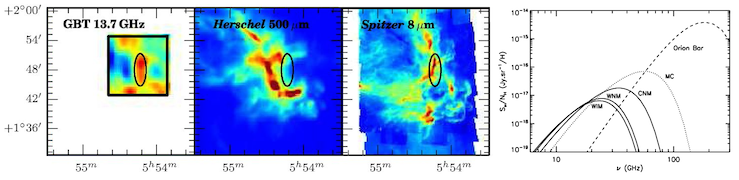}
  \caption{\label{fig:skalb_4_3_5_Fig_1} {\em Left}: maps of LDN1622 from GBT 13.7~GHz, Herschel 250~$\mu$m, and Spitzer 8~$\mu$m observations. {\em Right}: theoretical spinning dust emissivity for various interstellar environments from diffuse (CNM, WNM, WIM with increasing gas temperature) to dense (MC) and for a photodominated region (Orion Bar). Maps adapted from Harper et al. (2015) and spectra from Ysard et al. (2010).}
\end{figure}

\smallskip
\noi {\bf Anomalous Microwave Emission}

\smallskip
\noi In the late 1990s, a new emission component was detected between 10 and 50~GHz, which could not be explained by the traditional emission mechanisms known in this frequency range (CMB, free-free, synchrotron, and thermal dust). As a result, it was baptised the ``Anomalous Microwave Emission'' (AME). Since then, AME was detected in various interstellar environments: diffuse cirrus, dense clouds, compact \hii~regions, star forming regions (e.g., Ghosh et al. 2012; Harper et al. 2015; Dickinson et al. 2009; Scaife et al. 2010). At the arcmin scale, AME intensity is correlated with dust emission at mid-IR and far-IR wavelengths (Fig.\,\ref{fig:skalb_4_3_5_Fig_1}), the carriers of which are ultra-small grains (PAHs, a few nm in size) and bigger grains ($\sim 100-150$~nm), respectively. A number of models have been proposed to explain the AME and so far, even if still questioned (Hensley \& Draine 2016), the most plausible is electric dipole radiation from rapidly spinning PAHs (a.k.a. spinning dust, Draine \& Lazarian 1998; Ali-Ha\"imoud et al. 2009; Ysard \& Verstraete 2010). However, only a few studies have been able to tentatively show a better correlation between AME and PAHs than AME and bigger grains in a handful of clouds (Ysard et al. 2010; Harper et al. 2015). It is also difficult to tackle the environment physical properties responsible for the thermal and rotational excitation of PAHs, leading to mid-IR emission and AME, respectively. A large part of these difficulties arises from the lack of angular resolution of the existing facilities. With a frequency coverage up to 24~GHz and an angular resolution down to a few arcsecs, the SKA1-MID is perfectly suited to study AME and has the potential to deliver unique science for AME. Sensitive and high resolution SKA observations of dense clouds would give a clear picture of AME. AME studies would also benefit from the SKA bands 3 and 4 at lower frequency since they would allow us to distinguish AME from free-free and synchrotron emissions. Combined with higher frequency data (e.g., ALMA band 1 at 30~GHz and Spitzer IRAC in the mid-IR), it would be possible to establish the full spinning dust SED for a variety of targets. Once the spinning dust hypothesis would be confirmed, it would be possible to use this original tracer of PAHs as a new probe of the ISM. Indeed, the rotational excitation of PAHs depend on a number of parameters otherwise difficult to constrain: PAH size and electric dipole moment distributions, interstellar radiation field, H$^+$ and C$^+$ density, photoelectric emission, and H$_2$ formation. All these parameters are significant inputs when one tries to understand the chemical and dynamical properties of interstellar clouds.


\begin{figure}[!ht]
  \centering
  \includegraphics[width=0.95\linewidth]{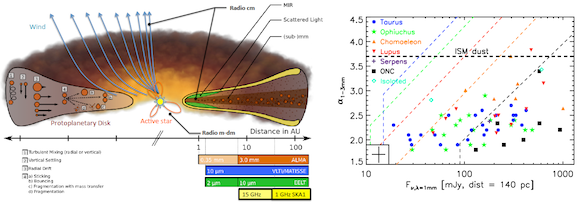}
  \caption{\label{fig:skalb_4_3_5_Fig_2} {\em Left panel}: sketch of a protoplanetary disk giving the main sources of continuum emission from the infrared to radio domains, the processes for grain growth, and the scales probed by various facilities in nearby star forming regions. {\em Right panel}: 1 to 3~mm spectral index $\alpha$ versus 1~mm flux of disks in nearby star forming regions and for the diffuse ISM. Low values of $\alpha$ indicate grain growth. Reproduced from Testi et al. (2014).}
\end{figure}

\smallskip
\noi {\bf Dust in protoplanetary disks}

\smallskip
\noi Protoplanetary disks, rich in gas and dust, are the birthplace of planets and therefore of life itself. Dust grains play an important role in disk evolution: their settling and growth are the first steps towards the formation of the rocky cores of planets. It is thus quite central to characterise their size and structure as a function of their location in the disks. The main observational proof of grain growth is the decrease of the mm-cm mass opacity spectral index from the diffuse ISM to very dense disks (Fig.\,\ref{fig:skalb_4_3_5_Fig_2}). Observations have revealed that grains can grow up to pebble-size (e.g., Ricci et al. 2010), which is difficult to explain theoretically due to the radial drift of solids towards the central star. One way to overcome this problem is dust trapping in pressure maxima (e.g., Fromang \& Nelson 2005). Observations also showed a radial dependence of the grain size in protoplanetary disks (Guilloteau et al. 2011). All these previous observational results imply that the only way to achieve breakthrough science in dust studies is to perform spatially resolved observations of disks at low frequency. For the first time, it will be possible to resolve disks in the microwave domain using SKA1-MID. SKA is also able to detect and spatially resolve the wind emission (see Fig.\,\ref{fig:skalb_4_3_5_Fig_2}) which has to be subtracted very accurately to study the dust properties. Combined with ALMA observations, SKA1-MID observations would allow us to radially trace grain growth down to a resolution of a few AU in nearby star forming regions. This kind of observational constraints are crucial to ascertain whether dust trapping is a realistic explanation for grain growth to the biggest sizes and also to determine timescale for the formation of the rocky cores of planets.
\noi Finally, mid-IR observations of disks around Herbig Ae/Be stars have shown that a big amount of mass can be trapped in nanometer-sized particles (e.g., Harbart et al. 2006). Detecting spinning dust emission from disks with SKA1-MID would allow us to better constrain their properties and the origin of this small grain population.\\

\parbox{0.9\textwidth}{
\noi{References:}\\
\noi{\scriptsize Ali-Ha\"imoud, Y., et al., 2009, ApJ, 395, 1055;
Dickinson, C., et al., 2009, ApJ, 690, 1585;
Draine, B.T. \& Lazarian, A., 1998, ApJ, 508, 157;
Fromang, S. \& Nelson R.P., 2005, MNRAS, 364, L81;
Ghosh, T., et al., 2012, MNRAS, 422, 3617;
Guilloteau, S., et al., 2011, A\&A, 529, 105;
Habart, E., et al., 2006, A\&A, 449, 1067;
Harper, S.E., et al., 2015, MNRAS, 453, 3375;
Hensley, B.S. \& Draine B.T., 2016, ApJ, 827, 45;
Ricci, L., et al., 2010, A\&A, 521, 66;
Scaife, A., et al., 2010, MNRAS, 405, L45;
Testi, L., \etal, 2014, AASKA14, 117;
Ysard, N. \& Verstraete L., 2010, A\&A, 509, 12;
Ysard, N., et al., A\&A, 509, L1
}}\\

\subsubsection{Faraday tomography}\label{sci:FarTom}
\vspace{0cm}

\noi In the past few years, Galactic magnetism has become a recognised and rapidly
growing field of research in astrophysics --  an evolution that was made possible 
by the great improvements in observational capabilities and in computational power. 
However, observations of the Galactic magnetic field face several important difficulties.
Indeed, the classical methods relying on Faraday rotation, synchrotron emission, 
and dust polarisation are neither direct (they involve the density of free electrons,
cosmic-ray electrons, or dust) nor complete (they give only the line-of-sight 
or plane-of-sky component of the magnetic field). 
In addition, they probe different (ionised, cosmic-ray filled, or dusty) regions 
of the multi-phase interstellar medium (ISM), such that they cannot easily 
be combined to yield the full magnetic field vector. 
But the major problem resides in the lack of access to the line-of-sight dimension,
since the above methods provide only line-of-sight integrated quantities, with no
information on how the integrant (Faraday rotation rate, synchrotron emissivity,
or dust emissivity) varies along the line of sight.
One way to overcome the latter problem, and thus gain access to the 3D structure 
of the Galactic magnetic field, is to perform Faraday tomography, 
also known as rotation measure synthesis (RMS).
The idea of this new technique is 
to exploit the wavelength-dependent Faraday rotation of the Galactic diffuse 
synchrotron emission to separate the different Faraday-rotating and
synchrotron-emitting regions and locate them along the line of sight.$\!$
\smallskip

\noi 
Remember that Faraday rotation is the rotation of the polarisation direction
of a linearly-polarised radio wave that propagates through a magneto-ionic medium,
where it interacts with the free (thermal) electrons gyrating about magnetic
field lines.
When the radio source is a background source,
i.e., when the regions of radio emission and Faraday rotation
are spatially separated,
the polarisation direction rotates by an angle

\begin{equation} 
\label{eq_Delta_theta}
\Delta \theta = {\rm RM} \ \lambda^2 \ ,
\end{equation}
where $\lambda$ is the observing wavelength
and RM is the so-called rotation measure, given by
\begin{equation}
\label{eq_RM}
{\rm RM} \, = \, {\cal C} \ \int _0 ^L n_e \ B_\parallel \ ds \ ,
\end{equation}

\smallskip

\noi with ${\cal C}$ a numerical constant, $n_e$ the thermal-electron density,
$B_\parallel$ the line-of-sight component of the magnetic field
(positive/negative when the field points toward/away from the observer),
and $L$ the path length between the source and the observer.
In practice, RM can be determined by measuring the polarisation direction 
of the incoming radiation at at least two different wavelengths.
Clearly, RM is a purely observational quantity,
which can be meaningfully measured only for a background radio source
and which can then be related to the physical properties
of the foreground Faraday-rotating medium through Eq.~(\ref{eq_RM}).
\smallskip

\noi 
In contrast, when the radio source is the Galaxy itself,
the regions of radio emission and Faraday rotation are spatially mixed.
In that case, $\Delta \theta$ no longer increases linearly with $\lambda^2$
(as in Eq.~(\ref{eq_Delta_theta})),
and the very concept of RM becomes meaningless.
However, one may resort to the more general concept of Faraday depth (FD),
defined as

\begin{equation}
\label{eq_FD}
\Phi(z) \, = \, {\cal C} \ \int _0 ^z n_e \ B_\parallel \ ds \ ,
\end{equation}

\smallskip

\noi where ${\cal C}$, $n_e$, and $B_\parallel$ have the same meaning
as in Eq.~(\ref{eq_RM}) and $z$ is the line-of-sight distance from the observer
(Burn 1966; Brentjens \& de Bruyn 2005).
$\Phi(z)$ has basically the same formal expression as RM (Eq.~\ref{eq_RM}),
but it differs from RM in the sense that it is a truly physical quantity,
which can be defined at any point of the Galaxy,
independent of any background radio source.
$\Phi(z)$ simply corresponds to the line-of-sight depth, $z$, measured in terms
of Faraday rotation -- in much the same way as optical depth corresponds
to line-of-sight depth measured in terms of opacity.
\smallskip

\noi 
When radio emission and Faraday rotation are mixed along the line of sight,
the polarised intensity measured at a given wavelength $\lambda$
is the superposition of the polarised emission produced at all line-of-sight
distances $z$, i.e., at all Faraday depths $\Phi$,
and Faraday-rotated by an angle $\Delta \theta = \Phi \ \lambda^2$:

\begin{equation}
\label{eq_PI}
P(\lambda^2) \, = \, \int _{-\infty} ^{+\infty} F(\Phi) \ 
e^{2 i \Phi \lambda^2} \ d\Phi \ ,
\end{equation}

\smallskip

\noi where $P(\lambda^2)$ is the complex polarised intensity ($P = Q + i U$)
at $\lambda$, and $F(\Phi)$ is the complex polarised intensity per unit $\Phi$,
also called complex Faraday dispersion function, at $\Phi$ (Burn 1966).
Since the Faraday rotation angle varies with wavelength,
the polarised intensities measured at different wavelengths
correspond to different combinations of all the line-of-sight contributions
and, therefore, provide different pieces of information.
Thus, the idea is to measure the polarised intensity at a large number
of different wavelengths and to convert its variation with $\lambda^2$
into a variation with $\Phi$.
Mathematically, this can be done inverting Eq.~(\ref{eq_PI}),
i.e., by taking the Fourier-like transform of $P(\lambda^2)$ to obtain $F(\Phi)$.
\smallskip

\noi 
\begin{figure}[h!]
\includegraphics[width=0.5\textwidth]{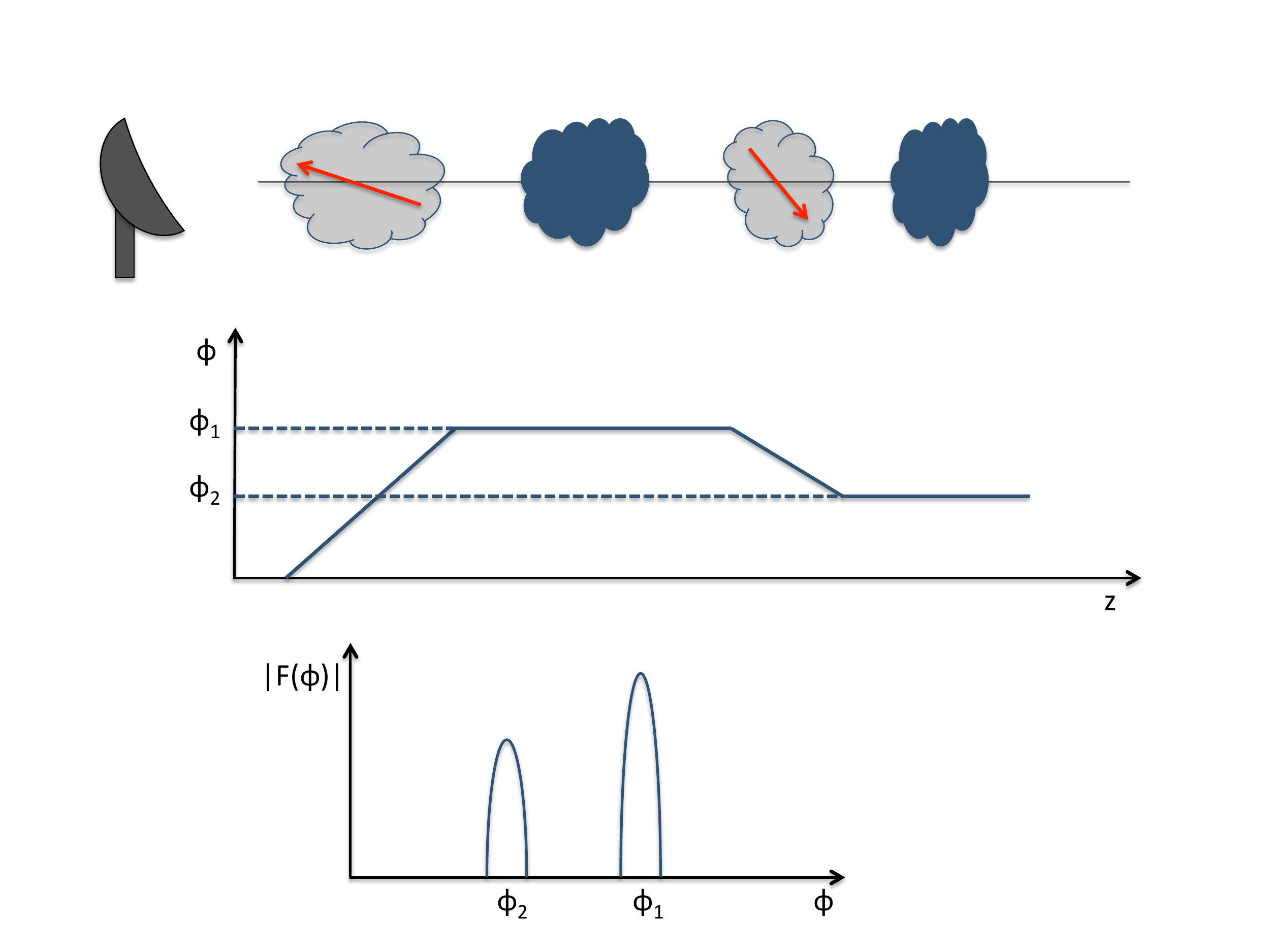}
\hfill
\includegraphics[width=0.5\textwidth]{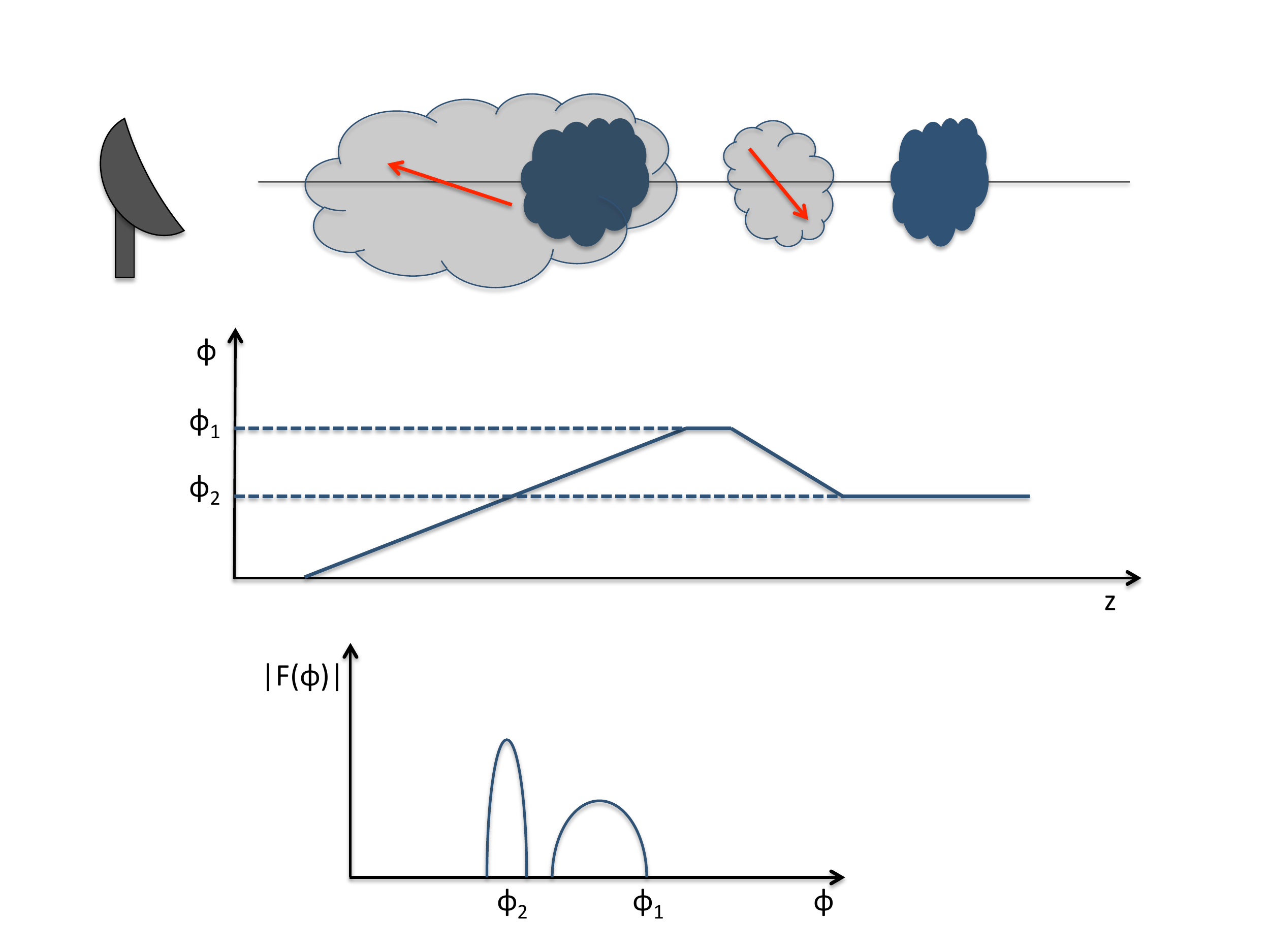}
\vspace*{0.1truecm} \\
\hspace*{0.21\textwidth} (a) \hfill (b) \hspace*{0.25\textwidth}
\caption{Schematics illustrating the concept of Faraday tomography.
The {\em top} panel in both (a) and (b) pictures the spatial configuration of the
system:
two Faraday-rotating clouds (light grey shading, with a red arrow representing 
the interstellar magnetic field)
and two synchrotron-emitting clouds (dark blue shading),
with the observer on the far left.
The {\em middle} panel shows how the Faraday depth, $\Phi$ (given by Eq.~(\ref{eq_FD})),
varies with line-of-sight distance from the observer, $z$.
The {\em bottom} panel provides the Faraday dispersion spectrum, $|F(\Phi)|$.
}
\label{figure_RMS}
\end{figure}
\smallskip

\noi
The method is illustrated in Fig.\,\ref{figure_RMS}, which depicts a situation
where the line of sight intersects two Faraday-rotating clouds (shaded in light grey),
across which $\Phi$ increases or decreases according to Eq.~(\ref{eq_FD}),
and two synchrotron-emitting clouds (shaded in dark blue).
The {\em top} panel in both Fig.\,\ref{figure_RMS}a and \ref{figure_RMS}b
indicates the positions of the four clouds along the line of sight
with respect to the observer placed on the far left, as well as the directions
of the interstellar magnetic field (red arrows) in the two Faraday-rotating clouds:
in the closer/farther cloud, the field points toward/away from the observer,
so that $B_\parallel$ is positive/negative and $\Phi$ increases/decreases
with increasing $z$.
The corresponding run of $\Phi$ with $z$ is plotted in the {\em middle} panel,
where $\Phi_1$ denotes the Faraday thickness of the closer cloud
and $\Phi_2$ the cumulated Faraday thickness of both Faraday-rotating clouds.
The {\em bottom} panel displays the Faraday dispersion spectrum, $|F(\Phi)|$,
with the two peaks representing the polarised emissions
from the two synchrotron-emitting clouds.
In Fig.\,\ref{figure_RMS}a, where the Faraday-rotating and synchrotron-emitting
clouds are spatially separated, the closer and farther synchrotron-emitting clouds
lie at FDs $\Phi_1$ and $\Phi_2$, respectively.
In Fig.\,\ref{figure_RMS}b, the closer synchrotron-emitting cloud
is embedded inside the closer  Faraday-rotating cloud,
so it has a finite Faraday thickness, i.e., it extends over a range of FDs
(up to nearly $\Phi_1$).
\smallskip

\noi
In practice, Faraday tomography can be used to separate synchrotron-emitting
regions located at different FDs along the line of sight
and to estimate their respective polarised synchrotron intensities,
which in turn can lead to the strength and the orientation
of their ${\boldvec B}_\perp$.
Faraday tomography can also be used to uncover intervening Faraday screens
and to estimate their Faraday thicknesses,
which in turn can lead to their $B_\parallel$.
The method is particularly interesting when the uncovered Faraday screens
can be identified with known gaseous structures,
because it then offers a new way of probing their magnetic field.
\smallskip

\noi
The SKA will be ideally suited to perform Faraday tomography,
because, in addition to its high sensitivity and fine angular resolution,
it will operate over a broad range of low radio frequencies.
The low frequencies will enhance the effects of Faraday rotation 
(which increase as $\lambda^2$) and hence improve sensitivity in FD space,
while the broad frequency coverage will make it possible to achieve 
fine resolution in FD space.
More quantitatively, if the wavelength range extends 
from $\lambda_{\rm min}$ to $\lambda_{\rm max}$,
the FD resolution is given by 
$\displaystyle \delta \phi \approx \frac{2 \sqrt{3}}{\Delta \lambda^2}$,
where $\Delta \lambda^2 = \lambda_{\rm max}^2 - \lambda_{\rm min}^2$,
and the maximum detectable Faraday thickness is given by 
$\displaystyle \Delta \phi_{\rm max} \approx \frac{\pi}{\lambda_{\rm min}^2}$
(Brentjens \& de Bruyn 2005).
Thus, with the frequency range 50 -- 350~MHz
SKA-LOW will reach $\delta \phi \approx 0.10~{\rm rad~m^{-2}}$
and $\Delta \phi_{\rm max} \approx 4.3~{\rm rad~m^{-2}}$, 
and with the frequency range 0.35 -- 15~GHz,
SKA-MID will reach $\delta \phi \approx 4.7~{\rm rad~m^{-2}}$
and $\Delta \phi_{\rm max} \approx 7\,850~{\rm rad~m^{-2}}$.
It then appears that SKA-LOW will be able to probe the local ISM 
with a finer FD resolution than LOFAR currently does 
($\delta \phi \approx 0.80~{\rm rad~m^{-2}}$,
for the presently used frequency range 115 -- 189~MHz),
in addition to its finer angular resolution, 
while SKA-MID will be sensitive to Faraday-thick objects
and able to probe the internal structure of objects like supernova remnants
($\Delta \phi \sim 100~{\rm rad~m^{-2}}$) at sub-arcsec resolution.\\

\parbox{0.9\textwidth}{
\noi{References:}\\
\noi{\scriptsize 
Burn, B. J., 1966, MNRAS, 133, 67;
Brentjens, M. A. \& de Bruyn, A. G., 2005, A\&A, 441, 1217
}}\\

\subsubsection{Magnetic fields in star formation regions: Zeeman effect of RRLs}\label{sci:zeem}
\vspace{0cm}

\noi Magnetic fields play a crucial role in the formation and evolution of interstellar objects. With typical strengths of a few microGauss in the diffuse interstellar medium (ISM) to a few milliGauss in the denser star formation regions (see e.g. Crutcher et al. 2010), magnetic fields can be as important for the gas dynamics as turbulent motions and gravity forces. One way to assess how magnetic fields control the flow of matter in the ISM is to quantify the variation of the field strength with gas density: this relation is different if, for instance, gas compression happens parallel or perpendicular to the magnetic field lines. Such studies rely on the Zeeman effect, which is the only available technique for directly measuring magnetic field strengths in the ISM of the Milky Way and other galaxies (e.g. Crutcher et al. 1993; Robishaw et al. 2015). Despite the continuous work, observations are still sparse and so is our understanding of the impact of magnetic fields in the different interstellar environments. 

\begin{SCfigure}[][!ht]
  \centering
  \includegraphics[width=0.4\linewidth]{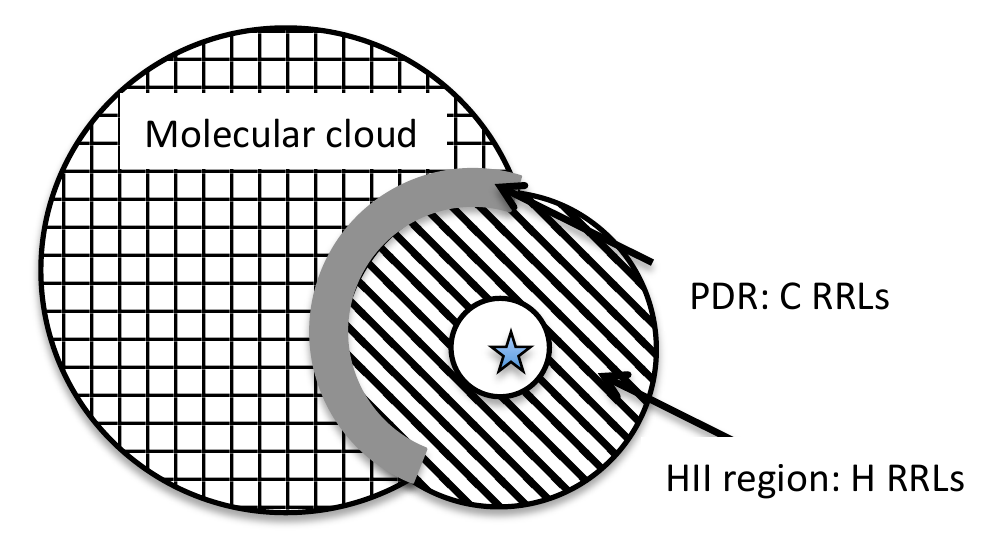}  \hspace*{1cm}
 \caption{\label{fig1} Cartoon showing the different media in a star formation region and where the H and C RRLs originate. Typical values of gas density and temperature in PDRs are $n_{\rm H} \approx 10^{4}$--$10^{6}$\,cm$^{-3}$ and $T\approx100$--300\,K; the gas in H{\sc ii} regions is much warmer, $T\approx10^4$\,K, and can be fairly dense, $n_{\rm H} \approx 10^{3}$--$10^{4}$\,cm$^{-3}$, if one considers compact ($\lsim1$\,pc) and ultra-compact ($\lsim 0.1$\,pc) H{\sc ii} regions (Tielens 2005).}
\end{SCfigure}

\smallskip
\noi 
Regions of massive star formation are the richest objects of the ISM for they comprise different environments. Notably, around the stars we find regions of ionised (H{\sc ii}) gas, which are separated from the parent molecular cloud by a thin layer of neutral gas, called the photodissociation region (PDR) (see Fig.\,\ref{fig1}). As a result of their different physical conditions, these media will emit different types of radiation. Of interest to this chapter are hydrogen and carbon radio recombination lines (RRLs): H RRLs sample H{\sc ii} regions, where the hydrogen is fully ionised, whereas C RRLs sample PDRs, where the hydrogen is atomic but the carbon is singly ionised. 
RRLs are powerful diagnostics of the physical conditions of the emitting gas (see e.g., Gordon \& Sorochenko 2009; Oonk et al. 2015) and potentially equally strong probes of the magnetic field through their Zeeman effect. 

\smallskip
\noi 
The Zeeman effect is the splitting of a spectral line into three or more components (linearly or circularly polarised).
The splitting between two consecutive sub-lines is $\Delta\nu_{\rm Z}=(b/2)|B|$, where $B$ is the magnetic field strength and $b$ is the splitting coefficient that depends on the spectral transition. When the frequency shift is much lower than the spectral line width, $\Delta\nu_{\rm Z}\ll \Delta\nu$ (which is the case for all Zeeman detections except in OH masers), the splitting is only detectable in the Stokes $V$ spectrum. The $V$ spectrum is the difference between the two circular polarisations. It has the (horizontal $S$-) shape of the derivative of the Stokes $I$ spectrum and is proportional to $B_{||}$, the line-of-sight component of the magnetic field. For a Gaussian line profile having central intensity $I_{\rm max}$, $|V_{\rm max}| = 1.43 I_{\rm max} b |B_{||}|/\Delta\nu$. The splitting coefficient for H and C RRLs is the same as for the H{\sc i} line, $b=2.8$\,Hz/$\mu$G (Troland \& Heiles 1977, Greve \& Pauls 1980).

\smallskip
\noi
Despite several attempts, no Zeeman detection of non-masing H RRLs has been claimed to date. Thum \& Morris (1999) reported the only Zeeman measurement of a H RRL maser transition at 1.3\,mm (H30$\alpha$). Measuring the Zeeman splitting of H RRLs is therefore exploratory work, which will be enabled with SKA1. As for C RRLs, Heiles \& Robishaw (2008) used the Green Bank Telescope to measure magnetic fields of $\sim 1000\,\mu$G (at $3\sigma$) towards a couple of PDRs. 
These results are suggestive of high magnetic field strengths in PDRs. They also point out the need to observe such small sources with an interferometer, to minimise the effects of beam averaging and depolarisation. Zeeman observations of C RRLs are ongoing with LOFAR (R. Oonk private communication). They will be pursued first with SKA1 at the same low frequencies; later with SKA2 we will be able to explore a higher frequency range at improved angular resolutions (see below).

\smallskip
\noi 
Fig.\,\ref{fig2} shows our estimates of the maximum intensity of the $V$ spectrum, $V_{\rm max}$, for H and C$n\alpha$ transitions ($n+1\rightarrow n$) as a function of quantum number $n$ and frequency. Note that there are more than 400 $\alpha$ transitions within the SKA1 frequency range, which can be stacked to increase the sensitivity to the Zeeman effect. We use gas densities and temperatures of $5\times10^{3}$\,cm$^{-3}$ and $10^{4}$\,K for the H{\sc ii} region and $6.7\times10^{5}$\,cm$^{-3}$ and $300$\,K for the PDR. In the H{\sc ii} region, 
 $n_{\rm H} = n_{\rm H^{+}} = n_{\rm e}$, where $n_{\rm e}$ is the electron density; in the PDR we assume $n_{\rm C^{+}}=n_{\rm e}=100$\,cm$^{-3}$, for a C abundance of about $10^{-4}$ (Tielens 2005). Moreover, we assume path lengths of 0.5\,pc and 0.05\,pc through the H{\sc ii} region and PDR, respectively. 
We take into account thermal, radiation, and collisional broadening of the lines (Gordon \& Sorochenko 2009; Salgado et al. 2017b). 
We assume that the H RRLs are emitted in local thermal equilibrium (LTE); detailed modelling of non-LTE effects shows that H line intensities can actually be enhanced at the highest frequencies ($\gsim 1$\,GHz). For C RRLs we include the departure coefficients from LTE, from the recent work of Salgado et al. (2017a, 2017b).

\begin{figure}[!h]
  \includegraphics[width=0.5\linewidth]{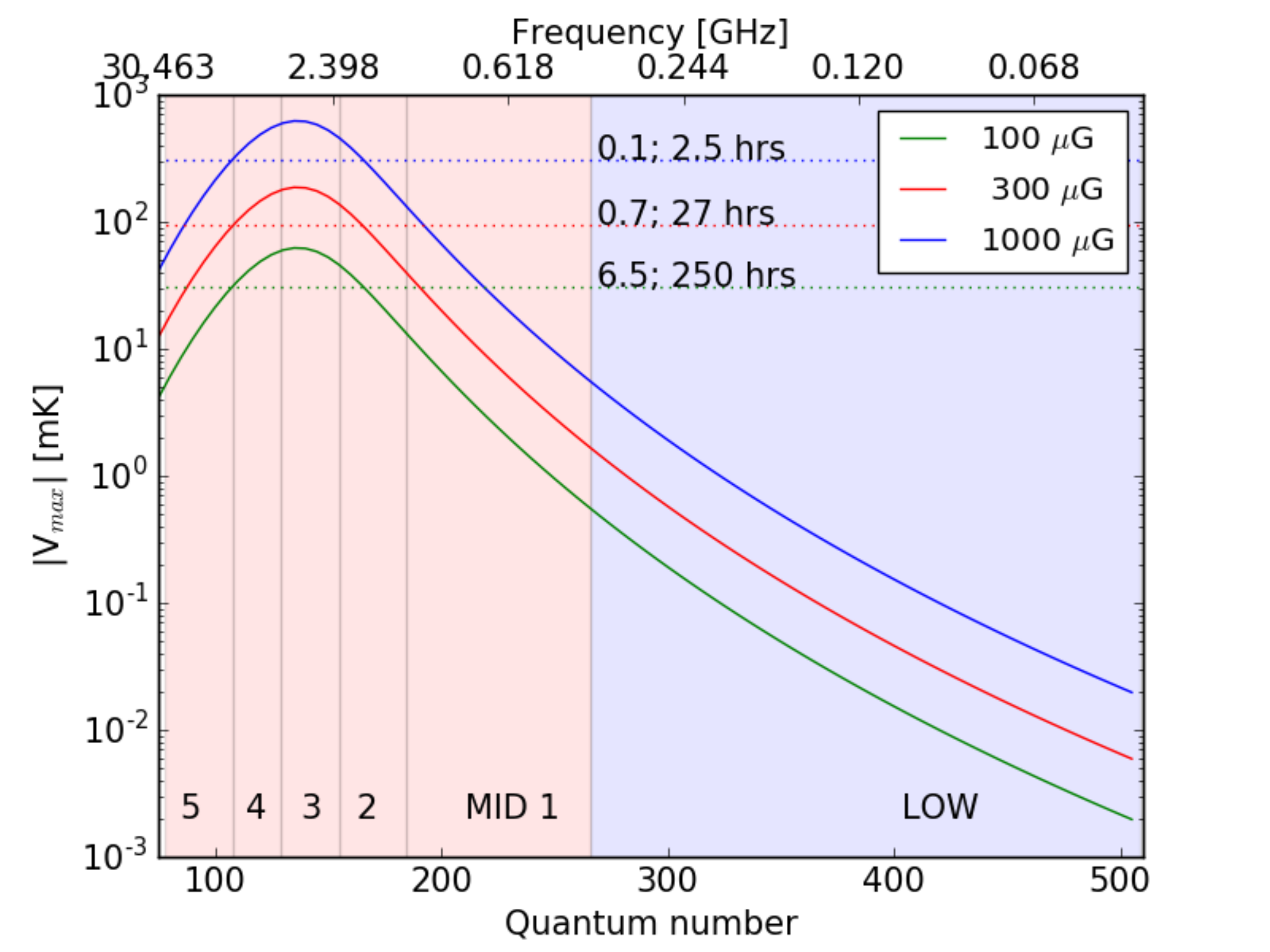}
  \includegraphics[width=0.5\linewidth]{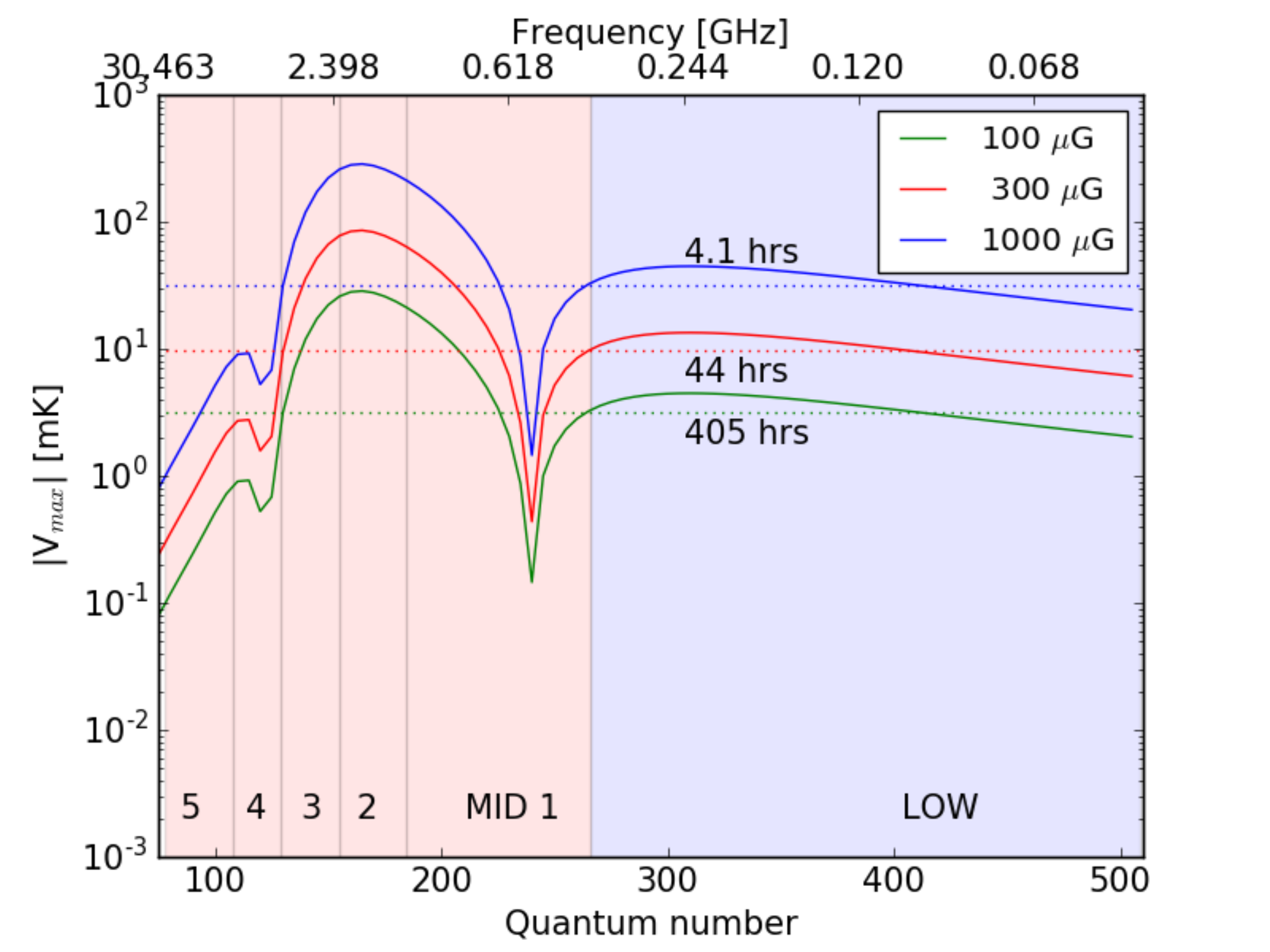}
  \vspace{-0.8cm}
 \caption{\label{fig2} Maximum (absolute) Stokes $V$ as a function of quantum number for H$n\alpha$ (\emph{left}) and C$n\alpha$ (\emph{right}) RRLs in the SKA1 frequency range (50\,MHz to 14\,GHz). The LOW and MID bands are identified by the blue and pink areas, respectively. The horizontal dotted lines show the integration time needed to detect the corresponding $V_{\rm max}$ at $5\sigma$, depending on angular resolution (see Table \ref{tab1}). The colours correspond to different values of $B_{||}$.
 }
\end{figure}

\begin{table}[!h]
\centering
\caption{Integration time needed to achieve a $5\sigma$ detection of the $V$ signal of H and C RRLs (calculated using equation 9.38 of Wilson, Rohlfs \& H\"{u}ttemeister 2009), for different values of $B_{||}$ and angular resolution, with SKA1. We take the mean $V_{\rm max}$ from Fig.\,\ref{fig2} within bands MID 2--4 for H RRLs and in the LOW band for C RRLs, and we consider (stack) the tens/hundreds of lines within those frequencies. We assume a spectral resolution of 2\,km/s to fully resolve the $V$ spectra of the narrowest RRLs ($\Delta V\gsim20$\,km/s). We use a mean sensitivity of 4 and 200\,m$^{2}$/K for a single SKA1-MID antenna and LOW station, respectively. We also take into account the number of antennas and stations (baseline-dependent) needed to reach a given angular resolution.}
\label{tab1}
\begin{tabular}{cccc}
 & \multicolumn{3}{c}{Time [hrs]}  \\ 
 \hline
$B_{||}$ [$\mu$G]  & H RRLs -- $0.\!^{\prime}5$ & HRRLs -- $0.\!^{\prime}2$ &  C RRLs -- $1.\!^{\prime}$5  \\
\hline 
10000  &  0.1 &  2.5  & 4.1    \\
300      &  0.7 &  27    & 44      \\
100      &  6.5 & 250  & 405      \\
\hline
\end{tabular}
\end{table}

\smallskip
\noi 
Table \ref{tab1} shows that the Zeeman effect of H and C RRLs is readily detectable with SKA1, for typical compact/ultra-compact H{\sc ii} regions and PDRs, and for $B_{||}\gsim100\,\mu$G. Such $B_{||}$ values have been measured in PDRs (e.g. Troland et al. 2016) but are higher than what is typically observed in large, low-density ($n_{\rm H}\sim10$\,cm$^{-3}$) H{\sc ii} regions (e.g. Harvery-Smith et al. 2011). However, the magnetic field could indeed be larger in the much denser compact/ultra-compact H{\sc ii} regions considered here. Angular resolutions of $0.\!^{\prime}2-0.\!^{\prime}5$ with MID are sufficient to observe such H{\sc ii} regions up to distances of a few kpc. As for PDRs with SKA1-LOW, angular resolutions better than $1.\!^{\prime}5$, which at the distance of Orion is equivalent to 0.2\,pc, imply too long integration times. More detailed studies of PDRs will be possible with SKA2, when the sensitivity will increase tenfold. Moreover, SKA2 will also allow us to explore C RRLs in the MID bands; even though they are as bright as the lines at lower frequencies, they are a factor of 10 narrower.
\\

\parbox{1.0\textwidth}{
\noi {\scriptsize Acknowledgements: We thank R. Oonk for kindly providing the results from non-LTE models and T. Robishaw for the discussions.}}
\\

\parbox{0.9\textwidth}{
\noi{References:}\\
\noi{\scriptsize 
	Crutcher R.M., et al., 1993, ApJ, 407, 175; 
	Crutcher R.M., 2010, ApJ, 725, 466;
	Gordon M.A. \& Sorochenko R.L., 2009, Radio recombination lines, ASSL, vol. 282; 
	Greve A. \& Pauls T., 1980, A\&A, 82, 388;
	Harvey-Smith L., et al., 2011, ApJ, 736, 83;
	Heiles, C. \& Robishaw T., 2008, IAUS, 259, 579;
	Oonk J.B.R, et al., 2015, AASKA14, 139;
	Robishaw T., et al., 2015, AASKA14, 110;
	Salgado F., et al., 2017a, ApJ, 837, 141;
	Salgado F., et al., 2017b, ApJ, 837, 142;
	Thum C. \& Morris D., 1999, A\&A, 344, 923; 
	Tielens A.G.G.M., 2005, The physics \& chemistry of the interstellar medium, Cambridge Uni. Press; 
	Troland T.H. \& Heiles C., 1977, ApJ, 214, 703;
	Troland T.H., et al., 2016, ApJ, 825, 2;
	Wilson T.L., Rohlfs K. \& H\"uttemeister S., 2009, Tools of radio astronomy, Springer-Verlag
}}\\

\subsubsection{Jets, outflows and young stellar objects}\label{sci:YSO}
\vspace{0cm}

\noi Young Stellar Objects' (YSOs) are stars in the early stages of their evolution, before they arrive on the main sequence. Most of them drive jets and bipolar outflows, that can harbour masers or Herbig-Haro objects. Their intense far UV field can also be at the origin of {\hii} regions. Finally, they are sometimes associated with protoplanetary disks. YSOs hence represent an important feedback form of  stars on galaxies, as supernova remnants (SNRs) do. YSOs are less spatially extended but more numerous than SNRs. Their injection of energy to the interstellar medium (ISM) is mostly in the form of shocks and energetic radiation, and probably cosmic rays (CRs) in lower proportions. They modify the dust and gas properties and composition. As such they play an important role in the chemistry hence in the cycle of matter in galaxies. They also play a role in the energetic balance of galaxies. Their study can shed light on star formation scenarios, specially on the evacuation of momentum from forming protostars. Finally, they could be acceleration sites for CRs.

\smallskip
\noi {\bf Young stellar objects}

\smallskip
\noi YSOs are often classified according to the mass of the forming object: from low-mass ($M <  3 M_{\rm sun}$) to high-mass ($M >  9 M_{\rm sun}$), through intermediate-mass (in between), due to the different formation scenarios.  Low-mass protostars are often isolated, sometimes in binary systems (e.g., BHR71, see Bourke et al. 1997), whereas their massive counterparts occur in clusters (e.g., G5.89--0.39, e.g., Hunter et al. 2008). The SKA contribution to the study of YSOs will be related to two questions. First, the high angular resolution and high sensitivity of the instrument will make the detection of multiple sources possible, as their separation within a binary system or a cluster. This is an important first step in the identification of sources driving outflows in complex systems such as G5.89--0.39. Additionally, upcoming SKA observations will contribute to the age determination of the sources (see G\"udel 2002 for a review on radio emission from stellar sources), a key parameter in understanding the evolution of clusters or binary systems.

\smallskip
\noi {\bf Jets and outflows}

\smallskip
\noi Following source identification, the separation of multiple flows and shock structures is also a topic where the high angular resolution of SKA will produce an impact. Indeed, it will contribute to disentangle the various outflow structures in a star-forming region (massive, like G5.89--0.39, or not, like the NGC1333 cloud, e.g., Dionatos \& G\"udel 2017). After this first separation step, isolating shock structures in outflows is also necessary before their characterisation. Indeed, the propagation of a shock in the ISM is naturally associated with several shocked layers (see e.g., Hollenbach 1997), which are generally mixed in a single-dish beam. Isolating at least those emitting in the radio range will enable a better characterisation. 

\smallskip
\noi
The shock structures thus isolated can then be characterised. This is done by comparing observations with models of shock propagation. 1D models such as the Paris-Durham model are used to interpret single-dish observations (e.g., Gusdorf et al. 2015). These observations usually trace the molecular and atomic shocked gas in the sub-mm/IR ranges of the spectrum. Few shock tracers fall within the SKA frequency ranges, with the notable exception of masers. OH, H$_2$O, CH$_3$OH and NH$_3$ masers all place constraints on dense gas properties (Frail 2011), and probe the local kinematics and magnetic field strength. All these masers will be observed with SKA with great sensitivity and will allow us to place constraints on shock models (see Gusdorf et al. 2012).

\smallskip
\noi
Based on these additional constraints and on complementary observations of molecular and atomic species (e.g., Bachiller et al. 2001), progress can be made in our understanding of shocks. This question encompasses the gas-phase and dust processes. In the gas phase, the observation of OH and H$_2$O masers will allow us to a better understanding of the chemistry of oxygen for instance, and the observation of complex organic molecules will also be a step forward in our understanding of astrochemical processes (e.g., Codella et al. 2015). Regarding dust, SKA observations could shed light on the anomalous microwave emission problem (e.g., Dickinson et al. 2015). This could open a new spectral window, whose combination with data at other wavelengths could make a better characterisation of dust properties and abundance possible. 

\smallskip
\noi
On the question of energetics, SKA observations will directly enable us to measure the energy and momentum injected into the ISM at low frequencies, for instance through the observation of recombination lines with spectral resolution (which are probes of abundance and velocity of certain gas components). Such observations could be used to benchmark simulations of star formation (e.g., Matsumoto et al. 2017 for a recent example). Additionally, the constraints placed on shock models will allow us to indirectly measure the momentum and energy injection at sub-mm to IR frequencies. 

\smallskip
\noi
More generally, SKA observations will allow us to better understand the process of star formation in various ways. First, imaging observations of the outflow systems at very high resolution and sensitivity will be compared to more detailed shock models than 1D ones in order to probe the role of outflows in extracting momentum from accreting systems. To this aim, pseudo-3D models such as Tram et al., submitted to MNRAS, or even more sophisticated ones, e.g., the magnetised disk wind models (e.g., Panoglou et al. 2012; Yvart et al. 2016) will be constrained by observations that probe the configuration of the jet/outflow systems. Moreover, the magnetic field measurements enabled by the SKA will shed light on the role of the magnetic field in the formation of stars and their associated outflows (see Li et al. 2014 for a recent review). Finally, SKA will enable the monitoring and characterisation of episodic phenomena associated with YSOs, in particular those associated with ejection of matter from the protostar (see Anglada et al. 1998). 

\smallskip
\noi {\bf {\hii} regions}

\smallskip
\noi This paragraph mostly concerns the late stages of massive YSOs, which have carved {\hii} regions in the ISM by their far UV radiation. The size of the {\hii} region depends on its origin: hypercompact when it is due to a photoevaporating disk, ultracompact when diskless protostars ionise their surroundings, and compact and classical when it is caused by several sources (e.g., Zinnecker \& Yorke 2007). Beyond the {\hii} region shell, one can often find a layer of dense gas where the physics and chemistry are driven by the far UV photons emanating from the {\hii} region, constituting a photon dominated or photodissociation region (PDR). This is the case in G5.89--0.39, a high-mass star forming region where a central O star has carved an ultra compact {\hii} region surrounded by a cocoon of dense gas (Hunter et al. 2008; Su et al. 2009). SKA is the right instrument to characterise both the {\hii} region and its PDRs, the signposts of stellar feedback. Indeed it will make the observation of typical tracers of {\hii} regions possible, as well as the study of their influence on the environment: hydrogen, helium, and carbon recombination lines (e.g., Oonk et al. 2015; Thompson et al. 2015; Beswick et al. 2015).

\smallskip
\noi {\bf Cosmic Ray acceleration}

\smallskip
\noi Synchrotron radio emission has been detected in several YSOs, such as IRAS~16547 (Garay et al. 2003), HH80/81 (Marti et al. 1993; Carrasco-Gonzalez et al. 2010), the triple radio source in Serpens (Rodriguez-Kamenetzky et al. 2016), DG~Tau (Ainsworth et al. 2014), and W3(0$_2$H) (Wilner et al. 1999) among others. The detection of synchrotron photons indicates the presence of relativistic electrons in the emitter. In all cases, the number density of non-thermal electrons required to explain the measured synchrotron fluxes ($\sim 1$~mJy) in the GHz domain is larger than the average CR electron density which would result from the propagation of background CR electrons in molecular clouds, and therefore local acceleration of electrons is needed. SKA with its improved sensitivity should increase substantially the number of detected YSOs emitting non-thermal radio radiation. In particular, SKA-LOW detection of several sources will complement first YSOs detection by the GMRT (Ainsworth et al. 2014) and will help in constraining the modeling of the relativistic electron distribution (Padovani et al 2016).

\smallskip
\noi
Energetic particles can be accelerated up to relativistic energies from the thermal pool by diffusing back and forth internal and termination shocks in protostellar jets (e.g., Araudo et al. 2007; Bosch-Ramon et al. 2010; Padovani et al. 2016). The magnetic field and its perturbations play an important role in the Fermi acceleration process. The value of the magnetic field in equipartition with non-thermal particles and in the one-zone approximation is between 0.1 and 100~mG, depending on the distance of the synchrotron emitter to the central protostar. However, the spatial resolution of the current facilities in the GHz domain is not enough to develop an accurate model of the magnetic field distribution in the shock downstream region. In this concern, future observations with SKA-MID at sub-arcsecond resolution will allow us to identify the acceleration and cooling regions, and to go in depth on the microphysics of the jet plasma by studying the magnetic field amplification and damping downstream of the shock. SKA high resolution observations of non-thermal YSO jets will open a new window to study CR acceleration in low velocity shocks and high density plasmas. \\

\parbox{0.9\textwidth}{
\noi{References:}\\
\noi{\scriptsize Ainsworth, R. E., et al., 2014, ApJ, 792, 18;}
\noi{\scriptsize Anglada, G., et al., 1998, AJ, 116, 2953;}
\noi{\scriptsize Araudo, A. T., et al., 2007, A\&A, 476, 1289;}
\noi{\scriptsize Bachiller, R., et al., 2001, A\&A, 372, 899;}
\noi{\scriptsize Beswick, R. J., et al., 2015, AASKA14, 70;}
\noi{\scriptsize Bosch-Ramon, V., et al., 2010, A\&A, 511, A8;}
\noi{\scriptsize Bourke, T. L., et al., 1997, ApJ, 476, 7812;}
\noi{\scriptsize Carrasco-Gonzalez, C., et al., 2010, AJ, 139, 2433;}
\noi{\scriptsize Codella, C., et al., 2014, AASKA14, 123;}
\noi{\scriptsize Dickinson, C., et al., 2014, AASKA14, 124;}
\noi{\scriptsize Dionatos, O. \& G\"udel, M., 2017, A\&A, 597, 64;}
\noi{\scriptsize Frail, D. A., 2011, MmSAI, 82, 703;}
\noi{\scriptsize Garay, G., et al., 2003, ApJ, 587, 739;}
\noi{\scriptsize G\"udel, M., 2002, ARA\&A, 40, 217;}
\noi{\scriptsize Gusdorf, A., et al., 2012, A\&A, 542, L19;}
\noi{\scriptsize Gusdorf, A., et al., 2015, A\&A, 575, 98;}
\noi{\scriptsize Hollenbach, D. J., 1997, IAU, 182, 181;}
\noi{\scriptsize Hunter, T. R., et al., 2008, ApJ, 680, 1271;}
\noi{\scriptsize Li, H. B., et al., 2014, PPV, 101;}
\noi{\scriptsize Marti, J., et al., 1993, ApJ, 416, 208;}
\noi{\scriptsize Matsumoto, T., et al., 2017, ApJ, 839, 69;}
\noi{\scriptsize Oonk, R., et al., 2015, AASKA14, 139;}
\noi{\scriptsize Padovani, P., et al., 2016, A\&A, 590, 8;}
\noi{\scriptsize Panoglou, D., et al., 2012, A\&A, 538, 2;}
\noi{\scriptsize Rodriguez-Kamenetzky, A., et al., 2016, ApJ, 818, 27;}
\noi{\scriptsize Su, Y.-N., et al., 2009, ApJ, 704, L5;}
\noi{\scriptsize Thompson, M., et al., 2015, AASKA14, 126;}
\noi{\scriptsize Wilner, D. J., et al., 1999, ApJ, 513, 775;}
\noi{\scriptsize Yvart, W., et al., 2016, A\&A, 585, 74;}
\noi{\scriptsize Zinnecker, H. \& Yorke, H. W., 2007, ARAA, 45, 281}
}

\subsubsection{Supernova remnants}\label{sci:SNR}
\vspace{0cm}

\noi At the latest stages of supernova (SN) explosions, supernova remnants (SNRs) represent an important feedback form of stars on galaxies. Indeed, SNRs re-distribute large amounts of energy to the interstellar medium (ISM) in the form of shocks, energetic radiation and cosmic rays. They modify both the dust and gas properties and compositions and as such play an important role in the chemistry, and hence in the cycle of matter in galaxies. They also play a vital role in the energetic balance of galaxies. Additionally, the shocks they generate have the potential to trigger a second generation of star formation. Finally, SNRs trap cosmic rays (CRs) that have been accelerated at the earlier stages of the SN explosion. SNRs are hence ideal laboratories to address numerous questions at the crossroads between ISM astrophysics and high-energy astrophysics. 

\smallskip
\noi
The first step towards a global understanding of SNRs consists of the identification and separation of the various shock components that propagate in relation with the SNR. Indeed, shocks in SNRs can be driven by the primary SN shock after its encounter with a molecular cloud, and by young stars (in the form of protostellar outflows) whose recent formation was triggered in the molecular cloud by either the SN shock wave or the stellar wind of the SNR's progenitor. This is the case for instance in the G clump of the IC443 SNR (Xu et al. 2011). Separating these various shock structures is a first step preliminary to all other scientific objectives, and it can only be achieved through high-angular and high-spectral resolution observations of the region (see Louvet et al. 2016 for an example of such a separation in a dense filament). Thanks to its high angular resolution, SKA will play a role in this crucial first step.

\smallskip
\noi
In a second step, the local ISM must be characterised, especially its shocked components. The underlying questions and methodology are exactly similar to what is done in the context of shocks associated to YSOs jets and outflows (see Sect.\,\ref{sci:YSO}): a shock model is first used to interpret sub-mm/IR observations (e.g., Anderl et al. 2014). Few shock tracers can be observed by SKA, contrary to maser emission that can be used to a posteriori validate this modelling (see Frail et al. 2011 and Gusdorf et al. 2012). Based on these additional constraints and on complementary observations of molecular and atomic species (e.g., van Dishoeck et al. 1993), further progress can be made in our understanding of SNR shocks. This question encompasses the gas-phase and dust processes (see Sect.\,\ref{sci:YSO} for details). Finally, SKA observations will directly enable us to measure the energy and momentum injected into the ISM at low frequencies through the observation of recombination lines with spectral resolution. Such observations could be used on the longer term to benchmark simulations of ISM evolution regulated by SN explosions (Hennebelle \& Iffrig 2014). Additionally, the constraints placed on shock models will allow us to indirectly measure the momentum and energy injection at sub-mm to IR frequencies. 

\smallskip
\noi
In relation with the topic of studying star formation potentially triggered by the passage of a SN shock, and beyond the separation between shock structures made possible by SKA, radio observations will make it possible to characterise ejection flares from protostars (Anglada et al. 1998) and their winds or even chromospheres for the more evolved ones (G\"udel 2002). The subsequent evolutionary stage determination will help to understand the causality between the SN shock and the local star formation. 
Finally, on the subject of CR-related questions, discussed in other sections (below and in Sect.\,\ref{sci:YSO}), it could be noted that the thorough assessment of the content of the ISM described here (magnetic field measurements included) will support the development and application of models of gamma-ray emission resulting from the interaction of CRs accelerated in the past stages of the SN explosion and trapped in its shock fronts, with the dense (shocked or not) medium (Gabici et al. 2009). SKA will thus also serve to study CR composition, acceleration, and propagation mechanisms. 

\smallskip
\noi {\bf Source identification}

\smallskip
\noi During the last two decades, the imaging atmospheric Cherenkov telescopes like H.E.S.S., MAGIC or VERITAS have discovered more than one hundred new gamma-ray sources of different types. However, in the case of H.E.S.S. a third of these new sources are still unidentified. Similarly, among the thousands of new sources detected by the gamma-ray LAT detector aboard Fermi space telescope, a third do not have a clear counterpart at any other wavelengths. The future at gamma-ray energies will be provided by the Cherenkov Telescope Array (CTA) which will start its Galactic plane survey in the early 2020's. CTA will consist of a northern and a southern site providing an order of magnitude gain in sensitivity and allowing the detection of 20-70 SNRs among which 7-15 would be resolved with CTA (Renaud 2011). This means that a large fraction of the detected shell-type SNRs will not be identified as such based solely on their TeV morphology. In addition, although CTA will bring an order of magnitude in sensitivity, the spatial resolution will be of the order of 3 arcmin (r$_{\rm 68 \%}$) creating source confusion in the plane. Surveys at other wavelengths will therefore be key to identify new SNRs. Historically most SNRs known in our Galaxy were discovered by radio surveys (Green 2009). The SKA1-LOW, with frequency range between 50 MHz and 350 MHz and a field of view of 30 deg$^2$, hosts great potential to discover a large number of SNRs and will be of crucial importance to identify gamma-ray sources or to constrain their non-thermal morphology. With a field of view of only 0.5 deg$^2$ but an excellent angular resolution of $0.22^{\prime\prime}$ (with respect to $11^{\prime\prime}$ for SKA1-LOW), the SKA1-MID will be ideal to detect and resolve young SNRs ($<$ 1000 years) which are the best candidates for the acceleration of PeV protons.

\begin{figure}[!ht]
  \centering
  \includegraphics{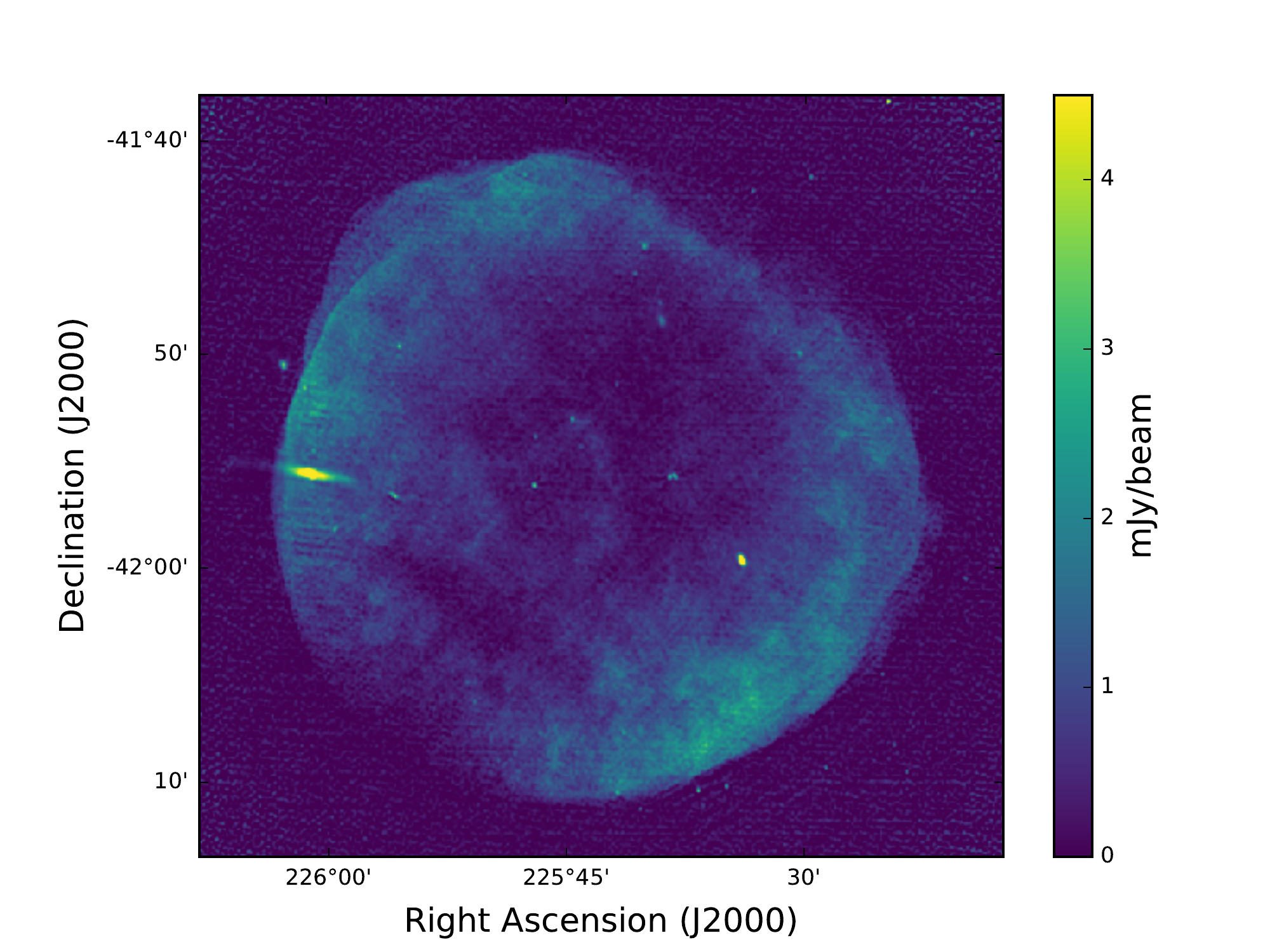}
  \caption{\label{fig:SNR} Total intensity image of the remnant of the historical event SN 1006 observed at 1.4 GHz. The beam size is 10 arcsec. The bright source on the left is an unrelated extra-galactic source. Adapted from Reynoso et al. 2013.}
\end{figure}

\smallskip
\noi {\bf Cosmic ray acceleration physics}

\smallskip
\noi SNRs have been known to be sources of relativistic particles and non-thermal radiation since the beginning of the 50's (Shklovskii 1953). Non-thermal radio emission results from synchrotron radiation of GeV electrons. Radio observations provide information about the remnant morphology, its magnetic field strength and orientation and particle acceleration in strong shock waves that pervade the ISM (Reynolds 2008). Recent X-ray arc-second angular resolution images of several historical SNRs by the American satellite Chandra show the presence of thin filaments with a typical thickness of a few percent of the SNR radius (Hwang et al. 2002). This, combined with the detection of SNRs at gamma rays, confirms that SNRs can accelerate energetic particles at least up to TeV energies (H.E.S.S. Collaboration 2011). The filament thickness implies that the magnetic field at the forward shock can reach strengths up to two orders of magnitude above mean interstellar values (Parizot et al. 2006). Magnetic field amplification seems to be a key process to explain the acceleration of hadrons up to PeV energies and the origin of cosmic rays (Bell 2004). Yet, the process of particle acceleration in strong SNR shock waves remains largely elusive. 

\smallskip
\noi SKA with its waveband coverage, its improved sensitivity and its angular resolution down to sub arc-second scales can bring important constraints on the process of particle acceleration in SNRs. In parallel to CTA, several X-ray facilities are expected to overlap with the science phase of SKA. This is likely for XMM-Newton and Chandra covering the 0.2 - 12 keV band as well as NuSTAR which operates in the 10 -- 80 keV energy range, and finally Athena+, due for launch in 2028, with a superb sensitivity and spectral resolution in the 0.1 -- 12 keV energy range. These missions will help to study the high energy TeV particles through their synchrotron emission while low frequency radio emission is linked to the GeV energy particles. High sensitivity observations at different wave bands combined with X-ray observations can be conducted to produce radio spectral maps of historical SNRs visible in the southern hemisphere like SN1006, RXJ 1713-3945 or Vela Jr. The full wavelength high sensitivity coverage from a few hundred MHz to X-rays can be used to test any particular spectral curvature related to the particle acceleration process. In parallel, it should be noted that the energy spectrum of secondary electrons produced via pp collisions starts to deviate from a pure power law below 1 GeV. This implies that the radio synchrotron emission from secondary electrons will start to diverge from a pure power-law at frequencies below 100 MHz for a magnetic field of 10~$\mu$G (Aharonian et al. 2013). This should be detectable by SKA1-LOW thus allowing us to trace high energy protons in SNRs.

\smallskip
\noi Finally, high-angular resolution observations can test the existence of radio filaments, the profile of radio synchrotron radiation behind the forward shock, and possibly the existence of a radio precursor associated with the streaming of low energy cosmic rays ahead the shock front. 
Covering SNR spectrum over a wide wavelength range will provide a complete view of particles accelerated in SNR shocks allowing us to better understand this long standing CR origin issue.\\

\parbox{0.9\textwidth}{
\noi{References:}\\
\noi{\scriptsize 
Aharonian, F. A., et al., 2013, arXiv:1301.4124;
H.E.S.S. Collaboration, 2011, A\&A, 531, A81;
Anderl, S., et al., 2014, A\&A, 569, 81;
Anglada, G., et al., 1998, AJ, 116, 2953;
Bell, A. R., 2004, MNRAS, 353, 550;
Frail, D. A., 2011, MmSAI, 82, 703;
Gabici, S., et al., 2009, MNRAS, 396, 1629;
Green, D. A., 2009, Bull. Astr. Soc. India, 37, 45; 
G\"udel, M., 2002, ARA\&A, 40, 217;
Gusdorf, A., et al., 2012, A\&A, 542, L19;
Hennebelle, P., \& Iffrig, O., 2014, A\&A, 570, 81;
Hwang, U., et al., 2002, ApJ, 581, 1101;
Louvet, F., et al., 2016, A\&A, 595, 122;
Parizot, E., et al., 2006, A\&A, 453, 387;
Renaud, M., 2011, MmSAI, 82, 726;
Reynolds, S. P., 2008, ARAA, 46, 89;
Reynoso, E. M., et al., 2013, AJ, 145, 104;
Shklovskii, I. S., 1953, Dokl.Akad.Naut. SSSR, 91, 475;
van Dishoeck, E. F., et al., 1993, A\&A, 279, 541;
Xu, J.-L., et al., 2011, ApJ, 727, 81
}}\\

\subsubsection{Pulsar census and probe of the interstellar medium} \label{sci:pulG}
\vspace{0cm}

\noi The SKA will have a wide field-of-view, high sensitivity, multi-beaming and sub-arraying capabilities, and it will be equipped with advanced pulsar search
backends. These design characteristics make this new radio telescope an ideal instrument to survey efficiently the whole sky accessible 
from the South African and Australian sites. In particular, SKA-MID will be the main tool for discovering distant pulsars in the Galactic plane, 
whereas SKA-LOW can both go deeper and survey faster at higher Galactic latitudes. 
\smallskip

\noi \textbf{Future populations studies} 
 
\smallskip

\noi The combined Milky Way surveys at multiple frequencies will increase the number of known pulsars by more than an order of magnitude, revealing both the young population (ages 10$^3$-10$^7$ years) strongly confined to the Galactic plane and the much older millisecond pulsars (ages of $\geq$ 10$^8$ yr) much more spread across the sky. Simulations show that one expects up to 9000 normal pulsars and 1500 millisecond recycled ones with the first step SKA1, while SKA2 will potentially find all of the Galactic radio-emitting pulsars in the SKA sky which are beaming in our direction (Bates et al. 2014; Keane et al. 2015). In terms of Galactic studies and stellar populations, SKA will offer a complete picture of the neutron star birth properties, of the interstellar matter (ionised gas) distribution and of the magnetic field structure, thanks to dispersion measures and Faraday rotation measures. Searches in the vicinity of the Galactic Centre should reveal the 1000 pulsars we infer to be closely orbiting Sgr A*, help constrain the mass and spin of the central object, and characterise its nature (massive Kerr black hole, boson star or black hole coupled with a scalar field, e.g. Vincent et al. 2016). Among the most exciting systems to be discovered, the SKA's instantaneous high sensitivity will allow efficient search for highly relativistic binaries, i.e. pairs of neutron stars or neutron star - white dwarf systems orbiting each other in less than a couple of hours. If not the dreamed neutron star - black hole association, or a few other triple systems similar to PSRJ0337+1715 to test directly the strong equivalence principle, this population of close compact binaries will provide us with a bunch of unprecedented tests for gravity theories (see Sect.\,\ref{sci:GT}). 

\smallskip

\noi Targeted searches towards globular clusters will help fully characterise the old pulsar population and provide unique tools to probe the structure, gas content, magnetic field, and formation history of the clusters (Hessels et al. 2015). With their high stellar densities, globular clusters are also the place of exotic gravitational associations, thanks to the unique environment for stellar collisions and interactions. The variety of individual systems, in terms of masses and nature of the components, will help constrain the equation-of-state of neutron-rich matter at super-nuclear densities and to test gravitational theories, and it will also provide novel insights for understanding stellar evolution and accretion physics. The timing of millisecond pulsars next to the cluster core can also reveal and constrain the presence of an intermediate mass black hole (Perera et al. 2017). Important results will also come from pulsar supernova remnant (SNR) and pulsar wind nebulae (PWN) associations, in particular thanks to the synergy with high energy astrophysics. SKA should measure the radio morphology, spectrum, and polarization properties of up to 100 PWNe (10 times more than today) and increase the number of eclipsing binary pulsars from 50 to 300 (Gelfand et al. 2015). Such samples will allow us in particular to determine how a neutron star's spin-down luminosity, age, and environment affect the evolution of its PWN. Finally, SKA will be the first telescope to probe efficiently the bright pulsar population in neighbouring galaxies and allow comparison in terms of average metallicity and environment. 

\smallskip

\noi  \textbf{Probing the interstellar medium using pulsar timing} 
\smallskip

\noi As the pulsar radio emission propagates through the interstellar medium (ISM), it is affected and modified by the ISM and its inhomogeneities. 
How severely the signal is affected mostly depends on the observing frequency and can usually be described by a power-law. The Dispersion Measure (DM) informs on the free electron content along each pulsar line of sight, its long term variation and the observed scintillation time and frequency scales reflect directly the characteristic scale of ISM turbulence and plasma cells.
At low frequencies, it is also possible to measure the small rotation measure (RM) values of the nearby population of pulsars. 
This provides an unprecedented tool to study the local magnetic field structure (Stappers et al. 2011). SKA, with its broad frequency coverage (up to 9 octaves!) 
and its high sensitivity at low frequencies where these effects are the most pronounced, will be a fantastic tool to study the ISM. The French pathfinder NenuFAR (10-80MHz), associated with LOFAR-HBA (110-250 MHz) and the Nan\c cay Radio Telescope (NRT;1.0-3.5GHz) will allow us to do the spadework for this program.

\smallskip

\noi \textbf{On going observational programs and collaborations in France} 

\smallskip

\noi The \href{https://mode.sciencesconf.org/}{\color{blue} \myul[blue] {MODE annual workshop and collaboration}} has been the place of fruitful discussions between French theoreticians and observers, about neutron stars and their environment. It regularly gathers scientists studying pulsar multi-wavelength emission, neutron star interiors, their magnetospheres and wind properties, and their associations with PWNe and SNRs. Those people (at e.g. CENBG, IRAP, LUPM) are all very interested in the SKA outcome in terms of population statistics.
The Nan\c cay-Orl\'eans group has a long term experience in pulsar timing and pulsar search with NRT and LOFAR. 
Two surveys of the Galactic plane have been made so far with the NRT, at the end of the 90's with the old receiver and currently with the powerful 512 MHz bandwidth NUPPI backend. Some targeted searches in unidentified Fermi sources were also made and led to the discovery of a few objects.
This group participates in the LOTAAS survey with LOFAR and leads the associated key science project with NenuFAR, to investigate the low frequency end of pulsar emission spectrum, as a pilot project for SKA. In particular, the aim here is to characterise the spectral turn-off observed below 100 MHz, understand the frequency evolution of the emission beam structure, and interpret these results in terms of radio emission processes. Finally, the same people are involved in the TRAPUM project, which is one of the key programs of MeerKAT, the South African precursor of SKA-MID.

\smallskip
\noi The French pulsar community has played a key role in the synergies between radio and gamma observations, with a large accompanying program of Fermi-LAT observations started already in mid-2006, before the satellite launch. This project has been extremely fruitful, with tens of common publications and $\sim$150 pulsars detected at high energies thanks to radio ephemerides. Equivalent programs at smaller scales were led in the same period with INTEGRAL, XMM-Newton, and HESS-I/II. The expected prolongation of Fermi's operation and the deployment of CTA will perpetuate this combined effort on the long term. Similarly, combined studies of timing radio observations with next-generation X-ray telescopes such as eROSITA  and ATHENA+ will help constrain the super-dense matter equation-of-state.\\

\begin{figure}[!ht]
  \centering
  \includegraphics[width=0.95\linewidth]{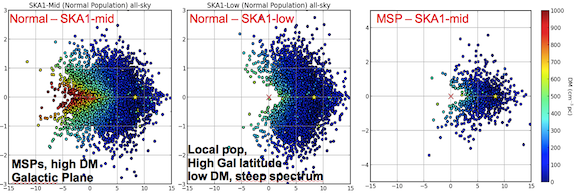}
  \vspace{-0.2cm}
  \caption{From Janssen et al. 2015: simulation of the ``normal'' and ``millisecond'' pulsar populations which will be detected by the SKA1-MID and SKA1-LOW. Projection onto the Galactic Plane in kpc scale.}
\end{figure}

\parbox{0.9\textwidth}{
\noi{References:} \\
\noi{\scriptsize
Bates, S. D., et al., 2014, MNRAS, 439, 2893;
Gelfand, J., et al., 2015, AASKA14, 46;
Hessels, J., et al., 2015, AASKA14, 47; 
Keane, E., et al., 2015, AASKA14, 40;
Perera, B. B. P., et al., 2017, MNRAS, 468, 2114;
Stappers, B. W., et al., 2011, A\&A, 53, A80;
Janssen, G., et al., 2015, SKA KSWG workshop, Stokholm;
Vincent, F. H., et al., 2016, PhysRevD, 94, 084045
}}\\

\subsubsection{Distance determination}
    \vspace{0cm}
    
\noi Despite the new era of high precision astrometry and  proper motions in the  Gaia era,  SKA will provide with its  high sensitivity the capability to obtain  microarcsec  astrometric precision for astronomical objects such as maser stars, pulsars in the Milky Way as well as in the local group universe. Combining the SKA with VLBI arrays, SKA-VLBI, will allow us to measure parallaxes as well as the trajectories of the
stars in the Galactic plane very precisely, specifically in molecular clouds where interstellar extinctions hampers Gaia to deliver precise parallaxes. An area of high interest will be clearly
the inner kpc of the Milky Way, the so-called ``Galactic Bulge'', a major component in our galaxy together with the Central Molecular Zone (CMZ) in the  innermost 200\,pc region  of the Milky Way, which
is a giant molecular cloud complex and contains about 10\% of the molecular gas of our Galaxy. The CMZ shows signs of active star formation and can provide us with a template to compare with the  quiescent galactic nuclei in the present-day Universe. Precise distances and proper motions would be essential to understand the origin of this high-complex zone as well as the interaction with the supermassive black hole and constraining the galactic
potential.

\begin{figure}[!ht]
  \centering
  \includegraphics[width=0.77\linewidth]{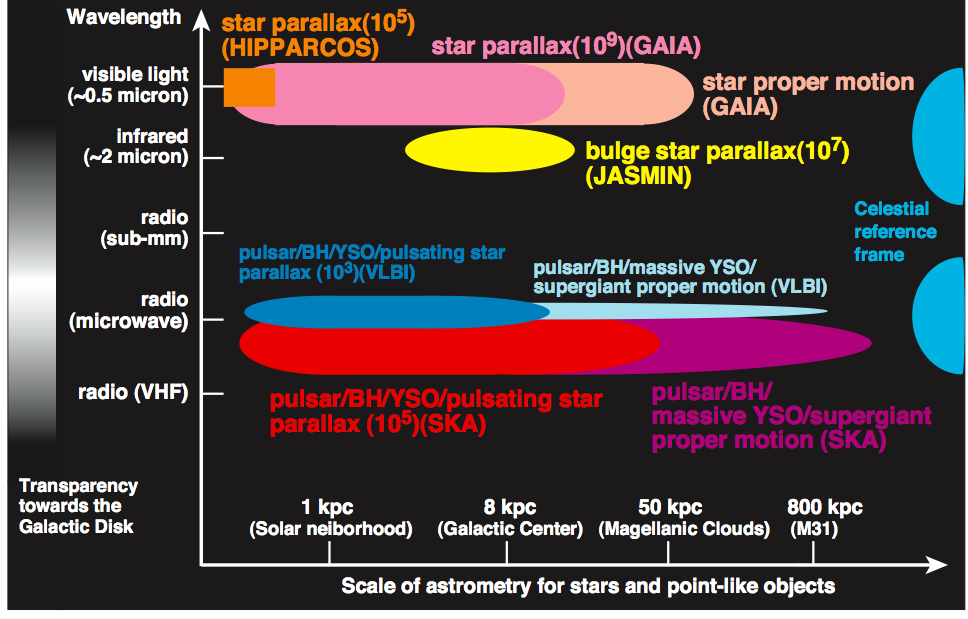}
  \caption{Present to future view of astrometric missions from optical to radio which target different sorts of celestial objects (from Imai et al. 2017).}
\end{figure}
 
\smallskip
\noi{\bf SKA and VLBI synergy}

\smallskip
\noi Due to its extremely high angular  resolution and sensitivity SKA-VLBI will be able to measure parallaxes and proper motions to better than 10 $\rm \mu arcsec$, allowing us  to establish a direct comparison with the the Gaia (ESA) mission (see also Ros et al. 2015), and favouring  a direct comparison with Gaia and a cross-correlation among both reference frames (International Celestial Reference Frame (ICRF)  and Gaia).
SKA-VLBI will be very complementary to Gaia as it traces stars in regions of high interstellar extinction (e.g.  massive star forming regions,  Galactic Centre and the CMZ, dark molecular clouds, etc..) where Gaia will not be able to penetrate. In addition, stellar objects with strong radio emission such as supernova remnants, pulsars and late-type stars such as OH/IR stars can be easily monitored by SKA-VLBI.
 According to Fomalont \& Reid (2004), all stars in our galaxy
with flux densities higher than 0.5 mJy can be detected by SKA-VLBI spaced every six months over a two-year period.
Proper motions of objects in the Milky Way can be obtained within a period of a few years. A typical example would be the proper motion
of $\rm Sgr A^{*}$  around the Galactic Centre and its supermassive black hole. Measuring the direct parallax of the Galactic Centre would give us
for the first time the fundamental parameter $\rm R_{0}$ (Distance to the Galactic Centre) to an accuracy of 1\% which will have a big impact in galactic astronomy.
Similarly, precise distances to stellar clusters  can be measured,  especially for low-mass stars, which will  contribute to extending the distance determinations of
stellar clusters. Masses of pre-main sequence objects (PMS) are calibrated from stellar mass-luminosity relations (e.g. Close et al.  2005) where SKA astrometry
will clearly refine measurements  of the distance and the orbital motions for several hundreds of stars (Guirado et al. 2015). This will lead to the  derivation of accurate masses and ages of the stars, which will give strong constraints to stellar evolutionary models (Baraffe et al. 2009). 

\smallskip
\noi{\bf The Galactic Centre and bulge : Synergy between radio and near-infrared astrometry}

\smallskip
\noi While the ESA Gaia mission will get precise parallaxes for stars in the different Milky Way
populations, the inner kpc of our Milky way will be not covered due
the high interstellar extinction.  The Japanese ``JASMINE'' (Japan Astrometry Satellite Mission for Infrared Exploration, planned for 2020) mission 
will measure positions and motions of stars around the nuclear bulge in the near-infrared. There will be  common targets between JASMINE and SKA, which are mainly circumstellar OH maser sources,  and their astrometry will be cross- checked. For stars in the innermost part of the nuclear disk, where it is difficult for JASMINE to access due to severe interstellar extinction, the SKA astrometry will compensate the sample of those stars.
One of the major questions to be addressed is the  formation mechanism of the super massive black hole (SMBH) in the centre of our galaxy. Was the  SMBH formed by gas accretion
 or by a merger process? What is the link between the SMBH, the nuclear central star cluster,  and  the nuclear disk (Genzel et al. 2010; Fritz et al. 2014)  ? The answer to these  questions is crucial and  will give us  a better estimate of   the galactic potential in the central region of the Milky Way. Precise stellar orbits with JASMINE and SKA will allow  us to better constrain the gravitational potential as well as the physical processes of the SMBH growth.

\smallskip
\noi {\bf Maser stars}

\smallskip
\noi In the Milky Way, maser emission is observed in  young stellar objects  and late type stars. The strongest maser emission comes
from SiO, OH and $\rm H_{2}O$. OH/IR stars show show strong maser emission at 1612 MHz with a typical double peak feature originating from the front and backside of the
circumstellar shell (Sevenster et al. 2001). Due to the variability of the maser, only a fraction of these maser stars has been detected.  The number of detected OH/IR stars in our Galaxy
depends on the distribution of the mass-loss rates of the AGB stars. A complete detection of the OH/IR  star population give us contraints on the total integrated
mass-loss rates in our Galaxy, a crucial parameter for chemical evolution models (Matteucci 2014). The kinematics of the maser stars allow us to study stellar kinematics in various galactic
components (thin disc, thick disc, bulge, halo).
We expect to find  weak maser emission in planetary nebulae or post-AGB stars where the maser emission is probably related to shock excitation in the interface between a fast post-AGB wind  and  a slow AGB star
wind, which produces low-luminosity maser emission (Etoka et al. 2015). There, SKA will clearly open a  new window  as it is expected
to find many maser in stars.\\

\parbox{0.9\textwidth}{
\noi{References:}\\
\noi{\scriptsize 
	Baraffe, I., et al., 2009, ApJ, 702, 27;
	Close, L.M., et al., 2005, Nature, 433, 286;
	Etoka S., et al., 2015, arXiv:1501.06153; 
	Fomalont E. \& Reid M., 2004, New Astronomy Reviews, 48, 1473;
	Fritz T.K., et al., ApJ, 821, 42;
	Genzel, R., et al., 2010, Reviews of Modern Physics, 82, 3121;
	Guirado, J.C., et al., 2015, Spanish SKA White Book;
	Imai, H., et al., 2017, Japan SKA Consortium;
	Matteucci, F., 2014, ``Chemical Evolution of the Milky Way and Its Satellites", in The Origin of the Galaxy and Local Group, Saas-Fee Advanced Course, Volume 37.~ISBN 978-3-642-41719-1.~Springer-Verlag Berlin Heidelberg, 2014, p.~145;
	Ros, E., et al., 2015, Spanish SKA White Book;
	Sevenster, M.N., et al., 2001, A\&A, 366, 481
}}\\

\newpage

\subsection{Planets, Sun, Stars and Civilizations}

\noindent {\normalsize Contributors of this section in alphabetic order: }

\smallskip

\noi {\sffamily \scriptsize
{\sffamily \bf {\bf E.~Blanc}} [\ceadam],
{\bf C.~Briand} [\lesia],
{\bf S.~Celestin} [\lpcee],
{\bf R.~Courtin} [\lesia],
{\bf J.~F.~Donati} [\irap],
{\bf T.~Farges} [\ceadam],
{\bf J.~Girard} [\irfu;\aim],
{\bf J.~M.~Griessmeier} [\lpcee;\usn],
{\bf M.~Janvier} [\ias],
{\bf E.~Josselin} [\lupm;\irap],
{\bf L.~Lamy} [\lesia],
{\bf T.~Le~Bertre} [\lermasorb],
{\bf S.~Masson} [\lesia;\usn],
{\bf M.~Pandey-Pommier} [\cral],
{\bf P.~Petit} [\irap],
{\bf N.~Vilmer} [\lesia],
{\bf P.~Zarka} [\lesia;\usn]
}

\subsubsection{Solar system planets}
\vspace{0cm}

\noi The following sections demonstrate the interest of survey observations by SKA1-LOW of solar system planets, for which strong interest and expertise exist in the French community.

\paragraph{Jupiter's magnetosphere and radio emissions}\label{science:jupiter}
\vspace{0cm}

\noi Jupiter is the solar system planet with the strongest magnetic field (up to 14 Gauss at the surface, $>$10 times larger than for any other planet. Its giant magnetosphere, than extends toward the Sun up to 100 Jovian radii (7 million km) from the planet, is a very effective particle accelerator. Magnetospheric plasma fed by the moon Io, Jupiter's ionosphere, and the solar wind, is accelerated to several keV energies in the auroral (high-latitude) regions, and up to several MeV energies in the radiation belts that are trapped along magnetic field lines with apex up to 4--6 Jovian radii above the cloud tops (Bagenal et al. 2014). Auroral keV electrons produce extremely intense cyclotron emissions at frequencies up to 40 MHz (Zarka 1998), that are below the frequency range of SKA1-LOW. But MeV radiation belt electrons produce intense synchrotron radiation, the broad spectrum of which covers frequencies from $\le$100 MHz to $\ge$ 10 GHz. The emission has strong linear polarisation, its flux peaks at $\sim$6.5 Jy between 100 MHz and 1 GHz and it decreases towards higher frequencies until merging into the thermal emission of the planetary disk above 10 GHz (de Pater et al. 2003).

\smallskip

\begin{figure}[!ht]
  \centering
  \includegraphics[width=0.67\linewidth]{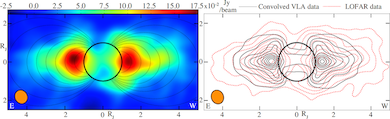}\hspace{0.2 cm}
  \includegraphics[width=0.31\linewidth]{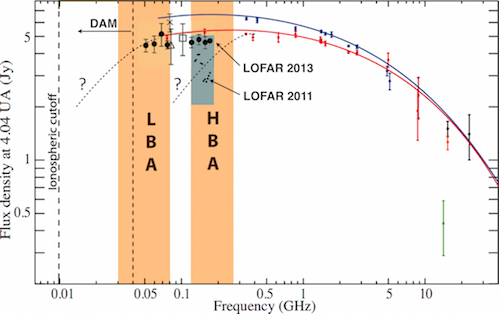}
  \caption{\label{fig_synchrotron}
{\em Left:} resolved brightness distribution of Jupiter's synchrotron emission with LOFAR at 127-172 MHz, with dipolar field lines superimposed.
{\em Center:} LOFAR image contours (red) compared to VLA ones at $\sim$5 GHz, by 10\% steps.
{\em Right:} measurements of Jupiter's synchrotron spectrum, obtained with the VLA between 1991 and 2004, and with LOFAR in 2011 and 2013 (Girard et al. 2016a,b). Blue and red curves were fitted by de Pater et al. (2003). Decametre emission (DAM -- not shown) dominates the spectral range below 40 MHz.}
\end{figure}

\noi  The belts being optically thin to synchrotron emission, but not homogeneous, their projection on the sky creates two asymmetric and variable ``ears'' surrounding the planetary disk (Fig.\,\ref{fig_synchrotron}). The extent, temporal and spectral variations of the belts contains the signatures of the physical processes (plasma sources, acceleration, sinks) that shape them. The Salammb\^o-3D code incorporate all these physical processes as well as radiative modeling of the belt emission (Sicard \& Bourdarie 2004, Nenon et al. 2017). LOFAR recently allowed us to obtain the first resolved image of the emission below 300 MHz (Girard et al. 2016a), taking into account the rapid source motion and wobbling, as well as direction-dependent calibration (Smirnov \& Tasse 2015). A 7-hour integrated image at $\sim$150 MHz demonstrates the broadest extent of electron belts of lower energy (emitting lower frequency synchrotron emission) as compared to higher energy electron belts observed with the VLA. Hourly sub-images compare well qualitatively with corresponding Salammb\^o-3D simulations. By contrast, the spectral variations and turnover are difficult to constrain by non-simultaneous observations in various spectral ranges. But we found a signal down to 50 MHz with an overall shape compatible with observations at higher frequencies.

\smallskip

\noi  SKA, covering the entire range of Jupiter's synchrotron emission with resolved images, will bring unique constraints to the distribution of energetic electrons in Jupiter's inner magnetosphere in all energy ranges (100s of keV to 10 MeV) simultaneously. It will also allow us to derive a reference instantaneous spectrum of the emission. It could also provide tomographic constraints on the topology of the near-planet low-latitude magnetic field, but following measurements by Juno we should already have a very accurate Jovian magnetic field model that will be a strong basis for the study of the belts structure and dynamics. Provided that the scheduling of survey observations by SKA in the ecliptic plane includes pointing at Jupiter at intervals of a few days, we will also derive the time variability of the belt processes at all timescales longer than this recurrence interval.\\

\parbox{0.9\textwidth}{
\noi{References:}\\
\noi{\scriptsize 
Bagenal, F., \etal , 2014, Space Sci. Rev., doi:10.1007/s11214-014-0036-8;
de Pater, I., \etal , 2003, Icarus, 163, 434;
Girard, J., \etal , 2016a, A\&A, A3, 587;
Girard, J., \etal , 2016b, SF2A, http://sf2a.eu/semaine-sf2a/2016/posterpdfs/237\_186\_46.pdf;
Nenon, Q., \etal , 2017, JGR, 122, 5158;
Sicard, A. \& Bourdarie, S., 2004, JGR, 109, 2216;
Smirnov, O. M. \& Tasse, C., 2015, MNRAS, 449, 2668;
Zarka, P., 1998, JGR, 103, 20159
}}\\

\paragraph{Planetary atmospheres}
\vspace{0cm}

\noi SKA will be a powerful instrument for studying planetary atmospheres thanks to its high sensitivity, angular resolution, spectral flexibility and resolution, that will help not only to detect faint thermal emissions down to the micro-Jansky level, but also resolve their morphology. These capabilities will enable significant advances in our understanding of the lower atmospheres of the terrestrial planets, mostly in the case of Venus. They will also further our knowledge of the deep atmospheres of the outer planets.

\smallskip

\noi On Venus, the total emission is the sum of the surface and atmospheric contributions. The atmospheric opacity in the microwave range is mainly determined by the abundance of the sulfur compounds, SO$_2$ and H$_2$SO$_4$.  While the distribution of these  compounds is fairly well understood, recent JVLA observations at 250 and 500 MHz (Butler 2014) have shown that there are still unknowns in what influences the long wavelength spectrum. The improved sensitivity of SKA in that domain will help resolve this issue. 

\smallskip

\noi The other area of interest concerns the deep atmospheres of the four outer planets. Their bulk composition is still poorly constrained in terms of the main cosmogonical elements, especially C, N, and O (Fletcher et al. 2009). Our incomplete knowledge is due to the difficulty to probe the atmospheres of these planets below the condensation levels of the volatiles containing N and O (NH$_3$, NH$_4$SH, H$_2$O) which are located much below the levels probed by infrared and microwave instruments (Atreya et al. 1999). As a consequence, the interior models of these planets are also poorly constrained because of the uncertainty in the abundance of heavy elements, which is dominated by that of H$_2$O (Nettelmann et al. 2013). As the atmospheric temperature increases with depth and the optical depth decreases with decreasing frequency, the thermal emission from the outer planets increases towards long wavelengths.

\smallskip

\noi At Jupiter, the strong magnetospheric synchrotron emission (see 4.1.1) dominates below $\sim$1 GHz, preventing to probe the deep atmosphere (except from an observer inside the radiation belts like the Juno spacecraft). But monitoring Jupiter's thermal emission above 1 GHz with sub-arcsecond resolution will reveal fine details of the spatial distribution of NH$_3$ and deep cloud structure, increasing our understanding of tropospheric dynamics.  

\smallskip

\begin{figure}[!ht]
  \centering
  \includegraphics[width=0.4\linewidth]{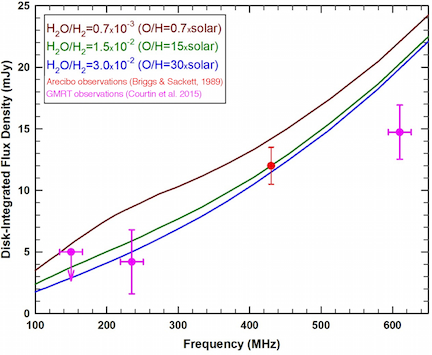}\hspace{0.5 cm}
  \includegraphics[width=0.4\linewidth]{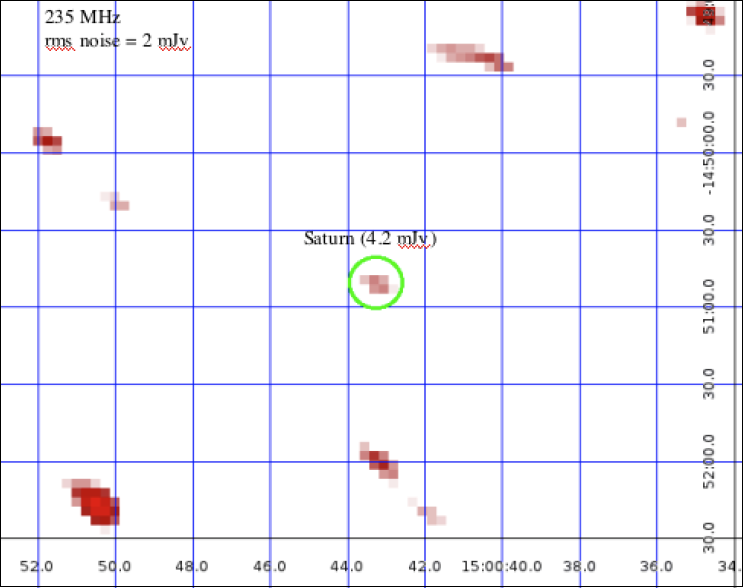}
  \caption{\label{fig:saturn} 
  {\em Left:} flux density measurements of Saturn, performed with the Arecibo (red) and GMRT (pink) radiotelescopes, compared with three model spectra assuming various abundances of H$_2$O in the deep atmosphere. The reference value for H$_2$O (Visscher \& Fegley 2005) corresponds to an O/H ratio 3.4 times the solar value of Lodders (2003). The range above $\sim$400 MHz is mostly influenced by NH$_3$, whereas H$_2$O vapour controls the slope of the 100-400 MHZ spectrum. Below 100 MHz free electrons become the dominant source of opacity.
  {\em Right:} detection (2$\sigma$) of Saturn at GMRT on 16/8/2014 at 235 MHz, with a flux density of 4.2 $\pm$ 2.6 mJy ($T_b$= 404 $\pm$ 249 K). The size of Saturn is consistent with a 18" disk convolved with a beam of FWHM 20". From (Courtin et al. 2015).}
\end{figure}

\noi At Saturn, no significant magnetospheric emission is expected because of electrons absorption by the rings. Observations at 428 MHz with the Arecibo radiotelescope permitted Briggs \& Sackett (1989) to constrain the NH$_3$ mole fraction within a factor of 2, but they were only marginally sensitive to the abundance of H$_2$O. Recent measurements with the Giant Meterwave Radio Telescope in India, at 150, 235, and 610 MHz, were compared with models of Saturn's thermal radiation assuming different water vapour concentrations (Courtin et al. 2015), favouring water-rich models with an O/H ratio of at least 15 (Fig.\,\ref{fig:saturn}, {\em left} panel). Contemporary LOFAR observations were heavily contaminated by extragalactic sources so that the thermal emission below $\sim$200 MHz remains undetected so far. SKA measurements will bring decisive constraints in the 100-600 MHz range, where Saturn's predicted  thermal spectrum strongly depends on the H$_2$O abundance. 
At 240 MHz, Saturn's thermal emission mostly originates from pressure levels between 20 and 400 bar, well below the H$_2$O-NH$_3$ condensation cloud, and is characterised by a brightness temperature of 490-800 K depending on the assumed H$_2$O concentration (0.2--10 times the value estimated by Visscher \& Fegley (2005), i.e. 0.7--34 times the O/H solar ratio of Lodders (2003)). Gas opacity decreases with decreasing frequency, so that higher brightness temperatures (550-1000 K) are expected at 180 MHz, corresponding to emission from around the 1 kbar level. The frequency range 180-240 MHz is most adapted to measuring the H$_2$O abundance since the spectrum in this range is primarily influenced by that quantity. At lower frequencies, we will probe the weakly ionised region of the atmosphere, never explored before and whose theoretical modeling is more uncertain (a brightness temperatures of 850-1650 K $\pm$20\% is predicted at 110 MHz). 
SKA observations between 150 and 600 MHz should reach a signal-to-noise ratio between 300 and 2500 with a 12 MHz bandwidth and 1 hour integration time, providing the first definitive test of Saturn's deep atmospheric abundance of water, and a first glimpse on the deeper weakly ionised layers. The H$_2$O abundance in the deep atmosphere will in turn allow us to constrain the O/H ratio relative to the solar abundance, which has important cosmogonical implications. It also strongly influences Saturn's J4 moment of inertia, which characterises the distribution of mass inside Saturn, with implications for its internal structure, figure, and rotation.

\smallskip

\noi Due to its smaller apparent diameter (3.7" instead of 17" for Saturn), Uranus' thermal spectrum is expected to be a few percent of that of Saturn in the same frequency range, assuming similar brightness temperatures. Hence, flux densities of about 0.2--0.6 mJy are predicted, that are within the reach of SKA with a theoretical SNR of 15--100 between 150 and 600 MHz (with 12 MHz bandwidth $\times$ 1 hour integration). Neptune, with flux densities one order of magnitude lower than Uranus, is not a realistic target for SKA. \\

\parbox{0.9\textwidth}{
\noi{References:}\\
\noi{\scriptsize 
Atreya, S.K., \etal , 1999, Planet. Space Sci., 47, 1243; 
Briggs, F.H. \& Sackett, P.D., 1989, Icarus, 80, 77;
Butler, B., 2014, AAS-DPS meeting \#46, id.416.04;
Courtin, R.,  \etal , 2015, SF2A proceedings, S. Boissier \etal eds., Paris, France; 
Fletcher, L.N., \etal , , 2009, Icarus, 199, 351;
Lodders, K., 2003, ApJ, 591, 1220; 
Nettelmann, N.,  \etal , 2013, Icarus, 225, 548;
Visscher, C. \& Fegley Jr., B., 2005, ApJ, 623, 1221 
}}\\

\paragraph{Planetary lightning}\label{science:plan_light}
\vspace{0cm}

\noi Lightning is a transient, high-current electrostatic discharge resulting from macroscopic electric charge separation (by convection and gravitation combined), following small-scale particle electrification (via collisions and charge transfer). A large-scale electric field builds up, which may eventually lead to breakdown and ionise the ambient medium, causing a lightning stroke. A lightning discharge consists of many consecutive strokes of a few $\mu$s duration and lasts typically 1-100 ms. It may have an important role in the atmospheric chemistry (production of non-equilibrium trace organic constituents potentially important for biological processes). Electromagnetic signatures include optical, VLF and LF radio emissions. The radio emission generated by lightning is broadband in nature. For example, the spectrum of terrestrial lightning peaks at tens of kHz, but extends to the VHF range (30-300 MHz) and above. Satellite observations show that the lightning phenomenon also takes place on other solar system planets (Zarka et al. 2004, and references therein). First ground-based detection occurred recently, limited by the low flux density of the emission. Its sensitivity and the low emission frequency (typically 
$<$ 100 MHz) make SKA1-LOW a promising instrument with potential for significant discoveries in this field, for terrestrial as well as giant planets of the Solar system.

\smallskip

\noi Existence of lightning on Venus remains controversial (e.g. Zarka et al. 2008). Venus lightning, if it exists, is either extremely rare (1 flash/hour, with long intervals of atmospheric electrical inactivity), very weak (10$^{2-3}$ times weaker than terrestrial lightning), or restricted to low frequencies (e.g. slow strokes with a very steep high frequency spectrum) which are unable to cross Venus' ionosphere. SKA will be able to probe the intensity and occurrence rate of Venus lightning. On Mars, despite a thin atmosphere, atmospheric electric discharges can happen in large scale dust storms or ``dust devils" (localised dust storms) which are thought to be able to generate substantial charge via triboelectric processes, i.e. contact electrification (Farrell et al. 1999). Observations were attempted, but reports on detection remain controversial (Ruf et al. 2009, Anderson et al. 2012). The radio flux is expected to increase with frequency, so that SKA should be able to either detect this emission, or prove that it does not exist. 

\smallskip

\noi  At Jupiter, no HF radio emission from lightning was detected despite the observation of optical lighting flashes. SKA1-LOW's sensitivity and resolution may allow us to detect rare intense events just above its LF limit of 50 MHz. On Saturn, radio emission caused by lightning has been observed from satellites as well as using ground-based radio telescopes such as UTR-2 (Zakharenko et al. 2012, Konovalenko et al. 2013), WSRT, and during test observations with LOFAR (Grie$\ss$meier et al. 2010, 2011). The flux density of Saturn's lightning detected at Earth is up to 100--1000 Jy (Zarka et al. 2004). SKA will provide the sensitivity permitting to monitor lightning from Saturn, addressing questions such as the discharge timescales, spectral profile, and its energy budget, the electrification processes compared to Earth's, the breakdown electric field (which depends on atmospheric pressure, composition and electron density), atmospheric dynamics and cloud structure, and geographical and seasonal variations. In addition, high resolution imaging will give access to spatial information. Similarly to Saturn's case, lightning-generated radio emission from Uranus was detected by Voyager 2 (Zarka and Pedersen 1986), with a corresponding flux density of 0.4--40 Jy at Earth.  Uranus' lightning is thus detectable with the SKA. At Neptune, only 4 weak events were detected during the Voyager 2 fly-by (Kaiser et al. 1991), too rare and weak to be an interesting target for SKA. However, occasional monitoring has the potential to discover Neptune's lightning if periods of strong activity occur (or to set upper limits on the emission duty cycle).

\smallskip

\noi  No lightning from exoplanets is expected to be detectable, in spite of some recent fanciful claims that do not need to be advertised here. \\

\parbox{0.9\textwidth}{
\noi{References:}\\
\noi{\scriptsize 
Anderson, M.M., \etal , 2012, ApJ, 744, 15; 
Farrell, W.M., \etal , 1999, GRL, 26, 2601;
Grie$\ss$meier, J.-M., \etal , 2010, ISKAF2010, PoS, 022, http://pos.sissa.it/cgi-bin/reader/conf.cgi?confid=112;
Grie$\ss$meier, J.-M., \etal , 2011, in PRE VII, H.O. Rucker \etal eds., Austrian Acad. Sci. Press, Vienna, 145;
Kaiser, M.L., \etal , 1991, JGR 96, 19043;
Konovalenko, A.A., \etal , 2013, Icarus 224, 4;
Ruf, C., \etal , 2009, GRL, 36, L13202;
Zakharenko, V., \etal , 2012, Planet. Space Sci., 61, 53;
Zarka, P., \etal , 2004, Planet. Space Sci., 52, 1435;
Zarka, P., \etal , 2008, Space Sci. Rev., 137, 257;
Zarka, P. \& Pedersen, B.M., 1986, Nature, 323, 605
}}\\

\paragraph{Electrical activity in Earth's atmosphere}\label{science:atmosphere}
\vspace{0cm}

\begin{figure}[!ht]
  \centering
  \includegraphics[width=0.95\linewidth]{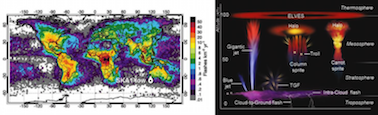}
  \caption{\label{fig_lightning} {\em Left:} annual lightning activity observed by the Optical Transient Detector and the Lightning Imaging Sensor on the TRMM mission (adapted from Christian et al. 2003). {\em Right:} sketch of the various TLE and TGF in relation with the discharges in the thundercloud which produces them (adapted from Blanc \& Farges 2012).}
\end{figure}

\noi Earth's thunderstorm activity consists of $\sim$45 flashes per second (Christian and al., 2003). 75\% of the discharges are intra-cloud, whereas 25\% are cloud-to-ground discharges. Lightning activity is inhomogeneously distributed in space and time: 90\% occurs over continents, mainly in the tropical belt, whereas 10\% occurs above the oceans (Fig.\,\ref{fig_lightning}, {\em left}). The activity takes place mainly in the summer and fall. In the 1990's, it was discovered that energetic phenomena occur above storm clouds: transient luminous events (TLE) (e.g. Pasko et al. 2012) and Terrestrial Gamma-ray Flashes (TGF) (Fishman et al. 1994) (Fig.\,\ref{fig_lightning}, {\em right}). SKA1-LOW may bring an original contribution to the understanding of several aspects of atmospheric electricity:

\begin{itemize}

\item start of the lightning discharge: Rison et al. (2016) showed that specific discharges, the so-called Narrow Bipolar Pulses, occur inside cumulonimbus at the very beginning of the lightning discharge process. These NBP are very energetic and radiate in the HF - VHF range (1 - 100\,MHz). Measurements below 100 MHz will strongly contribute to the understanding of these phenomena. More generally, the spectral slope of lightning radio emission is not well characterised above 10 MHz;

\item a new category of TLE has been observed at the top of tropical clouds from the International Space Station (Chanrion et al. 2017). They are reminiscent of blue jets, which remain poorly known because not easily accessible. These topside discharges, like the NEB, probably produce HF - VHF radiation that can be detected with SKA. Their quantitative study will improve our understanding of the formation of NOx components involved in the assessment of climate warming;

\item production of TGF requires the presence of electric fields high enough to accelerate electrons to relativistic energies. These electric fields are produced in intra-cloud discharges, and they accelerate secondary electron/positron beams that escape into space, and may produce radio and visible signals detectable from ground and space (Xu et al. 2017). The study of the radio signal with SKA will constrain the production mechanisms of TGF.

\end{itemize}

\smallskip

\noi Pioneering radio imaging measurements of lightning has been conducted with the LOFAR in order to locate electric discharges in 3D (Scholten et al. 2017). Additional location information will be brought by facilities such as the Weather Zone Total Lightning detection Network in Australia, as well as satellite observations from geostationary orbit. Several space experiments dedicated to the study of lightning, TLE, and TGF are (soon to be) launched: ISS--LIS (3/2017) and ASIM (2018) on board the ISS, and the satellite TARANIS (2019). Having the possibility of Target of opportunity Observations on short notice and taking care of not suppressing lightning radio signals (brief, broadband) in SKA1-LOW data will be important assets.\\

\parbox{0.9\textwidth}{
\noi{References:}\\
\noi{\scriptsize 
Blanc, E. \& Farges, T., 2012, PLS, 416, 48;
Chanrion, O., \etal , 2017, GRL, 44, 496; 
Christian, H.J., \etal , 2003, JGR, 108(D1), 4005;
Fishman, G.J., \etal , 1994, Science. 264, 1313;
Pasko, V.P.,, \etal , 2012, Space Sci. Rev., 168, 475;
Rison, W., \etal , 2016, Nature Comm., 7, 10721;
Scholten O., \etal , 2017, EPJ Web of Conf. 135, 03003, doi:10.1051/epjconf/201713503003;
Xu, W., \etal , 2017, GRL, 44, 2571
}}\\

\paragraph{Ionospheric physics \& Meteor showers} \label{science:meteorshowers}
\vspace{0cm}

\begin{figure}[!ht]
  \centering
  \includegraphics[width=0.375\linewidth]{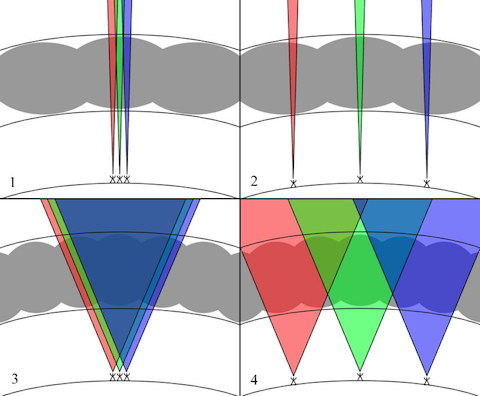}\hspace{0.2 cm}
  \includegraphics[width=0.375\linewidth]{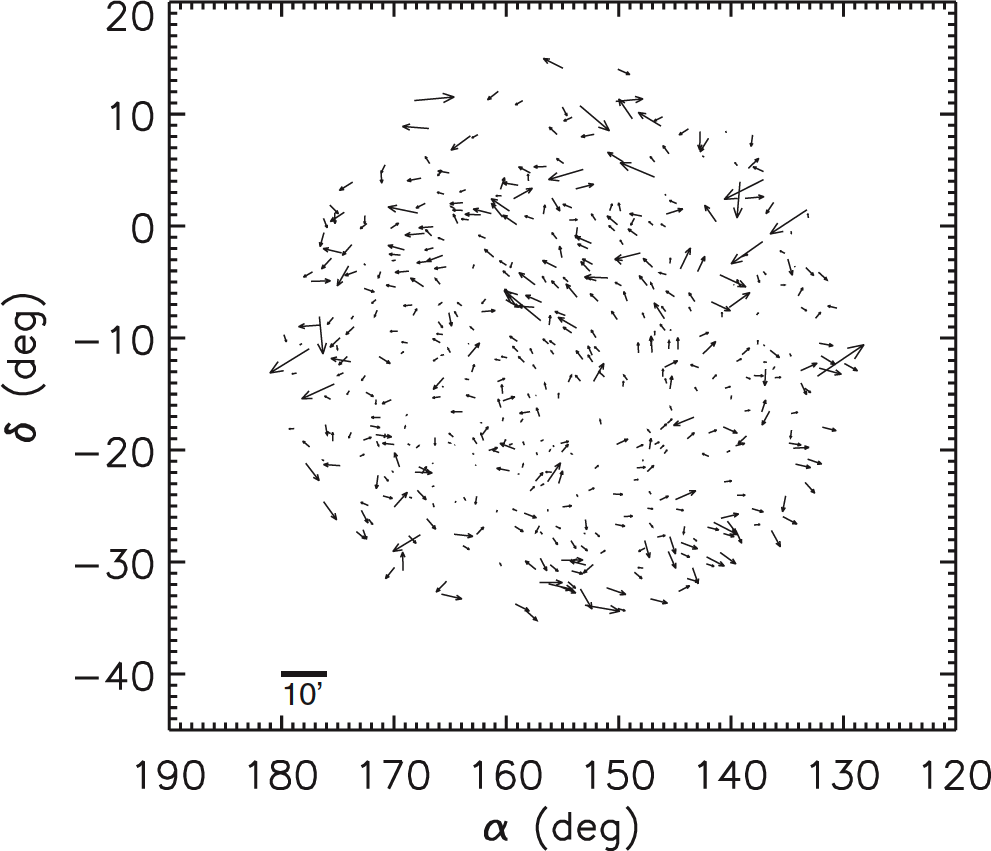}\hspace{0.2 cm}
  \includegraphics[width=0.21\linewidth]{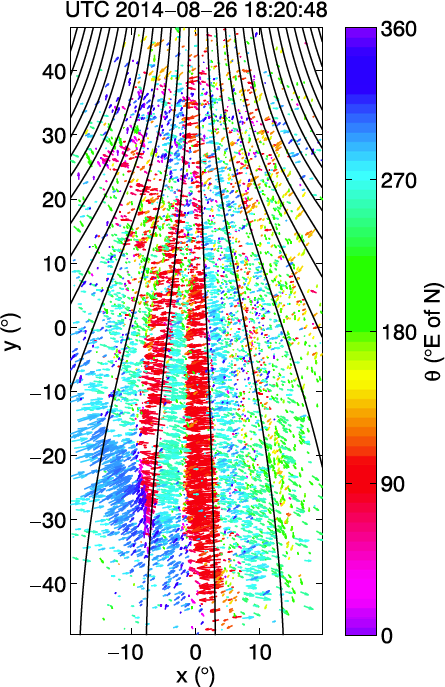}
  \caption{\label{fig_iono} 
{\em Left:} sketch of ionospheric electron density structure (grey) for a compact (1,3) and extended (2,4) array with narrow (1,2) and wide (3,4) FoV (red, green and blue). The compact array ``sees" the same ionosphere, not the extended one. A broad FoV implies ionospheric variations inside the beam (from Intema et al. 2009).
{\em Centre:} spatial offsets of source positions introduced by the ionosphere, as imaged by the MWA (wide FoV) and compared to  reference catalogue positions (from Williams et al. 2012).
{\em Right:} perpendicular TEC gradient direction deduced from MWA images (from Loi et al. 2016).}
\end{figure}

\noi The calibration of low frequency radio data (below $\sim$300 MHz) acquired by new generation radio interferometers like MWA, LOFAR and SKA imposes severe challenges due to ionospheric propagation combined with their large field of view (Fig.\,\ref{fig_iono}, {\em left}). The ionosphere is inhomogeneous in time and space (Davies 1990). The free electron column density along the line of sight also referred to as total electron content (TEC) and its organised (waves) and stochastic fluctuations cause dispersion, propagation delays, Faraday rotation, as well as distortion and phase shifts in the radio signal from astronomical sources. The resulting time- and direction-dependent phase and amplitude calibration errors may introduce source displacement (Fig.\,\ref{fig_iono}, {\em centre}) and deformation, false flux variability, and add spurious faint sources in the image or smear existing ones leading to their disappearance. The effect of the ionosphere is stronger at low frequencies, the phase shift being inversely proportional to the wavelength. These effects must be corrected in the imaging process. Algorithms have been developed to solve and correct for ionospheric phase errors from a pre-defined sky model, from SPAM (Intema et al. 2009), that produces a model of the ionospheric phase screen, to the fully time- and direction-dependent KillMS/DDfacet pipeline developed in Paris Observatory (Smirnov \& Tasse 2015). Besides drastically improving the image quality, these algorithms allow one to derive maps of TEC fluctuations (Fig.\,\ref{fig_iono}, {\em right}) and consequently study the ionosphere itself (Herne et al. 2013, Loi et al. 2016). These studies will be a permanent side product of SKA1-LOW observations.

\smallskip

\noi Meteoroids of various sizes (single objects, dust clusters or asteroid trails) frequently penetrate Earth's atmosphere at tens of km/s. Friction with air molecules produces ablation and ionization, forming a plasma cloud that reflects radio waves. Signals from ionospheric radars, AM/FM/TV broadcast and other narrowband transmitters reflect on these clouds. High time--frequency observations ($\le 0.1$ sec$\times0.1$ kHz) within narrow frequency bands (a few kHz) around transmitters' frequencies of reflected echoes allow to track the plasma cloud dynamics along the meteor entry path up to several seconds after full ablation. Estimation of the location and motion of the meteor and its trail may allow to estimate shower paths and strengths (densities of meteors in showers), trail reflectivity, and to classify automatically meteors by ``radar signature" analysis (Azarian \& Vaubaillon 2014). Maximum reflectivity should occur at the LF end of SKA1-LOW (peak around 50 MHz). Even in the super-clean environment of SKA1-LOW, narrowband RFI sources will exist (Expert Panel Report on RFI 2011) than will constitute illuminators of opportunity. SKA1-LOW can then be used as a passive radar receiver with high sensitivity, large FoV (possibly multibeam), and possibly fast imaging capability with high angular resolution fast imaging.\\
 
\parbox{0.9\textwidth}{
\noi{References:}\\
\noi{\scriptsize
Azarian, S. \& Vaubaillon, J., 2014, in NenuFAR: instrument description and science case, https://nenufar.obs-nancay.fr/, 118;
Davies, K., 1990, Ionospheric Radio, IET, UK;
Expert Panel Report on RFI, 2011, http://skatelescope.org/;
Intema, H.T., \etal , 2009, A\&A, 501, 1185; 
Herne, D., \etal , 2013, Proc. ASSC 13, Sydney, Australia, 129;
Loi, S.T., \etal , 2016, JGR, 121, 1569;
Smirnov, O. M. \& Tasse, C., 2015, MNRAS, 449, 2668;
Williams, C.L., \etal , 2012, ApJ, 755, 47
}}\\

\subsubsection{Exoplanets and star-planet interactions}\label{science:exop}
\vspace{0cm}

\begin{figure}[!ht]
  \centering
  \includegraphics[width=0.6\linewidth]{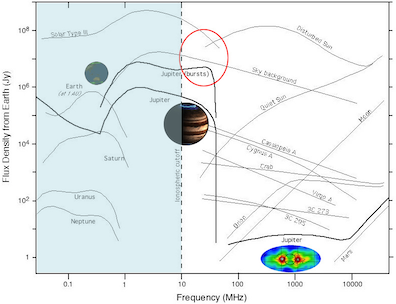}
  \caption{\label{radio_spectra} Spectra of astronomical radio sources detected from the Earth's vicinity. Auroral planetary spectra lie below the 
  Earth's ionospheric cutoff, except Jupiter's decametric (CMI, auroral and Io-induced) and metric-decimetric (synchrotron) emissions. 
  Normalized to the same observer distance of 1 AU, the Jupiter spectrum must be upscaled by a factor 20 (exceeding Solar bursts), Saturn by $\times$100, Uranus by $\times$400, and Neptune by $\times$900, so that all are grouped within 2-3 orders of magnitude. From (Zarka 2015).}
\end{figure}

\begin{figure}[!ht]
  \centering
  \includegraphics[width=0.8\linewidth]{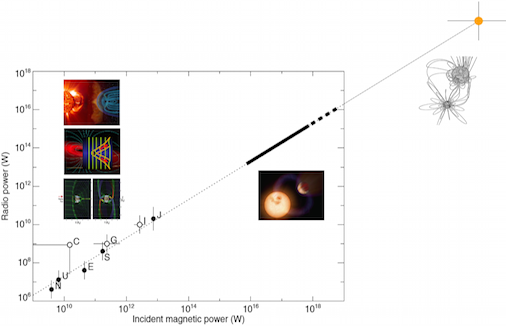}
  \caption{\label{scaling_law} Radio-magnetic scaling law relating magnetospheric (E,J,S,U,N = Earth, Jupiter, Saturn, Uranus and Neptune) and satellite-induced (I,G,C = Io, Ganymede, Callisto) radio power to incident Poynting flux of the plasma flow on the obstacle. Dashed line has slope 1, with a proportionality coefficient $2\times 10^{-3}$. The thick bar extrapolates to hot jupiters solar wind-magnetosphere interactions (solid) and satellite-planet electrodynamic interactions (dashed). The orange dot illustrates the case of the interacting magnetic binary RS CVn V711 $\tau$ (Budding et al. 1998). Insets sketch the types of interaction (solar wind-magnetosphere, Jupiter-Io, Jupiter-Ganymede). From (Zarka 2015).}
\end{figure}

\noi All magnetised solar system planets are sources of strong low-frequency radio emissions (Zarka 1998). These emissions are produced by electrons accelerated to a few keV energy via by solar wind-magnetosphere interaction (compression or magnetic reconnection), magnetosphere-ionosphere or magnetosphere-satellite coupling. When traveling along magnetic field lines above planetary polar regions, these electrons generate radio emissions at the local cyclotron frequency ($f_{ce}=eB/2 \pi m_e$, up to 40 MHz at Jupiter, $\le$1.5 MHz at all other planets), via a well-understood and well-documented non thermal coherent process, the Cyclotron Maser Instability (CMI -- Treumann 2006). These emissions can be as intense as solar ones (kJy to MJy -- cf. red circle on Fig.\,\ref{radio_spectra}), hence the idea to detect their exoplanetary counterparts from stellar distances. However, the LF sky background, (plus ionospheric scintillation and RFI) prevent us to detect them from these distances. Scaling laws have been derived that predict that hot jupiters may be emitters $10^3$ to $10^5$ times more intense than Jupiter (Fig.\,\ref{scaling_law}), and thus detectable at a few tens of pc range (Zarka et al. 2001, Zarka 2007, Grie$\ss$meier et al. 2007). Other studies favour fast rotating giant planets orbiting at a few AU from stars with a large X-UV brightness (Nichols, 2011).

\smallskip

\noi As hot jupiters on circular orbit are likely to be spin-orbit synchronised, their relatively slow sidereal rotation ($\sim$ days) may be unfavourable to sustain a strong planetary dynamo. Thus their magnetic field may be too weak to provide electron cyclotron frequencies $\ge$ a few tens of MHz. However, even if weakly magnetised or unmagnetised, these planets orbiting in a sub-Alfv\'enic stellar wind will magnetically interact with their parent star similarly to Io with Jupiter. The planet may thus induce (via unipolar inductor, Alfv\'en waves or magnetic reconnection) electron acceleration and thus intense radio emission, up to $10^{6-7}$ times Jupiter's, in the magnetic field of their parent star (Fig.\,\ref{scaling_law}). The study of star-planet plasma interactions was placed in the more general theoretical framework of the interaction of a magnetised plasma flow with an obstacle (Zarka 2017, and references therein). It also applies for example to the case of terrestrial planets orbiting a white dwarf, as analysed by Willes and Wu (2004, 2005) who predicted strong CMI emission at several GHz.

\smallskip

\begin{figure}[!ht]
  \centering
  \includegraphics[width=0.8\linewidth]{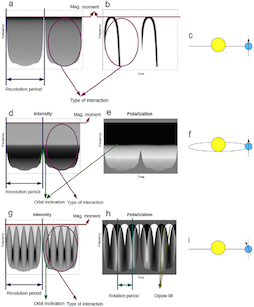}
  \caption{\label{exo_simu} Simulated dynamic spectra in intensity (a,b,d,g) and circular polarisation (e,h) and associated parameters of the system that can be determined: (a,b,c) exoplanetary magnetic field aligned with the rotation axis and 0$^\circ$ orbit inclination, for emission (a) of a full exoplanet's auroral oval and (b) an auroral active sector fixed in local time. (d,e,f) aligned exoplanetary magnetic field and 15$^\circ$ orbit inclination, for emission of a full auroral oval. (g,h,i) exoplanetary magnetic field tilted by 15$^\circ$ and 0$^\circ$ orbit inclination, for emission of a full auroral oval. From (Hess \& Zarka 2011).}
\end{figure}

\noi Based on our detailed theoretical understanding of magnetospheric emissions from solar system planets, it has been shown that detection (in the time--frequency plane, because no resolved imaging is possible for extrasolar hot jupiters) of the intensity and circular polarisation of the radio emission from an exoplanetary system will allow us to obtain unique information on the planetary magnetic field (strength, orientation, tilt) and thus the planetary dynamo process (that depends itself on the thermal state, composition, and dynamics of their interior -- very difficult to probe by other means), and on the physics of star-planet plasma interactions (topology, energetics). Detailed simulations are discussed in (Hess \& Zarka 2011) and illustrated in Fig.\,\ref{exo_simu}. Not only such a detection will open the broad new field of comparative exo-magnetospheric physics, but it will also permit to measure the planetary rotation (testing spin-orbit synchronisation) and orbit inclination, detect exo-moons, and provide a precious constraint on the planetary habitability (large-scale planetary magnetic fields play an important role in protecting the planet's surface and atmosphere). 

\smallskip

\noi Observational searches started even before the confirmed discovery of the first exoplanet (Zarka 2011). No confirmed detection has been 
reached yet (Zarka et al. 2015), but only a few tens of candidates were observed (by UTR-2, the VLA, the GMRT, and LOFAR). SKA1-LOW will give access to the range $\ge$ 50 MHz with a sensitivity $\sim$30 times better than LOFAR's. To the extent that Jupiter's magnetic field is not exceptionally strong, and that exoplanetary radio emission thus exists above 50 MHz, SKA1-LOW will be able to detect a signal equivalent to Jupiter's emitted from a distance up to $\sim$ 5 pc. Distinguishing a planetary signal from a stellar one should be facilitated by the expected strong circular polarisation of planetary CMI emission, and recurrence at the planetary orbital period. SKA will operate in large surveys (the one of interest for exoplanets being focused on low Galactic latitudes), so that systematic sensitive observation of all known exoplanets in the southern hemisphere ($\sim$800 today -- cf. \href{http://www.exoplanet.eu}{\color{blue} \myul[blue] {www.exoplanet.eu}}) will be carried out. Within 10 pc distance, there are 200 known stars and $\sim$50 currently known exoplanets, and this number should increase substantially with coming space missions dedicated to transits and powerful ground-based instruments. The accessible volume will be much increased if scaling laws derived in our solar system can be reliably extrapolated to exoplanetary systems, permitting to measure lower mass planets' dynamos and magnetospheres. Scheduling regular re-observations of the exhaustive set of planetary targets will increase the chances of radio burst detection and possibly allow us to address original questions such as the variability of the emission with the season or stellar cycle. SKA observations will be complemented at frequencies $<$50 MHz by NenuFAR observations (in the northern hemisphere -- Zarka et al. 2012). Note that the higher frequency synchrotron emission from high energy electrons in radiation belts is several orders of magnitude weaker than the CMI emission and thus should not be detectable at stellar distances. \\

\parbox{0.9\textwidth}{
\noi{References:}\\
\noi{\scriptsize
Budding, E., \etal , 1998, PASA, 15, 183;
Grie$\ss$meier, J.-M., \etal , 2007, A\&A, 475, 359;
Hess, S.L.G. \& Zarka, P., 2011, A\&A, 531, A29;
Nichols, J. D., 2011, MNRAS, 414, 2125;
Treumann, R., 2006, Astron. Astrophys. Rev, 13, 229;
Willes, A. J. \& Wu, K., 2004, MNRAS, 348, 285;
Willes, A. J. \& Wu, K., 2005, A\&A, 432, 1091;
Zarka, P., 1998, JGR, 103, 20159;
Zarka, P., 2007, Planet. Space Sci., 55, 598;
Zarka, P., 2011, in Planetary Radio Emissions VII, H. O. Rucker \etal eds., Austrian Acad. Sci. Press, 287;
Zarka, P., \etal , 2001, Astrophys. Space Sci., 277, 293;
Zarka, P.,  \etal , 2012, SF2A-2012, S. Boissier \etal eds., 687;
Zarka, P., \etal , 2015, AASKA14, 120;
Zarka, P. 2017, Handbook of Exoplanets, J. A. Belmonte \etal eds., Springer, in press
}}\\

\subsubsection{Solar radio emissions}\label{science:sun}
\vspace{0cm}

\begin{figure}[!ht]
  \centering
  \includegraphics[width=0.8\linewidth]{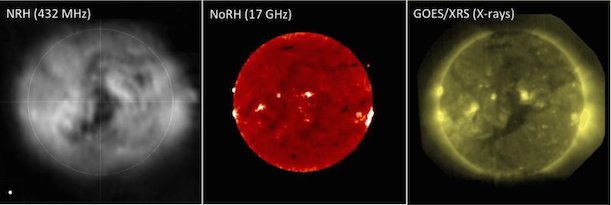}
  \caption{\label{fig_solar_images} The Sun as observed with the Nan\c{c}ay radioheliograph at 432 MHz ({\em left}), the Nobeyama one at 17 GHz ({\em centre}) and the GOES satellite ({\em right}). SKA will provide a missing link in understanding the thermal structure of the quiet corona and coronal holes at frequencies from 50\,MHz to 14\,GHz (adapted from Mercier \& Chambe 2012).}
\end{figure}

\begin{figure}
  \centering
  \includegraphics[width=0.5\linewidth]{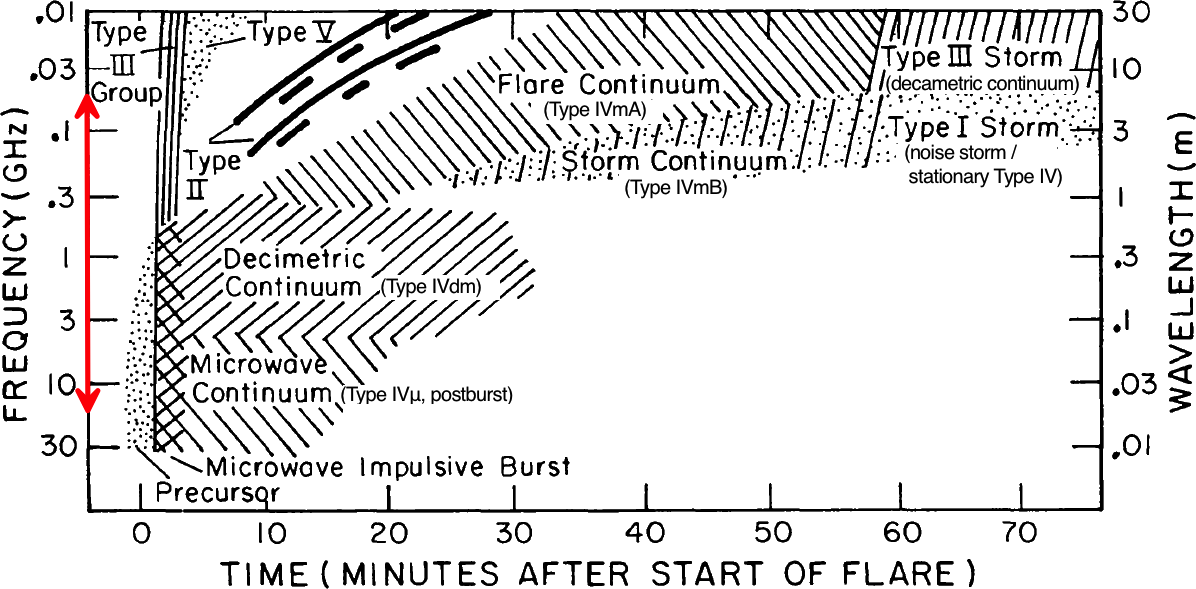}\hspace{0.3 cm}
  \includegraphics[width=0.20\linewidth]{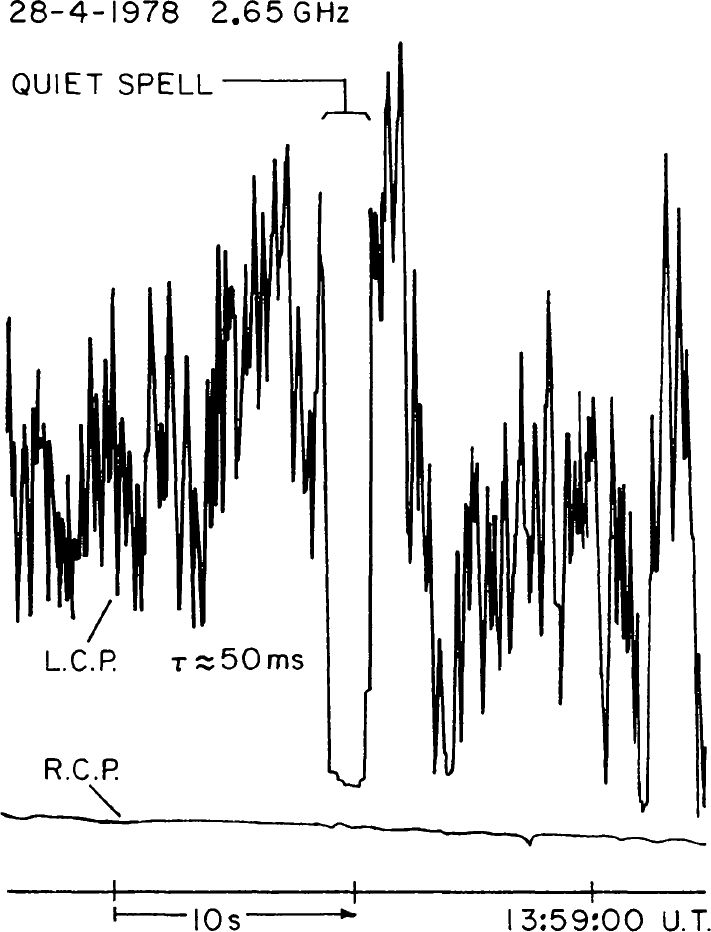}
  \caption{\label{fig_solar}
{\em Left:} schematic dynamic spectrum of a solar radio outburst such as might be produced by a large flare.  Outbursts often vary considerably from this ``typical spectrum''.  SKA will cover the spectral range indicated
by the red arrow.  This corresponds to altitudes ranging from $\lesssim$ 1.1 R$_S$  from the SunÕs centre, up to 1.5 R$_S$  (or higher in exceptional plasma conditions (Morosan et al. 2014)). {\em Right:} microwave spike bursts from the Sun at 2.65 GHz, recorded with a time constant of 10 ms. In the displayed event, spikes appeared only in LH polarisation, some with variations on a 50 ms time scale. From Dulk 1985.
}
\end{figure}

\noi The Sun is a strong radio emitter in the frequency range from $<$1\,MHz to $>$50\,GHz (e.g. Dulk 1985, Pick \& Vilmer 2008). Radio observations provide dynamic spectra (intensity, polarisation) from 10\,GHz to 1\,MHz and imaging information with solar dedicated radio-heliographs (Nobeyama at 35 and 17 GHz, Nan\c{c}ay (NRH) at several frequencies between 435 and 150\,MHz) or with radio interferometers developed for astronomy (VLA, LOFAR, \dots). The quiet Sun can be imaged at various frequencies with e.g. the above radio-heliographs (Dulk 1985, Mercier \& Chambe 2012; Fig.\,\ref{fig_solar_images}). Higher frequencies are produced closer to the solar surface. As a result, while quiet Sun images at high frequencies tend to match Extreme UV and X-ray images of the solar corona, at lower frequencies (e.g. 150\,MHz) they reveal coronal holes. 

\begin{figure}
  \centering
  \includegraphics[width=0.9\linewidth]{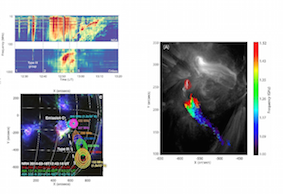}
  \caption{\label{fig_bougeret}
{\em Top left:} dynamic spectrum of a solar radio event observed with the spectrograph Orfees (in Nan\c{c}ay) and the Nan\c{c}ay Decameter Array (NDA), covering the frequency range between 1000 MHz to 10 MHz.  Type III, Type II and continuum (emission C) are observed at different frequencies and time intervals in the event, such as might be produced by a large flare.  {\em Bottom left:} Solar Dynamics Observatory/AIA three-colour image of 94, 131, and 335 \AA~channels showing a highly twisted structure with a jet. NRH contours at 445-150\,MHz are overplotted with the maximum and minimum brightness temperature indicated on the extreme frequencies. The radio sources occur above the northwest loop of the rope. The dashed lines indicate the frequencies of Type III emission and Emission ``C'' showing up during the interval 2 of the radio dynamic spectra above. SKA images will cover the altitude range between the eruptive rope observed in the EUV and the highest altitude radio sources observed with the NRH (Carley et al. 2016). {\em Right:} VLA imaging spectroscopy of a decimetric Type III burst (electron beam) between 1 and 1.5\,GHz (Chen et al. 2013).}
\end{figure}

\smallskip

\noi Radio emissions of the active Sun (radio bursts) display a very complex morphology in the time-frequency plane with drifting structures, bands, zebra (Fig.\,\ref{fig_solar}, {\em left} and Fig.\,\ref{fig_bougeret}, {\em top left}). Radio bursts associated with flares, shocks and coronal mass ejections (CMEs) are produced in the solar atmosphere by energetic electrons either through various plasma mechanisms (narrowband emissions observed in particular below 1 GHz) or by gyrosynchrotron process (large broadband continuum emission observed predominantly above 1\,GHz). For plasma mechanisms, the emitted frequency is related to the local plasma frequency ($f_{pe} = (1/2\pi)~(N_e e^2/\epsilon_0 m_e)^{1/2}$) or its double (harmonic) and therefore decreases with height in the solar corona. Measuring radio emissions at different frequencies (either with spectroscopy or through imaging) thus allows us to probe phenomena at different heights of the solar atmosphere as well as to follow the propagation of disturbances from e.g. the low corona to the interplanetary medium. In addition, some specific radio emissions (long duration radio continua called type IV) can be used as tracers of the development in the low corona of CMEs, well before they become detectable in the field of view of coronographs.

\smallskip

\noi The most frequent bursts produced in solar flares are Type III bursts associated to the propagation of relativistic (0.1-0.5c) electron beams in the corona (see e.g. Saint-Hilaire et al. 2013). Type II bursts are associated to shock waves travelling outward through the corona (Bougeret et al. 1998). Both beams and shock waves can propagate through the interplanetary medium to and beyond 1 AU (Fig.\,\ref{fig_bougeret}, {\em top left}). At frequencies in the 450-150\,MHz range, radio bursts have been routinely imaged with the NRH (Fig.\,\ref{fig_bougeret}, {\em bottom left}). With the LOFAR array, new observation modes combining (pseudo-)imaging and high-resolution dynamic spectra have been implemented (Morosan et al. 2014), allowing to localise type III sources in the high corona (in the 30-50\,MHz range). LOFAR observations also revealed circularly polarised drifting bursts which were attributed to CMI emission from the feet or magnetic loops (Morosan et al. 2015). At higher frequencies, some microwave spike bursts have been also attributed to the CMI on highly-magnetised magnetic loops. Fig.\,\ref{fig_solar} ({\em right}) displays an example of such circularly polarised spikes.

\smallskip

\noi The main interests of SKA for the study of the radio Sun will be its broad spectral coverage that will allow to observe a large portion of the solar radio spectrum with imaging capability in the frequency range from 14\,GHz to 50\,MHz. Its expected capability of providing simultaneously images and dynamic spectra will allow in particular to study in details the propagation and radiation of electron beams very close to the flare site (as demonstrated with the first observations of Type IIIdm bursts in the 1-2\,GHz range obtained with the recently upgraded Karl G. Jansky VLA; cf. Fig.\,\ref{fig_bougeret}, {\em right}) as well as from the acceleration region up to the higher corona (Fig.\,\ref{fig_bougeret}, {\em bottom left}).

\smallskip

\noi SKA will of course not produce routine solar observations, but regular and Target of Opportunity campaigns may bring significant progress in the understanding of e.g. acceleration, propagation and radiation of electron beams in the solar corona. These observations will complement those from solar dedicated instruments such as the MUSER radio-heliograph in China and EOVSA in the USA. Eventually, similar to the extrapolation of our knowledge of solar system planetsÕ magnetospheres to the study of exoplanets, our understanding of solar activity and bursts should form a basis for stellar radio bursts and activity studies (below).\\

\parbox{0.9\textwidth}{
\noi{References:}\\
\noi{\scriptsize
Bougeret, J.-L., \etal , 1998, GRL, 25, 2513;
Carley, E. et al., 2016, ApJ, 833, 87;
Chen, B. et al., 2013, ApJ, 703, L21;
Dulk, G.A., 1985, Ann. Rev. Astron. Astrophys., 23, 169;
Mercier, C., Chambe, G., 2012, A\&A, 540, A18;
Morosan, D.E., \etal , 2014, A\&A, 568, A67;
Morosan, D.E., \etal , 2015, A\&A, 580, A65;
Pick, M. \& Vilmer, N., 2008, A\&Arv, 16, 1;
Saint-Hilaire, P., \etal , 2013, ApJ, 762, 60
}}\\

\subsubsection{Stellar activity and environment}\label{science:stars}
\vspace{0cm}

\noi Stellar activity can be studied with SKA along with exoplanet search and study, with the same observational setups. SKA can provide a bridge between the study of solar and stellar bursts, as between that of solar system planets and exoplanets.

\paragraph{Stellar flares and analogues to Solar radio bursts}
\vspace{0cm}

\begin{figure}[!ht]
  \centering
  \includegraphics[width=0.55\linewidth]{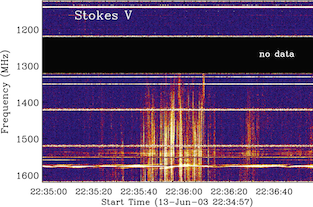}\hspace{0.5 cm}
  \includegraphics[width=0.39\linewidth]{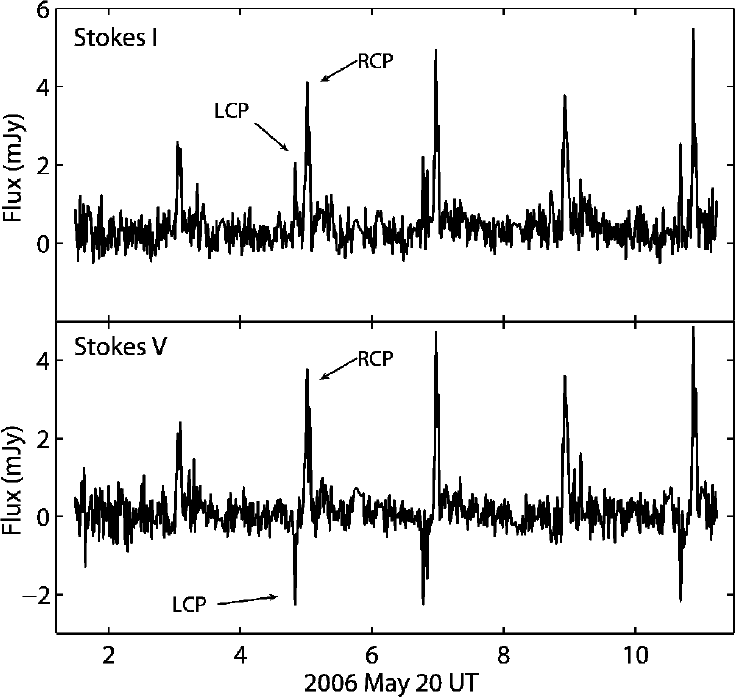}
  \caption{\label{fig_flares}
{\em Left:} dynamic spectrum of fast-drifting behavior seen on AD Leo with the Arecibo Observatory, around 1.5 GHz at 10 ms time resolution. Circularly polarised intensity is shown. From (Osten et al. 2006).
{\em Right:} light curves of the total intensity (Stokes I) and the circularly polarised (Stokes V) radio emission detected at 8.44 GHz from the M9 dwarf TVLM 513. RH (resp. LH) circular polarisation is represented by positive (resp. negative) values in the Stokes V light curve. Bursts of both 100\% RH circular emission (example highlighted as RCP) and 100\% LH circular emission (highlighted as LCP) are detected with a periodicity of 1.96 hr. From (Hallinan et al. 2007).}
\end{figure}

\noi Multi-wavelength observations (X, EUV, optical and radio) have shown that solar-like flares are a common feature of magnetically active stars. In particular, radio emissions during stellar flaring events have been reported continuously along the years (see reviews and references in Bastian 1990, Guedel 2002, White 2004 -- Fig.\,\ref{fig_flares}).

\smallskip

\noi Overall, stellar flares present a similar phenomenology with solar flares: light curves generally show an impulsive phase with an abrupt rise followed by a longer decay phase, as for solar flares, and the ratios of radio to UV emission are similar (Osten et al. 2004). However, radio emissions are not always present during stellar flares: Tingay et al. (2016) showed that the Kepler K2 field radio survey did not find any radio emission where optical flares had been detected. When radio emissions are detected alongside other wavelengths, the lack of amplitude correlation (e.g. between optical and radio, see Spangler 1976) poses the question of how and where energy is partitioned during those flares. Furthermore, compared with the Sun, some star types such as UV Ceti are radio-overluminous with regard to their associated X-ray emission (Guedel 1996). Hence, even though similar generic mechanisms may be at play at the Sun and on other stars during flares, the role of atmospheric compositions and magnetic fields need to be better understood. Sky surveys performed with large fields of view with SKA will help assess the link between emissions at different wavelengths and probe into the underlying mechanisms of flares.

\smallskip

\noi Different types of radio emissions from stars have been reported during stellar flares in the MHz-GHz range (see review by Guedel 2002). On the one hand, gyrosynchrotron emissions, related to relativistic electrons (Dulk 1985), provide a diagnostic of non-thermal electron populations. On the other hand, coherent emissions inform on the plasma processes at play. Study of the polarisation can help distinguish between the two possible coherent mechanisms, CMI or plasma emission (see e.g. the observations made at 154 MHz with the MWA by Lynch et al. 2017). Dynamic spectral analyses offered with the SKA capabilities will also be of interest in understanding the location of particle acceleration and drift of electron beams in the stellar coronae. SKA will provide a better understanding of the particle acceleration history, non-thermal energy release, coherent emission, and influence of the stellar atmosphere, as answering those questions require both broadband coverage and high sensitivity.

\smallskip

\noi Assuming that stellar flares are powered by mechanisms similar to those found at the Sun, we may wonder whether the equivalent of Coronal Mass Ejections (CMEs) exist. It is believed that these large amounts of ejected stellar material can be an important mass loss for the star, with possible impact on planetary disks and planetary habitability (Khodhachenko et al. 2013, Osten \& Wolk 2015, Drake et al. 2016). While CMEs observed at the Sun are associated with radio sources (e.g. Bastian 2001, 2007), their stellar counterparts have remained elusive. Ongoing programs (e.g. Villadsen et al. 2016) are promising, as well as future dynamic spectra obtained in the low frequencies ($\sim$100 MHz, for these frequencies will reveal outward-propagating CMEs) with SKA.

\smallskip

\noi Overall, by providing new observations in the field of stellar flares, SKA will help to better understand transients detected at stars (including stars equivalent to a young Sun, constraining scenarios of solar system evolution), and to assess planetary habitability, a field that has gained momentum with the exponential increase of detected exoplanets (\href{http://www.exoplanet.eu}{\color{blue} \myul[blue] {www.exoplanet.eu}}). To complement this section, the reader is referred to the review by Osten (2008), where interesting questions regarding radio observations of stellar transients with wide-field radio surveys such as SKA's are discussed in more details.\\

\parbox{0.9\textwidth}{
\noi{References:}\\
\noi{\scriptsize
Bastian, T.S., 1990, Solar Phys., 130, 265;
Bastian, T.S., \etal , 2001, ApJ, 558, 65;
Bastian, T.S., 2007, ApJ, 665, 805;
Drake, J.J., \etal , 2016, Proc. IAU Symp. 320;
Dulk, G.A., 1985, Ann. Rev. Astron. Astrophys., 23, 169;
Guedel, M., \etal , 1996, ApJ, 471, 1002;
Guedel, M., 2002, Ann. Rev. Astron. Astrophys., 40, 217;
Hallinan, G., \etal , 2007, ApJ, 663, L25;
Khodachenko M., \etal , 2013, Proc. IAU Symp. 300;
Lynch, C.R., \etal , ApJ Lett., 836, L30;
Osten R., \etal , 2005, ApJ, 621, 398;
Osten R., \etal , 2006, ApJ, 637, 1016;
Osten, R., Wolk, S., 2015, ApJ, 809, 79;
Spangler, S.R. \& Moffett, T.J., 1976, ApJ, 203, 497;
Tingay, S.J., \etal , 2016, Astron. J., 152, 82;
Villadsen J., \etal , 2016, Proc. IAU Symp. 320;
White S.M., 2004, New Astron. Rev., 48, 1319
}}\\

\paragraph{Cool dwarfs}
\vspace{0cm}

\begin{figure}[!ht]
  \centering
  \includegraphics[width=0.49\linewidth]{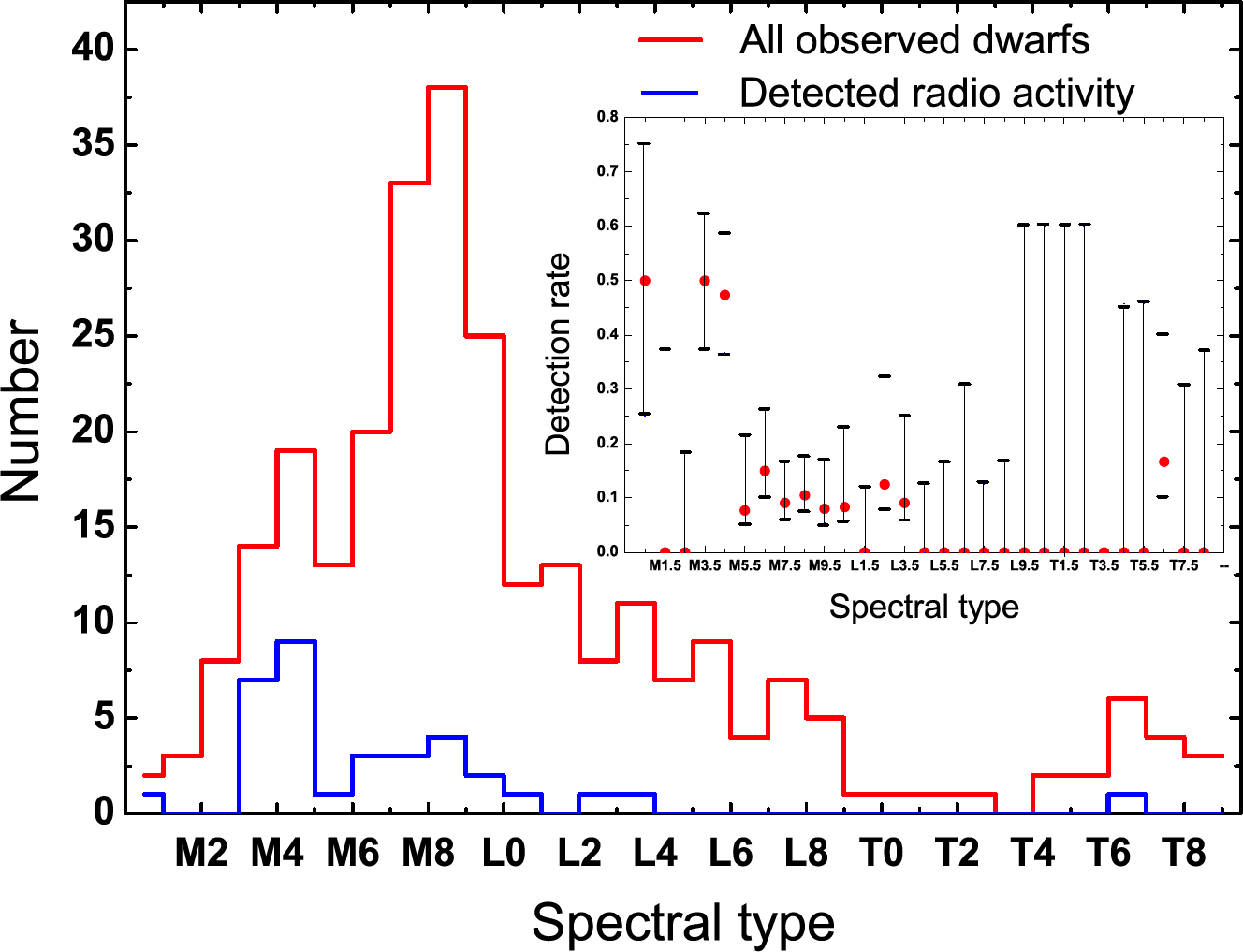}\hspace{0.2 cm}
  \includegraphics[width=0.49\linewidth]{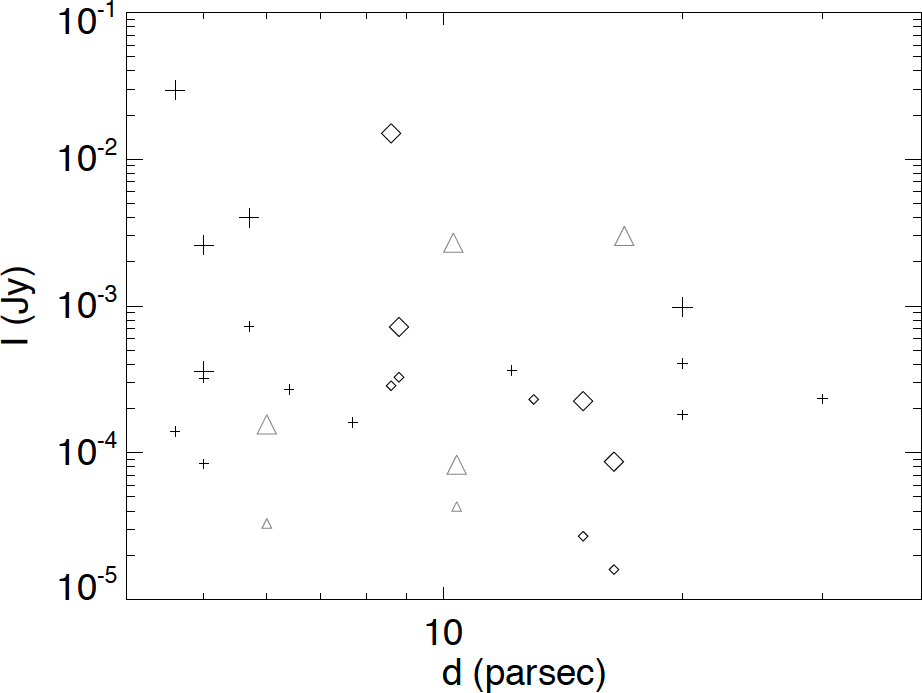}
  \caption{\label{fig_brown_dwarfs}
{\em Left:} number distribution of observed dwarfs (red line) and dwarfs with detected radio signal (blue line) with respect to their spectral type. Inset shows the detection rate (and its uncertainty). In the spectral type range M0-T8, 273 dwarfs have been observed, 34 of which with detected radio signal. The zero detection rate around spectral type M2 is most likely due to the limited number of observations. (Extracted from Antonova et al. 2013).
{\em Right:} intensity versus distance of the 20 brown dwarfs with detected radio emission (generally with a high circular polarisation fraction) at 4.2-8.5 GHz. Spectral types are represented by symbols (plus = M diamond = L, triangle = T). Large symbols represent the bursts, small ones the quiescent emissions.}
\end{figure}

\noi A limited number of bodies more massive than planets have been recently found to host similar non-thermal radio emissions, intense enough to be detected from the ground in the GHz range with peak flux densities up to a few mJy (Fig.\,\ref{fig_flares}, {\em right}). These include brown dwarfs and very low mass stars of spectral type later than M7, often referred to altogether as ultracool dwarfs. Fig.\,\ref{fig_brown_dwarfs}, {\em left} shows the rate and number of detections as a function of the spectral type, that provides a constraint on their internal structure with their capacity to generate and maintain a large scale magnetic field. Fig.\,\ref{fig_brown_dwarfs}, {\em right} displays the distribution of radio intensities versus the distance.

\smallskip

\noi The detected radio sources generally display two types of radiations: quiescent emissions are persistent, broadband, faint, and display a low degree of circular polarisation; bursts are variable at timescales of minutes, broadband, intense, with a high degree of circular polarisation (Berger et al. 2002, 2006, 2009; Burgasser \& Putman 2005; Hallinan et al. 2006, 2015; Phan-Bao et al. 2007; Antonova et al. 2008; MacLean et al. 2011, 2012; Route \& Wolszczan 2012 ; Lynch et al. 2015). Quiescent faint emission has been early thought to be produced by gyrosynchrotron emission from mildly relativistic electrons.

\smallskip

\noi Conversely, the electron-cyclotron maser instability (CMI) has been identified as the most plausible emission mechanism for most intense bursty emissions. This instability enables to amplify elliptically polarised radio waves from non-maxwellian weakly relativistic electrons close to the relativistic electron cyclotron frequency with a typical 1\% energy conversion efficiency. The CMI was originally developed to explain the terrestrial Auroral Kilometric Radiation (Wu \& Lee 1979, Treumann 2006) and it is now thought to be more generally responsible for all planetary auroral radio emissions detected in the solar system, ranging from decametric (40 MHz at Jupiter) to  kilometric ($\sim$100 kHz at Earth, Jupiter, Saturn, Uranus \& Neptune) wavelengths (Zarka 1998). The validity of the CMI has then been checked with in situ measurements at Earth (Roux et al. 1993; Ergun et al. 2000), at Saturn (Lamy et al. 2010; Mutel et al. 2010) and at Jupiter (Louarn et al. 2017). Overall, CMI-driven planetary auroral radio emissions provide a powerful diagnostic of magnetospheric processes: dynamics, coupling with the solar wind and/or moons, energy budgets, current systems, particle acceleration processes. For other CMI radio sources (exoplanets, stars), comparison to emission models (as in Hess \& Zarka, 2011, or Leto et al. 2016) gives access to key parameters such as the magnetic field strength at the source, and various parameters. 

\smallskip

\noi Applied to brown dwarfs, CMI emission implies magnetic field amplitudes of several kGauss at the source. Monitoring polarisation switches of the CMI bursts of ultracool dwarfs may reveal the existence and properties of magnetic cycles of very low-mass stars and brown dwarfs (Route 2016). Very little is currently known about the dynamo magnetic fields and cyclic behavior of these fully-convective objects. Monitoring them simultaneously with SKA and SPIROU (Infrared Spectropolarimeter at CFHT) should unveil their detailed magnetic properties, guide theory towards more predictive dynamo models and bridge the observational gap between magnetic fields of M dwarfs (Donati et al. 2008; Morin et al. 2008, 2010) and those of giant and terrestrial planets (Christensen et al. 2009; Morin et al. 2011).

\smallskip

\noi Observing ultracool dwarfs with SKA1-LOW will enable us to detect a wide variety of sources in the range 50-350 MHz yet unexplored with high sensitivity (hence probing source magnetic fields from 18 to 125 G). Such observations will provide a unique frame for comparative ``dwarfology" that will rely on our current understanding of solar system planetary auroral emissions. Interestingly, SKA observations of ultracool dwarfs may additionally provide a path toward the detection of (likely fainter) exoplanets, or at least hints to understand their non-detection.\\

\parbox{0.9\textwidth}{
\noi{References:}\\
\noi{\scriptsize
Antonova, A., \etal , 2013, A\&A, 549, A131;
Antonova, A., \etal , 2008, A\&A, 487, 317;
Berger, E. \etal , 2002, ApJ, 572, 503;
Berger, E., 2006, ApJ, 648, 629;
Berger, E., \etal , 2009, ApJ, 695, 310;
Burgasser, A. J. \& Putman, M. E., 2005, ApJ, 626, 486;
Christensen, U., \etal , 2009, Nature 457, 167;
Donati, J.-F., \etal , 2008, MNRAS 390, 545;
Ergun, R.E., \etal , 2000, ApJ, 538, 456;
Hallinan, G., \etal , 2006, ApJ, 653, 690;
Hallinan, G., \etal , 2015, Nature, 523, 568;
Hess, S.L.G. \& Zarka, P., 2011, A\&A 531, A29;
Lamy, L., \etal , 2010, GRL, 37, L12104;
Leto, P., \etal , 2016, MNRAS, 459,1159;
Louarn, P., \etal , 2017, GRL, in press;
Lynch, C., \etal , 2015, ApJ, 802,106;
McLean, M., \etal , 2011, ApJ, 741, 27;
McLean, M., \etal , 2012, ApJ, 746, 23;
Morin, J., \etal , 2008, MNRAS 390, 567;
Morin, J., \etal , 2010, MNRAS 407, 2269;
Morin, J., \etal , 2011, MNRAS 418, L133;
Mutel, R.L., \etal , 2010, GRL, 37, L19105;
Phan-Bao, N., \etal , 2007, ApJ, 658,  553;
Route, M. \& Wolszczan, A., 2012, ApJ, 747, L22;
Route, M., 2016, ApJ 830, L27;
Roux, A., \etal , 1993, JGR, 98, A7, 11657;
Treumann, R., 2006, Astron. Astrophys. Rev, 13, 229;
Trigilio, C., \etal , 2011, ApJ, 739, L10;
Wu, C.S. \& Lee, L.C., 1979, ApJ, 230, 621;
Zarka, P., 1998, JGR, 109, E9, 20159
}}\\

\paragraph{Circumstellar HI}

The \hi~line at 21 cm allows us to probe the kinematics and the physical 
properties of circumstellar shells over large sizes, and especially 
the properties of their external regions where stellar outflows interact 
with the ambient medium (Villaver et al. 2002; Hoai et al. 2015). 
Previous observations, in particular with the VLA and the NRT, 
have shown that the line widths are narrow (FWHM~$\leq$~5\,km/s), 
and that in some cases the centroid velocities are drifting with position  
on the sky along directions opposite to the stellar motions  
(Matthews et al. 2008). 

Thanks to the high spectral resolution available 
in the radio range, the \hi~study of galactic circumstellar shells with
the SKA can provide unique information on the mass loss history of stars, 
on the injection of stellar matter in the ISM, and thus on the cycle 
of matter in our Galaxy. We stress here the interest to map with a high spectral resolution 
(corresponding to $\sim$~0.1\,km/s) the \hi~line emission in the
directions of a variety of sources representative of the population 
of galactic mass losing stars. A spatial resolution of $\sim$~1\,\arcsec 
would allow us to resolve the kinematical structures which, in 3D
numerical simulations, are predicted to develop at the interfaces with 
the ISM, in nearby sources (typically 200-500\,pc, e.g. Mohammed et al. 2012).  
This project could be extended to the study of the cycle of matter 
in nearby galaxies, such as the Magellanic Clouds. 

The requirements are a good sensitivity in a large field of view 
($\sim$~one~degree), a good response to large scale structures 
(up to $\sim$~1/2~degree), as well as an ability to filter out 
strong background \hi~components originating in the ISM located 
on the same lines of sight. A radio interferometer, made of 12-m antennas, 
with a compact core and outer components providing long baselines, 
up to 100\,km, and a good coverage of the uv plane, 
such as the one foreseen for SKA1 in South Africa, 
seems very well adapted. However, the sensitivity requirements
are demanding, and the full potential of this programme might only be
revealed by the complete SKA (i.e. with a collecting area of 1\,km$^2$). \\

\parbox{0.9\textwidth}{
\noi{References:}\\
\noi{\scriptsize
Hoai, D.T., \etal, 2015, MNRAS, 449, 2386;
Matthews, L.D., \etal, 2008, ApJ, 684, 603;
Mohammed, S., \etal, 2012, A\&A, 541, 1;
Villaver, E., \etal, 2002, ApJ, 571, 880}}\\

\subsubsection{Planets, Stars and Commensal SETI observations} \label{science:seti}
\vspace{0cm}

\noi As discussed above, CMI emission is the most intense radio emission from planets, and has been detected from brown dwarfs and low-mass stars. Its detection is expected to bring unique information about exoplanets and star-planet plasma interactions (involving FGKM stars as well as white dwarfs), opening the news fields of comparative exo-magnetospheric physics and extrasolar space weather, and about auroral-like radio emissions from brown dwarfs and low-mass stars. For exoplanets, detection will give access to unique physical information (described above). For both exoplanets and low-mass stars, detection will permit to elucidate the type or ``engine" in operation : star-planet interactions (stellar wind control, triggering by CME, magnetic reconnection, unipolar induction), auroral-like currents in stellar environments. The expected sensitivity will also permit to detect radio emissions from stellar coronal mass ejections (CME, analog to Solar type II bursts), that may in turn trigger exoplanetary radio emissions, and from magnetic binaries with star-planet-like interaction.

\smallskip

\noi It is thus essential to promote an ambitious survey with SKA allowing us to search and study exoplanets, cool dwarfs and stars. It will allow us to study the evolution of the objects' properties (rotation periods, large-scale magnetic field amplitudes and topologies, atmospheres--ionospheres, radio fluxes), bringing new constraints on dynamo theories and radio emission scaling laws, and possibly enabling the discovery of low-mass (habitable?) planets around M dwarfs (Zarka et al. 2015). We have proposed such a survey, focused on low-frequencies and low galactic latitudes, in the frame of the SKA working group ``Cradle of Life". This survey should benefit from strong synergies with observations at other wavelengths.

\smallskip

\noi We named this proposal : Key Science Project ``Stars, Planets and Civilisations". Indeed, provided that beamformed raw voltages can be exported in parallel to imaging or beamformed spectral data, the proposed targets of this program are ideally adapted to a SETI commensal search, probing leakage radiation from Kardashev 0-1 civilisations of nearby planets and stars (Siemion et al. 2015). After interference elimination, sensitivities to $\sim 10^{10-11}$ W Equivalent Isotropic Radiated Power can be reached, nearing the highest power leakage transmitters from Earth. A particularly interesting idea is the possible existence of beamed unintentional signals that would be used for radio communication within a transiting multi-planet system (i.e. seen along the ecliptic plane), between planets and/or space probes, the detection of which would require transmitted powers of $\sim 10^{4-5}$ W only.

\smallskip

\noi The search and study of Planets, Sun, Stars, and Civilisations with SKA is a broad new field to explore (actually including many connected sub-fields). The theoretical frame is much advanced, and we can be very optimistic about the prospects and outcomes.\\

\parbox{0.9\textwidth}{
\noi{References:}\\
\noi{\scriptsize
Siemion, A.P.V., \etal , 2015, AASKA14, 116;
Zarka, P., \etal , 2015, AASKA14, 120
}}

\newpage

\subsection{Transient Universe}

\noindent {\normalsize Contributors of this section in alphabetic order: }

\smallskip

\noi {\sffamily \scriptsize
{\sffamily \bf {\bf M.~G.~Bernardini}} [\lupm],
{\bf C. Boisson} [\luth],
{\bf M. Cerruti} [\lpnhe],
{\bf E.~Chassande-Mottin} [\apc],
{\bf S.~Chaty} [\irfu;\aim;\iuf],
{\bf I.~Cognard} [\lpcee],
{\bf S.~Corbel} [\irfu;\aim;\usn],
{\bf M.~Coriat} [\irap],
{\bf R.~Dallier} [\subatech],
{\bf A.~Djannati-Ata\"i} [\apc],
{\bf L.~Guillemot} [\lpcee;\usn],
{\bf A.~Marcowith} [\lupm],
{\bf L.~Martin} [\subatech;\usn],
{\bf F.~Mottez} [\luth],
{\bf J.~P\'etri} [\stras],
{\bf B.~Revenu} [\subatech],
{\bf J.~Rodriguez} [\irfu;\aim],
{\bf G.~Theureau} [\lpcee;\usn;\luth],
{\bf S.~D.~Vergani} [\gepi],
{\bf N.~Webb} [\irap]
}

\subsubsection{Slow transients}

\paragraph{Accreting transients}
\vspace{0cm}

\noi The following text is adapted from the chapter \textit{Incoherent transient radio emission from stellar-mass compact objects in the SKA era} from the Proceedings of \textit{Advancing Astrophysics with the Square Kilometre Array} (Corbel et al. 2015).

\smallskip
\noi {\bf Introduction and context}

\smallskip
\noi Stellar-mass compact objects provide important laboratories for studying the fundamental coupling between the accretion process and the launching of energetic outflows, which often take the form of highly-collimated jets (Hughes 1991). A theoretical picture has been developed where the jets are composed of an electron/positron or electron/proton plasma (e.g. Bonometto \& Rees 1971), which is magnetically collimated as it flows away from the compact object. These jets may be powered either by tapping the energy of a rotating black hole (Blandford \& Znajek 1977), or by extracting energy from the accretion flow (Blandford \& Payne 1982). However, many basic aspects of jet physics are uncertain, including their composition and structure, as well as the mechanisms that power and collimate them. Relativistic jets are relevant in almost all fields of astrophysics, and in some cases may be the dominant output channel for the accretion power from black holes (Fender et al. 2003). They provide an important source of feedback to the surrounding environment, being able to either trigger or suppress star formation, accelerate cosmic rays, and seed the surrounding medium with magnetic fields. They have even been suggested to play a role in the reionisation of the Universe (e.g. Mirabel et al. 2011).
Despite their relative proximity, the lower masses of stellar-mass compact objects imply that they are observed at lower angular resolution (in terms of gravitational radii) than nearby AGN, yet they evolve through their duty cycles on human timescales, typically undergoing entire outbursts over periods of days to months. Thus, they provide unique insights into the coupling between accretion and outflow. Furthermore, comparative studies of the different classes of compact object can provide important insights into the necessary and sufficient ingredients for the jet launching processes.
The non-thermal radio emission from stellar-mass compact objects typically arises via synchrotron emission from relativistic particles spiraling around the magnetic field lines of the jets. While significant progress has been made over the past few decades in understanding the nature of the jets and their coupling to the accretion flow (see Fender 2006, for a review), investigations have been hampered by the limited sensitivity of past and current facilities (see also Fender et al. 2015). We detail in this chapter how the sensitivity improvement brought about by the SKA will allow us to study black hole outbursts throughout the Galaxy and out into the Local Group, determine the role of jets in the low-luminosity quiescent state, and extend our existing studies of black holes to the analogous yet fainter neutron star and white dwarf systems.

\smallskip
\noi {\bf Stellar mass black holes}

\smallskip
\noi From an observational standpoint, black hole transients spend most of their time in a quiescent state, at very low mass accretion rates. They occasionally undergo outbursts that last from a few months to $\sim$ a year, during which the flux rises by several orders of magnitude across the whole electromagnetic spectrum (McClintock \& Remillard 2006). These outbursts are associated with global changes in three main components: the jets, the accretion disk and the corona. The luminous outburst phase, with a luminosity $>10-30\%$ of the Eddington luminosity (the "soft state"), is dominated by thermal emission from the accretion disk. During the rise and decay phases of the outburst (the "hard" state), the bolometric luminosity of the source is dominated by non-thermal emission (synchrotron or inverse Compton emission from either the jets or the corona) extending up to the hard X-ray band. Following recent results with high-resolution X-ray spectroscopy, accretion disk winds are now recognised as ubiquitous in black hole X-ray binaries, and may carry away a significant fraction of the inflowing energy (Miller et al. 2006; Neilsen \& Lee 2009; Ponti et al. 2012; Diaz Trigo et al. 2013).
Observations of typical Galactic black holes have indicated two forms of jets associated with these primary accretion states; the slowly-varying, partially self-absorbed compact jets (with radio emission usually $\leqslant 30-50$ mJy for distances of a few kpc, and a flat or slightly inverted radio spectrum) observed in the hard state (Corbel et al. 2000; Fender et al. 2000; Dhawan et al. 2000; Stirling et al. 2001), and the bright ($0.1-10$ Jy, with an optically thin radio spectrum), strongly variable transient jets (occasionally showing apparent superluminal motion) detected during the transition from the hard to the soft state (Corbel et al. 2004; Fender et al. 2004b). The core radio emission is then strongly quenched during the soft state (Fender et al. 1999; Coriat et al. 2011). Relic radio emission can also be detected in some cases when the jets interact with the ambient medium, either as large-scale lobes (Gallo et al. 2005), or as faint, transient hot spots (Corbel et al. 2002), depending on the duty cycle of the central black hole.

\begin{figure}[!ht]
  \centering
  \includegraphics[scale=0.65]{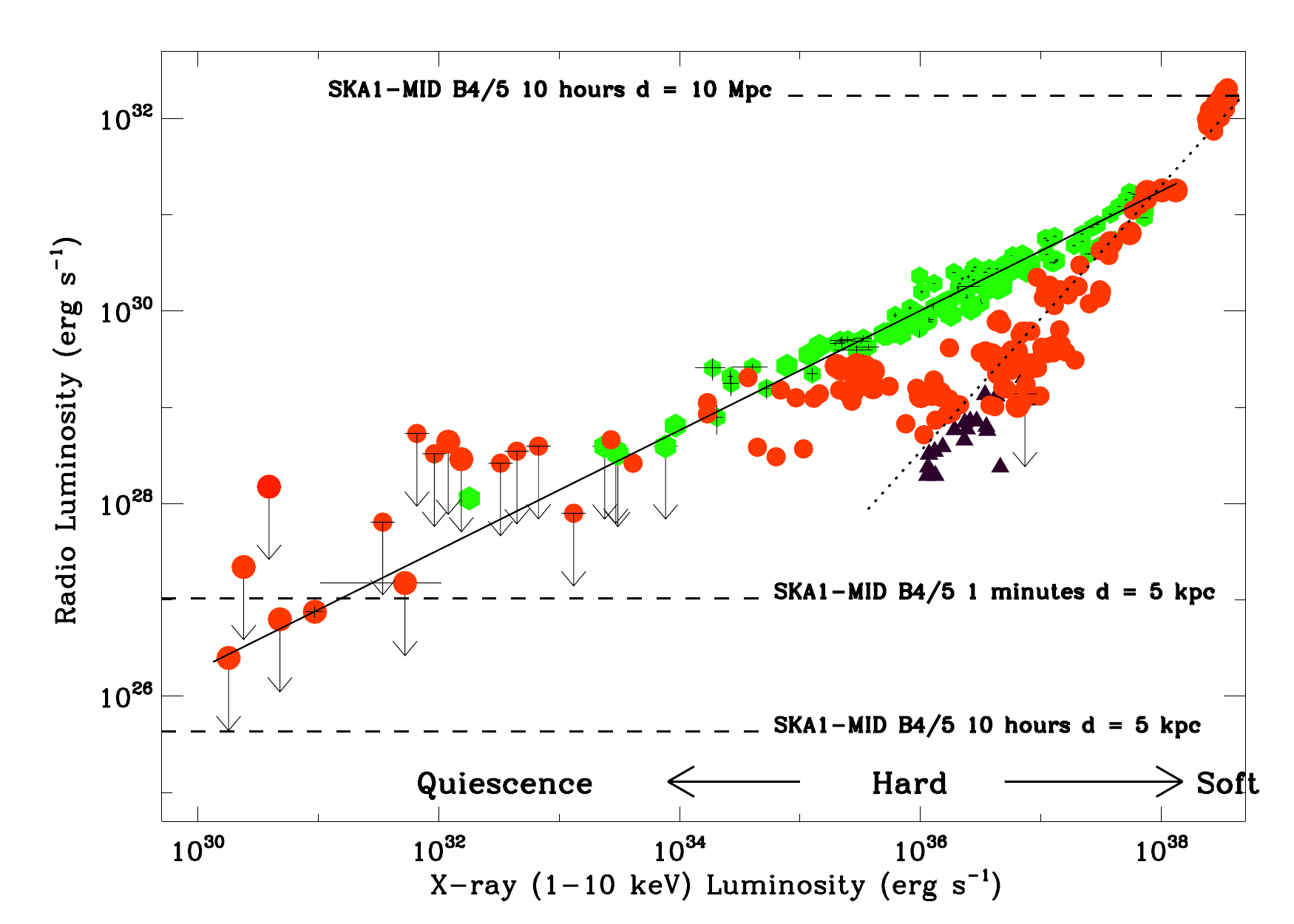}
  \caption{\label{fig:1A} Radio and X-ray (1-10 keV) luminosities for Galactic accreting binary black holes in the hard and quiescent states (see arrows at bottom for approximate luminosity levels). It illustrates  the standard correlation (defined by sources such as GX339--4 or V404 Cyg with index $\sim$ 0.6; green points) and the new correlation for the so-called ``outliers'' (defined by e.g.  H~1743$-$322 or Swift J1753.5$-$0127 with index $\sim$ 1.4; a significant fraction of the red points). Black triangles represent neutron star binaries. The solid line illustrates the fit to the whole 1997--2012 sample of GX339--4 (Corbel et al. 2013) with an extrapolation to the quiescent state. The dotted line corresponds to the fit to the data for H~1743$-$322, one of the representatives for the outliers (Coriat et al. 2011). Upper limits are plotted at the 3$\sigma$ confidence level. The horizontal dashed lines represent some sensitivity levels for SKA1-MID as discussed in the text. Figure adapted from Corbel et al. (2015).}
\end{figure}

\smallskip
\noi There is a strong correlation between the radio and X-ray emission in the hard and quiescent states (Corbel et al. 2003, 2013b; Gallo et al. 2003, 2012), indicating a possible coupling between the launch of the jets and the dynamics of the flow close to the accreting black hole (Fig.\,\ref{fig:1A}). It can be used to determine the expected source behaviour at the lowest radio luminosities, which remain relatively poorly explored owing to the limited sensitivity of current instruments (Miller-Jones et al. 2011). It is, for example, not clear whether quiescent black holes do host jets as in the brighter states. Current observations seem to indicate weak jet activity even in quiescence (Gallo et al. 2006, 2014), but it is not known what fraction of the liberated accretion power is carried away in the jets rather than being advected across the black hole event horizon (Fender et al. 2003). Furthermore, understanding the dichotomy in the radio/X-ray correlation (Coriat et al. 2011; Gallo et al. 2012) could provide a new way to explore the possibility of different couplings between accretion and ejection in black hole transients. With a sensitivity as good as 30 $\mu$Jy s$^{1/2}$ in band 4/5, SKA1-MID is the ideal instrument to monitor all detected black hole outbursts (up to a few tens per year) all the way down to quiescence. Furthermore, a long observation (10 hrs) with SKA1-MID could detect all quiescent black hole binaries up to a distance of 5 kpc. These transient black holes could either be discovered by targeted monitoring or a blind radio survey with SKA (Fender et al. 2015), or by any active multi-wavelength monitoring facility.
The brightest phases of black hole outbursts (the hard to soft state transitions) can easily be studied via snapshot observations, even for the most distant systems in our Galaxy, meaning that SKA (even in phase 1) will probe a significant fraction of the outburst activity for almost all black holes in our Galaxy, including the very faint, sub-Eddington transients, which have rarely been observed at radio frequencies. With a resolution on the order of a few tenths of an arcsecond,  the bright radio emission associated with the transient jets (Mirabel \& Rodriguez 1994) could also be resolved from a week after the onset of the radio flare, as typical proper motions are on the order of 15-20 mas day$^{-1}$. Good polarisation measurements (linear and circular) from SKA1-MID, associated with good signal purity, will provide key probes of the composition and geometry of the jets, and the structure of their magnetic fields.
The high sensitivity and ability to dump the visibilities on rapid ($\lesssim$1 s) timescales would allow us to probe any short-timescale variability of the jets, should it be present (shortest timescales of $\sim$10 minutes up to now, e.g. Corbel et al. (2000); Middleton et al. (2013)). Casella et al. (2010) detected rapid (sub-second) infrared variability from the jets of GX 339--4, which was correlated with the observed X-ray variability. Observing these variations at multiple frequencies as they propagate downstream in the jets would allow us to directly measure the speeds and sizes of the compact jets, providing the best observational constraints on their Lorentz factors.

\smallskip
\noi {\bf Neutron stars}

\smallskip
\noi 
Comparisons between accreting neutron stars and black holes are ideal for determining which effects seen from accreting black holes are fundamentally related to the presence of an event horizon, instead of being generic to accretion onto objects with deep gravitational potential wells. Largely speaking, the phenomenology of accretion onto neutron stars is similar to that onto black holes (e.g. Psaltis 2006). Many of the known differences can be explained in a straightforward manner by the presence of a solid surface for neutron stars and the lack of one for black holes.
Studies of accreting neutron stars will be bolstered by the enhanced sensitivity of the SKA. Systematic studies of the radio luminosities of accreting neutron stars as a function of their X-ray luminosities are far sparser than those for black holes. This is in part because the characteristic timescales on which accretion disks change scales inversely with the accretor mass, so that "typical" outbursts of accreting neutron stars are shorter than those for accreting black holes; and partly because both the peak X-ray luminosities of neutron star transients and the ratio of radio to X-ray flux for neutron stars are both lower than for black holes (Fender \& Kuulkers 2001; Wu et al. 2010; Migliari \& Fender 2006). As a result of these factors, most soft X-ray transients with neutron star accretors peak at flux densities of about 1 mJy or less (for typical distances of several kpc). Thus, it has traditionally been difficult to span more than a factor of 10 in radio luminosity for neutron stars. Furthermore, since the outbursts progress quickly, scheduling a large number of epochs places strong pressure on the ToO scheduling of existing arrays. Finally, since many of these sources are located in the Galactic Bulge, or elsewhere in the Southern Hemisphere, often only short observations are possible. Despite these problems, some progress has still been made on understanding the radio emission of neutron star X-ray binaries. A few things seem clear: the high magnetic field X-ray pulsars are not strong radio emitters (Migliari \& Fender 2006) and the radio and X-ray fluxes are correlated (see Fig.\,\ref{fig:1A}) for neutron star X-ray binaries that emit at less than about 10\% of the Eddington luminosity (Migliari et al. 2003; Tudose et al. 2009). The enhanced sensitivity of the SKA will allow more quantitative statements on these topics. For example, Migliari et al. (2003) find that $L_R \propto L_X^{1.4}$ for the neutron star 4U 1728-34 in hard spectral states, while Tudose et al. (2009) looked at Aql X-1 and found $L_R \propto L_X^{0.4}$, albeit while including data from a range of X-ray spectral states. Developing a sample of sources that has both many objects, and which spans a few orders of magnitude in radio luminosity would  bring the data quality into line with that from the black hole transients at the present time  (Fig.\,\ref{fig:1A}). 

\smallskip
\noi Additional studies of the thermal X-ray states of neutron stars in the radio may also hold important clues to understanding how jets are launched. In these systems, by analogy with the black holes, the accretion disks themselves are unlikely to supply much power to an accretion flow. On the other hand, the boundary layers, where the neutron star's accretion disk dissipates its excess rotational energy, should have the large scale height thought to be needed to power jets, and may interact with the magnetic field of the neutron star itself. Studies of a large sample of these soft states rather than merely the two that have been detected already (Migliari et al. 2004), may help us to understand jet production in this environment.

\smallskip
\noi A final class of systems that can probe the relationship between the accretion flow and the launching of jets are the recently-discovered class of transitional binary pulsars, which switch between accretion-powered and rotation-powered states on timescales of just weeks (e.g. Archibald et al. 2009; Papitto et al. 2013; Bassa et al. 2014). These systems have been shown to emit flat or inverted-spectrum radio emission in their accretion-powered states, which is consistent with partially-self absorbed synchrotron jets (Papitto et al. 2013; Bassa et al. 2014). Intensive radio monitoring of the transitions between these two states would provide unique insights into how the jets are formed and destroyed, and, together with simultaneous X-ray monitoring, would highlight the connection between the jets and the evolving accretion flow.

\smallskip
\noi {\bf White dwarfs}

\smallskip
\noi The occurrence of jets and fast collimated outflows is by no means restricted to accreting black holes or neutron stars. Through sensitive and timely observations in the last decade, numerous accreting white dwarfs (in cataclysmic variables, symbiotic stars and supersoft X-ray sources) have shown strong evidence for jets and jet-like shocked, collimated outflows, observed at radio frequencies and interpreted as synchrotron emission (e.g. Coppejans et al. 2015, 2016). However, the brightness temperatures of the radio emission from cataclysmic variables are small enough that thermal and/or cyclotron emission are still viable possibilities and given the multiple possible mechanisms, there may be some heterogeneity in emission processes. Better spectral measurements, and searches for circular polarisation, which should be possible with SKA, can help resolve these issues.
Perhaps the most striking (and encouraging) aspect is that these transient radio jets have all occurred in the prototypes of subclasses of accreting white dwarfs (Brocksopp et al. 2004; Rupen et al. 2008; Kording et al. 2008), suggesting they are more common in white dwarf accretors than previously assumed. 

\smallskip
\noi In close analogy to transient jets in X-ray binaries, the non-magnetic dwarf nova SS Cyg repeatedly exhibits radio outbursts associated with its disc outburst cycle (Kording et al. 2008).  Even though radio luminosities are low compared to X-ray binaries -- peak flux densities of $\sim$ 1 mJy at 1--10 GHz for a system at 114 pc (SS Cyg; Miller-Jones et al. 2013) -- these systems provide an important link in understanding how accretion is coupled to the outflow of matter across a range of compact accretors. With the expected sensitivities achieved by SKA1- MID we can extend the sample of dwarf novae observed at radio frequencies out to kiloparsec distances, where optical transient surveys such as CRTS and iPTF (and LSST in the SKA era) are finding thousands of new dwarf novae (e.g. Drake et al. 2014). The sheer numbers of systems, the accessible time scales of the disc instability cycle in cataclysmic variables (weeks to months), and the reasonably well-understood accretion discs around white dwarfs provide an excellent laboratory for the SKA to study accretion physics with targeted (and target-of-opportunity) observations.
Whilst thermal emission is the dominant component of radio emission in novae (Seaquist \& Bode 2008; O'Brien et al. 2015b), a significant number of novae exhibit non-thermal (synchrotron) emission associated with collimated bipolar and jet-like outflows (e.g. O'Brien et al. 2006; Woudt et al. 2009; O'Brien et al. 2015). As recurrent novae are prime candidates for the progenitors of type Ia supernovae, questions surrounding the nature and energetics of the outflow of material during a nova outburst are at the core of the debate as to whether a white dwarf grows in mass during successive nova cycles. A representative census of Galactic novae observed at the sensitivity of SKA1-MID (including band 4 or 5) is required to determine what fraction of novae show evidence for synchrotron emission and to fully understand the processes that lead to the formation of collimated jets following a thermonuclear runaway on the surface of the white dwarf. \\

\parbox{0.9\textwidth}{
\noi{References:}\\
\noi{\scriptsize 
Archibald, A., M., \etal, 2009, Science, 324, 1411; 
Bassa, C. G., \etal, 2014, MNRAS, 441, 1825;
Blandford, R. D. \& Znajek, R. L., 1977, MNRAS, 179, 433;
Blandford, R. D. \& Payne, D. G., 1982, MNRAS, 199, 883;
Bonometto, S. \& Rees, M. J., 1971, MNRAS, 152, 21;
Brocksopp, C., \etal, 2004, MNRAS, 347, 430;
Casella, P., \etal, 2010, MNRAS, 404, L21;
Coppejans, D. L., \etal, 2015, MNRAS, 451, 3801;
Coppejans, D. L., \etal, 2016, MNRAS, 463, 2229;
Corbel, S., \etal, 2000, A\&A, 359, 251;
Corbel, S., \etal, 2002, Science, 298, 196;
Corbel, S., \etal, 2003, A\&A, 400, 1007;
Corbel, S., \etal, 2004, ApJ, 617, 1272;
Corbel, S., \etal, 2013a, MNRAS, 431, L107;
Corbel, S., \etal, 2013b, MNRAS, 428, 2500;
Corbel, S., \etal, 2015, AASKA14, 62;
Coriat, S., \etal, 2011, MNRAS, 414, 677;
Dhawan, V., \etal, 2000, ApJ, 543, 373;
Diaz Trigo, M., \etal, 2013, Nature, 504, 260;
Drake, A. J., \etal, 2014, MNRAS, 441, 1186;
Fender, R. P., \etal, 1999, ApJ, 519, L165;
Fender, R. P., \etal, 2000, MNRAS, 312, 853;
Fender, R. P., \etal, 2003, MNRAS, 343, L99;
Fender, R. P., \etal, 2004b, MNRAS, 355, 1105;
Fender, R. P., 2006, Compact stellar X-ray sources (Cambridge University Press), 381;
Fender, R. P., \etal, 2015, AASKA14, 62;
Fender, R. P. \& Kuulkers, E. 2001, MNRAS, 324, 923;
Gallo, E., \etal, 2003, MNRAS, 344, 60;
Gallo, E., \etal, 2005, Nature, 436, 819;
Gallo, E., \etal, 2006, MNRAS, 370, 1351;
Gallo, E., \etal, 2012, MNRAS, 423, 590;
Gallo, E., \etal, 2014, MNRAS, 455, 290;
Hughes, P.A., 1991, Beams and jets in astrophysics (Cambridge University Press);
Kording, E., \etal, 2008, Science, 320, 1318;
McClintock, J. E. \& Remillard, R. A., 2006, Black hole binaries (Compact stellar X-ray sources), 157;
Middleton, M. J., \etal, 2013, Nature, 493, 187;
Migliari, S., \etal, 2003, MNRAS, 342, L67;
Migliari, S., \etal, 2004, MNRAS, 351, 186;
Migliari S. \& FenderR. P., 2006, MNRAS, 366, 79;
Miller, J. M., \etal, 2006, Nature, 441, 953;
Miller-Jones, J. C. A., \etal, 2011, ApJ, 739, L18;
Miller-Jones, J. C. A., \etal, 2013, Science, 340, 950;
Mirabel, I. F. \& Rodriguez, L. F., 1994, Nature, 371, 46;
Mirabel, I. F., \etal, 2011, A\&A, 528, 149;
Neilsen, J. \& Lee, J. C., 2009, Nature, 458, 481;
O'Brien, T., \etal, 2006, Nature, 442, 279;
O'Brien, T., \etal, 2015, AASKA14, 62;
Papitto, A., \etal, 2013, Nature, 501, 517;
Psaltis, D., 2006, Accreting neutron stars and black holes: a decade of discoveries (Cambridge University Press), 1;
Ponti, G., \etal, 2012, MNRAS, 422, L11;
Rupen, M. P., \etal, 2008, ApJ, 688, 559;
Seaquist, E. R., \& Bode, M. F., 2008, Cambridge University Press (Classical Novae), 141;
Stirling, A. M., \etal, 2001, MNRAS, 327, 1273;
Tudose, V., \etal, 2009, MNRAS, 400, 2111;
Woudt, P. A., \etal, 2009, ApJ, 706, 738;
Wu, Y. X., \etal, 2010, ApJ, 718, 620}}

\paragraph{Ultra Luminous X-ray sources}\label{sci:ULXS}
\vspace{0cm}

\noi Ultra Luminous X-ray sources (ULXs) are defined very simply as off-nuclear X-ray sources that have X-ray luminosities exceeding the Eddington luminosity for a stellar mass black hole of reasonable ($\sim$10 M$_\odot$) proportions (L$\sim$10$^{39}$ erg s$^{-1}$). For a long time they have been supposed to contain accreting black holes, which were thought to perhaps be of intermediate mass (10$^{2-5}$ M$_\odot$), which would explain the very high luminosities without invoking complicated scenarios such as super-Eddington accretion. However, thanks to the first results with NuSTAR (Bachetti et al. 2013; Walton et al. 2013), it seemed likely that the systems contained stellar mass black holes accreting above the Eddington limit. More recently, three ULXs have been shown to contain accreting neutron stars (Bachetti et al. 2014; F\"urst et al. 2016; Israel et al. 2017), thanks to the detection of X-ray pulsations of these systems, thus confirming that these objects do undergo super-Eddington accretion. A number of authors have also proposed that many ULXs may therefore contain neutron stars accreting at highly super-Eddington rates (e.g. Koliopanos et al. 2017; King, Lasota \& Klu\'zniak et al. 2017; Mushtukov et al. 2017). In addition, the extreme ULX ESO 243-49 HLX-1 (e.g. Farrell et al. 2009; Webb et al. 2010) has been shown to contain an intermediate mass black hole (IMBH) of $\sim$10$^{4}$ M$_\odot$ (e.g. Webb et al. 2012; Godet et al. 2012). This means that ULXs are a very heterogeneous collection of objects. Further observations are required to elucidate their nature. 

\smallskip

\begin{figure}[!ht]
  \centering
  \includegraphics[scale=0.285]{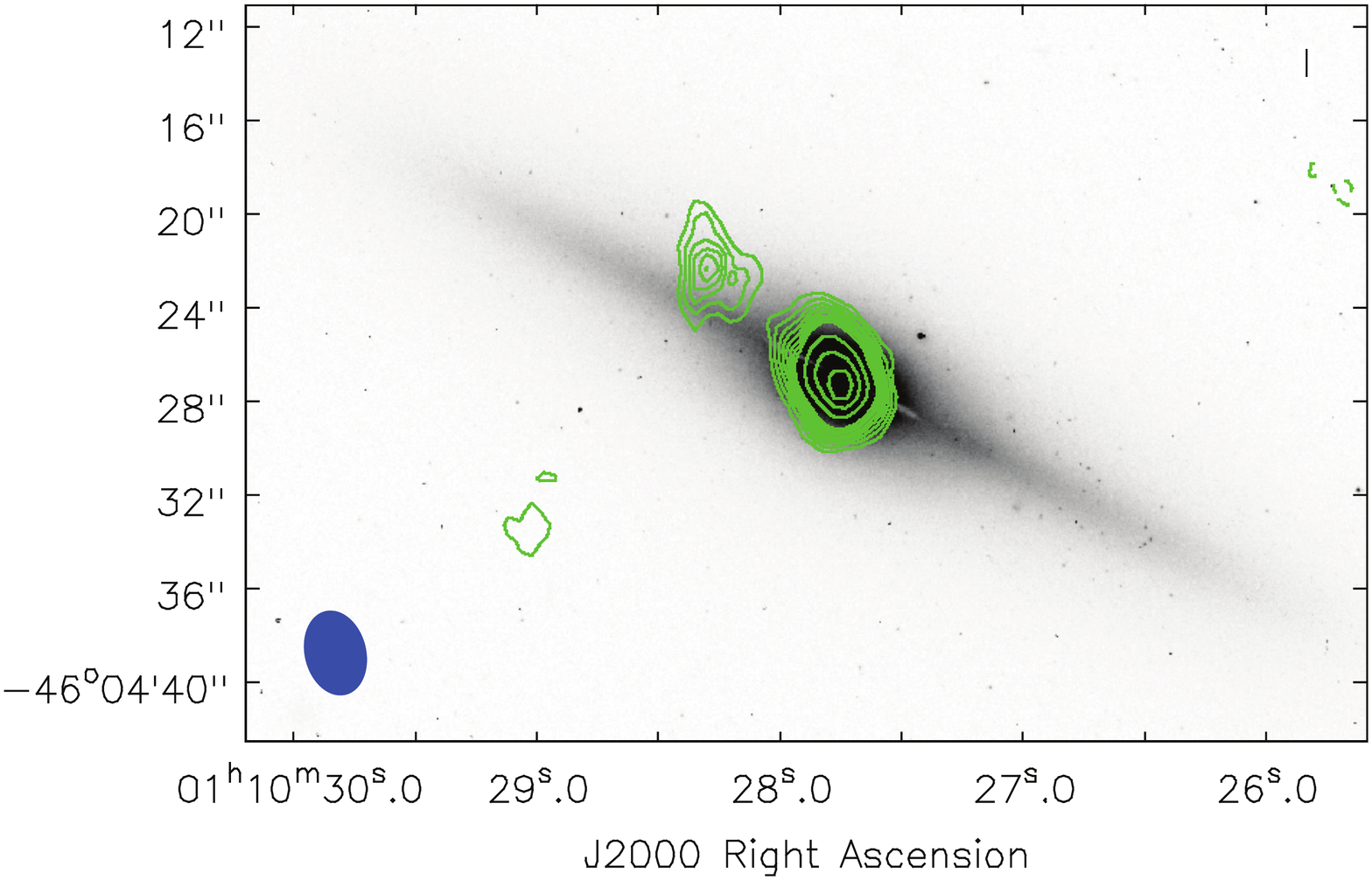}
  \includegraphics[scale=0.285]{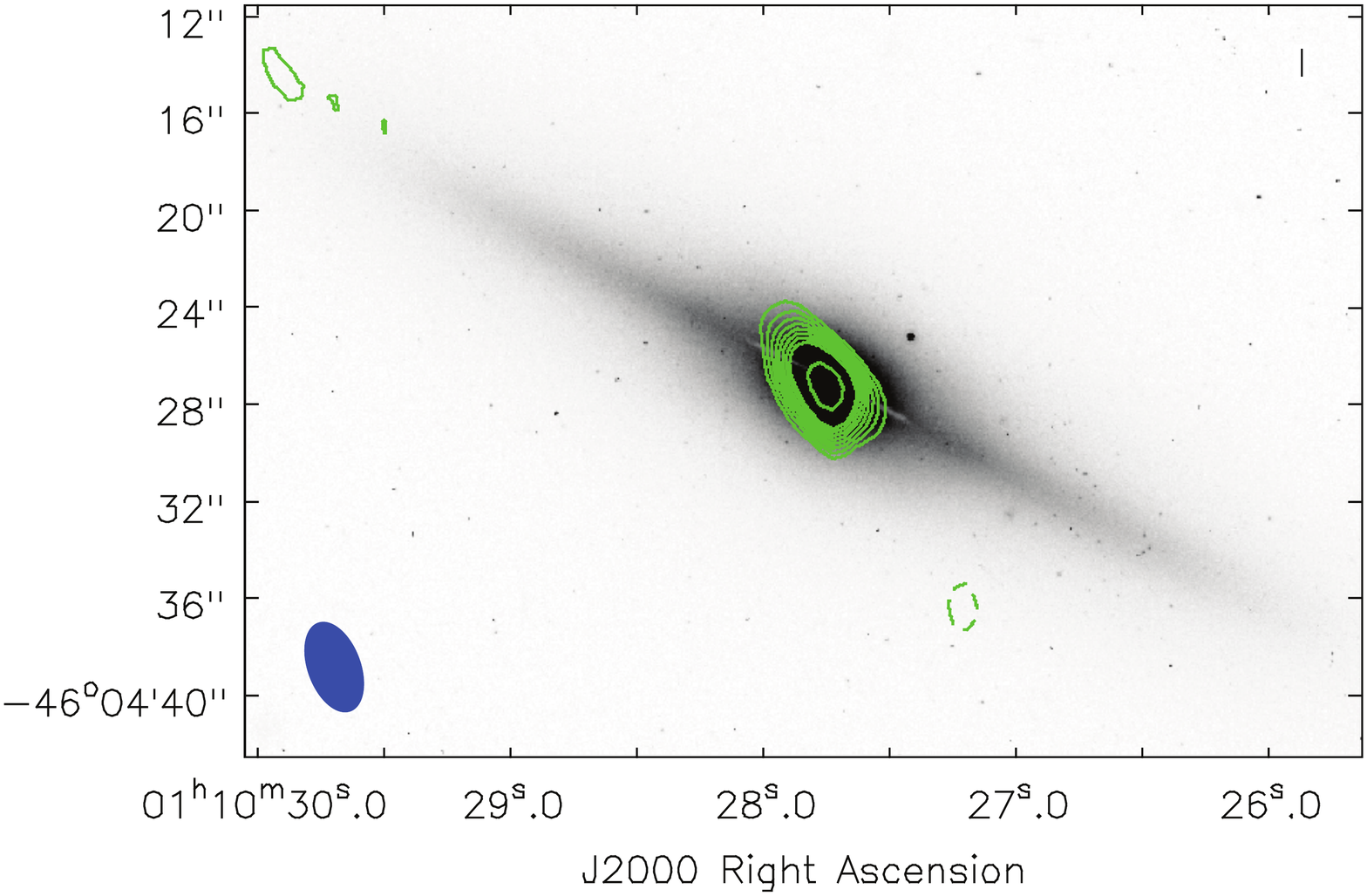}
  \caption{\label{fig:1B} ATCA  5 and 9 GHz combined radio observations (green contours) of HLX-1 and its host galaxy ESO 243-49. {\em Left panel}: 8$\sigma$ radio detection of HLX-1 at a flux density of 45 $\mu$Jy. {\em Right panel}: HLX-1 is undetected down an rms noise level of 7.0 $\mu$Jy/beam. Only ESO 243-49 is visible. See Webb \etal (2012) for details.}
\end{figure}

\noi However, whatever the nature of ULXs, they are extremely useful objects to understand the growth of supermassive black holes. For those ULXs accreting above the Eddington limit, studying them should help us uncover the physical processes behind these extreme accretion rates which are thought to play an essential role in the formation of supermassive black holes that appear early on in the Universe, (e.g. Mortlock et al. 2011). There are many other open questions concerning super-Eddington accretion outlined recently by Narayan, Sadowski \& Soria (2017), which could also be studied through observations of ULXs. These questions include (i) How viable is super-Eddington accretion ?  (ii) What is the geometry of the accretion flow? How does it impact observations as a function of inclination angle?  (iii) How luminous are super-Eddington systems? Are they radiatively efficient?  (iv) How much mechanical energy do super-Eddington disks produce in outflows? What role do the outflows play in feedback? (v) How often do super-Eddington disks produce relativistic jets? How do these jets compare with blazar jets?  Alternatively, supermassive black holes could grow from accretion onto intermediate mass black holes (e.g. Greene  2012, Volonteri 2012). Observing the size and the distribution of the population of IMBHs will help us probe open questions such as how are IMBHs formed and how do they evolve (Miller \& Colbert 2004)?

\smallskip

\noi Observations with the {\em SKA} will be essential to elucidate the nature of ULXs.  Firstly, the intrinsic power of the ULX can be estimated from the optical emission-line nebulae, created from shock-ionised driven jets, outflows or disc winds and/or because of photo-ionisation from the X-ray and UV emission around the black hole (Pakull \& Mirioni, 2002, 2003), where some of these show a radio counterpart. These radio nebulae can also be used to understand how outflows and photo-ionisation can play a role in the behaviour of the ULX and on its surrounding environment (Pakull \& Mirioni, 2002, 2003; Roberts et al. 2003; Kaaret et al. 2004). The sensitivity (and resolution) of SKA1-MID will allow us to detect and resolve these nebulae in galaxies out to 10s of Mpc with few hours of integration time. Phase 2 of the SKA will double the distance of detecting ULX bubbles of a given luminosity with only modest telescope time per target.  We will also be able to search for the presence of any compact radio source on the sub-arcsecond scale to determine if these nebulae are blown by compact jets emanating from the compact object. With quasi-simultaneous X-ray observations, using for example SVOM or Athena (see also Sect.\,\ref{science:synergies}), we can also estimate the mass of the black hole. This can be done using the fundamental plane of black hole activity, a correlation relating the X-ray luminosity, the radio luminosity and the mass of the black hole (e.g. Merloni et al. 2003; K\"ording et al. 2006; Plotkin et al. 2012).

\smallskip

\noi HLX-1 was the first ULX to show jets in the same way as X-ray binaries, although they are too faint to detect in the hard state and can only be detected when discrete jet ejection events occur during the transition from the hard to the soft state (Webb et al. 2012; Cseh et al. 2015). It is very likely that HLX-1 will be detectable in the hard state with SKA1-MID, extrapolating from the radio flares detected during the hard to soft state transitions, but closer ULXs ($<$95 Mpc away) will be much easier to detect than HLX-1 in the hard state. Other nearby ULXs have now also been seen to show radio jets close to the soft state, like XMMU J004243.6+412519 (Middleton et al. 2013)  which shows radio variability on a timescale of tens of minutes (arguing that the source is highly compact). Therefore, any X-ray (using for exemple SVOM or Athena, see also Sect.\,\ref{science:synergies}) and radio observations of this type of sources should be contemporary in order to constrain the mass of the compact object and understand the emission mechanisms. In addition, low state/quiescent IMBHs should also have steady jets. Madau \& Rees (2001) propose that there may be many IMBHs in a Milky Way type galaxy. SKA1-MID will be able to detect almost any quiescent IMBH in our Galaxy (assuming radio flux densities of $\mu$Jy level). Plotting the radio luminosities against quasi-simultaneous X-ray luminosities (the fundamental plane of black hole accretion) will indicate their IMBH nature (Maccarone  2004). Identifying where IMBHs reside will give us clues as to how they were formed (Miller \& Colbert 2004).

\smallskip

\noi In order to identify neutron stars as the compact objects in ULXs, the clearest cut way is to detect pulsations. ULXs are by definition very bright because of the high accretion rates and to date all of the pulsations have been detected in X-ray. However, we know that ULXs can vary by factors of 100-1000 or more in X-ray luminosity (e.g. the accreting neutron star ULX, M 82 X-2, Brightman et al. 2016). Such variability could be monitored using the SVOM satellite, for example. Once the neutron star is no longer accreting at high (super-Eddington) rates, it may be possible to detect the radio pulsations, in the same way as has been done for the accreting pulsars e.g. PSR 1023+0038 (Archibald et al. 2009) or PSR J1824-2452I (Papitto et al. 2013).  The wide field of view, great time resolution and sensitivity of SKA1-MID and SKA1-LOW will be essential tools to identify (or put limits on) pulsations from known ULXs and help find new pulsating neutron stars that may become ULXs later in their lives.  \\

\parbox{0.9\textwidth}{
\noi{References:}\\
\noi{\scriptsize 
Archibald, A. M., \etal, 2009, Science, 324, 1411; 
Bachetti, M., \etal, 2014, Nature, 514, 202;
Bachetti, M., \etal, 2013, ApJ, 778, 163;
Brightman, M., \etal, 2016, ApJ, 816, 60;
Cseh, D., \etal, 2015, MNRAS, 446, 3268;
Farrell, S. A., \etal, 2009, Nature, 460, 73;
F\"urst, F., \etal, 2016, ApJ, 831, L14;
Godet, O., \etal, 2012, ApJ, 752, 34;
Greene, J. E. 2012, Nature Communications, 3, 1304;
Israel, G. L., \etal, 2017, Science, 355, 817;
Kaaret, P., Ward, M. J., \& Zezas, A. 2004, MNRAS, 351, L83;
King, A., et al., 2017, MNRAS, 468, L59;
Koliopanos, F., \etal, 2017, A\&A;
Kording, E., et al., S. 2006, A\&A, 456, 439;
Maccarone, T. J., 2004, MNRAS, 351, 1049;
Madau, P. \& Rees, M. J. 2001, ApJ, 551, L27;
Merloni, A., et al., 2003, MNRAS, 345, 1057;
Middleton, M. J., \etal, 2013, Nature, 493, 187;
Miller, M. C. \& Colbert, E. J. M., 2004, International Journal of Modern Physics D, 13, 1;
Mortlock, D. J., \etal, 2011, Nature, 474, 616;
Mushtukov, A. A., \etal, 2017, MNRAS, 467, 1202;
Narayan, R., et al., 2017, arXiv:astro-ph/1702.01158;
Pakull, M. W. \& Mirioni, L., 2002, arXiv:astro-ph/0202488;
Pakull, M. W. \& Mirioni, L., 2003, RMxAC, 15, 197;
Papitto, A., \etal, 2013, Nature, 501, 517;
Plotkin, R. M., \etal, 2012, MNRAS, 419, 267;
Roberts, T. P., \etal, 2003, MNRAS, 342, 709;
Volonteri, M., 2012, Science, 337, 544;
Walton, D. J., \etal, 2013, ApJ, 779, 148;
Webb, N., \etal, 2012, Science, 337, 554;
Webb, N. A., \etal, 2010, A.J., 712, L107
}}\\

\paragraph{Gamma-ray bursts}\label{sci:GRB}
\vspace{0cm}

\smallskip
\noi {\bf Introduction} 

\smallskip
\noi Gamma-ray burst (GRB) afterglow studies in the radio domain provide complementary and sometimes unique diagnostics on GRB explosions and their environments. They may also allow the identification of very high redshift GRBs, possibly associated with PopIII stars. The sensitivities of MeerKAT and then SKA will hugely increase the number of radio afterglow detections. Furthermore, thanks to the wide field of view of the new and future radio arrays, it will be possible to identify GRB orphan afterglows and establish the association between gravitational waves and short GRBs.

\smallskip
\noi{\bf GRBs and their radio afterglows}

\smallskip
\noi GRBs are among the most energetic events in the universe. Thanks to their exceptional brightness they can be used as a unique tools to retrieve information on the high redshift universe. Forty years after their discovery, their origin and physics are still to be fully understood. There are two main phases of the gamma-ray burst phenomenon: the prompt emission (occurring primarily at gamma-ray energies and lasting at most few minutes), and the afterglow (a long-lasting, multi-wavelength emission which follows the main GRB event).

\begin{figure}[!ht]
  \centering
  \includegraphics[width=0.5\linewidth]{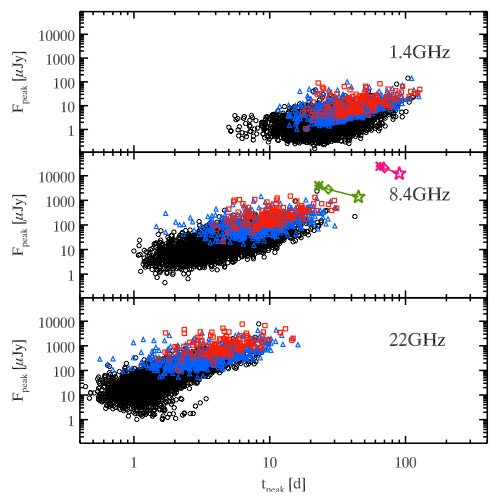}
  \caption{\label{fig:GRBaft} Peak flux versus time of the peak (observer frame) of the radio afterglow light-curve of the  population of GRBs at three frequencies (from Ghirlanda et al. 2013).}
\end{figure}

\smallskip
\noi Studies of the afterglow in the radio domain provide complementary and sometimes unique diagnostics on GRB explosions and their environments. Thanks to radio observations it is possible to add information to assess which are the radiative processes at play, to determine the total kinetic power of the jet, to test the jet expansion and to determine the interstellar medium structure and density. The radio emission has the advantage (compared to optical) to peak on timescales of few days and generally to keep bright up to several tens of days after the burst (Fig.\,\ref{fig:GRBaft};  the peak fluxes are weaker for larger wavelengths). Nonetheless only a few percent of radio GRB afterglows are detected with present radio instrumentation (see Chandra \& Frail 2012 for a review). This is mainly due to the fact that radio GRB afterglows are usually weak (flux at peak less than 1mJy). With the new generation of radio arrays it will be possible to easily detect much weaker fluxes. Part of the {\it ThunderKAT} KSP of the SKA precursor MeerKAT is dedicated to the detection and study of GRB radio afterglows, and SKA will be able to detect systematically GRB afterglows and also their transition to the non-relativistic phase.
SKA GRB afterglow observations will play a particular important role in the context of the Sino-French space mission {\it SVOM} (see Sect.\,\ref{syn:SVOM}). Indeed, taking also advantage of the synergy with CTA (see Sect.\,\ref{syn:CTA}), it will be possible to cover the emission of SVOM detected GRB from the very high (GeV/TeV) to the very low (radio) frequencies, and therefore obtain a complete knowledge of the emission processes  occurring both in the prompt and in the afterglow phases.

\smallskip
\noi {\bf Very high-z GRBs}

\smallskip
\noi Since the peak time of radio afterglow light-curves is larger for very high-z GRBs, radio afterglow detections can pinpoint to high redshift GRBs even without optical/near-infrared afterglow observations. Moreover radio detections can also allow the identification of PopIII GRBs. This should be possible also at LOFAR frequencies. In fact, a strong radio signature is expected for GRBs originating from PopIII stars, with high peak fluxes and much larger peak times (e.g.: Toma et al. 2011), allowing us to uniquely identify them. Radio afterglows of PopIII stars (or of very bright GRBs) can also in principle be used as background sources for 21 cm absorption line detections. 
At very high redshift, SKA afterglow observations will play also an important role in synergy with the {\it Athena} (see Sect.\,\ref{science:athena-erosita}) capabilities to constrain the chemical composition of Pop-III-GRB close environments, confirming the Pop-III origin of afterglows.  
The improved sensitivity of the future radio arrays will also make possible the detection of the radio emission of a much larger number of GRB host galaxies, allowing the direct determination of their star formation rate, important information as GRBs can be considered as direct tracer of star formation.

\smallskip
\noi {\bf Orphan afterglows} 

\begin{figure}[!ht]
  \centering
  \includegraphics[width=0.4\linewidth]{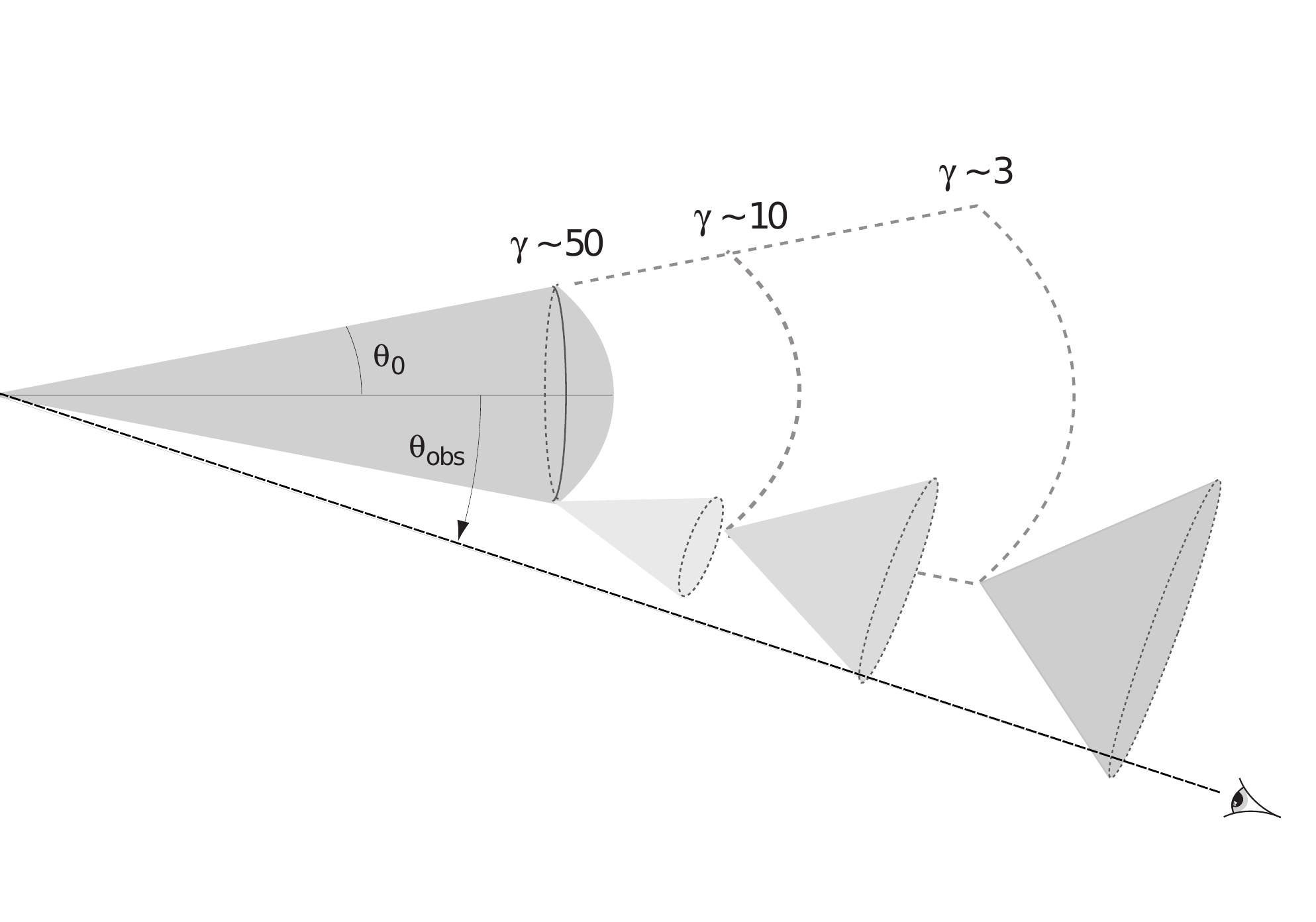}
  \caption{\label{fig:GRBorph} Schematic view of the jet deceleration (Granot et al. 2005). Due to the deceleration of the jet, the beaming angle of the emission within the jet is becoming wider with time. Even if the observer's line-of-sight is initially outside the jet cone - and therefore the prompt GRB emission cannot be observed - the effect might allow the observer to detect the afterglow - orphan afterglow - at later times.}
\end{figure}

\smallskip
\noi The GRB emission is beamed in the direction of the jet motion. To observe the prompt GRB emission, our line-of-sight has to lie within the jet cone. However, as the jet is gradually decelerated over time, the beaming angle of the emission becomes wider (see Fig.\,\ref{fig:GRBorph}) allowing the afterglow detection even for an observer not placed along the jet cone. Such an afterglow is called an {\it orphan afterglow}. In fact, on-axis GRBs are only the tip of the iceberg of the GRB population. For each GRB seen on axis, there should be hundreds of GRBs for which only the orphan afterglows can be detected. The detection of an orphan afterglow will bring independent confirmation of the jetted GRB outflow and the detailed analysis of their light curves will enable to differentiate between different jet models. The peak of the orphan afterglow will happen some tens of days after the GRB explosion. At these epochs, the afterglow spectra peak towards mm/radio frequencies. The future radio wide field surveys, especially with the ASKAP and SKA arrays, will be therefore very powerful ways to detect orphan afterglows and establish the association between gravitational waves and short GRBs (see Ghirlanda et al. 2014).\\

\parbox{0.9\textwidth}{
\noi{References:}\\
\noi{\scriptsize 
Chandra, P. \& Frail, D., 2012, ApJ, 746, 156;
Ghirlanda, G., \etal, 2013, MNRAS, 435, 2543;
Ghirlanda, G. \etal, 2014, PASA, 31, 22;
Granot, J., \etal, 2005, ApJ, 630, 1003;
Toma, K., \etal, 2011, ApJ, 731, 127
}}\\

\paragraph{High Mass X-ray Binaries}
\vspace{0cm}

\noi ``High-Mass X-ray Binaries'' (HMXB) are composed of a compact object (neutron star --NS-- or black hole --BH--) orbiting a luminous and massive early OB type companion star ($\geq 10 \msol$). We distinguish 3 types of HMXB by the process of accretion, ordered here by decreasing number of systems of each class: (i) Be X-ray binaries (BeHMXB), (ii) supergiant X-ray binaries (sgHMXB), and (iii) Roche Lobe Overflow systems (RLO).

\begin{figure}[!ht]
  \centering
  \includegraphics[width=0.54\linewidth]{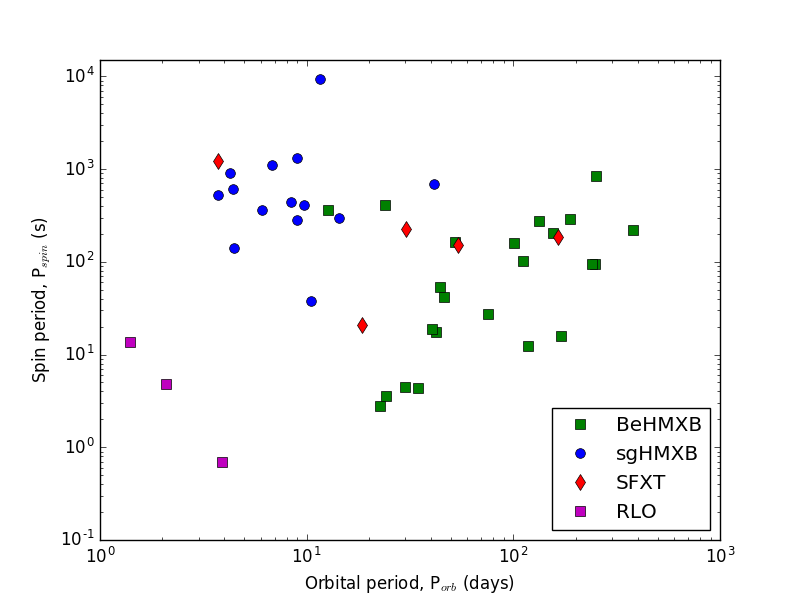}
  \includegraphics[width=0.44\linewidth]{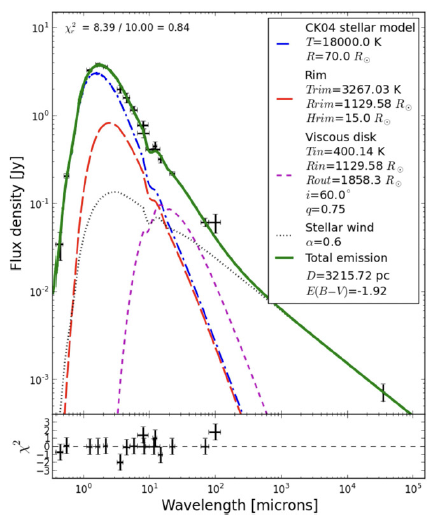}
  \caption{\label{fig:corbet} {\em Left}: Corbet diagramme showing the 3 different populations of HMXB. {\em Right}: SED of GX\,301-2 fitted with a stellar photosphere model, a rimmed disk, and a power law due to stellar wind (Servillat et al. 2014).}
\end{figure}

\smallskip

\noi (i) Be X-ray binaries (BeHMXB) host a main sequence donor star of spectral type B0-B2e III/IV/V, a rapid rotator surrounded by a circumstellar (so-called ``decretion'') disc of gas, as seen by the presence of a prominent H$\alpha$ emission line. This disc is created by a low velocity/high density stellar wind of $\sim 10^{-7} \msol / yr$. Transient and bright (``Type I'') X-ray outbursts periodically occur, each time the compact object (usually a NS on a wide and eccentric orbit) approaches periastron and accretes matter from the decretion disc (see Charles \& Coe 2006; Tauris \& van den Heuvel 2006). These systems exhibit a correlation between the spin and orbital period, as shown by their location from the lower left to the upper right of the Corbet diagram (see Fig.\,\ref{fig:corbet}), due to efficient transfer of angular momentum at each periastron passage: rapidly spinning NS corresponding to short orbital period systems. Apart from MWC\,656 hosting a BH (Casares et al. 2014), most BeHMXB seem to host a NS.

\smallskip

\noi (ii) Supergiant X-ray binaries (sgHMXB) host a supergiant star of spectral type O8-B1 I/II, characterised by a slow and dense, radiatively steady and highly supersonic stellar wind, radially outflowing from the equator. There are $\sim 16$ ``classical'' sgHMXB, most of them being close systems, with a compact object on a short and circular orbit, directly accreting from the stellar wind, through e.g. Bondy-Hoyle-Littleton process. Such wind-fed systems exhibit a luminous and persistent X-ray emission ($L_X = 10^{36-38} \ergs$), with superimposed large variations on short timescales, and a cut-off (10-30 keV) power-law X-ray spectrum. Located in the upper left part of the Corbet diagram (small orbital period $P_\mathrm{orb} \sim 3-10$\,days vs long spin periods $P_\mathrm{spin} \sim 100-10000$\,s, Fig.\,\ref{fig:corbet}), they do not show any correlation, since there is no net transfer of angular momentum. Nearly half of sgHMXB ($\sim 8$) exhibit a substantial intrinsic and local extinction $\nh \geq 10^{23} \cmmoinsdeux$, with a compact object deeply embedded in the dense stellar wind (such as the highly obscured IGR~J16318-4848, Chaty \& Rahoui 2012). Likely in transition to Roche Lobe Overflow, these systems are characterised by slow winds causing a deep spiral-in of the compact object, and leading to Common Envelope Phase. There exists also the possibility, for sgHMXB such as Cyg\,X-1, to accrete both through Roche lobe overflow and stellar wind accretion.

\smallskip

\noi A significant subclass of sgHMXB is constituted of 17 (+5 candidate) Supergiant Fast X-ray Transients (SFXT, see Negueruela et al. 2006). These systems, characterised by a compact object orbiting with $P_\mathrm{orb} \sim 3.3 - 100 $\,days on a circular or excentric orbit, and by $100-1000$\,s spin periods, span on a vast location in the Corbet diagram, mostly in between BeHMXB and sgHMXB (Fig.\,\ref{fig:corbet}). They exhibit short and intense X-ray outbursts, an unusual characteristic among HMXB, rising in tens of minutes up to a peak luminosity $L_X \sim 10^{35-37} \ergs$, lasting for a few hours, and alternating with long ($\sim 70$\,days) quiescence at $L_X \sim 10^{32-34} \ergs$, with an impressive variability factor $\frac{L_{max}}{L_{min}}$ going up to $10^2-10^5$. Various processes have been invoked to explain these flares, such as wind inhomogeneities, magneto/centrifugal accretion barrier, transitory accretion disc, etc (see Chaty et al. 2013 and references therein).

\smallskip

\noi (iii) Roche Lobe Overflow systems (RLO) host a massive star filling its Roche lobe, where accreted matter flows via inner Lagrangian point to form an accretion disc (similarly to LMXB). These systems, also called beginning atmospheric Roche lobe overflow, constitute the classical ``bright'' HMXB (such as Cen X-3, SMC X-1 and LMC X-4), with accretion of matter occurring through the formation of an accretion disc, leading to a high X-ray luminosity ($L_X \sim 10^{38} \ergs$) during outbursts. There are only a few sources, located in the lower left of the Corbet diagram (Fig.\,\ref{fig:corbet}), characterised by short orbital and spin periods.

\smallskip

\noi A number of 114 HMXB are reported in Liu et al. (2006), and 117 in Bird et al. (2016). By cross-correlating both catalogues, we find that the total number of HMXB currently known in our Galaxy amounts to 152 (see Fortin, Chaty \& Coleiro, in prep). Therefore, adding all known LMXB and HMXB systems, HMXB represent $\sim 43\%$ of the total number of high energy binary systems. Among HMXB, there are 70 BeHMXB, 35 sgHMXB and 56 HMXB of unidentified nature. sgHMXB can be further divised in 25 ``classical'' sources and 10 SFXT. Thus, HMXB can be divided respectively in 43\% of BeHMXB, 22\% of sgHMXB and 35\% of unidentified HMXB.

\smallskip

\noi To get an idea of the flux we would detect with SKA1, we now estimate the level of emission of HMXB in the radio domain, likely dominated by the intense stellar winds emanating from early type massive stars. To take an example, let us estimate the flux of the persistent HMXB hosting an hypergiant star GX\,301-2 (see discussion within Servillat et al. 2014, and Fig.\,\ref{fig:corbet} {\em right}). Using Equation (4) in Scuderi et al. (1998), we estimate a radio flux of 2.5 mJy at 8.6 GHz, while Pestolazzi et al (2009) have detected a flux around 1 mJy at 4.8 and 8.6 GHz with ATCA, in 1.5 hour integration. This discrepancy between the estimate and the detection level might be due either to rough understanding of the stellar wind emission, or even variability proper to this emission. Looking at the SED of GX\,301-2 in Servillat et al. (2014), an extrapolation at lower frequency suggests that this source will not be detectable within SKA1-LOW bands. However, the frequencies between 4.8 and 8.6 GHz will be covered by SKA1-MID, and we estimate that we should be able to detect this signal level in a few minutes.

\smallskip

\noi Such a detection with SKA1 with a continuous coverage will allow us to constrain the nature of the emission (thermal, free-free, non-thermal, synchrotron emission), and most important where it comes from: either from the stellar wind, or surrounding dust, or even from a transient accretion disk accumulating around the compact object. This integration time represents a lower limit for HMXB, as GX\,301-2 is one of the brightest HMXB.\\

\parbox{0.9\textwidth}{
\noi{References:}\\
\noi{\scriptsize 
  Bird, A. J., et al., 2016, ApJSS, 223, 15;
  Casares, J., et al., 2014, Nature, 505, 378;
  Charles, P. A. \& Coe, M. J., 2006, in {\em Compact stellar X-ray sources}, Cambridge University Press, 215;
  Chaty, S. \& Rahoui, F., 2012, ApJ, 751, 150;
  Chaty, S., 2013, Advances in Space Research, 52, 2132;
  Coleiro, A., \& Chaty, S., 2013, ApJ, 764, 185;
  Grimm, H.J., et al., 2003, MNRAS, 339, 793;
  Liu, Q.Z., et al. 2006, A\&A, 455, 1165;
  Liu, Q.Z., et al. 2007, A\&A, 469, 807;
  Negueruela, I., et al., 2006, in {\em ESA Special Publication};
  Scuderi, S., et al., 1998, A\&A, 332, 251;
  Servillat, M., et al., 2014, ApJ, 797, 114;
  Tauris, T. \& van den Heuvel E.P.J., 2006, in {\em Compact stellar X-ray sources}, Cambridge University Press, 623
}}

\paragraph{AGN variability}\label{science:AGNvar}
\vspace{0cm}

\smallskip
\noi Active galactic nuclei (AGN) are supermassive black holes at the centre of galaxies that are thought to 
be powered by accretion (e.g. Urry \& Padovani 1995). Radio-loud AGN typically exhibit relativistic 
outflows of matter, the so-called jets. 

\smallskip
\noi Variability of nuclear flux density and polarisation is a common characteristic of AGN, and
occurs with multiple characteristic timescales in individual objects, from hours to days to years. Variations between different
bands of the electromagnetic spectrum are often related to each other, with  emission at short wavelengths 
(X-rays or $\gamma$-rays) eventually propagating to longer wavelengths, as radio, with longer timescales. 
Emission at different frequencies probe different physical processes occurring in different regions of the
AGN central engine. Thus correlating variability over a range of radio frequencies with that seen across the electromagnetic
spectrum, can provide a powerful probe of the connection between disks and jets, inflow and outflow, turbulence and shocks, and activity and quiescence.

\smallskip
\noi Indeed AGN intrinsic variability exhibits both red and white noise power spectrum, as well as quasi-periodic 
oscillations in some cases (e.g. King et al. 2013). This may correspond to variations in
accretion rate (Chatterjee et al. 2011), flares and shocks in disks and jets  (Marsher 2016),  
or changes in Doppler boosting and jet precession (Lister et al. 2009), as well as reflect transitions 
between high and low states (Krauss et al. 2016). 

\smallskip
\noi These questions will be addressed specially through analysis of the massive data sets arising
from the deep all-sky surveys  with SKA1, including both total and polarimetric flux.

\smallskip
\noi {\bf Long term monitoring}

\smallskip
\noi  Long-term monitoring observations are required to sample different timescales and to allow correlations to observations at other frequencies. 
Day to year timescales for AGN variability correspond to the light crossing, viscous, and infall timescales of
the black hole and accretion disk. The delay between X-ray and radio variations can be hundreds of days implying that 
long duration coordinated campaigns are required for reasonable sampling (at least several years).  
Due to the limited survey rate of existing radio telescopes, AGN variability has been studied in detail only for a thousand of relatively 
bright objects (e.g.  Hovatta et al. 2007), or in only a few epochs for larger numbers of sources  down to lower flux density limits. 

\smallskip
\noi AGN dominate the radio sky at flux densities above a milli-Jansky (Seymour et al. 2008), and synoptic surveys with SKA1 
will provide  precision flux measurements as a function of time with a cadence of days or less for hundreds of thousands of AGN, 
and thus produce an archive of variability information for a large range of AGN, alleviating the biases inherent to small sample selection.

\smallskip
\noi  {\bf Radio loud}

\smallskip
\noi Variability at radio wavelengths is most marked in powerful radio-loud AGN. The largest and most rapid variations observed in nuclear non-thermal continuum 
emission is seen in BL Lacs and core-dominated quasars.  
These variations are explained primarily by Doppler boosting of the nuclear emission by highly relativistic jets viewed at small 
angles to the line of sight. Flares in the radio light curves might 
correspond to ejection of new relativistic components or emergence of shocks in 
the underlying flow (e.g. Valtaoja et al. 2002; Arlen et al. 2013; Feng et al. 2017). 

\smallskip
\noi Population studies of gamma-ray loud and quiet AGN was possible thanks to Fermi-LAT that monitors since 2008 
the entire gamma-ray sky every three hours (Atwood et al. 2009). Comparison of the gamma-ray emission to their radio properties showed that  gamma-ray detected
AGN have larger variability amplitudes in radio (Richards et al. 2011), faster apparent jet speeds
(Lister et al. 2009), and higher Doppler beaming factors (Savolainen et al. 2010) than non-detected
objects. Some studies have also reported that radio and gamma-ray luminosities are correlated
(Marscher et al. 2008; Ackermann et al. 2011; Nieppola et al. 2011), but it is difficult to assess 
the significance of such correlations as a whole without larger samples even though this implies a connection between
radio and gamma-ray variations in the sources. Moreover, the next generation very high energy Cherenkov
Telescope Array (CTA) (see Sect.\,\ref{syn:CTA}), in combination with SKA, will open the possibility to study the
connection between TeV and radio emission in a large number of sources, probing the acceleration
of particles up to the highest energies.

\smallskip
\noi As a matter of fact, many existing surveys have focused on blazars, where relativistic beaming and the compact
sizes of emitting regions produce more variability than in sources oriented further from the line of
sight. The wide field monitoring capabilities of SKA, however, will enable light-curves from all classes of AGN to be studied, 
refining unification schemes and our understanding of the geometry of the emitting regions.

\smallskip
\noi {\bf Radio quiet}

\smallskip
\noi Seyfert galaxies, classified as radio quiet (though not radio silent), are known to show variability in their continuum and line 
emission at X-ray, optical, and UV wavelengths, providing constraints on the variation of nuclear 
absorbing column, photon reprocessing, structure and dynamics of the broad-line 
region and ultimately the mass of the central object (ref here to put). But, although ten time 
more common than their radio-loud counterparts, only a small number of interesting Seyferts 
have been systematically monitored  at radio frequencies, with some studies motivated by serendipitous discovery of 
radio flares (Falcke et al. 2004). Radio flares seen in III Zw 2 and Mrk 348 suggest that jets in Seyferts can
accelerate from non-relativistic to relativistic speeds during an outburst, but
larger samples are required to confirm this (Mundell et al. 2009). Monitoring of broad absorption line BAL-QSOs and other radio quiet objects,
such as Mrk 231, has already revealed variations similar to blazars. Although large scale jets and
lobes do not develop in these objects, highly energetic flares can still occur close to the core (Reynolds et al. 2013). 

\smallskip
\noi Less than 10\% of Narrow-Line Seyfert I galaxies (NLS1s) are formally radio loud (Komossa
et al. 2006). However several of them are detected in gamma rays by Fermi (Abdo et al. 2009), confirming that they indeed can also emit high
energy electrons, and have prominent relativistic jets. Those also show
fast variations in their radio light curves, similar to blazars (Foschini et al. 2011; D'Ammando et al. 2012). Modeling these variations
can probe jet acceleration mechanisms and central engine physics (Kudryavtseva et al. 2011).

\smallskip
\noi Note that in some cases (e.g. Thyagarajan et al. 2011) variable radio emission is the only indicator of AGN activity, and can provide a powerful
probe of buried AGN which show no evidence of an accreting black hole at other wavelengths.

\smallskip
\noi But with few examples, it is difficult to establish if the flaring is a common property in so-called radio quiet 
AGN. With SKA, the number of low luminosity AGN monitored will be extremely large, allowing detailed studies of the variability and comparison to blazars. This
will give insights into the jet formation processes and bridge the gap between variations in jetted systems around stellar-mass black holes and the most extreme blazars.

\smallskip
\noi {\bf VLBI}

\smallskip
\noi VLBI observations provide a direct measure of relativistic motion in AGN jets, allowing 
to measure intrinsic jet parameters (as velocity, Doppler factor, opening
angle, inclination angle, magnetic field strength) and  have the potential to locate and constrain the size of emission regions in conjunction with
high energy and intermediate energy observations. 

\smallskip
\noi Radio observations with milli-arcsecond resolution have resolved the radio-emitting regions and measured 
outflow velocities from some well known AGN (e.g. Lister et al. 2013). Multi-wavelength monitoring of a 
few bright sources from radio to the Fermi GeV band, strongly suggests that the emission 
takes place at distances of several pc from the central engine, VLBI observations showing that (some) major $gamma$-ray
flares take place at the same time with the ejection of superluminal components from the radio core,
located several pc downstream of the central engine (e.g. Marscher et al. 2010; Arlen et al. 2013), or even co-spatial
with VLBI components downstream of the core (e.g. Agudo et al. 2011). At (very)high energies the angular resolution is insufficient 
and we must infer the size and location of the emission regions from flux variations. 
If gamma-ray and radio emission are triggered by shocks propagating along a relativistic jet, the time delay between flares in the two bands depends
on their separation. Several studies have found time-lagged correlation between these two energy bands, but without a large sample
with well-sampled light curves it is difficult to assess the significance of the correlations.  Flares and superluminal blobs may be connected 
to changes in magnetic field on timescales of years, or to current-driven instabilities, tangled magnetic fields, 
or reconnection events, and the VLBI extension of SKA can help probe this (Agudo et al. 2015).

\smallskip
\noi Also if flares in radio light curves correspond to ejection of new relativistic components
or emergence of shocks in the underlying flow, improved monitoring and high-resolution imaging using VLBI
techniques are required to confirm that radio jets are intrinsically non-relativistic during quiescence but that
Seyferts, as black-hole-driven active galactic nuclei (AGN), have the capacity to accelerate relativistic jets during
radio flares. 

\smallskip
\noi Current VLBI surveys reach milli-arcsecond resolutions, corresponding to parsec scales in AGN. Resolution of the baseline SKA-MID 
is an order of magnitude or so poorer but SKA2 (see Paragi et al. 2015) would enable the
expansion of existing monitoring programs to much larger number of sources. Polarisation capabilities will 
enable identification and tracking of jet components,
characterising changes in magnetic fields and their connection to flares, thus shedding light on jet
composition and the interaction of jets with the ambient medium. 

\smallskip
\noi More elements pointing towards the strength of SKA monitoring for time domain studies of AGN can be found in Bignall et al. 2015.\\

\parbox{0.9\textwidth}{
\noi{References:}\\
\noi{\scriptsize Abdo, A. A., et al., 2009,  ApJL, 707, L142;   
Ackermann, M. et al. 2011, ApJ, 741, 30;  
Agudo, I., et al. 2011, ApJ, 743, L10;   
Agudo, I.,  et al. 2015, AASKA14, 93;  
Arlen, T., et al. 2013, ApJ, 762, 92;   
Atwood, W. B., et al. 2009, ApJ, 697,1071;   
Bignall, H., et al., 2015, AASKA14, 58;  
Chatterjee, F., , R., et al. 2011, ApJ, 734, 43;    
D'Ammando et al. 2012, MNRAS, 426, 317;  
Falcke, H., et al. 2004,  New Astronomy Reviews, 48, 1157; 
Feng, Q., et al. 2017, arXiv 1708.06386;
Foschini, L., et al. 2011, MNRAS, 413, 1671;   
Hovatta, T., et al. 2007, A\&A, 469, 899;  
King, O. G., et al. 2013, MNRAS, 436, L114;  
Komossa, S., et al. 2006, AJ, 132, 531;  
Krauss, F., et al. 2016, A\&A, 581,130;  
Kudryavtseva, N.A., et al. 2011, MNRAS, 415, 1631;  
Lister, M. L., et al. 2009, AJ, 138, 1874;  
Lister, M. L., et al. 2013, AJ, 146, 120;  
Marscher, A. P., et al. 2008,  Nature, 452, 966;   
Marscher, A. P., et al., 2010, ApJ,  710, L126;   
Mundell, C., et al. 2009, ApJ, 703, 802;  
Nieppola, E.,  et al. 2011, A\&A, 535, 69;  
Paragi, Z., et al. 2015, AASKA14, 143;  
Reynolds, C.,  et al. 2013, ApJL, 776, L21;  
Richards, J. L., et al. 2011, MNRAS, 438, 3058;  
Savolainen, T., et al. 2010, A\&A, 489, L33;   
Seymour, N., et al. 2008, MNRAS, 386, 1695;  
Thyagarajan, N., et al. 2011,  ApJ, 742, 49;   
Valtaoja, E., et al. 2002, PASA, 2002, 19, 117;  
Urry, M., \& Padovani, P., 1995, PASP, 107,803}}\\

\paragraph{Supernov\ae}
\vspace{0cm}

\noi Supernov\ae \ (SNe) are among the most energetic events in the universe. They can release up to a few $10^{51}$ ergs in the interstellar medium in form of fast stellar ejecta and radiation. There are two types of supernova (SN): Type Ia SN, events produced by the thermonuclear explosion of an accreting white dwarf, core-collapse SNe, events associated with the death of stars more massive than 8 $M_{\odot}$. Radio observations are essential for many aspects of supernova physics especially to what concerns the investigation of the circum-stellar medium (CSM) properties, SN shock dynamics, energetic particles acceleration.

\smallskip
\noi {\bf Type Ia supernov\ae}

\smallskip
\noi Actually no type Ia SN has been detected yet at radio wavelength. Hence, the high sensitivity of SKA should allow to discover and follow in time the radio emission of such objects. SKA-MID configuration already reaches flux limits below some most luminous type Ia SNe like SN 2014J or SN 2011fe. SNIa radio detection will help to identify the configuration of the stellar system prior to the explosion: either a double or a single degenerate. Radio emission is expected from a single degenerate case as a probe of the CSM matter deposited by the companion star (Wang et al. 2015). 

\smallskip
\noi {\bf Core-collapse supernovae-CCSNe}

\smallskip
\noi CCSNe present a vast variety of sub-types from type IIP SNe (about half of the SNe), type IIL, IIb, Ib/c, IIn (see Smith 2014 for details). At present, about 50 SNe have been detected and then follow in time at radio wavebands (Weiler et al. 2002; Perez-Torres et al. 2014). SKA with its improved sensitivity at different wavelengths and combining detection in its pointed and survey modes is expected to detect several thousands of CCSNe (Perez-Torres et al. 2014) up to a redshift $Z\sim 0.25$ for the most intense IIn SNe. With the advent of SKA the community will be able to start statistical analysis of CCSNe properties depending of their subclass type.

\begin{figure}[!ht]
  \centering
  \includegraphics[width=0.4\linewidth]{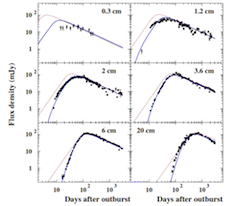}
  \caption{\label{fig:SNCh} VLA Radio data by Weiler et al. 2007: solid: best fit, dotted red: synchrotron self absorption model. Extracted from Tatischeff 2009.}
\end{figure}

\smallskip
\noi 
SKA will also be able to monitor the closest and most intense events as it was already done with the VLA for emblematic objects like SN1987A and SN1993J (see Fig.\,\ref{fig:SNCh}). Time evolution of the radio spectrum is of particular importance to understand shock dynamics, magnetic field generation, particle acceleration in different CSM environments. These issues are intimately related to the problem of the origin of cosmic rays and make these observations particularly relevant to trigger multi-wavelength observation campaigns involving high-energy instruments like the X-ray satellites XMM-Newton or Chandra and gamma-ray telescopes like the future Cerenkov Telescope Array.\\

\parbox{0.9\textwidth}{
\noi{References:}\\
\noi{\scriptsize 
Perez-Torres, M., et al., 2015, AASKA14, 60;
Smith, N., 2014, ARA\&A, 52, 478;
Tatischeff, V., 2009, A\&A, 499, 191;
Wang, L., et al., 2015, AASKA14, 64;
Weiler, K. W., et al., 2002, ARA\&A, 40 387;
Weiler, K. W., et al., 2007, ApJ 671 1959
}}\\

\subsubsection{Fast transients}

\paragraph{Fast Radio Bursts}
\vspace{0cm}

\noi Fast Radio Bursts (hereafter FRBs) are bright, millisecond-duration transient phenomena detected in the radio. The first FRB was detected in 2007 in archival data recorded as part of a 1.4-GHz survey for pulsars with the 64-m Parkes radio telescope, in Australia (Lorimer et al. 2007). A little more than 20 have now been discovered in data from several radio telescopes around the world, at 1.4 and 0.8 GHz\footnote{A catalog of FRBs is available at: \href{http://www.astronomy.swin.edu.au/pulsar/frbcat/}{\color{blue} \myul[blue] {http://www.astronomy.swin.edu.au/pulsar/frbcat/}}.} (Petroff et al. 2016). Similarly to radio signal from pulsars, observed bursts are dispersed by the ionised plasma following the classical DM$ / f^2$ frequency-time relation, where DM (the ``Dispersion Measure'') represents the integrated density of free electrons along the line-of-sight. The large DM values of FRBs seen so far range from about 300 cm$^{-3}$ pc to well above 1000 cm$^{-3}$ pc. Among other arguments, this suggests an extragalactic origin for the FRBs, although the exact phenomenon producing them is still actively debated. Proposed theories invoke superflares from soft gamma repeaters (e.g. Popov \& Postnov 2007), supergiant pulses from distant pulsars (e.g. Cordes \& Wasserman 2016), Alfv\'en waves from bodies orbiting a pulsar (Mottez \& Zarka 2014), to name but a few examples (see e.g. Katz 2016 for a review). The identification of FRB progenitors is further complicated by the fact that, unlike bursts from pulsars, FRBs do not appear to repeat. One notable exception is FRB~121102, seen to repeat by several radio telescopes (e.g. Spitler et al. 2016) including the Nan\c{c}ay Radio Telescope (see below). This enabled the association of the FRB with a dwarf galaxy at a distance of $\sim$1 Gpc (Chatterjee et al. 2017). However, it is unclear whether FRB~121102 is typical of other FRBs, since it is the only one seen to repeat until now. In any case, the firm identification of an FRB with a known object would certainly dramatically improve our understanding of this phenomenon. 

\smallskip
\noi Independently of their exact nature, the brightness and short durations of FRBs make them very useful astrophysical tools. For example, the dispersion measures of a large number of FRBs could be used to address the ``missing'' baryon problem, in the low redshift Universe. The scattering-induced smearing of FRB signals could be used to probe the turbulent intergalactic medium. Another potential application of the detection of a population of FRBs with high measured redshifts, is the study of the geometry of the Universe, via a model for the average electron density as a function of redshift. We refer the reader to the review by Macquart et al. (2015) and references therein, for additional details on these various potential studies and further possible applications. In addition to being extremely interesting phenomena in themselves, FRBs also are valuable cosmic probes. \\

\newpage

\noi \textbf{Current and planned efforts to observe FRBs at Nan\c{c}ay}
\smallskip

\noi The majority of FRBs seen thus far have been detected by analysing data from 1.4-GHz observations conducted with the Parkes radio telescope, often as part of the High Time Resolution Universe (HTRU) pulsar survey (see for instance Champion et al. 2016). FRB~121102 (discussed above), discovered at 1.4-GHz at Arecibo, was the first FRB not detected at Parkes (Spitler et al. 2014); and the discovery of FRB~110523 with the Green Bank Telescope at about 0.8-GHz demonstrated that FRBs can be seen at these lower frequencies (Masui et al. 2015). Recently, Caleb et al. (2017) reported the discovery of three FRBs in a survey of the southern sky at 0.8-GHz with the Molonglo Observatory Synthesis Telescope. It will be particularly interesting to see how many FRBs CHIME, currently in construction in Canada, will detect between 0.4 and 0.8-GHz, and also if instruments such as MWA and LOFAR will be able to detect FRBs despite the strongly dispersed and scattered signals at these very low radio frequencies (note that a search for FRBs at 145-MHz using LOFAR stations in France and in the UK resulted in no detection, see Karastergiou et al. 2015 for details). The estimated FRB rate based on the population seen so far is of several thousands FRBs per sky and per day (see e.g. Champion et al. 2016). 

\begin{figure}[!ht]
  \centering
  \includegraphics[width=0.53\linewidth]{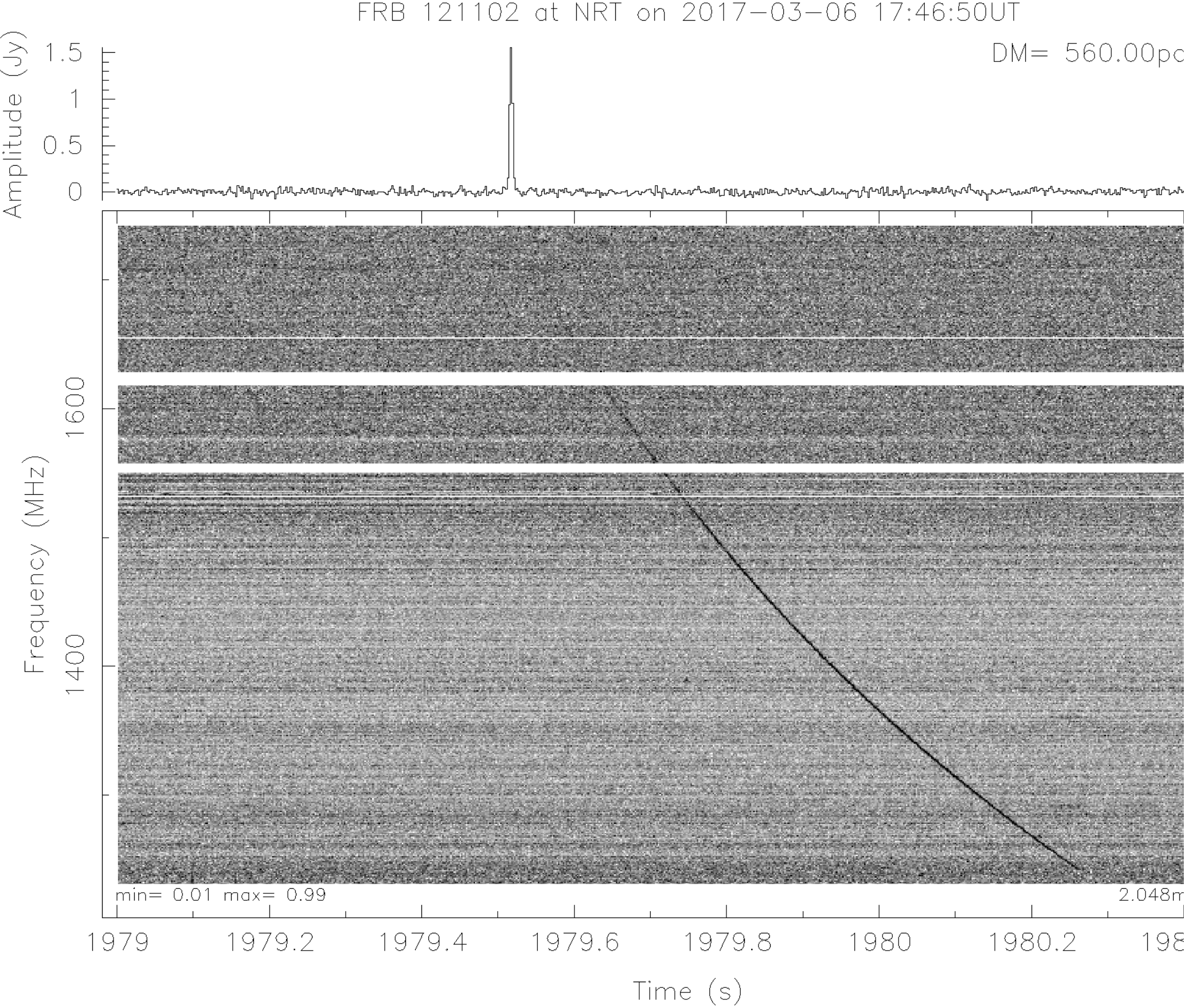}
  \caption{\label{fig:FRB121102} Detection of the repeating FRB~121102 at 1.4 GHz, with the Nan\c{c}ay Radio Telescope. \textit{Lower panel:} dynamic spectrum of the Nan\c{c}ay data, not corrected for the frequency-dependent dispersion. \textit{Upper panel:} integrated amplitude of the signal, corrected for the dispersion. Adapted from Cognard et al., in preparation.}
\end{figure}

\smallskip
\noi A survey for pulsars similar to HTRU is being conducted with the Nan\c{c}ay decimetric Radio Telescope (NRT) in France, at 1.4-GHz (Desvignes et al. 2013). No FRB has yet been found in the data from this pulsar survey. However, observations of the repeating FRB~121102 have revealed bright detections (see Fig.\,\ref{fig:FRB121102} for an example). At lower radio frequencies (below 100 MHz), Zarka \& Mottez (2016) simulated observations of FRBs with NenuFAR, a new telescope currently under construction at Nan\c{c}ay recognised as a pathfinder to SKA, and argued that it should be able to detect these transients provided they exist at these frequencies. If successful, such observations will give crucial information on FRB emission spectra and on the intergalactic medium. They will also be very useful for estimating the expected detection rate of future instruments operating at these low frequencies, such as SKA1-LOW (see below). \\

\smallskip

\noi \textbf{FRBs in the SKA era}
\smallskip

\noi With its dramatically increased sensitivity compared to present day radio telescopes, its large field-of-view and multi-beaming capabilities, and its continuously active transient detection systems, the SKA will be a wonderful FRB monitor. While current FRB searches yield a few detections per year, searches with SKA1-MID at GHz frequencies could for instance enable the discovery of several FRBs every week (see e.g. Macquart et al. 2015). Searches with SKA1-LOW below 350 MHz, although hampered by the much higher dispersion of the radio emission, may also yield large numbers of FRB detections. A much increased FRB population uncovered by the SKA will likely make it possible to characterise the FRB phenomenon in detail and use them as cosmological probes. The Orl\'{e}ans-Nan\c{c}ay group has a long experience in pulsar searches and is now conducting FRB searches with the NRT. The LUTh group at the Paris Observatory has been exploring possible theories for the nature and emission properties of FRBs. NenuFAR observations will be valuable for exploring FRB very low frequency emission. The French community is therefore ideally placed to contribute to FRB observation efforts with the SKA. \\

\parbox{0.9\textwidth}{
\noi{References:}\\
\noi{\scriptsize Lorimer, D. R., et al., 2007, Science, 318, 777; 
Petroff, E., et al., 2016, PASA, 33, 45; 
Popov, S. B., \& Postnov, K. A., 2007, arXiv:0710.2006; 
Cordes, J. M. \& Wasserman, I., 2016, MNRAS, 457, 232; 
Mottez, F. \& Zarka, P., 2014, A\&A, 569, 86; 
Katz, J. I., 2016, Modern Physics Letters A, 31, 14;
Spitler, L. G., et al., 2016, Nature, 531, 202; 
Chatterjee, S., et al., 2017, Nature, 541, 58;
Macquart, J. P., et al., 2015, AASKA14, 55;
Champion, D. J., et al., 2016, MNRAS, 460, 30;
Spitler, L. G., et al., 2014, ApJ, 790, 101;
Masui, K., et al., 2015, Nature, 528, 523;
Caleb, M., et al., 2017, MNRAS, 468, 3746;
Karastergiou, A., et al., 2015, MNRAS 452, 1254;
Desvignes, G., et al., 2013, Proc. IAU 291, 375;
Zarka, P. \& Mottez, F., 2016, Proc. SF2A meeting
}}

\paragraph{GW event follow-up}\label{sci:GW}
\vspace{0cm}

\noi {\bf Introduction} 

\smallskip
\noi SKA and its precursors can play a unique role in identifying the electromagnetic counterparts of gravitational wave (GW) events. An optimised observational strategy is fundamental to this purpose.

\smallskip
\noi {\bf Electromagnetic counterparts of GWs form merging of neutron stars}

\smallskip
\noi Even if to date GWs have been detected only from the merging of black hole binary systems, it is predicted that the Advanced LIGO/Virgo interferometers will also detect GWs coming from the merging of neutron stars. To fully exploit the huge scientific outcome of such detections it is necessary to identify the electromagnetic (EM) counterpart emission associated with the GW event. Many are the hints that the class of {\it short} GRBs (i.e.: GRBs having a duration of their prompt emission of less than 2\,s; see Sect.\,\ref{sci:GRB}) is associated with the merging of neutrons stars. Indeed, there are some evidences for the detection of kilonova explosions (predicted for neutron star merging) after short GRB explosions. Short GRBs and kilonova emission are the best candidates as GW EM counterparts.

\begin{figure}[!ht]
  \centering
  \includegraphics[width=0.6\linewidth]{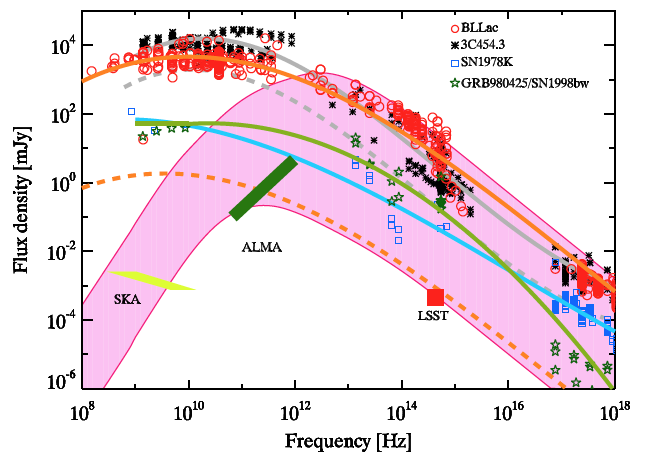}
  \caption{\label{fig:GRBaft2} Spectral energy distribution of the OA that can be detected by the LSST(pink filled region). The (5?) limits for a 12\,h continuum observation with the SKA, allowing to discriminate between different classes of transients, is shown by the yellow shaded region. See Ghirlanda et al. 2015 for further information.}
\end{figure}

\smallskip
\noi {\bf The detection of EM counterparts by SKA and SKA precursors}

\smallskip
\noi The error regions associated with a GW detection are and will be very large (from hundreds of degree$^2$ to $\sim1$ degree$^2$). This situation makes the identification of the EM counterpart particularly challenging, also because of the high number of transient phenomena in the Universe. The radio emission of the candidate counterparts can play a key role in disentangling which is the event associated with the GW detection (see e.g. Fig.\,\ref{fig:GRBaft2}).
The future radio wide field surveys, especially with the ASKAP and SKA arrays, will make a step forwards in detecting the orphan afterglows (see Sect.\,\ref{sci:GRB}) also of short GRBs. As orphan afterglow are believe to be hundreds of time more frequent than on-axis GRBS (see Sect.\,\ref{sci:GRB}), their detection is a key to find EM counterparts of GWs. Due to its frequency range, it is not clear if LOFAR and NenuFAR can also play a role in this sense (see Vergani \& Chassande-Mottin 2015).
The {\it ThunderKAT} group of the SKA precursor MeerKAT is focussing its effort also to the search of EM counterparts of GW events, with an observational strategy aiming at detecting the emission coming from GRB afterglows and kilonov\ae \,(see also Hotokezaka et al. 2016).
The information coming from the GW detection itself can be particularly important to set the observational strategy. Indeed, as shown by Salafia et al. 2017, such optimised strategy is able to reduce by a factor of ten the time needed to observe and detect the source, and increase the success of detections as, following this strategy, the observation of the counterparts will take place at epochs at which the source is still enough bright to be detected.\\

\parbox{0.9\textwidth}{
\noi{References:}\\
\noi{\scriptsize Ghirlanda, G., et al., 2015, A\&A, 578, 71;
Hotokezaka, K., et al., 2016, ApJ, 831, 190;
Salafia, O.S., et al., 2017, ApJ submitted;
Vergani, S.D. \& Chassande-Mottin, E., 2015, SF2A-2015: Proceedings of the Annual meeting of the French Society of Astronomy and Astrophysics. Eds.: F. Martins, S. Boissier, V. Buat, L. Cambr\'esy, P. Petit, pp.253-256
}}\\

\paragraph{Neutron stars magnetosphere and wind} 
\vspace{0cm}

\noi
With its unprecedented sensitivity, SKA will increase the number of detected radio-pulsars by one order of magnitude. We expect to observe about 10.000~pulsars or more, offering a high quality sample of neutron star population to investigate the still unsolved fundamental physics problem of their coherent radio emission. SKA will be able to diagnose the plasma regime given useful insight into the magnetosphere and identify the possible sites of radio wave generation.

\smallskip
\noi Amazingly, even after almost fifty years of intensive activity, our current understanding of the pulsar machinery stays on a rather low basic level. Apart from being a strongly magnetised rotating neutron star, further claims remain almost pure speculation. This is widely due to the lack of modeling based on a self-consistent use of first principles of electrodynamics, leaving our knowledge in an unsatisfactory state.

\smallskip
\noi According to current wisdom, the low-frequency spectrum of pulsed emission is required to be a coherent emission mechanism able to radiate very efficiently in the radio-band because the brightness temperature is much too high, larger than $10^{25}$~K which is unphysical. The details of this radio-wave generation are still unclear and highly debated within the community. The location of the emission region is also loosely constrained by observations so far. However, due to its large bandwidth and sensitivity, extending from 50~MHz up to 14~GHz (albeit the 1.65 - 5GHz band remains uncovered), the theory of radio emission will be underpinned by SKA measurements and eventually fully understood. Detailed analysis of single pulses observations achievable with SKA should unequivocally unveil the radio mechanism and open a new window on pulsar variability, beyond stationary models of their magnetospheres.

\smallskip
\noi  \textbf{Synergies with high energy emission physics}

\smallskip
\noi The discovery of more than 200 gamma-ray emitting pulsars, including the doubling of the number of known millisecond pulsars, has opened a new era in the field. The abundance of measurements in the broadband spectrum, e.g., light curves and pulse profiles, radio polarisation of single pulses, time lag between radio and high energy pulses,  have brought unprecedented constraints on high-energy emission models. The acceleration and emission zones can no longer be confined to the polar caps, but have to be pushed towards the outer magnetosphere, close to, or even beyond the light-cylinder in the wind. In addition, the magnetosphere can no more be considered as purely vacuum or force-free, but resistive (Kalapotharakos et al. 2014).  Significant contributions have been made by the French community to these recent advances. On the observational side, the synergy between radio and gamma-ray astronomers, working with the Nan\c cay radio telescope and the Fermi-LAT (e.g. Abdo \etal 2009, 2013), has been instrumental in the discoveries. In the phenomenological field, important results have been achieved through both analytical and numerical modeling works, e.g., the striped wind model (e.g., P\'etri 2016a) and the PIC\footnote{Particle In Cell} simulations (e.g., Cerruti \etal 2016).
These efforts should continue in the SKA era, in conjunction with the Fermi-LAT (of which the extension on the long term is likely), and the CTA observatory, where the French researchers are heavily involved.

\smallskip
\noi \textbf{Giant Radio Pulses (GRP)} 

\smallskip
\noi Giant radio pulses have an amplitude typically 1000 times above regular radio pulses, but their duration is much shorter from a few nanoseconds to a few microseconds. They  are seen only in a handful of pulsars and they are (so far) not correlated with optical, X, and gamma ray observations, neither with any properties of their host, apart from a high magnetic field at the light cylinder (Cognard \etal 1996). Theories include solitons (Mikhailovskii \etal 1985), scattering of photon beams of different frequencies (Petrova 2004), magnetic reconnection near the light cylinder (Lyutikov 2007), modulational instability  (Weatherall 2001). Thanks to its high sensitivity and wide spectral range, SKA will allow a characterisation of the instantaneous spectra of galactic GRPs and help for the first time to discriminate the various models.  For instance, (Lyutikov 2007) has shown that the Crab GRPs observations (e.g. Hankins 2007) are compatible with L-O mode triggered by magnetic reconnection in the equatorial region of the light cylinder. Because the L-O mode has a lower cut-off frequency, the extension toward lower frequencies of the emission band spectrum could confirm or invalidate this theory. Beyond the problem of waves in the magnetosphere, SKA observations will give clues to the important question of the dissipation of magnetic energy (here through reconnection) in the pulsar magnetosphere.  SKA will also allow for the first time search for giant pulses in other Local Group galaxies. The expected better statistics will help tracing the history of massive star formation in other galaxies (Kondratiev \etal 2013), and in the same time provide constraints on the inter-galactic medium through the analysis of propagation properties (see e.g. Sect.\,\ref{sci:pulG}).

\smallskip
\noi  \textbf{Magnetic field topology}

\smallskip
\noi The great precision and the detailed observations coming from current telescopes forces our modeling of pulsars to be increasingly refined, including general-relativistic and quantum electrodynamics effects. Such corrections largely underestimated in past works, become more popular to theoreticians because these effects clearly demonstrate their inevitable usefulness to understand pulsar magnetospheres. 

\smallskip
\noi A fundamental issue about neutron star magnetosphere concerns the question of the magnetic field topology. Until very recently, pulsar light-curves have been exclusively computed according to a dipole magnetic field structure, which is moreover located exactly at the centre of the star. Such assumption is highly restrictive and difficult to justify as it is a very special case very unlikely to occur in nature. This led P\'etri (2016b) to relax this assumption by assuming a off-centred dipole. Such geometry induces higher order magnetic multipoles that can be computed analytically (P\'etri 2015). Small scale magnetic field structures (Harding \& Muslimov 2011) as well as general relativity (Philippov \etal 2015; Belyaev \& Parfrey 2016) are required to enhance the pair creation rate around the polar caps. Such distortions from the standard view of a dipole impact strongly on the phase-resolved polarisation properties and are under scrutiny by French theoreticians.

\smallskip
\noi \textbf{Polarisation}

\smallskip
\noi Guided by near future measurements of X-ray polarisation emanating from neutron stars as performed by the IXPE mission (Imaging X-ray Polarimetry Explorer), for a launch expected in~2020, in addition to observations available in optical polarisation and the advent of the SKA project, the French community is highly motivated by a thorough multi-wavelength study of the phase-resolved polarisation properties of pulsed emission. Accurate fitting of pulse profiles polarisation have already been performed for the Crab in optical (P\'etri 2005) and predictions made in X-rays (P\'etri 2013) in the frame work of the striped wind. polarisation can severely constrain competing emission models (polar cap, slot gap, striped wind) as demonstrated by (Dyks \etal 2004). In the radio band, the plethora of observations concerning polarised emission of hundreds of pulsars still suffers from a lack of convincing explanations by a unique and reliable model. The quantity and complexity of the data collected in radio does not allow a synthetic description of these observations. Nevertheless, radio polarisation is a crucial and complementary aspect of optical/X-ray/gamma-ray observations to constrain the polar cap geometry (Radhakrishnan \& Cooke 1969). Radio astronomy is at the dawn of a new era with SKA. An off-centred dipole generates a clear and distinctive radio-polarisation signature (P\'etri 2017) very different from the traditional rotating vector model (Radhakrishnan \& Cooke 1969). Single pulse observations acquired by SKA will alleviate the degeneracy in the determination of the magnetosphere geometry. \\

\parbox{0.9\textwidth}{
\noi{References:}\\
\noi{\scriptsize 
Abdo, A. A., \etal, 2009, Science, 325, 848;
Abdo, A. A., \etal, 2013, ApJS, 208, 17;
Belyaev, M. A. \& Parfrey, K., 2016, ApJ, 830, 119;
Cerutti, B., \etal, 2016, MNRAS, 457, 2401;
Cognard, I., \etal, 1996, ApJL, 457, L81;
Dyks, J., \etal, 2004, ApJ, 606, 1125;
Hankins, T. H. \& Eilek, J. A., 2007, ApJ, 670, 693;
Harding, A. K. \& Muslimov, A. G., 2011, ApJ, 726, L10;
Kalapotharakos, C., \etal, 2014, ApJ, 793, 97;
Kondratiev, V., \etal, 2013, IAUS, 291, 431;
Lyutikov, M., MNRAS, 2007, 381, 1190;
Mikhailovskii, A. B., \etal, 1985, Soviet Astronomy Letters, 11, 78;
Petrova, S. A., 2004, A\&A, 424, 227;
Philippov, A. A., \etal, 2015, ApJL, 815, L19;
P\'etri, J. \& Kirk, J. G., 2005, ApJL, 627, L37;
P\'etri, J., MNRAS, 2011, 412, 1870;
P\'etri, J., MNRAS, 2013, 434, 2636 ;
P\'etri, J., MNRAS, 2015, 450, 714;
P\'etri, J., J.Pl.Ph., 2016a, 82, 635820502;
P\'etri, J., MNRAS, 2016b, 463, 1240;
P\'etri, J., MNRAS, 2017, 466, L73;
Radhakrishnan, V. \& Cooke, D. J., 1969, ApL, 3, 225;
Weatherall, J. C., 2001, ApJ, 559, 196
}}\\

\paragraph{Cosmic rays}
\vspace{0cm}

\noi Extensive atmospheric air showers (EAS)  induced by cosmic rays (CRs) emit radio pulses with a duration of tens of nanoseconds which illuminate areas of km$^2$ scale on the ground.  These radio signals can be used to infer the CR arrival direction, energy and mass composition.  The mass composition in particular is a very important quantity to study the origin and propagation of high energy CRs, which in spite of many decades of research are still among the great open questions in astroparticle physics.

\smallskip
\noi Several experiments are currently exploring the radio signal of EAS. At LOFAR, high-precision mass composition measurements based on CR radio detection are currently being performed in the 10$^{17}$--10$^{18}$~eV range (Buitink et al. 2016).  The Auger Engineering Radio Array (AERA), located at the Pierre Auger cosmic ray observatory in Argentina, performs radio measurements up to energies of $\approx 10^{19}$~eV, but with a much sparser antenna spacing than LOFAR, making the determination of the particle mass much more challenging (Gat\'e et al. 2016). CODALEMA at Nan\c cay is exploring various solutions to enrich and to improve the radio detection technique such as detecting on very large frequency band (2 - 200 MHz), pining down a new low frequency signal, improving the CR self-triggering system (Dallier et al. 2015). In the future, due to the large area of its core and its wide frequency coverage of 50-350 MHz, SKA-LOW will be very well-suited to perform high-precision CR composition measurements in the full energy range of 10$^{17}$--10$^{19}$~eV (Huege et al. 2016).  This is the energy range in which the transition from galactic to extragalactic cosmic rays is expected.  High-precision CR measurements with SKA-LOW can significantly improve our understanding of CR physics in this transition region, e.g. by disentangling individual CR mass groups such as proton, helium, silicon and heavier particles.  Finally, combining the expertise on CR detection and the tremendous amount of antennas, SKA might be a unique opportunity to search for and exploit in a beam-formed mode the radio counterpart of the gamma-ray induced air showers of identified gamma-ray sources.

\smallskip
\noi As it has already been shown at LOFAR, on a digital radio telescope observations of atmospheric air showers can run fully concurrently with other observation modes such as imaging.  However, for this it is necessary to shortly buffer all raw data of the SKA-LOW core antennas in ring buffers.  Upon an external trigger provided by a particle detector array, a few microseconds of the raw (waveform) data of the individual antennas need to be stored for analysis.  As the arrival direction of CRs is unknown a priori, pre-beamformed data is not suitable for CR studies through atmospheric air showers. The low-RFI environment of the SKA might also allow triggering from the pulsed radio signals themselves without the need to deploy a particle detector array.

\smallskip
\noi The authors of this proposal are the promoters of the CODALEMA experiment at the Nan\c cay radio observatory, dedicated since 2002 to the radio detection of cosmic ray air showers (Torres-Machado, 2013). Since 2014, together with an European team of researchers from Germany and Netherlands, they propose to use the NenuFAR telescope, completely surrounded by the CODALEMA detectors and which could be externally triggered on a cosmic ray signal detected by CODALEMA, to serve as a pathfinder for the SKA-EAS operating mode (Dallier, 2015). They are involved in the ``SKA Focus Group'' on cosmic ray physics, and have submitted several contributions to the future SKA design and programs including the possibility to test the concepts in Nan\c cay (Huege et al. 2014). The physics goals of this observing program have been extensively presented in several conferences (Huege 2014, 2015 and 2016).\\

\parbox{0.9\textwidth}{
\noi{References:}\\
\noi{\scriptsize 
Buitink, S., \etal, 2016, Nature 531, 70;
Gat\'e F., for the Pierre Auger Collaboration, 2016, 7$^{th}$ International Conference on Acoustic and Radio EeV Neutrino Detection (ARENA 2016), Groningen, The Netherlands;
Dallier, R., \etal, 2015, 34$^{th}$ ICRC, The Hague, Netherlands;
Huege, T., \etal, 2016, 7$^{th}$ International Conference on Acoustic and Radio EeV Neutrino Detection (ARENA 2016), Groningen, The Netherlands;
Torres-Machado, D., \etal, 2013, (CODALEMA Coll.), 33$^{rd}$ ICRC, Rio de Janeiro, Brazil;
Huege, T., \etal, 2015, AASKA14, 148;
Huege, T., \etal, 2015, 34$^{th}$ ICRC, The Hague, Netherlands
}}\\

\newpage
\subsection{Fundamental physics}

\smallskip

\noi {\sffamily \scriptsize
{\sffamily \bf {\bf S.~Babak}} [\apc],
{\bf L.~Blanchet} [\iapsorb],
{\bf P.~Charlot} [\lab],
{\bf I.~Cognard} [\lpcee],
{\bf A.~F.~Fantina} [\iaa;\ganil],
{\bf L.~Guillemot} [\lpcee;\usn],
{\bf F.~Gulminelli} [\lpc],
{\bf S.~Lambert} [\syrte],
{\bf A.~Le Tiec} [\luth]
{\bf J.~Margueron} [\usa;\ipn],
{\bf J.~Novak} [\luth],
{\bf M.~Oertel} [\luth],
{\bf A.~Petiteau} [\apc],
{\bf G.~Theureau} [\lpcee;\usn;\luth],
}

\subsubsection{Pulsar timing arrays as gravitational wave detectors}
\vspace{0cm}

\noi The timing of an array of stable millisecond pulsars (PTA) works as a Galactic sensor for gravitational wave direct detection 
in the frequency regime from nHz to $\mu$Hz. 
It is complementary to ground based (LIGO-Virgo, Hz-kHz) and future spatial (LISA, mHz) interferometers. 
This technique is the unique one probing the emission of the supermassive black hole binary (SMBHB) population that was formed in the hierarchical large scale 
structure and galaxy building scenario. Thanks to the tremendous results from LIGO-Virgo announced in 2016, we now know that gravitational waves exist. 
Even though PTAs have yet produced only upper limits on the nHz background or individual sources gravitational radiation, the recent results start to constrain 
models of structure formation, the growth of the galaxy central black hole and  the migration rate of the black hole binaries formed in mergers towards the 
gravitational wave emission regime. With improving radio telescope detectors and data analysis techniques, a detection is expected within the coming 
5-10 years with the current generation of instruments, while we will have to wait for SKA1 to fully confirm preliminary detections, identify and characterise 
the sources and their spectrum.

\smallskip

\noi \textbf{NRT, EPTA and IPTA.}

\smallskip

\noi The "pulsar group" in Orl\'eans (LPC2E, CNRS), associated with the scientific exploitation of the decimetric Nan\c cay Radio Telescope (NRT) 
and with the construction of a new low frequency compact array NenuFAR in France, has been one of the founding members of the European Pulsar Timing Array (EPTA). 
This consortium was created in 2006 in a kick off meeting in Paris Observatory, and it gathers scientific teams associated with the five 100-m 
class radio telescopes of the continent: Jodrell Bank, UK, Westerbork, NL, Effelsberg, Ger, Cagliari, It, Nan\c cay Radio Telescope (NRT), Fr. 
Pulsar timing observations represent a large fraction of telescope time at all these facilities, with a total amount of more than 
3000 hours dedicated per year, since the start of the project.
EPTA members are collaborating for upgrading dedicated pulsar backends with state of the art techniques and they participate 
actively to receiver developments at each telescope. In this respect, NRT is equipped with one of the most productive 
(60\% of European timing data in the period 2006-2017) and most  precise timing instrumentation in the world (NUPPI back-end, Desvignes et al. 2011). 
Its pioneering development of coherent pulsar dedispersion on wide frequency band in L and S domains (currently 512 MHz, 
2 GHz of bandwidth in preparation) make the French group one of the leaders in the field. 
The European collaboration has also been the framework for the LEAP-ERC experiment, 
demonstrating the feasibility of a continental phased array of the five 100-m class telescopes and its optimisation for pulsar timing. 
This composite radio telescope reaches the same sensitivity as the Arecibo dish, but on a much greater area on the sky (Bassa et al. 2016). 

\begin{figure}
 \centering
    \begin{subfigure}[c]{0.6\textwidth}
        \includegraphics[width=\textwidth]{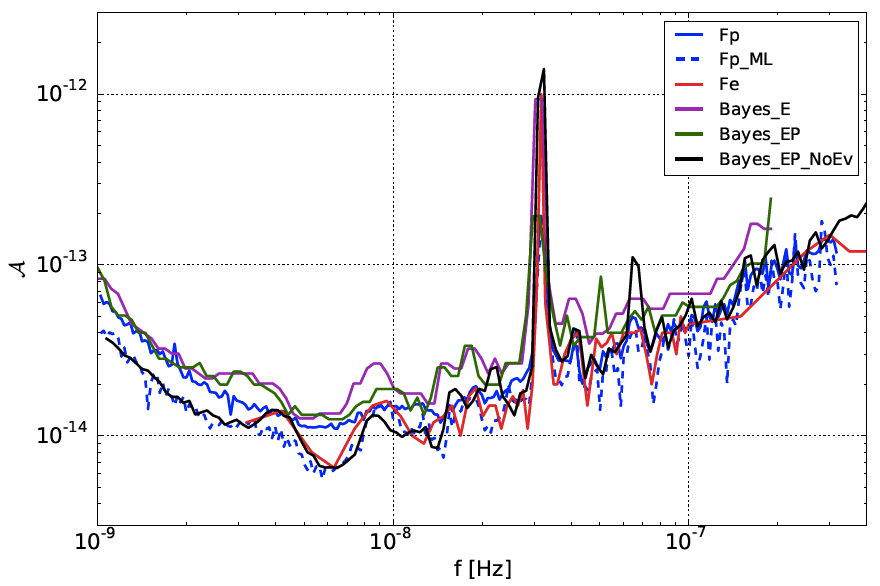}
        \caption{From Babak et al. 2016: first 95\% upper limits on the gravitational wave strain from a super massive black hole binary, as a continuous single source (for 3 frequentist methods and 3 bayesian methods). 
        This results exclude the existence of SMBHBs with separation $<$ 0.01pc and chirp mass M$_c$ $>$ 10$^9$ M$_{\odot}$ out to a distance of about 25Mpc (well beyond Virgo), and with M$_c$ $>$ 10$^{9.5}$ M$_{\odot}$ out to a distance of about 200 Mpc (twice the distance to Coma).}
        \label{fig:PTAFig1a}
    \end{subfigure}
    \quad
     \begin{subfigure}[c]{0.33\textwidth}
        \includegraphics[width=\textwidth]{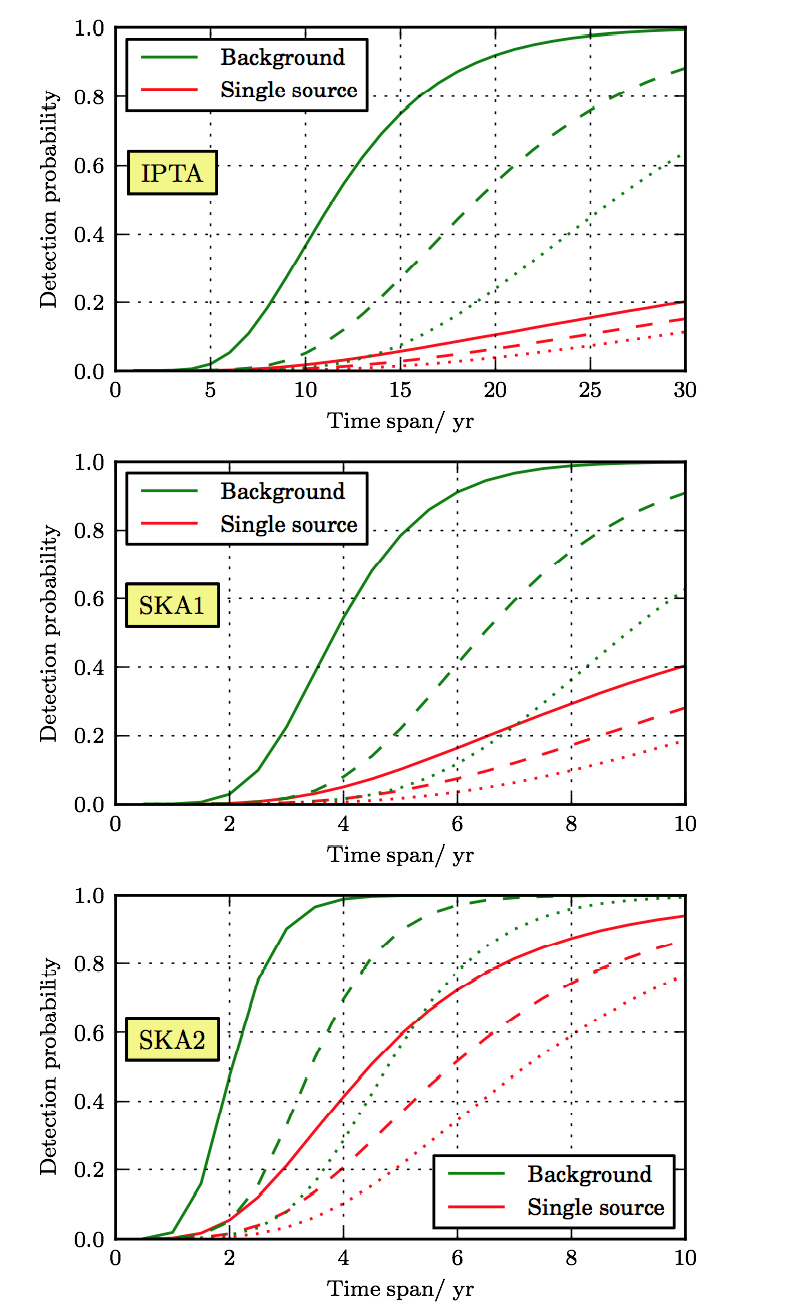}
        \caption{Probability of gravitational wave signal detection with IPTA,  SKA1 and SKA2 (Rosado, Sesana, Gair 2015)}
        \label{fig:PTAFig2a}
    \end{subfigure}
\end{figure}

\smallskip

\noi At the world wide level, the International Pulsar Timing Array (IPTA) is presently made of the contributions of the three continental consortia: 
EPTA in Europe, PPTA in Australia and NanoGRAV in the US. With the development of FAST in China, of MeerKAT in South Africa, 
and the recent evolutions of the GMRT in India, it will rapidly include new groups around the planet.
This consortium, with its long term program and world wide organisation, is the cornerstone of 
the pulsar SKA Science Working Group (SSWG), and make it one of the most structured Science Groups 
of the SKA organisation. The authors of this section are involved at different level in this international structure:
members of the SSWG, members of the EPTA steering committee and IPTA board, Co-Is of the TRAPUM and MeerTime
key projects with the SKA precursor MeerKAT.

\smallskip

\noi \textbf{Gravitational astronomy with PTA in the SKA context.}

\smallskip

\noi In 2015 and 2016, in phase with a first massive data release (11 years times series, Desvignes et al. 2016; Caballero et al. 2016), the French
NRT team and EPTA/IPTA collaborations have produced a first set of long term results,
with a major impact on the gravitational emission in the nHz regime and on the population of supermassive black hole binaries 
(e.g. Lentati et al. 2015; Babak et al. 2016; Taylor et al. 2015; Verbiest et al. 2016; Lentati et al. 2016). In the meantime, the results from a large set 
of short term projects have been published, led by the young scientists among the collaboration and based on individual source astrophysics 
and gravity tests in binaries (5-10 publications/yr; see Sect.\,\ref{sci:GT}).

\smallskip

\noi For high precision timing, SKA will both bring a leap in sensitivity and versatility.  The SKA survey will provide us with a large number of new stable millisecond pulsars
(1500 MSPs expected with SKA1). The gain in signal to noise ratio  (S/N) being directly convertible in timing accuracy, the predicted 10 times better sensitivity 
with respect to current 100-m class telescopes shall bring up to 200 the number of pulsars timed better than 500 ns over 10 years, with 10-20 around or better 
than 100 ns rms. The wide frequency domain accessible with the SKA (50MHz-13.8 GHz) will allow fine studies of the interstellar medium imprints on the signal 
and a better monitoring of the associated red noise component which perturbs the gravitational signal detection. In this respect, the multi-telescope Nan\c cay 
facility will provide us with pioneering results, thanks to the possibility of simultaneous observation campaigns with NenuFAR, LOFAR-FR606 and the NRT in 
the range 15MHz-3.5 GHz. Pulse jittering is also a critical source of noise in the timing residuals, typically at the 100 ns rms level, it produces a higher threshold 
with shorter integrations, and it can thus limit the detection of gravitational waves with a highly sensitive radio telescope like the SKA, where a natural tendency 
would be to reduce observing time of the brightest sources. The two ways to address this problem are: 1) using the sub-arraying capability of SKA and doing multiple parallel 
long exposures observations; 2) using the existing 100-m class telescopes to monitor on the long term most of the bright millisecond pulsars used in PTAs. 
In such a context, the low declination limit (-39$^{\circ}$) of the NRT would make it a precious complement to SKA pulsar observations.

\smallskip

\noi While we expect to detect the first signal from the stochastic gravitational background or the first single super massive black hole binary with the IPTA, SKA1 
should definitely confirm and characterise the sources or populations associated to this detection. On a longer time scale, full SKA will be the only instrument allowing 
to really exploit gravitational astronomy in the nHz-$\mu$Hz domain.\\

\parbox{0.9\textwidth}{
\noi{References:}\\
\noi{\scriptsize 
Babak, S., \etal, 2016, MNRAS,  455, 1665;
Bassa, C. G., \etal, 2016, MNRAS 456, 2196;
Caballero, R. N., \etal, 2016, MNRAS,  457,  4421;
Desvignes, G., \etal, 2011, AIP Conference Proceedings, Volume 1357, pp. 349-350;
Desvignes, G., \etal, MNRAS, 2016, 458,  3341;
Lentati, L., \etal, MNRAS, 2015, 453,  2576;
Lentati, L., \etal, MNRAS, 2016, 458, 2161;
Rosado, P. A., \etal, 2015, MNRAS 451, 2417;
Taylor, S. R., \etal, PhRvL, 2015, 115,  041101;
Verbiest, J. P. W., \etal, MNRAS, 2016, 458,  1267
}}\\

\subsubsection{Binary pulsars as natural laboratories to test Gravitation theories} \label{sci:GT}

\vspace{0cm}

\noi With orbital periods of only a few hours, pulsar/white-dwarf or pulsar/neutron-star compact binary systems are natural laboratories for testing Gravity theories in the strong field regime. In the most interesting systems, the timing precision reached by state of the art radio observations allows us to measure all Keplerian and several post-Keplerian orbital parameters in only a few years. Beyond the accurate measurement of both masses, which is crucial for constraining neutron star internal structure models (see Sect.\,\ref{sci:NS}), one can accurately measure the relativistic precession of the periastron, a combination of the gravitational redshift and the second order Doppler effect, the Shapiro time delay (both its range and shape), and the gravitational damping of the orbit due to gravitational wave emission (see e.g. Kramer et al. 2017). One can also measure the spin-orbit coupling effect implying geodetic orbital precession and hence get information on the spins. Finally, one can test for possible variations of the gravitational constant, produce the most stringent test of scalar-tensor gravity, and of some alternative theories motivated by dark matter (Freire et al. 2012). Such tests of alternative theories with binary pulsars are very important for the direct detection of gravitational waves by LIGO/VIRGO, since they permit to restrict the number of alternative theories to be considered for building gravitational-wave templates to be used in the data analysis of these detectors.

\smallskip

\noi \textbf{Recent results obtained with NRT and other 100-m class radio telescopes} 

\smallskip

\noi The Nan\c{c}ay decimetric Radio Telescope (NRT) is one of the rare instruments on Earth able to provide both timing quality and data rate for this kind of scientific application, in particular for densely and correctly sampling the orbital phase of such systems. Since 2006, 
90 binary pulsars are then monitored on a regular basis, with time series extending beyond a decade. Combining these NRT timing data with those of other antennae (Effelsberg, GBT, Arecibo, Parkes, Westerbork, Jodrell Bank) the Nan\c cay group obtained several results : 

\begin{figure}
 \centering
    \begin{subfigure}[c]{0.54\textwidth}
        \includegraphics[width=\textwidth]{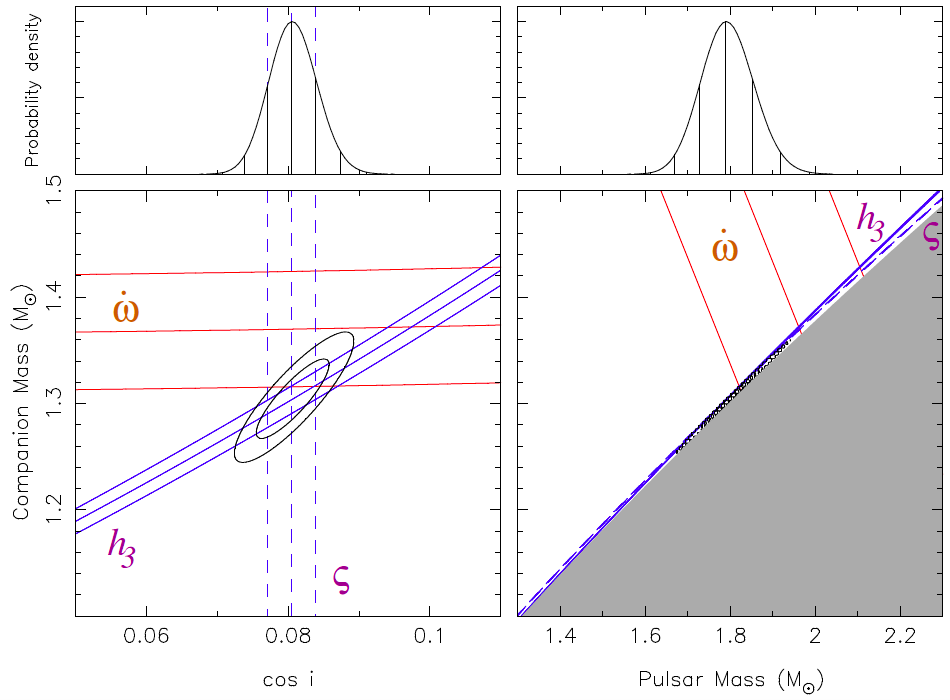}
        \caption{From Cognard et al. 2017: Current constraints from timing of PSR J2222$-$0137. Each triplet of lines corresponds to the nominal and 1 $\sigma$ uncertainties of the post-Keplerian parameters: rate of advance of periastron $\dot{\omega}$ (solid red lines) and, in blue, orthometric parameters for the Shapiro delay  (dashed and solid). The black contour lines include 68.3 and 95.4\% of the 2-D probability distribution function derived from $\chi^2$ maps.}
        \label{fig:PTAFig1}
    \end{subfigure}
    \quad
     \begin{subfigure}[c]{0.36\textwidth}
        \includegraphics[width=\textwidth]{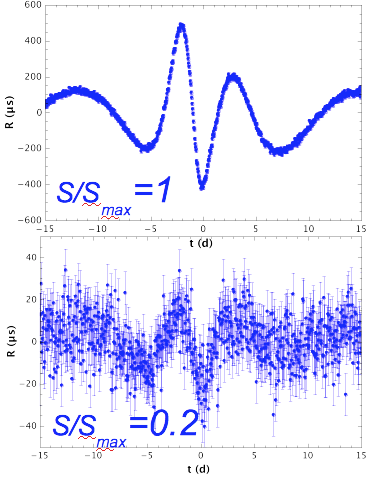}
        \caption{From Psaltis, Wex \& Kramer 2016: characteristic timing residuals from quadrupole moment with spin 100\% ({\em top}) or 20\% ({\em bottom}) of maximum}
        \label{fig:PTAFig2}
    \end{subfigure}
\end{figure}

\begin{itemize}

\item first Shapiro delay measurement and masses for the neutron-star/white-dwarf (NS/WD) system PSR J1802$-$2124 (Ferdman et al. 2010);  
\item constraints on PSR J1012+5307's eccentricity, variation of semi-major axis and orbital period, upper limit on dipolar emission and on the variations of the gravitational constant G (Lazaridis et al. 2009);
\item evidence of a secular variation of PSR J2051-0827's projected semi-major axis and interpretation in terms of spin-orbit coupling, tidal effects and associated variations of the quadrupole moment (Lazaridis et al. 2011); 
\item discovery of a strong magnetic field around the supermassive black hole at the centre of the Galaxy (Eatough et al. 2013); the measurement of five post-Keplerian parameters in the double neutron star PSR J1756$-$2251 and a series of tests of General Relativity (Ferdman et al. 2014); 
\item evidence for geodetic precession in PSR J1906+0746 (157 years) and accurate measurement of both neutron star masses (Desvignes et al. 2008, Van Leeuwen et al. 2015, Desvignes et al. 2017); 
\item pulsar timing evidence for an intermediate-mass black hole in the globular cluster NGC 6624 (Perera et al. 2017); 
\item accurate mass measurements for both components of the PSR J2222$-$0137 binary and of the rate of advance of periastron, constraints on dipolar radiation and spontaneous scalarization (Cognard et al. 2017).
\end{itemize}

\smallskip

\noi \textbf{Testing Gravity in the SKA era}

\smallskip

\noi Current radio timing observations of binary pulsars enable tests of the quasi-stationary strong-field regime in the first order post-Newtonian approximation and of gravitational wave damping at leading order (i.e. second-and-a-half post-Newtonian order). With its higher sensitivity and higher timing accuracy, SKA will revolutionize the field and allow us to measure other post-Keplerian parameters and higher order effects. We expect in particular more than a hundred of double neutron star (NS) systems, with phase resolved short period orbits. At 50 ns of timing precision, we know that the Shapiro effect will be detectable at inclinations down to 40$^{\circ}$ and lead to 5 times more NS masses and also new geodetic precession measurements. This precision will also allow to observe frame dragging from apparent semi-major axis variations in tight binaries with large spin/orbit angle, and to measure the Lense-Thirring effect through its contribution to $\dot{\omega}$ thanks to the combination of accurate Shapiro delay, variation of the orbital period $\dot{P_b}$ and VLBI distance. These data will also be unique for the measurement of NS moments of inertia and contraints on the Equation of State (see Sect.\,\ref{sci:NS}). 

\smallskip

\noi The high timing precision reached by SKA will allow exploring new tests of the strong equivalence principle, and gravitational dipole radiation, e.g. due to a scalar field component of gravitation.  The discovery of new highly asymmetrical NS-WD systems  such as PSR J1738+0333 and PSR J0348+0432 will offer a unique test of the effacement property in so-called scalar tensor theories (see e.g. Wex 2014, Berti et al. 2015)
or allow to constrain the differential free fall in the Milky Way gravitational potential (the so-called Damour-Sch\"afer test). Finally, the possible discovery of a triple system such as the recently discovered PSR J0337+1715 (Ramson et al. 2014), but with an external neutron star (instead of a white dwarf) would considerably improve the test of the universality of free fall in a strong external field. Such triple systems are also interesting for checking Newtonian resonant orbital effects such as the Kozai-Lidov mechanism.

\smallskip

\noi All gravitational astronomers know that the holy grail would be the discovery of a neutron-star/black-hole system. Liu et al. (2012, 2014) showed for example that with a fast millisecond pulsar orbiting a 10-30 M$_{\odot}$ black hole (BH) like those discovered with LIGO, SKA could measure the quadrupole moment with enough precision to constrain the BH mass to better than 0.1\% and its spin better than 1\%. Such a discovery will also allow us a direct test of the cosmic censorship conjecture (there exists a maximum spin, S $<$ G M$^2$/c). Liu et al. (2012) and Psaltis, Wex \& Kramer (2016) also showed that with a pulsar orbiting the central Milky Way black hole Sgr A* in 0.1 year, we would be able to test the no-hair theorem to about 1\% precision, thanks to a high amplitude (a few 100 $\mu$s !) and the characteristic periodic effect of the quadrupole moment on the timing residuals. \\

\parbox{0.9\textwidth}{
\noi{References:}\\
\noi{\scriptsize 
Cognard, I., \etal, 2017, ApJ, submitted; 
Berti, E., \etal, 2015, CQG, 32, 243001;
Desvignes, G., \etal, in preparation; 
Eatough, R. P., \etal, 2013, Nature, 501, 391; 
Ferdman, R. D., \etal, 2010, ApJ, 711, 764; 
Ferdman, R. D., \etal, 2014, MNRAS, 443, 2183; 
Freire, P. C. C., \etal, 2012, MNRAS, 423, 3328;
Kramer, M., \etal, 2017, in preparation; 
Lazaridis, K., \etal, 2009, MNRAS, 400, 805;
Lazaridis, K., \etal, 2011, MNRAS, 414, 3134;
van Leeuwen, J., \etal, 2015, ApJ, 798, 118;
Liu, K., \etal, 2012, ApJ, 747, 1;
Liu, K., \etal, 2014, MNRAS, 445, 3115; 
Perera, B. B. P., \etal, 2017, MNRAS, 468, 2114; 
Psaltis, D., \etal, 2016, ApJ, 818, 121;
Ransom, S. M., \etal, Nature, 2014, 505, 520;
Shao, L., \etal, 2015, AASKA14, 42;
Wex, N., 2014, eprint arXiv:1402.5594
}}\\

\subsubsection{Neutron star equation of state}\label{sci:NS}
    
\vspace{0cm}
    
\begin{figure}[!ht]
    \includegraphics[width=.5\textwidth]{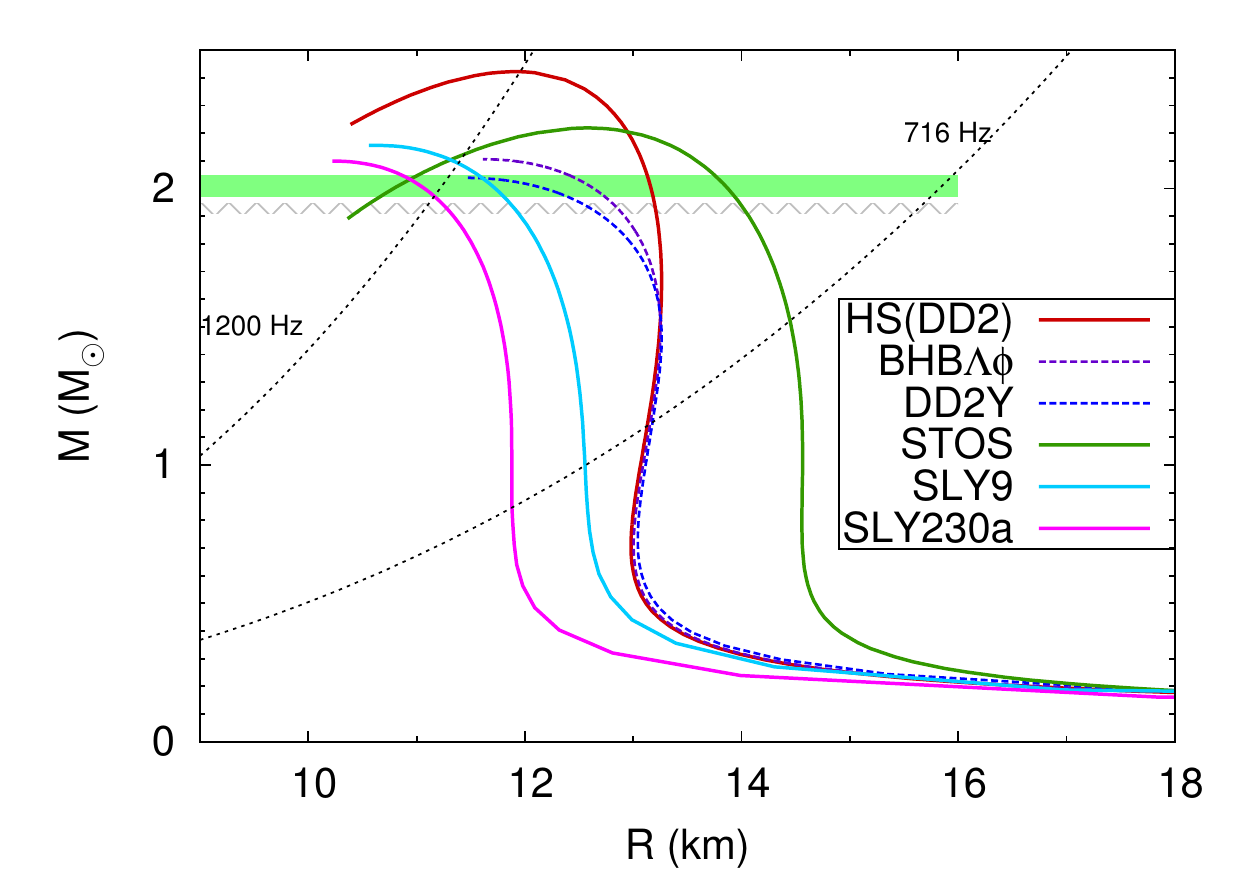}\hfill
    \includegraphics[width=.5\textwidth]{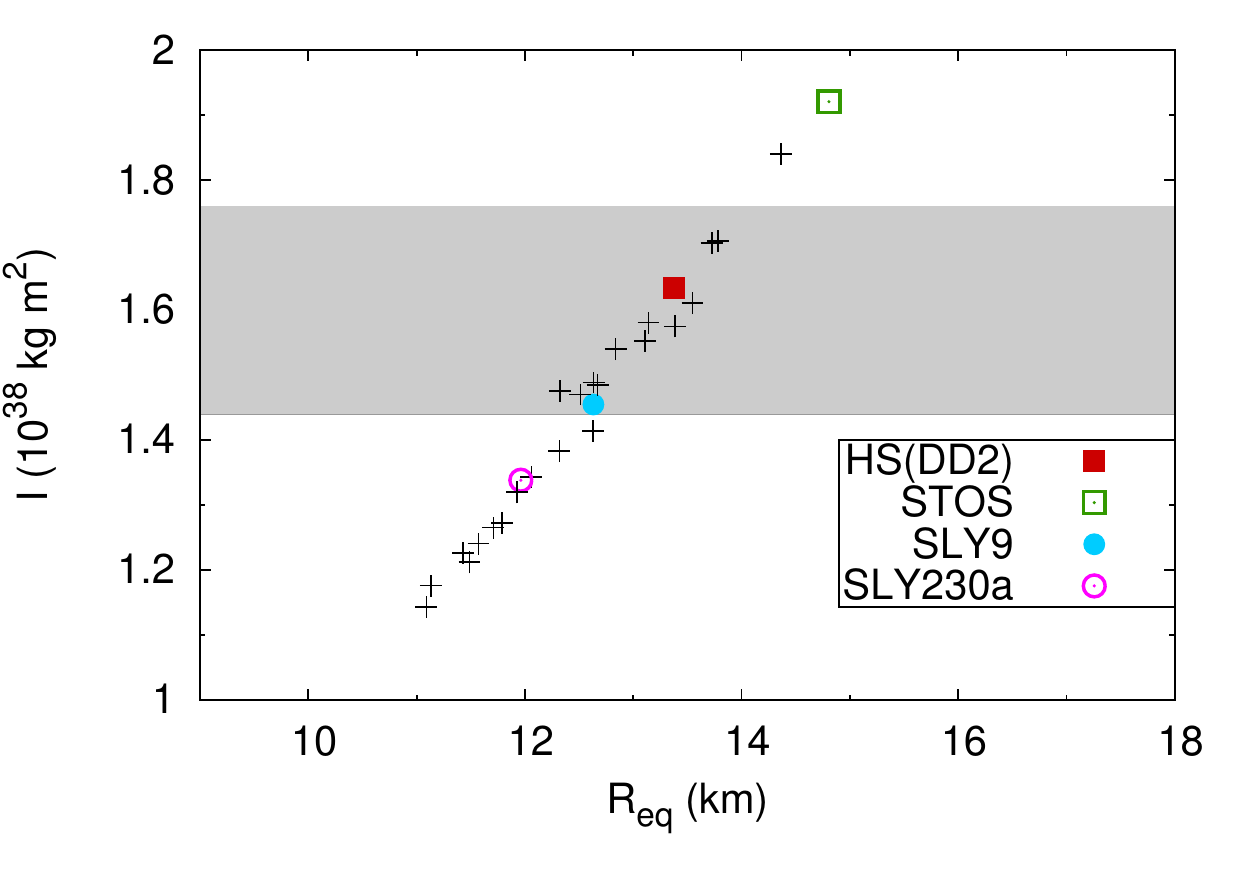}
  \caption{\label{fig:mr} {\em Left}: Gravitational mass ($M$)-radius ($R$)
    relation for non-rotating NSs employing different EoSs. Models
    producing NSs with a purely nucleonic core are indicated by solid
    lines and those with a core containing hyperons at high density by
    dashed lines. The two horizontal bars indicate the masses of the
    current most massive observed NSs. Dotted lines indicate the
    minimum radius for a given mass assuming rotation at a Kepler
    frequency of 716 Hz (the fastest spinning known pulsar) and 1200
    Hz, respectively. {\em Right}: Moments of inertia $I$ calculated with
    mass and rotation frequency of PSR J0737-3039A for different EoSs
    vs the equatorial NS radius ($R_{\rm eq}$). The horizontal grey
    bar indicates a possible measurement of the moment of inertia at
    10\% accuracy foreseen for SKA. Different EoS models are taken
    from the \textsc{Compose} data base (Typel et al. 2013) and calculations
    have been performed with the \textsc{LORENE} library (Gourgoulhon
    et al. 2016). }
\end{figure}

\noi Matter in the neutron star (NS) interior is compressed to
densities exceeding that in the centre of atomic nuclei, opening the
possibility to probe the nature of the strong interaction under
conditions dramatically different from those in terrestrial
experiments and to clarify the composition of compact stars. Several
models are considered and these stars can be either standard NSs
composed of nucleons and nuclei, or stars containing, in addition to
nucleons, strange baryons (hyperons) or mesons at the centre, or
possibly ``hybrid stars'', i.e., with a quark matter core, or even
pure strange quark stars. Via the equation of state (EoS), matter
properties uniquely relate mass and radius of the star, see
Fig.\,\ref{fig:mr} ({\em top-left} panel). The ultimate key to constraining the
NS EoS would thus be to measure masses and radii of NSs. The latter
are, however, difficult to determine and current values are subject to
large systematic uncertainties.

\smallskip
\noi \textbf{Masses} 

\smallskip
\noi Since, for a given EoS, there is a limit in the
mass for a NS (due to general-relativistic effects), the observation
of a massive NS can already exclude some EoS models, put constraints
on the strong interaction at high densities and answer questions about
the composition. In particular, pulsar timing measurements in double
NS systems and millisecond pulsars (MSPs) in binary systems can
deliver high precision NS masses (see also Sec. 4.6.2). The
SKA is expected to increase the total number of mass determinations by
roughly a factor 10 (Watts et al. 2015). The improved timing precision
will in addition greatly reduce uncertainties on already measured
masses. The discovery of a NS mass with a mass only 10\% higher
than the current record holders, PSR J 1614-2230 (Demorest et
al. 2010; Fonseca et al. 2016) and PSR J0348+0432 (Antoniadis et al
2013) would exclude many EoSs and thus seriously challenge our
understanding of dense matter, see Fig.\,\ref{fig:mr}, {\em left} panel.

\newpage
\noi  \textbf{Radii and  moments of inertia.}  
  
\smallskip
\noi Current standard techniques to determine NS
radii rely on \textit{i)} spectroscopic measurements of accreting
neutron stars in quiescence (Heinke et al. 2014) or \textit{ii)} on
the study of thermonuclear (type I) X-ray bursts (\"Ozel \& Freire
2016 and references therein), or \textit{iii)} timing observations of
surface inhomogeneities of rotating stars (see Miller \& Lamb 2016;
Haensel et al. 2016). Determining the moment of inertia $I$ for a NS
with well known mass would be a fascinating alternative way (Morrison
et al. 2004). As can be seen from Fig.\,\ref{fig:mr} ({\em right} panel), a
10\% precision on $I$ for the double pulsar, possible with the SKA in
less than 20 years (Watts et al. 2015), would constrain the radius to
roughly 10\%, too, better than present determinations with standard
techniques. In addition, SKA1(2) should discover up to 100 (180) new
double neutron star systems, and we can hope that some of them present
more favorable geometries for measuring $I$ with high precision, see
also Sec. 4.6.2.

\smallskip
\noi \textbf{Rotation frequencies} 

\smallskip
\noi The NS spin frequency $f$ must be
lower than the mass shedding limit, the Kepler frequency $f_K$. To a
very good approximation,
$f_K = C (M_g/m_\odot)^{1/2} \, (R/10\, \mathrm{km})^{-3/2}$, with the
constant $C = 1.08 $kHz for NSs (Haensel et al. 2009). More compact
stars support higher rotation rates and the discovery of a sub-ms
pulsar would severely constrain the EoS, see Fig.\,\ref{fig:mr},
{\em left} panel. The dotted lines indicate the minimum radius for a
given mass assuming rotation at a Kepler frequency of 716 Hz (the
fastest spinning known pulsar, PSR J1748-2446ad, Hessels et al. 2006)
and 1200 Hz, respectively. Even if SKA1-MID and SKA2 do not find
sub-millisecond pulsars, SKA1-MID will certainly discover at least
few MSPs spinning faster than PSR J17482446ad, while SKA2 will
deliver a full census of the population of the of the nearby, thus
unaffected by scattering and dispersion delays, ultra rapidly spinning
MSPs (Watts et al. 2015).

\smallskip
\noi \textbf{Glitches} 

\smallskip
\noi The high precision timing provided by the SKA
(see Sec. 4.6.1) might not only constrain the EoS, but could
improve greatly our understanding of dense matter superfluidity in the
interior of NSs through the study of glitches, small irregularities in
the pulsar's rotation rate. Several hundreds of glitches have been
observed during almost fifty years in more than 150 pulsars
(\href{http://www.jb.man.ac.uk/pulsar/glitches.html}{\color{blue} \myul[blue] {www.jb.man.ac.uk/pulsar/glitches.html}}, Espinoza et al. 2011), mainly trough long term monitoring programs. The slow post-glitch relaxation on time scales up to months or years is a clear
indication for a superfluid component in the NS interior (Baym et
al. 1969). A dedicated observational campaign on $\sim$20-30 objects
could explore fully the glitch size distribution and track the
recovery of all detected glitches. It might be possible, too, to
observe the glitch rise time, which then gives insight into the
superfluid NS properties (Sourie et al. 2017). All these can
contribute in an important way to a quantitative understanding of the
glitch mechanism.

\smallskip
\noi The French community contributes considerably to these efforts via

\begin{enumerate}
\item[a)] the development of the numerical library \textsc{LORENE} (\url{http://lorene.obspm.fr}) which allows to compute (rotating) NS models in general relativity and thus relate dense matter properties to NS observables (Gourgoulhon et al. 2016),\\
\item[b)] the development of new EoS models (Oertel et al. 2017; Gulminelli \& Raduta 2015; Tan et al. 2016; Fantina et al. 2016),\\
\item[c)] the development of the EoS data base \textsc{Compose} (\url{http://compose.obspm.fr}), which aims in collecting microphysics data for modeling compact stars,\\
\item[d)] the development of a collaboration between theoreticians and the experimental nuclear-physics group in GANIL; specifically, an experiment to measure properties of neutron-rich nuclei relevant for compact-star modelling has been accepted and will be performed in the next months (Bastin et al. 2016).
\end{enumerate}

\smallskip
\noi  This shows that strong links exist with the French experimental and theoretical nuclear-physics community.\\

\parbox{0.9\textwidth}{
\noi{References:}\\
\noi{\scriptsize 
Antoniadis, J., et al., 2013, Science, 340, 448;
Bastin, B., et al., 2016, JYFL experiment I220;
Baym, G., et al., 1969, Nature, 224, 673;
Demorest, P., et al., 2010, Nature, 457, 1081;
Espinoza, C.M., et al., 2011, MNRAS, 414, 1679;
Fantina, A.F., et al., 2016, Phys. Rev. C, 93, 015801;
Fonseca, E., et al., 2016, ApJ, 832, 167;
Gourgoulhon, E., et al., 2016, Astrophysics Source Code Library, 1608.018;
Gulminelli, F. \& Raduta, A.R., 2015, Phys. Rev. C, 92, 055803;
Haensel, P., et al., 2009, A\&A, 502, 605;
Haensel, P., et al., 2016 EPJA, 52, 59;
Heinke, C., et al., 2014, MNRAS, 444, 443;
Hessels, J., et al., 2006, Science, 311, 1901;
Miller, M. C. \& Lamb, F. K., 2016, EPJA, 52, 63;
Morrison, I. A., et al., 2004, ApJ, 617, L135;
Oertel, M., et al., 2017, Rev. Mod. Phys., 89, 015007;
\"Ozel, F. \& Freire, P., 2016, ARA\&A, 54, 401;
Sourie, A., et al., 2017, MNRAS, 464, 4641;
Tan, N.~H., et al., 2016, Phys. Rev. C, 93, 035806; 
Typel, S., et al., 2015, Phys. Part. Nucl., 46, 633;
Watts, A., et al., 2015, AASKA14, 43
}}\\

\subsubsection{Reference frames}\label{sci:RF}
\vspace{0cm}

\noi The determination of the motion of the celestial bodies through repeated measurements of their position using astrometric techniques constitutes the foundation of our present understanding of the Universe. For consistency in time and space, and hence comparisons for different celestial bodies, such determinations must be established in well-defined coordinate (or reference) systems. In practice, coordinate axes for celestial systems are not defined directly but only implicitly through the adoption of a set of fiducial directions, precisely identified and highly stable over long timescales. Specific bodies possessing the required properties are used to materialise such directions. On the celestial sphere, these form a grid of points whose two-dimensional coordinates define a reference frame.

\smallskip
\noi For the past 20 years, fundamental reference frames have been based on extragalactic sources, namely quasars or more generally active galactic nuclei (AGN). These objects possess highly compact central emission (of synchrotron origin), ensuring well-defined fiducial directions, and have no detected proper motions due to their location at cosmological distances. Furthermore, their positions can be determined with exquisite accuracies, below the milliarcsecond (mas) level, using Very Long Baseline Interferometry (VLBI), a radio interferometric technique utilising antennas spread over the entire Earth, hence forming intercontinental baselines (see Sect.\,\ref{sci:vlbi}).

\begin{figure}[h]
  \centering
  \includegraphics[angle=-90,width=0.85\linewidth,clip=true,trim= 40mm 0 40mm 0]{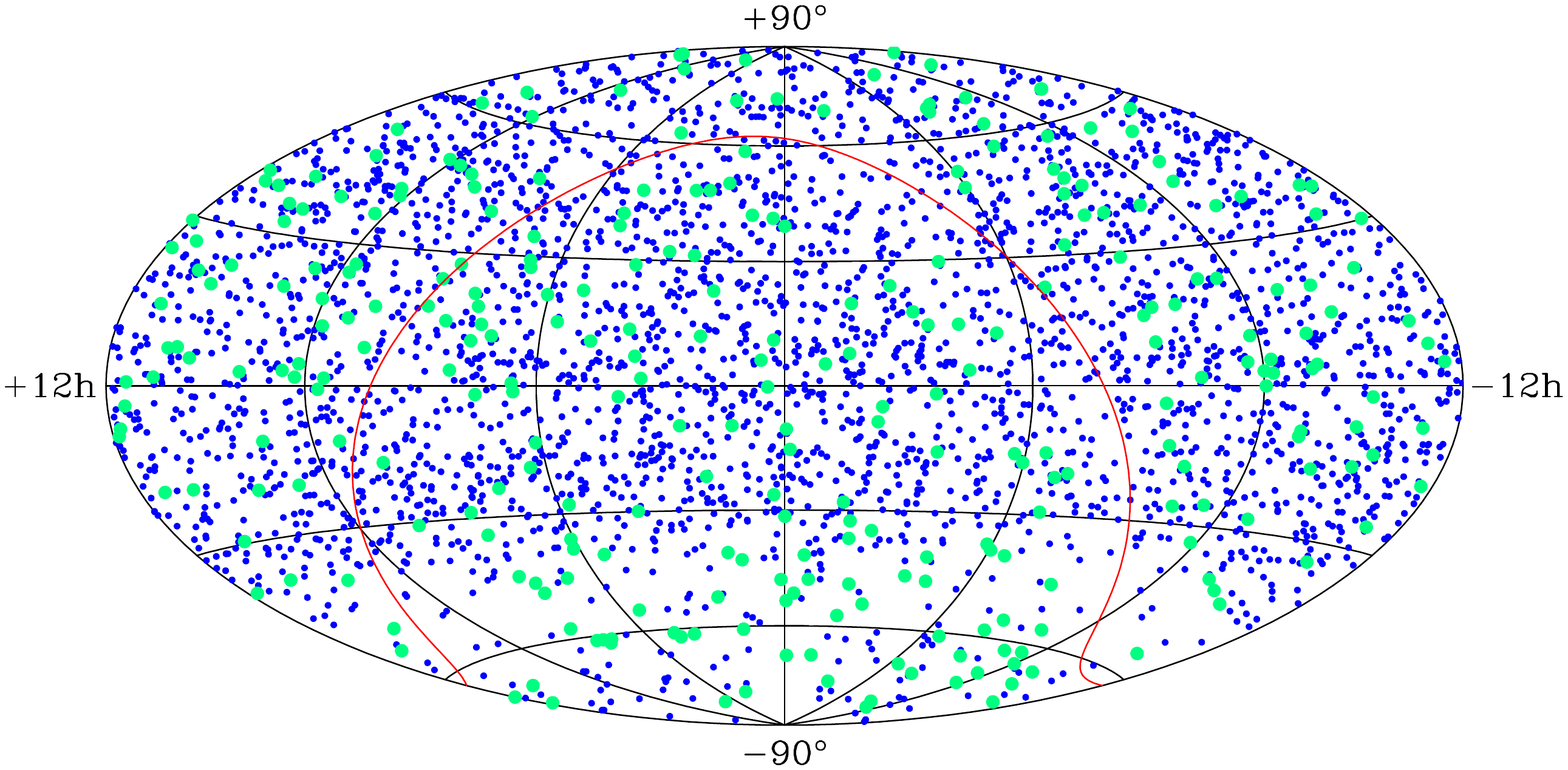}
  \caption{Sky distribution of the 3414 radio sources of ICRF2 (Ma et al. 2009; Fey et al. 2015). Green dots identify the 295 defining sources. The red line represents the Galactic equator.}
  \label{icrf2}
\end{figure}

\smallskip
\noi The current best realisation of the extragalactic frame is the second version of the International Celestial Reference Frame (ICRF2), recognised as the primary reference frame as per a resolution of the IAU adopted in 2009 (Ma et al. 2009; Fey et al. 2015). The ICRF2 is made up of 3414 extragalactic radio sources covering the entire sky (Fig.\,\ref{icrf2}). Of these, 295 are used to define the axes of the frame with an accuracy of $\sim$0.01\,mas. Positional accuracy for the individual objects reaches 0.04\,mas for the most compact and best observed sources in ICRF2. The next generation celestial reference frame is expected for 2018 with the presentation of ICRF3 to the IAU for adoption at the upcoming IAU General Assembly in summer 2018. ICRF3 is built by a \href{https://www.iau.org/science/scientific\_bodies/working\_groups/192/}{\color{blue} \myul[blue] {Working Group of the IAU}} chaired by P. Charlot (LAB) and with 20\% of its membership from the French community. Compared to ICRF2, ICRF3 will have a larger number of sources ($>$~4000), be more accurate and have a more uniform precision in position. It should also comprise positions at three frequencies, 8 GHz, 24 GHz and 32 GHz, another new feature with respect to ICRF2, where positions were limited to the standard 8\,GHz radio frequency.

\smallskip
\noi The upcoming ICRF3, with its increased positional accuracy, but also thanks to an observing span approaching 40\,years (1979--2017), will allow the scientific community to tackle new questions in astronomy and fundamental physics. The long time base now makes possible the detection of Galactic aberration, a secular effect introduced by the rotation of the Solar System barycentre around the Galactic centre. This effect is now emerging from the current VLBI data with a magnitude of 0.005\,mas/yr through apparent long-term proper motions of the radio sources (Titov \& Lambert 2013). In a similar way, the low-frequency ($<10^{-9}$\,Hz) gravitational wave background, although not detected at present, may be revealed in the future through such quasar proper motions (Gwinn et al. 1997; Titov et al. 2011). VLBI is also essential for testing General Relativity, e.g.~through the determination of the relativistic parameter $\gamma$ (Lambert \& Le-Poncin Lafitte 2011) or for trying alternate theories (Le-Poncin Lafitte et al. 2016). Incorporating SKA1-MID as an element of world-wide VLBI arrays, as described in Sect.\,\ref{sci:vlbi}, will largely improve the sensitivity of such arrays, hence permitting to expand considerably the celestial frame, which is essential to further constrain or reveal the above effects. With its location in the southern hemisphere, SKA1-MID will also help to correct the currently uneven sky distribution which comprises much less sources below $-45^{\circ}$~declination (see Fig.\,\ref{icrf2}) due to the sparseness of VLBI telescopes in the South.

\begin{figure}[t]
  \centering
  \includegraphics[width=0.245\linewidth,clip=true,trim=5mm 50mm 10mm 30mm]{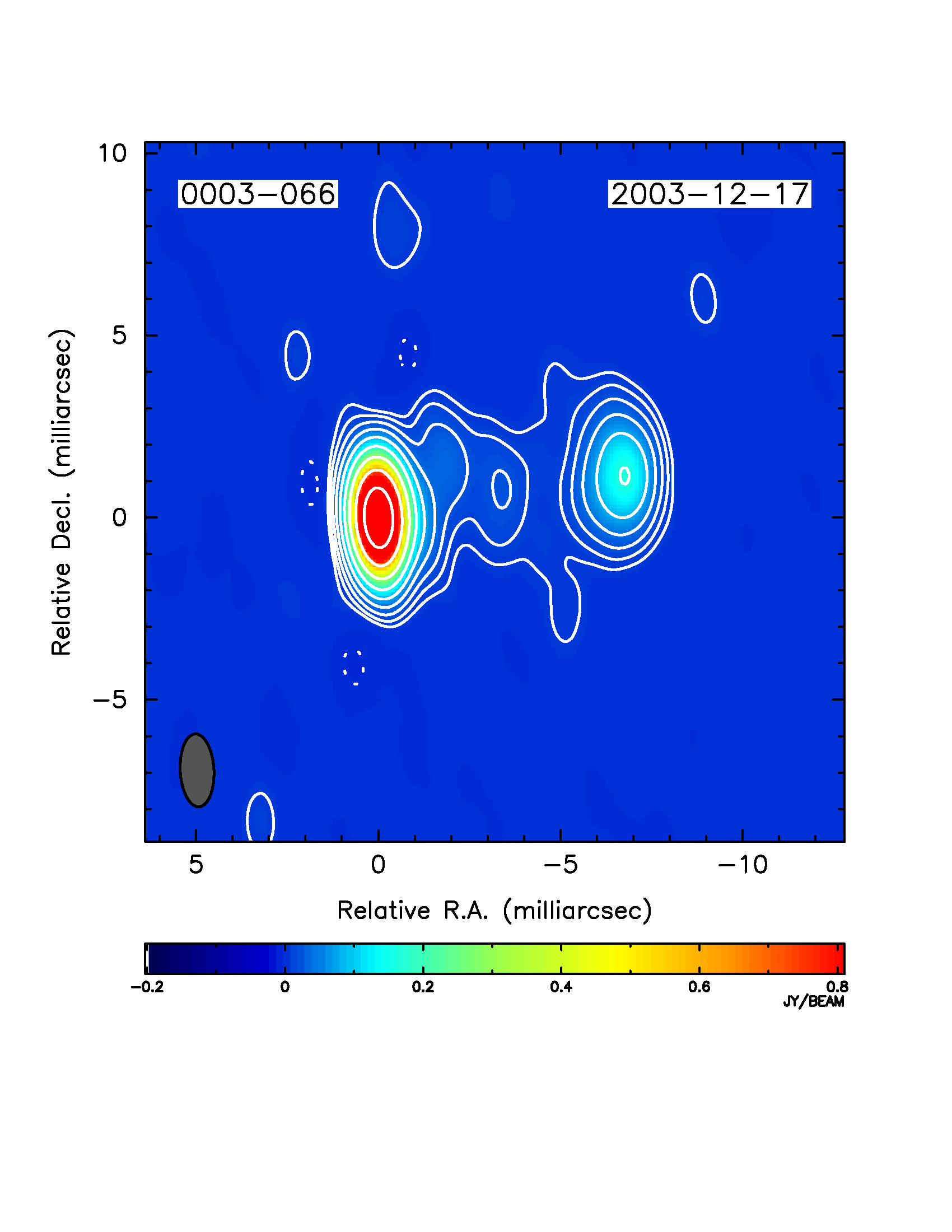}
  \includegraphics[width=0.245\linewidth,clip=true,trim=5mm 50mm 10mm 30mm]{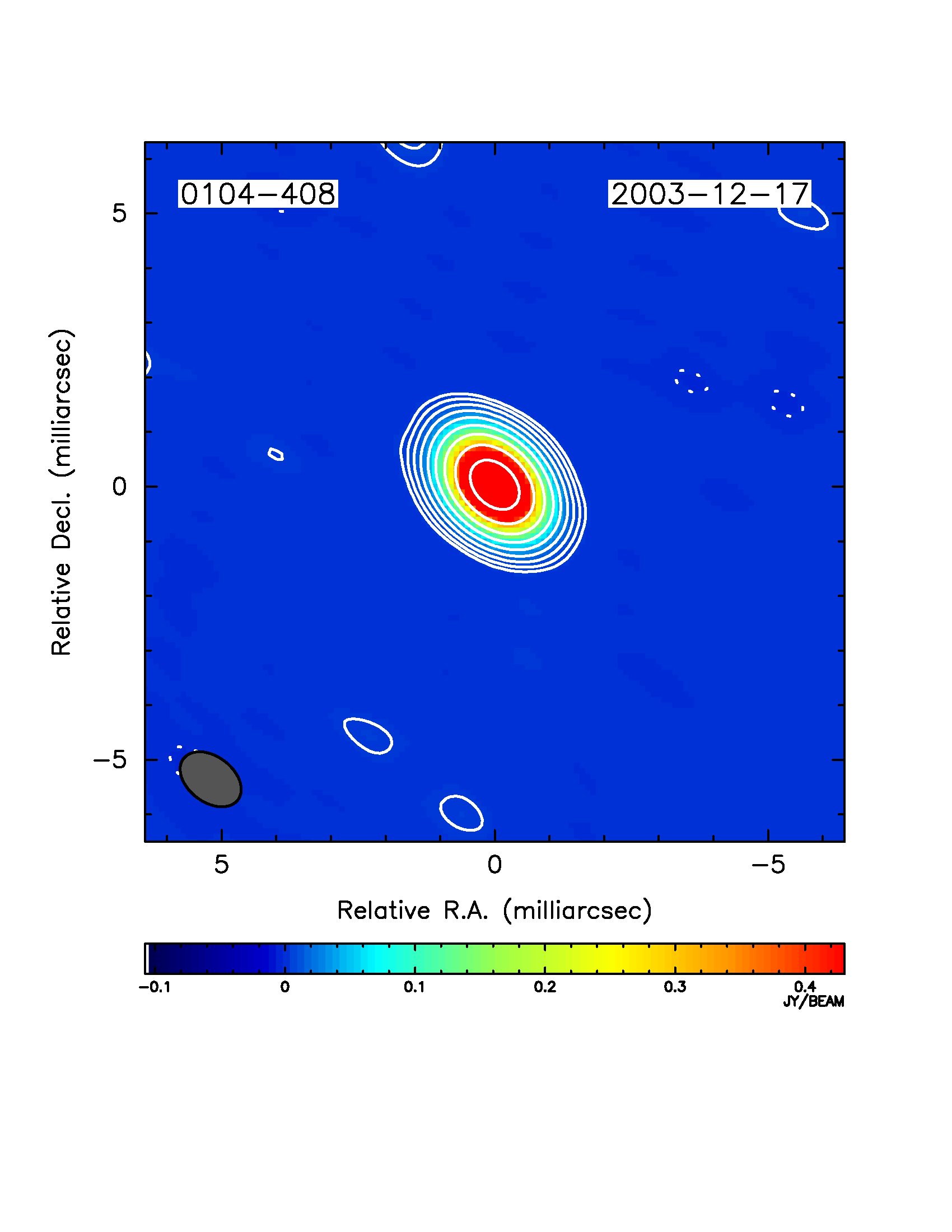}
  \includegraphics[width=0.245\linewidth,clip=true,trim=5mm 50mm 10mm 30mm]{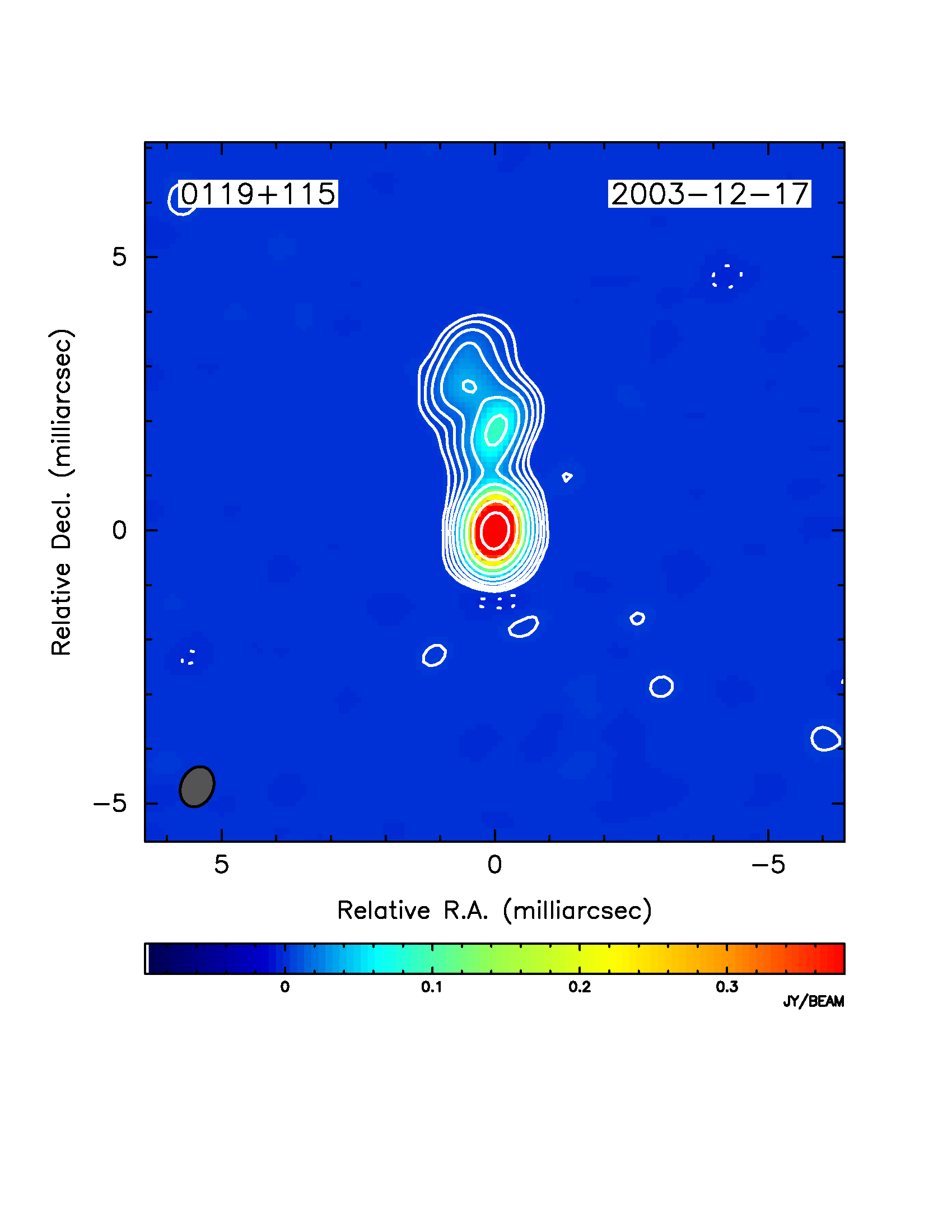}
  \includegraphics[width=0.245\linewidth,clip=true,trim=5mm 50mm 10mm 30mm]{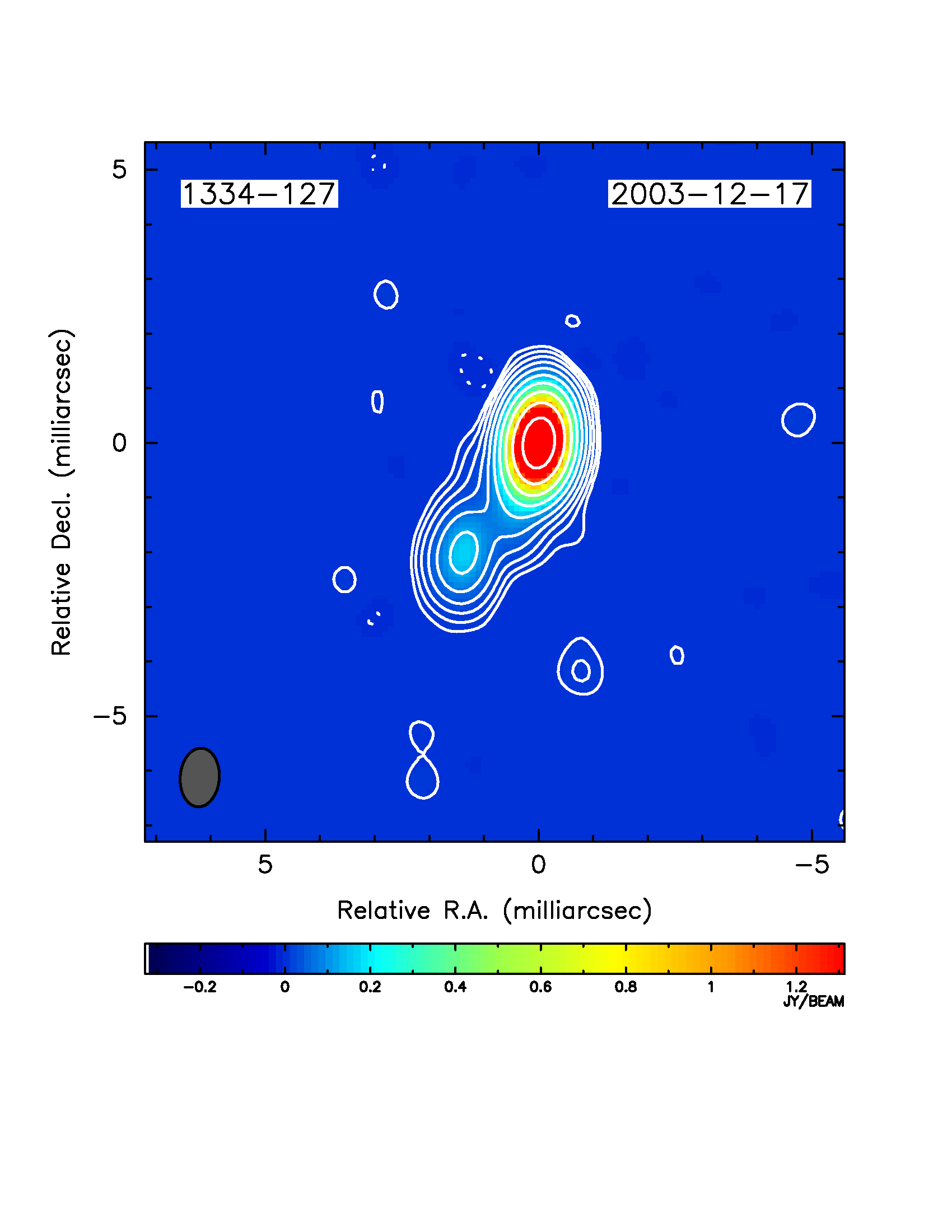}
  \caption{A sample of VLBI maps from the Bordeaux VLBI Image Database (BVID) showing the predominantly core-jet structure of those reference frame sources on milliarcsecond scales.}
  \label{bvid}
\end{figure}

\smallskip
\noi The multi-frequency positional information in ICRF3 together with the optical positions derived with the Gaia space mission (Mignard et al. 2016) will also provide new insights into the physics of AGN. At radio frequencies, these objects generally feature a bright compact core and a single-sided relativistic jet with blobs of emission moving away from the core on time scales of months to years (Fig.\,\ref{bvid}). Future reference frames will have to account for such time-varying extended internal structures for the highest accuracy, a perspective that motivates the systematic VLBI imaging program\footnote{All VLBI images produced through this program are publicly available through the \href{http://www.astrophy.u-bordeaux.fr/BVID/}{\color{blue} \myul[blue] {Bordeaux VLBI Image Database (BVID)}}.} of the Bordeaux group to monitor the structure of the ICRF~sources. In the light of such extended emission, comparison of the ICRF3 and Gaia positions will be essential in understanding whether the radio emission and optical emission are superimposed in these objects. Estimates of such radio-optical ``core shifts'' indicate that they amount to 0.1\,mas on average (Kovalev et al. 2008), which is significant considering the anticipated ICRF3 and Gaia position accuracies. While potentially affecting the alignment between the two frames, the ICRF3-Gaia positional differences will also offer a unique opportunity to directly determine those core-shifts and probe the geometry of quasars in the framework of unified AGN theories. In particular, such measurements may help to locate the optical region relative to the relativistic radio jet and determine whether the dominant optical emission originates from the accretion disk or the inner portion of the jet. As noted above, incorporation of SKA1-MID in VLBI arrays will permit to vastly augment the number of sources in the radio frame, thereby permitting to tackle such studies for complete source samples including thousands of objects.

\smallskip
\noi Beyond fundamental physics, astronomy and astrophysics, celestial reference frames have also applications in many other areas. This includes navigation of spacecrafts in the Solar System through differential VLBI astrometry relative to angularly-close quasars. Additionally, VLBI contributes to geodesy and the monitoring of the Earth's rotation. In particular, it determines uniquely the Earth's nutation, which depends on the geophysical properties of the Earth, hence providing constraints on the Earth's interior structure (Mathews et al. 2002; Rosat et al. 2017). Expansion of the celestial frame through SKA-VLBI measurements will thus reverberate on all those areas as well.\\

\parbox{0.9\textwidth}{
\noi{References:}\\
\noi{\scriptsize Fey, A. L., et al., 2015, AJ, 150;
Gwinn, C. R, et al., 1997, ApJ, 485, 87;
Kovalev, Y.~Y., et al., 2008, A\&A, 483, 759;
Lambert, S.B. \& Le Poncin-Lafitte, C., 2011, A\&A, 529, 70;
Le Poncin-Lafitte, C., et al., 2016, Phys. Rev. D, 94, 125030;
Ma, C., et al., 2009, IERS Technical Note No. 35, Eds. A. Fey and D. Gordon, Verlag des Bundesamsts f\"ur Kartographie und Geod\"asie, Frankfurt am main;
Mathews, P. M., et al., 2002, JGR, 107(B4), ETG 3-1;
Mignard, F., et al., 2016, A\&A, 595, 5;
Rosat, S., et al., 2017, GJI, 208, 211;
Titov, O. \& Lambert, S., 2011, A\&A, 559, A95;
Titov, O., et al., 2011, A\&A, 529, A91
}}

\newpage
\section{Scientific synergies with other instruments}\label{science:synergies}

\noindent {\normalsize Contributors of this section in alphabetic order: }

\smallskip

\noi {\sffamily \scriptsize
{\sffamily\bf E.~Armengaud} [\irfu],
{\bf A.~J.~Banday} [\irap],
{\bf P.~Charlot} [\lab],
{\bf E.~Chassande-Mottin} [\apc],
{\bf N.~Christensen} [\artemis],
{\bf N.~Clerc} [\irap],
{\bf B.~Cordier} [\irfuSAp],
{\bf F.~Hammer} [\gepi],
{\bf P.~Guillard} [\iapsorb],
{\bf A.~Gusdorf} [\lermasorb],
{\bf O.~Hervet} [\stcruz],
{\bf C.~Lachaud} [\apc],
{\bf S.~Lambert} [\syrte],
{\bf O.~Le~F\`evre} [\lam],
{\bf M.~Lehnert} [\iapsorb],
{\bf C.~Magneville} [\irfu],
{\bf S.~Maurogordato} [\lagrange],
{\bf Y.~Mellier} [\iapsorb],
{\bf N.~Nesvadba} [\ias],
{\bf N.~Palanque-Delabrouille} [\irfu],
{\bf G.~W.~Pratt} [\irfu;\aim],
{\bf M.~Puech} [\gepi],
{\bf J.~Richard} [\cral],
{\bf H.~Sol} [\luth],
{\bf M.~Tagger} [\lpcee],
{\bf S.~D. Vergani} [\gepi],
{\bf N.~Webb} [\irap],
{\bf P.~Zarka} [\lesia;\usn]
}

\subsection{Euclid}\label{science:euclid}
\vspace{0cm}

\noi{\bf Mission description}

\smallskip
\noi Euclid is a M-class space mission led by ESA, aimed to understand the nature of dark energy, and study the distribution of dark matter. The Euclid mission is composed of the following main elements: a telescope 1.2m in diameter, a visible imager (VIS), a near infrared imaging spectrograph and photometer (NISP), VIS and NISP both with a field of view of 0.5 square degree, and a science ground segment (SGS). Euclid will run two surveys, a wide survey over 15\,000 square degrees and a deep survey over 40 square degrees. The mission duration is set for about 6 years with a launch not earlier than 2020.

\smallskip
\noi {\bf Wide cosmology survey}

\smallskip
\noi The Euclid cosmology survey will cover 15\,000 square degree of high galactic latitude, both in imaging and in spectroscopy (Euclid {\it Red book}, Laureij et al. 2011). Deep images will be obtained with the VIS instrument in a broad visible {\it r+i+z} band covering  from  550 to  900 nm, providing two key elements to the mission: detection of faint sources and shape measurements to analyse weak lensing properties. The VIS images will reach a limiting magnitude V=24.5 (10$\sigma$) for sources with extent $\sim$0.3 arcsec. The NIR imaging spectrograph (NISP) will provide Y, J and H band images down to H$_{AB}\sim$24 (5$\sigma$) over the same field of view of the VIS imaging. In total more than 1.5 billion sources will be imaged in the wide survey. Along with the imaging, the NISP will provide slitless spectroscopy over the same area. The spectroscopic survey aims to obtain redshift information based on the H$\alpha$ line for over 25 million galaxies with $0.9<z<1.8$,  reaching a limiting flux 2$\times 10^{-16}$ erg.cm$^2$.s$^{-1}$. With this large sample of galaxies with accurate spectroscopic redshifts it will be possible to analyse the clustering of galaxies  at a cosmic time when sensitivity to a dark energy component is maximum.

\smallskip
\noi The complementarity between Euclid and SKA for Cosmology has been
reported by Kitching et al. (2015).  They emphasise that
the two  experiments are  synergistic in several respects, both
through the scientific objectives and through the control of
systematic effects on computing cosmological parameters. On the weak
lensing side Euclid and SKA will be complementary in measuring the
weak lensing shear and magnification, together with the positions of
galaxies, projected on the plane of the sky as well as in
redshift. Cross and auto correlation statistics will allow to identify
systematic effects affecting measurements, and ideally minimise them
to an unprecedented accuracy and precision.

\smallskip
\noi As identified in Kitching et al. (2015), the combination of SKA with Euclid will improve the accuracy of Baryon Acoustic Oscillations (BAO), and redshift space distortions (RSD) measured from clustering analysis. Improved constraints on massive neutrinos become possible. Combined analysis of different populations with different biases with the dark matter distribution, can yield a further improvement on the measurement of cosmological parameters, especially from constraints on large scales.  

\smallskip
\noi {\bf Galaxy formation and evolution}

\smallskip
\noi With Euclid delivering images and spectrophotometric information probing mostly the stellar emission in galaxies, and SKA informing mostly about the gas content, combining the two will lead to a comprehensive information content for each galaxy in common. It will become possible to investigate the evolution of the star formation rate, gas and stellar mass content, over a range of galaxy properties, including their environment, in a range of redshift up to z=2 when combined with the Euclid wide survey. For the 25 million galaxies with Euclid spectra and H$\alpha$ measurement, it will become possible to compare the instantaneous star formation rate to the gas content. 

\smallskip
\noi The Euclid deep survey over 40 deg$^2$ will give a unique opportunity to probe the massive end of the galaxy mass function up to the reionisation epoch z$\sim$7. The deepest SKA surveys should ideally be set overlapping with the Euclid deep survey areas. Here again, and depending on the depth of the SKA surveys, the joint information about stellar and gas content will be invaluable to investigate the build up of the massive galaxy population from the end of reionisation to the peak in star formation activity.

\smallskip
\noi {\bf Galaxy clusters}

\smallskip
\noi Combining SKA and Euclid will open a new era in the understanding
 of the physics of galaxy clusters and their use as a new solid
 primary  cosmological probe. 
Euclid surveys are expected to deliver an unprecedented  catalog of
$\sim 10^5$ massive clusters spanning the redshift range [0,2]
(Laurejis et al. 2011).  Analysis of cluster counts as a function of
mass and redshift is a young and promising cosmological probe which is
deeply investigated in the context of next generation surveys. Current
forecasts show that its success will depend crucially on our capacity
to precisely determine and calibrate the observable-mass relation
(Sartoris et al. 2016), which is  critical at high redshift where mass
estimates are particularly difficult to obtain by most techniques.
Following a sample of these clusters with SKA1-MID will allow to
measure the SZ decrement produced by these systems (Grainge et
al. 2015) and will provide the calibration of the mass-richness relation for  Euclid clusters up to the highest redshifts ($z \sim 2$) where individual mass determination by weak-lensing techniques reaches its limit. This will then improve drastically the precision of the cosmological parameters inferred from the Euclid cluster catalog. 

\smallskip
\noi Comparison of the thousands of rich clusters detected through their synchrotron radiation in SKA1 surveys  (see Sect.\,\ref{science:clustersNT}) will be compared to those detected  from galaxy overdensities and weak lensing peaks by Euclid. Analysis of  the signatures of the non-thermal component of the ICM revealed by SKA1 surveys will then be addressed at the light of  state-of-the-art mass maps reconstructed from Euclid weak lensing analysis and galaxy density maps. This will give a revolutionary vision of the formation of these systems in the context of their surrounding mass and  galaxy distribution. \\

\parbox{0.9\textwidth}{
\noi{References:}\\
\noi{\scriptsize 
Grainge, K., et al., 2015, AASKA14, 170;
Kitching, T., et al., 2015, AASKA14, 146;
Laurejis, R., et al., 2011, arXiv1110.3193, Euclid Definition Study Report;
Sartoris, B., et al., 2016, MNRAS, 459, 1764
}}

\subsection{E-ELT}
\vspace{0cm}

\noi The European Extremely Large Telescope (E-ELT), with its 39m diameter
aperture, will provide Europe with the biggest ground-based optical
astronomical facility ever built as from 2024. During the very first
years of operation, it will be progressively equipped with a
instrumental suite largely covering the parameter space in term of
observational capabilities (imaging, integrated and 3D spectroscopy),
wavelength (from the visible to the NIR).

\smallskip

\noi \href{http://www-astro.physics.ox.ac.uk/instr/HARMONI/}{\color{blue} \myul[blue] {HARMONI}} will be one of the two first-light instruments for the E-ELT.
It is an integral field spectrograph operating in the optical and
near-infrared (0.5-2.45 $\mu$m), with spectral resolutions ranging
between R=3\,000 and R=20\,000 (Thatte et al. 2016). HARMONI is designed
to work together with various adaptive-optics systems (SCAO, GLAO, LTAO)
with a good spatial sampling (between 4 and 60 mas), albeit within a
small FoV (between 1 and 10 arcsec). With its numerous spatial and
spectral configuration, HARMONI will be a generic instrument for first
light, in particular for the study of stellar populations, exoplanets
and high redshift galaxies.

\smallskip

\noi The timeline for HARMONI (first light in 2024) makes it suitable for
following-up SKA sources. HARMONI can perform a detailed follow-up of
kinematics and chemical properties for individual galaxies at $z<5$
selected with SKA in 21 cm \hi~and/or radio continuum. At higher
redshift, HARMONI will have the spatial and spectral resolution to
study bright sources inside reionisation bubbles mapped out by SKA. It
will be possible to correlate the \hi~content of these sources with
their Ly-$\alpha$ properties, in particular the presence of extended
haloes.

\smallskip

\noi MOSAIC\footnote{\url{http://mosaic-elt.eu}} will the E-ELT
multi-object spectrograph as from 2026. It will work in the optical
and near-infrared (0.45-1.8 $\mu$m) with spectral resolutions
R=5\,000, and R=15\,000 in specific windows (Hammer et al. 2016),
providing integrated spectroscopy of $\sim$100-200 objects in a
$\sim$40 $arcmin^2$ patrol field as well as multi-object
spatially-resolved spectroscopy with 10 IFUs -- a unique capability
amongst all ELTs. MOSAIC will be optimised for survey speed and is
aimed to become a workhorse spectrograph for the E-ELT. These
capabilities will allow tackling fundamental questions in cosmology
such as how matter is distributed in distant galaxy haloes, and how
present-day galaxies assembled. This includes the detection of nearby
primordial stars, the very first galaxies at the EoR, the most
exhaustive dynamical survey of distant galaxies ever undertaken, and
the detailed study of the stellar populations (Evans et al. 2015).
Most of these science cases are expected to provide very strong
synergies with SKA.

\smallskip

\noi One of the prominent synergies with MOSAIC will be related to the
distribution of dark and baryonic matter in distant galaxy haloes. In
particular, MOSAIC will take advantage of both the (MOAO) adaptive
optics corrections and the optimized surface brightness sensitivity
provided by its (multiple) $\sim$ 75 mas IFUs to sample the ionized
gas extending beyond the optical radius of distant galaxies (Contini
et al. 2016). This will allow constructing accurate rotation curves
and dark matter profiles in $z \sim 2-4$ galaxies (Puech et al. 2010),
while SKA1-MID will probably remain limited in providing \hi-based rotation
curves only up to $z \sim 0.2-0.5$ (Blyth et al. 2015). The redshifts
around $z \sim 2$ will be of particular interest since here SKA1 and
then SKA2 will be able to detect a large fraction of galaxies and
determine their \hi~ cold gas content.

\smallskip

\begin{figure}[!ht]
  \centering
  \includegraphics[width=0.65\linewidth]{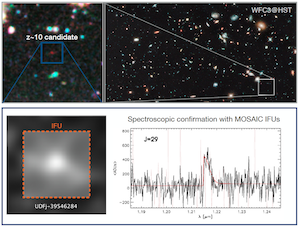}
  \caption{\label{fig:LAE} Upper panel: $z \sim 10$ candidate in the Hubble
    Ultra Deep Field (Bouwens et al. 2011). Lower panel: Planned
    follow-up of the $z \sim 10$ candidate with the MOSAIC IFU ({\em left}). The
    expected spectra of the candidate as observed by MOSAIC at E-ELT
    in $\sim$ 10 hr of integration time is shown in the {\em lower right}
    panel (Disseau et al. 2014). Adapted from Hubble Ultra Deep Field
    2009-2010 and UDFj-39546284. Credit: NASA, ESA, G. Illingworth
    (University of California, Santa Cruz), R. Bouwens (University of
    California, Santa Cruz, and Leiden University), and the HUDF09
    Team.}
\end{figure}

\noi MOSAIC will also target the low-luminosity sources that are
expected to be responsible for the reonisation of the Universe beyond
$z \sim 6$. MOSAIC will observe large samples of Lyman Break Galaxies
as faint as $J_{AB} = 30$ and measure the equivalent widths of their
Ly-$\alpha$ emission (see Fig.\,\ref{fig:LAE}; Disseau et al. 2014) to
study the evolution of the Ly-$\alpha$ Luminosity Function as a
function of redshift and better constrain the EoR. The higher spectral
resolution compared to JWST will be decisive to study the physics (from
the Ly-$\alpha$ line profile and other ISM lines) within these first
galaxies. The correlation between these sources and the \hi~ direct
imaging at scales of arc-minutes that SKA1-LOW will produce at $z \sim
6-28$ are expected to generate very strong synergies that will lift the veil
on the physics of EoR (Koopmans et al. 2015).\\

\parbox{0.9\textwidth}{
\noi{References:}\\
\noi{\scriptsize
  Blyth, S., \etal, 2015, AASKA14, 128;
  Bouwens, R.J., \etal, 2011, Nature, 469, 504;
  Contini, T., \etal, 2015, \aap, 591, 49;
  Disseau, K., \etal, 2014, SPIE, 9147, 91
  Evans, C., Puech M. (Eds), \etal, 2015, ArXiv:1501.04726;
  Hammer, F., \etal, 2016, SPIE, 9908, 24;
  Koopmans, L. V. E., \etal, 2015, AASKA14, 1;
  Puech, M., \etal, 2010, \mnras, 402, 903;
  Thatte, N., \etal, 2016, SPIE, 9908, 1
}}\\

\subsection{4MOST Cosmology Redshift Survey}

\vspace{0cm}

\noi 4MOST is a major new wide-field high-multiplex spectroscopic survey facility under development for the VISTA telescope of ESO. 4MOST has a broad range of science goals ranging from galactic archaeology and stellar physics to the high-energy physics, galaxy evolution, and cosmology. It is a natural 
follow-up instrument for other massive space-based and ground-based projects, such as GAIA, Euclid, e-Rosita or SKA. Starting in 2022, 4MOST will 
deploy 2436 fibres in an hexagonal 4.1 square degree field-of-view using a positioner based on the tilting spine principle. The fibres will feed 
one high-resolution (R~20,000) and two medium-resolution (R~5000) spectrographs with fixed 3-channel designs and identical 6K x 6K CCD detectors. 
4MOST will have a unique operation concept, in which 5-year public surveys both from the 4MOST Consortium and the ESO community will be combined 
and observed in parallel during each exposure. At least 15 surveys, both galactic and extra-galactic, will be run in parallel. The 4MOST Facility 
Simulator (4FS) was developed to demonstrate the feasibility of this observing concept, showing that we can expect to observe above 50 million 
objects in the entire survey period and will eventually be used to plan and conduct the actual survey (de Jong. et al. 2016). 

\smallskip

\begin{figure}[!ht]
  \centering
  \includegraphics[width=0.95\linewidth]{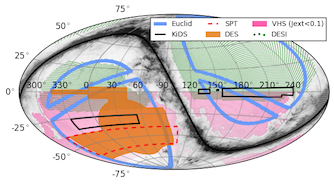}
  \caption{\label{fig:skycoverage} Overview of the sky area covered by the 4MOST Cosmology Redshift Survey compared to other cosmology experiments. 
	The 4MOST CRS aims towards a homogenous 
	target density over the full southern sky hemisphere ($\delta<0$) based on target selection with VHS apart from regions of high galactic 
	extinctions and/or high stellar density (pink area), with higher target densities achieved over the DES footprint (orange).}
\end{figure}

\noi The goal of the 4MOST Cosmology Redshift Survey (CRS) is to study the nature of gravity and dark energy, responsible for the structure formation in the 
Universe and its expansion, respectively. To this end, it will map the three-dimensional matter distribution in the Southern sky. We aim at 
performing precise measurements of baryonic acoustic oscillations (BAO) and redshift space distortions (RSD) from the three-dimensional distributions 
of galaxies with an unprecedented accuracy at redshifts up to about $z=1$. In particular the objectives are to measure redshifts for $\sim$8-10 
million (a) Bright Galaxies (BG) and (b) Luminous Red Galaxies (LRG), and for 10 million (c) Emission Line Galaxies (ELG). Additionally, this survey 
will measure the distribution of quasars and the Lyman-$\alpha$ forest between redshift 2.2-3.5 and the redshifts of $>200,000$ SNe 1a or their host galaxies using triggers from DES and LSST. The survey will cover 10-13K square degrees, concentrating on the area covered by the Dark Energy Survey (DES) where deep imaging will be available (see Fig.\,\ref{fig:skycoverage}), allowing for instance cross-correlations with weak lensing surveys and CMB experiments. Among all 4MOST surveys, the CRS provides the largest number of targets and represents the majority of all extragalactic targets. We describe below some science goals which will benefit from the combination of 4MOST and SKA.

\smallskip

\smallskip

\noi {\bf 21\,cm surveys}
\smallskip

\noi Intensity mapping surveys with the Square Kilometre Array pathfinders (MeerKAT in South Africa and ASKAP in Australia) and later with the SKA itself, will be a very powerful cosmological probe. Future intensity mapping experiments will be able to measure the rate of cosmic growth through redshift-space distortions of the power spectrum (Bull et al. 2015).  Measurements of the \hi~power spectrum will still be degenerate with the \hi~bias, but cross-correlation between the 21\,cm intensity maps and large-area optical surveys provides a useful cross-check of this parameter and the neutral hydrogen density.  This measurement has already been performed using radio data from the Green Bank telescope, and optical data from the WiggleZ Dark Energy Survey (Masui et al. 2013).  Cross-correlation between 4MOST and a large-area future intensity mapping experiment will be be a valuable probe of cosmology and galaxy evolution.

\smallskip

\noi{\bf Radio continuum surveys}
\smallskip

\noi The large-area radio continuum surveys such as ASKAP EMU (Evolutionary Map of the Universe, Norris et al. 2011) will cover the whole southern sky.  EMU will detect roughly 70 million radio galaxies out to redshift 5 and provide constraints on dark energy, modified gravity and primordial non-Gaussianity through measurements of the angular correlation function, cosmic magnification and Integrated Sachs Wolfe (ISW) effect.

\noi The radio spectrum is too smooth to measure redshifts, and all of these measurements will be enhanced by the addition of redshift information to allow division of the sample into redshift bins.  This will be contributed by the 4MOST CRS, either directly, or indirectly via photo-z calibration.

\smallskip

\noi{\bf Velocity power spectrum}
\smallskip

\noi Maximum-likelihood methods can constrain the amplitude of the auto- and cross-power spectra of density and velocity, which can trace the growth rate of structure as a function of scale . Originally proposed by Jaffe \& Kaiser (1995), it has seen recent success in its application to the peculiar velocity sample from the 6dFGS (Johnson et al. 2014; Adams et al. 2017).  Howlett et al. (2017) find that combining the peculiar velocity and density fields from the Taipan and ASKAP WALLABY surveys can constrain the growth rate to $\sim3\%$. The higher number density of 4MOST objects would greatly improve on this constraint, both separately, and in combination with other measurements of peculiar velocities, such as those from WALLABY, SKA or LSST Supernovae.\\

\parbox{0.9\textwidth}{
\noi{References:}\\
\noi{\scriptsize 
Adams, C., \etal , 2017, submitted;
Bull, P.,  \etal,  2015, ApJ, 803, 21; 
Howlett, C., \etal , 2017, MNRAS, 464, 2517;
Jaffe, A. H. \& Kaiser, N., 1995, APJ, 455, 26;
Johnson, A., \etal, 2014, MNRAS, 444, 3926;
de Jong, R., \etal, 2016, SPIE, 9908, 1;
Masui, K., \etal, 2013, ApJ, 763, 20;
Norris, R., \etal, 2011, PASA, 28, 215
}}\\

\subsection{JWST}
\vspace{0cm}

\noi 
The James Webb Space Telescope (JWST) is the large aperture, near-IR optimised successor of the \textit{Hubble} and \textit{Spitzer} space telescopes. JWST's launch date is October 2018 and it will be carried on an Ariane~5 rocket which will place it in a Lissajous orbit around the Earth-Sun L2 point. 
Covering the wavelength range $0.6 - 28\,\mu$m, with its large, segmented, $6.5$~m primary mirror (see Gardner et al. 2006 for a review of JWST's characteristics), will achieve a gain in sensitivity and spatial resolution of about two-orders of magnitude compared to ground-based facilities and \textit{Spitzer}. The four science instruments comprising the integrated science instrument module (ISIM) are:

\begin{description}
\item[\textbf{NIRCam: }] A Near-Infrared Camera ($0.6 - 5\,\mu$m), with two broad- and intermediate-band imaging modules, each with a $2.16 \times 2.16$ arcmin$^2$ field of view (Beichman et al. 2012). The modules have a short- and a long-wavelength channel, taking images simultaneously with light split by a dichroic at about $2.35 \,\mu$m. The short-wavelength channel, covering the wavelength range $0.6 - 2.3\,\mu$m, is  sampled at $4096 \times 4096$ pixels (0.0317 arcsec pixel$^{-1}$), the long-wavelength channel at $2048 \times 2048$ pixels (0.0648 arcsec pixel$^{-1}$).

\item[\textbf{NIRSpec: }] A Near-Infrared Spectrograph ($0.6 - 5\,\mu$m) with 3 observing modes: a low resolution $\mathcal{R} \sim 100$ prism mode, a $\mathcal{R} \sim 1000$ multi-object mode and a $\mathcal{R} \sim 3000$ Integral Field Unit (IFU) or long-slit spectroscopy mode (Bagnasco et al. 2007). 
In the $\mathcal{R} \sim 100$ and $\mathcal{R} \sim 1000$ modes, NIRSpec uses a micro-electromechanical system (MEMS) to provide dynamic aperture shutter masks (``microshutter array''). These masks allow us to obtain simultaneous spectra of more than 100 objects in a $>9$ arcmin$^{2}$ field of view ($\sim 3.4 \times 3.4$~arcmin$^{2}$ at 1~$\mu$m). At $\mathcal{R} \sim 100$ one prism spectrum covers the full $0.6 - 5 \,\mu$m wavelength range. At $\mathcal{R} \sim 1000$, three gratings are necessary to cover the full $1 - 5 \,\mu$m wavelength.

\item[\textbf{MIRI: }]  A Mid-InfraRed Instrument, developed by a European consortium (Wright et al. 2014), with imaging and spectroscopy covering the $5-27\,\mu$m wavelength range. Its design consists of two main modules, an imager (Bouchet et al. 2015), a low-resolution spectrometer (Kendrew et al. 2015) and an IFU medium resolution spectrograph (MRS; Wells et 2015).

\item[\textbf{FGS/NIRISS}]
The Fine Guidance System/Near-InfraRed Imager and Slitless Spectrograph, provided by the Canadian Space Agency. NIRISS is a wide-field, 2.2' $\times$ 2.2', imager capable of $1.0-2.5$~$\mu m$ slitless spectroscopy (Beichman et al. 2014).   It also includes a spectroscopic observing mode optimised for exploring exoplanets atmospheres.
\end{description}

\smallskip
\noi In the following we illustrate some Galactic and extra-galactic science topics where JWST and SKA have strong synergies. This is by no means intended to be a comprehensive review, but rather we highlight a few examples, biased towards interstellar medium studies, showcasing the power of combining JWST and SKA observations.

\smallskip
\noi {\bf Galactic science} 

\smallskip
\noi The complementarity between SKA and JWST wavelengths translates into the following Galactic science topics:
\begin{itemize}
\itemsep-0.2em
\item In shock regions (e.g., supernova remnant shocks, see Sect.\,\ref{sci:SNR}, or jets and outflows, see Sect.\,\ref{sci:YSO}), the local density can be constrained by H$_2$ observations with JWST. These types of constraints have proven to be robust using JWST's predecessors like ISO or Spitzer (see, e.g., Neufeld et al. 2009), because H$_2$ is significantly excited in the warm shocked gas, its transitions are optically thin, and also its abundance makes it chemistry-independent. Once the density of H$_2$ is known, especially at high angular resolution, the chemistry of complex organic molecules (COMs) will be better constrained, and the excitation of, for example, OH masers better understood. Both COMs and maser emissions can be probed through SKA observations. JWST will also allow us to more accurately characterise shocked regions through observations of ionised lines which trace fast or externally irradiated shocks, and of other lines of abundant molecular species such as water.
\item In photodissociation regions (PDRs) and \hii~regions, the study of the physics and chemistry will similarly benefit from observations with JWST and the constraints it will provide on H$_2$ density (e.g., Habart et al. 2011) complemented by COMs, masers, and also the radio recombination line observations by SKA (hydrogen, helium, and carbon). In such environments, the synergy between JWST and SKA will extend to the study of dust properties. Dust plays key roles in both the physical and chemical processes operating in PDRS and \hii~regions. The exact role they play depends on the detailed dust properties such the distribution of grain sizes, the detailed structure of the grains, and on their composition. It is well documented that continuum observations at multiple wavelengths are mandatory to robustly constrain these properties.
\item Beyond these phases of the interstellar medium (ISM), the study of dust will benefit from combined results from both SKA and JWST. To reinforce what is written in previous sections, the validation of the interpretation of anomalous microwave emission by means of spinning dust requires the combined observation of dust emission between 10 and 50~GHz with SKA, and of PAH emission at a few microns with JWST. The high angular resolution of both telescopes will allow us to obtain a deep understanding of this phenomenon. The high-angular resolution of both SKA and JWST will also allow us to study grain growth in protostellar disks, where these telescopes do not probe exactly the same layers. We will be able to ``peel back'' the layers and thus obtain a comprehensive characterisation of the whole disk structure.
\end{itemize}
\smallskip
\noi 

\smallskip
\noi {\bf Extragalactic science} 

\smallskip
\noi JWST will detect warm/hot dust emission, and ionised and molecular hydrogen lines out to very high redshifts, which will therefore allow us to complement spectral line diagnostics and provide independent dust and gas mass and energy budget estimates in the early universe. For instance: 
\begin{itemize}
\itemsep-0.2em
\item JWST and SKA observations will make it possible to probe sub-kpc scales and low column density gas of the multiphase ISM of nearby galaxies. JWST will complement SKA observations of the \hi~gas by allowing us to study the relationship between HI, H$_2$ and ionised gas, as well as the dust distribution. High-resolution multi-wavelength observations are the keys to studying the connection between star formation on small scales and global scaling laws.
\item JWST will allow us to unveil obscured AGN and extend the AGN luminosity function to similar redshifts as SKA ($z \approx 6$). SKA and JWST will also spatially resolve the morphology and the kinematics of the dusty gas torus around nearby AGN, thus extending the studies of the gas properties of their host galaxies.
\item JWST will extend the UV line emission and absorption to high-redshift galaxies, by building the faint-end of the UV luminosity function, thus providing unique and complementary constraints for SKA on how the universe was reionised. 
\item  Recent observations show that the gas in halos of galaxies is multiphase (e.g. Werk et al. 2014; Emonts et al. 2016). JWST and SKA have unique and complementary capabilities in imaging and characterising the ionised, atomic and molecular gas and dust in halos of galaxies, up to high redshifts ($z \approx 3$). This will potentially revolutionise our view of the gas content in the circum-galactic medium of galaxies.
\end{itemize}
\smallskip
\noi 

\noi Generally speaking, the synergy between JWST and SKA will be made even more efficient on all these subjects by including high-angular resolution observations at sub-millimetre wavelengths with ALMA (see Sect.\,\ref{science:alma}) filling the gap between the IR/MIR and radio ranges: the combined use of these three telescopes will ensure an optimal scientific return in all these active research areas for decades to come.\\

\parbox{0.9\textwidth}{
\noi{References:}\\
\noi{\scriptsize 
Bagnasco, G., et al., 2007, SPIE, 6692;
Beichman, C. et al., 2012, SPIE, 8442, 2;
Beichman, C., et al., 2014, PASP, 126, 1134;
Bouchet, P., et al., 2015, PASP, 127, 612; 
Emonts, B. H. C., et al., 2016, Science, 354, 1128;
Habart, E., et al., 2011, A\&A, 527, 122;
Kendrew, S. et al., 2015, PASP, 127, 623;
Neufeld, D. A., et al., 2009, ApJ, 706, 170;
Wells, M., et al., 2015, PASP, 127, 646;
Werk, J. K., et al., 2014, ApJ, 792, 8;
Wright, G. S., et al., 2015, PASP, 127, 595
}}\\

\subsection{DESI}
\vspace{0cm}

\noi The Dark Energy Spectroscopic Instrument (DESI) is a Stage IV ground-based survey dedicated to cosmology. It will be installed on the 4-m Mayall telescope in Kitt Peak, Arizona, in order to obtain more than 30 million galaxy and quasar redshifts. This will permit the study of BAO and the growth of structures with unprecedented precision on a wide area and over a large range of redshifts (Aghamoussa et al. 2016).

\smallskip
\noi The DESI robotically-actuated fiber-fed spectrograph is capable of taking up to 5,000 simultaneous spectra over the wavelength range $360-980$~nm. 
The fibers feed ten three-arm spectrographs with resolution $R=2,000-5,500$ depending on wavelength. The DESI instrument will conduct a five-year survey to cover 14,000 deg$^2$. Spectroscopic targets will be selected in four classes from imaging data: luminous red galaxies up to $z=1$, emission line galaxies up to $z=1.7$ and quasars both as tracers of the matter distribution and, in the redshift range $2.1<z<3.5$, for their Lyman-$\alpha$ forest absorption features. Several French laboratories from CEA, INSU and IN2P3 take an active part in the development of this DOE-led project, at the same time in instrumental developments (spectrographs), survey preparation (target selection, pipeline) and science. Of particular interest concerning SKA, french groups possess a key scientific expertise in the Lyman-$\alpha$ forest, from the development of quasar search algorithms to the reconstruction of the forest power spectrum at small scales and at the BAO scale, and its cosmological implications. French members strongly contributed and contribute to this research topic within the former BOSS and current eBOSS redshift surveys (e.g. Delubac et al. 2015).

\smallskip
\noi As is well known, the cross-correlations between different probes of large-scale structures is a nice tool in order to improve their sensitivities to cosmological parameters. This is due to several reasons : extra information from different cosmological dependences of the cross-correlations, improved knowledge of the tracer biases, and better constraints on single-probe systematics.  Although located in the Northern hemisphere, a large fraction of the DESI footprint will overlap with large-area ($3\pi$) SKA cosmological surveys, and in particular all of the DESI Southern Galactic Cap. Cross-correlations between different DESI and SKA probes will therefore be possible and provide significant improvements in terms of cosmological constraints - in a way very similar to what was described for 4MOST in the previous section. We note in particular that the DESI footprint will be complementary to the 4MOST one, which will only cover sky regions with negative declination.

\smallskip
\noi It is recognised that several sources of systematics, and in particular foreground contaminations, will be an important challenge to the SKA1 \hi~intensity mapping survey. For redshifts above $z=2.1$, a way to identify residual systematics in this survey, and to reduce foreground contaminations will be to cross-correlate the 21\,cm IM signal with Lyman-$\alpha$ forest spectra, as suggested in (Sarkar et al. 2011). While the 21\,cm signal is expected to originate mostly from damped Lyman-$\alpha$ (DLA) systems, the Lyman-$\alpha$ forest is associated to much smaller fluctuations of the neutral IGM. Simulations (Carucci et al. 2017) suggest that both emission and absorption signals should be largely anti-correlated. This should permit improvements by $\sim 30$~\% in the cosmological parameter estimations, and more importantly provide an observable less sensitive to foreground contaminations than the 21\,cm-alone power spectrum. Such a cross-correlation should be feasible already with existing BOSS [resp.~eBOSS] data, which have a density of $\sim 15$ [resp.~22] quasars/deg$^2$. The DESI survey will improve both the signal-to-noise and resolution with respect to BOSS/eBOSS, and increase the density of Lyman-$\alpha$ forests to $\sim 50/$deg$^2$. Note that a large number of individual DLA systems will also be identified from the DESI survey.\\

\vspace{0.1cm}
\parbox{0.9\textwidth}{
\noi{References:}\\
\noi{\scriptsize Aghamoussa, A., \etal \,(DESI Collaboration), 2016, arxiv:1611.00036;
Delubac, T., \etal\, (BOSS Collaboration), 2015, A\&A, 574, A59;
Sarkar, T. G. \etal, 2011, MNRAS, 410, 1130;
Carucci, I. P., et al., 2017, JCAP04, 001
}}\\

\subsection{LSST}
\vspace{0cm}

\noi The Large Synoptic Survey Telescope (\href{http://lsst.org}{\color{blue} \myul[blue] {LSST}}) is the major optical 
photometric survey instrument of the next decade (Ivezic et al. 2008).
The LSST telescope under construction on Cerro Pachon in Chile features an 8.4 meter diameter (6.7 m effective) primary 
mirror and a compact 3-mirror optical design providing a very wide field of view of $\sim 9 \, \mathrm{deg}^2$.
The LSST camera will cover the $\sim 64\, \mathrm{cm}$ diameter focal plane  with 189 16-MPixels $4\, \mathrm{k} \times 4\, \mathrm{k}$ CCDs,
for a total of 3.2 billion pixels that will be read in 2 s.
The camera will be equipped with six broadband optical filters (ugrizy)
covering the wavelength range from near UV  to near infrared (320-1050 nm). 

\smallskip
\noi LSST will carry out a ten year survey of the southern sky in the six filters, starting in 2022. 
Its total optical throughput (\'etendue = $\mathrm{ aperture \times field \, of \, view= 320 \, m^2 deg^2}$) 
will be two orders of magnitude greater than any existing facility, allowing a survey of the sky to unprecedented depth and size. 
The entire southern sky will be mapped every $\sim 4$ nights by LSST, with i-band magnitude depth reaching $m_i \sim 24.5$ 
in each single visit with an exposure time of only 40s, including dead time.  
At the end of the 10-year survey, the southern sky will be visited nearly a 1000 times using the six filters, 
with the i-band detection limit reaching $m_i \sim 27$ in the deep co-added images.

\smallskip
\noi LSST will provide a static census of nearly 40 billion objects, as well as dynamic, time domain mapping of 
variabilities for a huge number of faint objects. Many areas of research in astronomy and physics will 
benefit from the LSST survey, and especially the four following broad scientific topics which have driven the 
instrument design and survey strategy optimisation (Ivezic et al. 2013): 
moving objects in the solar system, 
structure and formation of the Milky Way, 
the transient sky and variable Universe, 
and the nature of the dark matter and dark energy. 

\smallskip
\noi A consortium of French scientists in CNRS-IN2P3 laboratories is contributing to the LSST construction effort and will 
have full access to the LSST data and catalog. The French R\&D and technical contributions relate to the LSST camera
electronics and mechanics (filter exchanger), calibration, camera control system and data processing. 
 
\smallskip
\noi Cosmology and more specifically dark energy probes are the main LSST science area in which the french LSST community
has been involved up to now. However, other science topics are likely to evolve and grow, as French researchers from INSU or CEA (outside the IN2P3 
laboratories) join the project. 
 
\smallskip
\noi As is the case for SKA, managing the LSST data flow (15 TB produced per night, 5 PB of raw image data per year) will be a major 
technical challenge (Juric et al. 2015). The French consortium is also responsible for handling the data processing for half of the data.
Sharing technical expertise, aiming for common design choices and technical synergies whenever possible, will likely benefit both projects. 

\smallskip
\noi LSST survey will provide shear measurement for weak lensing and accurate photometric redshift estimates 
for a few billion $(2-4 \times 10^9)$ galaxies in the southern sky, that is to say about 40-50 galaxies per arcmin$^2$ with $m_i \lesssim 25.3$ and redshift errors $(\sigma_z \lesssim 0.05 (1+z))$. Such a large sample will enable cosmological studies using nearly all of the cosmological probes and allow LSST to
measure with high precision the distance-redshift and growth-redshift relations. 
A precision of $\sim 0.5 \%$ on the distance and $ \sim 2 \%$ on the growth factor is forecast over the 
redshift range $0.5 < z < 3$ (Abell et al. 2009). 
In addition, LSST will discover of order $\sim 10^5$ type Ia supernovae annually, with a mean redshift $z \sim 0.45$, 
and maximum redshift $z \sim 0.7$. Combining distance-redshift constraints from supernovae with weak lensing, LSS, BAO 
as well as galaxy cluster studies will constrain the mean value of the dark energy parameter $w$ to about 1\%.  

\smallskip
\noi Future constraints on the cosmological model and its parameters are likely to be limited by systematics. 
Complementarity of SKA and optical surveys such as LSST for mitigating systematic errors, 
especially for BAO and weak lensing, is widely recognised (see e.g. Bacon et al. 2015) and is an area where joint SKA-LSST studies 
can improve the cosmological constraints. One forefront example is the ability of SKA \hi \, observations to yield estimates of intrinsic position angles of galaxies, thus providing a way to mitigate LSST shear systematics due to intrinsic alignment, arguably one of the main issues in LSST weak lensing programs.

\smallskip
\noi In addition, at least four other science areas will benefit from the complementarities and synergies between SKA and LSST: 

\begin{enumerate}
\item {\bf PhotoZ : } The determination of galaxy redshifts from photometric measurements is a key objective 
of the LSST survey, because this information will be exploited for all the extragalactic and cosmological studies. 
A number of studies have  demonstrated the importance of spectroscopic samples for training PhotoZ reconstruction codes, as 
well as calibration, in particular using cross-correlation methods  (Newman et al. 2013).
The \hi\ sources detected by SKA at large redshift will be a valuable addition to optical spectroscopic samples, 
as well as 3D \hi \, mass maps obtained through the Intensity Mapping technique.  \\
Moreover, SKA can use the LSST catalog to determine the redshift for the radio sources which it will detect.

\item {\bf Galaxy formation and evolution : } Next generation of instruments promise to revolutionize our understanding of galaxy evolution, especially at high redshift. With their comparable footprints (in several redshift bands, LSST will provide optical photometric measurements for nearly all the \hi \, sources that SKA will detect), LSST and SKA will play a key role in this respect.
For instance, SKA continuum survey will obtain unprecedented high-resolution, multi-scale 3D coverage of AGN and star forming regions, while LSST will provide redshift and other properties (mass, metallicity, age) of stellar populations within galaxies. LSST will also benefit from SKA detection of AGN activity in order to correctly model the galaxy spectrum at optical and near-infrared wavelengths.

\item {\bf Strong lenses : } SKA and LSST will each potentially detect several 10000s lensed systems. For SKA, these will include gas clouds, for which SKA will determine distance, but also continuum synchrotron emission from AGN or star forming regions, for which LSST data may provide distance information.
  Furthermore, joint optical and radio data will be powerful to fight false positive and to associate lensed systems to the lensing galaxies, paving the road to accurate mass distribution modelling and studies of galaxy formation  source populations at high magnification. Finally, multi-wavelength measurements and follow-ups of gravitational time delays will represent an important joint cosmological topic for both instruments and communities. 

\item {\bf Transients : }  LSST promises to revolutionise the area of Time Domain Astronomy. With a baseline cadence of 
$\sim 9 \, \mathrm{deg}^2$ field visit every $40\,$s (with 2 back-to-back $15\,$s exposures), and one nominal 
  revisit after 60 min when possible, LSST will be able to scan several time scales significantly shorter than the full-sky
  survey few-day time scale. Furthermore, LSST is committed to distribute in a lapse time of 60 seconds unrestricted alerts of 
  the multi-million transient and variable events detected each night. Long recognised as a key science driver for LSST,
  the transient Universe has also come to the fore in the case of SKA and its pathfinders and precursors, with key programmes
  in fast and slow transients.
  The transient science drivers for SKA have been extensively discussed in (Fender et al. 2015; Macquart et al. 2015; Corbel et al. 2015), ranging from the very high rate of AGN flaring detections to the prospects for TDE detection or FRB breakthroughs.

  Without going into these detailed analyses here, it is clear that the value of finding radio transients is severely
  reduced if counterparts at other wavelengths are not also identified, and among them the optical 
  or infrared bands are often most needed, especially to identify candidates for spectroscopic follow-up (and possibly redshift measurement) in the most exciting cases. While, in the era of parallel wide-field SKA and LSST observations looking at the same portion of the sky,
  the optical data of the quiet counterpart will be readily available for most fields, the actual real-time follow-up for either instrument
  based on alert from the other seems like a formidable goal to achieve. Indeed both LSST and SKA will have unprecedented rates of alerts,
  which will be extremely hard to efficiently triage without prior synchronisation. Thus, while it is important to investigate the best ways to enable target of opportunity observations for either or both instruments, some reflexion should also be devoted to the possibility to build campaigns where the same sky is observed at the same time.

\end{enumerate}

\parbox{0.9\textwidth}{
\noi{References:}\\
\noi{\scriptsize 
  Ivezic, Z., et al., 2008, 0805.2366v4; 
  Ivezic, Z., et al., 2013, http://ls.st/LPM-17; 
  Juric, M., et al., 2015, 1512.07914;  
  Abell, P. A., et al., 2009, 0912.0201; 
  Bacon, D., et al, 2015, AASKA14, 145; 
  Newman, J., et al., 2013, 1309.5384;
  Fender, R., et al., 2015, AASKA14, 51;  
  Macquart, J.P., et al., 2015, AASKA14, 55; 
  Corbel, S., et al., 2015, AASKA14, 53  
}
} 
\subsection{Athena and eROSITA}\label{science:athena-erosita}
\vspace{0cm}

\noi {\bf Galaxy Clusters}

\smallskip

\noi The combination of radio and X-ray observations gives insights into the interplay of thermal and non-thermal emission across a wide variety of astrophysical sources. In this context, the coupling of SKA observations with missions offering deep X-ray survey capability, or new X-ray instruments with unprecedented angular resolution and sensitivity, will lead to new insights. Exploitation of existing missions such as XMM-Newton and Chandra, will hopefully continue as SKA comes on-line. The present contribution discusses further possibilities in the context of galaxy cluster observations for the next-generation X-ray survey, eROSITA (2018), and ESA’s next-generation X-ray observatory, Athena (2028). 

\smallskip


\noi The extended ROentgen Survey with an Imaging Telescope Array (eROSITA, Merloni et al. 2012) is the primary instrument on the Russian Spektrum-Roentgen-Gamma (SRG) mission, to be launched in 2018. eROSITA has been conceived to perform a deep survey of the entire X-ray sky that will be about 20 times more sensitive than ROSAT in the soft energy (0.5-2 keV) band; in addition, the satellite will undertake the first true imaging survey in the hard energy (2-10 keV) band. The seven eROSITA telescopes combined will have an effective area of 0.14 m$^2$ at 1~keV, a field of view of  $1.03^\circ$, and an average survey-mode angular resolution of  $\sim 28^{\prime\prime}$ in the soft band and $\sim 40^{\prime\prime}$ in the hard band. Over the course of its four-year survey, eROSITA is expected to detect $\sim 10^5$ galaxy clusters up to $z > 1$, and $3 \times 10^6$ AGN across the extragalactic sky, in addition to Galactic objects such as accreting binaries, active stars, supernova remnants, and diffuse emission within and without the Galaxy.

\smallskip

\noi A fraction of the X-ray sources discovered in the eROSITA survey will also emit in radio wavelengths. Cross-comparison of X-ray and radio surveys will enable improved source identification and source classification (e.g. of AGN into radio-loud and radio-quiet classes). New discoveries of Galactic and extragalactic accreting black holes will enable the black hole fundamental plane (e.g. Merloni et al. 2003) to be examined in greater detail and at greater dynamic range, leading to new insights into the relationship between accretion and ejection through the disc-jet connection. 

\smallskip

\noi SKA is also expected to detect thousands of extended radio sources associated to the non-thermal (synchrotron) emission from galaxy clusters (haloes, relics and individual radio galaxies; e.g. Norris et al. 2011; Br\"uggen et al. 2012; Reiprich et al. 2013).  The origin of such diffuse radio emission within the intracluster medium (ICM) is unclear, although their properties correlate well with those of the hot gas observed in X-rays (e.g. Cassano et al. 2010)  suggesting a tight relationship between non-thermal and thermal components in a cluster. Extended halo and relic emission in particular is seen to be associated to cluster dynamical state; this information will be useful in combination with information on the X-ray morphology and global parameters obtained from the eROSITA survey to provide insights into the formation and astrophysics of haloes and relics, and the cluster cosmic ray population in general (e.g. Schuecker et al. 2001; see below). Observations of tailed radio galaxies out to large distances trace the ICM (e.g. Blanton et al. 2001) that can be studied via its X-ray emission. Furthermore, estimates of the magnetic field strengths in the intracluster medium (ICM) can be obtained from measurements of Faraday rotation through the ICM, in combination with an independent measurement of the ICM electron density (e.g. Clarke et al. 2001). The intrinsic polarisation of the source need not be known, as the effect can be observed as a characteristic wavelength-dependent rotation measure (RM) signature. Thus the combination of X-ray and radio observations gives unique insights into cluster-wide magnetic fields. The All-Sky SKA Rotation Measure Survey aims to increase by five orders of magnitude existing data sets by providing an all-sky grid of Faraday rotation measures at a spacing of $20-30^{\prime\prime}$ between sources. Combined with measurements of the electron density from eROSITA measurements, this rotation measure grid will improve immeasurably our knowledge of the spatial distribution, physical characteristics, and evolution of galaxy cluster magnetic fields.\\


\smallskip

\noi The Advanced Telescope for High ENergy Astropysics (Athena, Nandra et al. 2013)  is the European Space Agency's L2 mission, scheduled for launch in 2028.  Athena has been conceived to combine a large effective area X-ray telescope with state-of-the-art instrumentation for spatially-resolved X-ray imaging spectroscopy. The telescope, based on ESA’s Silicon Pore Optics technology, will have an effective area of 2~m$^2$ at 1~keV, coupled with an angular resolution of $\sim 5^{\prime\prime}$ (Willingale et al. 2013). The Athena Science Instrument Module will carry the X-ray Integral Field Unit (X-IFU) for high-resolution ($\Delta E \sim 2.5$~eV) imaging X-ray spectroscopy (Barret et al. 2016), and the Wide Field Imager (WFI) for X-ray spectro-imaging across a $40^{\prime}$ field of view (Rau et al. 2016).

\smallskip

\noi Athena is an observatory-class mission that will make detailed observations of many astrophysical objects, so the area of overlap with SKA is potentially vast. Areas in which combined Athena X-ray and SKA radio observations are expected to provide particular insight include:

\begin{itemize}
\item	The relationship between turbulence and bulk motions in the ICM and the presence of radio haloes (e.g. Cassano et al. 2010, 2013). These works have shown that Mpc-scale radio halo emission is associated almost exclusively with systems that present highly perturbed X-ray images, indicating significant dynamical activity, while objects with a relaxed X-ray morphology do not appear to host radio emission. Furthermore, the radio power is correlated to the cluster X-ray luminosity. The combination of SKA and Athena will allow these relationships, which are thus-far limited to the nearest, most massive objects, to be probed to down to far lower masses and up to higher redshifts, opening up the field to studies of evolution.

\item	The detailed physics of shock acceleration in galaxy cluster radio relics. These extended, filamentary radio emission regions appear to be linked to re-acceleration of fossil relativistic particles as a shock travels towards the cluster periphery (van Weeren et al. 2017). Robust measurement of the Mach number from several different methods is necessary to probe these shocks. These  can be determined both in radio (from the spectral injection index) and in X-rays (from application of the Rankine-Hugoniot shock conditions to density and temperature jumps). SKA will vastly increase the number of known relic systems with favourable viewing angles, while the high throughput of Athena will allow accurate pre- and post-shock densities and temperatures to be measured.

\item	The balance between heating and cooling in the centre of galaxy clusters and its relationship to the interaction between the central AGN and the ICM (e.g. Croston et al. 2013). In particular, Athena will allow measurement of the line profiles and centroid variations in the vicinity of the AGN jets, allowing the velocity field to be mapped to high precision. The thermal and non-thermal energy content of the X-ray cavities will also be mapped for the first time, enabling their composition to be established. Finally,  jet power measurements  can be compared to accretion rates of hot and cold material, enabling insights into the accretion process and black hole growth to be obtained.

\end{itemize}

\smallskip

\noi {\bf AGN}
\smallskip

\noi Complimentary  SKA and  Athena observations will help us understand many of the open questions regarding Active Galactic Nucleii (AGN), see Sect.\,\ref{sci:AGN}. In a similar way to the Ultra Luminous Sources (ULXs, see Sect.\,\ref{sci:ULXS}), X-ray and radio observations will help us investigate the mechanical  energy  carried  away by  jets  from AGN. This is believed  to  control  hot-gas  cooling  in  massive ellipticals  and  groups  and  clusters  of  galaxies  via  a  feedback  loop  in  which  jets  heat  the  hot  gas,  suppressing  star  formation and regulating their own fuel supply. Current X-ray observations have revealed compelling evidence for such AGN feedback (e.g. McNamara \& Nulsen 2007). However, the physics of how the balance between these processes is established and maintained, and how it evolves with time, is poorly understood.  Athena observations will allow us to measure the kinematics and energy injected into the environs and the  SKA observations will allow us to analyse, in detail, the nature and scale of the jets. This will help us understand how  the energy  input  from  jets  is  dissipated  and  distributed  throughout  the  intercluster medium,  and  how  the  energy  balance  between  cooling  and  heating is  maintained  and  established  in regions where the most massive galaxies are being formed.  Further, we will also be able to make estimates of the masses of black holes using the X-ray and radio observations, in order to understand the growth of black holes (Sect.\,\ref{sci:ULXS}), by using the fundamental plane of black hole activity.

\smallskip

\noi However, there are galaxies in which the central massive black hole is not accreting at a high rate, i.e. is not an AGN, but can become bright during a tidal disruption event (TDE). These occur when a star in a galaxy wanders too close to the central massive black hole. The star disrupts when the tidal forces exceed the self-gravity of the star and a previously undetected central supermassive black hole will become extremely bright, allowing it to be studied.  10$^{-4}$ – 10$^{-5}$ TDEs are expected per galaxy per year (Rees 1988). To date, only about 30 TDEs have been detected in X-ray (Komossa 2015) out of about 70 candidates proposed to date. As the tidal radius of the black hole must be outside of the Schwarzschild radius for us to observe the tidal disruption event, it is only for the lower mass black holes ($<$ 10$^8$ M$_\odot$) that we detect these events, which may allow us then to detect the extremely rare intermediate mass black holes (IMBH, $\sim$10$^{2-5}$ M$_\odot$, see Sect.\,\ref{sci:ULXS}).  Surveys with X-ray observatories such as SVOM can pick out such events and then they can be followed up rapidly in radio to hunt for jetted emission that has only been seen in 3 TDEs to date (e.g. Burrows et al. 2011; Cenko et al. 2012). Why only some TDEs show such emission is still unclear. Then with Athena the bright X-ray emission can be modelled to determine the mass of the black hole, the accretion mechanism and if the event is close and therefore very bright, to detect an iron line in the spectrum, allowing us to put limits on the black hole spin (e.g. Karas et al. 2014). 

\smallskip

\noi Combined observations with the SKA and Athena will be essential to elucidate the nature of Ultra Luminous X-ray sources (ULXs). Using Athena/X-IFU observations, we will be able to search for emission and absorption lines, evidence for a ULX wind (e.g. Pinto, Middleton \& Fabien 2016) that may be at the origin of the radio nebulae seen around some ULXs (Cseh et al. 2015). With quasi-contemporary SKA and Athena X-ray observations for example, necessary as stellar mass black holes are known to vary rapidly, see Sect.\,\ref{sci:ULXS}, and using the fundamental plane of black hole activity, an estimate of the mass of the black hole can be made. Radio survey observations and complimentary X-ray observations will also be able to identify the many IMBHs in the Milky Way. \\

\parbox{0.9\textwidth}{
\noi{References:}\\
\noi{\scriptsize 
Barret, D., et al., 2016, {\tt  arXiv:1608.08105};
Blanton, E., et al., 2001, AJ, 121, 2915;
Br\"uggen, M., et al., 2012, SSRv, 166, 187; 
Burrows, D.N., et al., 2011, Nature, 476, 421 
Cassano, R., et al. 2010, ApJ, 721, L82;
Cassano, R., et al. 2013, ApJ, 777, 141;
Cenko, S.B., et al. 2012, ApJ, 753, 77;
Clarke, T.E., et al. 2001, ApJ, 547, L111;
Croston, J.H., et al. 2013, {\tt arXiv:1306.2323};
Cseh, D., et al., 2015a, MNRAS, 452, 24;
Karas, V., et al., 2014, arXiv:1409.3746;
Komossa, S.\ 2015, Journal of High Energy Astrophysics, 7, 148;
McNamara, B.~R., \& Nulsen, P.~E.~J.\ 2007, ARA\&A, 45, 117;
Merloni, A., et al., 2012, {\tt arXiv:1209.3114};
Nandra, K., et al., 2013, {\tt arXiv:1306.2307};
Norris, R., et al., 2011, JApA 32, 599; 
Pinto, C., Middleton, M. J., \& Fabian, A. C. 2016, Nature, 533, 64; 
Rau, A., et al., 2016, {\tt  arXiv:1607.00878}; 
Rees, M.~J.\ 1988, Nature, 333, 523; 
Reiprich, T., et al., 2013, SSRv, 177, 195; 
Schuecker, P. et al., 2001, A\&A, 378, 408; 
van Weeren, R., et al., 2017, Nature Astronomy, 1, 0005; 
Willingale, R., et al., 2013, {\tt arXiv:1307.1709}
}}\\

\subsection{CTA} \label{syn:CTA}
\vspace{0cm}

\noi The Cherenkov Telescope Array (CTA) is currently the biggest project of very high energy astrophysics, gathering at this time more than 1500 members spread in 32 countries.
CTA represents the new generation of gamma-ray imaging Atmospheric Cherenkov telescope (IACTs), that will succeed to the current arrays H.E.S.S. (Namibia), VERITAS (Arizona - US), and MAGIC (La Palma - Canary islands).
CTA will be composed of two distinct arrays, one for each hemisphere for a full-sky coverage. The northern array which will take place on the MAGIC site, is expected to hold 19 telescopes, while the southern one in Chili (Paranal) will be much bigger with an nominal configuration of 99 telescopes.
Thanks to CTA, the far end of the electromagnetic spectra from 20 GeV up to 300 TeV will be observed with an unprecedented sensitivity and angular resolution.

\smallskip

\noi With a partial array expected to start operating in 2018-2019, routine user operation beginning at the horizon 2022, and a lifetime of 30 years, CTA will be fully operational during the future SKA era.
The scientific synergies between SKA and CTA will reveal the non-thermal universe by linking the both ends of the electromagnetic spectra for numerous common astrophysical topics, as referenced below.

\smallskip

\noi {\bf Radio-loud active galactic nuclei}

\smallskip

\noi Active galactic nuclei (AGN) are the archetypal sources able to provide powerful emission of radio and very high energy (VHE) gamma-rays with strong but often non-trivial correlation between them.
Many open questions about the nature of the non-thermal emission can only be answered by considering together these two energy domains.
SKA1 will deeply enhance the sample of large scale jet morphologies and polarisation, increasing our understanding of jet global energetics and long term processes of particle acceleration and cooling. 
The future VLBI extension of SKA will also unveil the pc-scale structure and kinematic of numerous new sources.

\smallskip

\noi CTA should boost the sample of AGN detected in the VHE range by about a factor 10, giving new sights from particle emission processes at super-massive black holes vicinity. Its sensitivity will also allow to catch extremely fast flares with variability under a min (Sol et al. 2013). Gamma-ray flares are known to have side effects on radio polarisation and VLBI kinematics still far to be understood which will be a main topic addressed to the future synergies with SKA.
Combining such SKA and CTA results will shed new light on AGN and should solve the long-standing issue of unification scenarios of extragalactic radiosources

\smallskip

\noi {\bf Transients: GRBs, gravitational waves and FRBs}

\smallskip

\noi CTA is developed to scan large portions of the sky dedicated to the search of VHE counterpart of transient phenomena, as gamma-ray bursts (GRBs) or gravitational waves (GWs) which present large uncertainty on their location. Its sensitivity allows much larger regions to scan than previous IACTs with promising detection results for covers as large as $\sim 1000$ deg$^2$ (Bartos et al. 2014). As for the radio GRB afterglows observations, SKA will significantly contribute to the search and characterisation of GWs.
Detecting the same event through the two messengers would allow a better identification of the sources and their environment, and a refinement of their parameters.

\smallskip

\noi Fast radio bursts (FRBs) are one of the most mysterious astrophysical phenomenon highlighted these last years, and are consequently one the main key science project of SKA. Following the progenitors of FRBs, strong gamma-ray emission is expected to be seen (Murase et al. 2016). CTA results from FRB observations will give important constraints on their nature, either by direct detections or upper limits.

\smallskip

\noi {\bf Cosmology: dark matter and reionisation}

\smallskip

\noi The most accepted dark matter hypothesis considers weakly interacting massive particles (WIMPs) which can self-annihilate, converting their large rest masses into other  particles, including gamma-rays. 
The search of a gamma-ray emission from dark matter halos is among the most critical research topics of CTA (Carr et al. 2015). SKA will be naturally associated to this quest by looking the possible indirect radio counterpart of such an annihilation. Indeed, a diffuse radio emission is expected to be emitted from secondary electrons interacting with the magnetised atmosphere of galaxy clusters and galaxies. Hence the two experiments would be able to provide independent cross-checks on the same astrophysical targets.

\smallskip

\noi The study of the reionisation period via the redshifted 21\,cm line will be one of the major SKA achievement, providing direct imaging over most of the redshift range z $\sim$ 6 - 28 with SKA1-LOW. 
Although CTA will not be able to access such a redshift, the AGN gamma-ray absorption by pair creation on the extragalactic background light (EBL) field was proven to give strong constraints on its density. The measurements of EBL until redshift $\sim$ 1 can thus provide important cross-checks of cosmic reionisation models that will be enhanced by SKA observations (Mazin et al. 2013).

\smallskip

\noi {\bf Galactic sources: pulsars, PWN and supernova remnants}

\smallskip

\noi SKA will increase by a large number the detected pulsar population, as well as the understanding of pulsar magnetosphere thanks to its wide bandwidth and exceptional sensitivity. On the other hand, a major breakthrough happened these last years with the direct pulse detection of the the Crab and Vela pulsars by IACTs (Aliu et al. 2008; Gadjus et al. 2015). The improvements that CTA will bring in term of lower energy threshold and sensitivity are expected to provide a spectral gamma-ray characterisation of several pulsars and enhance their multi-wavelength emission models.

\smallskip

\noi It is still unclear whether pulsar wind nebulae (PWN) and supernova remnants accelerate cosmic rays up to PeV energies. This fundamental question should have soon answers from both SKA and CTA.
SKA1, using both the MID and LOW elements, will already unveil dozens of new PWN providing vital new insight in the composition of the wind and the acceleration of leptons.
CTA will also increase the VHE detected number of PWN, but also define more precisely their spectral features (de O\~{n}a-Wilhelmi et al. 2013). And in a more general way, the angular resolution of CTA will lead to spectacular progresses to define the gamma-ray structure of supernova remnants (Nakamori et al. 2015).

\smallskip

\noi {\bf Cosmic particles}

\smallskip

\noi Both experiments plan to better define the cosmic ray population from ground detection and analysis of air-showers. The capabilities of SKA1-LOW from the radio signature of air-showers will be optimised for particles from $1.0 \times 10^{17}$ to $1.0 \times 10^{19}$ eV. CTA will be mainly focused on the electron spectrum, reaching energies up to $\sim$ $1.0 \times 10^{15}$ eV (Picozza \& Boezio 2013). The deep knowledge on air-showers properties gained by the CTA community will be a strong asset for future common works with the SKA collaboration.\\

\parbox{0.9\textwidth}{
\noi{References:}\\
\noi{\scriptsize 
Aliu, E., \etal , 2008, Science, 322, 1221;
Bartos, I., \etal , 2014, MNRAS, 443, 738;
Carr, J., \etal , 2015, PoS, (ICRC2015)1203;
de O\~{n}a-Wilhelmi, E., \etal , 2013, APh, 43, 287;
Gadjus,  M., \etal , 2015, PoS, (ICRC2015)841;
Mazin, D., \etal , 2013, APh, 43, 241;
Murase, K., \etal , 2016, MNRAS, 461, 1498;
Nakamori, T., \etal , 2015, PoS, (ICRC2015)774;
Picozza, P. \& Boezio M., 2013, APh, 43, 163;
Sol, H., \etal , 2013, APh, 43, 215
}}

\subsection{LIGO and Virgo}

\vspace{0cm}

\noi A new era for transient astronomy began with the first gravitational-wave
events detected by the Advanced LIGO detectors. One of the up-coming milestones is to
connect this new type of observation with that of conventional astronomy by the detection of an electromagnetic counterpart to a gravitational-wave event. To
this aim, an extensive electromagnetic follow-up program has been set up through
collaborative agreements signed between the LIGO and Virgo collaborations and
more than 80 astronomer teams around the world. This network of follow-up
instruments includes several SKA precursors such as ASKAP, MeerKAT and the
MWA. Alerts are generated from the gravitational-wave data and communicated to
those teams through a dedicated network.

\smallskip

\noi As of today, short-hard gamma-ray bursts are the most promising counterpart
candidate. There are strong indications that those gamma-ray bursts are
connected with compact binary mergers (including at least a neutron star). If
this is true, gravitational-wave events from such binary mergers would then be
associated with a prompt gamma-ray flash, and an afterglow emission that extends
from X-rays down to radio frequencies. Only few short gamma-ray burst afterglows
have been detected so far. Recent reviews (Chandra \& Frail 2012) mention the
afterglow detection for only two short gamma-ray bursts so far,
i.e. GRB050724 and GRB051221A; both were observed by the Very Large Array at 8.5 GHz.

\smallskip

\noi It is expected that the SKA will revolutionise this science. The SKA-MID survey
will improve those detection rates by many orders of magnitude
(Burlon et al. 2015). Furthermore, it is expected that the SKA will also detect many
``orphan'' afterglows associated to sources that are not pointing to Earth (see Sect.\,\ref{sci:GRB}).
Those future detections will be potentially associated to LIGO/Virgo events
if they occur within the horizon of the gravitational-wave detectors.

\smallskip

\noi The LIGO Scientific Collaboration and the Virgo Collaboration have produced a plan for the scientific
observations to be conducted with the current second generation ``advanced'' detectors over
the next 10 years, through 2025 (Abbott et al. 2016). Over this period, the Japanese KAGRA detector
and the LIGO India observatory will also start operation. The SKA development plan
is well in-line with this schedule.

\smallskip

\noi The size of the gravitational-wave error region is determined by the ratio
between the observed gravitational wavelength by the distance between
gravitational detectors in the network. For instance, the LIGO network resolves
the source in a very large sky region, that extends over 1000 square degrees,
typically. Thanks to its very large field-of-view, including in the GHz range,
the SKA will be a powerful and versatile observatory for the follow-up of
gravitational-wave candidates.\\

\parbox{0.9\textwidth}{
\noi{References:}\\
\noi{\scriptsize Chandra, P. \& Frail, D. A., 2012, ApJ 746, 2;
Burlon, D., et al., 2015, AASKA14, 52;
Abbott, B. P., et al., 2016, LRR, 19, 1
}}\\

\subsection{CMB polarisation experiments}

\vspace{0cm}

\noi Measurements of the cosmic microwave background (CMB) provide the
cleanest experimental approach to search for evidence of an
inflationary phase in the early Universe. In particular, the
primordial gravitational waves generated during that epoch induce a
specific signature in its polarisation properties.

\smallskip
\noi The CMB polarisation signal can be divided into even and odd
parity $E$- and $B$-modes (Kamionkowski et al. 1997). $E$-mode
polarisation, sourced by both scalar (curvature) and tensor
(gravitational wave) perturbations, has been detected at high
significance. However, the primordial $B$-mode polarisation signal,
arising only from the intrinsically weaker tensor perturbations, has
yet to be discovered. We quantify constraints on $B$-modes in terms of
the ratio, $r$, of the tensor fluctuations (gravitational waves) to
scalar (density) fluctuations, evaluated at a given spatial
wavenumber. The $B$-mode power spectrum has a peak at the horizon
scale at recombination ($\ell \sim 90$) with an amplitude proportional
to this ratio. Reionisation then introduces an additional peak at
low-$\ell$ ($\ell \sim 10$) with an amplitude that depends on the
optical depth of the Universe, $\tau$.

\smallskip 
\noi Current measurements of the Galactic foreground emission imply
that primordial $B$-modes will be sub-dominant on all angular scales
and over all observational frequencies in the microwave regime
(Krachmalnicoff et al. 2016; Choi et al. 2015).  Measurements of the
CMB signal at the nK level thus requires subtraction of the
foregrounds at sub-percent accuracy. 

\smallskip
\noi Analysis of the WMAP and Planck data has demonstrated
that the polarised Galactic emission is well-described by 
the combinaton of
synchrotron radiation and thermal dust emission (Planck Collaboration X
2016). Synchrotron emission, produced by cosmic ray electrons
spiraling in the Galactic magnetic field, 
dominates at low frequencies
($< 100$\,GHz),  while at higher frequencies polarised dust emission is the
dominant foreground component.  In principle, the CMB polarisation
signal can be distinguished from these Galactic foregrounds by their
different frequency spectra, and SKA should allow the study of the low
frequency foregrounds up to the 12--14\,GHz window.

\smallskip
\noi The measured synchrotron emission is dependent on the density of the
relativistic electrons along a given line-of-sight, and approximately
to the square of the plane-of-sky magnetic field component, and can be
strongly polarised in the direction perpendicular to the Galactic
magnetic field.  Its spectral energy distribution (SED) is
typically modeled as a power law, although there is no precise
determination of the spatial variation of the synchrotron spectral
index, either in intensity or polarisation. Some form of spectral
curvature is often considered relevant for the description of
observations at microwave frequencies (Kogut et al. 2007).  In detail,
the observed polarised emission is seen to arise mainly in a narrow
Galactic plane and well-defined filamentary structures 
that can extend over 100 degrees across the sky and 
be polarised at a level of
$\sim$40\% (Vidal et al. 2015). However, away from these features and
at high latitudes, the polarisation fraction 
is less than $\sim$15\%.

\smallskip 
\noi Although the synchrotron and thermal dust emission are 
the dominant contributors to the diffuse polarised Galactic foreground,
uncertainties in the current data may still allow other
components
to contribute at
fainter levels. The so-called "Anomalous" microwave emission (AME) has
been studied by Planck towards specific dust and \hii~regions in
intensity and defined the shape of its emission at frequencies above
$\sim$ 30\,GHz. If AME is solely due to spinning dust particles
(Draine et al. 1998), then we expect it to have a 
polarisation percentage of less than 1\%, and a polarisation
fraction that decreases with increasing frequency. Indeed, the most
recent theoretical work (Draine et al. 2016) predicts negligible
polarisation at frequencies above 10~GHz. However, AME might arise
from other physical mechanisms and it is important to confirm its
polarisation properties empirically. While recent measurements imply
that the level at which AME is polarised is low (Planck Collaboration XXV 2016),
a failure to account for it could bias the measurement of $r$ in
$B$-mode searches (Remazeilles et al. 2016). SKA could provide
detailed information about the low frequency emission of this
foreground component.

\smallskip
\noi Although these foregrounds are a particular obstacle to the extraction
of the $B$-mode signal on the largest angular scales, SKA high
resolution and sensitivity maps of smaller regions will still
contribute to an improved understanding of both their physical nature
and statistical properties over the sky. Furthermore, the development of
mosaicing techniques allowing matching between different
fields-of-view and large-scale calibration will allow the mapping of
the foregrounds on intermediate and large angular scales.
Such information is crucial for component separation strategies that
will benefit by extending the frequency range of various
foreground templates adopted in the analysis and/or the inclusion of
constraints about the spectral behaviour or statistical properties of
the different components. In this context, a particularly important
question is the extent to which the
emission from a specific component is coherent across
frequencies. As an example, the synchrotron emission along a given
line-of-sight depends on the distribution of sources and the magnetic
field structure. The detailed interplay of these physical structures,
combined with the extent to which they are correlated or otherwise,
can result in changes of both the polarisation fraction and angle with
frequency. Moreover, this behaviour could incur departures from
the power-law behaviour of either the frequency dependence of the
emission, or power-spectrum parametrisation, as adopted in various
component separation strategies. In addition, the extent to which
foregrounds can be treated as Gaussian random fields, an assumption of
other foreground cleaning approaches, will also be affected by this
complex foreground behaviour.

\begin{figure}[!ht]
  \centering
  \includegraphics[width=0.48\linewidth]{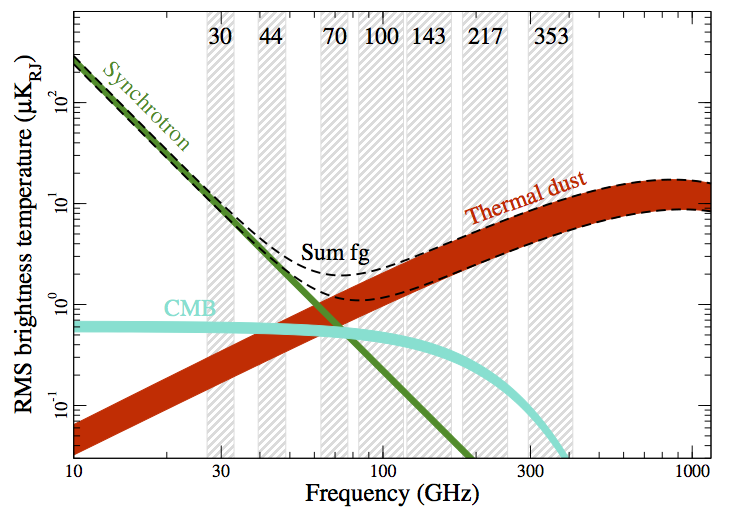}
  \includegraphics[width=0.45\linewidth]{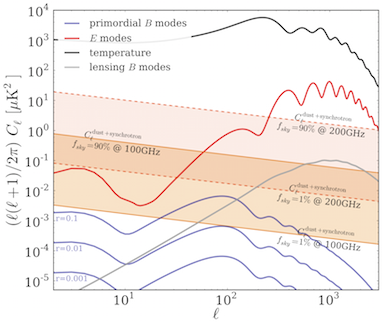}
  \caption{{\em Left}: spectral distribution of polarised foreground emission at 
1 degree resolution as estimated for between 80 and 90\% 
sky coverage. Units are r.m.s brightness temperature. Taken from
Planck Collaboration X (2016). {\em Right}: angular power spectra showing
primordial $B$ modes, lensing $B$ modes, total intensity, and $E$
modes, as well as the total contribution of polarised $B$-mode
foregrounds (dust plus synchrotron), expected on the cleanest 1\% to
90\% of the sky, at $100$ and $200$\,GHz. From Errard et al. (2016).
}

  \label{fig:CMBForegrounds} 
\end{figure}

\smallskip
\noi The presence of polarised extragalactic sources constitutes an
additional source of contamination.  In general, the compact source
contribution does not impact the large angular scales (near the
reionisation peak), but can play an important role on intermediate and
small angular scales. It is likely that the contamination will need to
be removed with a combination of masking or subtracting individual
sources, and modelling residuals at the power-spectrum level.
We can seek to improve the source detection process using as a proxy
deep catalogues of compact sources detected by SKA. It should also be
noted that a large fraction of the radio sources that can be detected
in CMB experiments are variable. One way of addressing this problem is
to plan follow-up observations of sources in such a way that these
observations are performed near simultaneously with the CMB
observations. \\

\parbox{0.9\textwidth}{
\noi{References:}\\
\noi{\scriptsize 
Choi, S. K. \& Page L.A., 2015, JCAP, 12, 20;
Draine, B. T. \& Lazarian A., 1998, ApJ, 508, 157;
Draine, B. T. \& Hensley B. S., 2016, ApJ, 831, 59;
Errard, J., \etal, 2016, JCAP, 03, 052;
Kamionkowski, M., \etal, 1997, PRL, 78, 2058;
Kogut, A., \etal, 2007, ApJ, 665, 355;
Krachmalnikoff, N., \etal, 2016, A\&A, 588, A65;
Planck Collaboration X, 2016, A\&A, 594, A10;
Planck Collaboration XXV, 2016, A\&A, 594, A25;
Remazeilles, M, \etal, 2016, MNRAS, 458, 2032;
Vidal, M., \etal, 2015, MNRAS, 452, 656
}}\\

\subsection{ALMA}\label{science:alma}
\vspace{0cm}

\noi The Atacama Large Millimetre Array (ALMA), located on the
Chajnator plateau at 5000~m altitude in the Chilean Atacama desert, is
a giant array of 66 radio antennas with baselines of up to 16~km. ALMA
operates in the millimetre radio regime, in ten frequency bands
between 40 and 960~GHz. This joint European/US and Japanese
interferometer, which is now nearing full commissioning, is the most
advanced millimetre radio facility in the world, with a strong focus
on the cold neutral medium and star formation in the Milky Way and
external galaxies. ALMA is groundbreaking in particular because it
opens up new observational windows in the sub-millimetre regime, and
because it offers the unique possibility of high-resolution
observations with minimum beam sizes of currently about
0.02\arcsec. Its main strengths are to characterise the cold
interstellar medium, molecular clouds, and star and planet formation
in the Milky Way, and to trace cold atomic and molecular gas and star
formation in external galaxies throughout cosmic history. ALMA can
image the gas kinematics in protostars and protoplanetary disks around
young Sun-like stars in the nearest molecular clouds, and detect
spectral line emission from CO or [CII] in a Milky-Way like galaxy at
redshift $z=3$. France is amongst the most competitive ALMA users,
having reached the highest European proposal success rate in 2015.

\smallskip
\noi Combining SKA and ALMA will boost our capabilities in studying
star formation even further. Neither facility alone covers the full
range of molecular transitions and star-formation tracers, which
together characterise the immense complexity in the main environments
where stars form, be it at the scales of planetary disks around young
stars, giant molecular clouds and star-forming regions, or entire
galaxies and galaxy clusters. Examples where SKA/ALMA synergies 
will be particularly fruitful range from the erosion of 
protoplanetary disks around nearby massive star to studies of
large samples of galaxies over wide redshift ranges (e.g. Carilli et al. 1999).

\smallskip
\noi For example, in protoplanetary disks, ALMA will trace the
molecular gas, whereas SKA will probe the photoionised gas flow from
the disk as well as the interaction zone between the gas and the
impinging stellar wind (e.g, Fuller et al. 2014). ALMA and SKA will
also be highly complementary in probing dust grain growth and the
formation of complex molecules within these disks (Fuller et
al. 2014), an important step towards understanding the origin of life on
Earth, and potentially in other planetary systems. For extragalactic
studies, SKA will quantify AGN and star formation activity from the
synchrotron and free-free components of the centimetre continuum, and
probe neutral atomic Hydrogen through HI, whereas ALMA will trace
dense molecular clouds and warm dust heated by star formation. The
tight empirical correlation between far-infrared and centimetre radio
continuum (FRC) and radio polarimetry will open up novel avenues to
probe star-formation histories and AGN content, and magnetic fields,
cosmic ray budgets, and the internal structure of the interstellar
medium in galaxies out to high redshift, which are hitherto very
little explored for all but the most nearby galaxies. The FRC and HI
surveys will provide redshifts for large samples of star-forming
galaxies over wide redshift ranges, and enable targeted follow-up with
ALMA of the molecular gas and dust. The SKA/ALMA synergy will be
transformational to study how diffuse, ambient gas seen in HI
condenses into molecular clouds, unimpeded by dust, and from the Milky
Way to the most intensely star-forming galaxies at the peak in cosmic
star formation history. Understanding the link between star formation
and galaxy growth is one of the fundamental open questions of
extragalactic astronomy today.

\smallskip
\noi Another fundamental question is why star formation and baryon
cooling are so inefficient, in individual molecular clouds abd distant
galaxies, alike. Only about 1\% of the available gas forms stars
within a free-fall time (e.g. Shimajiri et al. 2017). Theory suggests
that turbulence and magnetic fields both play a role, but only
turbulence can be probed directly today with ALMA (e.g. Canameras et
al. 2017). Magnetic fields are probably also important (e.g.
Hennebelle et al. 2011), but can be studied directly and at high
resolution only when SKA will become operational.
 
\smallskip
\noi On top a wealth of complementary scientific themes, ALMA
illustrates also how a novel world-class facility can become the basis
of a new, transformative field at the forefront of current
research. The success of ALMA, one of the most competitive
observatories in the world today, is therefore an excellent example to
highlight the immense foreseeable scientific and community-building
potential of SKA.\\

\parbox{0.9\textwidth}{
\noi{References:}\\
\noi{\scriptsize 
Canameras, R., et al., 2017, A\&A accepted, astro-ph/1704.5853;
Carilli, C. L. \& Yun, M.S., 1999, ApJL, 513, 13; 
Fuller, G., et al., 2015, AASKA14, 152;
Hennebelle, P., et al., 2011, A\&A, 528, 72; 
Shimajiri, Y., et al., 2017, A\&A accepted, astro-ph/1705.00213
}}\\

\subsection{NenuFAR} \label{syn:nenufar}
\vspace{0cm}

\noi NenuFAR (the New extension in Nan\c{c}ay upgrading LOFAR) is a low-frequency SKA pathfinder presently in construction and commissioning in Nan\c{c}ay (Zarka et al. 2012, 2015a). Made of $\sim$100 hexagonal arrays of 19 antennas, it will be altogether and simultaneously a  beamformer more sensitive than the LOFAR core (with $\sim1^\circ$ instantaneous resolution) and a spectropolarimetric imager in the range 10-85 MHz. The imager will work in fast (1 sec) low resolution ($\sim1^\circ$) mode, and in a slower mode ($\sim$8 hours) providing 10' resolution. Being connected to the LOFAR array, NenuFAR will also be a LOFAR super LBA station. All the different modes are usable together in parallel, with the only constraint that the analog beam of each elementary 19-antenna array can point at a single direction of the sky at any given time (bit different arrays can point to different directions and the instantaneous analog FoV is broad, of order of several 10's degrees. NenuFAR will be equipped of various backends: the LOFAR station backend in super station mode, an autonomous beamformer (LaNewBa) capable of synthesising up to 768 beamlets of 200-kHz bandwidth independently steerable, and a LOFAR-like correlator. The beamformer will be connected to a pulsar receiver (computing in real-time dedispersion + dynamic spectra), and hopefully to a SETI-machine (computing super-high spectral resolution maps).

\begin{figure}[!ht]
  \centering
  \includegraphics[width=0.8\linewidth]{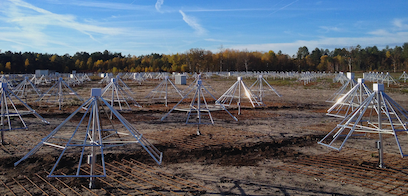}
  \caption{\label{NenuFAR}
The NenuFAR array in construction in Nan\c{c}ay.}
\end{figure}

\smallskip

\noi Beyond the technological experience gained through the development of NenuFAR, that can be partly applied to SKA (see Sect.\,\ref{tech:NenuFAR}), NenuFAR will also serve to prepare the French community, through the analysis of LF radioastronomy products (dynamic spectra and image cubes, in intensity and polarisation), to the exploitation of SKA. A large fraction of the French SKA community is interested in its pathfinder NenuFAR.

\smallskip

\noi Being located in the northern hemisphere (at a latitude of 47.4$^\circ$ North and a longitude of 2.2$^\circ$ East), NenuFAR can cover the latitude range from 90$^\circ$ to about -20$^\circ$ latitude. SKA-LOW will be located in Western Australia at a latitude about 26.5$^\circ$ South and a longitude of 116$^\circ$ East, and will observe up to +20$^\circ$ latitude. The two instrument will thus share the equatorial sky. Being separated by $\sim$7.6 hours in longitude, they will be able to observe simultaneously any given target about 1 hour per day quite far from zenith, but conversely they can provide a complementary coverage for $\sim$16 hr per day of sources within $\pm20^\circ$ latitude. NenuFAR covers the spectral range 10-85 MHz, whereas SKA-LOW will cover the range 50-350 MHz. Thus besides ensuring the above-mentioned coverage in the band 50-85 MHz, NenuFAR will considerably extend the SKA-LOW range toward the lowest frequencies observable from the ground.

\smallskip

\noi SKA and NenuFAR will be very complementary for studying several scientific subjects. While SKA will be the best instrument to detect the Epoch of Reionisation signal, NenuFAR may be a serious competitor to SKA for the detection of the dark ages/cosmic dawn signal, due to its greater compacity (filling factor) (Koopmans et al. 2015). SKA will be a formidable pulsar detection machine, but NenuFAR may allow to study their spectra down to 10 MHz, identifying turnover frequencies, measuring DM with exquisite accuracy and addressing in a complementary way interstellar propagation effects and the physics of pulsar magnetospheres (Zakharenko et al. 2013). LOFAR detected far less non-pulsar transients than expected in imaging mode (Stewart et al. 2016), so that the outcome of SKA in that domain in difficult to predict. Conversely, the Ukrainian UTR-2 array detected many transient in its blind beamformed search of the northern sky (Zakharenko et al. 2015; Vasylieva 2015). NenuFAR being a sensitive beamformer, it is expected to be an efficient transients detector. The blind detection of transients by an instrument will motivate comparisons and follow-up studies with the other one.

\smallskip

\noi SKA will study stellar emissions (flares, cool dwarfs -- Sect.\,\ref{science:stars}) and search for emissions induced by star-planet plasma interactions (Sect.\,\ref{science:exop}), radio emissions from planetary magnetospheres themselves may be at too low frequencies for SKA (like Jupiter's decametric auroral and satellite-induced emissions). They will be accessible with NenuFAR, down to 10 MHz. Jupiter itself, will have its synchrotron radiation studied with SKA (Sect.\,\ref{science:jupiter}) but its decametre emission accessible from NenuFAR only. For Solar studies (Sect.\,\ref{science:sun}), SKA and NenuFAR will be very complementary, covering respectively the altitude range (for plasma emissions) from 1. to $\sim$1.5 R$_S$ and from $\sim$1.2 to 3--5 R$_S$. This also means that scintillation of radio sources through the Solar wind can be studied up to that distance with NenuFAR. NenuFAR and SKA will cover the different frequency ranges where planetary lightning (Saturn, Uranus, Venus? -- Sect.\,\ref{science:plan_light}), terrestrial lightning and transient luminous events (Sect.\,\ref{science:atmosphere}), meteors showers (Sect.\,\ref{science:meteorshowers}), and Radio Recombination Lines (Asgekar et al. 2013; Oonk et al. 2014) can be usefully studied. It is also desirable that SETI searches be conducted in ``piggyback" mode both on SKA (Sect.\,\ref{science:seti}) and on NenuFAR. The science of NenuFAR has been described in a reference document (NenuFAR-France consortium 2014) that should be published soon.\\

\parbox{0.9\textwidth}{
\noi{References:}\\
\noi{\scriptsize 
Asgekar, A.,  \etal , 2013, A\&A, 551, L11;
Koopmans, \etal , 2015, AASKA14, 1;
NenuFAR-France consortium, 2014, https://nenufar.obs-nancay.fr/Argumentaire-scientifique.html;
Oonk, J.B.R., \etal , 2014, MNRAS, 437, 3506;
Stewart, A.J., \etal , 2016, MNRAS, 456, 2321;
Vasylieva, I., 2015, PhD thesis, https://tel.archives-ouvertes.fr/tel-01246634;
Zakharenko, V.V., \etal , 2013, MNRAS, 431, 3624;
Zakharenko, V.V., \etal , 2015, OAP, 28, 252;
Zarka, P., \etal , 2012, SF2A-2012, S. Boissier \etal eds., 687;
Zarka, P., \etal , 2015a, Int. Conf. Antenna Theory \& Techniques, Kharkiv, Ukraine, 13;
Zarka, P., \etal , 2015, AASKA14, 120
}}

\subsection{SVOM}\label{syn:SVOM}
\vspace{0cm}

\noi Scheduled for a launch in 2021, SVOM (``Space-based multi-band astronomical Variable Objects Monitor'') is a Sino-French space mission dedicated to the study of the transient sky, focusing on the Gamma-Ray Bursts (GRBs).
 
\smallskip
\noi The satellite payload encompasses a coded-mask telescope operating in the 4-150 keV energy range for real-time detection and localisation of all known types of GRBs, a non-imaging gamma-ray monitor extends the GRB spectroscopy up to MeV energies, and two narrow-field follow-up telescopes to refine the GRB positions and to study their afterglow emission in the X-ray and visible bands. The pointing strategy of the satellite has been optimised to favour the detection of GRBs and transients located in the night hemisphere, in order to enhance ground-based observations in the first minutes and to facilitate GRB redshift measurements by the largest telescopes. The SVOM ground segment combines a wide-field camera to catch the GRB prompt emission in the visible band and two robotic telescopes to measure the photometric properties of the early afterglow in the NIR/visible band. 

\begin{figure}[htpb]
  \centering
  \includegraphics[trim=60 80 60 80,scale=0.7,clip=true]{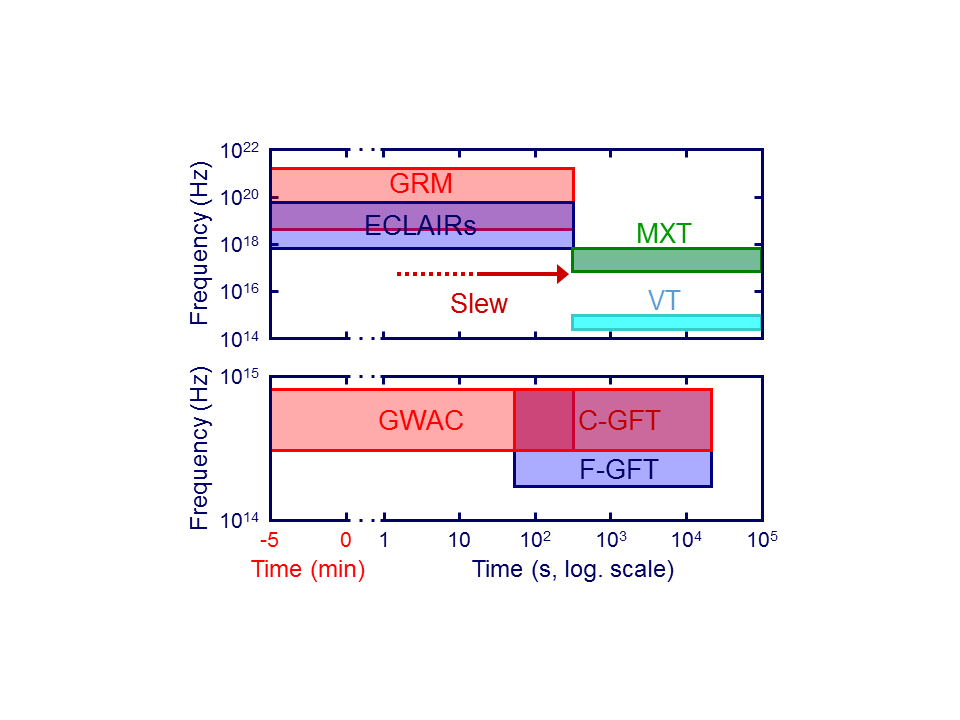}
 \caption{Spectral coverage of the GRB prompt emission and its afterglow with SVOM instruments, as a function of time (the burst is detected at time t = 0). The upper panel refers to space-born instruments (GRM, ECLAIRs, MXT, VT), the lower panel to ground-based instruments (GWAC, C-GFT, F-GFT).}
 \label{SVOM_Instru}
\end{figure}

\smallskip
\noi All details of the SVOM mission and science perspectives are given in Wei et al. 2016. Fig.\,\ref{SVOM_Instru} summaries the spectral coverage of the different SVOM instruments.

\begin{figure}[!ht]
\centering
\includegraphics[trim=52 250 52 250,scale=1.,clip=true]{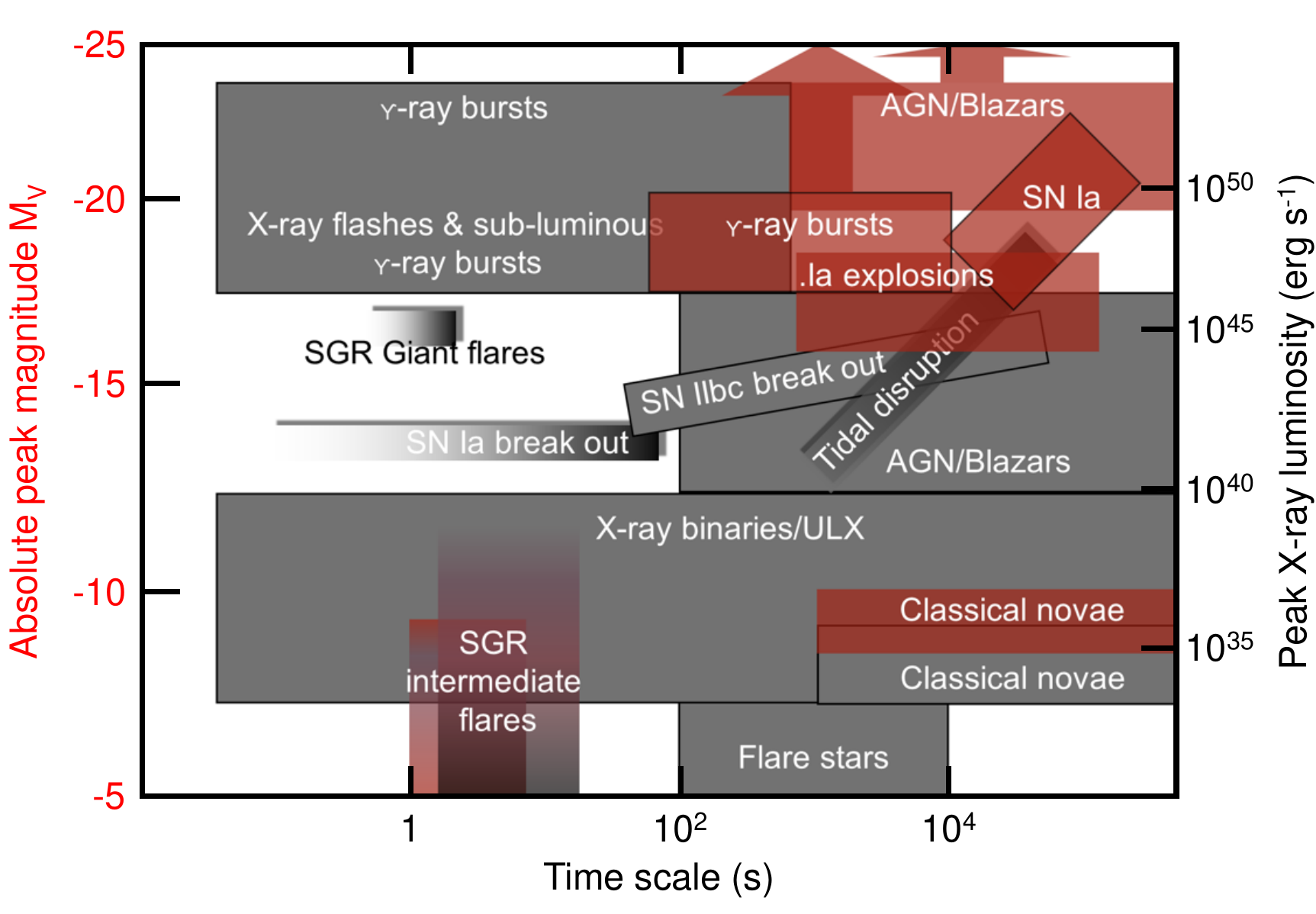}
 \caption{\label{fig:SVOM_transients} The characteristic absolute optical magnitude (red) and X-ray luminosity (grey) versus characteristic timescale for various classes of object which could be studied by SVOM. The optical/X-ray scales are on the left/right axes. Sources/phenomena that have been relatively poorly studied are plotted without border and with fading shading. Some object classes have a large range in characteristic timescales due to different physical processes. Adapted from Jonker et al. 2013.}
\end{figure}

The observation program of SVOM is split in three part :

\begin{itemize}

\item CP : the Core Program is dedicated to the GRB science. Sect.\,\ref{sci:GRB} describes the GRB science with SKA. SVOM will provide GRB alerts to the community and is expected to play a central role in GRB science.

\item GP : the General Program consists in the targets the satellite will follow while waiting for a GRB. The General Program will allow the survey of interesting targets compliant with the satellite attitude law (Virgo cluster for example). Joint observations with SKA could be planned.

\item ToO : the Target of Opportunity program consists in the target programmed from the ground after the reception of an alert (there are different kind of ToO depending on the delay we accept for the start of the observations). The ToO program is of first interest for transient events detected by SKA. Fig.\,\ref{fig:SVOM_transients} shows interesting transient events that can be studied with SVOM. The nominal ToO program (ToO-NOM) allows one ToO per day of one orbit with a typical 48h delay (5 ToO-NOM per day after 3 years). With the ToO-NOM program we will be able to follow SKA transients events to characterise them both in X-rays and optical (with space and ground telescopes). The exceptional ToO-EX allows to reduce the delay to 12h in case of an interesting event (once per week in average) and permits to extend the observation to one day. Finally the multi-messenger ToO-MM will allow the follow-up of multi-messenger events whose localisation is uncertain by tiling a part of the error box region (gravitational wave will be of first importance (see Sect.\,\ref{sci:GW}) as well as HE neutrinos).

\end{itemize}

\smallskip 

\noindent
SKA and SVOM will be major instruments. At the interface we find the transient events observed in multi-wavelength (GRBs,...) and the multi-messenger astrophysics. Synergies have to be identified and dedicated programs built, in the following years, to be ready for the launch of SVOM.\\

\parbox{0.9\textwidth}{
\noindent
{References:}\\
\noindent
{\scriptsize 
Wei, J., et al., 2016, arXiv:1610.06892 : The Deep and Transient Universe in the SVOM Era: New Challenges and Opportunities - Scientific prospects of the SVOM mission;
Jonker, P., et al., 2013, arXiv:1306.2336
}}\\

\subsection{VLBI}\label{sci:vlbi}
\vspace{0cm}

\noi VLBI is a unique tool in astronomy to study the compact radio emission of celestial bodies in extreme details and to pinpoint their direction in the sky with unprecedented accuracy. The technique has been used in fundamental astronomy for establishing celestial reference frames and in astrophysics to investigate the non-thermal continuum and line emission of galactic and extragalactic objects. The main such targets studied with VLBI are the black-hole powered active galactic nuclei and their relativistic jets, disks, shells, and the atomic and molecular gas outflows in star forming regions, young stellar objects, stars, supernovae and their descendants. Additionally, ultra-precise relative VLBI astrometry has made possible the determination of distances and transverse velocities of Galactic objects out to a distance of tens of kpc through measurements of proper motions and parallaxes.

\smallskip
\noi The current VLBI arrays which have the capability to observe in the SKA1-MID frequency range (350~MHz--14~GHz) are the European VLBI Network (EVN), which comprises antennas in Europe, China and South Africa, the Very Long Baseline Array (VLBA) in the United States, and the Australian Long Baseline Array (LBA). At higher frequencies (22 GHz and above), the VERA (VLBI Exploration of Radio Astrometry) array in Japan and the Korean VLBI Network (KVN) are also available. All such arrays are run independently, based on pier-reviewed scientific proposals submitted by the community. They can also conduct joint observations if needed.
On the geodesy and astrometry side, VLBI observing is coordinated by the International VLBI Service for geodesy and astrometry (IVS). The primary mission of IVS is to monitor the Earth's rotation and establish fundamental terrestrial and celestial reference frames. The Bordeaux (LAB) and Paris (SYRTE) VLBI groups are partners of IVS and contribute as analysis and data centres. Of particular interest is the next generation IVS~system, namely VGOS (VLBI Global Observing System), designed to observe 24 hours a day, ultimately all year round, and in the process of deployment. VGOS will be made up of fast-moving 12-m antennas similar to those of the SKA DISH array and will observe a broad frequency band (2--14 GHz), largely overlapping with Bands~4 and~5 of SKA.

\begin{figure}[t]
  \centering
  \includegraphics[width=0.70\linewidth,clip=true,trim=0 16mm 0 15mm]{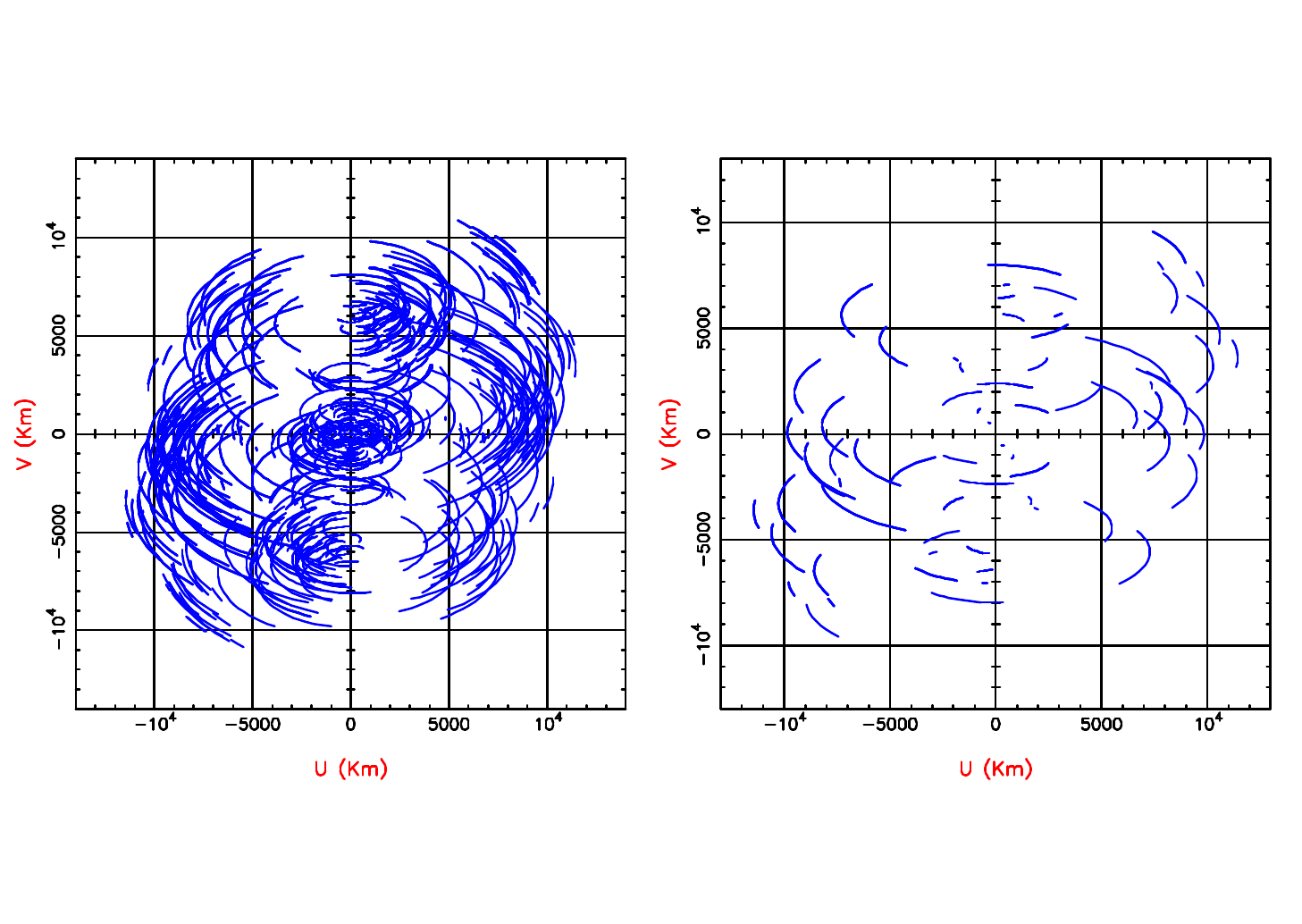}
  \caption{{\em Left}: $u$-$v$ coverage at the Galactic centre for a 12-hour observation with 24 telescopes from the EVN, LBA and AVN with SKA1-MID. {\em Right}: $u$-$v$ coverage for a more typical 4-hour observation for a source at $-20^{\circ}$~declination with 9~telescopes from the above configuration (reproduced from Paragi et al. 2015).}
  \label{uv-coverage}
\end{figure}

\smallskip
\noi In Europe, the Joint Institute for VLBI in Europe (JIVE), an ERIC entity funded by 10~partners, including France, is charged with the processing of the observations acquired with the EVN using its high-level correlator. Additionally, JIVE provides assistance to the users at all stages of the observing process, from proposal preparation to data analysis. In the coming years, the technical capabilities and services offered by the institute will be further enhanced through the JUMPING JIVE (2017--2020) H2020~project, of which CNRS (through the~LAB) is a partner. The enhancements include adding a geodetic/astrometric capability to the correlator, making the observation scheduling smoother, and preparing for the future, especially for SKA-VLBI. Work will also be directed towards incorporating new telescopes and developing new partnerships, notably in Africa, along with outreach.

\smallskip
\noi The inclusion of a high-resolution component with baselines longer than 1000~km has long been considered an essential part of the SKA concept. Such a component will greatly broaden the science delivered by SKA by providing unparalleled flux density sensitivity on compact angular scales probed by VLBI. Different approaches will be considered for SKA1 and SKA2 to achieve high resolution. In SKA1, the SKA core will be incorporated as an additional element in existing VLBI networks, hence boosting their sensitivity, with correlation accomplished using current VLBI correlators. Key to the success of this approach will be the addition of a few (2--4) remote telescopes in Africa to provide the short and medium length baselines to the SKA core required to obtain good $u$-$v$ coverage (Fig.\,\ref{uv-coverage}). Ideal locations for these stations would be the developing African VLBI Network (AVN) stations in Zambia, Ghana, Kenya, and Madagascar (Gaylard et al. 2011). In the current era of rapidly expanding electronic-VLBI (e-VLBI) array, the organisation of such an array should not present any logistical problems. Implementation of the high-resolution component in SKA2 will be accomplished by having the remote stations an integral part of the array, with a significant portion of the collecting area (e.g. 25\%) extending thousands of kilometres from the core, and with processing carried out with the SKA processor. This second-stage approach will make SKA a true array of transcontinental dimensions providing milliarcsecond resolution at sub-\muJy\  sensitivities. An overview of the science to be achieved with such an instrument may be found in Godfrey et al. (2011) and Paragi et al. (2015). As for the reference frame area, further details are also given in Sect.\,\ref{sci:RF} of this book.\\

\parbox{0.9\textwidth}{
\noi{References:}\\
\noi{\scriptsize Gaylard, M. J., et al., 2011, Proceedings of SAIP2011, the 56th Annual Conference of the South African Institute of Physics, Eds. I.~Basson and A.~E. Botha, p.~425; 
Godfrey, L., et al., 2011, SKA Memo No. 135;
Paragi, Z., \etal, 2015, AASKA14, 143
}}\\

\newpage
\section{Technological developments}\label{technology}

\noindent {\normalsize Contributors of this section in alphabetic order: }

\smallskip

\noi {\sffamily \scriptsize
{\sffamily\bf M.~Allen} [\stras],
{\bf R.~Ansari} [\lal],
{\bf R.~Ammanouil} [\lagrange],
{\bf J.~Bobin} [\irfu;\aim],
{\bf S.~Bosse} [\usn],
{\bf R.~Boyer} [\lss]
{\bf M.~Caillat} [\lermasorb],
{\bf J.~Cohen-Tanugi} [\lupm],
{\bf B.~Da~Silva} [\usn],
{\bf L.~Denis} [\usn],
{\bf W.~van Driel} [\gepi],
{\bf M.~N.~El~Korso} [\leme],
{\bf A.~Ferrari} [\lagrange],
{\bf C.~Ferrari} [\lagrange],
{\bf R.~Flamary} [\lagrange],
{\bf N.~Gac} [\lss],
{\bf S.~Gauffre} [\lab],
{\bf F.~Genova} [\stras]
{\bf J.~Girard} [\irfu;\aim],
{\bf P.~Larzabal} [\satie],
{\bf D.~Mary} [\lagrange],
{\bf D.~Mourard} [\lagrange],
{\bf J.-F.~Nezan} [\insarennes],
{\bf V.~Ollier} [\satie;\lss],
{\bf B.~Quertier} [\lab],
{\bf S.~Rakotozafy Harison} [\usn],
{\bf C.~Richard} [\lagrange],
{\bf J.-L.~Starck} [\irfu;\aim],
{\bf M.~Tagger} [\lpcee],
{\bf C.~Tasse} [\gepi],
{\bf S.~Torchinsky} [\apc],
{\bf P.~Zarka} [\lesia;\usn]
}

\subsection{SKA pathfinders and prototypes}\label{tech:pp}

\subsubsection{LOFAR and NenuFAR}\label{tech:NenuFAR}
\vspace{0cm}

\noi When they joined the international LOFAR collaboration and installed the LOFAR FR606 station in Nan\c{c}ay, French scientists and engineers started developing NenuFAR (the New extension in Nan\c{c}ay upgrading LOFAR), an original instrument aimed at pushing performances in the lowest frequency range of the telescope (10-85 MHz). The initial goal was to create a massive LOFAR  ``superstation'', replacing each of the 96 low-band antennae of FR606 by a mini-array of 19 analog-phased antennae, thus promising to enhance considerably and in many ways  the performances of  LOFAR. As the design progressed it became clear that, when used by LOFAR in single-station mode, this led to an instrument with unique capabilities. It was thus decided to add to NenuFAR a local dedicated back-end, that permits it to be operated {\em simultaneously} and {\em independently}, as two instruments: the LOFAR SuperStation (LSS) and a stand-alone fully autonomous instrument (Zarka et al. 2014). The only constraint is that the stand-alone observations must be done within the wide (10's of degrees) station beam defined by LOFAR when the european network uses the superstation. More recently, it was decided to make the stand-alone NenuFAR an imaging instrument by adding a few ($\sim 6$) mini-arrays a few km from the core, enabling it to produce images with a strongly reduced confusion limit. When completed, NenuFAR will include 96 mini-arrays of 19 dual-polarised elements, densely covering a disk of 400 m in diameter, plus the $\sim 6$ remote ones. Its construction and operations are fully  modular, so that it  is presently under construction with funding secured for $\sim 60$ mini-arrays plus the imager, and commissioning is in progress. NenuFAR (see also Sect.\,\ref{syn:nenufar} for the scientific synergies) has been labelled as a pathfinder of SKA. 

\begin{figure}[!ht]
  \centering
  \includegraphics[width=0.8\linewidth]{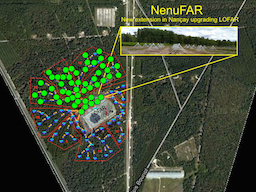}
  \caption{\label{fig:NenuFARLay}
Layout of the 96 NenuFAR tiles, with green circles indicating the part of the array that will be completed by the end of 2017.}
\end{figure}

\smallskip
\noi The technological developments have involved: 
\begin{itemize}
\item The antennae radiators: optimised parameters
include a broad and smooth beam (nearly isotropic,
albeit with extinction below 20 deg elevation) and
maximum efficiency (related to electrical and ground
losses), over a large frequency bandwidth; this implies
an antenna radiation resistance and (low) reactance as
constant as possible over the band of interest and versus
time; cost effectiveness and compatibility with
LOFAR strongly favoured linearly polarised crossed
dipoles; the optimal compromise is a thick inverted-
V dipole similar to the LWA Fork (see Fig.\,\ref{NenuFAR}), with a
metallic ground screen. Since the optimisation study resulted in a design extremely close to the one used for the LWA,  the LWA radiator model was chosen.
\item The antennae contain each an ASIC preamplifier, whose design (made in France) ensures a very flat response over the whole (10-85 MHz) frequency range with a noise ~10 dB below the sky noise level (Girard et al. 2012).
\item The phasing within the mini-arrays is performed by delay lines. In order for this to
be cost-effective, they must be mutualised for
groups of antennas, e.g. by arranging antennas with a regular spacing in two orthogonal directions. This has lead to a choice of 19 antennas per mini-array in a 3 /
4 / 5 / 4 / 3 pattern with each line shifted by 1/2 inter-
antenna spacing relative to its neighbours.
The absolute value of inter-antenna spacing was set to
5.5 m in order to maximise the effective area without
overlap at LF, while keeping the NenuFAR extent
compatible with its hosting at the Nan\c{c}ay station.
\item The 96 tiles are arranged in a relatively dense
layout (see Fig.\,\ref{fig:NenuFARLay}), providing
a smooth overall beam with a low side lobe
level.
The optimal distribution of the 96 mini-arrays was
computed using the ``Boone algorithm'' (Boone 2001), taking into account
a site mask of the Nan\c{c}ay station including its limits
and forbidden areas (the station FR606 itself and
other antennas of the site). It gives NenuFAR a smooth,
Gaussian distribution of visibilities in the (u,v) plane,
and thus a near-Gaussian  beam pattern.
The layout of trenches and cables connecting the 
mini-arrays to the FR606 backend was optimised using a
reasonable cost ratio per unit length of trench/cable in
input to a specific optimisation algorithm. 
\item 
Pointing will be controlled by a
dedicated LCU (Local Command Unit) connected to
96 electronic modules, one in each LF tile, specifying
the phasing scheme to be applied at any given time;
these modules are designed to be completely
radio-quiet outside pointing time; pointing will occur
at intervals from 10 to 60 sec, ensuring low gain
variation in the main beam direction. The NenuFAR
LCU will be connected to the LOFAR LCU of station
FR606. When NenuFAR is used in the SuperStation mode pointing commands will come from the LOFAR 
operations centre via the LOFAR LCU and be translated
into phasing commands of the mini-arrays by the NenuFAR LCU; data recorded by the local LOFAR backend will
be sent to the LOFAR central correlator. In Standalone
mode, pointing orders will come from a local
command computer, and the data will be recorded
locally.
\item
An autonomous beamformer (LaNewBa) has been developed for use in the stand-alone mode, LaNewBa can synthesise up to 768 independently steerable beamlets of 200-kHz bandwidth. A  LOFAR-like correlator for the imager will be implemented.
Furthermore, the beamformer will be connected to UnDySPuTeD, a pulsar receiver, computing in real-time dedispersion and dynamic spectra over the full 10-85 MHz band, with variable spectral and temporal resolution.
Hopefully a SETI-machine (computing super-high spectral resolution maps) will be added.
\end{itemize}

All documentation can be found at the \href{https://nenufar.obs-nancay.fr/}{\color{blue} \myul[blue] {instrument website}}.\\

\parbox{0.9\textwidth}{
\noi{References:}\\
\noi{\scriptsize Boone, F., 2001, A\&A, 377, 368;
Girard, J., et al., 2012, C.R. Phys. 13, 33;
Zarka, P., et al., 2014, \href{http://nenufar.obs-nancay.fr/IMG/pdf/nenufar-science-case-v5_2014_10_10_pz.pdf}{\color{blue} \myul[blue]{NenuFAR : instrument description and science case}}
}}\\
\subsubsection{EMBRACE}\label{tech:embrace}
\vspace{0cm}

\begin{figure}[ht!]
 \centering \includegraphics[width=0.95\linewidth,clip]{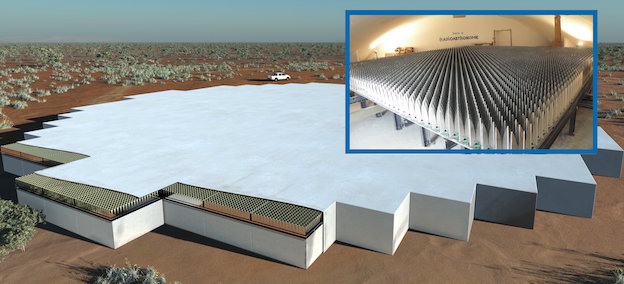}
\caption{This artist's conception shows a single station of a mid-frequency
 aperture array instrument proposed for the Square Kilometre Array
 (SKA).  Over two hundred stations will be required for the full
 SKA. The inset shows the currently operational EMBRACE array at
 the \nancay\ Radio Observatory in France.  \enancay\ is composed of
 4608~Vivaldi antenna elements separated from each other by 12.5~cm,
 making it a dense array for frequencies above 1200~MHz.  \enancay\
 measures 8.42~m~$\times$~8.42~m for a total area of
 70.8~m$^2$. } \label{fig:embrace}
\end{figure}


\noi Modern digital technology has sparked a revolution in radio astronomy
techniques, resulting in a fundamental change in how telescopes are
built and operated.  Instead of large mechanical structures, aperture
arrays can be built which are composed of many small antennas with
signals combined together to form the equivalent of a telescope with
large collecting area.  The design philosophy puts the hard work on
the digital electronics, while the mechanical parts, such as dishes
and antennas, are made as simply as possible.  The result is an
instrument built from many small mechanical components and
electronically combined together with the power of fast digital
processors.  Not only is such a system easier to build and maintain,
it also provides distinct operational advantages.  Perhaps most
importantly, aperture arrays have an extremely large field of view for
fast surveys of the sky.  This is the technology which makes the
billion galaxy catalogue a feasible goal for the SKA, and in turn,
will make a fundamental advance in our understanding of Dark Energy
(Rawlings et al. 2004).

\smallskip
\noi The challenge to move this design philosophy to higher frequencies is
mainly driven by the large number of components and associated
electronics.  At frequencies around 1 GHz, the goal is to maintain the
large field of view and observational flexibility made possible by the
aperture array technology while surpassing the sensitivity of large
parabolic dishes.  This can only be done by fully sampling the
aperture with antennas that are closely spaced together.  The result
is an instrument that collects all the available signal, as would do a
dish of the same size, but having all the advantages of an aperture
array, including the possibility for full sky imaging.

\smallskip
\noi Observatoire de Paris has been a major partner in the development of
dense phased arrays for radio astronomy since 2005, working closely
with The Netherlands Foundation for Radio Astronomy (ASTRON).  The
joint project is called EMBRACE (Electronic MultiBeam Radio Astronomy
Concept, Kant et al. 2011).  With significant funding from European Commission FP6
project SKADS, an EMBRACE prototype was built and is operational at
Nan\c{c}ay since 2010 (see Fig.\,\ref{fig:embrace}, Torchinsky et al, 2016).  The Observatoire de
Paris developed the analogue integrated circuit responsible for the
first stage of beam forming (Bosse et al. 2010), and also provided the
sophisticated Monitoring and Control software (Taffoureau et al. 2011)
Observatoire de Paris continues to be responsible for the
characterisation and operation of the EMBRACE prototype with an
on-going observing programme.

\smallskip
\noi \enancay\ has demonstrated its capability as a radio astronomy instrument,
including astronomical observations of pulsars and spectroscopic
observations of galaxies (Torchinsky et al. 2016).  The multibeam
capability has also been demonstrated.  For several years until 2017
when the prototype was ordered to shutdown, EMBRACE operated in the
manner of a facility instrument, doing regularly scheduled
observations showing that dense aperture array technology is stable
and reliable in the long term which is a key operational advantage for
the future Square Kilometre Array.\\

\parbox{0.9\textwidth}{
\noi{References:}\\
\noi{\scriptsize
  Bosse, S., et al., 2010, Proc. EuMC, pp. 106--109;
  Kant, G.W., et al., 2011, IEEE Trans. A\&P, 59, 1990;
  Rawlings, S., et al., 2004, NewAR, 48, 1013;
  Taffoureau, C., et al., 2011, ADASS XXI, ASP Conf. Series, 461, 209;
  Torchinsky, S.A., et al., 2016, A\&A 589, 77
  }}\\

\subsection{Participation in design studies}

\subsubsection{SKADS and PrepSKA} \label{PrepSKA}
\vspace{0cm}


\noi More than fifty years ago, Jean Heidmann at the Observatoire de Paris
pointed out that a radiotelescope with nearly a square kilometre of
collecting area was the next logical step in radioastronomy (Heidmann
1966).  The Nan\c{c}ay Radio Telescope was just beginning operations
and was the world's largest collecting area for neutral hydrogen observations,
but Heidmann already understood that a much bigger instrument was
necessary in order to make cosmological surveys.  He mentioned the
possibility of cataloguing tens of millions of sources.  This was
visionary, and today the catalogue of a billion galaxies for
cosmological studies remains a driving key project for the SKA.

\smallskip
\noi When the International Astronomical Union (IAU) created the working
group for a large radio telescope in 1993, French astronomers were
active participants, and this effort continued through to the formal
creation of the SKA Project by the IAU in 2000.

\smallskip
\noi The most significant mobilisation of French effort for SKA began in
2004-2005 with the submission and start of the ``Square Kilometre Array
Design Studies'' (SKADS).  SKADS was funded by the European Commission
Framework~Programme~6 and had 26~international partners from
13~countries.  Observatoire de Paris was a major partner in the SKADS
consortium, having responsibility for the workpackage ``Continuous assessment and critical reviews'', the
vice-chairmanship of the Board, and a seat in the
Management Team with the Project Scientist provided by Observatoire de
Paris.

\smallskip
\noi SKADS focussed on technological development for an aperture array
implementation of SKA and also included science simulations to help
inform the design process.  The resulting ``SKADS Simulated Skies''
was the work of a number of PhD theses, including in France, and the
papers produced by SKADS continue to be cited today (see for example,
Baek et al. 2009; Levrier et al. 2009).  The technology development led
to the first astronomically capable dense aperture array for radio
astronomy called EMBRACE (Kant et al. 2011; Torchinsky et al. 2016; see also Sect.\,\ref{tech:embrace}).  In particular, the EMBRACE prototype and the future Mid Frequency
Aperture Array (MFAA) for SKA relies on integrated analogue
beamforming technology developed at the Nan\c{c}ay Radio Observatory
(Bosse et al. 2010; see Sect.\,\ref{tech:bossegauffre}).
An overview of the SKADS project and results is published in the
proceedings of the final conference (Torchinsky et al. 2009).

\smallskip
\noi When the EC-FP6 funding ended for SKADS at the end of 2009, the EC-FP7
project PrepSKA was already underway, and SKADS partners naturally
joined this effort.  Much of the SKADS technical development evolved
into the Aperture Array Verification Programme and was integrated to
PrepSKA.  The Observatoire de Paris maintained its leadership with the
EMBRACE prototype testing, and developed EMBRACE into a functioning
astronomically capable radio telescope with a long term observational
programme (Torchinsky et al. 2016).  The scientific efforts also
continued, and French efforts in PrepSKA were instrumental in laying
the groundwork for the next major step towards the SKA.

\smallskip
\noi In 2011, France could not join for programmatic reasons as a full member of the new SKA Organisation.  Nevertheless, thanks to the experience acquired during SKADS and PrepSKA, Observatoire de Paris provides the scientific leadership for the SKA MFAA consortium, with the Project Scientist having this r\^ole since more than a decade, and engineers from Nan\c{c}ay and Bordeaux still provide key technology for analogue integrated receiver, beamforming and digitisation (Sect.\,\ref{tech:bossegauffre}). The rest of this book details further and more recent SKA-related developments in France.\\


\parbox{0.9\textwidth}{
\noi{References:}\\
\noi{\scriptsize
  Baek, S., et al., 2009, A\&A 495, 389;
  Bosse, S., et al,. 2010, Proc. EuMC, pp. 106--109;
  Heidmann, J., 1966, L. Astr. 80, 157;
  Kant, G.W., et al, 2011, IEEE Trans. A\&P, 59, 1990;
  Levrier, F., et al, 2009, PoS, 132, 5;
  Torchinsky, S.A., et al., (eds). 2009, PoS, 132;
  Torchinsky, S.A., et al,. 2016, A\&A 589, 77
  }}\\

\subsubsection{Interoperability Frameworks and Design Studies}\label{tech:interoperability}
\vspace{0cm}

\noi The SKA will lead astronomy into a new era of Big Data where astrophysics research will rely on interoperable e-Infrastructures for the sharing and analysis of data. The realisation of multi-wavelength, multi-messenger and time domain astronomy requires that the SKA and other large observational and computational instruments of the discipline work together in an integrated way with common solutions. France is making significant contributions to build the necessary e-Infrastructures for astronomy and astroparticle physics that will directly benefit the SKA. This work is pursued via participation in the European H2020 projects ``Astronomy ESFRI and Research Infrastructure Cluster Astronomy ESFRI and Research Infrastructure Cluster'' (\href{https://www.asterics2020.eu}{\color{blue} \myul[blue] {ASTERICS}}), ``Research Data Alliance'' (\href{https://www.rd-alliance.org}{\color{blue} \myul[blue] {RDA}}) and ``Advanced European Network of E-infrastructures for Astronomy with the SKA'' (AENEAS), and builds on long term experience of developing the Virtual Observatory framework at the European and international levels(``European Virtual Observatory'', \href{http://www.euro-vo.org}{\color{blue} \myul[blue] {Euro-VO}}, Genova et. al 2015; ``International Virtual Observatory Alliance'', \href{http://ivoa.net}{\color{blue} \myul[blue] {IVOA}}), and the provision of reference data services such as the ``Centre de Donn\'{e}es Astronomiques de Strasbourg'' (\href{http://cdsweb.u-strasbg.fr}{\color{blue} \myul[blue] {CDS}}).

\smallskip
\noi {\bf ASTERICS}

\smallskip
\noi The rationale of ASTERICS is to address cross-cutting synergies and common challenges shared by astronomy ESFRI facilities (CTA-SKA-KM3NeT-ELT). It brings together the astronomy, astrophysics and particle astrophysics communities, in addition to other related research infrastructures. Its major objectives are to support and accelerate the implementation of the ESFRI telescopes, to enhance their performance beyond the current state-of-the-art, and to see them interoperate as an integrated, multi-wavelength (MW) and multi-messenger (MM) facility. The French community is very active within ASTERICS.

\smallskip
\noi CDS leads the ``Data Access, Discovery and Interoperability'' (DADI) work package of ASTERICS. The goal is to make the ESFRI and pathfinder project data available for discovery and usage by the whole astronomical community, interoperable in a homogeneous international framework, and accessible with a set of common tools. More specifically, the project is based on three actions: {\em a)} train and support ESFRI project staff in the usage and implementation of the VO framework and tools, {\em b)} train and support the wider astronomical community in scientific use of the framework, and {\em c)} adapt the VO framework and tools to the ESFRI project needs, and make sure European astronomers remain lead actors in the IVOA. The DADI work package also ensures a liaison with the RDA on data sharing initiatives.

\smallskip
\noi The Laboratoire d'Annecy-le-Vieux de Physique de Particules (LAPP) leads the ``OBservatory E-environments LInked by common ChallengeS'' (OBELICS) work package. The objectives are to enable interoperability and software re-use for the data generation, integration and analysis of the ASTERICS ESFRI and pathfinder facilities. An essential ingredient is the creation of an open innovation environment for establishing open standards and software libraries for multiwavelength/multi-messenger data. Furthermore, development of common solutions for streaming data processing and extremely large databases is required. Study of advanced analysis algorithms and software frameworks for data processing and quality control is the third focus area. The specific objectives are: {\em a)} train researchers and data scientists in the ASTERICS ESFRI and pathfinder projects to apply state-of-the-art parallel software programming techniques, to adopt big-data software frameworks, to benefit from new processor architectures and e-science infrastructures, {\em b)} to maximise software re-use and co-development of technology for the robust and flexible handling of the huge data streams generated by the ASTERICS ESFRI and pathfinder facilities, {\em c} to adapt and optimise extremely large database systems to fulfil the requirements of the ASTERICS ESFRI projects, and {\em d} to study and demonstrate data integration across ASTERICS ESFRI and pathfinder projects using data mining tools and statistical analysis techniques on Petascale data sets.

\smallskip
\noi Thanks to the links with ASTRONET, we are also chairing the Asterics Policy Forum, part of the management work package. The goal of the policy forum is to study how to harmonise joint and efficient scheduling, operation and interoperability of the various MW/MM telescopes. This global question will be addressed through 4 main strategic topics: 1/ Joint time allocation, 2/ observing strategies for MM campaigns, 3/ data access and sharing, and 4/ general policies of common interest (towards next generation ESFRI RI). The study is based on the four ESFRI RI involved in ASTERICS (CTA, E-ELT, KM3NET and SKA) but will also involve many other European astronomical facilities. 

\smallskip
\noi {\bf AENEAS}

\smallskip
\noi The AENEAS project is developing a concept and design for a distributed, federated European Science Data Centre (ESDC) to support the astronomical community in achieving the scientific goals of the SKA.  The approach taken by AENEAS is to leverage existing products, technologies, services, best practices and standards offered by European e-Infrastructures and integrated e-Infrastructures worldwide. 

\smallskip
\noi AENEAS, which is coordinated by the Dutch institute ASTRON, includes participants from all the major European countries who are either official members or interested in the development of the SKA, as well as from the SKA Organisation and its international non-European partners. France is represented by CNRS/INSU, with four institutes taking part to two Work Packages (WP): WP2 (``Development of ESDC Governance Structure and Business Models'') and WP5 (``Access and Knowledge Creation''). Within this latter, particularly relevant is the role of CDS, which is contributing high level expertise on the development of interoperable systems for astronomical data services and the use of the Virtual Observatory framework. Also Fran\c{c}oise Genova (CDS) is mandated by RDA Europe to be the liaison with the AENEAS project. 

\smallskip
\noi Data from the SKA Observatory will have to be transported from South Africa and Australia to the regional data centres. Agreements with the network providers in the different host countries of these centres will have to be established. In Europe, a working group coordinated by the G\'EANT infrastructure operates within the framework of the AENEAS project (WP4: ``Analysis of Global SKA Data Transport and Optimal Storage Topologies''). Its objective is to assess the feasibility of transporting data from the SKA Observatory to Europe, as well as between the ESDC nodes. In order to prepare the transport of the data for a possible node in France, the coordination SKA~France made contact with RENATER. Discussions related to specific challenges of these data centres are also underway within SKA France with experts of the field, both from industry (e.g. Bull-ATOS, DDN Storage, TAS) and from other academic institutes (e.g. INRIA).

\smallskip
\noi {\bf The Strasbourg Astronomical Data Centre (CDS)}

\smallskip
\noi The CDS is an astronomy  data centre that adds value to published and reference data, making them available via services that are widely used by the international astronomical community: SIMBAD, the reference database for astronomical objects, VizieR, for astronomical catalogues and associated data published in journals, the Aladin interactive sky atlas and Virtual Observatory portal, the CDS X-Match large astronomical catalogue cross-correlation service, and the CDS Portal. 

\smallskip
\noi  The continuing role of CDS as a reference data centre in the era of SKA will include adaption of CDS services in support of the SKA. The CDS intends to host appropriate reference data generated by the SKA, and to support a high level of interoperability with remotely distributed data.  The CDS will promote the application of HiPS technologies (Fernique et. al 2015) as a ``hierarchical approach to Big Data'' for the publishing, visualisation and analysis of SKA science-ready data, supporting models of distributed data hosting and remote computing. The CDS X-Match service and algorithms (Pineau et. al 2017) will also be developed for fast cross-correlation of the largest astronomical catalogues.  \\

\parbox{0.9\textwidth}{
\noi{References:}\\
\noi{\scriptsize
Fernique, P., Allen, M.G., Boch, T., et al.\ 2015, \aap, 578, A114; 
Pineau, F.-X., Derriere, S., Motch, C., et al.\ 2017, \aap, 597, A89; 
Genova, F., Allen, M.~G., Arviset, C., et al.\ 2015, Astronomy and Computing, 11, 181 
 }}\\

\subsection{Engagements in pre-construction phase}\label{tech:preconstr}

\subsubsection{Signal processing} \label{texh:TdS}
\vspace{0cm}

\smallskip 

\noi In an instrumental context such as that of the SKA, the processing of the signal recorded during the observations is an unprecedented challenge because of the complexity of the operations to be performed and the amount of acquired data that must be transmitted, stored and analysed. As described in Sect.\,\ref{intro:techno}, two SKA consortia are setting in place the operations that will transform the electric signals resulting from the array of thousands of antennas to the data product ready for science analysis: while the ``Central Signal Processor'' (CSP) is in charge of designing the hardware and software for the time-domain analysis and the correlation of interferometric data, the ``Signal Data Processing'' (SDP) element will transform raw visibilities in the deepest and highest resolution radio images of the sky.

\noi The SDP pipeline will include complex mathematical operations for performing the essential steps of calibration and imaging. Sitting on an elegant and general formulation (the Radio Interferometry Measurement Equation; Hamaker \& Bregman 1996; Hamaker et al. 1996), modern interferometry allows to create complex post-processing corrections, more powerful than the adaptive optics. But these new interferometers generate very large amounts of data (SKA will produce $\sim10$ times more of data than all internet), and the inversion of a large ($10^{9-12}$) system of equation poses new and sometimes profound problems at the crossroads of mathematics, physics and algorithmic. Estimating the so-called ``direction-dependent'' terms in the measurement equation is a difficult and open problem (also known as ``third generation calibration'', see Noordam \& Smirnov 2011). Indeed, the system of many trillions of equations is non-linear, and sometimes ill conditioned. 

\smallskip

\noi As detailed in the next sections, the main contributions to the SKA data processing in France are related to the development of new algorithms for the radio interferometric calibration and imaging steps, and on hardware and software optimisation. Aware of this worldwide recognised French signal processing expertise and of the complexity of the SKA ``Big Data'' challenge, SKA~France has undertaken an effort to coordinate the necessary know-how in France, both at academic and industry level. Several meetings dedicated to radio signal processing, in both its theoretical and applied aspects, have allowed to identify a group of French experts who are currently working on algorithmic developments for the SKA, including joint software and hardware tests that are unique in the international panorama and are based on the collaboration between industry and academic partners (see also Sect.\,\ref{industry:HPC}). A press release of the SKA Organisation (``French researchers push forward radio image quality in view of the SKA telescope''\footnote{The following links are available for more information: \href{https://ska-france.oca.eu/images/SKA-France-Media/Goa-FR-signal-processing-SKA.pdf}{\color{blue} \myul[blue] {text of the press release}}, \href{https://skatelescope.org/news/highlights-from-the-ska-science-conference-in-goa/} {\color{blue} \myul[blue] {SKAO Science Highlights}}, \href{https://twitter.com/SKA_telescope/status/796959266435280896} {\color{blue} \myul[blue] {interview about this topic of the SKA~France coordinator}}}.), published during the 2016 Science Conference in Goa (\href{https://indico.skatelescope.org/event/391/}{\color{blue} \myul[blue] {``SKA 2016: Science for the SKA generation''}}), testifies the great interest with which the international community is following the signal processing developments of the French community.

\smallskip
\noi {\bf Image reconstruction}

\smallskip
\noi An ideal radio interferometer composed of co-planar and identical antennas samples the sky in the Fourier domain. Reconstruction of the sky brightness consists of solving the hill-posed problem associated to these measures. However, for such arrays to provide meaningful images accurate, gain calibration which consists of solving for the unknown complex antenna directional gains, using a known (prior, or iteratively constructed) model of the sky is of critical importance.

\begin{figure*}[ht!]
\begin{center}
\includegraphics[width=0.6\textwidth]{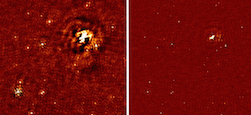}
\end{center}
\caption{\label{fig:LOFAR} Direction-dependent effects distort the
  electromagnetic wavefront, disturb the polarization of the signal, and vary in the
field of view. An algorithm exploiting a mathematical shortcut unveiled by the
"Wirtinger" formalism allows to efficiently solve a large system
of equations. Implementation of a post-processing adaptative optics
then becomes possible. Here the difference between LOFAR images at
$\sim130$ MHz,
uncorrected and corrected for the direction dependent effect ({\em left} and {\em right} respectively).}
\end{figure*}

\begin{figure*}[ht!]
\begin{center}
\includegraphics[width=.6\textwidth]{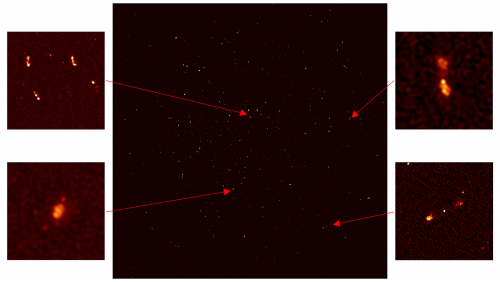}
\end{center}
\caption{\label{fig:MeerKAT} MeerKAT is the SKA precursor telescope in South Africa. Its 64 antennas will integrate SKA1-MID. In collaboration with Oleg Smirnov and his group in South
Africa, the final first light MeerKAT image was made thanks to
the Wirtinger third-generation algorithm, on a computation node of the radio astronomy station of
Nan\c{c}ay. This image was presented to the Minister of
Naledi Pandor for the inauguration of MeerKAT (see \href{https://www.obspm.fr/premiere-image-pour-meerkat.html}{\color{blue} \myul[blue] {original press release}}).}
\end{figure*}

\smallskip
\noi Two researchers of Paris Observatory and Rhodes University / SKA South Africa (Cyril Tasse and Oleg Smirnov, respectively) have discovered that using recent theoretical results on complex optimisation, the bilinear structure of equation of the measure w.r.t. the unknown gains made it possible to obtain Jacobians and Hessians having exceptional properties. Compact but non-trivial structures appear in these matrices, and mathematical shortcuts can be used to solve the non-linear calibration problem (see Tasse 2014, and Smirnov \& Tasse 2015 for more details). It then becomes possible to solve all the directions of the simultaneously and in full-polarization. The processing gain is of the order of the squared number of antenna ($\sim10^{3-4}$ for LOFAR and $\sim10^{6}$ for SKA). A new imaging and deconvolution framework adapted to the specificities of Wirtinger's algorithm (DDFACET, see Figs. \ref {fig:LOFAR} and \ref{fig:MeerKAT}) has also been developed (Tasse et al. 2017 in prep). This new family of complex optimisation algorithm are being applied to reduce the tens of Peta Bytes of the \href{http://www.lofar.org/astronomy/surveys-ksp/surveys-ksp}{\color{blue} \myul[blue] {LOFAR surveys}}, see Fig.\,\ref{fig:LOFAR}. The \href{https://www.obspm.fr/meerkat-first-light-image.html}{\color{blue} \myul[blue] {first light MeerKAT image}} has been produced using that method, see Fig.\,\ref{fig:MeerKAT}.

\smallskip
\noi In parallel, two groups in France have started to develop sparse reconstruction techniques for radio interferometric (RI) and image deconvolution for SKA and pathfinder telescopes (such as LOFAR). 

\begin{figure}[!ht]
  \centering
  \includegraphics[width=0.85\linewidth]{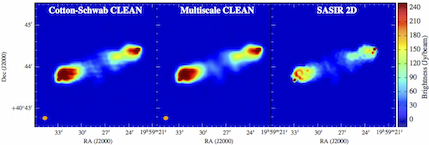}
  \caption{\label{fig:ReconsSASIR} Reconstructed images of Cygnus A from real LOFAR observation. Images are obtained using CLEAN-Based methods and SASIR. SASIR reconstruction presents a higher angular resolution and a lower residual level than that obtained with the two other methods. Adapted from Girard et al. 2015.}
\end{figure}

\smallskip
\noi In the scope of Frontier Project FP3 (Labex UnivEarthS), researchers at CEA-AIM demonstrated with SASIR (``Sparse Aperture Synthesis Image Reconstruction'') the feasibility and the application to real LOFAR data, of using new imaging methods based on the latest mathematical frameworks of Compressed Sensing. They demonstrated that, in a ``controlled'' context (simulation and real data of good quality), SASIR brings better image reconstruction with low residuals and higher angular resolution compared to classical CLEAN-based methods (Garsden et al. 2015; Girard et al. 2015), see Fig.\,\ref{fig:ReconsSASIR}. The current phase of the development is to extend SASIR for multi-channel data: 1) in a ``Time-agile'' Sparse deconvolution algorithm (Girard et al., to be submitted) or ``2D-1D'' and 2) in a ``Frequency-agile'' Sparse deconvolution algorithm (Jiang et al., in press and to be submitted) or ``fGMCA''. Accounting for Direction-Dependent Effects is performed using the imaging framework DDFACET described in the previous paragraph.

\smallskip
\noi The first extension performs image reconstruction of radio transients associated with high-energy, catastrophic events involving particle/plasma/radiation interactions. They are difficult to image because they are either diluted (in time integrated image), or undetectable (low SNR of snapshot images). To cope with this problem, SASIR 2D1D deconvolution method was implemented by adding the temporal axis to the reconstruction (Girard et al., to be submitted). Sources in the data are deconvolved both spatially and temporally using 2D wavelet transforms (IUWT for spatial wavelets and 1D Haar or Daubechies wavelet as temporal in time). SASIR 2D1D brings an improvement (compared to snapshot imaging with CLEAN-based methods) of a factor of 5 in residuals in the images (root mean square) and factor of 2 improvement in the temporal profile reconstruction error on simulated data, thus increasing the probability of transients detection. A real 1 second VLA dataset containing a single pulsar event demonstrated the capability of temporal profile reconstruction.

\smallskip
\noi The second extension of the CEA-AIM code relies on the formulation of wideband image deconvolution (i.e. multiple frequency channels) as a blind source separation problem. The sky is modelled as an unknown (spatial and frequency) mixture of unresolved and extended sources. Current methods rely on ``serial'' single-channel deconvolution or ``joint'' deconvolution. The fGMCA method combines: 1) frequency dependent PSF deconvolution using ForWard (Neelamani et al. 2004) and 2) ``Blind Source Separation'' solutions based on source morphology (gMCA, Bobin et al. 2007). fGMCA can not only reconstruct the data spatially, but also provide the independent frequency spectra of each recovered source. Main results of this theoretical work is the capability to disentangle the spectrum of Cygnus A (a complex extended source with filamentary structure) with a gaussian source blended with the central core of Cygnus A, below 1\% reconstruction error. Application to real multi-channel RI data shows promising results in the context of the upcoming data from SKA1-MID and LOW.

\begin{figure}[!ht]
  \centering
  \includegraphics[width=0.85\linewidth]{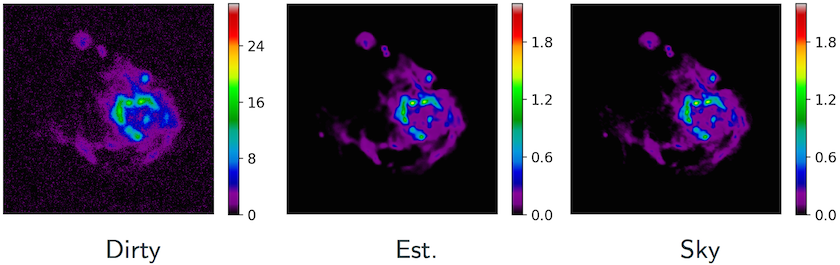}
  \caption{\label{fig:ReconsMUFFIN} Multi-wavelengths reconstruction of a radio emission of an \hii~region in M31 galaxy using MUFFIN. The size of the  data cube is $(256\times 256)$ pixels $\times 100$ wavelengths. The figure shows the dirty image, the estimated sky and the true sky at the higher frequency. Adapted from Ammanouil et al. 2017.}
\end{figure}

\smallskip
\noi The Observatoire de la C\^ote d'Azur (OCA) has a long term experience in inverse problems for image reconstruction and actively develops algorithms for RI image deconvolution. The first main contribution is MORESANE (Dabbech et al. 2015), a greedy algorithm for narrow-band deconvolution which has been tested on SKA1 simulated observations of galaxy clusters (Ferrari et al. 2015). Since 2015, this research activity is developed within the ANR Project MAGELLAN (2015-18, PI OCA) which aims to investigate the potential of machine learning methods for the RI imaging. With regards to image deconvolution, they develop MUFFIN (Multi-Frequency Sparse Radio Interferometric imaging) which is a wide-band deconvolution algorithm based on sparse analysis priors and a state-of-the-art primal-dual optimisation algorithm (Deguignet et al. 2016). The spectral dimension critically blows up the size of this inverse problem. A major advantage of MUFFIN is that the most computing demanding steps (e.g. spatial wavelet decompositions and adjuncts  of the decomposition operator)  are separable w.r.t. the wavelengths, leading to an efficient parallel implementation on shared-memory architectures. A particular effort has been made to derive a scalable method for self tuning the various parameters, such as the regularisation coefficients (Ammanouil et al. 2017), see Fig.\,\ref{fig:ReconsMUFFIN}. Note that computer simulations are performed using the SKA Amazon Web Services Astrocompute Programme, in collaboration with SKA-SA. A collaboration with C. Tasse at Observatoire de Paris aims to integrate MUFFIN in the minor loop of DDFACET.

\smallskip
\noi {\bf Calibration}

\smallskip
\noi Calibration is an essential challenge for the SKA, which is formed of a huge number of elementary antennas with a wide field of view, resulting in large collecting area and high resolution imaging. To achieve the theoretical optimal performance bounds, calibration, i.e., estimation of all unknown perturbation effects along the radiation propagation path, is a cornerstone of the imaging step. Indeed, lack of calibration results in severe distortions and artefacts during the image reconstruction step.

\smallskip
\noi Several researchers in ENS Cachan, L2S Univ. Paris-Sud and Paris-Ouest have been working for since several years on this field. In Ollier et al. (2016), they proposed a novel scheme based on a broad class of distributions, gathered under the so-called spherically invariant random noise modeling (e.g., this includes the Student's t and the K-distribution). With such flexible modeling, we can adaptively consider non-Gaussian heavy-tailed distributed noise in the presence of outliers, but also Gaussian noise when no outliers are present. The proposed algorithm is based on a step-wise numerical approach of the relaxed maximum likelihood estimation. It is worth noted that robust calibration can be even more enhanced if we exploit the known structure of some specific physical parameters across frequency.  Considering the variation of parameters w.r.t. frequency leads to a more accurate direction dependent calibration and is particularly relevant for the SKA, since data volumes are huge and multiple frequency sub-bands are present. Based on this, in Ollier et al. (2017), they proposed a robust distributed calibration of radio interferometers with direction dependent distortions using the so-called Consensus-ADMM. Specifically,  a set of computational agents is considered, each solving a local subproblem in a restricted frequency interval. Communication with a fusion center enforces consensus among agents thanks to known constraints, the goal being to solve a global constrained optimization problem while maintaining a low computational cost. Finally, numerical simulations highlighted the better estimation performance of the proposed technique w.r.t. non-robust and/or mono-frequency state-of-the-art cases (Ollier et al. 2016; Ollier et al. 2017). More specifically, the SIRP-based proposed algorithm reveals to be more robust than state-of-the-art calibration techniques, i.e., the classical non-robust Gaussian case which amounts to solve a least squares (Kazemi et al. 2011) and an alternative which proposes a Student's t with independent and identically distributed entries (Kazemi \& Yatawatta 2013) to model the noise and outlier contribution.

\smallskip
\noi {\bf Developments on architectures}

\smallskip
\noi Between 30\% and 40\% of the SKA budget will be spent for the computing, including both hardware and software costs on both sites in Australia (SKA1-LOW) and Southern Africa (SKA1-MID). SKA1 represents 10\% of the final system and approximately 1\% of the data to compute. As described in Sect.\,\ref{intro:techno}, three processing systems will be developed (remote station, CSP, SDP) and the data from each telescope has to be processed independently and then correlated between each others in order to compute very high-resolution images of the sky (imaging part). 

\smallskip
\noi It is impossible to store the data in memories due to their huge amount (157 terabytes per second for \href{http://skatelescope.org/australia/}{\color{blue} \myul[blue] {SKA1-LOW}} for instance, enough to fill up 35000 DVD every second). Furthermore part of the computations must be done on site, as close as possible to the antennas to decrease the cost of the data transfers. The computers have thus to be powered with local production systems accurately sized. For each specific computing the best hardware/software solution must be found in terms of number of instruction per watt and per second.  

\smallskip 
\noi The goal of a team of researchers at VAADER Rennes 1 University and L2S Univeristy Paris-Sud is to evaluate the best available technologies to provide this efficiency at low cost in a long-term perspective. This work is done with 2 laboratories already in SKA : the Auckland University of Technology (AUT - Andrew Ensor - New Zealand) and the Rhodes University (Oleg Smirnov - South Africa). Three technologies are especially investigated: FPGA (Field-Programmable Gate Array), GPU (Graphics Processing Unit) and the manycore MPPA (Massively Parallel Processor Array) from the French company Kalray. The transformation of the GPU to a manycore architecture has allowed a democratisation of its use since 10 years. It has been approved as a redoubtable powerful computing solution for massively parallel algorithms and experienced by the GPI team for several inverse problems (Gac et al. 2010). Nevertheless its energy efficiency is not always optimal and it could become in the future a barrier to its extensive use for big HPC project as SKA. Thats why, FPGA, known to be one of best energy efficient computation resources could be relevant in SKA context. FPGA will probably become more interesting with the emergence of new design tools known as HLS (High Level Synthesis). HLS tools reduce the time to market compared with the use of language description such as VHDL or Verilog. The Kalray MPPA has been designed from the beginning and exclusively as an efficient manycore architecture. MPPA is also a promising hardware accelerator for SKA data processing. All the CSP, SDP and SKA imaging algorithms are concerned by this work and the first to be studied are FFT, Cross Correlations, Gridding, Calibration algorithms. \\

\parbox{0.9\textwidth}{
\noi{References:}\\
\noi{\scriptsize 
Ammanouil, R., et al., 2017, EUSIPCO; 
Bobin, J., et al., 2007, IEEE Trans. on Im. Proc., 16, 11; 
Dabbech, A., et al., 2015, A\&A, 576, A7; 
Deguignet, J., et al., 2015, EUSIPCO; 
Ferrari, C., 2015, AASKA14, 75;
Gac, N., et al., 2010, proc. of CT meeting; 
Girard, J. N., et al., 2015, JINST 10 C08013; 
Garsden, H., et al., 2015, A\&A, 575, 90; 
Hamaker, J. P. \& Bregman, J. D., 1996, A\&AS, 117, 161;
Hamaker, J. P., et al., 1996, A\&AS, 117, 137;
Hascoet, J., et al., 2015, DASIP; 
Kazemi, S., et al., 2011, MNRAS, 414, 2; 
Kazemi, S., Yatawatta, S., 2013, MNRAS, 435, 1;
Neelamani, R., et al., 2004, IEEE Trans. on Sig. Proc., 52, 2; 
Noordam, J. E. \& Smirnov, O. M., 2010, A\&A, 524, 61;
Ollier, V., et al., 2017, IEEE TSP, DOI: 10.1109/TSP.2017.2733496;
Ollier, V., et al., 2017, submitted to IEEE SPL;
Smirnov, O. M. \& Tasse, C., 2015, MNRAS, 449, 3;
Tasse C., 2014, arXiv:1410.8706
}}

\subsubsection{Research and Development on radio-frequency technologies}\label{tech:bossegauffre}
\vspace{0cm}

\noi Both Paris Observatory and {\it Laboratoire d'Astrophysique de Bordeaux} (LAB) are involved on Research and Development (R\&D) of radio-frequency (RF) technologies related to the next-generation radio telescopes. This activity includes a wide range of applications, from Low Noise Amplifiers (LNA) to fast digitisation, with analogue and digital conditioning techniques.   In particular, Paris Observatory and its department {\it Unit\'e Scientifique de Nan\c cay} (USN) are specialised in RF related devices and sub-systems through the design of Application Specific Integrated Circuits (ASIC) and further application developments, while LAB develops the electrical cards (wide band digitiser and digital filtering cards) by using components off the shelf (COTS)  to establish the link between the RF part and the digital processing back-ends (correlator).

\smallskip
\noi The SKA-related R\&D projects developed or proposed by USN and LAB concern different receiving technologies for different frequency bands, from 50MHz to 24GHz.

\bigskip
\noi {\bf USN contribution for SKA Aperture Arrays}

\smallskip
\noi USN has been working on technological developments for dense aperture arrays since 2004 and is now the task leader of the Integrated Receiver within the AAMID consortium for the Mid-Frequency Aperture Array (MFAA).  These developments were supported by funds of Paris Observatory, CNRS/INSU, {\it R\'egion Centre} (more recently) and, from 2012 to 2016, by the French National Research Agency (ANR) through the AAIR project ``Aperture Array Integrated Receiver''. Over the years, the work done by USN for SKA has also been considered of importance for other SKA consortia, such as, very  recently, the AADC (Aperture Array Design and Construction) consortium for the Low Frequency Aperture Array (LFAA).

\smallskip
\noi The technical challenges of interest for USN can be grouped mainly in three areas, which will be detailed below: LNA, analogue signal conditioning, fast digitisation and smart phased arrays.

\smallskip
\noi \underline{\it LNA} 

\smallskip
\noi To observe the sky, we need both antennas and signal amplification which do not add a significant thermal noise. The MFAA frequency bandwidth is nearly 2 octaves around 1 GHz and the sky noise within the band is 3-7 Kelvins (K). The MFAA consortium requirement for 2025 (Phase 2 of the SKA) is to have an amplifier with a maximum added noise of 30 K.  With cooling, this can be easily achieved. But with the scale of MFAA, it is not possible to cryogenically cool the amplifier. This problem is mainly due to the cost (the number of antennas being extremely high). That is the reason why the 30 K value must be achieved without cooling, this being a very important performance parameter for the highest-frequency portion of the telescope. It is important to stress that the energy consumption is an important constraint, since the SKA antennas will be in the desert. Based on these elements, the technological specifications to realise the LNA must respect the  following requirements: massive production capability and low cost for each LNA, low power consumption, and low noise temperature. 

\smallskip
\noi The micro-electronic group in USN has a wide expertise in the design of ASIC and in the fabrication of the circuit boards for wide band frequency range, particularly developing low noise receivers for SKA projects by using low cost and mature Silicon-Germanium (SiGe) technologies. SiGe technology seems indeed to be a good compromise in terms of cost/energy/performance. The SiGe ASIC design is a very good technology for MFAA, but also for LFAA. USN started SiGe ASIC LNA studies in collaboration with the company \href{http://www.amstechnologies.com/company/}{\color{blue} \myul[blue] {AMS Technologies}} in 2004; at the end of that year, a partnership was developed with the company \href{http://www.nxp.com/about/about-nxp/about-nxp:ABOUT-NXP}{\color{blue} \myul[blue] {NXP}} and, in particular, its French laboratory NXP-Caen.

\smallskip
\noi The noise temperature of 30 K, without cooling, is very difficult to be achieved. The last ASIC LNA designed at USN has 45 K noise temperature at 1500 MHz. The requirement of 30 K being, as stated above, extremely important for the MFAA consortium, further development are under study at USN. Future designs could include a SiGe technology with a better optimisation and other topology (e.g. the use of a SiGe technology with a finer engraving, or of a GaAs process). However, with these solutions both cost and power consumption could be potentially higher. USN has recently also designed an LNA integrated on a time delay ASIC for the LFAA consortium. 

\smallskip
\noi \underline{\it Analogue signal conditioning}  

\smallskip
\noi The digital backend and processing have often lead to significant cost and power consumption. So, in order to achieve a better compromise on performance over cost, one solution could be not to  pursue a digital processing for each antenna. This solution could be of interest both for MFAA and LFAA (even if, for this latter case, it is not the currently adopted solution). In such a case, an analogue conditioning is potentially required by an analogue combination of many antennas while realising an analogue phase-shifting or time-delay. Within the community, USN has a leading expertise in this domain. For over 10 years, USN engineers have designed SiGe ASIC with phase-shifting or time-delay for MFAA and, only recently, for LFAA. With these ASIC, the cost and power consumption of phased array can be reduced. The first radio telescope prototype of the dense phased array, EMBRACE (see Sect.\ref{tech:embrace}), has been realised in 2009 with a phase-shifting ASIC conceived by USN. More recently, USN designed a time-delay ASIC for MFAA. The last time-delay based ASIC produced by USN, in which an LNA is integrated, has been for the LFAA consortium. This chip, called ``Low Noise Delay'' (LND) could  reduce the cost enormously, lowering also the power consumption. SPI or I2C bus are integrated on these ASIC to control all the phase-shifting or time-delay  states. The phase-shifting have 16 states on 360 degrees and the time-delay have 2.4 ns with plus or minus 20 ps of step for MFAA, and 40 ns with 150 ps of step for LFAA. Single-ended or differential mode with a rejection of common mode are important to decrease the power consumption. USN designs ASIC on these 2 modes. Differential mode is easily achieved with SiGe technology. USN can also designs ASIC to convert from a single-ended mode to a differential mode.

\smallskip
\noi \underline{\it Fast digitisation}

\smallskip
\noi The SKA will need to digitise the beams with a large frequency bandwidth and a high dynamic range. Consequently, the cost and the power consumption are very high. USN can design fast ASIC ADC for MFAA based on SiGe technology. Flash and Interpolated Flash folding has been designed using bipolar transistors. This will lead to low-cost fast ADC but with a very high power consumption.

\smallskip
\noi The expertise of USN on the analogue conditioning and the ability to design ASIC for analogue beamformer offers an option to reduce the overall cost on phased array. In addition, an ADC design with a lower power consumption is expected. This becomes a priority for the digitisation. USN will design ADC with a slow topology but a very low power consumption and a clock at 200 MHz for the moment. It will be based on CMOS technology for MFAA. Also, USN is interested in the demand from other consortia to design a very fast ADC on SiGe technology with the RF part on the same ASIC.
In reality, the ADC on the radio telescopes are away from antennas. The link between the RF part and the digitisation is realised by RF cable or more recently by analogue optical fibre. To digitise the beams closed to antennas, the objectives of USN, is to reduce the cost. Then a new problem arise to synchronise a fast clock, due to the repartition of ADC on dense phased array. USN studies the different technical solutions for synchronisation. In the future, these solutions could be integrated within ASIC ADC. 

\smallskip
\noi \underline{\it Smart phased arrays}

\smallskip
\noi Energy is potentially the major obstacle for a dense phased array under study by MFAA. These arrays being built in the desert, the distribution of energy will lead to high cost, thus pushing for the lowest power consumption. This is particularly true for dense phased arrays. The analogue summation or combination allows a reduction in the power consumption, which is though still very high. To realise a gigantic radio telescope in the desert, and in particular for mid-frequency aperture arrays, a breakthrough technology is necessary for energy friendly requirement.

\smallskip
\noi In this framework, USN is starting a first study to design new RF ASICs, which will allow a management of radio telescope power consumption. The electromagnetic environment is not the same over time during the day. Also, each scientific observation does not need the same technical performances. The aim of this breakthrough technology  is therefore to realise an adjustment of the power consumption from a particular electromagnetic environment and scientific observation. In fact, the power consumption is the same whatever the parasitic power level and scientific observations. To fix this problem, the ASIC should have an adaptability on power consumption, versus an adjustable gain level and mostly adjustable intermodulation level. In addition, to minimise the power consumption, a possible solution would be to cancel the parasitic emissions, an ideal electromagnetic environment. By introducing a phase-shift of 180 degrees in the RF signal path at the spare time of the instrument, the parasitic power level will be highly reduced as well as the intermodulation levels; consequently, the overall power consumption could be much lower.

\bigskip
\noi {\bf LAB contribution for SKA1-MID}

\begin{figure*}[ht!]
\begin{center}
\includegraphics[width=.6\textwidth]{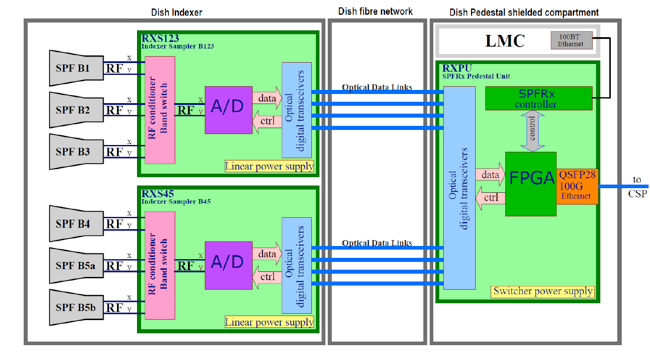}
\end{center}
\caption{\label{fig:LAB} Simplified block diagram of SPF (Single Pixel Feed) receivers.}
\end{figure*}

\smallskip
\noi The electronic group at LAB, a department of Bordeaux University, is specialised in both high speed analog and digital electronics. Over the past 10 years, LAB has acquired an expertise in digital electronics through its involvement in major ground-based and space projects. It has designed, fabricated, characterised and delivered the digitisation units for two major radio astronomy instruments: the ESA Herschel Space Observatory and the ALMA interferometer, including design, prototyping and manufacturing of the digital filtering sub-system of the ALMA correlator. In addition, LAB was involved in a number of academic and industrial R\&D programs aiming at designing wide band digitisation module in order to prepare the new generation of radio telescopes. In the framework of the SKA radio telescope, LAB leads the SKA1-MID Band 5 receiver (RXS45) developments within the DISH and the WBSPF (Wide Band Single Pixel Feed) consortia. This receiver must be production ready for the Phase 1 of SKA, i.e. in early 2019. The electronic group of LAB uses low-cost components off the shelf (COTS) for designing the RXS45 receiver (shown in Fig.\,\ref{fig:LAB}).

\smallskip
\noi The RXS45 receiver is split into two parts:

\begin{itemize}

\item the first one on the feed indexer platform, including the RF signal conditioner module, the digitiser and the optic data transmission boards;

\item the second one in the dish pedestal composed of the optic reception board and the Field-programmable gate array (FPGA) carrier board for data processing (DPU, Digital Process Unit). These boards are placed inside a shielded compartment.

\end{itemize}

\noi The optical data links between the digitiser board and the digital process unit ensure a galvanic isolation to prevent the ``pollution'' of the receiver by the Radio Frequency Interferences (RFI) generated by the digital circuits.

\smallskip
\noi For the RF signal conditioner module, a modular RF chain is developed to meet the input return loss (15 dB min) and ripple (1.5dBpp max) requirements for each frequency sub-band (4, 5a, 5b). This RF chain is composed of coaxial components, and, a more integrated modules with surface mount components, reported on 4 layers PCB (Printed Circuit Board) for cost and size savings. The digitiser board embeds two commercially available 3-bit, 10-level ADCÕs which allow to digitise both polarisation signals at 9 GSps for the SKA1 Band 5a (from 4.6 to 8.5 GHz) and at 16 GSps for the Band 5b (from 8.3 to 15.4 GHz).

\smallskip
\noi The electronic group of LAB pursues a market survey on the wide band ADC's for the SKA Phase 2 in order to implement two other frequency bands: Band 4 (from 2.8 to 5.18 GHz) and Band 5c (from 15 to 24 GHz). SKA2-MID requirements for the ADC are a 4-bit resolution with a minimum sampling of 25 GSps. LAB studies also the digital filter algorithm for extracting two 2.5 GHz sub-bands within the Band 5c for the logic programmable component (FPGA) selection and for the digital processing board design. This board allows to capture and process ADC data and to packetise and send the resulting data to CSP with an 100G link. The LAB contributions in SKA are funded by Bordeaux University, CNRS/INSU and {\it R\'egion Nouvelle Aquitaine}.


\newpage
\section{Industrial perspectives and solutions}\label{industry}

\smallskip

\noi {\sffamily \scriptsize 
{\sffamily \bf J.-T.~Acquaviva} [\ddn],
{\bf K.~Barriere} [\cnrs],
{\bf F.~Bellossi} [\ariane],
{\bf M.~Caillat} [\lermasorb],
{\bf J.-M.~Denis} [\atos],
{\bf C.~Ferrari} [\lagrange],
{\bf A.~Julier} [\tas],
{\bf G.~Marquette} [\cnrs],
{\bf C.~Mazauric} [\atos],
{\bf X.~Olive} [\tas],
{\bf Y.~Pennec} [\airliquide]
{\bf E.~Raffin} [\atos],
{\bf S.~Rawson} [\callisto],
{\bf X.~Vigouroux} [\atos]}

\subsection{Overview}
\vspace{0cm}

\subsubsection{Context}
\vspace{0cm}

\noi As detailed in the previous sections, the Square Kilometre Array radio telescope will be one of the largest physics instruments ever designed and built worldwide. It aims at tackling fundamental questions, both in astrophysics and fundamental physics. 

\smallskip
\noi In France, the community of astronomers and physicists is strongly involved since more than ten years in the preparation of the scientific program as well as in the technical design of the instrument. Today this community is growing rapidly and around 400 scientists are now directly or indirectly working on scientific topics related to the SKA in France. In addition, this huge concentration of efforts is accompanied by a continuously increasing involvement of experts from the industry, which is summarised in this section.

\smallskip
\noi In practice, about thirty French scientists are involved since the beginning in the SWG as core members or associate members, two of them being co-chairpersons. This number is rapidly increasing this last weeks. Industry and labs are jointly involved in five technical consortia through scientists and experts membership, as well as in Precursors (EMBRACE, Sect.\,\ref{tech:embrace}) and Pathfinders (LOFAR and NenuFAR, Sect.\,\ref{tech:NenuFAR}).

\smallskip
\noi Since 2016, large companies and Small- and Medium-size Enterprises (SMEs) are regularly meeting under the auspices of the SKA~France coordination, to prepare for the re-entrance of France in SKA organisation and to be in position to propose the best technological solutions that SKA will require. In this perspective, collaborative projects through top level private-public partnerships are put together aiming at producing these technological solutions with the double objective of:

\begin{itemize}

\item facing the main technical challenges brought by the SKA, still unsolved as per today,
\item contributing to enhance the performance of the instrument,

\end{itemize}

\noi while coping with the economic limitations, both at CAPEX and OPEX levels.

\subsubsection{The SKA challenges and French contribution}\label{industry:axes}
\vspace{0cm}

\noi The SKA will be deployed on 2 sites, in South Africa and West Australia, and in 2 phases, SKA1 and SKA2. SKA1 will represent around 10\% of the total capacity of the SKA, and will be realised by 2025, within a budget envelope capped to 674\,M\euro. Today, due to the instrument complexity, the SKA relies on proven technologies, but currently used at lower scale. Here the gigantism of the instrument, as the data volume to be treated (data traffic greater than todayÕs WWW internet traffic, already at SKA1 level), imposes to solve huge technical challenges to achieve the expected performances. 

\smallskip
\noi These challenges, accentuated by economic and environmental constraints, include:

\begin{itemize}

\item the manufacturing, implementation and control-command of tens of thousands of high-tech mechanical and electronic devices,

\item the energy supply of these measurement devices distributed in desert areas over thousands of kilometres range,

\item the collection, transport and processing of enormous volumes of data produced by the instrument (several Petabits/s).

\end{itemize}

\noi To apprehend them in the most efficient way, the SKA~France coordination has performed a joint comprehensive analysis of both the expertise available in France and the needs of the SKA project not yet covered by the partnering countries. In France, the expertise in low and mid-frequency radio astronomy lies mainly in the development of detectors, dense phased-array technologies, signal processing and data distribution, prototype design and manufacturing (NenuFAR, EMBRACE). Based upon this analysis, an ensemble of workshops grouping industry and academic experts has led to the identification of 4 key axes:

\begin{itemize}

\item high-frequency digitalisation technologies and HPC,
\item signal processing,
\item renewable energy production, distribution and storage,
\item system engineering.

\end{itemize}

\noi Companies and labs have organised themselves together in project consortiums and have started working on enhancing their respective capacities through collaborative projects aligned with SKA targets.

\smallskip
\noi Although this is an iterative process, the objective is to be in capacity to propose to the SKA technical solutions that are relevant, innovative, proven, and respectful of the environment, as well as to provide a complete cost-effective scenario for their implementation, operation and maintenance, all this in due time (i.e. before the accomplishment of the Critical Design Review).

\begin{figure}[!ht]
  \centering
  \includegraphics[width=0.8\linewidth]{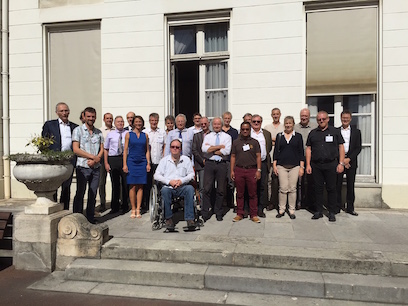}
  \caption{\label{fig:industry} Picture taken in one of the different technological workshops organised by the SKA~France coordination in the last months. Second SKA~France HPC Workshop (CNRS Headquarters in Paris; September 9, 2016).}
\end{figure}

\subsubsection{A systemic approach}
\vspace{0cm}

\noi Because of the interdisciplinary dimension of the SKA instrument and because of the transdisciplinary dimension presented by its challenges, no single solution will solve the ÒequationÓ posed by the SKA. It is also clear that a sequential approach leads to adding-up all constraints from each phase/technological development, both in terms of time and cost. 

\smallskip
\noi A systemic approach is thus needed to integrate in the same timeframe the objectives to be reached, the limited resources available and the technological performances available on due time. It makes re-thinking the whole process in particular at this crucial time of cost revision exercise by SKAO.

\smallskip
\noi What stated above is particularly true for the four key axes identified in Sect.\,\ref{industry:axes}: 

\begin{itemize}

\item Energy supply for Low- and Mid-frequency antennas: production, distribution and storage of renewable energy in highly remote areas. Hybrid systems and energy efficiency will be at the heart of the solution to be put in place.

\item HPC and Big Data: the super-computer that will be able to cope with the huge demands of the SKA does not exist yet, neither the data storage system, nor the data transmission system. These topics is totally transversal and of interest to the whole industry and research world. Big Data and the ``{\it Plan Super-Calculateur}'' are part of the French ``National Strategy for Research'' and strongly supported at ministry level. 

\item Energy and HPC: the massive computing performance is directly linked to the energy consumption (electricity, heat, etc.). French specialists from industry and labs work on the whole computing chain, hardware and software, from the processor (CPU, GPU) to the storage capacity and access speed, and to build a new algorithmic approach to optimise the computational time and performance. They work together with energy specialist to optimise the full energy chain.

\item Cryogenics and energy: leading companies in the domain of cooling and cryogenics work in partnership with research laboratories to develop multifunctional systems able to cool down to few tens of Kelvin, and at the same time feed, an energy device that produces the necessary electrical power for remote antennas.

\item System engineering: by definition, the global architect of the whole instrument. Worldwide leading companies, specialised in aerospace or marine projects, have already started working on the SKA as an ``industry project'' to bring the necessary system culture for the realisation of this giant radio telescope.

\end{itemize}

\smallskip
\noi Although it is not a science/technology domain, financial engineering is also part of the game within this systemic approach. In particular, countries are more and more cautious vis a vis the engagements they sign for a number of decades. Then CAPEX and OPEX together, aligned with industry standards, lead to new business models, which include all the above in terms of investment, maintenance costs and service/products associated on a mid-long term perspective. 

\smallskip
\noi The SKA is therefore not only a cost, but an opportunity to generate new businesses, then revenue, either directly though the technologies developed, or through other applications and products/services, and/or through the image gained in being associated to a worldwide acknowledged scientific and technological adventure.

\subsubsection{The benefits for the SKA and from the SKA}
\vspace{0cm}

\noi The SKA, because of its dimensions and its scientific perspectives, creates a fantastic momentum worldwide, at scientific and industry levels, but moreover at citizen level. As such it is a worldwide model to create scientific knowledge, to develop industry cooperation, and to enrich global citizens through jobs and value creation locally and for global application. 

\newpage
\smallskip
\noi {\bf SKA, aggregator of talents and competencies}

\smallskip
\noi The multidisciplinary aspect of the SKA requires having experts and scientists from different scientific fields, technology domains working together: electronics (digital and analogic), mechanics, automatics, HW and SW, energy, etc. It creates multidisciplinary teams, sharing and exchanging their respective knowledge, bringing new ways of thinking. The benefit is systematic, resulting in a mutual enrichment from different work cultures, business, practices, etc.

\smallskip
\noi {\bf SKA, international organisation, global player}

\smallskip
\noi Due to its global dimension and global perspectives, the playground of SKA is worldwide. All technologies developed in the scope of SKA can find other applications in other domains or in other geographical areas:

\begin{itemize}

\item the energy system to be developed in South Africa for MeerKAT, then SKA1-LOW will be replicable in other remote areas, where people will benefit from 3 technological jumps immediately with energy system well adapted to their needs,
\item the satellite data transmission system will find many other potential applications,
\item the cryogenics/energy system will be a first demonstrator of multi-application.

\end{itemize}

\smallskip
\noi {\bf SKA, use case for HPC and Big Data}

\smallskip
\noi It is crystal clear that what will be done for the SKA will be applied for many other Big Data consumers, such as meteorologists, geophysicists, biomedical scientists, etc. Then SKA is an opportunity to build a community of scientists and industry experts around this specific question and give a strong impetus to the whole HPC science and business. 

\smallskip
\noi {\bf SKA, a sustainable, ethical, environmentally-friendly project}

\smallskip
\noi The global realisation about climate change impact, thus the Paris agreement signature late 2015 on one side, and the digital revolution which have disrupted all societal value scales on the other side, SKA has a duty to be an emblematic example of a worldwide achievement respectful of the environment, knowledge-sharing oriented, technology-driven for the sake of humanity welfare, while guaranteeing the success of its gigantic scientific objectives.

\smallskip
\noi The SKA will be a fantastic opportunity for everyone, thanks to its global visibility, and the scientific and technology it brings, to benefit from the outcome of this unique and universal instrument.

\subsection{French industry capacity for the SKA}

\subsubsection{High performance computing}\label{industry:HPC}
\vspace{0cm}

\smallskip
\noi {\bf Hardware and application integration} 

\smallskip
\noi High performance computing is getting standardised on off the shelf hardware components. It could be X86 (Intel instruction set) general purpose processing devices, NVidia accelerators or rising ARM based solutions. Even more specialised, high level of integration between hardware and applications with very specialised devices, using FPGA boards, or ASICS compute engines could be used for reaching extreme performances and for minimising the power consumption.

\smallskip
\noi Whatever the chosen approach, some difficult technical points have to be addressed. The very first one if the adaptability of the software to the hardware platform, or the other way around, the adaptability of the hardware platform for the application, or algorithm. A wrong combination could drive to low performances, high power consumption and poor performance per watt ratio. It is today a challenge to identify that the chosen combination between hardware and software provides very bad results. It is even more difficult to obtain optimal results (high performance, low power consumption and best performance per watt ratio) when several combinations provide, at first touch, interesting results. One could say it is easy to answer this question: let's test all or at least a significant number of combinations and letÕs choose the optimal one! While this holistic approach is relevant at the theoretical level, it is not possible to implement it in the ``real and concrete life''. The very main reasons are time, resources, and cost.

\smallskip
\noi At first, porting one application to several platforms is extremely time consuming when the hardware platform is emerging or extremely difficult to use (ASICS simulators, FPGA programming languages, \dots) and requires a collaborative effort between the scientist developing the algorithm and the computer scientist porting and optimising the algorithm to the platform. In second place, getting relevant numbers and comparing the relevant combinations requires, as just mentioned, a significant effort from many different resources with different skills: mathematicians, physicists, computer scientists, developers, parallel programming experts and so on. Those resources are scarce and it is often extremely difficult to get them available at the same time to work as a team. As a direct consequence of this team work from specialised experts which has to be sustained through a significant amount of time and the several hardware platforms to be evaluated, the cost is significant.

\smallskip
\noi The SKA project as a whole, and specifically the SKA~France coordination is willing to address this challenge, thanks to the world class expertise of the French astronomers in signal and image analysis and thanks to the contribution the French SKA organisation wants to bring to the global SKA project.

\smallskip
\noi ATOS HPC community considers the problems the French SKA community is willing to address as representative of the future regular requests from the HPC and Big Data community (again, this is all about finding, for a given algorithm or a given class of algorithm the best possible combination between hardware and software to reach optimal performance for the minimal power consumption). Thus, through the collaboration with SKA France, ATOS expects to gain additional expertise, proves its expertise for addressing this general-purpose question and reach worldwide visibility as HPC and Big Data expert.

\smallskip
\noi ATOS has expressed the deepest interest on the HPC challenges of the SKA project and, more particularly, on code optimisation activities within the SKA~France coordination.

\smallskip
\noi In this framework, two working groups led by ATOS, and in particular by its Center for Excellence in Parallel Programming (CEPP), have been identified, focusing on Co-design and Parallelisation of codes for radio data processing. SKA~France is in charge of coordinating these activities, which include different partners. A wiki page has been set-up to organise and follow-up meetings and discussions. The main goal of this ATOS contribution is to probe CEPP skills and its interest to take part in the SKA~France effort, targeting the SKA Science Data Processing (SDP) consortium.

\smallskip
\noi As first experiments, Bull has worked on the most advanced code that has been developed in France, already widely used on radio astronomical data (e.g. LOFAR, MeerKAT, JVLA \dots), that is the imaging tool DDFacet by Cyril Tasse (see Sect.\,\ref{texh:TdS}) and collaborators. DDFacet has been tested with two different compiler suites: GNU and Intel. The idea was to dive into the DDFacet software stack to understand the different phases of processing and to extract their respective part of the total execution time. The main outcome of this collaboration has highlighted the necessity to build a strong collaboration with developers to improve DDFacet performance.

\smallskip
\noi {\bf Data storage, distribution and preservation} 

\smallskip
\noi In our digital century, mastering data has become a national sovereignty issue. This results in the requirement to develop and maintain a set of skills and the corresponding industrial tools. Data has to be captured at the fastest possible pace, manipulated in a safe and efficient way, shared among multiple actors despite geographical distribution. This is the daunting task facing all modern organisations. 

\smallskip
\noi Data has not to be considered as a static element anymore, but to be envisioned for its whole life cycle from production, to transformation, to usage  up to the production of value.

\smallskip
\noi Regarding this data life cycle, for each and every stage, the Square Kilometre Array is pushing the borders to unseen levels. Traditional dichotomy between HPC workloads which are oriented toward massive ingestion of data and cloud functionalities designed for capacity and data sharing is irrelevant for the SKA project.
\smallskip
\noi Therefore, if the quantitative metrics related to data for the SKA project are mind-blowing, the real breakthrough may rely on the qualitative process of its data  management. Considering the whole data value chain, SKA offers the ability to design dual technologies, which means designing and maturing to address the specific challenges of HPC, but in the longer term, being valuable to a broader, if not mainstream market.

\smallskip
\noi The HPC nature of the SKA project is acknowledged by the community, as illustrated by the SKA Keynote to be delivered at the next supercomputing conference\footnote{SC17, in Denver, November 2017. ``Supercomputing'' in the most important supercomputing conference in the world. The event takes place every year, always hosted in the United States.}. Similarly, the SKA project is specifically addressed in the Strategic Research Agenda of the ETP4HPC consortium. 

\smallskip
\noi The ability to ingest data at a rate of several TeraByte/s is in reach of current supercomputers when they are augmented by I/O accelerator such as burst buffers. An illustration being 1 of the 10 largest supercomputers in the world, where DDN operates a burst buffer at 1.5 TB/s.

\smallskip
\noi Such burst buffers are mandatory to deliver the ingestion bandwidth within a reasonable energy envelop, with the energy footprint being reduced by a factor of 20 compared to traditional technologies.

\smallskip
\noi However the SKA challenge is not only about the deployment of the widest burst buffer in the world, but also to provide the ability to tolerate failures, to self-heal and to implement predictive maintenance. All of these functionalities, which are mandatory for the SKA, are still in gestation in the HPC market. The SKA project will dramatically foster these innovations.

\smallskip
\noi In a movement, quite similar to what is observed at CERN, scaling simultaneously the volume of data and the degree of sharing bridge the gap between traditional HPC and cloud data centre. For instance the CERN file system named ÔEOSÕ presents many similarities with the Dropbox internal file system.

\smallskip
\noi Therefore, the question asked by the SKA project: ÒHow to distribute hundreds of PetaBytes of data products per year to thousands of scientists around the world?Ó This has its answer outside traditional HPC.

\smallskip
\noi Geographic distribution and replication of data is mostly observed in media streaming where there is a need to deliver content from the closest data silo to the end user. Disaster recovery, prior to quite a different problem, implies also to protect, synchronise and migrate huge amounts of data across organisations.

\smallskip
\noi As an example, to protect the company digital assets, DDN has deployed a trans-oceanic disaster recovery scheme for a major web actor leading to the movement of 15 PetaBytes per year.

\smallskip
\noi The SKA project has been planed by the community for many years, notably by the design and installation of a pathfinder radio telescope to prepare the SKA initiative: the Karoo Array Telescope in South Africa.

\smallskip
\noi The data backbone for the Karoo Telescope is The South African Very Large Database project. DDN has deployed geographical distributed data management solutions in both Johannesburg and Cap town for this project. Allowing scientists to share, replicate and manipulate data across sites. This previous experiment leads to a better understanding of the necessities and constraints of the astrophysicists.

\smallskip
\noi DDN has recently established a R\&D centre in Meudon, nearby Paris. This site hosts most of the emerging tech and Software Defined Storage department. The SKA project presents a major opportunity to showcase and develop DDN French engineering and at a broader extend to elevate the skills and expertise of its network of local partners.

\smallskip
\noi {\bf Data science for monitoring}

\smallskip
\noi Being a major actor of the European space industry, Thales Alenia Space (TAS) use data science approach on several domains to handle massive volume of data belonging to space system or coming from Earth Observation system. Overall, TAS data scientists are representing a workforce of 20+ people.

\smallskip
\noi Two major experiences are described hereafter.

\begin{itemize}

\item TAS collect, store, process and monitor thousands of telemetries from dozens of satellites and space platforms. These telemetries are lengthy time signals collected during the whole satellite lifetime, making them 15-year long or more for one satellite. TAS has developed analysis tools with user utility as the main objective like synthetic views of the parameters state, tools designed to extract valuable information from thousands of signals that cannot be manually explored. TASÕs approach relies on mathematical signal processing to provide a rich description of the signals. The semantic description of each signal is exploited for health monitoring (including predictive maintenance) and anomaly detection.

\item TAS has applied new technologies (big data, cloud architecture, \dots) to the Ground Mission Segment (GMS) data of Galileo. Two use cases have been implemented : global analysis of GMS' log message (dashboard providing global indicators and high level synthesis of flow); architecture based on COTS and big data storage, to collect, parse and store all data flow transiting into GMS, allowing user-defined processing execution, template based reports and advanced visualisation tools.

\end{itemize}

\noi TAS could support the definition, implementation and operational usage of such approach to monitor the SKA infrastructure (i.e. whole set of radio-telescopes), based on big data and information extraction technologies. TAS could also provide a more global support to SKA project, on big data computation through its own data scientist communities.

\subsubsection{Cryogenics systems}
\vspace{0cm}

\begin{figure}[!ht]
  \centering
  \includegraphics[width=0.95\linewidth]{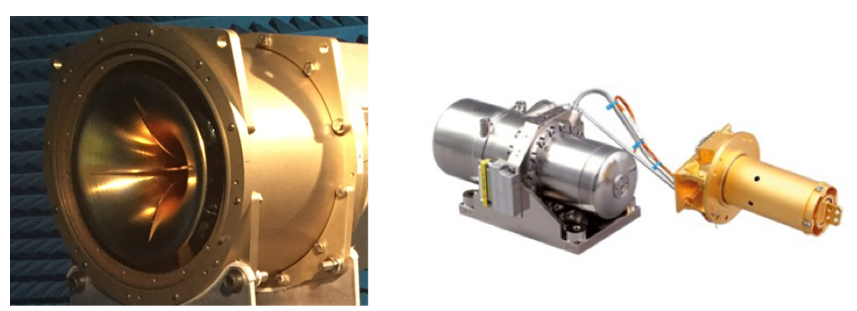}
  \caption{\label{fig:callisto} {\em Left panel:} Wideband Cryogenic Receiver (Callisto). {\em Right  panel:} Stirling PT Cryocooler (Air Liquide).}
\end{figure}

\noi A key challenge for the Square Kilometre Array is to reduce both the capital and operations costs of the facility.  Operations costs are a particular concern because over the lifetime of the facility the cost of ownership will exceed the original capital outlay by a large factor.  The cost of electrical power dominates the operations budget, with maintenance costs also being significant.

\smallskip

\noi To achieve the required sensitivity of the SKA mid-frequency dish array, the front-end receivers will require cryogenic cooling.  Traditional implementation of such cryogenic receivers requires high-vacuum cryostats and closed-cycle helium refrigerators.  The refrigerators draw a considerable amount of power.  The cryogenic and vacuum systems require frequent preventative maintenance as well as corrective maintenance to repair failures.  The SKA will require up to 10000 cryogenic receivers.  Clearly innovative new cooling solutions are required to implement the SKA receivers if the project is to be affordable and viable.

\smallskip

\noi Cryogenic refrigeration technology has been developed for more than 20 years for cooling instruments mounted on spacecraft which are required to be both energy efficient and highly reliable These technologies directly address the power consumption and maintenance challenges which are the weakness current cryogenic systems, which rely on G-M cycle refrigerators.   In France, there is a great deal of expertise in industry (Air Liquide, Thales Cryogenics, Absolut System) as well as in public research institutions (CEA) on space-borne cryocoolers using Stirling cycle and Pulse Tube technology.  This heritage can be levered to develop a design for high reliability, high efficiency cryocooler with performance suitable for the SKA, which requires operating temperatures below 30 Kelvin.   Callisto, an SME based near Toulouse, has been producing cryogenic receivers for satellite communications and radio astronomy for over 20 years and has the capabilities and expertise to integrate this new technology into receivers for the SKA.

\subsubsection{Energy production and storage}
\vspace{0cm}

\noi Energy is a global issue and SKA is a postcard example of the challenges at stakes. The telescope site of SKA-MID I requires more than 5 MW for powering the central computing facility as well as the dish antennas. The site needs and remoteness challenge the electrical grid network in terms of capacity, availability, efficiency and infrastructures.

\smallskip
\noi The weather conditions on the site are very favorable to Solar Photovoltaic as a primary source of Energy. Recent and rapid progresses in technologies have brought down the cost sufficiently to become competitive with the grid. Nevertheless, PV is inherently intermittent in nature both at short (day), and long (weeks) cycle. In turns, it is mandatory to implement large capacity storage to buffer supply fluctuations.

\smallskip
\noi Hydrogen is a well suited vector to address long cycle fluctuations as the storage capacity can be dimensioned separately from the power capacity. A typical Hydrogen Energy Storage includes an electrolyzer (PEM, SOFC, Alkalyne), a large capacity storage (Bag, Compressed, Liquid), and a power unit (PEM, SOFC, Turbine).

\smallskip
\noi Air Liquide has a long standing heritage as a manufacturer and user of Hydrogen technologies starting with the Pioneering days of powering the Ariane Launcher, to its current activities in commercial cars refueling station or autonomous electrical units. Focusing on SKA needs, we have built a first small scale breadboard of a complete Hydrogen Energy Storage system. We foresee this prototype to address industrial scale application. As a major stepping stone in this development, Air Liquide has already delivered a 1 MW commercial PEM power unit and all ancillary services. Air Liquide has large subsidiaries worldwide (including in South Africa) dealing with the production and distribution of industrial gases throughout the countries.

\smallskip
\noi SKA is a unique opportunity to deploy at full scale a Hydrogen Energy Storage. This energy vector can be applied both at the Dish site and at the Telescope site. We believe, this solution can answer the baseline requirement and improve on the system margins at a competitive cost with the grid. Looking forward, powering SKA with Hydrogen will pave the way for broad scale use of sustainable technologies for a green future.



\subsubsection{System engineering}
\vspace{0cm}

\noi The landmark international science SKA project aims at building and operating a radio astronomy instrument of unprecedented sensitivity and accuracy. The project presents numerous significant challenges over an extremely wide scope of technical domains and technological, production, operations and management issues. Technological development and breakthroughs are necessary but not sufficient to make SKA project successful: its success can only be guaranteed through the implementation of a harmonised system engineering and management capacity for mega-projects. 

\smallskip

\noi SKA is to be considered as a system of systems: it is composed of different complex and interconnected systems dealing with a very wide variety of functions and interfaces; it is largely distributed between different locations over the world; the project organisation involves numerous multinational consortia with significant autonomy level. Demand to develop and manage such systems is constantly increasing, in particular for Very Large Research Infrastructures (VLRI) as well as in worldwide institutional projects and in space projects.

\smallskip

\noi Early and continuous application of state-of-the-art system engineering and Assembly, Integration and Validation (AIV) process is a cornerstone to secure the on-time, on-quality and on-cost result of such programmes, and to guaranty final operational performances. In particular, SKA success deals with observation accuracy and sensitivity, overall availability and reliability, affordable and minimised CAPEX and OPEX, and with future evolution and upgrade capacity.

\smallskip
\noi Two main companies in France have an internationally recognised system engineering expertise and are involved in SKA~France activities: Ariane Group and TAS.

\smallskip
\noi Ariane Group developed an expertise in management, system engineering and AIV of such complex projects, for their core programmes Ariane and Navy Deterrence Forces including their ground support and operational bases. Through a ten year diversification program, this expertise and associated processes and tools are now used for the benefit of a large scope of worldwide mega-projects such as ITER (nuclear fusion), Space Surveillance and SESAR (Air Traffic Management).

\smallskip

\noi By its involvement and commitment on system engineering and AIV activities, Ariane Group could support the scientific community by securing the operational performances, the schedule and the good use of the fundings provided to the SKA project. 

\smallskip
\noi Thales has a recognised expertise and experience in the fields of complex system of system projects management, engineering, architecting and AIV.TAS has developed its own system capacities taking benefit of the Thales experience, tools and projects. Among more than 10 so called ``system'' programs under development, some significant examples are: a) Syracuse 3 \& 4 (TAS and Thales are responsible for the design and development of the entire system), b) Galileo (TAS is responsible for system design and Ground Mission Segment), c) ALMA - Atacama large Millimeter Array (provision of 25 giant antennas for the most complex and powerful ground-based astronomical observatory ever built). TAS aims at bringing into the SKA project its experience, background and skills in system engineering, AIV and security domain, as well as its unrivalled European experience on such projects, acquired through ALMA.

\subsubsection{Antenna design and Telecommunications}
\vspace{0cm}

\noi The antenna product portfolio of TAS covers most of the state of the art from UHF to W-band (10,GHz) : reflectors, feeds, arrays, passive, active, large bandwidth, low and high gain, \dots, supported by a unique expertise in radiation patterns of very complex antenna subsystems. TAS is the antenna supplier for CNES and DGA of active antenna arrays including anti-jamming capabilities. This category of antennas is very linked to the system, and includes digital processing and Beam Forming Network capability. TAS experience in system and complex antenna design could bring an important technical added-value to the SKA project. 

\smallskip
\noi A key challenge is the transmission of very large amounts of data from remote sites to user regional centres spread all over the world. Therefore, a wide range of telecommunication technologies shall be contemplated, from ground and deep-sea optical fibers to spatial transmissions. For SKA, Thales Alenia Space is ready to propose spatial solutions ranging from Ka-band, Q-band, V-band and above, as well as free space optical links which could bring unprecedented data rates (Terabits/s). Such solutions are based on single satellites or constellations, with or without inter-satellite links, and with various types of telecommunication payload and ground infrastructure. They may use existing satellites or dedicated one's.


\newpage
\section{Interdisciplinary challenges}

\smallskip

\noi {\sffamily \scriptsize
{\sffamily\bf C. Ferrari} [\lagrange],
{\bf G.~Marquette} [\cnrs],
{\bf M.~P\'erault} [\lermasorb],
{\bf J.-P.~Vilotte} [\ipgp]
}


\subsection{Introduction}

\noi The SKA is undoubtedly one of the biggest observational infrastructures ever conceived and, as such, it represents a fantastic opportunity to face a variety of challenges in a wide range of research fields. 

\smallskip
\noi As extensively described in the previous chapters of this book, as well as in the SKA Science Book produced by the international community (\href{https://pos.sissa.it/cgi-bin/reader/conf.cgi?confid=215}{\color{blue} \myul[blue] {``Advancing Astrophysics with the Square Kilometre Array''}}, AASKA14 reference in this volume), the SKA is interdisciplinary within the field of physical sciences itself, in view of the vast range of interlinked astronomical and fundamental physics questions that it will allow us to address. The investigations permitted by SKA will cover an unprecedented range of spatial scales and cosmological epochs, from the very initial phases of structure formation, to the study of planets and the search of extra-terrestrial life in the local Universe. In the meanwhile SKA will also turn our own Galaxy into the largest ``instrument'' capable of detecting gravitational waves through pulsar studies.

\smallskip
\noi This wealth of topics is allowed by the unique combination of a huge frequency coverage, of gigantic collecting areas, exquisite site quality, immense computing resources and advanced algorithmic developments, none of which did exist a few dozens of years ago, when the embryo idea of what is now the SKA project started to be conceived by a few visionary pioneers. At that time, technology was not ready to allow such an observatory, which still represents an engineering challenge in many respects, appealing to many technological domains, as summarised in Sect.\,\ref{mi:tech}. Particularly important in terms of interdisciplinary interests, the SKA is one of the largest ``Big Data'' project ever conceived: SKA1 will produce a huge amount of data compared to current standards and capabilities, generating a data traffic equivalent to today's global Internet traffic, and triggering ambitious developments in Data Science.

\smallskip
\noi The SKA is not only a big project in terms of telescope size: it is the international scientific facility that will benefit of the largest variety of cultures and expertise from all over the world. In Sect.\,\ref{mi:so} we will summarise some of the most interesting human and society aspects that the international organisation preparing and later operating the SKA Observatory will have to face.

\subsection{A wide variety of technical challenges}\label{mi:tech}
\vspace{0cm}

\noi Already in Phase 1 of the project, the SKA telescopes, with components distributed over several hundred kilometres in desert areas, will pose particularly difficult engineering problems, which need to be faced by joint interdisciplinary and multi-approach expertise. Aware of this, the approach of the SKA Organisation and, at national level of SKA France (see Sects.\,\ref{industry} and \ref{technology}), is to coordinate the activities of a wide range of both private companies and public research laboratories.

\smallskip
\noi Among the main technological challenges of the SKA, which trigger a global transdisciplinary search for integrated solutions (co-design), we recall: 

\begin{itemize}

\item distribution of uninterrupted electrical energy under strong constraints (absence of electromagnetic interference in particular) in a desert environment; 

\item minimising the costs, whilst maximising the reliability and ease of maintenance of all hardware components (antennas, receivers, computers, transmission lines); 

\item accurate real-time control of a many-elements large-scale infrastructure; 

\item complex problems of interferometric signal processing;

\item management of large data flows, with, for each of the 2 components of SKA1, a raw data rate of the order of 10\,Tb/s, a required computing power of 100\,Pflop/s, and an archiving capacity of hundreds of PB/year.

\end{itemize}

\noi The last point of the list above is of greatest interest in terms of interdisciplinary developments. This is not only true because of the confluence of signal propagation, processing and control issues with the fundamental astrophysical challenges. But this is also true because the expected outcomes in data and signal processing are likely to trigger similar developments in other fields of application, far away from the SKA project and its astrophysical applications. Indeed, SKA represents a particularly stimulating ``use case'' for a very large subset of data science and high-performance information technologies, including:

\begin{itemize}

\item onboard computing systems;

\item transport, storage and management of massive data sets;

\item co-design of massive processing and high-level statistical analysis of the signal, including advances in virtualisation technologies;

\item real-time and off-real-time control of particularly complex systems;

\item multi-scale analysis of data over a very wide frequency range, involving different antenna geometries and electrical signal processing; 

\item probabilistic inference methods, including recent advances in Bayesian methods for complex and large-scale physical models, as well as machine learning methods.

\end{itemize}

\noi In this framework, one of the current objectives of the SKA France Coordination is to initiate an interdisciplinary project in data science. This aim is motivated by the new technological (``container'': network, storage architecture, data-bases, computing architectures) and methodological (``content'': new methods of statistical analysis and probabilistic inference) challenges associated with the acquisition and exploitation of very large volumes of multimodal data (events, time series, images) that will be generated by the SKA.

\smallskip
\noi Motivated by the complexity and challenges of the SKA in the data science domain, our action should allow us to identify and boost new collaborations and synergies with other applicative domains that share the same data issues and research practices, as well as with various data science disciplines (e.g. mathematics, computer research) whose research is based on statistical description of data, mathematical tools, computing and data technologies, virtualised environments. Our aim is that experts from different scientific communities will be able to work along two lines:

\begin{itemize}

\item \textbf{Interdisciplinarity through applications:} identify SKA signal processing and analysis issues (scientific, methodological, technological) that are in common with other applicative domains (geophysics, meteorology, particle physics, medical imaging, etc.), allowing to share existing expertise and to construct multi- and inter-disciplinary collaborations.

\item \textbf{Interdisciplinarity within the SKA project:} identify how to respond to the challenges of the entire data processing and analysis chain by integrating the technological and methodological components.

\end{itemize}

\subsection{A scientific project with a big expected impact on society} \label{mi:so}
\vspace{0cm}

\noi Actors from all around the World are committed to build and operate two gigantic scientific infrastructures which will be located in desert areas of Western Australia and South Africa: this ambitious endeavour simultaneously meets fascinating problems of cooperation and governance at this global scale, as well as issues related to the impact of the infrastructure and to the relations with the indigenous populations, which in turn might feed several domains of social sciences with interesting {\it experiments}. 

\subsubsection{Local development}
\vspace{0cm}

\smallskip 
\noi The development of radioastronomy in Western Australia has given rise to a longstanding collaboration between the scientific community and Australian Aboriginal people (the Wajarri Yamatji), who are traditional custodians of the land on which the SKA1-LOW telescope will be built. For this reason, the Yamatji Marlpa Aboriginal Corporation is partner of the ``Indigenous Land Use Agreement'', whose partners include also the Australian Government, the Western Australia Government, the Western Australian State Minister for Lands and the Commonwealth Scientific and Industrial Research Organisation (CSIRO). It is important to stress that, through their collaboration with the SKA project, the Wajarri Yamatji people receive a mixture of monetary and non-monetary benefits aimed at providing long term economic benefit (more information at \href{http://www.ska.gov.au/Mid\%20West\%20Community/Pages/default.aspx}{\color{blue} \myul[blue] {SKA Australia web page}}).

\smallskip
\noi Due to a more recent development of astronomical facilities in South Africa, the collaboration of local populations with SKA South Africa is rapidly evolving in current years, with a fantastic educational effort of SKA SA, in collaboration with the Ministry of Science and Technology. The construction of the SKA precursor MeerKAT has provided not only economical benefits (about 40\% of the equipments come from local companies), but also the opportunity to train people. Many educational actions are organised with students, from primary schools up to university level. As announced at the \href{http://www.ska.ac.za/students/}{\color{blue} \myul[blue] {SKA South Africa web page}}, in February 2017, about 826 students had benefited from SKA South Africa bursaries and scholarships, including many young persons from other African countries. Thanks to the SKA project, top-level international researchers are now moving to South Africa from all over the world, boosting the research and educational level, as well as the international visibility of the country. The most recent example is the organisation of the international conference \href{http://www.astroinformatics2017.ska.ac.za}{\color{blue} \myul[blue] {``Astroinformatics 2017''}} in Cape Town on November 2017. As mentioned above, all these activities are not involving only South Africa, but the whole continent, since the SKA is expected to be extended in several different countries in its Phase 2 (Zambia, Namibia, Botswana, Kenya, Mozambique, Ghana, Madagascar). Nine countries are partners today of the Square Kilometre Array Africa organisation. An exciting recent event resulting from this collaboration happened at the beginning of July 2017, when the Ministries of Ghana and South Africa announced the successful conversion of a communications antenna in Ghana into a functioning Very Long Baseline Interferometry (VLBI) radio telescope.
 
\smallskip
\noi The interest of French researchers to join the educational effort around the SKA project is clear, with already an active participation, for instance, to the annual Postgraduate Bursary Conference that, since 2006, brings together postdoctoral fellows, students, scientists and engineers in South Africa to develop an interactive community around the SKA project.

\subsubsection{Science and Culture}
\vspace{0cm}

\begin{figure*}[ht!]
\begin{center}
\includegraphics[width=0.9\textwidth]{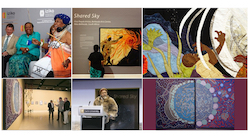}
\end{center}
\caption{\label{fig:sharedsky} Opening of ``Shared Sky'' in South Africa ({\em top}) and Australia ({\em bottom}) in the presence of Aboriginal Australian and South African artists, as well as SKA and ministry representatives.}
\end{figure*}

\noi In order to symbolically bring together ``under one sky'' all countries working within the SKA project, as well as the traditional populations of the lands that will host the telescopes, the SKA Organisation, in collaboration with SKA Australia and SKA South Africa, has organised \href{http://skatelescope.org/shared-sky/}{\color{blue} \myul[blue] {``Shared Sky''}}, a collaborative travel exhibition celebrating humanity's ancient cultural wisdom through artworks of both Aboriginal Australian and South African artists. The exhibition was officially opened by SKA Director General Prof. Phil Diamond during the 2014 SKA Engineering Conference and, since then, it is travelling all over the world, being generally associated with big scientific events related to the SKA project. In collaboration with the SKA Organisation, SKA France is trying to bring Shared Sky in France at the occasion of the \href{http://www.esof.eu/en/home.html}{\color{blue} \myul[blue] {``Euroscience Open Forum - ESOF 2018''}}, which will be held in Toulouse in July 2018. Two paintings of the exhibition are part of the cover image of this volume, a few other wonderful examples are shown in Fig.\,\ref{fig:sharedsky}. The indigenous populations have been so enthusiastic of this initiative that others are following, in particular in Australia: in 2015 the \href{http://www.ska.gov.au/Mid\%20West\%20Community/balagardibarnagardi/Pages/default.aspx}{\color{blue} \myul[blue] {``Balagardi Barnagardi''}} exhibition celebrated the unique relationship between the SKA project and the Wajarri community in Western Australia's Mid-West region, and new paintings have been produced through the collaboration of artists from \href{http://www.ska.gov.au/Mid\%20West\%20Community/Pages/SKA-inspires-new-Indigenous-art.aspx}{\color{blue} \myul[blue] {the Yamaji Art Centre with astronomers from CSIRO and Curtin University at the Murchison Radioastronomy Observatory}}.
  
\subsubsection{Global governance}
\vspace{0cm}

\noi In parallel to these varied connections in the economic, educational and cultural domains, the global nature of the project requires the development of inventive solutions to the many problems that such a large scale collaboration requires in terms of political engagement, of governance, of communication, of economic and scientific return as well as of intellectual property protection, under the overarching requirement of uncompromis efficiency and scientific performance. The fascinating problems arising might undoubtedly trigger original work in the matter of governance and cooperation models of partner countries from the five continents. 

\subsection{Conclusion}
\vspace{0cm}

\noi Already when Phase 1 construction will be complete, the scale of the instrument will make it a project which will uniquely address key scientific questions in astrophysics and fundamental physics, with major connections with the rapidly developing field of data science. The construction of the project, on the other hand, is a challenging industrial undertaking calling for radical innovations, strongly motivating technology companies, in a large range of technological fields. Besides, geopolitical and cultural diversity issues place the project in a pioneering position in global cooperation and development programs. The fascinating questions arising for the design, construction and operation phases of the project are likely to foster interesting studies in various domains of social sciences. 

\newpage
\section{Conclusions and Perspectives: towards the ``Maison SKA France''}\label{conclusions}
\smallskip

\noi {\sffamily \scriptsize
{\sffamily\bf C. Ferrari} [\lagrange],
{\bf G.~Marquette} [\cnrs],
{\bf M.~P\'erault} [\lermasorb]}

\subsection{France positioning assessment}
\vspace{0cm}
\noi Achievements and perspectives presented above result from the academic and industry partners' work realised together since more than a year under the auspices of the SKA France Coordination put in place mid-2016.  

\smallskip
\noi The first step of this informal structure has been to evaluate, by discussing with SKAO and with international colleagues, where the scientific and technological open challenges of the SKA project reside, and where French actors could bring an added value using a different and complementary approach to jointly address SKA performance and cost. Capitalising on the results obtained by all the partners strongly involved in the SKA project since years, the objective of SKA France is indeed to make the possible entry of France among the SKA member countries a positive factor for everybody, the SKA project in first row. After this first analysis, the SKA France Coordination has invited the main stakeholders -- scientific and industrials -- to gather in a succession of workshops dedicated to specific topics matching both French available expertise and SKA needs: energy, HPC, signal processing, joint HPC and energy challenges, system integration.

\smallskip
\noi This has led to the structuring of a huge scientific community, going far beyond astronomers: today more than 250 researchers are ``SKA-earmarked'' in France, with research fields going from radioastronomy to satellite communication, through Big Data, energy production and storage, as well as system integration. 
It has led also to the federation of companies and research institutes, resulting in the creation of ad-hoc consortia, grouping companies from different industry sectors and laboratories to prepare for the optimum scientific and technological solutions for SKA1 whenever possible, and for SKA2 on longer time-scales.

\smallskip
\noi A full multi-disciplinary approach is almost the golden rule to allow a systemic vision which will allow us to optimise both the scientific and technology performances, imposed by the challenge of several breakthroughs to reach the gigantic SKA aims. This is particularly true for enabling scientists to fully exploit the gold-mine of SKA data through the next supercomputers generation, Big Data management and transport, and multi-functional energy systems, as discussed previously.

\smallskip
\noi Finally the outstanding dimension of the SKA imposes the scientific instrument construction and operation to be treated as an industrial project. Hence the obligation to incorporate in the SKA industry landscape worldwide majors in the domain of project management. As such, the participation of international companies with strong experience in aerospace projects for instance is mandatory.

\smallskip
\noi One year after the SKA France Coordination settling, the road which has been run is huge in terms of structuring and mobilisation of all the actors. This white book is a proof of this achievement. In the light of all the results harvested and the definite will of scientists and industry to demonstrate to all French stakeholders the need to go for the SKA now, it has been concluded that it was time to move to the next step and to organise differently in a common structure -- The Maison SKA France -- to best prepare French actors to this big challenge.

\subsection{The ``Maison SKA France''}
\vspace{0cm}

\noi The SKA France Coordination, together with its private partners, has decided to create the ``Maison SKA France'' as a forum for its members to meet, exchange, share strategy and perspectives in order to organise their participation to the preparatory works of the SKA and its IGO. 

\subsubsection{Rationale}
\vspace{0cm}

\noi SKA France private and public partners have proposed to organise themselves within a light but efficient legal structure to:

\begin{itemize}
\item acknowledge the capacity and power of academic and industry partners to help achieve SKA's goals;
\item be an attractor of the French competencies and expertise to strengthen the potential of SKA;
\item be a recognised body to interface with the ministries in France and with the organisation SKA;
\item be a force of proposition to evolve the SKA program in the best interests of the international astronomical community.
\end{itemize}

\subsubsection{Objectives}
\vspace{0cm}

\begin{itemize}
\item Demonstrate to the French Research Ministry public-private mobilisation to bring France into the SKA Project Office as soon as possible, aiming at 2018;
\item promote and recognise the excellence of French scientific teams and industry to be at the fore-front of the science to be developed in SKA;
\item develop the expertise and technologies needed to meet the challenges of SKA through strong Private-Public Partnerships. Implementation through theses/postdocs and collaborative research projects financed by public and/or private funds through shared cost actions. 
\end{itemize}

\subsubsection{SKA France as an innovative financial model}
\vspace{0cm}

\noi Because of the perpetually increasing costs demand for Very Large Research Infrastructures, CAPEX and OPEX become monstrous and, even at global level, they cannot be borne anymore by public money only. This is particularly true in astronomy where the requests of the science is directly correlated to the technology performance. Therefore there is a need to put together new business models to cope with this stringency. 

\smallskip
\noi The ``Maison SKA France'' has been built in this perspective too, and has to be seen as a precursor of a new paradigm for industry and research relations, both having same calendar and performance agenda, but different final use perspectives. The multi-usage, transversal domain dimension of the SKA makes it a perfect case study for such new and innovative financial approach. 

\subsubsection{SKA for France}
\vspace{0cm}

\noi SKA, for France, is not only a fantastic scientific objective, not only one of the biggest technological challenge ever faced by scientific projects, not only an example for future research environmentally friendly, citizen-oriented tools, but also a precursor of sustainable, economic and societal instrument to develop technologies and business at global scale as well as worldwide knowledge sharing, education enhancement and cultural enrichment.

\smallskip
\noi The ``Maison SKA France'' is ready to put this fantastic opportunity for France and SKA into action.

\newpage
\pagestyle{plain}
\cleardoublepage\phantomsection\addcontentsline{toc}{section}{List of contributors}{}{}

\noi {\bfseries\boldmath\LARGE \color{hbl} List of contributors}\\

\begin{ltabulary}{Lll}
\textcolor{hbl}{Fabio ACERO} & {\footnotesize \href{mailto:fabio.acero@cea.fr}{fabio.acero@cea.fr}} & {\footnotesize {\footnotesize Charg\'e de recherche} } \\
{\scriptsize \slshape \irfu; \aim} & & \\
\multicolumn{3}{l}{} \\
\textcolor{hbl}{Jean-Thomas ACQUAVIVA} & {\footnotesize \href{mailto:jacquaviva@ddn.com}{jacquaviva@ddn.com}} & {\footnotesize {\footnotesize Senior Software at DataDirect Networks} } \\
{\scriptsize \slshape \ddn} & & \\
\multicolumn{3}{l}{} \\
\textcolor{hbl}{R\'emi ADAM} & {\footnotesize \href{mailto:remi.adam@oca.eu}{remi.adam@oca.eu}} & {\footnotesize Post-doctorant}  \\ 
{\scriptsize \slshape \lagrange} & & \\
\multicolumn{3}{l}{} \\
\textcolor{hbl}{Nabila AGHANIM} & {\footnotesize \href{mailto:nabila.aghanim@ias.u-psud.fr}{nabila.aghanim@ias.u-psud.fr}} & {\footnotesize Directrice de recherche}  \\
{\scriptsize \slshape \ias} & & \\
\multicolumn{3}{l}{} \\
\textcolor{hbl}{Mark ALLEN} & {\footnotesize \href{mailto:mark.allen@astro.unistra.fr}{mark.allen@astro.unistra.fr}} & {\footnotesize Directeur de recherche}  \\
{\scriptsize \slshape \stras} & & \\
\multicolumn{3}{l}{} \\
\textcolor{hbl}{Marta ALVES} & {\footnotesize \href{mailto:marta.alves@irap.omp.eu}{marta.alves@irap.omp.eu}} & {\footnotesize Post-doctorante}  \\
{\scriptsize \slshape \irap} & & \\
\multicolumn{3}{l}{} \\
\textcolor{hbl}{Rita AMMANOUIL} & {\footnotesize \href{mailto:rita.ammanouil@oca.eu}{rita.ammanouil@oca.eu}} & {\footnotesize Post-doctorante}  \\
{\scriptsize \slshape \lagrange} & & \\
\multicolumn{3}{l}{} \\
\textcolor{hbl}{R\'eza ANSARI} & {\footnotesize \href{mailto:ansari@lal.in2p3.fr}{ansari@lal.in2p3.fr}} & {\footnotesize Professeur des universit\'es}   \\
{\scriptsize \slshape \lal} & & \\
\multicolumn{3}{l}{} \\
\textcolor{hbl}{Anabella ARAUDO} & {\footnotesize \href{mailto:aaraudo@gmail.com}{aaraudo@gmail.com}} & {\footnotesize Post-doctorante}  \\
{\scriptsize \slshape \lupm} & & \\
\multicolumn{3}{l}{} \\
\textcolor{hbl}{Eric ARMENGAUD} & {\footnotesize \href{mailto:eric.armengaud@cea.fr}{eric.armengaud@cea.fr}} & {\footnotesize Ing\'enieur-Chercheur CEA}  \\
{\scriptsize \slshape \irfu} & & \\
\multicolumn{3}{l}{} \\
\textcolor{hbl}{Bego\~na ASCASO} & {\footnotesize \href{mailto:ascaso@apc.univ-paris7.fr}{ascaso@apc.univ-paris7.fr}} & {\footnotesize Post-doctorante}  \\
{\scriptsize \slshape \apc} & & \\
\multicolumn{3}{l}{} \\
\textcolor{hbl}{Evangelie ATHANASSOULA} & {\footnotesize \href{mailto:evangelie.athanassoula@lam.fr}{evangelie.athanassoula@lam.fr}} & {\footnotesize Astronome}  \\
{\scriptsize \slshape \lam} & & \\
\multicolumn{3}{l}{} \\
\textcolor{hbl}{Dominique AUBERT} & {\footnotesize \href{mailto:dominique.aubert@unistra.fr}{dominique.aubert@unistra.fr}} & {\footnotesize Ma\^itre de conf\'erences}   \\
{\scriptsize \slshape \stras} & & \\
\multicolumn{3}{l}{} \\
\textcolor{hbl}{Stanislav BABAK} & {\footnotesize \href{mailto:stanislav.babak@apc.in2p3.fr}{stanislav.babak@apc.in2p3.fr}} & {\footnotesize Directeur de recherche}  \\
{\scriptsize \slshape \apc} & & \\
\multicolumn{3}{l}{} \\
\textcolor{hbl}{Aurore BACMANN} & {\footnotesize \href{mailto:aurore.bacmann@univ-grenoble-alpes.fr}{aurore.bacmann@univ-grenoble-alpes.fr}} & {\footnotesize Astronome-adjointe}  \\
{\scriptsize \slshape \univgren} & & \\
\multicolumn{3}{l}{} \\
\textcolor{hbl}{Anthony BANDAY} & {\footnotesize \href{mailto:abanday@irap.omp.eu}{abanday@irap.omp.eu}} & {\footnotesize Directeur de recherche}  \\
{\scriptsize \slshape \irap} & & \\
\multicolumn{3}{l}{} \\
\textcolor{hbl}{K\'evin BARRIERE} & {\footnotesize \href{mailto:kevin.barriere@cnrs-dir.fr}{kevin.barriere@cnrs-dir.fr}} & {\footnotesize Ing\'enieur} \\
{\scriptsize \slshape \cnrs} & & \\
\multicolumn{3}{l}{} \\
\textcolor{hbl}{Fr\'ed\'eric BELLOSSI} & {\footnotesize \href{mailto:frederic.bellossi@ariane.group}{frederic.bellossi@ariane.group}} & {\footnotesize Program manager}  \\
{\scriptsize \slshape \ariane} & & \\
\multicolumn{3}{l}{} \\
\textcolor{hbl}{Jean-Philippe BERNARD} & {\footnotesize \href{mailto:Jean-Philippe.Bernard@irap.omp.eu}{Jean-Philippe.Bernard@irap.omp.eu}} & {\footnotesize Directeur de recherche}  \\
{\scriptsize \slshape \irap} & & \\
\multicolumn{3}{l}{} \\
\textcolor{hbl}{Maria Grazia BERNARDINI} & {\footnotesize \href{mailto:bernardini@lupm.in2p3.fr}{bernardini@lupm.in2p3.fr}} & {\footnotesize Post-doctorante}  \\
{\scriptsize \slshape \lupm} & & \\
\multicolumn{3}{l}{} \\
\textcolor{hbl}{Matthieu B\'ETHERMIN} & {\footnotesize \href{mailto:matthieu.bethermin@lam.fr}{matthieu.bethermin@lam.fr}} & {\footnotesize Astronome-adjoint}  \\
{\scriptsize \slshape \lam} & & \\
\multicolumn{3}{l}{} \\
\textcolor{hbl}{Elisabeth BLANC} & {\footnotesize \href{mailto:elisabeth.blanc@cea.fr}{elisabeth.blanc@cea.fr}} & {\footnotesize Directeur de recherche CEA}  \\
{\scriptsize \slshape \ceadam} & & \\
\multicolumn{3}{l}{} \\
\textcolor{hbl}{Luc BLANCHET} & {\footnotesize \href{mailto:luc.blanchet@iap.fr}{luc.blanchet@iap.fr}} & {\footnotesize Directeur de recherche}  \\
{\scriptsize \slshape \iapsorb} & & \\
\multicolumn{3}{l}{} \\
\textcolor{hbl}{J\'er\^ome BOBIN} & {\footnotesize \href{mailto:jerome.bobin@cea.fr}{jerome.bobin@cea.fr}} & {\footnotesize Ing\'enieur-Chercheur CEA}  \\
{\scriptsize \slshape \irfu; \aim} & & \\
\multicolumn{3}{l}{} \\
\textcolor{hbl}{Samuel BOISSIER} & {\footnotesize \href{mailto:samuel.boissier@lam.fr}{samuel.boissier@lam.fr}} & {\footnotesize Charg\'e de recherche}  \\
{\scriptsize \slshape \lam} & & \\
\multicolumn{3}{l}{} \\
\textcolor{hbl}{Catherine BOISSON} & {\footnotesize \href{mailto:catherine.boisson@obspm.fr}{catherine.boisson@obspm.fr}} & {\footnotesize Astronome}  \\
{\scriptsize \slshape \luth} & & \\
\multicolumn{3}{l}{} \\
\textcolor{hbl}{Alessandro BOSELLI} & {\footnotesize \href{mailto:alessandro.boselli@lam.fr}{alessandro.boselli@lam.fr}} & {\footnotesize Directeur de recherche} \\
{\scriptsize \slshape \lam} & & \\
\multicolumn{3}{l}{} \\
\textcolor{hbl}{Albert BOSMA} & {\footnotesize \href{mailto:albert.bosma@lam.fr}{albert.bosma@lam.fr}} & {\footnotesize Directeur de recherche} \\
{\scriptsize \slshape \lam} & & \\
\multicolumn{3}{l}{} \\
\textcolor{hbl}{St\'ephane BOSSE} & {\footnotesize \href{mailto:stephane.bosse@obs-nancay.fr}{stephane.bosse@obs-nancay.fr}} & {\footnotesize Ing\'enieur de recherche} \\
{\scriptsize \slshape \usn} & & \\
\multicolumn{3}{l}{} \\
\textcolor{hbl}{Sandrine BOTTINELLI} & {\footnotesize \href{mailto:sbottinelli@irap.omp.eu}{sbottinelli@irap.omp.eu}} & {\footnotesize Ma\^itre de conf\'erences} \\
{\scriptsize \slshape \irap} & & \\
\multicolumn{3}{l}{} \\
\textcolor{hbl}{Fran\c{c}ois BOULANGER} & {\footnotesize \href{mailto:francois.boulanger@lra.ens.fr}{francois.boulanger@lra.ens.fr}} & {\footnotesize Directeur de recherche} \\
{\scriptsize \slshape \lermasorb; \ias} & & \\
\multicolumn{3}{l}{} \\
\textcolor{hbl}{R\'emy BOYER} & {\footnotesize \href{mailto:remy.boyer@l2s.centralesupelec.fr}{remy.boyer@l2s.centralesupelec.fr}} & {\footnotesize Maitre de conf\'erences} \\
{\scriptsize \slshape \lss} & & \\
\multicolumn{3}{l}{} \\
\textcolor{hbl}{Andrea BRACCO} & {\footnotesize \href{mailto:andrea.bracco@cea.fr}{andrea.bracco@cea.fr}} & {\footnotesize Post-doctorant} \\
{\scriptsize \slshape \apc} & & \\
\multicolumn{3}{l}{} \\
\textcolor{hbl}{Carine BRIAND} & {\footnotesize \href{mailto:carine.briand@obspm.fr}{carine.briand@obspm.fr}} & {\footnotesize Astronome} \\
{\scriptsize \slshape \lesia} & & \\
\multicolumn{3}{l}{} \\
\textcolor{hbl}{Martin BUCHER} & {\footnotesize \href{mailto:bucher@apc.univ-paris7.fr}{bucher@apc.univ-paris7.fr}} & {\footnotesize Directeur de recherche} \\
{\scriptsize \slshape \apc} & & \\
\multicolumn{3}{l}{} \\
\textcolor{hbl}{Veronique BUAT} & {\footnotesize \href{mailto:veronique.buat@lam.fr}{veronique.buat@lam.fr}} & {\footnotesize Professeur des universit\'es} \\
{\scriptsize \slshape \lam} & & \\
\multicolumn{3}{l}{} \\
\textcolor{hbl}{Laurent CAMBRESY} & {\footnotesize \href{mailto:laurent.cambresy@astro.unistra.fr}{laurent.cambresy@astro.unistra.fr}} & {\footnotesize Astronome} \\
{\scriptsize \slshape \stras} & & \\
\multicolumn{3}{l}{} \\
\textcolor{hbl}{Michel CAILLAT} & {\footnotesize \href{mailto:michel.caillat@obspm.fr}{michel.caillat@obspm.fr}} & {\footnotesize Ing\'enieur de recherche} \\
{\scriptsize \slshape \lermasorb} & & \\
\multicolumn{3}{l}{} \\
\textcolor{hbl}{Jean-Marc CASANDJIAN} & {\footnotesize \href{mailto:jean-marc.casandjian@cea.fr}{jean-marc.casandjian@cea.fr}} & {\footnotesize Ing\'enieur-Chercheur CEA} \\
{\scriptsize \slshape \irfu; \aim} & & \\
\multicolumn{3}{l}{} \\
\textcolor{hbl}{Emmanuel CAUX} & {\footnotesize \href{mailto:Emmanuel.Caux@irap.omp.eu}{Emmanuel.Caux@irap.omp.eu}} & {\footnotesize Directeur de recherche} \\
{\scriptsize \slshape \irap} & & \\
\multicolumn{3}{l}{} \\
\textcolor{hbl}{SŽbastien C\'ELESTIN} & {\footnotesize \href{mailto:sebastien.celestin@cnrs-orleans.fr}{sebastien.celestin@cnrs-orleans.fr}} & {\footnotesize Ma\^itre de conf\'erences} \\
{\scriptsize \slshape \lpcee} & & \\
\multicolumn{3}{l}{} \\
\textcolor{hbl}{Matteo CERRUTI} & {\footnotesize \href{mailto:mcerruti@lpnhe.in2p3.fr}{mcerruti@lpnhe.in2p3.fr}} & {\footnotesize Post-doctorant} \\
{\scriptsize \slshape \lpnhe} & & \\
\multicolumn{3}{l}{} \\
\textcolor{hbl}{Patrick CHARLOT} & {\footnotesize \href{mailto:patrick.charlot@u-bordeaux.fr}{patrick.charlot@u-bordeaux.fr}} & {\footnotesize Directeur de recherche} \\
{\scriptsize \slshape \lab} & & \\
\multicolumn{3}{l}{} \\
\textcolor{hbl}{Eric CHASSANDE-MOTTIN} & {\footnotesize \href{mailto:ecm@apc.in2p3.fr}{ecm@apc.in2p3.fr}} & {\footnotesize Directeur de recherche} \\
{\scriptsize \slshape \apc} & & \\
\multicolumn{3}{l}{} \\
\textcolor{hbl}{Sylvain CHATY} & {\footnotesize \href{mailto:chaty@cea.fr}{chaty@cea.fr}} & {\footnotesize Professeur des universit\'es}   \\ 
{\scriptsize \slshape \irfu; \aim; \iuf} & & \\
\multicolumn{3}{l}{} \\
\textcolor{hbl}{Nelson CHRISTENSEN} & {\footnotesize \href{mailto:nelson.christensen@oca.eu}{nelson.christensen@oca.eu}} & {\footnotesize Directeur de recherche}  \\ 
{\scriptsize \slshape \artemis} & & \\
\multicolumn{3}{l}{} \\
\textcolor{hbl}{Laure CIESLA} & {\footnotesize \href{mailto:laure.ciesla@cea.fr}{laure.ciesla@cea.fr}} & {\footnotesize Post-doctorante} \\ 
{\scriptsize \slshape \irfu; \aim} & & \\
\multicolumn{3}{l}{} \\
\textcolor{hbl}{Nicolas CLERC} & {\footnotesize \href{mailto:nicolas.clerc@irap.omp.eu}{nicolas.clerc@irap.omp.eu}} & {\footnotesize Charg\'e de recherche}  \\ 
{\scriptsize \slshape \irap} & & \\
\multicolumn{3}{l}{} \\
\textcolor{hbl}{Johann COHEN-TANUGI} & {\footnotesize \href{mailto:johann.cohen-tanugi@umontpellier.fr}{johann.cohen-tanugi@umontpellier.fr}} & {\footnotesize Charg\'e de recherche}   \\
{\scriptsize \slshape \lupm} & & \\
\multicolumn{3}{l}{} \\
\textcolor{hbl}{Isma\"el COGNARD} & {\footnotesize \href{mailto:icognard@cnrs-orleans.fr}{icognard@cnrs-orleans.fr}} & {\footnotesize Directeur de recherche}   \\
{\scriptsize \slshape \lpcee} & & \\
\multicolumn{3}{l}{} \\
\textcolor{hbl}{Fran\c{c}oise COMBES} & {\footnotesize \href{mailto:francoise.combes@obspm.fr}{francoise.combes@obspm.fr}} & {\footnotesize Astronome; Professeur des universit\'es}   \\
{\scriptsize \slshape \colfr} & & \\
\multicolumn{3}{l}{} \\
\textcolor{hbl}{Barbara COMIS} & {\footnotesize \href{mailto:barbara.comis@lpsc.in2p3.fr}{barbara.comis@lpsc.in2p3.fr}} & {\footnotesize Post-doctorante} \\
{\scriptsize \slshape \lpsc} & & \\
\multicolumn{3}{l}{} \\
\textcolor{hbl}{St\'ephane CORBEL} & {\footnotesize \href{mailto:stephane.corbel@cea.fr}{stephane.corbel@cea.fr}} & {\footnotesize Professeur des universit\'es}  \\
{\scriptsize \slshape \irfu; \aim; \usn} & & \\
\multicolumn{3}{l}{} \\
\textcolor{hbl}{Bertrand CORDIER} & {\footnotesize \href{mailto:bcordier@cea.fr}{bcordier@cea.fr}} & {\footnotesize Ing\'enieur-Chercheur CEA}  \\
{\scriptsize \slshape \irfuSAp} & & \\
\multicolumn{3}{l}{} \\
\textcolor{hbl}{Mickael	CORIAT} & {\footnotesize \href{mailto:mickael.coriat@irap.omp.eu}{mickael.coriat@irap.omp.eu}} & {\footnotesize Astronome-adjoint}  \\
{\scriptsize \slshape \irap} & & \\
\multicolumn{3}{l}{} \\
\textcolor{hbl}{R\'egis	COURTIN} & {\footnotesize \href{mailto:regis.courtin@obspm.fr}{regis.courtin@obspm.fr}} & {\footnotesize Charg\'e de recherche}  \\
{\scriptsize \slshape \lesia} & & \\
\multicolumn{3}{l}{} \\
\textcolor{hbl}{Helene COURTOIS} & {\footnotesize \href{mailto:helene.courtois@univ-lyon1.fr}{helene.courtois@univ-lyon1.fr}} & {\footnotesize Professeur des universit\'es}   \\ 
{\scriptsize \slshape \ucb} & & \\
\multicolumn{3}{l}{} \\
\textcolor{hbl}{Bruno DA SILVA} & {\footnotesize \href{mailto:bruno.dasilva@obs-nancay.fr}{bruno.dasilva@obs-nancay.fr}} & {\footnotesize Ing\'enieur de recherche} \\
{\scriptsize \slshape \usn} & & \\
\multicolumn{3}{l}{} \\
\textcolor{hbl}{Emanuele DADDI} & {\footnotesize \href{mailto:emanuele.daddi@cea.fr}{emanuele.daddi@cea.fr}} & {\footnotesize Ing\'enieur-Chercheur CEA}  \\
{\scriptsize \slshape \irfu; \aim} & & \\
\multicolumn{3}{l}{} \\
\textcolor{hbl}{Richard	DALLIER} & {\footnotesize \href{mailto:richard.dallier@subatech.in2p3.fr}{richard.dallier@subatech.in2p3.fr}} & {\footnotesize Ma\^itre-Assistant IMT Atlantique}  \\
{\scriptsize \slshape \subatech} & & \\
\multicolumn{3}{l}{} \\
\textcolor{hbl}{Emmanuel DARTOIS} & {\footnotesize \href{mailto:emmanuel.dartois@u-psud.fr}{emmanuel.dartois@u-psud.fr}} & {\footnotesize Directeur de recherche}  \\
{\scriptsize \slshape \ias} & & \\
\multicolumn{3}{l}{} \\
\textcolor{hbl}{Karine DEMYK} & {\footnotesize \href{mailto:karine.demyk@irap.omp.eu}{karine.demyk@irap.omp.eu}} & {\footnotesize Directrice de recherche}  \\
{\scriptsize \slshape \irap} & & \\
\multicolumn{3}{l}{} \\
\textcolor{hbl}{Jean-Marc DENIS} & {\footnotesize \href{mailto:jean-marc.2.denis@atos.net}{jean-marc.2.denis@atos.net}} & {\footnotesize Head of Strategy, Big Data} \\ 
{\scriptsize \slshape \atos} & & \\
\multicolumn{3}{l}{} \\
\textcolor{hbl}{Laurent DENIS} & {\footnotesize \href{mailto:laurent.denis@obs-nancay.fr}{laurent.denis@obs-nancay.fr}} & {\footnotesize Ing\'enieur de recherche}  \\ 
{\scriptsize \slshape \usn} & & \\
\multicolumn{3}{l}{} \\
\textcolor{hbl}{Arache DJANNATI-ATA\"I} & {\footnotesize \href{mailto:djannati@in2p3.fr}{djannati@in2p3.fr}} & {\footnotesize Directeur de recherche}  \\ 
{\scriptsize \slshape \apc} & & \\
\multicolumn{3}{l}{} \\
\textcolor{hbl}{Jean-Fran\c{c}ois DONATI} & {\footnotesize \href{mailto:jean-francois.donati@irap.omp.eu}{jean-francois.donati@irap.omp.eu}} & {\footnotesize Directeur de recherche}  \\
{\scriptsize \slshape \irap} & & \\
\multicolumn{3}{l}{} \\
\textcolor{hbl}{Marian DOUSPIS} & {\footnotesize \href{mailto:marian.douspis@ias.u-psud.fr}{marian.douspis@ias.u-psud.fr}} & {\footnotesize Astronome}  \\ 
{\scriptsize \slshape \ias} & & \\
\multicolumn{3}{l}{} \\
\textcolor{hbl}{Wim van DRIEL} & {\footnotesize \href{mailto:wim.vandriel@obspm.fr}{wim.vandriel@obspm.fr}} & {\footnotesize Astronome}  \\ 
{\scriptsize \slshape \gepi} & & \\
\multicolumn{3}{l}{} \\
\textcolor{hbl}{Mohammed Nabil EL KORSO} & {\footnotesize \href{mailto:m.elkorso@parisnanterre.fr}{m.elkorso@parisnanterre.fr}} & {\footnotesize Ma\^itre de conf\'erences}   \\ 
{\scriptsize \slshape \leme} & & \\
\multicolumn{3}{l}{} \\
\textcolor{hbl}{Edith FALGARONE} & {\footnotesize \href{mailto:edith.falgarone@ens.fr}{edith.falgarone@ens.fr}} & {\footnotesize Directrice de recherche}  \\ 
{\scriptsize \slshape \lermasorb} & & \\
\multicolumn{3}{l}{} \\
\textcolor{hbl}{Anthea F. FANTINA} & {\footnotesize \href{mailto:anthea.fantina@ganil.fr}{anthea.fantina@ganil.fr}} & {\footnotesize Charg\'e de recherche}  \\ 
{\scriptsize \slshape \iaa; \ganil} & & \\
\multicolumn{3}{l}{} \\
\textcolor{hbl}{Thomas FARGES} & {\footnotesize \href{mailto:thomas.farges@cea.fr}{thomas.farges@cea.fr}} & {\footnotesize Expert senior CEA}  \\ 
{\scriptsize \slshape \ceadam} & & \\
\multicolumn{3}{l}{} \\
\textcolor{hbl}{Andr\'e FERRARI} & {\footnotesize \href{mailto:andre.ferrari@unice.fr}{andre.ferrari@unice.fr}} & {\footnotesize Professeur des universit\'es}   \\ 
{\scriptsize \slshape \lagrange} & & \\
\multicolumn{3}{l}{} \\
\textcolor{hbl}{Chiara FERRARI} & {\footnotesize \href{mailto:chiara.ferrari@oca.eu}{chiara.ferrari@oca.eu}} & {\footnotesize Astronome}  \\ 
{\scriptsize \slshape \lagrange} & & \\
\multicolumn{3}{l}{} \\
\textcolor{hbl}{Katia FERRI\`ERE} & {\footnotesize \href{mailto:katia.ferriere@irap.omp.eu}{katia.ferriere@irap.omp.eu}} & {\footnotesize Directrice de recherche}  \\ 
{\scriptsize \slshape \irap} & & \\
\multicolumn{3}{l}{} \\
\textcolor{hbl}{R\'emi FLAMARY} & {\footnotesize \href{mailto:remi.flamary@unice.fr}{remi.flamary@unice.fr}} & {\footnotesize Ma\^itre de conf\'erences}   \\ 
{\scriptsize \slshape \lagrange} & & \\
\multicolumn{3}{l}{} \\
\textcolor{hbl}{Nicolas GAC} & {\footnotesize \href{mailto:nicolas.gac@lss.supelec.fr}{nicolas.gac@lss.supelec.fr}} & {\footnotesize Ma\^itre de conf\'erences}  \\
{\scriptsize \slshape \lss} & & \\
\multicolumn{3}{l}{} \\
\textcolor{hbl}{St\'ephane GAUFFRE} & {\footnotesize \href{mailto:Stephane.Gauffre@obs.u-bordeaux1.fr}{Stephane.Gauffre@obs.u-bordeaux1.fr}} & {\footnotesize Ing\'enieur de recherche}  \\
{\scriptsize \slshape \lab} & & \\
\multicolumn{3}{l}{} \\
\textcolor{hbl}{Fran\c{c}oise GENOVA} & {\footnotesize \href{mailto:francoise.genova@astro.unistra.fr}{francoise.genova@astro.unistra.fr}} & {\footnotesize Directrice de recherche}  \\
{\scriptsize \slshape \stras} & & \\
\multicolumn{3}{l}{} \\
\textcolor{hbl}{Julien GIRARD} & {\footnotesize \href{mailto:julien.girard@cea.fr}{julien.girard@cea.fr}} & {\footnotesize Ma\^itre de conf\'erences}  \\
{\scriptsize \slshape \irfu; \aim} & & \\
\multicolumn{3}{l}{} \\
\textcolor{hbl}{Isabelle GRENIER} & {\footnotesize \href{mailto:isabelle.grenier@cea.fr}{isabelle.grenier@cea.fr}} & {\footnotesize Professeur des universit\'es}  \\ 
{\scriptsize \slshape \irfu; \aim} & & \\
\multicolumn{3}{l}{} \\
\textcolor{hbl}{Jean-Mathias GRIESSMEIER} & {\footnotesize \href{mailto:jean-mathias.griessmeier@cnrs-orleans.fr}{jean-mathias.griessmeier@cnrs-orleans.fr}} & {\footnotesize Astronome-adjoint}  \\
{\scriptsize \slshape \lpcee; \usn} & & \\
\multicolumn{3}{l}{} \\
\textcolor{hbl}{Pierre GUILLARD} & {\footnotesize \href{mailto:guillard@iap.fr}{guillard@iap.fr}} & {\footnotesize Ma\^itre de conf\'erences}  \\
{\scriptsize \slshape \iapsorb} & & \\
\multicolumn{3}{l}{} \\
\textcolor{hbl}{Lucas GUILLEMOT} & {\footnotesize \href{mailto:lucas.guillemot@cnrs-orleans.fr}{lucas.guillemot@cnrs-orleans.fr}} & {\footnotesize Astronome-adjoint} \\ 
{\scriptsize \slshape \lpcee; \usn} & & \\
\multicolumn{3}{l}{} \\
\textcolor{hbl}{Francesca GULMINELLI} & {\footnotesize \href{mailto:francesca.gulminelli@unicaen.fr}{francesca.gulminelli@unicaen.fr}} & {\footnotesize Professeur des universit\'es} \\ 
{\scriptsize \slshape \lpcee; \usn} & & \\
\multicolumn{3}{l}{} \\
\textcolor{hbl}{Antoine GUSDORF} & {\footnotesize \href{mailto:antoine.gusdorf@lra.ens.fr}{antoine.gusdorf@lra.ens.fr}} & {\footnotesize Charg\'e de recherche}   \\ 
{\scriptsize \slshape \lermasorb} & & \\
\multicolumn{3}{l}{} \\
\textcolor{hbl}{Emilie HABART} & {\footnotesize \href{mailto:emilie.habart@ias.u-psud.fr}{emilie.habart@ias.u-psud.fr}} & {\footnotesize Ma\^itre de conf\'erences} \\ 
{\scriptsize \slshape \ias} & & \\
\multicolumn{3}{l}{} \\
\textcolor{hbl}{Fran\c{c}ois HAMMER} & {\footnotesize \href{mailto:francois.hammer@obspm.fr}{francois.hammer@obspm.fr}} & {\footnotesize Astronome}  \\ 
{\scriptsize \slshape \gepi} & & \\
\multicolumn{3}{l}{} \\
\textcolor{hbl}{Patrick HENNEBELLE} & {\footnotesize \href{mailto:patrick.hennebelle@cea.fr}{patrick.hennebelle@cea.fr}} & {\footnotesize Ing\'enieur-Chercheur CEA}  \\
{\scriptsize \slshape \irfu; \aim} & & \\
\multicolumn{3}{l}{} \\
\textcolor{hbl}{Fabrice HERPIN} & {\footnotesize \href{mailto:fabrice.herpin@u-bordeaux.fr}{fabrice.herpin@u-bordeaux.fr}} & {\footnotesize Astronome}  \\ 
{\scriptsize \slshape \lab} & & \\
\multicolumn{3}{l}{} \\
\textcolor{hbl}{Olivier HERVET} & {\footnotesize \href{mailto:ohervet@ucsc.edu}{ohervet@ucsc.edu}} & {\footnotesize Post-doctorant}  \\ 
{\scriptsize \slshape \stcruz} & & \\
\multicolumn{3}{l}{} \\
\textcolor{hbl}{Annie HUGHES} & {\footnotesize \href{mailto:Annie.Hughes@irap.omp.eu}{Annie.Hughes@irap.omp.eu}} & {\footnotesize Astronome-adjointe}  \\ 
{\scriptsize \slshape \irap} & & \\
\multicolumn{3}{l}{} \\
\textcolor{hbl}{Olivier ILBERT} & {\footnotesize \href{mailto:olivier.ilbert@lam.fr}{olivier.ilbert@lam.fr}} & {\footnotesize Astronome-adjoint}  \\ 
{\scriptsize \slshape \lam} & & \\
\multicolumn{3}{l}{} \\
\textcolor{hbl}{Miho JANVIER} & {\footnotesize \href{mailto:miho.janvier@ias.u-psud.fr}{miho.janvier@ias.u-psud.fr}} & {\footnotesize Astronome-adjoint}  \\ 
{\scriptsize \slshape \ias} & & \\
\multicolumn{3}{l}{} \\
\textcolor{hbl}{Eric JOSSELIN} & {\footnotesize \href{mailto:eric.josselin@obs-mip.fr}{eric.josselin@obs-mip.fr}} & {\footnotesize Ma\^itre de conf\'erences}  \\ 
{\scriptsize \slshape \lupm; \irap} & & \\
\multicolumn{3}{l}{} \\
\textcolor{hbl}{Alain JULIER} & {\footnotesize \href{mailto:alain.julier@thalesaleniaspace.com}{alain.julier@thalesaleniaspace.com}} & {\footnotesize Head of Future Projects Department}  \\ 
{\scriptsize \slshape \tas} & & \\
\multicolumn{3}{l}{} \\
\textcolor{hbl}{Cyril LACHAUD} & {\footnotesize \href{mailto:cyril.lachaud@in2p3}{cyril.lachaud@in2p3}} &  {\footnotesize Ma\^itre de conf\'erences}  \\ 
{\scriptsize \slshape \apc} & & \\
\multicolumn{3}{l}{} \\
\textcolor{hbl}{Guilaine LAGACHE} & {\footnotesize \href{mailto:guilaine.lagache@lam.fr}{guilaine.lagache@lam.fr}} & {\footnotesize Astronome}  \\ 
{\scriptsize \slshape \lam} & & \\
\multicolumn{3}{l}{} \\
\textcolor{hbl}{Rosine LALLEMENT} & {\footnotesize \href{mailto:rosine.lallement@obspm.fr}{rosine.lallement@obspm.fr}} & {\footnotesize Directrice de recherche}  \\ 
{\scriptsize \slshape \gepi} & & \\
\multicolumn{3}{l}{} \\
\textcolor{hbl}{S\'ebastien LAMBERT} & {\footnotesize \href{mailto:sebastien.lambert@obspm.fr}{sebastien.lambert@obspm.fr}} & {\footnotesize Astronome-adjoint}  \\ 
{\scriptsize \slshape \syrte} & & \\
\multicolumn{3}{l}{} \\
\textcolor{hbl}{Laurent LAMY} & {\footnotesize \href{mailto:laurent.lamy@obspm.fr}{laurent.lamy@obspm.fr}} & {\footnotesize Astronome-adjoint}  \\ 
{\scriptsize \slshape \lesia} & & \\
\multicolumn{3}{l}{} \\
\textcolor{hbl}{Mathieu LANGER} & {\footnotesize \href{mailto:mathieu.langer@ias.u-psud.fr}{mathieu.langer@ias.u-psud.fr}} & {\footnotesize Ma\^itre de conf\'erences}  \\
{\scriptsize \slshape \ias} & & \\
\multicolumn{3}{l}{} \\
\textcolor{hbl}{Pascal LARZABAL} & {\footnotesize \href{mailto:pascal.larzabal@satie.ens-cachan.fr}{pascal.larzabal@satie.ens-cachan.fr}} & {\footnotesize Professeur des universit\'es}  \\
{\scriptsize \slshape \satie} & & \\
\multicolumn{3}{l}{} \\
\textcolor{hbl}{Guilhem LAVAUX} & {\footnotesize \href{mailto:guilhem.lavaux@iap.fr}{guilhem.lavaux@iap.fr}} & {\footnotesize Charg\'e de recherche}  \\
{\scriptsize \slshape \iapsorb} & & \\
\multicolumn{3}{l}{} \\
\textcolor{hbl}{Thibaut LE BERTRE} & {\footnotesize \href{mailto:thibaut.lebertre@obspm.fr}{thibaut.lebertre@obspm.fr}} & {\footnotesize Directeur de recherche}  \\
{\scriptsize \slshape \lermasorb} & & \\
\multicolumn{3}{l}{} \\
\textcolor{hbl}{Olivier LE F\`EVRE} & {\footnotesize \href{mailto:olivier.lefevre@lam.fr}{olivier.lefevre@lam.fr}} & {\footnotesize Astronome}  \\
{\scriptsize \slshape \lam} & & \\
\multicolumn{3}{l}{} \\
\textcolor{hbl}{Alexandre LE TIEC} & {\footnotesize \href{mailto:letiec@obspm.fr}{letiec@obspm.fr}} & {\footnotesize Charg\'e de recherche}  \\
{\scriptsize \slshape \luth} & & \\
\multicolumn{3}{l}{} \\
\textcolor{hbl}{Bertrand LEFLOCH} & {\footnotesize \href{mailto:bertrand.lefloch@univ-grenoble-alpes.fr}{bertrand.lefloch@univ-grenoble-alpes.fr}} & {\footnotesize Directeur de recherche}  \\
{\scriptsize \slshape \univgren} & & \\
\multicolumn{3}{l}{} \\
\textcolor{hbl}{Matthew LEHNERT} & {\footnotesize \href{mailto:lehnert@iap.fr}{lehnert@iap.fr}} & {\footnotesize Directeur de recherche}  \\
{\scriptsize \slshape \iapsorb} & & \\
\multicolumn{3}{l}{} \\
\textcolor{hbl}{Marianne LEMOINE-GOUMARD} & {\footnotesize \href{mailto:lemoine@cenbg.in2p3.fr}{lemoine@cenbg.in2p3.fr}} & {\footnotesize Charg\'ee de recherche}  \\
{\scriptsize \slshape \unib} & & \\
\multicolumn{3}{l}{} \\
\textcolor{hbl}{Fran\c{c}ois LEVRIER} & {\footnotesize \href{mailto:francois.levrier@ens.fr}{francois.levrier@ens.fr}} & {\footnotesize Ma\^itre de conf\'erences}  \\
{\scriptsize \slshape \lermasorb} & & \\
\multicolumn{3}{l}{} \\
\textcolor{hbl}{Marceau LIMOUSIN} & {\footnotesize \href{mailto:marceau.limousin@lam.fr}{marceau.limousin@lam.fr}} & {\footnotesize Charg\'e de recherche}  \\
{\scriptsize \slshape \lam} & & \\
\multicolumn{3}{l}{} \\
\textcolor{hbl}{Darek LIS} & {\footnotesize \href{mailto:darek.lis@obspm.fr}{darek.lis@obspm.fr}} & {\footnotesize Professeur des universit\'es}  \\
{\scriptsize \slshape \lermasorb} & & \\
\multicolumn{3}{l}{} \\
\textcolor{hbl}{Andr\'e L\'OPEZ-SEPULCRE} & {\footnotesize \href{mailto:lopez@iram.fr}{lopez@iram.fr}} & {\footnotesize Charg\'e de recherche}  \\
{\scriptsize \slshape \iram} & & \\
\multicolumn{3}{l}{} \\
\textcolor{hbl}{Juan MACIAS-PEREZ} & {\footnotesize \href{mailto:macias@lpsc.in2p3.fr}{macias@lpsc.in2p3.fr}} & {\footnotesize Directeur de recherche}  \\
{\scriptsize \slshape \lpsc} & & \\
\multicolumn{3}{l}{} \\
\textcolor{hbl}{Christophe MAGNEVILLE} & {\footnotesize \href{mailto:christophe.magneville@cea.fr}{christophe.magneville@cea.fr}} & {\footnotesize Ing\'enieur-Chercheur CEA}  \\
{\scriptsize \slshape \irfu} & & \\
\multicolumn{3}{l}{} \\
\textcolor{hbl}{Alexandre MARCOWITH} & {\footnotesize \href{mailto:Alexandre.Marcowith@umontpellier.fr}{Alexandre.Marcowith@umontpellier.fr}} & {\footnotesize Directeur de recherche}  \\
{\scriptsize \slshape \lupm} & & \\
\multicolumn{3}{l}{} \\
\textcolor{hbl}{J\'er\^ome MARGUERON} & {\footnotesize \href{mailto:j.margueron@ipnl.in2p3.fr}{j.margueron@ipnl.in2p3.fr}} & {\footnotesize Charg\'e de recherche}  \\
{\scriptsize \slshape \usa; \ipn} & & \\
\multicolumn{3}{l}{} \\
\textcolor{hbl}{Gabriel MARQUETTE} & {\footnotesize \href{mailto:gabriel.marquette@cnrs-dir.fr}{gabriel.marquette@cnrs-dir.fr}} & {\footnotesize D\'el\'egu\'e aux relations industrielles} \\
{\scriptsize \slshape \cnrs} & & \\
\multicolumn{3}{l}{} \\
\textcolor{hbl}{Douglas MARSHALL} & {\footnotesize \href{mailto:douglas.marshall@cea.fr}{douglas.marshall@cea.fr}} & {\footnotesize Ma\^itre de Conf\'erences} \\
{\scriptsize \slshape \irfu; \aim} & & \\
\multicolumn{3}{l}{} \\
\textcolor{hbl}{Lilian MARTIN} & {\footnotesize \href{mailto:lilian.martin@subatech.in2p3.fr}{lilian.martin@subatech.in2p3.fr}} & {\footnotesize Charg\'e de recherche} \\
{\scriptsize \slshape \subatech; \usn} & & \\
\multicolumn{3}{l}{} \\
\textcolor{hbl}{David MARY} & {\footnotesize \href{mailto:david.mary@unice.fr}{david.mary@unice.fr}} & {\footnotesize Professeur des universit\'es}  \\
{\scriptsize \slshape \lagrange} & & \\
\multicolumn{3}{l}{} \\
\textcolor{hbl}{Sophie MASSON} & {\footnotesize \href{mailto:Sophie.Masson@obspm.fr}{Sophie.Masson@obspm.fr}} & {\footnotesize Astronome-adjointe}  \\
{\scriptsize \slshape \lesia; \usn} & & \\
\multicolumn{3}{l}{} \\
\textcolor{hbl}{Sophie MAUROGORDATO} & {\footnotesize \href{mailto:sophie.maurogordato@oca.eu}{sophie.maurogordato@oca.eu}} & {\footnotesize Directrice de recherche} \\
{\scriptsize \slshape \lagrange} & & \\
\multicolumn{3}{l}{} \\
\textcolor{hbl}{Cyril MAZAURIC} & {\footnotesize \href{mailto:cyril.mazauric@atos.net}{cyril.mazauric@atos.net}} & {\footnotesize HPC Applications and performances expert}\\ 
{\scriptsize \slshape \atos} & & \\
\multicolumn{3}{l}{} \\
\textcolor{hbl}{Yannic MELLIER} & {\footnotesize \href{mailto:mellier@iap.fr}{mellier@iap.fr}} & {\footnotesize Astronome}\\ 
{\scriptsize \slshape \iapsorb} & & \\
\multicolumn{3}{l}{} \\
\textcolor{hbl}{Marc-Antoine MIVILLE-DESCH\^ENES} & {\footnotesize \href{mailto:mamd@ias.u-psud.fr}{mamd@ias.u-psud.fr}} & {\footnotesize Directeur de recherche} \\
{\scriptsize \slshape \ias} & & \\
\multicolumn{3}{l}{} \\
\textcolor{hbl}{Ludovic MONTIER} & {\footnotesize \href{mailto:ludovic.montier@irap.omp.eu}{ludovic.montier@irap.omp.eu}} & {\footnotesize Ing\'enieur de recherche} \\
{\scriptsize \slshape \irap} & & \\
\multicolumn{3}{l}{} \\
\textcolor{hbl}{Fabrice MOTTEZ} & {\footnotesize \href{mailto:fabrice.mottez@obspm.fr}{fabrice.mottez@obspm.fr}} & {\footnotesize Directeur de recherche} \\
{\scriptsize \slshape \luth} & & \\
\multicolumn{3}{l}{} \\
\textcolor{hbl}{Denis MOURARD} & {\footnotesize \href{mailto:denis.mourard@oca.eu}{denis.mourard@oca.eu}} & {\footnotesize Astronome} \\
{\scriptsize \slshape \lagrange} & & \\
\multicolumn{3}{l}{} \\
\textcolor{hbl}{Nicole NESVADBA} & {\footnotesize \href{mailto:nicole.nesvadba@ias.u-psud.fr}{nicole.nesvadba@ias.u-psud.fr}} & {\footnotesize Charg\'ee de recherche}   \\ 
{\scriptsize \slshape \ias} & & \\
\multicolumn{3}{l}{} \\
\textcolor{hbl}{Jean-Franois NEZAN} & {\footnotesize \href{mailto:jnezan@insa-rennes.fr}{jnezan@insa-rennes.fr}} & {\footnotesize Professeur des Universit\'es}   \\ 
{\scriptsize \slshape \insarennes} & & \\
\multicolumn{3}{l}{} \\
\textcolor{hbl}{Pasquier NOTERDAEME} & {\footnotesize \href{mailto:noterdaeme@iap.fr}{noterdaeme@iap.fr}} & {\footnotesize Charg\'e de recherche}  \\ 
{\scriptsize \slshape \iapsorb} & & \\
\multicolumn{3}{l}{} \\
\textcolor{hbl}{J\'er\^ome NOVAK} & {\footnotesize \href{mailto:jerome.novak@obspm.fr}{jerome.novak@obspm.fr}} & {\footnotesize Directeur de recherche}  \\ 
{\scriptsize \slshape \luth} & & \\
\multicolumn{3}{l}{} \\
\textcolor{hbl}{Pierre OCVIRK} & {\footnotesize \href{mailto:pierre.ocvirk@astro.unistra.fr}{pierre.ocvirk@astro.unistra.fr}} & {\footnotesize Astronome-adjoint}  \\ 
{\scriptsize \slshape \stras} & & \\
\multicolumn{3}{l}{} \\
\textcolor{hbl}{Micaela OERTEL} & {\footnotesize \href{mailto:micaela.oertel@obspm.fr}{micaela.oertel@obspm.fr}} & {\footnotesize Charg\'ee de recherche}  \\ 
{\scriptsize \slshape \luth} & & \\
\multicolumn{3}{l}{} \\
\textcolor{hbl}{Xavier OLIVE} & {\footnotesize \href{mailto:xavier.olive@thalesaleniaspace.com}{xavier.olive@thalesaleniaspace.com}} & {\footnotesize Head Ground Segment \& Value-added Data section}  \\ 
{\scriptsize \slshape \tas} & & \\
\multicolumn{3}{l}{} \\
\textcolor{hbl}{Virginie OLLIER} & {\footnotesize \href{mailto:virginie.ollier@satie.ens-cachan.fr}{virginie.ollier@satie.ens-cachan.fr}} & {\footnotesize Doctorante}  \\ 
{\scriptsize \slshape \satie; \lss} & & \\
\multicolumn{3}{l}{} \\
\textcolor{hbl}{Nathalie PALANQUE-DELABROUILLE} & {\footnotesize \href{mailto:nathalie.palanque-delabrouille@cea.fr}{nathalie.palanque-delabrouille@cea.fr}} & {\footnotesize Directeur de Recherche CEA}  \\ 
{\scriptsize \slshape \irfu} & & \\
\multicolumn{3}{l}{} \\
\textcolor{hbl}{Mamta PANDEY-POMMIER} & {\footnotesize \href{mailto:mamtapan@gmail.com}{mamtapan@gmail.com}} & {\footnotesize Post-doctorante} \\ 
{\scriptsize \slshape \cral} & & \\
\multicolumn{3}{l}{} \\
\textcolor{hbl}{Yan PENNEC} & {\footnotesize \href{mailto:yan.pennec@airliquide.com}{yan.pennec@airliquide.com}} & {\footnotesize Scientist and Business Developer} \\ 
{\scriptsize \slshape \airliquide} & & \\
\multicolumn{3}{l}{} \\
\textcolor{hbl}{Michel P\'ERAULT} & {\footnotesize \href{mailto:michel.perault@ens.fr}{michel.perault@ens.fr}} & {\footnotesize Directeur de recherche} \\ 
{\scriptsize \slshape \lermasorb} & & \\
\multicolumn{3}{l}{} \\
\textcolor{hbl}{Celine PEROUX} & {\footnotesize \href{mailto:celine.peroux@lam.fr}{celine.peroux@lam.fr}} & {\footnotesize Charg\'ee de recherche}  \\ 
{\scriptsize \slshape \lam} & & \\
\multicolumn{3}{l}{} \\
\textcolor{hbl}{Pascal PETIT} & {\footnotesize \href{mailto:ppetit@irap.omp.eu}{ppetit@irap.omp.eu}} & {\footnotesize Astronome}  \\ 
{\scriptsize \slshape \irap} & & \\
\multicolumn{3}{l}{} \\
\textcolor{hbl}{J\'er\^ome P\'ETRI} & {\footnotesize \href{mailto:jerome.petri@astro.unistra.fr}{jerome.petri@astro.unistra.fr}} & {\footnotesize Ma\^itre de conf\'erences}  \\ 
{\scriptsize \slshape \stras} & & \\
\multicolumn{3}{l}{} \\
\textcolor{hbl}{Antoine  PETITEAU} & {\footnotesize \href{mailto:petiteau@apc.univ-paris7.fr}{petiteau@apc.univ-paris7.fr}} & {\footnotesize Ma\^itre de conf\'erences}  \\ 
{\scriptsize \slshape \apc} & & \\
\multicolumn{3}{l}{} \\
\textcolor{hbl}{J\'er\^ome PETY} & {\footnotesize \href{mailto:pety@iram.fr}{pety@iram.fr}} & {\footnotesize Astronome}  \\ 
{\scriptsize \slshape \iram; \lermasorb} & & \\
\multicolumn{3}{l}{} \\
\textcolor{hbl}{Gabriel W. PRATT} & {\footnotesize \href{mailto:gabriel.pratt@cea.fr}{gabriel.pratt@cea.fr}} & {\footnotesize Ing\'enieur-Chercheur CEA}  \\ 
{\scriptsize \slshape \irfu; \aim} & & \\
\multicolumn{3}{l}{} \\
\textcolor{hbl}{Mathieu PUECH} & {\footnotesize \href{mailto:mathieu.puech@obspm.fr}{mathieu.puech@obspm.fr}} & {\footnotesize Astronome-adjoint}  \\ 
{\scriptsize \slshape \gepi} & & \\
\multicolumn{3}{l}{} \\
\textcolor{hbl}{Benjamin QUERTIER} & {\footnotesize \href{mailto:benjamin.quertier-dagorn@u-bordeaux.fr}{benjamin.quertier-dagorn@u-bordeaux.fr}} & {\footnotesize Ing\'enieur de recherche}  \\ 
{\scriptsize \slshape \lab} & & \\
\multicolumn{3}{l}{} \\
\textcolor{hbl}{Erwan RAFFIN} & {\footnotesize \href{mailto:erwan.raffin@atos.net}{erwan.raffin@atos.net}} & {\footnotesize HPC Applications and performances expert}\\ 
{\scriptsize \slshape \atos} & & \\
\multicolumn{3}{l}{} \\
\textcolor{hbl}{Sangitiana RAKOTOZAFY HARISON} & {\footnotesize \href{mailto:sangitiana@obs-nancay.fr}{sangitiana@obs-nancay.fr}} & {\footnotesize Ing\'enieur de recherche}  \\ 
{\scriptsize \slshape \usn} & & \\
\multicolumn{3}{l}{} \\
\textcolor{hbl}{Stephen RAWSON} & {\footnotesize \href{mailto:steve.rawson@callisto-space.com}{steve.rawson@callisto-space.com}} & {\footnotesize CEO}   \\ 
{\scriptsize \slshape \callisto} & & \\
\multicolumn{3}{l}{} \\
\textcolor{hbl}{Matthieu RENAUD} & {\footnotesize \href{mailto:matthieu.renaud@umontpellier.fr}{matthieu.renaud@umontpellier.fr}} & {\footnotesize Charg\'e de recherche}  \\ 
{\scriptsize \slshape \lupm} & & \\
\multicolumn{3}{l}{} \\
\textcolor{hbl}{Beno\^it REVENU} & {\footnotesize \href{mailto:benoit.revenu@subatech.in2p3.fr}{benoit.revenu@subatech.in2p3.fr}} & {\footnotesize Directeur de recherche}  \\ 
{\scriptsize \slshape \subatech} & & \\
\multicolumn{3}{l}{} \\
\textcolor{hbl}{C\'edric RICHARD} & {\footnotesize \href{mailto:cedric.richard@unice.fr}{cedric.richard@unice.fr}} & {\footnotesize Professeur des universit\'es}  \\ 
{\scriptsize \slshape \lagrange} & & \\
\multicolumn{3}{l}{} \\
\textcolor{hbl}{Johan RICHARD} & {\footnotesize \href{mailto:johan.richard@univ-lyon1.fr}{johan.richard@univ-lyon1.fr}} & {\footnotesize Astronome-adjoint} \\ 
{\scriptsize \slshape \cral} & & \\
\multicolumn{3}{l}{} \\
\textcolor{hbl}{Fran\c{c}ois RINCON} & {\footnotesize \href{mailto:francois.rincon@irap.omp.eu}{francois.rincon@irap.omp.eu}} & {\footnotesize Charg\'e de recherche}  \\ 
{\scriptsize \slshape \irap} & & \\
\multicolumn{3}{l}{} \\
\textcolor{hbl}{Isabelle RISTORCELLI} & {\footnotesize \href{mailto:isabelle.ristorcelli@irap.omp.eu}{isabelle.ristorcelli@irap.omp.eu}} & {\footnotesize Charg\'ee de recherche}  \\ 
{\scriptsize \slshape \irap} & & \\
\multicolumn{3}{l}{} \\
\textcolor{hbl}{Jerome RODRIGUEZ} & {\footnotesize \href{mailto:jrodriguez@cea.fr}{jrodriguez@cea.fr}} & {\footnotesize Ing\'enieur-Chercheur CEA}  \\ 
{\scriptsize \slshape \irfu; \aim} & & \\
\multicolumn{3}{l}{} \\
\textcolor{hbl}{Mathias SCHULTHEIS} & {\footnotesize \href{mailto:mathias.schultheis@oca.eu}{mathias.schultheis@oca.eu}} & {\footnotesize Astronome-adjoint}  \\
{\scriptsize \slshape \lagrange} & & \\
\multicolumn{3}{l}{} \\
\textcolor{hbl}{Carlo SCHIMD} & {\footnotesize \href{mailto:carlo.schimd@lam.fr}{carlo.schimd@lam.fr}} & {\footnotesize Ma\^itre de conf\'erences}   \\
{\scriptsize \slshape \lam} & & \\
\multicolumn{3}{l}{} \\
\textcolor{hbl}{Benoit SEMELIN} & {\footnotesize \href{mailto:benoit.semelin@obspm.fr}{benoit.semelin@obspm.fr}} & {\footnotesize Professeur des universit\'es}  \\
{\scriptsize \slshape \lermasorb} & & \\
\multicolumn{3}{l}{} \\
\textcolor{hbl}{H\'el\`ene SOL} & {\footnotesize \href{mailto:helene.sol@obspm.fr}{helene.sol@obspm.fr}} & {\footnotesize Directeur de Recherche}  \\
{\scriptsize \slshape \luth} & & \\
\multicolumn{3}{l}{} \\
\textcolor{hbl}{Jean-Luc STARCK} & {\footnotesize \href{mailto:jstarck@cea.fr}{jstarck@cea.fr}} & {\footnotesize Directeur de Recherche CEA}  \\
{\scriptsize \slshape \irfu; \aim} & & \\
\multicolumn{3}{l}{} \\
\textcolor{hbl}{Michel TAGGER} & {\footnotesize \href{mailto:michel.tagger@cnrs-orleans.fr}{michel.tagger@cnrs-orleans.fr}} & {\footnotesize Directeur de recherche}  \\
{\scriptsize \slshape \lpcee} & & \\
\multicolumn{3}{l}{} \\
\textcolor{hbl}{Cyril TASSE} & {\footnotesize \href{mailto:cyril.tasse@obspm.fr}{cyril.tasse@obspm.fr}} & {\footnotesize Astronome-adjoint}  \\
{\scriptsize \slshape \gepi} & & \\
\multicolumn{3}{l}{} \\
\textcolor{hbl}{Gilles THEUREAU} & {\footnotesize \href{mailto:theureau@cnrs-orleans.fr}{theureau@cnrs-orleans.fr}} & {\footnotesize Astronome}  \\
{\scriptsize \slshape \lpcee; \usn; \luth} & & \\
\multicolumn{3}{l}{} \\
\textcolor{hbl}{Stephen TORCHINSKY} & {\footnotesize \href{mailto:steve.torchinsky@obspm.fr}{steve.torchinsky@obspm.fr}} & {\footnotesize Ing\'enieur de recherche}   \\
{\scriptsize \slshape \apc} & & \\
\multicolumn{3}{l}{} \\
\textcolor{hbl}{Charlotte VASTEL} & {\footnotesize \href{mailto:charlotte.vastel@irap.omp.eu}{charlotte.vastel@irap.omp.eu}} & {\footnotesize Astronome-adjointe}  \\
{\scriptsize \slshape \irap} & & \\
\multicolumn{3}{l}{} \\
\textcolor{hbl}{Susanna D. VERGANI} & {\footnotesize \href{mailto:susanna.vergani@obspm.fr}{susanna.vergani@obspm.fr}} & {\footnotesize Charg\'ee de recherche}  \\
{\scriptsize \slshape \gepi} & & \\
\multicolumn{3}{l}{} \\
\textcolor{hbl}{Laurent VERSTRAETE} & {\footnotesize \href{mailto:laurent.verstraete@ias.u-psud.fr}{laurent.verstraete@ias.u-psud.fr}} & {\footnotesize Professeur des Universit\'es} \\ 
{\scriptsize \slshape \ias} & & \\
\multicolumn{3}{l}{} \\
\textcolor{hbl}{Xavier VIGOUROUX} & {\footnotesize \href{mailto:xavier.vigouroux@atos.net}{xavier.vigouroux@atos.net}} & {\footnotesize Head of CEPP} \\ 
{\scriptsize \slshape \atos} & & \\
\multicolumn{3}{l}{} \\
\textcolor{hbl}{Nicole VILMER} & {\footnotesize \href{mailto:nicole.vilmer@obspm.fr}{nicole.vilmer@obspm.fr}} & {\footnotesize Directrice de recherche}  \\
{\scriptsize \slshape \lesia} & & \\
\multicolumn{3}{l}{} \\
\textcolor{hbl}{Jean-Pierre VILOTTE} & {\footnotesize \href{mailto:vilotte@ipgp.jussieu.fr}{vilotte@ipgp.fr}} & {\footnotesize Professeur} \\ 
{\scriptsize \slshape \ipgp} & & \\
\multicolumn{3}{l}{} \\
\textcolor{hbl}{Natalie WEBB} & {\footnotesize \href{mailto:natalie.webb@irap.omp.eu}{natalie.webb@irap.omp.eu}} & {\footnotesize Astronome}  \\
{\scriptsize \slshape \irap} & & \\
\multicolumn{3}{l}{} \\
\textcolor{hbl}{Nathalie YSARD} & {\footnotesize \href{mailto:nathalie.ysard@ias.u-psud.fr}{nathalie.ysard@ias.u-psud.fr}} & {\footnotesize Charg\'ee de recherche}  \\
{\scriptsize \slshape \ias} & & \\
\multicolumn{3}{l}{} \\
\textcolor{hbl}{Philippe ZARKA} & {\footnotesize \href{mailto:philippe.zarka@obspm.fr}{philippe.zarka@obspm.fr}} & {\footnotesize Directeur de recherche}  \\
{\scriptsize \slshape \lesia; \usn} & & \\
\multicolumn{3}{l}{} \\
\end{ltabulary}

\newpage
\pagestyle{plain}
\cleardoublepage\phantomsection\addcontentsline{toc}{section}{List of additional French supporters to the SKA project}{}{}

\noi {\bfseries\boldmath\LARGE \color{hbl} List of additional French supporters to the SKA project}\\

\noi Members of the French community who are not co-authors of this White Book, but expressed their interest in the SKA project by completing the form at the web page \href{http://artemix.obspm.fr/le-projet-ska-m-interesse}{\color{blue} \myul[blue] {``Le projet SKA m'int\'eresse''}}. Follow \href{http://artemix.obspm.fr/le-projet-ska-m-interesse/supporters}{\color{blue} \myul[blue] {this link}} for a summary of their scientific interests. \\

\begin{ltabulary}{Ll}
\textcolor{hbl}{Laurence ALSAC} & {\footnotesize \href{mailto:laurence.alsac@obs-nancay.fr}{laurence.alsac@obs-nancay.fr}} \\
{\scriptsize \slshape \usn} & \\
\multicolumn{2}{l}{} \\
\textcolor{hbl}{Philippe AMRAM} & {\footnotesize \href{mailto:philippe.amram@lam.fr}{philippe.amram@lam.fr}} \\
{\scriptsize \slshape \lam} & \\
\multicolumn{2}{l}{} \\
\textcolor{hbl}{Slaheddine ARIDHI} & {\footnotesize \href{mailto:slah@sensoriaanalytics.com}{slah@sensoriaanalytics.com}} \\
{\scriptsize \slshape Sensoria Analytics} & \\
\multicolumn{2}{l}{} \\
\textcolor{hbl}{Monique ARNAUD} & {\footnotesize \href{mailto:monique.arnaud@cea.fr}{monique.arnaud@cea.fr}}  \\
{\scriptsize \slshape \irfu; \aim} & \\
\multicolumn{2}{l}{} \\
\textcolor{hbl}{Hakim ATEK} & {\footnotesize \href{mailto:hakim.atek@iap.fr}{hakim.atek@iap.fr}} \\
{\scriptsize \slshape \iapsorb} & \\
\multicolumn{2}{l}{} \\
\textcolor{hbl}{Jonathan AUMONT} & {\footnotesize \href{mailto:jonathan.aumont@ias.u-psud.fr}{jonathan.aumont@ias.u-psud.fr}} \\
{\scriptsize \slshape \ias; \irap} & \\
\multicolumn{2}{l}{} \\
\textcolor{hbl}{David BARATOUX} & {\footnotesize \href{mailto:david.baratoux@ird.fr}{david.baratoux@ird.fr}} \\
{\scriptsize \slshape Institut de Recherche pour le D\'eveloppement, France} & \\
\multicolumn{2}{l}{} \\
\textcolor{hbl}{Bruno BARELAUD} & {\footnotesize \href{mailto:bruno.barelaud@xlim.fr}{bruno.barelaud@xlim.fr}} \\
{\scriptsize \slshape XLIM UMR CNRS 7252, Universit\'e de Limoges, France} & \\
\multicolumn{2}{l}{} \\
\textcolor{hbl}{Pierre BARGE} & {\footnotesize \href{mailto:pierre.barge@lam.fr}{pierre.barge@lam.fr}} \\
{\scriptsize \slshape \lam} & \\
\multicolumn{2}{l}{} \\
\textcolor{hbl}{Severin BARTH} & {\footnotesize \href{mailto:severin.barth@obs-nancay.fr}{severin.barth@obs-nancay.fr}} \\
{\scriptsize \slshape \usn} & \\
\multicolumn{2}{l}{} \\
\textcolor{hbl}{Alain BAUDRY} & {\footnotesize \href{mailto:alain.baudry@u-bordeaux.fr}{alain.baudry@u-bordeaux.fr}} \\
{\scriptsize \slshape \lab} & \\
\multicolumn{2}{l}{} \\
\textcolor{hbl}{G\'erard BEAUDIN} & {\footnotesize \href{mailto:gerard.beaudin@obspm.fr}{gerard.beaudin@obspm.fr}} \\
{\scriptsize \slshape \lermasorb} & \\
\multicolumn{2}{l}{} \\
\textcolor{hbl}{Alexandre BEELEN} & {\footnotesize \href{mailto:alexandre.beelen@ias.u-psud.fr}{alexandre.beelen@ias.u-psud.fr}} \\
{\scriptsize \slshape \ias; \lam} & \\
\multicolumn{2}{l}{} \\
\textcolor{hbl}{Christophe BENOIST} & {\footnotesize \href{mailto:christophe.benoist@oca.eu}{christophe.benoist@oca.eu}} \\
{\scriptsize \slshape \lagrange} & \\
\multicolumn{2}{l}{} \\
\textcolor{hbl}{Matthieu BERTHOMIER} & {\footnotesize \href{mailto:matthieu.berthomier@lpp.polytechnique.fr}{matthieu.berthomier@lpp.polytechnique.fr}} \\
{\scriptsize \slshape LPP, \'Ecole Polytechnique, Sorbonne Universit\'es, Observatoire de Paris, Paris-Scalay} & \\
\multicolumn{2}{l}{} \\
\textcolor{hbl}{Florian BOLGAR} & {\footnotesize \href{mailto:florian.bolgar@obspm.fr}{florian.bolgar@obspm.fr}} \\
{\scriptsize \slshape \lermasorb} & \\
\multicolumn{2}{l}{} \\
\textcolor{hbl}{Sylvain BONTEMPS} & {\footnotesize \href{mailto:sylvain.bontemps@u-bordeaux.fr}{sylvain.bontemps@u-bordeaux.fr}} \\
{\scriptsize \slshape \lab} & \\
\multicolumn{2}{l}{} \\
\textcolor{hbl}{Nicolas BOUCHE} & {\footnotesize \href{mailto:nicolas.bouche@irap.omp.eu}{nicolas.bouche@irap.omp.eu}} \\
{\scriptsize \slshape \irap} & \\
\multicolumn{2}{l}{} \\
\textcolor{hbl}{Fran\c{c}ois BOUCHET} & {\footnotesize \href{mailto:bouchet@iap.fr}{bouchet@iap.fr}} \\
{\scriptsize \slshape \iapsorb} & \\
\multicolumn{2}{l}{} \\
\textcolor{hbl}{G\'eraldine BOURDA} & {\footnotesize \href{mailto:geraldine.bourda@u-bordeaux.fr}{bouchet@iap.fr}} \\
{\scriptsize \slshape \lab} & \\
\multicolumn{2}{l}{} \\
\textcolor{hbl}{Jean-Claude BOURET} & {\footnotesize \href{mailto:jean-claude.bouret@lam.fr}{jean-claude.bouret@lam.fr}} \\
{\scriptsize \slshape \lam} & \\
\multicolumn{2}{l}{} \\
\textcolor{hbl}{Jonathan BRAINE} & {\footnotesize \href{mailto:jonathan.braine@u-bordeaux.fr}{jonathan.braine@u-bordeaux.fr}} \\
{\scriptsize \slshape \lab} & \\
\multicolumn{2}{l}{} \\
\textcolor{hbl}{Denis BURGARELLA} & {\footnotesize \href{mailto:denis.burgarella@lam.fr}{denis.burgarella@lam.fr}} \\
{\scriptsize \slshape \lam} & \\
\multicolumn{2}{l}{} \\
\textcolor{hbl}{R\'emi CABANAC} & {\footnotesize \href{mailto:rcabanac@irap.omp.eu}{rcabanac@irap.omp.eu}} \\
{\scriptsize \slshape \irap} & \\
\multicolumn{2}{l}{} \\
\textcolor{hbl}{Sylvie CABRIT} & {\footnotesize \href{mailto:sylvie.cabrit@obspm.fr}{sylvie.cabrit@obspm.fr}} \\
{\scriptsize \slshape \lermasorb} & \\
\multicolumn{2}{l}{} \\
\textcolor{hbl}{Benoit CARRY} & {\footnotesize \href{mailto:benoit.carry@oca.eu}{benoit.carry@oca.eu}} \\
{\scriptsize \slshape \lagrange} & \\
\multicolumn{2}{l}{} \\
\textcolor{hbl}{Baptiste CECCONI} & {\footnotesize \href{mailto:baptiste.cecconi@obspm.fr}{baptiste.cecconi@obspm.fr}} \\
{\scriptsize \slshape \lesia} & \\
\multicolumn{2}{l}{} \\
\textcolor{hbl}{Benjamin CENSIER} & {\footnotesize \href{mailto:benjamin.censier@obs-nancay.fr}{benjamin.censier@obs-nancay.fr}} \\
{\scriptsize \slshape \usn} & \\
\multicolumn{2}{l}{} \\
\textcolor{hbl}{St\'ephane CHARLOT} & {\footnotesize \href{mailto:charlot@iap.fr}{charlot@iap.fr}} \\
{\scriptsize \slshape \iapsorb} & \\
\multicolumn{2}{l}{} \\
\textcolor{hbl}{Bertrand CHAUVINEAU} & {\footnotesize \href{mailto:chauvineau@oca.eu}{chauvineau@oca.eu}} \\
{\scriptsize \slshape \lagrange} & \\
\multicolumn{2}{l}{} \\
\textcolor{hbl}{Thierry CONTINI} & {\footnotesize \href{mailto:thierry.contini@irap.omp.eu}{thierry.contini@irap.omp.eu}} \\
{\scriptsize \slshape \irap} & \\
\multicolumn{2}{l}{} \\
\textcolor{hbl}{Vincent COUD\'E DU FORESTO} & {\footnotesize \href{mailto:vincent.foresto@obspm.fr}{vincent.foresto@obspm.fr}} \\
{\scriptsize \slshape \lesia} & \\
\multicolumn{2}{l}{} \\
\textcolor{hbl}{Audrey COUTENS} & {\footnotesize \href{mailto:audrey.coutens@u-bordeaux.fr}{audrey.coutens@u-bordeaux.fr}} \\
{\scriptsize \slshape \lab} & \\
\multicolumn{2}{l}{} \\
\textcolor{hbl}{Karol DESNOS} & {\footnotesize \href{mailto:kdesnos@insa-rennes.fr}{kdesnos@insa-rennes.fr}} \\
{\scriptsize \slshape IETR, INSA Rennes, UMR CNRS 6161, UBL, France} & \\
\multicolumn{2}{l}{} \\
\textcolor{hbl}{Pierre-Alain DUC} & {\footnotesize \href{mailto:pierre-alain.duc@astro.unistra.fr}{pierre-alain.duc@astro.unistra.fr}} \\
{\scriptsize \slshape \stras} & \\
\multicolumn{2}{l}{} \\
\textcolor{hbl}{Laurent DUGOUJON} & {\footnotesize \href{mailto:laurent.dugoujon@st.com}{laurent.dugoujon@st.com}} \\
{\scriptsize \slshape STMicroelectronics} & \\
\multicolumn{2}{l}{} \\
\textcolor{hbl}{Jean-Baptiste DURRIVE} & {\footnotesize \href{mailto:jean.baptiste.durrive@e.mbox.nagoya-u.ac.jp}{jean.baptiste.durrive@e.mbox.nagoya-u.ac.jp}} \\
{\scriptsize \slshape Nagoya University, Japon} & \\
\multicolumn{2}{l}{} \\
\textcolor{hbl}{Evan EAMES} & {\footnotesize \href{mailto:evan.eames@obspm.fr}{evan.eames@obspm.fr}} \\
{\scriptsize \slshape \lermasorb} & \\
\multicolumn{2}{l}{} \\
\textcolor{hbl}{Beno\^it EPINAT} & {\footnotesize \href{mailto:benoit.epinat@lam.fr}{benoit.epinat@lam.fr}} \\
{\scriptsize \slshape \lam} & \\
\multicolumn{2}{l}{} \\
\textcolor{hbl}{St\'ephanie ESCOFFIER} & {\footnotesize \href{mailto:escoffier@cppm.in2p3.fr}{escoffier@cppm.in2p3.fr}} \\
{\scriptsize \slshape Centre de Physique des Particules de Marseille; CNRS/IN2P3 \& AMU} & \\
\multicolumn{2}{l}{} \\
\textcolor{hbl}{Gilles ESPOSITO-FAR\`ESE} & {\footnotesize \href{mailto:gef@iap.fr}{gef@iap.fr}} \\
{\scriptsize \slshape \iapsorb} & \\
\multicolumn{2}{l}{} \\
\textcolor{hbl}{Alexandre FAURE} & {\footnotesize \href{mailto:alexandre.faure@univ-grenoble-alpes.fr}{alexandre.faure@univ-grenoble-alpes.fr}} \\
{\scriptsize \slshape \univgren} & \\
\multicolumn{2}{l}{} \\
\textcolor{hbl}{Thierry FOGLIZZO} & {\footnotesize \href{mailto:foglizzo@cea.fr}{foglizzo@cea.fr}} \\
{\scriptsize \slshape \irfuSAp} & \\
\multicolumn{2}{l}{} \\
\textcolor{hbl}{Maurice GHEUDIN} & {\footnotesize \href{mailto:maurice.gheudin@obspm.fr}{maurice.gheudin@obspm.fr}} \\
{\scriptsize \slshape \lermasorb} & \\
\multicolumn{2}{l}{} \\
\textcolor{hbl}{Olivier GODET} & {\footnotesize \href{mailto:ogodet@irap.omp.eu}{ogodet@irap.omp.eu}} \\
{\scriptsize \slshape \irap} & \\
\multicolumn{2}{l}{} \\
\textcolor{hbl}{Paolo GOLDONI} & {\footnotesize \href{mailto:goldoni@apc.in2p3.fr}{goldoni@apc.in2p3.fr}} \\
{\scriptsize \slshape \apc} & \\
\multicolumn{2}{l}{} \\
\textcolor{hbl}{Christian GOUIFF\`ES} & {\footnotesize \href{mailto:christian.gouiffes@cea.fr}{christian.gouiffes@cea.fr}} \\
{\scriptsize \slshape CEA, DRF/IRFU/SAP/LEPCHE, Saclay, France} & \\
\multicolumn{2}{l}{} \\
\textcolor{hbl}{Oliver HAHN} & {\footnotesize \href{mailto:oliver.hahn@oca.eu}{oliver.hahn@oca.eu}} \\
{\scriptsize \slshape \lagrange} & \\
\multicolumn{2}{l}{} \\
\textcolor{hbl}{Bernard JARRY} & {\footnotesize \href{mailto:bernard.jarry@xlim.fr}{bernard.jarry@xlim.fr}} \\
{\scriptsize \slshape CNRS UMR\,7252, Universit\'e de Limoges, France} & \\
\multicolumn{2}{l}{} \\
\textcolor{hbl}{Thierry LANZ} & {\footnotesize \href{mailto:thierry.lanz@oca.eu}{thierry.lanz@oca.eu}} \\
{\scriptsize \slshape \lagrange} & \\
\multicolumn{2}{l}{} \\
\textcolor{hbl}{Damien LE BORGNE} & {\footnotesize \href{mailto:leborgne@iap.fr}{leborgne@iap.fr}} \\
{\scriptsize \slshape \iapsorb} & \\
\multicolumn{2}{l}{} \\
\textcolor{hbl}{Amandine LE BRUN} & {\footnotesize \href{mailto:amandine.le-brun@cea.fr}{amandine.le-brun@cea.fr}} \\
{\scriptsize \slshape \aim} & \\
\multicolumn{2}{l}{} \\
\textcolor{hbl}{Elena LEGA} & {\footnotesize \href{mailto:elena.lega@oca.eu}{elena.lega@oca.eu}} \\
{\scriptsize \slshape \lagrange} & \\
\multicolumn{2}{l}{} \\
\textcolor{hbl}{James LEQUEUX} & {\footnotesize \href{mailto:james.lequeux@obspm.fr}{james.lequeux@obspm.fr}} \\
{\scriptsize \slshape \lermasorb} & \\
\multicolumn{2}{l}{} \\
\textcolor{hbl}{Alan LOH} & {\footnotesize \href{mailto:alan.loh@obspm.fr}{alan.loh@obspm.fr}} \\
{\scriptsize \slshape \lesia} & \\
\multicolumn{2}{l}{} \\
\textcolor{hbl}{Benoit LOTT} & {\footnotesize \href{mailto:lott@cenbg.in2p3.fr}{lott@cenbg.in2p3.fr}} \\
{\scriptsize \slshape Centre Etudes NuclŽaires de Bordeaux Gradignan, CNRS, IN2P3, Bordeaux, France} & \\
\multicolumn{2}{l}{} \\
\textcolor{hbl}{Sofian MAABOUT} & {\footnotesize \href{mailto:sofian.maabout@u-bordeaux.fr}{sofian.maabout@u-bordeaux.fr}} \\
{\scriptsize \slshape Laboratoire Bordelais de Recherche en Informatique (LaBRI), Universit\'e de Bordeaux, France} & \\
\multicolumn{2}{l}{} \\
\textcolor{hbl}{Suzanne MADDEN} & {\footnotesize \href{mailto:suzanne.madden@cea.fr}{suzanne.madden@cea.fr}} \\
{\scriptsize \slshape CEA, DRF/IRFU/SAP/LFEMI, Saclay, France} & \\
\multicolumn{2}{l}{} \\
\textcolor{hbl}{Anna MANGILLI} & {\footnotesize \href{mailto:anna.mangilli@irap.omp.eu}{anna.mangilli@irap.omp.eu}} \\
{\scriptsize \slshape \irap} & \\
\multicolumn{2}{l}{} \\
\textcolor{hbl}{J\'er\^ome MARGUERON} & {\footnotesize \href{mailto:j.margueron@ipnl.in2p3.fr}{j.margueron@ipnl.in2p3.fr}} \\
{\scriptsize \slshape Institut de Physique Nucl\'eaire de Lyon, France} & \\
\multicolumn{2}{l}{} \\
\textcolor{hbl}{Simona MEI} & {\footnotesize \href{mailto:simona.mei@obspm.fr}{simona.mei@obspm.fr}} \\
{\scriptsize \slshape \lermasorb} & \\
\multicolumn{2}{l}{} \\
\textcolor{hbl}{Anne-Laure MELCHIOR} & {\footnotesize \href{mailto:anne-laure.melchior@obspm.fr}{anne-laure.melchior@obspm.fr}} \\
{\scriptsize \slshape \lermasorb} & \\
\multicolumn{2}{l}{} \\
\textcolor{hbl}{Jean-Baptiste MELIN} & {\footnotesize \href{mailto:jean-baptiste.melin@cea.fr}{jean-baptiste.melin@cea.fr}} \\
{\scriptsize \slshape CEA, DRF/IRFU/SPP, Saclay, France} & \\
\multicolumn{2}{l}{} \\
\textcolor{hbl}{Daniel MENARD} & {\footnotesize \href{mailto:daniel.menard@insa-rennes.fr}{daniel.menard@insa-rennes.fr}} \\
{\scriptsize \slshape INSA, Rennes, France} & \\
\multicolumn{2}{l}{} \\
\textcolor{hbl}{Patrick MICHEL} & {\footnotesize \href{mailto:michelp@oca.eu}{michelp@oca.eu}} \\
{\scriptsize \slshape \lagrange} & \\
\multicolumn{2}{l}{} \\
\textcolor{hbl}{H\'elo\"ise M\'EHEUT} & {\footnotesize \href{mailto:heloise.meheut@oca.eu}{heloise.meheut@oca.eu}} \\
{\scriptsize \slshape \lagrange} & \\
\multicolumn{2}{l}{} \\
\textcolor{hbl}{Thierry NOEL} & {\footnotesize \href{mailto:thierry.noel@st.com}{thierry.noel@st.com}} \\
{\scriptsize \slshape STMicroelectronics} & \\
\multicolumn{2}{l}{} \\
\textcolor{hbl}{D\'eborah PARADIS} & {\footnotesize \href{mailto:deborah.paradis@irap.omp.eu}{deborah.paradis@irap.omp.eu}} \\
{\scriptsize \slshape \irap} & \\
\multicolumn{2}{l}{} \\
\textcolor{hbl}{Maxime PELCAT} & {\footnotesize \href{mailto:mpelcat@insa-rennes.fr}{mpelcat@insa-rennes.fr}} \\
{\scriptsize \slshape \insarennes} & \\
\multicolumn{2}{l}{} \\
\textcolor{hbl}{Valeria PETTORINO} & {\footnotesize \href{mailto:valeria.pettorino@cea.fr}{valeria.pettorino@cea.fr}} \\
{\scriptsize \slshape \aim} & \\
\multicolumn{2}{l}{} \\
\textcolor{hbl}{Sandrine PIRES} & {\footnotesize \href{mailto:sandrine.pires@cea.fr}{sandrine.pires@cea.fr}} \\
{\scriptsize \slshape \aim} & \\
\multicolumn{2}{l}{} \\
\textcolor{hbl}{Simon PRUNET} & {\footnotesize \href{mailto:prunet@cfht.hawaii.edu}{prunet@cfht.hawaii.edu}} \\
{\scriptsize \slshape CFHT} & \\
\multicolumn{2}{l}{} \\
\textcolor{hbl}{Delphine RUSSEIL} & {\footnotesize \href{mailto:delphine.russeil@lam.fr}{delphine.russeil@lam.fr}} \\
{\scriptsize \slshape \lam} & \\
\multicolumn{2}{l}{} \\
\textcolor{hbl}{Quentin SALOM\'E} & {\footnotesize \href{mailto:quentin.salome@ias.u-psud.fr}{quentin.salome@ias.u-psud.fr}} \\
{\scriptsize \slshape \ias} & \\
\multicolumn{2}{l}{} \\
\textcolor{hbl}{Philippe SALOM\'E} & {\footnotesize \href{mailto:philippe.salome@obspm.fr}{philippe.salome@obspm.fr}} \\
{\scriptsize \slshape \lermasorb} & \\
\multicolumn{2}{l}{} \\
\textcolor{hbl}{Youssouf SAMEUT BOUHAIK} & {\footnotesize \href{mailto:youssouf.sameut\_bouhaik@hotmail.com}{youssouf.sameut\_bouhaik@hotmail.com}} \\
{\scriptsize \slshape \lss} & \\
\multicolumn{2}{l}{} \\
\textcolor{hbl}{Norma SANCHEZ} & {\footnotesize \href{mailto:norma.sanchez@obspm.fr}{norma.sanchez@obspm.fr}} \\
{\scriptsize \slshape \lermasorb} & \\
\multicolumn{2}{l}{} \\
\textcolor{hbl}{Aurore SAVOY-NAVARRO} & {\footnotesize \href{mailto:aurore.savoy-navarro@apc.univ-paris7.fr}{aurore.savoy-navarro@apc.univ-paris7.fr}} \\
{\scriptsize \slshape \apc} & \\
\multicolumn{2}{l}{} \\
\textcolor{hbl}{Daniel SCHAERER} & {\footnotesize \href{mailto:daniel.schaerer@unige.ch}{daniel.schaerer@unige.ch}} \\
{\scriptsize \slshape CNRS \& Observatoire de Gen\`eve} & \\
\multicolumn{2}{l}{} \\
\textcolor{hbl}{Hayato SHIMABUKURO} & {\footnotesize \href{mailto:hayato.shimabukuro@obspm.fr}{hayato.shimabukuro@obspm.fr}} \\
{\scriptsize \slshape \lermasorb} & \\
\multicolumn{2}{l}{} \\
\textcolor{hbl}{Eric SLEZAK} & {\footnotesize \href{mailto:eric.slezak@oca.eu}{eric.slezak@oca.eu}} \\
{\scriptsize \slshape \lagrange} & \\
\multicolumn{2}{l}{} \\
\textcolor{hbl}{David SMITH} & {\footnotesize \href{mailto:smith@cenbg.in2p3.fr}{smith@cenbg.in2p3.fr}} \\
{\scriptsize \slshape CENBG/IN2P3/CNRS, France} & \\
\multicolumn{2}{l}{} \\
\textcolor{hbl}{Genevieve SOUCAIL} & {\footnotesize \href{mailto:gsoucail@irap.omp.eu}{gsoucail@irap.omp.eu}} \\
{\scriptsize \slshape \irap} & \\
\multicolumn{2}{l}{} \\
\textcolor{hbl}{Philippe STEE} & {\footnotesize \href{mailto:philippe.stee@oca.eu
}{philippe.stee@oca.eu}} \\
{\scriptsize \slshape \lagrange} & \\
\multicolumn{2}{l}{} \\
\textcolor{hbl}{Antoine STRUGAREK} & {\footnotesize \href{mailto:antoine.strug@cea.fr}{antoine.strug@cea.fr}} \\
{\scriptsize \slshape CEA} & \\
\multicolumn{2}{l}{} \\
\textcolor{hbl}{Christian SOURACE} & {\footnotesize \href{mailto:christian.surace@lam.fr}{christian.surace@lam.fr}} \\
{\scriptsize \slshape \lam} & \\
\multicolumn{2}{l}{} \\
\textcolor{hbl}{Christophe TAFFOUREAU} & {\footnotesize \href{mailto:ctaffoureau@obs-nancay.fr}{ctaffoureau@obs-nancay.fr}} \\
{\scriptsize \slshape \usn} & \\
\multicolumn{2}{l}{} \\
\textcolor{hbl}{Lidia TASCA} & {\footnotesize \href{mailto:lidia.tasca@lam.fr}{lidia.tasca@lam.fr}} \\
{\scriptsize \slshape \lam} & \\
\multicolumn{2}{l}{} \\
\textcolor{hbl}{Cedric VIOU} & {\footnotesize \href{mailto:Cedric.Dumez-Viou@obs-nancay.fr}{Cedric.Dumez-Viou@obs-nancay.fr}} \\
{\scriptsize \slshape \usn} & \\
\multicolumn{2}{l}{} \\
\textcolor{hbl}{Valentine WAKELAM} & {\footnotesize \href{mailto:valentine.wakelam@u-bordeaux.fr}{valentine.wakelam@u-bordeaux.fr}} \\
{\scriptsize \slshape \lab} & \\
\multicolumn{2}{l}{} \\
\end{ltabulary}

\end{document}